\documentclass{article}
\usepackage[T1]{fontenc}
\usepackage{amsmath,amssymb,amsthm}
\usepackage{booktabs,array}
\usepackage{graphicx}
\usepackage{xcolor}
\usepackage[hypertexnames=false,colorlinks=true,linkcolor=blue!50!black,citecolor=blue!50!black,urlcolor=blue!50!black]{hyperref}

\providecommand{\ICMLOPTIONS}{\usepackage[preprint]{icml2027}}
\ICMLOPTIONS
\hypersetup{pdfauthor={Kundan Baliga, Ryan Kim},pdfsubject={Research preprint}}
\usepackage[capitalise,noabbrev,nosort]{cleveref}
\theoremstyle{plain}
\newtheorem{theorem}{Theorem}
\newtheorem{proposition}[theorem]{Proposition}
\newtheorem{lemma}[theorem]{Lemma}
\newtheorem{corollary}[theorem]{Corollary}
\newtheorem{conjecture}[theorem]{Conjecture}
\theoremstyle{definition}
\newtheorem{remark}[theorem]{Remark}
\newtheorem{example}[theorem]{Example}
\newtheorem{definition}[theorem]{Definition}
\newcommand{\jmlrQED}{\hfill$\square$}

\newif\ificmlextended
\newif\ifoldpullback
\oldpullbacktrue
\newif\ificml % venue switch: false for focs-anon.tex and preprint.tex; icml.tex says \icmltrue right after this file
\newcommand{\vsw}[2]{\ificml#1\else#2\fi}
\newenvironment{icmltight}{\ificml\def\qquad{\hskip0.6em\relax}\thinmuskip=2mu\thickmuskip=2.5mu plus 1mu\medmuskip=2mu plus 1mu minus 1mu\fi}{}
\newenvironment{mbfigure}{\ificml\begin{figure*}[t]\else\begin{figure}[t]\fi}{\ificml\end{figure*}\else\end{figure}\fi}
\usepackage{tikz}
\usetikzlibrary{positioning}
\newcommand{\Acon}{12.78}
\newcommand{\BtwoNmax}{8}
\newcommand{\CAexact}{3981.99}
\newcommand{\CJ}{633.23}

\newcommand{\NtwoChar}{7/8}
\newcommand{\NtwoDisc}{7/8}
\newcommand{\NtwoS}{\{1\}}
\newcommand{\NzeroLHS}{320}
\newcommand{\NzeroRHS}{318.31}
\newcommand{\Sigcon}{4.51}
\newcommand{\TAone}{48}
\newcommand{\TRa}{28.3}
\newcommand{\TRb}{14.8}

\newcommand{\clSeq}{2.93, 2.93, 3.27, 3.46, 3.58, 3.63, 3.65, 3.62, 3.58, 3.51, 3.43}
\newcommand{\clTmax}{11}

\newcommand{\expMel}{0.36}
\newcommand{\expMelRange}{5\le T\le 11}
\newcommand{\expSpec}{0.36}
\newcommand{\expSpecRange}{4\le T\le 13}
\newcommand{\famTmax}{13}
\newcommand{\gadgetExp}{0.29}
\newcommand{\melTmaxAll}{8}

\newcommand{\necI}{0.22859}
\newcommand{\necLarge}{0.19}
\newcommand{\necSigMax}{0.252}
\newcommand{\necSigMin}{0.211}
\newcommand{\necTmax}{14}
\newcommand{\necTmin}{6}
\newcommand{\necTthr}{11}
\newcommand{\necWTfirst}{6}
\newcommand{\necWTlast}{12}
\newcommand{\necWfirst}{0.133}
\newcommand{\necWlast}{0.028}
\newcommand{\necc}{0.2099}
\newcommand{\necerr}{21.6}
\newcommand{\neckappa}{0.35}
\newcommand{\propBNmax}{4}
\newcommand{\sidonBmax}{3}
\newcommand{\sidonConst}{32}
\newcommand{\sidonNhi}{7.8\cdot10^{7}}
\newcommand{\sidonNlo}{256}

\newcommand{\specSpreadPct}{0.2}

\newcommand{\throne}{179.3}
\newcommand{\thrtwo}{63.81}


\crefname{conjecture}{Conjecture}{Conjectures}
\crefname{example}{Example}{Examples}
\crefname{definition}{Definition}{Definitions}
\crefname{remark}{Remark}{Remarks}
\crefname{lemma}{Lemma}{Lemmas}
\crefname{proposition}{Proposition}{Propositions}
\crefname{corollary}{Corollary}{Corollaries}
\crefname{theorem}{Theorem}{Theorems}
\crefname{appendix}{Appendix}{Appendices}
\newcommand{\E}{\mathbb{E}}
\newcommand{\R}{\mathbb{R}}
\newcommand{\C}{\mathbb{C}}
\newcommand{\Wt}{\widetilde W}
\newcommand{\disc}{\operatorname{disc}}
\newcommand{\sgn}{\operatorname{sgn}}

\icmltrue
\ifdefined\ICMLEXTENDED\icmlextendedtrue\else\oldpullbackfalse\fi
\let\cite\citep
\usepackage{etoolbox}
\robustify{\cref}
\robustify{\Cref}

\icmltitlerunning{The Scratchpad Complexity of Autoregressive Multiplication}

\begin{document}
\twocolumn[
  \icmltitle{The Scratchpad Complexity of \texorpdfstring{Autoregressive\\Multiplication}{Autoregressive Multiplication} with Hard Attention}
  \begin{icmlauthorlist}
    \icmlauthor{Kundan Baliga}{uiuc}
    \icmlauthor{Ryan Kim}{uiuc}
  \end{icmlauthorlist}
  \icmlaffiliation{uiuc}{University of Illinois Urbana-Champaign, Urbana, IL, USA}
  \icmlcorrespondingauthor{Kundan Baliga}{kbaliga2@illinois.edu}
  \icmlkeywords{transformers, chain of thought, multiplication, hard attention, expressivity, lower bounds,
    pseudorandomness, finite-state models, branching programs}
  \vskip 0.3in
]
\printAffiliationsAndNotice{}

\newcommand{\icmlcotpartialstatement}{%
Use the operands and architecture of \cref{thm:cot-avg}. Let $n=5N$, $b=\lfloor\eta n/4\rfloor$,
$s=\lfloor\sigma n\rfloor$ for positive constants $\eta,\sigma$ from the proof, and
$I=\{N-b+1,\ldots,N\}$.
For each $0\le t\le s$, let $A_t$ be the first $t$ symbols actually generated
by the model for total length $|x|+t+2N$. At digit $i\in I$, supply the
context $xA_tM_1\cdots M_{i-1}$. Decode its readout as $1$ for token $1$
and $0$ for every other token. Let $C_{t,i}$ be $+1$ if the decoded bit
equals $M_i$ and $-1$ otherwise. Write $K_t$ for the number of positions
in $I$ whose original readout token is exactly the required bit.

\begin{corollary}[Teacher-forced agreement after a short trace]
\label[corollary]{cor:icml-cot-partial}
Let $U_b$ be uniform on $\{-1,+1\}^b$, independent of $A_t$.
For all sufficiently large $N$ and every $0\le t\le s$,
\[
 d_{\rm TV}\bigl(\mathcal L(A_t,C_t),\mathcal L(A_t,U_b)\bigr)
 \le e^{-\eta n/2}.
\]
Consequently, for $0<\gamma\le 1/2$,
\begin{align*}
 &\Pr\bigl[\exists\, t\le s:\ K_t\ge(1/2+\gamma)b\bigr]\\
 &\hspace{1em}\le (s+1)\bigl(e^{-2\gamma^2 b}+e^{-\eta n/2}\bigr).
\end{align*}
For fixed $\gamma>0$ this probability is exponentially small in $N$,
including when the auxiliary length is selected from the operands.
\end{corollary}

}

\newcommand{\icmlcotpartialproof}{%
\subsection{Joint correctness on a teacher-forced block}
\label[appendix]{app:icml-cot-partial}

\icmlcotpartialstatement

\begin{proof}[Proof of \cref{cor:icml-cot-partial}]
Fix $t$ and a realized string $\tau$. Put
$p_\tau=\Pr[A_t=\tau]$ and let $\mu_\tau(v)=\Pr[A_t=\tau,C_t=v]$.
For nonempty $J\subseteq I$, the final correlation estimate in
\cref{app:icml-cot-joint} gives
\[
 \left|\sum_v\mu_\tau(v)\prod_{i\in J}v_i\right|
 =\left|\mathbb E\left[W_\tau\prod_{i\in J}\widehat M_i g_i\right]\right|
 \le 2e^{-\eta n}.
\]
The signed mass $d_\tau(v)=\mu_\tau(v)-2^{-b}p_\tau$ has zero sum
and the same nonempty Walsh moments. Cauchy--Schwarz and Parseval give
\[
 \sum_v |d_\tau(v)|
 \le \Bigl(\sum_{J\subseteq I}
       \bigl|\textstyle\sum_v d_\tau(v)\prod_{i\in J}v_i\bigr|^2\Bigr)^{1/2}
 \le 2^{b/2+1}e^{-\eta n}.
\]
Sum over $\tau$ and divide by two. Since
$|\mathcal A_t|\le e^{\eta n/4}$ by \eqref{eq:icml-linear-traces} and
$b\le\eta n/4$, this gives the claimed total variation distance.
Each exactly correct token has $C_{t,i}=+1$. Under $U_b$ the count of
positive signs is binomial with parameters $b$ and $1/2$.
Apply its Hoeffding upper bound and sum over the $s+1$ candidate lengths.
\end{proof}

The total variation statement concerns a fixed total-length model. For
example, take two models with constant answer tokens $0$ and $1$,
respectively. Selecting the length whose constant equals $M_N$ makes the
selected correctness sign at $N$ always positive. The tail bound accounts
for this choice. It evaluates each position on its true product prefix.
}

\newcommand{\icmlrecordstatement}{%
\begin{corollary}[Information in the readout records]\label[corollary]{cor:icml-records}
Fix the architecture and a binary MSB-first target in the
linear-exponent range of \cref{thm:topk-depth}. Let $R$ be the tuple of top records at up to $cN$ positions
fixed independently of the operands in the true-prefix run, for sufficiently small $c>0$. Write $U$ for an independent
fair bit. Then $d_{\rm TV}((M_T,R),(U,R))$ and $I(M_T;R)$ are exponentially
small in $N$. Any probabilistic readout from $R$ incurs expected log loss
at least $\ln2-e^{-\Omega(N)}$.
\end{corollary}
A small linear block of future digits is jointly close to uniform given records
strictly before the block. Both results allow a short trace and a bounded adaptive
record reader (\cref{app:icml-records}).
}

\newcommand{\icmlrecordproof}{%
\subsection{Posterior bias and probabilistic readouts}\label[appendix]{app:icml-records}
\icmlrecordstatement

Use the grouped pullback \eqref{eq:icml-linear-approximant} with the
applicable discrepancy exponent at the target (\cref{thm:main}). For some sufficiently
small constants $\delta,\eta>0$, any probe $F(R)$ with $|F|\le1$ has
$|\E[\widehat M_TF(R)]|\le e^{-\eta n}/2+e^{-3\delta n/4}$.
Choose $F$ as the sign of the posterior bias
$B(R)=2\Pr[M_T=1\mid R]-1$. This gives $\E|B(R)|\le e^{-\Omega(N)}$.
The joint total variation distance in \cref{cor:icml-records} is
$\E|B(R)|/2$. Concavity of binary entropy gives
$h_2((1+b)/2)\ge1-|b|$ for $-1\le b\le1$, in bits.
Thus $I(M_T;R)\le 1-H_2(M_T\mid R)\le\E|B(R)|$ in bits.
Any probabilistic forecast has conditional log loss at least the
posterior entropy, by nonnegativity of relative entropy. A scalar
Brier loss has conditional minimum $(1-B(R)^2)/4$.
More generally, for a binary proper loss the excess of the fair forecast
over the optimal risk is convex in the posterior probability and vanishes
at $1/2$. If its values at $0$ and $1$ are finite, their maximum times
$|B(R)|$ bounds this excess by convexity.

\paragraph{Auxiliaries and adaptive record access.}
Fix a total-length model with a specified sufficiently short auxiliary
length $t$. Reveal the emitted trace $A_t$ and allow a reader to query
a sufficiently small linear number of record positions at or before
the target readout. Its only input-dependent observations are $A_t$
and the records it queries. Independent randomization is allowed by
conditioning. For each possible trace $\tau$, verify it with the
$t$ top records defining $W_\tau$ in \cref{app:icml-cot-joint}.
For each proposed visited set $S$, simulate the reader from the
records in $S$ on this literal context. Reject if it queries outside
$S$ or finishes with a different visited set. The probe times this
acceptance indicator and $W_\tau$ is a bounded function of at most
$t+|S|$ specified records. The expansion over traces and visited
sets is exact. By \eqref{eq:icml-linear-traces} and the subset entropy
bound, the product of the two counts has arbitrarily small exponential
rate as the two budgets decrease. Summing the correlation bounds
still gives exponentially small posterior bias for the complete
observation, including the trace and the ordered query transcript.

\paragraph{A block of future digits.}
Let $I$ be a consecutive block of $b\le cN$ digits around the midpoint,
with every index in $[N/2,N]$, and take $c>0$ sufficiently small.
Let $R$ consist of the fixed-position records, or the short trace and
adaptive transcript just described, with every record computed on
the context strictly before the first target digit. Then
\[
 d_{\rm TV}((M_I,R),(U_I,R))\le e^{-\Omega(N)},\qquad
 H_2(M_I\mid R)\ge b-e^{-\Omega(N)},
\]
where $U_I$ is an independent uniform bit vector and $H_2$ is entropy
in bits. The same entropy is a lower bound on joint expected log loss
when the logarithm is base two.

For a nonempty $J\subseteq I$ and $u=\max J$, use the probe that is the
sign of the conditional expectation of $\chi_J=\prod_{i\in J}\widehat M_i$
given $R$. All its observations occur before the first target. The proof
above approximates the probe in the formal cube for $u$ and charges
any trace and visited-set counts. Multiply each approximant by
$\prod_{i\in J\setminus\{u\}}\widehat M_i$ without changing its Walsh norm.
Correlation at $u$ bounds $\E|\E[\chi_J\mid R]|$ by $e^{-\kappa N}$ for
a uniform $\kappa>0$. Fourier inversion and the triangle inequality give
expected conditional total variation at most $2^{b-1}e^{-\kappa N}$.
For each value of $R$, couple $M_I$ to a uniform vector with disagreement
probability equal to the conditional total variation $d$. The conditional
entropy deficit is at most $h_2(d)+bd$, by bounding the entropy of the
uniform vector given the product vector and using the disagreement event.
Jensen's inequality keeps the expected deficit exponentially small.

\paragraph{Fixed alphabets.}
Under the fixed-alphabet correlation hypothesis of
\cref{thm:icml-transfer}, use the complex pullback in \cref{lem:position-union} and test
the phase of each nontrivial posterior character. Fourier inversion gives
exponentially small expected total variation from uniform on the $q$ symbols.
For a posterior law $p$, $H(p)\ge-\ln\max_z p(z)$ and
$\max_z p(z)\le1/q+d_{\rm TV}(p,U_q)$, so the expected log loss is
at least $\ln q-e^{-\Omega(N)}$.
}

\newcommand{\icmlrelatedmain}{%
\section{Related work}
\paragraph{Attention and circuit complexity.}
\citet[Thm.~1]{HAF22} simulate generalized unique hard-attention
encoders, which select one key per head, by polynomial-size,
constant-depth Boolean circuits. We use their simulation for that
model; our bounds for arbitrary top-$k$ scores and row updates
follow directly from position-union approximation.
\citet[Lemmas~7 and~10]{LMN93} give low-degree Fourier approximation
for such circuits in $L_2$ under the uniform cube measure. At the
midpoint, the genuine prefix graph has density $2^{-(N-1)}$ in that
cube. Restriction to the graph can amplify the $L_2$ error by
$2^{(N-1)/2}$. The classical tail estimate at sublinear degree gives
insufficient uniform error for this change of measure at depth
at least two. It does yield a short-prefix prediction bound
(\cref{app:icml-circuit-comparison}). Our position-union lemma
controls error pointwise, including on the genuine prefix graph.

\citet[Thm.~3]{RemscrimSipser13} use bounded-independence
pseudorandomness~\citep{Braverman10} to fool $\mathrm{AC}^0$ tests on
the operands and a retained central product block with
$2^{-N^{\Omega(1)}}$ error. Comparing a prediction to a target in
that block also bounds next-bit advantage when only the earlier
central digits are supplied. In our task the leading digits are
also genuine, and marked replay accounts for the generated trace.

\paragraph{Lemma C.6 and scratchpad bounds.}
Conditional on Conjecture~C.5, \citet[arXiv v3, Lemma~C.6]{AHRH25}
assert $\Omega(N)$ average sensitivity for every Boolean predictor
correct on genuine contexts.
Write $G$ for the graph of the genuine midpoint prefix. Let $F$ be
the midpoint sign on $G$ and $+1$ elsewhere. Every cube edge where
$F$ changes touches $G$, which has density $2^{-(N-1)}$. Hence its
uniform-cube average sensitivity is at most
$2(3N-1)2^{-(N-1)}$, contradicting the claimed $\Omega(N)$.
The proof would need a bound on the low-degree Fourier mass of the
Boolean extension under the uniform cube measure. The moments in
Eq.~(29) are taken on $G$ (\cref{app:c6}).
The deduction of the lemma's full-product UHAT consequence
loses its sensitivity premise. Theorem~4.5 has a separate proof
for the single-bit, worst-case task. Our full-product result follows
from position-union approximation and marked replay
(\cref{tab:icml-prior-scratch}).

\citet[Thm.~B.17 and Cor.~B.6]{MKSH26} bound the low-degree operand
Fourier mass of one central product bit and the information a successful
trace must reveal under their model-size conditions. Their bit has index $N-1$ from the least significant
end (starting at zero). Our $M_N$ has index $N$ in that convention,
and our correlation test may use the entire genuine earlier product
prefix. Marked replay then bounds the number of auxiliary \emph{tokens}
for joint generation, even when a token's information capacity grows.
\paragraph{Learning and expressivity.}
\citet{AbbeGlobality24} use globality to analyze efficient weak
learning on a cycle task under specified training assumptions.
Their inductive scratchpads use supervised intermediate steps.
\citet{HahnRofin24} connect input sensitivity to parameter norms,
layer normalization and the local loss landscape. Our bounds
concern what a fixed-resource model can represent after arbitrary
training. They do not give optimization bounds, predict trained
model accuracy, or prove length generalization.
\citet{SHT23} use communication complexity to bound the representation
size for triple detection. Our pullback controls the low-degree mass
of a prediction on genuine prefixes, and marked replay counts
auxiliary tokens even when individual hidden states carry unrestricted
information. \Cref{tab:icml-related} gives a broader comparison.
}

\newcommand{\icmlrelatedtable}{%
\begin{table*}[t]
\caption{Results on multiplication and on transformer scratchpad complexity.
The setting column identifies the resources or tests to which each result applies.}
\label{tab:icml-related}
\centering
\small
\begin{tabular}{@{}>{\raggedright\arraybackslash}p{.23\textwidth}
                  >{\raggedright\arraybackslash}p{.28\textwidth}
                  >{\raggedright\arraybackslash}p{.44\textwidth}@{}}
\toprule
Work & Setting & Result \\
\midrule
\citet{AHRH25}
& Unique hard attention, fixed depth and heads, growing scratch vocabulary allowed
& Worst-case $\Omega(N)$ scratchpad lower bound for one middle bit
with operands as input. An $O(N\log N)$ scratchpad with growing vocabulary writes the full product.
Conjecture C.5 concerns prediction with the true prefix. \\
\citet{RemscrimSipser13}
& $\mathrm{AC}^0$ distinguishers on operands and retained central digits
& Stretched-exponential pseudorandomness for ordinary and carryless
multiplication after discarding $N^\alpha$ extreme digits in total. \\
\citet{Tomlinson26}
& Bounded attention prefix oracle with bounded information flow
& $\Omega(n)$ reasoning tokens for binary majority, triplet matching
and graph reachability. \\
\citet{MerrillSabharwalCoT24}
& Log-precision transformer decoders with chain of thought
& Language-class bounds by reasoning length, including a polynomial-time
characterization with polynomially many steps and generalized pre-normalization. \\
\citet{MKSH26}
& Low-degree operand tests and information in chains of thought
& Fourier-mass bound for one central product bit and a
model-size-dependent lower bound on information revealed by the trace. \\
This work
& Separate bounded operand tests and full-prefix parities.
Fixed-resource causal top-$k$ for generation, growing scratch vocabulary allowed
& Proves the conjectured true-prefix bound. Average-case lower
bounds for full-product generation, including a linear deletion budget.
$\Theta(N)$ auxiliary tokens with $O(N)$ vocabulary entries. \\
\bottomrule
\end{tabular}
\end{table*}

}

\newcommand{\icmlrelatedappendix}{%
\section{Further related work}\label[appendix]{app:icml-related}
\icmlrelatedtable

\subsection{Circuit simulation with a limited prefix}
\label[appendix]{app:icml-circuit-comparison}
Fix a model that predicts the binary midpoint bit $M_N$ from both
operands and $m$ prescribed earlier product bits. Suppose its signed
readout $F$ on the formal cube has a polynomial-size circuit of
depth $d$ in $\mathrm{AC}^0$. The simulation of \citet[Thm.~1]{HAF22}
applies to a fixed-depth, fixed-head unique hard-attention readout.
The proof enumerates activation records at each length. Fixed
separators and a Boolean next-bit readout can be built into the
circuit at that length.

Let $P$ be a low-degree Fourier truncation and let $U$ and $G$ denote
the uniform formal and genuine-context laws. The latter is uniform
on a fraction $2^{-m}$ of the cube. Thus
\[
  \|F-P\|_{L_2(G)} \le 2^{m/2}\|F-P\|_{L_2(U)}.
\]
The factor is attained for errors supported on the graph, for example
an indicator requiring $X=0$ and all $m$ supplied bits to be zero.
Fix $0\le\alpha<\beta<1/d$ and take $m\le N^\alpha$. By the
Fourier tail estimate of \citet[Lemma~7]{LMN93}, a cutoff
$O(N^{d\beta})=o(N)$ gives uniform-cube $L_2$ error at most
$e^{-N^\beta}$ with a sufficiently large cutoff constant.
Parseval and Cauchy--Schwarz give $A(P)\le e^{o(N)}$, since the
number of retained coefficients is $\exp(o(N))$. Every formal
character times the midpoint sign is controlled by
\cref{prop:finite-fourier}. Applying this to $P$ and charging
the $L_2(G)$ error gives
\[
 |\E_G[\widehat M_N F]|
 \le 2^{-N/2560+o(N)}+2^{m/2}e^{-N^\beta}
 = e^{-\Omega(N^\beta)}.
\]
This comparison uses a single prediction. The full-prefix bound
and the generation result use uniform approximation of joint
readouts and the vocabulary-independent count of replay histories.

\paragraph{Attention models.}
Hahn~\cite{Hahn20} studies sensitivity and hard-attention restrictions. Hao, Angluin
and Frank~\cite{HAF22} give a circuit simulation for unique hard attention; Barcel\'o et al.\ and
Yang et al.~\cite{BKLP24,YCA24,YSCA24} relate masked hard attention to logical characterizations
under specific positional assumptions. Strobl et al.~\cite{StroblSurvey24} survey the different
models. Merrill and Sabharwal~\cite{MerrillSabharwal23} give a threshold-circuit upper bound at
logarithmic precision. Sanford et al.~\cite{SHT23,SHT24} study communication and parallel computation
in transformers. \citet{PNP24} use communication complexity to study
single-layer limitations for iterated function composition, with
bounds depending on head count, dimension, and precision.
\citet{CPW24} give a depth--dimension lower bound for sequential
composition. Our bound concerns prediction advantage for fixed-depth top-$k$ routing with the
true product prefix and unrestricted states. Buss~\cite{Buss92} establishes worst-case hardness
for checking multiplication by showing that its graph is $\mathrm{AC}^0$-equivalent to majority. Kozachinskiy et al.~\cite{KSW26} use sensitivity for one-layer PARITY
lower bounds, and Sato et al.~\cite{SKK26} study generation order empirically for multiplication.

\paragraph{Multiplication in branching programs.}
Ordered binary decision diagrams require exponentially many nodes
to compute the lower middle bit exactly under every variable order
\citep{Bryant91,Woelfel05}. \citet{Ponzio98} obtain a lower bound for
read-once branching programs, which may choose their next input
variable adaptively. \citet{SW03} give lower bounds for computing
the lower middle bit at small error and an inverse-polynomial-error
read-once approximation with polynomial size under uniform inputs.
These results count nodes in operand-query programs. Here an adaptive
attention head scores the entire visible context, including the
genuine prefix in the prediction setting, and the generation
bound counts emitted auxiliary tokens.

\paragraph{Digit estimates and scratchpad simulation.}
\citet{MauduitRivat09,MauduitRivat10} develop digital truncation methods.
\citet{Bourgain11} applies these ideas to weighted bilinear Walsh sums.
The elementary first-half proof uses his two-digit grouping
(\cref{app:finite-fourier}).
\citet{FengCoT23} construct traces for fixed-field tasks.
\citet{LiSvete2026} simulate Word RAMs with polylogarithmic overhead.
Our packed transform has input-independent addresses, which permit one
head to execute each bounded-arity word operation in a constant number
of tokens. \citet{Tomlinson26} derive token lower bounds from access
to prefix information. Our averaging is over uniform operands and
covers free runs with errors in the predicted product.

\paragraph{Serial computation and arithmetic training.}
\citet{LiLiuZhouMa24} use chain of thought to simulate serial
computation on dynamic-programming tasks. Our upper bounds use
fixed arithmetic schedules, and our lower bound counts the tokens
needed to produce a product on uniform inputs. \citet{Malach24}
studies the length complexity of learned traces and gives efficient
Turing-machine simulations using trained linear next-token predictors.
\citet{EGKZ22} give bounded-norm attention constructions and
sample-complexity bounds for functions of sparse subsets of tokens.
\citet{LeeArithmetic24}
study how intermediate targets affect learned arithmetic.
\citet{ChoArithmetic25} combine task-specific scratchpads with
position coupling to improve length generalization for multiplication
and multi-operand addition. Their scratchpads focus each prediction on
few relevant tokens. Our construction prescribes such selections and
gives explicit token and vocabulary bounds for exact multiplication.
}

% Focused ICML main text.
\begin{abstract}
How many auxiliary tokens does a causal hard-attention transformer
need to multiply two $N$-bit integers? First, we resolve the
true-prefix parity conjecture of Amiri Bavandpour et al. on
independent uniform $N$-bit operands. The correlation bound permits
any bounded test of each operand and every earlier product-bit parity.
Second, we give a task-independent transfer from low-degree
correlation bounds to causal hard-attention lower bounds.
Position-union approximation verifies which keys win, and marked
replay counts possible scratch traces using their positions.
Third, at fixed depth, head count, and number of selected keys,
the probability of generating the product with a sufficiently small
linear scratch and edit budget is exponentially small even with a
growing vocabulary. A one-head, one-layer packed FFT attains linear
scratch with linear vocabulary and constant hidden dimension.
\end{abstract}

\section{Introduction}\label{sec:intro}
A decoder that knows both operands can compute any product digit exactly.
What changes if each attention head can return only a bounded number of
records, even though it can score the entire prefix? In top-$k$
attention, each head returns only the $k$ highest-scoring records.
We study two tasks.
In \emph{true-prefix prediction} the decoder receives the operands and all
correct earlier digits. In \emph{free-running generation} it reads its
own outputs and can use intermediate tokens as a scratchpad.

Even a restricted view of the inputs can contain substantial information.
Let $X,Y$ be independent uniform integers in $\{0,\ldots,2^N-1\}$,
$N\ge2$, and let $M_1,\ldots,M_{2N}$ be their zero-padded product
digits, most significant first. Given the first $N-1$ product bits
$M_{<N}$, all of $Y$, and $X\bmod 8$, an unrestricted decoder predicts
$M_N$ with probability at least $7/8$. It divides the half-open
interval for $XY$, of width $2^{N+1}$, by $Y$ when $Y\ge2^{N-2}$.
This restricts $X$ to an interval of width at most
eight, containing only one integer of the given residue. This event
has probability $3/4$. A fair guess suffices otherwise. A
fixed-depth top-$k$ transformer with fixed head count and $k$ has
exponentially vanishing advantage even with both complete operands and the
genuine prefix.

\paragraph{Arithmetic pseudorandomness.}
\citet[Conjecture~C.5]{AHRH25} asked whether any parity of the operands
and the genuine earlier digits has appreciable correlation with the
next bit. We resolve the conjecture and allow arbitrary bounded tests
of each operand separately in place of its parity (\cref{thm:main}).
One exponential rate holds uniformly over all prefix masks. An
additive Fourier estimate and an integer-grid counting argument give
the proof (\cref{sec:setting,app:finite-fourier}).

\paragraph{Transfer for bounded attention.}
The transfer applies to any task whose output symbols have small
low-degree correlations with the input and genuine earlier symbols
(\cref{thm:icml-transfer}). Approximation controls the input dependence
of the selected positions. Replay reconstructs generated scratch
tokens from a short description of which positions the network used.
For multiplication, this gives exponentially small next-digit
advantage throughout the interior, in either digit order and every
fixed radix (\cref{thm:topk-depth}).

\paragraph{Maximal scratchpad complexity.}
Generation requires the model to write the entire product on a free
run. We show that this needs linear scratch on independent uniform
operands, even allowing some incorrect or missing digits. A model with
unrestricted local computation and vocabulary can always store one
longer input prefix per token. This gives a universal linear scratch
upper bound for deterministic tasks on a fixed input layout and
prescribed output length (\cref{app:icml-cot-circuits}). Multiplication
requires this maximal order. The packed FFT achieves it with $O(N)$
token types and constant hidden dimension, while the general
accumulator can use exponentially many types. \Cref{tab:icml-prior-scratch} compares
the token bounds with the earlier worst-case single-bit result.

\begin{samepage}
\begin{theorem}[Exact generation]\label{thm:cot-avg}\label{thm:cot-packed}
\textbf{Lower bound for any finite vocabulary.}
Fix $L,H,k\ge1$ and an input layout with $O(1)$ fixed separators.
There are $c_0,c_1>0$ such that the following holds for all sufficiently
large $N$. Fix one causal top-$k$ model of depth $L$ and $H$ heads per
layer for each candidate output length, with any finite vocabulary per
model. For independent uniform $N$-bit operands, the probability that any
candidate generates the exact $2N$-bit product after $t\le c_0N$
auxiliary tokens is at most $e^{-c_1N}$. The choice of $t$ may
depend on the operands.

\textbf{Upper bound.} A one-layer, one-head model with
fixed routing writes the exact product on every input using
$O(N)$ auxiliary tokens. The direct packed FFT uses constant
hidden dimension, $O(N)$ vocabulary entries and $O(N)$ scalar
parameters, including the positional table. All parameters have
$O(\log(N+1))$ bits and are constructible in polynomial time
given the input layout (\cref{app:icml-cot-packed}).
\end{theorem}
\end{samepage}

Candidate models can use separate parameters and vocabularies at
each length. For one top-$1$ head in one layer, binary MSB-first
output, and at most $5N$ context rows per candidate,
$c_0=1/4096$ and $c_1=(\ln 2)/2048$ suffice for $N\ge8192$
(\cref{sec:icml-cot}).

\begin{corollary}[Generation with edits]\label[corollary]{cor:icml-generation-edits}
Under the lower-bound hypotheses of \cref{thm:cot-avg}, for sufficiently
small $c_0>0$ and some $c_1>0$, the probability that any candidate's
entire continuation can be aligned to the product by deleting at most
$c_0N$ tokens from each string is at most $e^{-c_1N}$. An auxiliary
token uses one deletion on the generated side. A substitution can be
represented by one deletion on each side.
\end{corollary}

In the same architecture and layout, the edit bound permits
$\lfloor N/32768\rfloor$ deletions per string with success at most
$2^{-N/32768}$ for $N\ge32768$ (\cref{app:icml-cot-edit}).

\begin{figure}[t]
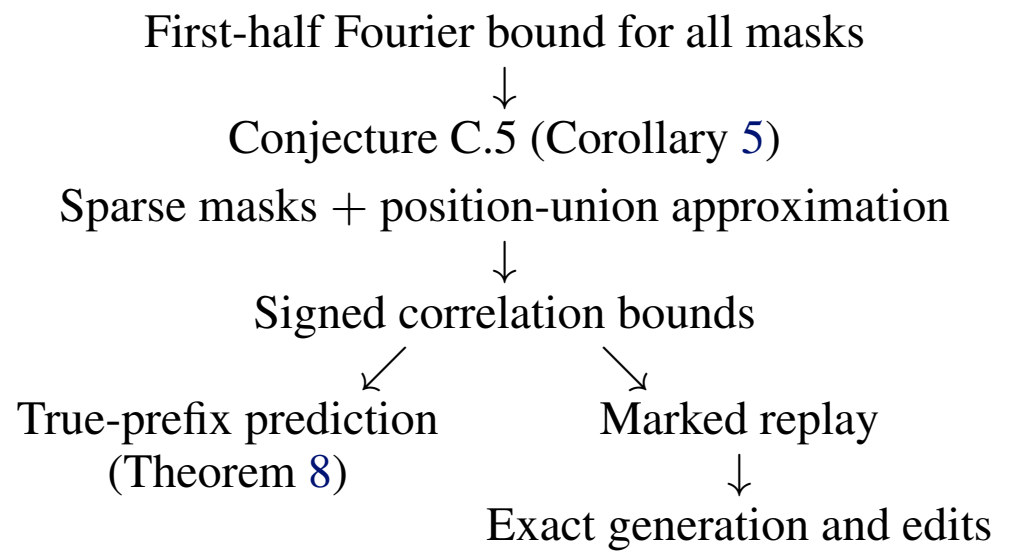

\centering\small
\begin{tabular}{@{}c@{}}
First-half Fourier bound for all masks \\
$\downarrow$ \\
Conjecture C.5 (\cref{cor:C5}) \\[.7ex]
Sparse masks $+$ position-union approximation \\
$\downarrow$ \\
Signed correlation bounds \\
$\swarrow\qquad\qquad\searrow$ \\
\begin{tabular}{@{}c@{\hspace{1em}}c@{}}
True-prefix prediction & Marked replay \\
(\cref{thm:topk-depth}) & $\downarrow$ \\
& Exact generation and edits
\end{tabular}
\end{tabular}
\caption{Proof dependencies. Approximation checks guessed winners
against losing keys. Generation also replays the scratch tokens from
marked positions. A packed FFT supplies the upper bound.}
\label{fig:icml-proof-dependencies}
\end{figure}

Read \cref{fig:icml-proof-dependencies} from the arithmetic estimates
to the two applications. The full proof order, including the
independent FFT construction, is given in \cref{app:icml-core}.

The transfer covers both digit orders and every fixed radix, including
decimal. A generated token designated as scratch can occur between
product digits, and the lower bound charges these tokens together
with the edits (\cref{thm:icml-radix-edit}). The appendix gives
extensions to growing resources and packed words.
With $w$ bits per input and product word, the lower bound still
requires $\Omega(N/w)$ auxiliary tokens in either order when $N/w\to\infty$.
Ordinary packed words attain this count with unrestricted vocabulary
(\cref{thm:icml-packed-replay}).

\begin{table}[t]
\centering\small
\caption{Binary scratchpad token bounds at fixed architectural
resources. Both lower bounds allow a growing vocabulary.}
\label{tab:icml-prior-scratch}
\begin{tabular}{@{}>{\raggedright\arraybackslash}p{.16\columnwidth}
                   >{\raggedright\arraybackslash}p{.33\columnwidth}
                   >{\raggedright\arraybackslash}p{.39\columnwidth}@{}}
\toprule
Bound & \citet{AHRH25} & This work \\
\midrule
Lower & $\Omega(N)$ in the worst case for one midpoint bit from the operands
& $\Omega(N)$ for nonvanishing success on the full product of uniform operands,
including a small edit budget \\
Upper & $O(N\log N)$ for the full product with a growing alphabet
& $O(N)$ for the full product with $O(N)$ symbols and one head in one layer \\
\bottomrule
\end{tabular}
\end{table}

\newcommand{\icmlsoftmaxsection}{%
\section{Softmax attention}\label{sec:icml-softmax}
\Cref{tab:icml-scope} summarizes the architectural hypotheses.
\paragraph{Stable softmax with sampled decoding.}
Fix depth and head count, and allow arbitrary attention weight profiles.
On certified reference trajectories assume a fixed-radius stability tube,
a uniform bound on attention-weighted value variance, and uniform
Lipschitz bounds on layer maps, row updates, and individual logits.
The categorical decoder samples at a temperature bounded below by a
positive constant. No attention-score or output-logit gap is required.
For true contexts of length $n$ in the range of \cref{thm:topk-depth}, set
$d=2^L-1$ and let $\beta$ be the mass of uncertified contexts.
By \cref{prop:icml-sampled-softmax}, sampled next-token accuracy is at most
\[
  B_n=\tfrac12+\beta+O\bigl(n^{-1/(2d)}\ln(2n)\bigr),
\]
and expected natural-log loss is at least $-\ln B_n$. For a sufficiently
small linear deletion budget, free-running success is at most
$\beta_{\rm edit}+e^{-\Omega(N)}$, where $\beta_{\rm edit}$ is the
probability of success with no certified successful tape. These
conclusions permit growing vocabularies. Sampling each head by a fixed
table of exponential races yields a hard-attention simulation. On a
certified tape, a uniform per-logit error controls relative likelihood
even for rare output tokens. The appendix gives sufficient dot-product
operator-norm bounds for certifying every tape
(\cref{app:icml-softmax-sampling}).

\begin{table}[t]
\caption{Scope of the architectural results.}
\label{tab:icml-scope}
\centering
\small
\begin{tabular}{@{}>{\raggedright\arraybackslash}p{.26\columnwidth}
                  >{\raggedright\arraybackslash}p{.68\columnwidth}@{}}
\toprule
\textbf{Attention} & \textbf{Conditions and conclusion} \\
\midrule
Hard top-$k$ & Fixed depth, head count and fan-in yield exponentially
small advantage and linear scratchpad lower bounds
(\cref{thm:topk-depth,thm:cot-avg}). \\
Diffuse softmax & Uniform mass cap and bounded values. Sharp gain
and noise bounds for first-half prediction
(\cref{prop:icml-sharp-diffuse}). \\
Stable softmax & Variance and local stability bounds. Greedy decoding
also needs margins. Sampling needs a temperature floor.
Bounds charge uncertified mass
(\cref{prop:icml-stable-softmax,prop:icml-sampled-softmax}). \\
Unrestricted softmax & Exact multiplication is possible at constant
depth (\cref{rem:averaging}). \\
\bottomrule
\end{tabular}
\end{table}

\paragraph{Sharp threshold for diffuse softmax.}
Take $\alpha N\le T\le N$ with fixed $\alpha>0$ and a layout that preserves each operand's
most-significant-first order. In a fixed-depth causal softmax model, bound layer-$\ell$ value
norms by $B_\ell$ and attention weights at query position $i$ by $\rho_\ell/i$.
Use the global Lipschitz constants $R$ for the per-logit readout, $U_\ell$ for row updates
in the head outputs, and $A_\ell$ for full layers (\cref{app:icml-softmax-sampling}). Set
\[
 G=R\sum_\ell U_\ell B_\ell\rho_\ell\prod_{j>\ell}A_j .
\]

\begin{theorem}[Sharp gain for diffuse softmax]\label{prop:icml-sharp-diffuse}
Fix $a\in(0,1/2)$. Next-bit success at least $1/2+a$ with the true prefix requires
$G/\tau=\Omega_{\alpha,a}(N)$ at sampling temperature $\tau>0$, and
$G/\gamma=\Omega_{\alpha,a}(N)$ for greedy decoding if the top-two logit gap is at least
$\gamma>0$ except on input mass at most $a/4$. For adversarial head-output errors
bounded by $u B_\ell\rho_\ell$, robust success at least $1/2+a$ requires
$u=O_{\alpha,a}(1/N)$. A single attention layer with one uniform head of linear
value width and a constant-depth ReLU readout with polynomial parameter count attains
both gain orders and radius $\Theta(1/N)$.
\end{theorem}

For any fixed $\eta>0$, the proof matches at least a $1-\eta$ fraction of inputs
in pairs sharing the true prefix and having opposite target bits. Each pair
differs in $O_\eta(1)$ operand bits at positions $\Omega(N)$, so diffuse heads
assign them $O(1/N)$ mass.
The matching construction uses a constant-depth majority circuit for multiplication
and a globally Lipschitz ReLU mask.
Under uniform rowwise bounds, positive-temperature sampling keeps
full-product success exponentially small at any prescribed
scratchpad length. Causal influence decays fast enough that summing
squared changes in token log probabilities costs only $O(1/N)$
(\cref{prop:icml-diffuse-transcript}).
}

\newcommand{\icmlfiniteonelayerstatement}{%
\begin{corollary}[Explicit one-layer scratchpad budget]\label[corollary]{cor:icml-one-layer-linear}
In the setting of \cref{thm:cot-avg}, take one layer, $h=Hk$, and context length
at most $4N$. If $N\ge4096h$ and $s\le\lfloor N/(64h)\rfloor$, the probability
of exact generation with an input-selected auxiliary length $0\le t\le s$
is at most
\[
 2(s+1)\left(2^{-N/(4096h)}+2^{-N/8192}\right).
\]
\end{corollary}
At $N=1024$ with one top-$1$ head and at most one auxiliary token,
\cref{tab:icml-finite-scratchpad} gives a bound of $0.014\%$.
}

\newcommand{\icmlimpact}{Multiplication gives a controlled setting for
studying how attention gathers information across intermediate tokens.
The bounds concern specified mathematical architectures and input laws.
The finite certificates apply to the stated bounded-selection models.
They do not quantify the reasoning requirements of deployed systems.}

% The trace bounds are rounded upward by exact integer comparisons.
% The parity numerator is divided by 4^N in both the plot and table.
\newcommand{\midpointdata}{%
 \midpointrow{2}{10}{0.775975}%
 \midpointrow{3}{22}{0.578702}%
 \midpointrow{4}{56}{0.464120}%
 \midpointrow{5}{150}{0.387790}%
 \midpointrow{6}{348}{0.300366}%
 \midpointrow{7}{972}{0.229750}%
 \midpointrow{8}{2360}{0.184240}%
}
\ExplSyntaxOn
\newcommand{\midpointparity}[2]{\fp_eval:n { #2 / 4 ^ #1 }}
\ExplSyntaxOff
\newcommand{\icmlmidpointfigure}{%
\begin{figure}[t]
\centering
\begin{tikzpicture}[x=.84cm,y=3.25cm,font=\footnotesize]
\foreach \y/\tick in {0/0,.2/0.2,.4/0.4,.6/0.6,.8/0.8}{
  \draw[black!17] (1.75,\y) -- (8.25,\y);
  \node[left] at (1.75,\y) {\tick};
}
\draw (1.75,0) -- (8.25,0);
\foreach \x in {2,...,8}{\node[below] at (\x,0) {\x};}
\node[anchor=south west] at (1.65,.82) {Absolute correlation};
\node[below] at (5,-.083) {Operand bits $N$};
\def\midpointrow##1##2##3{%
  \edef\low{\midpointparity{##1}{##2}}
  \draw[black!48,line width=.8pt] (##1,\low) -- (##1,##3);
  \fill[blue!62!black] (##1,\low) circle[radius=2.4pt];
  \draw[orange!65!black,fill=white,line width=1.1pt]
       (##1,##3) ++(-2pt,-2pt) rectangle ++(4pt,4pt);
}
\midpointdata
\end{tikzpicture}
\caption{Midpoint discrepancy on independent uniform operands.
Filled circles give the exact maximum absolute correlation over operand
parities and product masks ending at $T=N$. Open squares certify an
upper bound for the larger class of separate bounded tests. Each
vertical segment brackets the discrepancy at that width. Both endpoints measure
arithmetic correlation. No model was trained for this figure.}
\label{fig:midpoint-discrepancy}
\end{figure}
}

% Shared main-body piece (section: the pseudorandomness theorem); input by main-body.tex (FOCS and named preprint) and by icml.tex (ICML 2027).
\section{The pseudorandomness theorem}\label{sec:setting}

\ificml
A parity records whether a chosen set of bits contains an even or odd
number of ones. A parity test predicts a target bit appreciably better
than chance only if it has appreciable correlation with that bit.
Conjecture C.5 asks for one bound that works for every choice of input
and earlier product bits, including masks that grow with the prefix.
\fi

Write $X_i$ and $Y_j$ ($i,j\in[N]=\{1,\dots,N\}$) for the binary digits of $X$ and $Y$ (their order plays no role) and
$\hat Z=2Z-1\in\{\pm1\}$ for a bit $Z$. The conjecture of~\cite[App.~C.2.2, Eq.~(23)]{AHRH25} bounds the absolute correlation.

\begin{conjecture}[{\cite[Conjecture C.5]{AHRH25}}]\label[conjecture]{conj:C5}
There is $D>0$ such that for all $1\le T\le N$, $A,B\subseteq[N]$ and $C\subseteq[T-1]$,
\[
\Bigl|\E\Bigl[\hat M_T\prod_{i\in A}\hat X_i\prod_{j\in B}\hat Y_j\prod_{k\in C}\hat M_k\Bigr]\Bigr|\le2^{-DT}.
\]
\end{conjecture}

\noindent For $N=1$ the absolute bound fails because the leading product digit is always $0$
and the empty operand masks give absolute correlation one. At $N=2$
the leading bit is one only at $X=Y=3$. Its sign has absolute
correlation $14/16=\NtwoChar$ with the constant test. \Cref{thm:main} below proves it
with arbitrary bounded $f(X)$ and $g(Y)$ in place of the input parities. The real
content is uniformity: one rate $2^{-DT}$ for all $2^{T-1}$ sets $C$ of earlier bits.

The notation is collected in \vsw{\cref{app:icml-core}}{\cref{tab:notation}}. Throughout, $\log$ is the natural logarithm and
$\log_2$ the binary one. The bit $M_k(x,y)$ is the $k$-th binary digit of $xy/4^N\in[0,1)$. For a nonempty
$S\subseteq[T]$ with $\max S=T$ let $\chi_S(x,y)=\prod_{k\in S}(-1)^{M_k(x,y)}$; with the binary digits $b_k(u)$ of
$u\in[0,1)$ and the Walsh functions $W_S=\prod_{k\in S}(-1)^{b_k}$, $\chi_S(x,y)=W_S(xy/4^N)$. Let
\vsw{\begin{multline*}
\disc_N(S)=\max\{|\E f(X)g(Y)\chi_S(X,Y)|:\\ f,g:\{0,\dots,2^N-1\}\to\C,\ |f|,|g|\le1\};
\end{multline*}}{\[
\disc_N(S)=\max\{|\E f(X)g(Y)\chi_S(X,Y)|:\ f,g:\{0,\dots,2^N-1\}\to\C,\ |f|,|g|\le1\};
\]}
the maximum is attained, because the expression is continuous on a compact set. Since $\hat M_k=2M_k-1=-(-1)^{M_k}$, the
quantity in \cref{conj:C5} is $\pm\E[\chi_S(X,Y)\prod_A\hat X_i\prod_B\hat Y_j]$ with $S=C\cup\{T\}$ (the sign is
$(-1)^{|S|}$); the input characters are functions of $X$ alone and of $Y$ alone with modulus $1$. Hence \cref{conj:C5}
follows from the bound $\disc_N(S)\le2^{-DT}$.

\begin{theorem}[Product-bit discrepancy]\label{thm:main}
\ificml
For $N\ge2$ and every mask with $T=\max S\le N$,
$\disc_N(S)\le\min\{2^{-T/2560},46\cdot2^{-T/40}\}$.
\else
With the explicit constant $D=1/512$, $\disc_N(S)\le2^{-DT}$ for all $N\ge2$ and all $S$ with $\max S=T\le
N$. For $T\le N$, the stronger bound $\disc_N(S)\le2^{-T/16}$ holds if $T\ge157$, or if $T\ge64$ and
either $T\le4N/5$ or both $N\ge8$ and $T\ge16\log_2N+180$. Further, $\disc_N(S)\le2^{-T/8}$ for all
$N\ge402$ and $96\le T\le N$, and $\disc_N(S)\le2^{-T/6}$ for all $N\ge951$ and $144\le T\le N$.
For every $\eta\in(0,1/3)$ there are $D_\eta>0$ and $N_\eta$ such that
$\disc_N(S)\le2^{-D_\eta T}$ for every $N\ge N_\eta$ and every mask with $N<T\le(4/3-\eta)N$.
\fi
\end{theorem}

\ificml
\paragraph{An arithmetic proof of the conjecture.}
The full proof is in \cref{app:finite-fourier}. A two-digit grouping
argument controls the Fourier coefficients uniformly over masks
\citep[Lemma~3]{Bourgain11}. For example, on four equal
cells a mask containing both digits has signs $(+,-,-,+)$. At frequency
one the unsigned cell phases are $(1,-i,-1,i)$, so the signed sum
is $2+2i$ with normalized magnitude $1/\sqrt2$.
Each product-frequency kernel has a Plancherel bound, and a count
of digit-boundary crossings transfers the resulting estimate to the
integer grid. \Cref{sec:icml-fourier} sketches this proof and the
sparse-mask estimate used for prediction at all interior positions.
\fi

\begin{corollary}\label[corollary]{cor:C5}
\Cref{conj:C5} holds for every $N\ge2$, with a constant $D$ independent of $N$ (take $S=C\cup\{T\}$,
$f=\prod_A\hat X_i$ and $g=\prod_B\hat Y_j$ in \cref{thm:main}). For $N=1$ the absolute form fails.
\end{corollary}

\ificml\else
\noindent By the discrepancy method, \cref{thm:main} also bounds the communication complexity of
$\bigoplus_{k\in S}M_k$ across $X\,|\,Y$ by $DT-O(1)$ bits from below (\cref{rem:cc}).

\fi
\ificml
\else
\noindent The transformer consequences (\cref{sec:transformers}\vsw{}{, \cref{tab:transformers}}) apply \cref{thm:main} through
inequality~\eqref{eq:walsh} below; the headline is \cref{thm:topk-depth} for every fixed depth. The proof of \cref{thm:main}
combines the Mellin--Hankel bound (\cref{lem:hankel}) with two results that may be of independent interest: the decay
of a Mellin symbol below, and a transfer from the continuum to integers (\cref{thm:transfer}, described in
\cref{sec:discrete} and stated in full in \cref{app:transferstmt}). In them $K_S$ is the integral operator on $L^2[0,1]$ with
kernel $W_S(st)$, $\Wt_S(\xi)=\int_0^1W_S(u)u^{-1/2+i\xi}\,du$ is the Mellin transform of $W_S$ on the critical line,
and $\mu_S=\sup_{\xi\in\R}|\Wt_S(\xi)|$ (\cref{sec:overview}).
\fi

\newcommand{\mellinsymbolstatement}{%
\begin{theorem}[decay of the Mellin symbol]\label{thm:symbol}
Let $T\ge1$, $S\subseteq[T]$ with $T\in S$, and $\xi\in\R$.
\textup{(a)} $|\Wt_S(\xi)|\le3982\cdot2^{-T/4}\min\{1,(2^T/|\xi|)^{3/4}\}$; in particular
$\|K_S\|\le\mu_S\le3982\cdot2^{-T/4}$.
\textup{(b)} \textup{(computer-assisted)} With $\nu=-\frac12\log_20.6646=0.29472\ldots$,
$|\Wt_S(\xi)|\le17077.1\cdot2^{-\nu T}\min\{1,(2^T/|\xi|)^{1-\nu}\}$.
\end{theorem}
}
\ificml\else\mellinsymbolstatement\fi

\subsection{Why the arithmetic estimate works}\label{sec:icml-fourier}

\paragraph{The obstacle.}
A product digit flips repeatedly as the product crosses digit
boundaries. An arbitrary parity of product digits combines many such
patterns. The operand tests can reweight rows and columns of the
product table, so balancing the signs over all entries does not suffice.
We need cancellation even against these separate reweightings.

\paragraph{Pairing digit factors.}
Imagine evaluating the product of two continuous numbers in $[0,1]$.
The parity sign is constant on intervals determined by the highest
selected digit. Its Fourier expansion expresses that sign in oscillatory
product kernels. For each frequency, the digit choices factor into a
sine or cosine for each position. Pairing positions gives the same
strict contraction for every pair of choices. The highest selected digit
also suppresses low frequencies, while integration across a digit
interval controls high frequencies. Together these facts give a
bound with one decay rate for every mask. Uniformity matters because
a network can combine many parities in the same predictor.

\paragraph{Returning to exact operands.}
The continuous estimate applies to integer operands after thickening
each grid point into a short interval. The signs can disagree only
when the product crosses a digit boundary. The continuum norm is
$\mu\le64\,2^{-\gamma T}$ with $\gamma\approx0.0571$.
Intervals of relative width $\epsilon$ cost $\mu/\epsilon$ in the
continuum estimate, and the residue count costs $O(\epsilon)$
through the midpoint. Balancing the two terms yields an exponent
$\gamma/2$, rounded down to $1/40$ in the bound with prefactor $46$.
Direct pairings at small $T$ allow the weaker $1/2560$ exponent
with prefactor one (\cref{app:finite-fourier}).

\paragraph{Past the midpoint.}
A mask selecting few product digits has modest
absolute Fourier mass. After suppressing very low frequencies using
its last selected digit, we can bound each remaining product-frequency
sum by geometric cancellation on the operand grid. This suffices for
the bounded-selection results in both digit orders
(\cref{app:icml-interior}). The extended archive develops dense-mask
and nonuniform-operand estimates. The theorems here use neither extension.

% Shared main-body piece (section: hard-attention transformers); input by main-body.tex (FOCS and named preprint) and by icml.tex (ICML 2027).
\ificml
\section{Bounded selection and correlation transfer}\label{sec:transformers}
\else
\section{Application: hard-attention transformers}\label{sec:transformers}
\fi

\paragraph{The model.}\label{sec:model}
\vsw{}{The headline is \cref{thm:topk-depth}; \cref{tab:transformers}
(\cref{app:transfull}) lists the main transformer results, with proofs in \cref{app:transfull,app:transformers,app:topk}. }
We use the unique-hard-attention transformers (UHATs) of~\cite[\S2]{AHRH25}
\vsw{}{(details in \cref{app:model})}. Layer $0$ is $y_i^{(0)}=\mathrm{emb}(\text{token}_i)+p_i$. In layer $\ell$, each of $H$
heads at position $i$ scores the positions $j\le i$ (causal masking) by
$a_{i,j}=(y_j^{(\ell-1)})^\top\mathbf K_h^\top\mathbf Q_hy_i^{(\ell-1)}$ and attends to the single best one, ties
being broken by a fixed priority order on positions; then $y_i^{(\ell)}=f^{\rm MLP}(y_i^{(\ell-1)}+\sum_hV_hy_{j_h}^{(\ell-1)})$
with an arbitrary map $f^{\rm MLP}$. Width and precision are unrestricted, and there is one transformer per total length.
\ificml
In radix $q$, the fixed input layout has one token per digit of each
of two $N$-digit operands and $O(1)$ fixed separators. Step $T$ reads
the first $T-1$ padded product digits in the specified output order.
The context length is $n=2N+T-1+O(1)$, and the readout is taken at its
last position. A prediction succeeds only when the emitted token is
the correct radix-$q$ digit.
\else
The input $x$ lists the $2N$ bits of $X$ and $Y$ in any fixed order, with possibly $O(1)$ separators at fixed positions.
Step $T$ is correct on $(X,Y)$ if the output at the last position of $xM_1\cdots M_{T-1}$ (the \emph{true} prefix) is
$M_T$. With the signed output $\hat O_T$, equal to $1$, $-1$ or $0$ for output token $1$, $0$ or any other
token, the advantage is at most $\frac12|\E\hat M_T\hat O_T|$.

\fi

\begin{definition}[top-$k$ attention]\label[definition]{def:topk}
In a \emph{top-$k$} layer each head at position $i$ orders the positions $j\le i$ by a score, which may be an arbitrary
function of the two states and the two positions, and then by a fixed priority order on positions, and selects the first
$\min(k,i)$ of them. The new state at $i$ is an arbitrary function of the old state at $i$ and, for every head, the
ordered list of the selected positions and their states. Width, precision and positional encodings are unrestricted,
and the model may differ at every length. A UHAT is the case $k=1$ with bilinear scores.
\end{definition}

\ificml
\noindent A \emph{record} consists of a position and its state at a
given layer. A \emph{probe} is a bounded function of specified records.

\noindent A top-$k$ head may average its $k$ winning records. Averaging
hard attention with unbounded ties and ordinary dense softmax may
read all the eligible keys.
\ificml
% Shared main-body piece (section: technical overview); input by main-body.tex (FOCS and named preprint) and by icml.tex (ICML 2027).
\subsection{Correlation transfer for autoregressive tasks}\label{sec:overview}

\paragraph{Literal contexts and genuine prefixes.}
A \emph{literal context} fixes the positions of raw inputs, supplied
output symbols, and auxiliary tokens. Its \emph{formal cube} lets the
raw inputs and supplied outputs vary independently, with any given
auxiliary tokens held fixed. The model has a readout on every string
in this cube. To measure accuracy we evaluate the readout under the
task's joint law, which supplies a genuine prefix determined by the
inputs in multiplication.

In the binary case the basis functions are parities of subsets of
these slots. The degree is the number of slots in a parity. A formal
parity evaluated on the true context $Z=(X,Y,M_{<T})$ factors into
one parity of each operand and one of the genuine prefix.
Multiplying it by the target sign adds digit $T$ to the product mask.
Thus the discrepancy bound controls each term. If $A(f)$ is the sum
of the absolute Fourier coefficients of a real readout $f$, then in the
binary MSB-first setting with applicable exponent $D$,
\begin{equation}\label{eq:walsh}
|\E[\hat M_T f(Z)]|\le 2^{-DT}A(f).
\end{equation}
An arbitrary readout may have too much coefficient mass for this
estimate to help. Bounded selection gives an approximant with both
small degree and controlled mass. Small degree is enough because
the arithmetic input can then be restricted to sparse product masks.
Over a fixed finite abelian alphabet, a character here is a product
of one-symbol characters. Degree counts its nonconstant slots.

\newcommand{\icmlgrowingstatement}{%
\begin{theorem}[Growing-resource scratchpad transfer]
\label[theorem]{thm:icml-growing-scratchpad}
Use the joint input and output law and the correlation hypothesis of
\cref{thm:icml-transfer}, with at least $\beta N$ qualifying targets
and a fixed finite abelian alphabet of size $q\ge2$. Let $n=\Theta(N)$ be a padded
context-length bound allowing a linear budget of extra generated symbols,
and write $a=1+Hk$. If
\[
  2^L\ln\big(a\ln(2n)\big)=o(\ln n),
\]
there is a radius $r=n^{1-o(1)}$ for which the probability that a free
run admits an alignment after at most $r$ deletions from each string is
at most $\exp(-n^{1-o(1)})$. The model may depend on the generated
length and the finite scratch vocabulary may grow with $N$.
\end{theorem}
}

\newcommand{\icmlsecondarytransfer}{%
\begin{corollary}[Systematic encoding with large dual distance]\label[corollary]{cor:icml-dual-code}
Let $U$ be uniform on $\mathbb F_2^N$ and output the $p=\Theta(N)$ parity
bits of a systematic binary linear code in any prescribed order. If
the dual distance is $\Omega(N)$, any fixed-depth, fixed-head, fixed-$k$
predictor has at most exponentially small advantage at every parity
bit given the true prefix. The probability of aligning a free run
after at most $cN$ deletions on each side is exponentially small for
some $c>0$.
\end{corollary}
Dual distance makes all low-degree target correlations zero even
though every parity bit is a parity of the input itself. The transfer uses approximate
rather than exact position unions at the raw layer to control degree
(\cref{lem:position-union-degree,app:icml-transfer-task}).
}

\paragraph{Position-union approximation.}
To express a probe of $m$ upper-layer records in terms of lower-layer records,
guess the union $S$ of all their winning keys. The query and candidate
records determine the proposed lists. Reject a guess if its lists
have the wrong size or do not use exactly $S$. For each outside key,
test whether it loses to the last proposed winner at every eligible
query and head. Taking the conjunction accepts exactly the true union.
One selected record can encode the global condition that all the
others lost, so counting only selected edges cannot bound what an
adaptive head computes.

\smallskip
\noindent\textbf{Example.} Take two top-$1$ queries at rows $4$ and $5$.
The eligible keys and their scores on one input are
\[
\begin{array}{c|rrrrr}
\text{key}&1&2&3&4&5\\ \hline
\text{query }4&1&6&4&2&\text{--}\\
\text{query }5&3&2&8&1&5
\end{array}
\]
Guess $S=\{2,3\}$. Inside $S$ the winners are $2$ and $3$ with
scores $6$ and $8$. Outside keys $1,4,5$ each lose wherever eligible,
so the guess passes. Guess $S'=\{2,5\}$ instead. Its proposed
winners are $2$ and $5$ with scores $6$ and $5$. Key $3$ loses at
query $4$ but beats key $5$ at query $5$, so the guess fails.
For either guess let $I_j$ indicate that outside key $j$ beats a
proposed winner at some eligible query. With $r=\sum_j I_j$, the
exact test is $\prod_j(1-I_j)$. The quadratic
$P(r)=(2(r-2)^2-1)/7$ equals $1$ at $r=0$ and has magnitude $1/7$
on $r=1,2,3$. It gives $1$ on $S$ and $1/7$ on $S'$.
For example, under the incorrect guess the monomial $I_1I_3$ reads
only outside records $1$ and $3$ once the queries and proposed
winners are given.

For a context of length $n$, use a polynomial of degree
$O(\sqrt{n\log(1/\varepsilon)})$ and uniform error $\varepsilon$
for the shared conjunction~\citep{deWolf08}. Each monomial
introduces at most that many outside records. The number of guessed
unions and the coefficient mass are controlled by subset entropy
and Parseval. Repeating for a fixed number of layers gives
the following joint bound.

\smallskip\noindent\textbf{Position-union lemma}
(\cref{lem:position-union}).
Fix $L,H,k$ and a finite abelian raw alphabet. For any
$\delta,\rho>0$ there is an $\eta>0$ such that for all large $n$,
every probe $F$ of modulus at most one of at most $\eta n$ final-layer
records admits a product-character polynomial $P$ on the formal
context with
\[
 \begin{gathered}
 \|F-P\|_\infty\le e^{-\eta n}/2,\\
 A(P)\le e^{\delta n/4},\qquad \deg P\le\rho n.
 \end{gathered}
\]
The bound is uniform on the formal context cube, including strings
whose supplied prefix disagrees with their operands.

\paragraph{Autoregressive transfer.}
Suppose the task has an exponentially small correlation bound for
every low-degree character times a nontrivial target character.
On its genuine contexts, that rate absorbs the coefficient mass of
the approximant. For generation, the same argument treats a small
linear collection of readouts as one probe. Replay adds the readouts
at the scratch positions to this probe. Its position code bounds the
number of histories we have to consider (\cref{sec:icml-cot}).

\begin{theorem}[Low-degree autoregressive transfer]\label{thm:icml-transfer}\label{thm:icml-scratchpad-transfer}
Let $G$ be a fixed finite abelian alphabet of size $q\ge2$. Given $N$
input symbols and $O(N)$ output symbols in a fixed linear layout,
let $\epsilon_t$ bound the absolute correlation of every nontrivial
target character with every input and true-prefix character of
total degree at most $\rho N$, for fixed $\rho>0$.
Fix $L,H,k$ as in \cref{def:topk}. An alignment is an order-preserving match after deletions from each string.
\begin{enumerate}
\item At every $t$ with $\epsilon_t\le e^{-\gamma N}$ for fixed
$\gamma>0$, true-prefix next-symbol success is at most
$1/q+e^{-\Omega(N)}$.
\item If the condition holds at $\beta N$ positions for fixed
$\beta>0$, there are $c,c'>0$ such that with probability at most
$e^{-c'N}$ some free run aligns to the target after at most $cN$
deletions from each string. The causal model may depend on the
proposed length and the finite scratch vocabulary may grow with $N$.
\end{enumerate}
\end{theorem}

For a systematic binary linear code with linear dual distance,
the input bits and generated parity bits obey no parity relation
of sublinear size. The low-degree target correlations in the transfer
theorem are therefore zero. For a linear number of output parity
bits this gives the same edit-tolerant generation lower bound in
any fixed order (\cref{cor:icml-dual-code}).

\else
% Shared main-body piece (section: technical overview); input by main-body.tex (FOCS and named preprint) and by icml.tex (ICML 2027).
\section{Technical overview}\label{sec:overview}

\paragraph{From transformers to correlations.} At step $T$ a transformer reads $2N+T-1$ bits: the operands and the
prefix. Call them \emph{formal} bits $\mathbf z\in\{0,1\}^m$, $m=2N+T-1$, allowed to take any values; the model is defined
on every such string, consistent with a product or not. Expand $f:\{0,1\}^m\to\R$ in Walsh characters,
$f=\sum_U\widehat f(U)\chi_U$, and let $A(f)=\sum_U|\widehat f(U)|$ be its Walsh norm; $A(fg)\le A(f)A(g)$, and a
conjunction of literals has $A\le1$. On the true context $Z=(X,Y,M_{<T})$ a character splits as
$f_A(X)\,g_B(Y)\prod_{k\in C}(-1)^{M_k}$: a parity of bits of $X$, one of bits of $Y$, and one of a set
$C\subseteq[T-1]$ of earlier product bits. Its correlation with $\hat M_T$ is therefore
$\pm\E f_A(X)g_B(Y)\chi_{C\cup\{T\}}(X,Y)$, at most $2^{-DT}$ by \cref{thm:main}. Summing term by term, in the ranges of \cref{thm:main} with the corresponding exponent $D$,
\begin{equation}\label{eq:walsh}
|\E[\hat M_Tf(Z)]|\le2^{-DT}A(f)\vsw{\quad}{\qquad}\text{for every }f:\{0,1\}^m\to\R .
\end{equation}
Two features matter. The sets $C$ that appear are arbitrary, so the bound must be uniform over masks. And
\eqref{eq:walsh} says nothing about inconsistent strings: $f$ is expanded on the formal cube, but only its values on true
contexts enter the expectation; this is where the uniform-measure Parseval step in the proof of Lemma~C.6 of~\cite{AHRH25} does not apply as stated
(\cref{app:c6}).
So a model is useless at step $T$ once its output is uniformly close, on the whole formal cube, to a function of Walsh
norm $2^{o(T)}$. For one UHAT layer the output is exactly a signed sum of at most $2(2n)^H$ conjunctions: the keys are raw tokens, so
``every other key loses to the winner'' is a condition on one token at a time (\cref{thm:onelayer}). In layer two the
keys are first-layer states, which depend on many tokens; the same condition becomes an AND of $n$ predicates of
first-layer states, whose exact Walsh norm can be exponential (\cref{rem:twolayer}). We replace that AND by a uniform
low-degree approximation (\cref{fig:idea}(b), \cref{sec:transformers}).

\begin{mbfigure}
\centering
\begin{minipage}[c]{0.44\linewidth}\centering
\includegraphics[width=\linewidth]{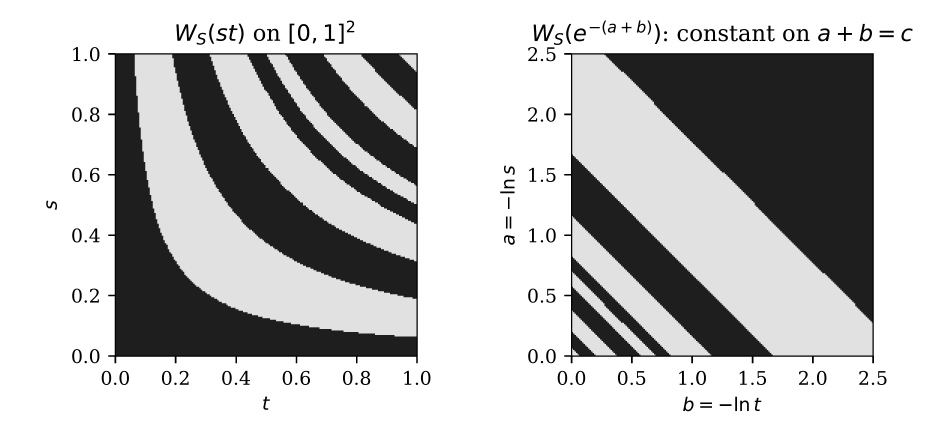}\\[-1pt]{\small(a)}
\end{minipage}\hfill
\begin{minipage}[c]{0.53\linewidth}\centering\footnotesize
\begin{tikzpicture}[box/.style={draw,rounded corners=2pt,align=left,inner sep=3pt,text width=0.93\linewidth},
  node distance=2.2mm]
\node[box](a){\textbf{Layer 1 (top-$1$).} A winner is a conjunction of one-token literals; every predicate of a first-layer
state has $A\le A_0=2(2n)^{H_1}$.};
\node[box,below=of a](b){\textbf{Layer 2.} $\hat O_T=\sum_{\text{tuples}}o\cdot\mathrm{AND}(v_1,\dots,v_n)$, with $A(v_j)\le A_0$
(exact norm can be exponential).};
\node[box,below=of b](c){\textbf{Approximate.} Replace each AND by a polynomial with error at most $\vartheta/B$ (\cref{lem:nor-small}),
where $B=\max(1,\sum_{\text{tuples}}|o|)$. For $\vartheta=e^{-aN/\log^2N}$ and small $a$, $\log A(P)\le DT\ln2/2$.};
\node[box,below=of c](d){\textbf{Correlate.} By \eqref{eq:walsh} with \cref{thm:main} for all masks:
$|\E\hat M_T\hat O_T|\le\vartheta+2^{-DT}A(P)$.};
\draw[->,thick](a)--(b);\draw[->,thick](b)--(c);\draw[->,thick](c)--(d);
\end{tikzpicture}\\[1pt]{\small(b)}
\end{minipage}
\caption{The two transfers. (a) The kernel $W_S(st)$, $S=\{1,3,4\}$ (black $=+1$), changes sign along hyperbolas
$st=\text{const}$; after $s=e^{-a}$, $t=e^{-b}$ it is constant on the lines $a+b=c$, so the operator is a Hankel
operator. (b) The top-$1$ argument of \cref{thm:full-two}: from the model to a function of small Walsh norm that is
uniformly close to its output on the whole formal cube, then to \cref{thm:main}.}
\label{fig:idea}
\end{mbfigure}

\paragraph{Where parity enters.} As in the key idea above, $\|K_S\|\le\mu_S$ (\cref{lem:hankel}), and bounding
$\Wt_S$ is a one-dimensional problem. Self-similarity of $W_S$ under $u\mapsto2u$ splits $\Wt_S$ into dyadic shells, and
integration by parts handles all frequencies outside a resonance band $2^{T/2}<\xi<2^{5T/4}$ (\cref{prop:warmup}).
That uses only the cell structure of $W_S$, which is not enough inside the band (\cref{prop:necessity},
\cref{fig:mellin}). There we use that the Fourier transform of a parity is a product of one cosine or sine per
bit (\cref{lem:walshft}); averaged over the frequencies a shell samples, each of about $T/2$ active factors contributes
at most $2^{-1/2}$ (\cref{lem:transfer}), which gives $2^{-T/4}$ (\cref{sec:symbol}). \Cref{sec:discrete} passes from
the continuum to integers.

\fi

\subsection{Interior prediction for multiplication}
\else
\noindent This covers averaging the values of the $k$ best keys, but not \emph{averaging hard attention}, which averages
all keys tied at the maximum however many there are, nor softmax attention without score margins
(\cref{app:topkmodel,cor:softmax,rem:averaging}).
\fi

\ificml\else
\noindent For one UHAT layer the bound is direct: the output is a signed sum of $2(2n)^H$ conjunctions, so
$|\E[\hat M_T\hat O_T]|\le2(2n)^H2^{-DT}$ and advantage $2^{-DN/2}$ at $T=N$ needs $\Omega(N/\log N)$ heads
(\cref{thm:onelayer,cor:onelayer} in \cref{app:onelayer}).

% FOCS-MOVED2: thm:full-two -> app:twospectral; its proof-idea paragraph is covered there
\noindent A \vsw{direct proof}{warm-up} for two layers, with one explicit Chebyshev-type polynomial, is \cref{thm:full-two} in
\cref{app:twospectral}. The theorem below covers every fixed depth.\vsw{ In words: at fixed depth the advantage is at
most $e^{-t}$ with $t$ of order $n/(\ln n)^{2^L-2}$, that is $\exp(-n^{1-o(1)})$; the explicit parameters make the
dependence on $L$, $H$ and $k$ precise.}{}

\fi
\begin{icmltight}\begin{theorem}[top-$k$ attention at every depth]\label{thm:topk-depth}
\ificml
Fix a radix $q\ge2$, a margin $0<\eta<1$, and $L,H,k$.
For independent uniform $N$-digit operands and either digit order,
at every true-prefix step $\eta N\le T\le(2-\eta)N$,
next-digit accuracy is at most $1/q+e^{-\Omega(N)}$.
The constants may depend on the fixed parameters.
\else
Fix $\alpha\in(0,1]$. Let $\alpha N\le T\le N$, or set $\alpha=1$
and fix $\eta\in(0,1/3)$ with $N<T\le(4/3-\eta)N$ and
$N\ge N_\eta$ from the post-midpoint estimate.
Consider any $L$-layer model of \cref{def:topk} with at most $H$
heads per layer, on the true context of length $n\le4N$.
Write $S=|\E[\hat M_T\hat O_T]|$.
For fixed $L,H,k$ there are $c_0>0$ and $n_0$ depending only on $\alpha,D,L,H,k$ and the stated size threshold such that
$S\le\exp(-c_0n/(\ln n)^{2^L-2})$ for $n\ge n_0$. In particular $S\le\exp(-n^\gamma)$
for every fixed $\gamma<1$ and all sufficiently large $n$.
\fi
\end{theorem}\end{icmltight}
\ificml
For a one-layer binary decoder at $N=T=1000$, a genuine-prefix
prediction with $Hk=8$ has advantage below $3.01\cdot10^{-15}$.
This is an upper bound on the accuracy gain over a fair guess at a
single teacher-forced step. The proof shares probes across heads
(\cref{app:midpoint-diagnostic}). Growing-resource bounds are in
\cref{app:icml-prediction-resources}.
\fi

\newif\ificmlresourcegeneral
\ificml\ifoldpullback\else\icmlresourcegeneraltrue\fi\fi
\newcommand{\topkgrowingdetails}{%
\ificmlresourcegeneral
\begin{theorem}[Low-degree transfer with growing resources]\label{thm:growing-resource-transfer}
On a fixed layout of $n$ positions, let $Z$ be a context over a
fixed finite abelian alphabet of size $q\ge2$ (fixed slots are allowed)
and let $Y$ be its target symbol. The joint law can be arbitrary.
Suppose that for some $\rho,\gamma>0$ every nontrivial target character
has absolute correlation at most $e^{-\gamma n}$ with every
formal character of degree at most $\rho n$.
For top-$k$ networks of \cref{def:topk}, allow $L,H,k$ to vary. Choose $0<\vartheta<\min(\rho,1/2)$
with $\Phi(\vartheta)+\vartheta\ln(q-1)\le\gamma$, where $\Phi$ is binary entropy in nats.
Put $a=1+Hk$, $K_{\mathrm{set}}=4a\ln(4n)$, $Q=2^{L-1}$ and
\[
 R_0=\max\{(9c^2/\vartheta)^2,4/\gamma,1\},\qquad
 t=\frac{n}{R_0^Q K_{\mathrm{set}}^{2Q-2}},
 \qquad c=3.34\sqrt{1+\ln 2}+1.
\]
For sufficiently large $n$ depending only on $q,\rho,\gamma,\vartheta$, if $t\ge K_{\mathrm{set}}+\ln(2L)$, accuracy is
at most $1/q+2e^{-t}$. Each nontrivial signed character correlation
is bounded by $2e^{-t}$.
\end{theorem}
\fi
\begin{remark}[Explicit fan-in and depth bounds for multiplication]\label[remark]{rem:topk-explicit}
\ificmlresourcegeneral
In \cref{thm:topk-depth}, the elementary sparse estimates provide
the hypothesis of \cref{thm:growing-resource-transfer} in either
order and every fixed radix. Use its definitions and bound, with
parameters depending only on the radix and margin.
\else
Under the binary MSB-first context assumptions of \cref{thm:topk-depth},
put $\alpha=\min(1,T/N)$ and write $S=|\E[\hat M_T\hat O_T]|$,
$a=1+Hk$, $K_{\mathrm{set}}=4a\ln(4n)$, $Q=2^{L-1}$, and put
\[
 \begin{aligned}
 \delta&=\min(1,D\alpha\ln2/4),\quad R_0=(\ifoldpullback16\else256\fi)c^2/\delta,\\
 t&=\frac{n}{R_0^Q K_{\mathrm{set}}^{2Q-2}},
 \end{aligned}
\]
with $D=\vsw{1/2560}{1/512}$ (or a stronger applicable exponent) for $T\le N$ and \ificml $D=c_\eta/2$ otherwise,\else $D=D_\eta$ otherwise,\fi and $c=3.34\sqrt{1+\ln2}+1$ \ificml as in the small-error conjunction bound.\else from \cref{lem:topk-poly}.\fi
For arbitrary positive integers $L,H,k$, if $t\ge K_{\mathrm{set}}+\ln(2L)$ then $S\le2e^{-t}$.
\fi
For fixed $L$ this condition holds for all large $n$ whenever
$a\le n^{1/[2^L-1]-\varepsilon}$, for any fixed $\varepsilon>0$. The exact finite condition retains
the logarithmic factors at the boundary exponent $1/[2^L-1]$.

For the uniform growing-depth bound let $u=\ln n$ and put
\[
 \begin{aligned}
 h(a)&=\lceil\log_2(8+4\log_2 a)\rceil,\\
 J&=\log_2 u-\log_2\ln u-h(a).
 \end{aligned}
\]
\ificmlresourcegeneral
For sufficiently large $n$ and $u\ge\max(4,8\sqrt{R_0})$, the digit
advantage is at most $e^{-n^{3/4}}$ at every $1\le L\le J$,
also when $H,k$ vary with $n$.
\else
If $u\ge\max(4,8\sqrt{R_0})$, then $S\le2\exp(-n^{3/4})$ for every $1\le L\le J$, also when $H,k$ vary with $n$.
\fi
\end{remark}

\begin{corollary}\label[corollary]{cor:topk-loglog}
In the setting of \cref{thm:topk-depth}, advantage is at most $e^{-n^{3/4}}$ at depths $L\le J$
under the condition of \cref{rem:topk-explicit}, including variable $H,k$. Thus for any fixed $p>0$, advantage at least $n^{-p}$
at step $T$ requires more than $\log_2\ln n-\log_2\ln\ln n-h(1+Hk)$ layers for all sufficiently large $n$, how large
\ificmlresourcegeneral
depending only on $p$ and the radix, margin and size thresholds above.
\else
depending only on $p$, $\alpha$, $D$ and the stated size threshold.
\fi
\end{corollary}
}
\ificml\else\topkgrowingdetails\fi

\ificml\else
\noindent At $L=2$ the fixed-head rate is that of \cref{thm:full-two}. At $L=3$ the theorem gives
\cref{cor:topk-three} (\cref{app:topkk}), in which $k$ may grow up to $\ln n$, and it implies the rate of
\cref{thm:three}\vsw{ for three unrestricted layers (\cref{app:three})}{}. The polynomial approximation holds for any input task.

\fi
\ificml
\newcommand{\icmlpredictionextensions}{%
\subsection{Prediction resources}\label[appendix]{app:icml-prediction-resources}
At fixed head count and fan-in, inverse-polynomial advantage needs
depth $\Omega(\log\log n)$. For fixed $L$, the necessary selection
budget is $1+Hk\ge n^{1/(2^L-1)-o(1)}$ in the range of
\cref{thm:topk-depth}. Fixed routing suffices for exact prediction
with exponent $1/L$. \Cref{fig:icml-k-gap,rem:topk-explicit}
show the open gap and the explicit resource bounds. A small linear
number of final-layer records also reveals exponentially little
about a binary interior target (\cref{cor:icml-records}).
\begin{figure}[!t]
\centering
\resizebox{.56\columnwidth}{!}{%
\begin{tikzpicture}[font=\small]
\begin{scope}[x=0.84cm,y=3.55cm]
\fill[blue!9] (1,0) -- plot[domain=1:6,samples at={1,2,3,4,5,6}](\x,{1/(pow(2,\x)-1)}) -- (6,0) -- cycle;
\fill[orange!16] (1,1) -- plot[domain=1:6,samples at={1,2,3,4,5,6}](\x,{1/\x}) -- (6,1) -- cycle;
\draw[->] (0.83,0) -- (6.23,0) node[right] {$L$};
\draw[->] (1,0) -- (1,1.09) node[above] {$\beta$};
\foreach \l in {1,...,6}{
  \draw (\l,0) -- ++(0,-.025) node[below] {\l};
}
\foreach \b/\lbl in {0/0,.5/{1/2},1/1}{
  \draw (1,\b) -- ++(-.07,0) node[left] {\lbl};
}
\draw[very thick,blue!65!black] plot[domain=1:6,samples at={1,2,3,4,5,6}](\x,{1/(pow(2,\x)-1)});
\draw[very thick,orange!70!black,dashed] plot[domain=1:6,samples at={1,2,3,4,5,6}](\x,{1/\x});
\node[anchor=west,align=left,text=blue!65!black] at (1.06,.095) {Negligible\\advantage};
\node[anchor=west,text=orange!70!black] at (2.15,.58) {Exact, fixed routing};
\fill[blue!65!black] (3,{1/7}) circle (1.3pt);
\node[font=\scriptsize,text=blue!65!black,anchor=south] at (3,{1/7+.015}) {$1/7$};
\node at (4.2,.151) {Open gap};
\end{scope}
\begin{scope}[shift={(1.55cm,2.41cm)},font=\scriptsize]
\node[align=left,anchor=south west,fill=white,draw=black!30,rounded corners=1pt,inner sep=3pt]
  at (0,0) {Depth with fixed $H,k$\\Adaptive lower: $\Omega(\log\log n)$\\Fixed routing (exact): $\Theta(\log n)$\\Adaptive upper (small error): $O(\log n/\log\log n)$};
\end{scope}
\end{tikzpicture}}
\caption{Fan-in exponents for interior prediction. The blue lower bound
holds in either digit order and every fixed radix.
With a fixed head count and $k=n^\beta$, the blue curve shows the necessary exponent
$1/(2^L-1)$ for inverse-polynomial advantage with adaptive routing, up to $o(1)$
(\cref{rem:topk-explicit}). The dashed curve shows the sufficient exponent $1/L$ for
exact prediction with fixed routing, attained when $k+1\ge n^{1/L}$. The gap leaves
room for both adaptive routing and weaker accuracy requirements. The adaptive upper bound in the depth inset uses
the binary MSB-first construction in the first half in \cref{prop:icml-adaptive-depth-upper}.}
\label{fig:icml-k-gap}
\end{figure}

\paragraph{An upper bound on the required $k$.}
For fixed depth $L$, one head per layer suffices for exact prediction when
$(k+1)^L\ge n$ in \cref{def:topk}. Let $b=k+1$. At layer $j$, position $i$
selects the distinct positive positions $i-hb^{j-1}$ for $1\le h\le k$,
using fixed scores and ignoring any fillers. Each state merges these
position-tagged records with its own. After layer $j$, position $i$ holds the
whole interval $[\max(1,i-b^j+1),i]$ by induction. The final update can
therefore compute the next product digit from the full context. This uses unrestricted
states and computation in the update. For $L>1$ it rules out a near-linear
necessary $k$ per head in this model.

\paragraph{A receptive-field bound for fixed routing.}
For binary MSB-first prediction, if fixed routing makes a signed readout a function of at most $m$ formal input bits,
Parseval and Cauchy--Schwarz give Walsh norm at most $2^{m/2}$. Thus \eqref{eq:walsh}
bounds advantage by $\tfrac12\,2^{m/2-DT}$. This also applies to sliding-window,
dilated, and block-sparse softmax with fixed support: the data-dependent weights are
functions of the states inside that support. For $L$ layers with $H$ lists of at most
$k$ fixed positions per query, including residual edges gives $m\le(1+Hk)^L$, so
\begin{equation}\label{eq:fixed-routing-lower}
 \Pr[\text{correct}]-\tfrac12 \le 2^{(1+Hk)^L/2-DT-1}.
\end{equation}
Here $D$ is the applicable discrepancy exponent from \cref{thm:main}. In the range of
\cref{thm:topk-depth}, inverse-polynomial advantage requires $(1+Hk)^L=\Omega(n)$:
for fixed $H,k$ this gives $L\ge\log_{1+Hk}n-O(1)$, and for fixed $L$ it gives
$1+Hk=\Omega(n^{1/L})$. For either output order and every fixed radix,
the sparse-mask estimate gives the same necessary receptive-field size
$m=\Omega(n)$. A bounded character of the readout on $m$ slots has
degree at most $m$ and coefficient mass at most $q^{m/2}$. When $m/n$
is sufficiently small, the uniform sparse correlation bound absorbs this mass.

}
\fi

\newcommand{\multiplicationtopktechnical}{%
\begin{lemma}[polynomial approximation of fixed-depth top-$k$ models]\label[lemma]{lem:topk-poly}
Let $a=1+Hk$, $K_{\mathrm{full}}=4a^L\ln(4n)$ and $c=3.34\sqrt{1+\ln2}+1<5.4$, the constant of \cref{lem:nor-small} at error
$e^{-s}$, $s\ge1$. For every $L$-layer top-$k$ transformer on $n$ positions with at most $H$ heads per layer and at most one
variable bit per position, and every $v$ with $K_{\mathrm{full}}+\ln(2L)\le v\le n$, there is a real polynomial $P$ on the whole input
cube with $|P-\hat O|\le e^{-v}/2$ everywhere and $\deg P\le4c^2nK_{\mathrm{full}}(v/(nK_{\mathrm{full}}^2))^{2^{-L}}$, where $\hat O$ is the signed
output at the last position. For $L\ge2$ there is also a possibly different $P$ with the same uniform error and $\ln A(P)\le4c^2nK_{\mathrm{full}}^2(v/(nK_{\mathrm{full}}^2))^{2^{-(L-1)}}$.
The same holds with $\hat O$ replaced by any function of the top state at the last position with values in $[-1,1]$.
\end{lemma}

\noindent With $v=K_{\mathrm{full}}+\ln(2L)$ the degree is $O_{L,H,k}((n\ln n)^{1-2^{-L}})$ at inverse-polynomial error.

\paragraph{Proof idea} (\cref{app:topkdepth}). We pull a polynomial backwards through the layers, starting from the
readout, on the whole cube of \emph{Cartesian} alphabets (at each position, every value the update takes on every
combination of lower states). A character of the upper states reads at most $d$ positions. Guessing the joint transcript
of those positions (query states, selected positions and their states) is correct exactly when a conjunction of $n$
tests holds, one per position, on a single lower state each; tests at one position are intersected. One small-error AND
polynomial (\cref{lem:nor-small}) replaces the conjunction at degree $c\sqrt{ns}$ for error $e^{-s}$, and Parseval charges
the $e^{O(K_{\mathrm{full}}d)}$ guesses to $s$, inside the square root. So degrees follow $d_{\rm new}\approx\sqrt{nK_{\mathrm{full}}d}$ and errors add.
At the bottom the conjunctions are products of literals, kept exact, and \eqref{eq:walsh} finishes. Approximating each
query separately would multiply degrees; the single joint conjunction is the shared-input composition of~\cite{BKT19}.

% FOCS-MOVED: Further results, Related work -> appendix

}
\ificml\else\multiplicationtopktechnical\fi

\newcommand{\icmlselectionsummary}{%
\paragraph{Beyond hard selection.}
The same bounds apply to softmax with a fixed support of at most $k$
keys (\cref{cor:fixed-support-softmax}). Dense softmax can aggregate
all operand bits in one head. Linear scratch is again necessary
under adversarial head-output error at a fixed positive relative radius or independent
Gaussian head noise at bounded signal-to-noise ratio, including
fresh recomputation (\cref{cor:icml-precision-main,prop:gaussian-scratchpad}).
The perturbation models and finite calibration are in
\cref{app:icml-dense-scope}.

\paragraph{Deterministically rounded dense attention.}
At logarithmic fixed-point precision, one constant-width layer can
multiply exactly with $O(N/\log\log N)$ scratch tokens using dot-product
scores. Relative-position bias reduces this to $\Theta(N/\log N)$
with sequential accumulation. Both constructions use length-dependent
positions and a polynomial untied vocabulary. In the latter model
the converse charges the memory of each row evaluation
(\cref{thm:icml-rounded-dense}). These guarantees use a specified
rounding order. An adversary may perturb the same outputs differently.

\begin{proposition}[Effective support for robust generation]
\label[proposition]{prop:icml-effective-robust}
Run the renormalized top-$k$ model. At every visited head let $t$
be the discarded softmax mass at its incoming states and let
$\mu_{\rm tail},\mu_{\rm keep}$ be the corresponding conditional means.
Suppose $\beta_k$ is the probability that in any candidate-length
rollout the displacement $t\|\mu_{\rm tail}-\mu_{\rm keep}\|_2$
exceeds the head radius, taking the displacement as zero if $t=0$.
Then robust generation success is at most
$b_k^{\rm gen}+\beta_k$, where $b_k^{\rm gen}$ is the hard top-$k$
generation bound for the same edit budget. With values bounded by
$B$, $2Bt$ upper-bounds the displacement at admissible states.
\end{proposition}

The certificate concerns the truncated rollout. The adversary
reproduces that rollout until a head fails the certificate
(\cref{app:icml-dense-scope}). Greedy agreement of the clean
dense and truncated runs is a different claim and needs
separate control of the output margins.
}

\newcommand{\icmlselectionappendix}{%
\subsection{Selection and aggregation}\label[appendix]{app:selection-coverage}
A top-$k$ head can compare every candidate score, but its update
sees only the $k$ winning records. This permits averaging those
winning values. A sparse softmax head with a fixed support also
sees only its support, although its weights can depend on the
states. A dense head can assign positive weight to every key.
For example, uniform attention can send each operand bit into
its own value coordinate and an unrestricted readout can multiply.
Unbounded tie averaging can do the same by tying every score.
Even one selected record can encode a global condition, since a
designated key wins only when it beats every competitor. Counting
the records reachable after selection therefore does not account for
what adaptive selection can reveal.

\begin{corollary}[Fixed-support softmax]\label[corollary]{cor:fixed-support-softmax}
The prediction and generation bounds in
\cref{thm:topk-depth,thm:cot-avg,cor:icml-generation-edits} apply to a causal
softmax model whose heads each use a fixed support of at most
$k$ keys. Each eligible set may depend on layer, head, row, and
total length, with pairwise scores and row-local values and updates.
This includes bounded sliding-window, dilated, and block-sparse masks
whose supports are input-independent.
\end{corollary}
\begin{proof}
Give each supported key a higher selection score than every
unsupported key and break ties by a fixed priority. Top-$k$ then
returns every supported record. The update discards fillers,
computes the original logits and the exact softmax mean on the
support, and applies the original update. Induction over layers
preserves states on every literal context. Induction over
generated tokens then preserves the entire run, including when
the auxiliary length varies within the stated model family.
\end{proof}

A query-dependent count bounded by $K$ is also covered: select
the top $K$ and use the incoming state to truncate the list.
A threshold known from that state works in the same way if it
retains at most $K$ keys. More generally, a normalization rule
$w_j=[a(s_j-\tau)]_+^r$ with fixed $a,r>0$ and $\sum_j w_j=1$
is covered when it always has at most $K$ positive weights:
the head can reconstruct the unique $\tau$ from the returned
scores alone. An extra statistic of the discarded scores or
an unbounded support is outside these simulations.

For dense heads the error model determines the generation bound.

\begin{table}[t]
\centering\small
\caption{Architectural coverage. All prediction statements use a
genuine prefix. Generation statements are free-running.}
\label{tab:selection-coverage}
\begin{tabular}{@{}p{.34\columnwidth}p{.60\columnwidth}@{}}
\toprule
Attention mechanism & Scope of lower bound \\
\midrule
Top-$k$ and bounded fixed-support softmax & Prediction and generation at fixed resources, including averaging over the selected keys. \\
Dense softmax, adversarial errors & Robust generation under simultaneous head-output errors (\cref{cor:icml-precision-main}). \\
Dense softmax, independent Gaussian noise & Cached or independently recomputed states. Linear scratch at bounded $B/\sigma$ and $N^{1-o(1)}$ at $1+B/\sigma=N^{o(1)}$ (\cref{prop:gaussian-scratchpad}). \\
Unbounded tie averaging & Can implement uniform aggregation. \\
\bottomrule
\end{tabular}
\end{table}
}

\newcommand{\icmldenseappendix}{%
\subsection{Robust dense-softmax generation}
\label[appendix]{app:icml-dense-scope}
Dense attention may use any weight profile from finite pairwise scores.
Require each head value to
have Euclidean norm at most $B$ on states reachable under allowed
perturbations. At every head and layer, for every row, an online
adversary may replace the mean at the actual incoming state by a
vector within Euclidean distance $\varepsilon>0$.

\begin{corollary}[Precision and robust scratchpad length]
\label[corollary]{cor:icml-precision-main}
Fix $L,H$ and use the length models and local updates of
\cref{app:dense-scratchpad} on independent uniform binary operands.
If $B/\varepsilon=O(1)$, robust success with at most $cN$ total
scratch tokens and edit operations is at most $e^{-\Omega(N)}$ for
some $c>0$. If $1+B/\varepsilon=N^{o(1)}$, there is a budget of
$N^{1-o(1)}$ with success at most $\exp[-N^{1-o(1)}]$.
\end{corollary}

Independent Gaussian head noise gives the same asymptotic budgets
and rates with $\sigma$ in place of $\varepsilon$, both with cached
states and with fresh recomputation (\cref{prop:gaussian-scratchpad}).
The power-law tradeoffs for growing $B/\varepsilon$ or $B/\sigma$
are in \cref{cor:asymp,prop:gaussian-scratchpad}.

\paragraph{Finite-length calibration.}
For $N=1000$ and a row budget of $5000$, take two layers, one head
per layer, and $B/\varepsilon=1024$. With per-site rejection
probability at most $1/20000$, the bounded-value adversarial
simulator uses $83{,}076{,}475$ draws per head site. The resulting
bound is vacuous at this scale
(\cref{lem:dense-sampling-domination,thm:dense-scratchpad-rates}).

\paragraph{How the error model enters.}
An exponential race selects one record with its softmax probability;
fixed clocks give a hard selector. In the adversarial model we resample
until the empirical mean lies in the allowed ball at the current state.
If each site accepts with probability at least $1-\delta$, the hard
sampler retains at least $(1-\delta)^M$ of the conditioned path mass
over $M$ sites (\cref{lem:dense-sampling-domination}).
For Gaussian head noise the simulator instead combines several sample
means to approximate the prescribed Gaussian density. For fresh
recomputation we change the law only at deleted emissions, then
compare correct predictions with a randomly tested cached pass
(\cref{prop:gaussian-scratchpad}).

\paragraph{Complement to rounded arithmetic.}
\citet{Feng24} prove size lower bounds under specified low-precision
internal arithmetic and give an exact constant-depth multiplier
of hidden dimension $O(N^2)$ at logarithmic arithmetic precision.
Here width and local computation can grow without restriction.
The constraint is simultaneous absolute error relative to the value
bound, evaluated on the perturbed run.
The companion studies achievable numerical robustness
radii~\cite{ArithmeticRobustnessCompanion}.
}

\section{Scratchpad lower bounds and matching constructions}\label{sec:icml-cot}
\paragraph{Exact generation with one top-one head.}
Take one layer and one head that selects a single key, with at most
$5N$ context rows. Fix a length with $t\le\lfloor N/4096\rfloor$
initial scratch tokens. Generate those tokens, then supply genuine
binary product bits in MSB-first order. Test the last
$b=\lfloor N/1024\rfloor$ predictions through the midpoint. A
successful free run passes every test because its history agrees
with the supplied bits. With signs $g_i$ for the predicted bits,
mapping any nondigit to zero before taking its sign, the agreement
indicator is $2^{-b}\prod_i(1+\widehat M_i g_i)$.

For a nonempty term, let $J$ be its tested set and $u=\max J$.
Guess the union of winning positions of the $m=t+|J|$ queries,
including the scratch queries. Guess the raw bits at those winning
positions and at the queries when they are ordinary input or
supplied-bit slots. Replay with selection restricted to the guessed
union. Each scratch token is recovered before any later query uses
it. Require that the replay uses the entire guessed union and
that every outside key loses under the model's scores and tie rule.
Each outside comparison has at most one unfixed raw bit, so its
condition is a literal, a constant, or impossible. The true branch
is the only passing one. The product of predicted signs is fixed
on that branch.

Its indicator is a conjunction of literals, with Fourier coefficient
mass at most one since each fixed sign contributes $(1\pm z)/2$.
Multiplying by the earlier true signs just shifts the Fourier
support. The remaining true sign at $u$ is the target. By
\cref{prop:finite-fourier}, each branch contributes at most
$46\,2^{-u/40}$ to the signed moment. Write $S(n,m)$ for the count of
subsets of at most $m$ of $n$ positions. There are at most
$2^{2m}S(5N,m)$ branches. The inequality
$S(5N,m)\le(5eN/m)^m$ and $m\le5N/4096$ bound this count by
$2^{5N/256}$. Since $u\ge1023N/1024$, the moment is at most
$2^{6-223N/40960}$.

The empty term contributes $2^{-b}$ and the nonempty terms have
total weight at most one. For $N\ge8192$, sum over lengths using
$1+\lfloor N/4096\rfloor\le2^{N/4096}$ to obtain
\[
 \Pr[\mathrm{success}]\le
  2^{1-6N/8192}+2^{6-213N/40960}\le2^{-N/2048}.
\]

\paragraph{Marked replay.}
Fix marked output slots to reconstruct. Mark their final-layer
readouts. At each earlier layer mark each query's own row and its
selected keys. For $a=1+Hk$ and $r$ designated
outputs, there are at most $a^{L-j}r$ marks in layer $j$. Encode only
the layerwise \emph{sets} of positions and the fixed-alphabet raw
symbols on the marked input and retained-answer slots.

\smallskip
\noindent\textbf{Replay and count with two layers.}
Take seven slots and one top-$1$ head. Fix deleted output slots $5$
and $7$, predicted from final-layer rows $4$ and $6$. Suppose these
queries select $2$ and $5$. Including the residuals, the layer-one
marks are $\{2,4,5,6\}$. If those rows select $1,2,4,5$, respectively,
the raw marks are $\{1,2,4,5,6\}$.
The code stores the two sets and the supplied bits, say $1,0,1,0$ at
slots $1,2,4,6$. Let layer one retain the pair (own token, selected
token). At row $4$ in layer two, the own pair $(1,0)$ and selected
pair $(0,1)$ have sum two. Let this readout emit a token $\tau_2$
tagging that sum at slot $5$. At row $6$ the selected layer-one pair
is now $(\tau_2,1)$. A readout that tests for the tag $\tau_2$ emits
$1$ at slot $7$. All original winners are present and any extra
eligible mark already lost, so replay recomputes the winners with
the fixed model.

For these set sizes and four supplied bits, there are at most
$\binom{7}{4}\binom{7}{5}2^4=11{,}760$ candidate codes, even when the
auxiliary vocabulary is larger. This count allows every pair of sets
of the prescribed sizes, including pairs that could never occur.

In general, chronological decoding reconstructs the entire token at
each marked readout, even if it has a large vocabulary. On binary
inputs and targets the number of codes is at most
$S(n,a^Lr)^L2^{a^Lr}$. At fixed $L,H,k$, this costs
$O(r\log(en/r))$ bits (\cref{lem:marked-position-code}).

\paragraph{Extending the argument to multiple layers.}
At each layer the losing tests depend on the preceding layer's states.
The position-union lemma approximates their conjunction and pulls a
probe of the resulting readouts back to the raw slots. For each
scratch string, the probe includes tests that verify its emissions.
The last nontrivial target in the agreement expansion then controls
the moment. The marked position code counts the scratch strings,
and its exponential rate shrinks with the scratch fraction
(\cref{app:icml-cot-joint}).

\paragraph{Allowing edits.}
Fix deletion masks on the generated and target strings. Pair
survivors in order, supply their paired true symbols, and generate
at the deleted slots. An alignment witness again agrees with this
replay. The same test loses only the deleted targets, and the
same position code now includes deleted emissions before the final
tested target. Counting mask pairs adds a term with arbitrarily
small exponential rate as the deletion fraction shrinks
(\cref{eq:icml-general-alignment}).
This also covers interleaved scratch and causal stopping policies
whose realized runs agree with admissible fixed-length continuations.

\paragraph{Finite certificates.}
For binary MSB-first generation with one top-$1$ head and one layer,
integer checks of analytic bounds certify exact success below
$3.170\%$ at $N=1000$ with one initial scratch slot and a fixed
length (\cref{app:midpoint-diagnostic}). A separate decision-tree
certificate gives one-deletion alignment success at most $0.495552$
at $N=490$ and $0.334588$ at $N=512$, and two-deletion success
at most $0.016752$ at $N=1024$ (\cref{app:decision-tree-edits}).
The analytic edit bound first gives a nonzero budget at $N=32768$
(\cref{app:icml-cot-edit}).

\paragraph{Matching constructive upper bounds.}
A one-head causal model implements a bounded-arity word program with
fixed read addresses. For a binary instruction a spacer reads and stages
one operand, then the staged row reads the other and emits the result.
The packed transform uses $O(N/\log N)$ words of $O(\log N)$ bits.
Two normalized fixed-point transforms, pointwise multiplication,
the inverse transform, and a carry pass cost $O(N)$ word
instructions~\citep[\S4.1]{Furer2014}. Each register can be
split into a fixed number of limbs of width
$\lfloor\log_2(N)/3\rfloor$ for large $N$.
Each word operation expands into a constant number of instructions
on at most three limbs. The tagged implementation therefore uses
$O(N)$ symbols and achieves linear token length with constant hidden
dimension (\cref{app:icml-cot-packed}). The $O(N)$ auxiliary tokens
carry $O(N\log N)$ bits in total.

The positional schedule, token bank, and weights are generated for
each operand length. Vocabulary tradeoffs and circuit simulations
are given in \cref{cor:icml-vocab,app:icml-cot-circuits}.
For scratchpad design in this model, the lower bound counts gathering
steps even with a large alphabet. The FFT schedule attains that
order by using each token to carry a logarithmic-size limb.

\icmlselectionsummary
\icmlrelatedmain
% Shared main-body piece (section: limitations); input by main-body.tex (FOCS and named preprint) and by icml.tex (ICML 2027).
\section{Limitations}\label{sec:limits}

\ificml
\paragraph{Finite lengths and non-uniformity.}
The linear lower bound is asymptotic at fixed depth, head count, and
fan-in. For example, the explicit one-head exact-generation certificate
at $N=1000$ gives a sub-$1/2$ success bound only through two initial
auxiliary tokens (\cref{app:midpoint-diagnostic}). The packed FFT is a
family of length-specific models. Its polynomial-time compiler builds
the vocabulary, positional table, and weights for each $N$. A single
fixed transformer that runs the same construction across lengths
remains open.

\paragraph{Vocabulary, access, and noise.}
With a constant alphabet, the lower bound is $\Omega(N)$ and the
fixed-routing construction uses $O(N\log N)$ auxiliary tokens
(\cref{cor:icml-vocab}). The logarithmic gap remains open.
For fixed routing and a fixed auxiliary length, closing it at
$O(N)$ would give linear-size bounded-fan-in Boolean circuits
for multiplication (\cref{app:icml-cot-circuits}).
For depth $L$, the necessary adaptive fan-in exponent $1/(2^L-1)$
and the sufficient fixed-routing exponent $1/L$ leave a gap.
The lower bound covers both monotone output orders and prescribed
permutations with linearly many interior targets whose earlier digits
avoid at least one extreme band (\cref{cor:icml-permutation}).
Unrestricted permutations, dependent sources, and stopping rules
with additional hidden state remain open. For dense softmax the error budgets concern
whole head outputs on the perturbed run. The finite calibration is
vacuous at $N=1000$ and $B/\varepsilon=1024$ under the specified
two-layer sampling bound (\cref{app:icml-dense-scope}). Relating
these budgets to specific floating-point kernels requires an
arithmetic error model.
\else
\paragraph{Explicit discrepancy.} The bounds $D=1/512$ and $D=1/16$ have the respective ranges in \cref{thm:main}. No uniform $D$ exceeds $\log_2(8/7)$ (\cref{ex:small}).

\paragraph{Finite input sizes.} At $T=N$ and $n=3N-1$, the analytic one-layer bound is below $1$ from
$N=163,219,278,338,594$ for $H=1,2,3,4,8$, respectively (\cref{thm:onelayer}).
With two top-$1$ layers and one head each, success at $N=65{,}536$ is at most $0.507017$ using the
certified discrepancy or $0.518380$ using the analytic bound (\cref{thm:finite-two}).
Neither guarantee covers $64$-bit operands. For most other applications the bounds have untuned exponents
and implicit constants.

\paragraph{Scope.} The average-case results concern teacher forcing on independent uniform $N$-bit operands, binary
output most significant bit first. The next-bit bound reaches $T\le(4/3-\eta)N$ for fixed $\eta>0$ and
sufficiently large $N$. Head counts and $k$ are fixed as $n$ grows except in \cref{thm:onelayer}, the head-count trade-off of
\cref{thm:full-two}, \cref{cor:topk-three}, in which $k$ may grow, and \cref{thm:topk-depth,cor:topk-loglog,cor:softmax},
which allow $H,k$ to vary with $n$. Softmax attention is covered only with margins (\cref{cor:softmax});
averaging hard attention and softmax without margins can predict $M_T$ exactly (\cref{rem:averaging}). The
chain-of-thought bounds are worst case and polynomial at every depth\vsw{ for unique hard attention}{} (\cref{thm:cot-worst}), and average case at every
fixed depth (\cref{thm:cot-avg}). Polynomial-size constant-depth circuits are open from depth three on
(\cref{sec:ac0}).\ificml\else\label{mainend}\fi

\fi

\label{mainend}

\noindent\begin{minipage}{\columnwidth}\section*{\ificml AI Disclosure Statement\else AI Disclosure\fi}
AI tools, including GPT-6.2 Astra (OpenAI) and Claude (Anthropic), assisted with mathematical reasoning and proofs, code, and manuscript preparation. The authors chose the research questions and directions, checked and revised all arguments, and are responsible for all claims. The supplement reproduces the computational checks.
\end{minipage}\par

\section*{Impact Statement}
\icmlimpact

\bibliography{refs,refs-poly,refs-topk,refs-ac0,refs-cot,refs-precision,refs-scan}
\bibliographystyle{icml2027}

\newpage
\appendix
\ExplSyntaxOn
\renewcommand{\thesection}{\int_to_Alph:n{\value{section}}}
\ExplSyntaxOff
\onecolumn
\crefalias{section}{appendix}
\sloppy
\section{Complete proofs of the main results}
\label[appendix]{app:icml-core}
\paragraph{Reading order.}
\begin{enumerate}
\item For Conjecture C.5, read the two-digit Fourier contraction
and integer-grid transfer in \cref{app:finite-fourier}. For prediction
throughout the interior, read the sparse-mask estimates in
\cref{app:icml-interior}.
\item For the task-independent transfer, start with the conjunction
approximation (\cref{lem:nor-small}), then the position-union lemma and
the last-character moment (\cref{lem:position-union,lem:icml-character-moment}).
Combining these with either arithmetic estimate gives next-symbol
hardness at its qualifying positions.
\item For exact or edit-tolerant generation, add the replay code and
alignment count (\cref{lem:marked-position-code,eq:icml-general-alignment}).
The finite generation certificates use the one-layer probe counts
in \cref{app:midpoint-diagnostic,app:decision-tree-edits}.
\ificmlextended
Additional rows in this archival version use the Mellin bounds
mapped in that section. The all-mask interior extension uses
Bourgain's Type II inequality.
\else
The bounded-selection generation proofs are included here.
The dense-mask Type II argument appears in the extended version.
\fi
\item The linear-vocabulary upper bound uses the limb FFT
(\cref{app:icml-cot-packed}). Only the smaller-vocabulary numerical
upper bounds use Harvey's circuit accounting through the bit-batching
simulation (\cref{cor:icml-vocab,app:icml-cot-circuits}).
\item For dense softmax, the sampling and transcript comparisons
reduce to the hard-attention transfer
(\cref{app:dense-scratchpad,app:gaussian-transcript}).
For deterministic rounding, a separate scan-memory reduction uses the
companion's balanced-prefix rank bound (\cref{app:icml-rounded-dense}).
\end{enumerate}

\subsection*{Notation}
\begin{center}
\begin{tabular}{ll}
\toprule
Symbol & Meaning \\
\midrule
$N,T,S$ & Operand length, target index, and selected product indices with $\max S=T$ \\
$\disc_N(S)$ & Maximum absolute correlation with a product of bounded operand tests \\
$n,L,H,k$ & Context length, layers, heads per layer, and keys selected per head \\
$Z,\mathbf z$ & Random genuine context and a point in the formal symbol cube \\
$A(f),\hat O_T$ & Walsh norm and signed output \\
\bottomrule
\end{tabular}
\end{center}
\subsection{A short Fourier proof through the midpoint}\label[appendix]{app:finite-fourier}

\begin{proposition}\label[proposition]{prop:finite-fourier}
For $N\ge2$ and $\max S=T\le N$,
\[
 \disc_N(S)\le\min\{2^{-T/2560},\,46\,2^{-T/40}\}.
\]
\end{proposition}

\begin{proof}
Put $m=2^N$, $q=2^T$, $Q=m^2/q$, and $e(t)=\exp(2\pi it)$.
We bound the continuum kernel $W_S(uv)$ by additive Fourier
expansion and compare it with the integer product on short portions
of the input cells.

\paragraph{Two binary digits at a time.}
Let $h_0(u)=|\cos\pi u|$, $h_1(u)=|\sin\pi u|$ and
$A_s(h)=\prod_{k=1}^s h_{\mathbf1_{k\in S}}(h/2^k)$. Write $c_h=\int_0^1 W_S(u)e(-hu)\,du$. Then $c_0=0$, since
$T\in S$. Integrating separately over the $T$-digit cells and the
offset within each cell gives, for $h\ne 0$,
\[
 |c_h|=A_T(h)\left|\frac{\sin(\pi h/q)}{\pi h/q}\right|
 \le A_T(h)\min\{1,q/(2|h|)\}.
\]
The first factor follows by summing independently over the binary
digits of the cell index. We use the two-digit grouping of
\citet[Lemma~3]{Bourgain11}, with the following explicit contraction.

Define $\kappa^2=\sqrt{2+\sqrt2}/4=\cos(\pi/8)/2$. We claim that
\[
 2^{-s}\sum_{h=0}^{2^s-1}A_s(h)\le 2\kappa^s.
\]
For $b\in\{0,1\}$ write $(P_bF)(x)=\bigl(h_b(x/2)F(x/2)+
h_b((x+1)/2)F((x+1)/2)\bigr)/2$. Splitting into preimages of doubling
expresses the dyadic average as the corresponding product of $P_b$'s applied
to $1$ at $0$. Each $P_b1$ equals
$U(x)=(\cos(\pi x/2)+\sin(\pi x/2))/2$ on $[0,1]$. The squares of the two
weights in $P_b$ sum to one, so Cauchy--Schwarz gives
\[
 P_bU(x)\le\tfrac12\sqrt{U(x/2)^2+U((x+1)/2)^2}
 =\tfrac14\sqrt{2+\sin(\pi x/2)+\cos(\pi x/2)}\le\kappa^2.
\]
Positivity gives a factor $\kappa^2$ per pair of letters. An unpaired letter
costs at most $1/\sqrt2<2\kappa$, proving the claim.
Put $\gamma=-\frac12\log_2\cos(\pi/8)$, so $(2\kappa)^T=q^{1/2-\gamma}$.
Then $1/20<\gamma<1/4$: indeed $\cos^2(\pi/8)<6/7$ and
$2(6/7)^5<1$, while $\cos(\pi/8)>1/\sqrt2$.

On positive frequencies the resulting band estimates are
\begin{align}
 \sum_{2^{s-1}\le h<2^s}|c_h|
 &\le 2\pi\,2^{s-T}(2\kappa)^s &&(1\le s\le T),\label{eq:finite-fourier-low}\\
 \sum_{q2^j\le h<q2^{j+1}} |c_h|
 &\le (2\kappa)^T &&(j\ge0).\label{eq:finite-fourier-high}
\end{align}
For $s<T$ retain the last-bit factor $|\sin(\pi h/q)|\le\pi 2^{s-T}$,
bound the intermediate factors by one, and sum $A_s$ over its period.
At $s=T$ apply the complete average directly. Each band above $q$ is
covered by $2^j$ periods of $A_T$ and pays the additional factor
$q/(2|h|)\le2^{-j-1}$. Negative frequencies have the same estimates.

\paragraph{The continuum product kernel.}
The kernel $e(huv)$ on $[0,1]^2$ has $L^2$ operator norm at most
$|h|^{-1/2}$ by Plancherel, extending the input by zero and rescaling
the output variable by $h$. Summing \eqref{eq:finite-fourier-low}
and \eqref{eq:finite-fourier-high} with this weight, including both
frequency signs, gives
\[
 \mu:=\|K_S\|_{2\to2}
 \le \sum_{h\ne0}\frac{|c_h|}{\sqrt{|h|}}
 \le\left(\frac{4\pi\sqrt{2}}{1-2^{-(1-\gamma)}}+
          \frac{2}{1-2^{-1/2}}\right)q^{-\gamma}
 <64q^{-\gamma}.
\]
Apply this bound first to the Fejér means, whose multipliers lie
in $[0,1]$. They are bounded by one and converge off the finitely
many cell boundaries, so dominated convergence applies against
every product of $L^2$ functions on $[0,1]$.

\paragraph{From short cell intervals to integer products.}
Give each integer $x$ the interval $[x/m,(x+\epsilon)/m)$ with
$0<\epsilon\le1$. For $|f|,|g|\le1$ put the values $f(x)/\epsilon$ and
$g(y)/\epsilon$ on these intervals, respectively, and zero elsewhere.
Each extended function has $L^2$ norm at most $\epsilon^{-1/2}$.
Their continuum correlation equals the expectation obtained by adding
independent uniform offsets in $[0,\epsilon)$ to the integers before
multiplying, while keeping $f(x)g(y)$ fixed. Its modulus is at most
$\mu/\epsilon$.

The offsets increase $xy$ by less than $2m\epsilon$. A boundary can
therefore be crossed only if $xy\equiv-d\bmod Q$ for some integer
$1\le d\le R=\lfloor2m\epsilon\rfloor$. The strictly positive $d$
takes account of the half-open cells when $xy$ is itself on a boundary.
For fixed $x$ put $g_x=\gcd(x,Q)$. At most $R/g_x$ choices of $d$ are
eligible, and each gives at most $1+mg_x/Q$ values of $y$ since the
solutions form one class modulo $Q/g_x$. This also holds at $x=0$,
where $g_x=Q$. Thus at most $R(m+m^2/Q)$ operand pairs can cross any
boundary. Each contributes at most two to the mismatch, giving
\[
 \disc_N(S)\le \mu/\epsilon+4\epsilon(1+q/m).
\]
Since $\mu\le1$, take $\epsilon=\sqrt{\mu/[4(1+q/m)]}$ (or take the
limit at $\mu=0$). This gives $\disc_N(S)\le4\sqrt{(1+q/m)\mu}$.
For $q\le m$, the bound is less than $46\,2^{-T/40}$, which also
implies $2^{-T/2560}$ for $T\ge256$.

\paragraph{The bounded range.}
If two columns agree on a fraction $a$ of rows, their combined
contribution, normalized per column, is at most
$\sqrt{a^2+(1-a)^2}$ by Cauchy--Schwarz applied to
$a|g(y)+g(y')|+(1-a)|g(y)-g(y')|$ and division by two.
For $2\le T\le255$, pair $m/4\le y<m/2$ with $2y$. These pairs cover
half the columns. Put $\lambda=qy/m$. In the coordinate $z=\lambda x/m$,
each full unit interval has constant first sign and opposite second signs
on its two halves, because the cell indices $2j,2j+1$ differ only in digit $T$.
Each half contains at least $\lfloor m/(2\lambda)\rfloor$ grid points.
Both agreement and disagreement therefore have fraction at least
$\lfloor\lambda\rfloor\lfloor m/(2\lambda)\rfloor/m$.
For $T\ge3$ we have $\lambda\ge2$ and $m/(2\lambda)\ge1$,
so the fraction is at least $(2/3)(1/2)/2=1/6$.
The discrepancy is at most $(1+\sqrt{26}/6)/2<13/14<2^{-1/10}$
(the last comparison follows by taking tenth powers), which suffices for $T<256$.
For $T=2$ the two arguments of the floors are at least one, giving fraction
$1/8$ and discrepancy at most $(1+\sqrt{50}/8)/2<63/64$.
For $T=1$ pair $y<m/4$ with $y+3m/4$. The low column is positive
throughout. The high column is positive for $x<m/2$ and negative for
$x\ge3m/4$. The resulting bound is at most $(1+\sqrt{10}/4)/2<63/64$.
Since $63/64<2^{-2/2560}$, this completes the bounded range.
\end{proof}

\paragraph{The weighted absolute Fourier mass.}
The same two dyadic band bounds give
\[
 \sum_{h\ne0}|c_h|\sqrt{\frac{1+\ln|h|}{|h|}}
 \ \ll\ \sqrt{1+T}\,2^{-\gamma T}.
\]
The extra square-root factor on band $s$ or $j$ is at most
$O(\sqrt{1+T})$ or $O(\sqrt{1+T+j})$, respectively, and the latter
is summable against $2^{-j/2}$.

\subsection{Finite midpoint correlations}\label[appendix]{app:midpoint-diagnostic}

\Cref{fig:midpoint-discrepancy} uses the following seven certified intervals. At $T=N$, we enumerated every prefix mask and used an integer
Walsh--Hadamard transform on the operand coordinates to compute
the exact maximum over operand parities. To bound discrepancy
over arbitrary bounded tests, let
$K_S(x,y)=\prod_{j\in S}\hat M_j(x,y)$ for $x,y\in\{0,\ldots,2^N-1\}$.
With $\disc_N(S)$ normalized by $4^N$, Cauchy--Schwarz gives
$\disc_N(S)\le 2^{-N}\|K_S\|_{2\to2}$. The matrix $K_S$ is real
symmetric, so $\|K_S\|_{2\to2}^{16}\le \operatorname{tr}(K_S^{16})$.
We evaluated the trace with integer arithmetic for every mask and
rounded the largest normalized sixteenth root upward by exact integer
comparisons.

\begin{center}
\def\midpointrow#1#2#3{#1 & $#2/4^{#1}$ & $#3$ \\}
\begin{tabular}{@{}rcc@{}}
\toprule
Operand bits $N$ & Exact maximum over parities &
Upper bound for $\max_{S:\max S=N}\disc_N(S)$ \\
\midrule
\midpointdata
\bottomrule
\end{tabular}
\end{center}

The parity column lower-bounds the discrepancy maximum because
parities are admissible bounded tests. The trace column upper-bounds
the same maximum even for complex tests. The finite calculation
covers binary midpoint masks for the listed operand lengths. The
supplement includes the code in
\texttt{midpoint-discrepancy/parity\_spectral.py}.

\paragraph{A separate finite prediction experiment.}
For the binary layout $X$, separator, $Y$, separator, and true prefix,
the unrestricted-score one-layer, one-head top-$1$ optimum at the
midpoint is $16/16$, $58/64$, $194/256$, and $659/1024$ for
$N=2,3,4,5$, respectively. These are counts of correct predictions
over all $4^N$ ordered operand pairs. To reproduce them, fix the
last query bit. The scores now order the possible
(position, raw bit) pairs, so the winner is the first true literal
in a decision list, with a default for fixed positions. Conversely
the allowed score and update functions implement any such list.
For a remaining set $D$ of examples let $m(D)$ be its majority count.
One can stop with $m(D)$ correct or test a literal, predict the
majority on the passing set and recurse on the failing set. Maximizing
over nonconstant splits and summing the optima for the two query bits
gives the counts above. The supplement's \texttt{one-head-exact}
code verifies the optima and replays the chosen heads.

\paragraph{Finite prediction bounds by shared probes.}
Take independent uniform $N$-bit operands and the indicated raw
binary layout with $2\le T\le N$, fixed tie priority, and one layer
of $H$ top-$k$ heads. Write $J=Hk$ and
$B(d,j)=\sum_{i=0}^{\min(d,j)}\binom di$. A cylinder fixes values on a
subset of the formal operand and product-prefix coordinates. Let $\delta$
be the maximum absolute expectation of the target sign times a
cylinder indicator, including cylinders that fix the final prefix bit. For a
cylinder fixing $a$ operand bits and $b$ prefix bits, its probability
is at most $\min(2^{-a},(2+b\ln 2)2^{-b})$. Thus a cylinder fixing $e$
coordinates has probability at most
\[
 u_e=\max_{a+b=e}\min(2^{-a},(2+b\ln 2)2^{-b}),
\]
where the maximum is over nonnegative integers. To prove the prefix
bound, put $m=2^N$, $K=2^{T-1}$ and $z=2^{-b}$. There are $Kz$ permitted
prefix words. Given $X=x>0$, each allows an interval of $Y$ of length
$m^2/(Kx)$. Their combined probability is at most $\min(1,zm/x)+Kz/m$.
The left Riemann sum of $\min(1,z/t)$ on $[0,1]$, with value one at zero,
is at most its integral $z(1+b\ln2)$ plus $1/m$. Now $K/m\le1/2$
and $1/m\le z/2$. The operand bound uses independence of its raw bits.

Fix the last query bit. Order each head's possible (position, token)
pairs by score and tie priority. Scan its list until $k$ present
records, or all available positions, are selected. Simulate the heads
in a fixed order and cache every raw bit discovered. Known and fixed
entries need no probes. A fresh probe asks whether the current
candidate is present and is a hit if it is. Every hit uses one
selection slot. Completed branches have at most $J$ hits and a branch
that still needs a probe has at most $J-1$. Truncate at $d$ fresh
probes, also processing all deterministic steps at the cutoff.
Prefix-free completed branch words inject into $d$-bit words with at
most $J$ ones by appending zeros. Unfinished paths have length $d$
and at most $J-1$ ones. Each path fixes the probed bits, since the
branch word determines the next probe. Restoring the query bit,
the completed paths contribute at most $2B(d,J)\delta$ to the absolute
signed correlation. The unfinished mass is at most $2B(d,J-1)u_{d+1}$.
Using zero for the signed readout on unfinished paths and dividing
the correlation bound by two gives
\begin{equation}\label{eq:icml-probe-advantage}
 \Pr[\mathrm{correct}]-\tfrac12
 \le B(d,J)\delta+B(d,J-1)u_{d+1}.
\end{equation}
If the decoder can emit a nondigit, assign that output signed value
zero. The accuracy is at most one half plus half the correlation. Independent model
randomness can be fixed before this argument.

\paragraph{Signed-cylinder estimate.}
We show
\begin{equation}\label{eq:icml-signed-cylinder}
 \delta\le 40(\lambda/\sqrt2)^{T/2},\qquad \lambda=37/30.
\end{equation}
In fact the operand cylinder can be replaced by any product of
separate bounded tests. Fix the prefix literals on $D\subseteq[T-1]$
and let $w(u)$ be their indicator times the sign of digit $T$ on
$[0,1)$, with the half-open binary cells. Put $q=2^T$.
The continuous additive Fourier coefficients satisfy $c_0=0$ and
\[
 |c_h|=\left|\frac{\sin(\pi h/q)}{\pi h/q}\right|
 |\sin(\pi h/2^T)|
 \prod_{j<T}p_j(h/2^j),\qquad
 p_j(u)=\begin{cases}
 1/2,&j\in D,\\
 |\cos(\pi u)|,&j\notin D .
 \end{cases}
\]
Indeed, integration within a cell gives the first factor. Averaging
the two possibilities for each digit gives the remaining factors.
A fixed digit contributes one term of modulus $1/2$.

For one-periodic $F$, define
$(\mathcal P_pF)(u)=p(u/2)F(u/2)+p((u+1)/2)F((u+1)/2)$.
The fixed-digit operator has supremum norm one. For $p=|\cos\pi u|$
and $M=2^r$ with $r\ge1$, repeated double angles give
\[
 (\mathcal P_p^r1)(u)
 =\frac{\sin(\pi u)}{M}\sum_{j=0}^{M-1}
     \frac1{\sin(\pi(u+j)/M)},\qquad 0<u<1 .
\]
To maximize it, expand the quotient
$\sin(M\theta)/\sin\theta=\sum_{a}e^{ia\theta}$, with
$a=-(M-1),-(M-3),\ldots,M-1$. At $\theta=\pi(u+j)/M$
the numerator is $(-1)^j\sin(\pi u)$. Thus summing
$(-1)^j e^{ia\pi(u+j)/M}$ over $j$ and pairing $a$ with $-a$ gives
\[
 (\mathcal P_p^r1)(u)=\frac1M\sum_a
   \frac{\cos(a\pi(u-1/2)/M)}{\cos(a\pi/(2M))}.
\]
The denominators are positive, so the maximum is at $u=1/2$.
Consequently $\|\mathcal P_p^r\|_{\infty\to\infty}$ is
$(2/M)\sum_{j<M/2}\csc((2j+1)\pi/(2M))$.
For $r\ge10$, the chord lower bound for sine and the integral
test bound this by $2+r\ln2\le\lambda^{r+1}$. The last
inequality holds at $r=10$, and the ratio of successive left
sides decreases and starts below $\lambda$. For $1\le r\le9$,
the exact rational sine lower bound
$(157/50)t-((22/7)t)^3/6\le\sin(\pi t)$ for $0\le t\le1/2$
gives upper bounds on $1000\|\mathcal P_p^r\|_{\infty\to\infty}$
of $1421,1861,2303,2745,3186,3628,4069,4511,4952$, respectively.
Each is below $1000\lambda^{r+1}$.

Summing the first $s$ factors over a period of length $2^s$
equals $(\mathcal P_{p_1}\cdots\mathcal P_{p_s}1)(0)$.
Each free run of length $r$ costs at most $\lambda^{r+1}$ and
the fixed digits cost one. There are at most one more free runs
than fixed digits, giving total at most $\lambda^{s+1}$. For
$2^{s-1}\le h<2^s$, $1\le s\le T$, the target sine is at most
$\pi 2^{s-T}$. Sum the first $s$ factors when $s<T$ and use two
periods of the first $T-1$ factors when $s=T$. This bounds the
sum of $|c_h|$ on the band by $2\pi 2^{s-T}\lambda^s$. Beyond $q$,
on $q2^j\le h<q2^{j+1}$, the sinc factor is at most $2^{-j-1}$
and the first $T-1$ factors cover $2^{j+1}$ periods. This
bounds the sum on each such band by $\lambda^T$.

Plancherel gives operator norm at most $|h|^{-1/2}$ for the
kernel $e^{2\pi i h u v}$ on the unit square. Apply this to
Fej\'er sums of $w$ and pass to the limit by dominated convergence.
The continuum operator with kernel $w(uv)$ therefore has norm
$\mu\le1$ and
\[
 \mu\le\sum_{h\ne0}\frac{|c_h|}{\sqrt{|h|}}
 \le\left(\frac{4\pi\sqrt2}{1-1/(\sqrt2\lambda)}
             +\frac2{1-1/\sqrt2}\right)(\lambda/\sqrt2)^T
 <50(\lambda/\sqrt2)^T .
\]
Here the two geometric sums come from the bands below $q$ and
at or above $q$, including both frequency signs.

For the integer transfer, offset each scaled operand $x/m$ uniformly
to its right by at most $\varepsilon/m$, with $0<\varepsilon\le1$.
The offset expectation against separate tests of modulus at most one
is at most $\mu/\varepsilon$, since the two continuum weighting
functions have $L^2$ norms at most $\varepsilon^{-1/2}$ each.
A change in $w$ requires crossing a product cell boundary with
unscaled spacing $Q=m^2/q$. The unscaled product increases by less
than $2m\varepsilon$. Hence $xy\equiv-d\pmod Q$ for an integer
$1\le d\le R=\lfloor2m\varepsilon\rfloor$. For fixed $x$ and
$g=\gcd(x,Q)$, at most $R/g$ values of $d$ qualify and each gives at
most $1+mg/Q$ values of $y$. The crossing fraction is at most
$R(1/m+1/Q)$, and the sign change costs at most two. The original
absolute expectation is thus at most
$\mu/\varepsilon+4\varepsilon(1+q/m)$. Minimize over
$\varepsilon$ to get $4\sqrt{(1+q/m)\mu}$, with a limit if $\mu=0$.
Because $q\le m$, this proves \eqref{eq:icml-signed-cylinder}.

The same bound with an extra factor $2^{s/4}$ holds if the cylinder
is multiplied by the parity of any $s$ strict-prefix digits. An
overlap with a fixed digit only changes a constant sign. The remaining
parity digits replace cosine factors by absolute sine factors. Their
operators have norm at most $\sqrt2$, since
$|\sin(\pi u/2)|+|\cos(\pi u/2)|\le\sqrt2$. Among the first $j$
factors, let $c,f,h$ count cosine, fixed and sine factors. There are
at most $f+h+1$ cosine runs, so the product norm is at most
$\lambda^{c+f+h+1}2^{h/2}=\lambda^{j+1}2^{h/2}$. Both band sums
thus gain at most $2^{s/2}$, yielding $2^{s/4}$ after the same
square-root grid transfer. Denote the resulting bound by
$\delta_s(T)=40(\lambda/\sqrt2)^{T/2}2^{s/4}$.

\paragraph{Exact certificates and scope.}
At $N=T=1000$, take $d=188$ in \eqref{eq:icml-probe-advantage}.
The supplement's \texttt{check\_finite\_top1\_cylinders.py} uses
$40(37/30)^{500}/2^{250}$ and the rational upper bound
$0.693147181$ on $\ln2$, certified by the degree-16 Taylor lower
bound on its exponential. With binomial sums in exact arithmetic,
it gives advantage below $3.01\cdot10^{-15}$ for $J=8$ and
$6.45\cdot10^{-6}$ for $J=16$. It also checks the trigonometric
bounds above with rational inequalities. These certificates concern
one prediction on a genuine prefix. The finite generation bound below uses a separate counting argument.

For growing $N=T$, take $d=\lceil T/5\rceil$ and $J\le T/60$.
Then $B(d,J)\le 11^J(12/11)^d\le2^{5d/12}$ by weighting the
binomial terms, and $u_{d+1}\le\sqrt{2+(d+1)\ln2}\,2^{-(d+1)/2}$
by the geometric mean bound. Using $\lambda^{16}<2^5$ in
\eqref{eq:icml-signed-cylinder}, both terms in
\eqref{eq:icml-probe-advantage} decay exponentially. Fixed routing can
gather both operands once $J\ge2N$, so the total one-layer selection
budget for constant advantage is linear in $N$.

\paragraph{Finite exact generation at a fixed auxiliary length.}
Use the layout $X$, separator, $Y$, separator, and fix one causal
one-layer, one-head top-$1$ model with $t$ designated initial
auxiliary slots and then $2N$ product slots. Its vocabulary may
be any finite set. Put $n=4N+t+2$ and
\[
 S(n,p)=\sum_{j=0}^{\min(n,p)}\binom nj,\qquad
 C_t=\sum_{j=0}^{\min(2N,t)} 2^j\binom{2N}{j}.
\]
For $1\le b<N$, the probability of generating the exact product is
at most
\begin{equation}\label{eq:icml-finite-exact-generation}
 2^{-b}\left[1+C_t\,\delta_{b-1}(N)
     \sum_{\ell=1}^b\binom b\ell 2^{2(t+\ell)}S(n,t+\ell)\right].
\end{equation}
This is a fixed-length bound for exact generation. It gives no
edit-tolerance guarantee.

To count the possible auxiliary traces, mark just the operand
positions selected while generating the $t$ auxiliary tokens. There
are at most $t$ such marks. Their positions and bits specify the
trace. In a chronological replay, every query is at a fixed separator
or an already reconstructed auxiliary token. Include all eligible
marked operand positions, fixed separators and already reconstructed
auxiliary tokens in each ranking. This includes the actual winner
without introducing any key that could beat it. Thus there are at
most $C_t$ distinct traces.

For each trace, freeze its symbols and supply the genuine product
digits. Exact generation implies correctness of the $b$ tested
readouts for the digits $N-b+1,\ldots,N$. Expand those equalities
using binary signs, mapping any nonbinary prediction to a fixed bit
to upper-bound success. The trivial term sums to $2^{-b}$ because
one trace is realized for each input. A term with $\ell>0$ tested
digits has last index $u\ge N-b+\ell$ and $\ell-1$ strict-prefix
signs. The indicator that the trace is realized, multiplied by the
predicted signs, is a bounded probe on at most $p=t+\ell$ readout
rows. To expand it exactly into cylinders, guess their union of
winning positions and assign the binary symbols of the queries
and guessed winners. This costs at most $S(n,p)$ guesses and
$2^{2p}$ assignments per guess. With these symbols fixed, test
outside keys against the proposed winners. For each outside
binary position all of its losing tests restrict just one bit.
Their conjunction is a cylinder (or empty), and the readouts
are constant on it. Requiring that the winners inside the guess
use exactly the guessed union makes the expansion exact. Apply
the signed-cylinder bound to each term. Since
$\lambda/\sqrt2<1<\lambda$,
\[
 \delta_{\ell-1}(u)\le
 \delta_{\ell-1}(N-b+\ell)
 =\lambda^{-(b-\ell)/2}\delta_{b-1}(N)
 \le\delta_{b-1}(N).
\]
Summing over the traces and tested subsets proves
\eqref{eq:icml-finite-exact-generation}. The same result holds
after averaging input-independent model or decoding randomness.

\begin{center}
\begin{tabular}{@{}rrrr@{}}
\toprule
Operand bits $N$ & Auxiliary slots $t$ & Tested digits $b$ & Exact success upper bound \\
\midrule
$1000$ & $0$ & $7$ & $0.007904$ \\
$1000$ & $1$ & $5$ & $0.03170$ \\
$1000$ & $2$ & $3$ & $0.125953$ \\
$2000$ & $2$ & $11$ & $0.000566$ \\
$5000$ & $10$ & $18$ & $0.000003950$ \\
$10000$ & $20$ & $36$ & $1.459\cdot10^{-11}$ \\
\bottomrule
\end{tabular}
\end{center}
The supplement's \texttt{check\_finite\_top1\_cylinders.py} verifies
these upper bounds by fourth powers and exact rational comparisons.
At $N=1000$, the listed $b$ uniquely minimize
\eqref{eq:icml-finite-exact-generation} for $t=0,1,2$. The last summand
of the correction, including the prefactor $2^{-b}$, increases by at
least $2^{5/4}$ per unit increase in $b$. Once that summand alone
exceeds the stated upper bound, it excludes every larger $b$.
For $t=3$, the trivial term at $b=1$ is $1/2$ and the last
summand at $b=2$ already exceeds $1/2$. At fixed $b$ the full
expression is nondecreasing in $t$, so two is the largest number
of initial auxiliary slots for which this certificate gives a
bound below $1/2$ at $N=1000$.

A fixed family with one model for each candidate length $t=0,1,2$
has probability below $0.16556$ of producing the exact product with
any permitted length, by a union bound over those three models.
The same cap holds if the choice of length depends on the operands
or on the decoder's randomness. These numerical guarantees are for
one-layer, one-head top-$1$ models in this initial-slot layout.

\icmlmidpointfigure
\ificml\subsection{Small-error Boolean polynomials}Let $\mathrm{NOR}_m(z)=\prod_{i\le m}(1-z_i)$. Only two facts about the \fi
Chebyshev polynomial $T_d$ are used: $T_d(\cos t)=\cos(dt)$, so $|T_d|\le1$ on $[-1,1]$; and
$T_d(\cosh t)=\cosh(dt)\ge e^{dt}/2$ for $t\ge0$. Also $\cosh t=1+2\sinh^2(t/2)$, and $\operatorname{asinh}$ is concave
on $[0,\infty)$ with $\operatorname{asinh}0=0$, so $\operatorname{asinh}u\ge(\operatorname{asinh}(u_0)/u_0)\,u$ for
$0\le u\le u_0$.

\begin{lemma}[plain Chebyshev]\label[lemma]{lem:nor-cheb}
For every $m\ge1$ and $0<\varepsilon\le1$ there is a multilinear polynomial $N$ in $m$ variables with \textup{(i)}
$N(0)=1$ and $|N(z)|\le\varepsilon$ for every $z\in\{0,1\}^m\setminus\{0\}$, so $|N-\mathrm{NOR}_m|\le\varepsilon$ on
the cube; \textup{(ii)} $|N|\le1$ on $\{0,1\}^m$; \textup{(iii)} $\deg N\le D_1(m,\varepsilon):=\min(m,\lceil
0.57\sqrt m\ln(2/\varepsilon)\rceil)$.
\end{lemma}

\begin{proof}
Let $d=\lceil0.57\sqrt m\ln(2/\varepsilon)\rceil$. If $d\ge m$ take $N=\mathrm{NOR}_m$ (this covers $m=1$). Otherwise
$m\ge2$; put $q(k)=T_d((m+1-2k)/(m-1))/T_d((m+1)/(m-1))$. For $1\le k\le m$ the argument of the numerator lies in
$[-1,1]$. Write $(m+1)/(m-1)=\cosh t_0$; then $\sinh(t_0/2)=1/\sqrt{m-1}\le1$, so by concavity
$t_0\ge2\operatorname{asinh}(1)/\sqrt{m-1}\ge2\operatorname{asinh}(1)/\sqrt m$ and
$T_d(\cosh t_0)\ge\frac12\exp(2\operatorname{asinh}(1)d/\sqrt m)\ge1/\varepsilon$, because
$0.57\ge1/(2\operatorname{asinh}1)=0.5672\ldots$. So $q(0)=1$ and $|q(k)|\le\varepsilon$ for $1\le k\le m$. Let $N$ be
the multilinear polynomial equal to $q(z_1+\dots+z_m)$ on the cube; its degree is at most $d$.
\end{proof}

\begin{lemma}[small-error NOR]\label[lemma]{lem:nor-small}
For every $m\ge1$ and $0<\varepsilon\le1$ there is a multilinear polynomial $N$ in $m$ variables with properties
\textup{(i)} and \textup{(ii)} of \cref{lem:nor-cheb} and
$\deg N\le D_2(m,\varepsilon):=\min\bigl(m,\lfloor3.34\sqrt{m\ln(2/\varepsilon)}\rfloor+1\bigr)$.
\end{lemma}

\begin{proof}
Put $L=\ln(2/\varepsilon)\ge\ln2$, $\lambda=\lceil L\rceil$ and $r=\lceil\sqrt{mL}/1.316\rceil$. If $\lambda\ge m$, take
$N=\mathrm{NOR}_m$; then $L>m-1$, so $m<\sqrt{mL}+1$. Let $\lambda\le m-1$. For $1\le i\le\lambda-1$ let
$d_i=\lceil(\pi/4)\sqrt{m/i}\rceil$, $\rho_i=\cos(\pi/(2d_i))$ (a root of $T_{d_i}$) and
$S_i(k)=T_{d_i}(1-(1-\rho_i)k/i)$. Then $S_i(0)=1$ and $S_i(i)=0$. Since $1-\rho_i\le\pi^2/(8d_i^2)\le2i/m$, the
argument stays in $[-1,1]$ for $0\le k\le m$, so $|S_i(k)|\le1$ there. Put
\[
q(k)=\frac{T_r((m-k)/(m-\lambda))}{T_r(m/(m-\lambda))}\prod_{i=1}^{\lambda-1}S_i(k).
\]
Then $q(0)=1$, $q(k)=0$ for the integers $1\le k\le\lambda-1$, and for $\lambda\le k\le m$ the first numerator has its
argument in $[0,1]$, so $|q(k)|\le1/T_r(m/(m-\lambda))$. Write $m/(m-\lambda)=\cosh t_1$; then
$\sinh(t_1/2)=\sqrt{\lambda/(2(m-\lambda))}\ge u:=\sqrt{\lambda/(2m)}$ with $u\le1/\sqrt2$, so by concavity
$t_1\ge2\sqrt2\operatorname{asinh}(1/\sqrt2)\,u\ge1.316\sqrt{\lambda/m}$ (as $2\operatorname{asinh}(1/\sqrt2)=
1.31695\ldots$). Hence $T_r(\cosh t_1)\ge\frac12\exp(1.316r\sqrt{\lambda/m})\ge\frac12\exp(\sqrt{mL}\sqrt{L/m})=
1/\varepsilon$, and $|q(k)|\le\varepsilon$ for all integers $1\le k\le m$. Using $\sum_{i\le t}i^{-1/2}\le2\sqrt t$,
$\deg q\le r+\sum_{i<\lambda}d_i\le\sqrt{mL}/1.316+1+(\lambda-1)+(\pi/2)\sqrt{m(\lambda-1)}$, and
$\lambda-1<L\le\sqrt{mL}$, so $\deg q<(1/1.316+1+\pi/2)\sqrt{mL}+1\le3.34\sqrt{mL}+1$. Let $N$ be the multilinear
polynomial equal to $q(|z|)$ on the cube.
\end{proof}

\noindent This is Sherstov's construction~\cite[Lemma~4.4 and Theorem~4.5]{Sherstov18}, stated there for AND with an
unspecified constant, mirrored by $k\mapsto m-k$; the proof above only makes the constants explicit. The bound
$\deg_\varepsilon(\mathrm{OR}_m)=O(\sqrt{m\log(1/\varepsilon)})$ goes back to Buhrman, Cleve, de Wolf and
Zalka~\cite{BCdWZ99}, and de Wolf~\cite[Thm.~1]{deWolf08} gives the general characterisation for symmetric functions;
those proofs go through quantum algorithms, and Sherstov's does not. Below $D(m,\varepsilon)$ stands for
$D_2(m,\varepsilon)$, $N_{m,\varepsilon}$ for the polynomial of \cref{lem:nor-small}, $N_{0,\varepsilon}=1$ and
$D(0,\varepsilon)=0$; $D$ is nondecreasing in $m$, and $D(m,\varepsilon)\le3.34\sqrt{m\ln(2/\varepsilon)}+1$.

\ificmlextended
\subsection{Additional scratchpad consequences}
\icmlfiniteonelayerstatement
\fi
\subsection{Proof of \cref{thm:cot-avg}}\label[appendix]{app:icml-cot-joint}

A literal context has fixed tokens and slots with one formal bit each.
The slots for fixed auxiliary symbols contribute no variable.
The asymptotic scratchpad proof uses only the weaker discrepancy exponent
$D=1/2560$ from \cref{prop:finite-fourier}. Put $n=5N$, $h=Hk$,
$a=1+h$, and $\delta=D\ln(2)/10$. Thus \eqref{eq:walsh} bounds the correlation of $2M_u-1$
with any function $P$ of the true context by $e^{-\delta n}A(P)$ for
$N/2\le u\le N$. All constants below may depend on $L,H,k$ and $D$.
Write $\Phi(x)$ for the binary entropy in nats at $\min(x,1/2)$, with
$\Phi(0)=0$. The number of subsets of $[n]$ of size at most $xn$ is at
most $e^{n\Phi(x)}$ for $x\ge0$. For $0<x\le1/2$ this follows by weighting
each such subset $S$ by $x^{|S|}(1-x)^{n-|S|}\ge e^{-n\Phi(x)}$.

\begin{lemma}[Position-union approximation]\label[lemma]{lem:position-union}\label[lemma]{lem:position-union-degree}
Fix $L,H,k$, a finite abelian alphabet $G$, and $\delta,\rho>0$.
There is $\eta\in(0,\min(1/10,\delta/4))$ such that, for large $n$,
every complex probe $F$ of modulus at most one on at most $\eta n$
top records of a literal context of length at most $n$ has a
product-character approximant $P$ on its formal raw coordinates with
\begin{equation}\label{eq:icml-linear-approximant}
 \|F-P\|_\infty\le e^{-\eta n}/2,\qquad
 \ln\max(1,A(P))\le\delta n/4,\qquad \deg P\le\rho n .
\end{equation}
Here $A(P)$ sums the moduli of the coefficients and degree counts
nontrivial coordinates.
\end{lemma}

\begin{proof}
\emph{Exact one-layer pullback.}
Represent a function at the current layer as a sum of terms,
each a bounded function (absolute value at most one) of at most $m$ records.
Let $B\ge1$ bound the sum of the absolute coefficients. Initially $B=1$.
For a term on a set $Q$ of queries, guess the union $S$ of all positions
selected across $Q$ and its heads. This gives at most
$\exp(n\Phi(hm/n))$ choices, since $|S|\le hm$.
The lower records in $Q\cup S$ suffice to sort the eligible positions of
$S$ for each query and head. Keep that head's prescribed number of positions
in the model's order, including its fixed tie priority. Discard the guess
if too few positions are eligible or if the union of the retained lists is
different from $S$. The resulting value of the term, set to zero for a
discarded guess, is a bounded function of these shared records. For each
position $i\notin S$, require its record to rank below the last retained
position for every head and query to which $i$ is eligible and whose list
is nonempty. This is one Boolean test of record $i$ and the shared records.
The conjunction of these tests verifies all lists. For an accepted union
the lists are exactly the global lists, so the union is unique. Conversely
the union of the true lists passes. Thus the sum over $S$ of the shared
term times its conjunction equals the original term pulled back one layer.
For a discarded guess the tests may be assigned arbitrary Boolean values.

\emph{Approximate one-layer pullback.}
At each pullback, approximate each conjunction of at
most $n$ tests by a polynomial of degree $O(\sqrt{n\log(1/\varepsilon)})$
at uniform error $\varepsilon$, the standard bound for AND
\citep{deWolf08}. The quantitative bounds use the explicit version
in \cref{lem:nor-small}. Set $\varepsilon=e^{-\lambda}$ with
\[
 \lambda=\eta n+\ln(2L)+\ln B+n\Phi(hm/n).
\]
The total added uniform error is at most $e^{-\eta n}/(2L)$. On its
Boolean test inputs the polynomial is bounded by two and has degree
$d\le c\sqrt{n\lambda}$ for an absolute $c$ and large $n$. Its Walsh
expansion has at most $\exp(n\Phi(d/n))$ characters and coefficient mass
at most $2\exp(n\Phi(d/n)/2)$ by Parseval and Cauchy--Schwarz. After
substitution each character is a bounded function of the shared records
and at most $d$ further records. Hence one can take
\[
 m'\le am+d,\qquad
 \ln B'\le \ln B+n\Phi(hm/n)+\ln2+\tfrac n2\Phi(d/n).
\]
\emph{Composition.} The exact step identifies shared records and losing
tests, and the polynomial step turns low-degree products of those tests
back into bounded record probes. Thus the only quantities passed to the
next layer are the fraction of records read per term and the log
coefficient mass per position. Bound these by $x_r$ and $y_r$,
respectively. Start $x_0=\eta$, $y_0=0$ and, for $0\le r<L$,
define
\[
 z_r=c\sqrt{2\eta+y_r+\Phi(hx_r)},\quad
 x_{r+1}=ax_r+z_r,\quad
 y_{r+1}=y_r+\Phi(hx_r)+\eta+\tfrac12\Phi(z_r).
\]
Here $z_r$ bounds the fraction of additional records used to check
the losing positions. For $n\eta\ge\ln(2L)$ the preceding bounds give $m/n\le x_r$ and
$(\ln B)/n\le y_r$ after $r$ pullbacks. Since $L$ is fixed and $\Phi$ is
continuous at zero, $x_r,y_r\to0$ as $\eta\downarrow0$ at every stage.
A bounded function of $m$ raw symbols has character mass at most
$|G|^{m/2}$ by Parseval and degree at most $m$.
Choose $\eta$ so that $x_L\le\rho$ and
$y_L+x_L\ln|G|/2\le\delta/4$. The errors sum across the layers.
\end{proof}

\topkgrowingdetails

\ifoldpullback\else
\begin{proof}[Proof of \cref{thm:growing-resource-transfer,rem:topk-explicit}]
Use the position-union expansion with total error $e^{-v}/2$, where
$v=t$ and $K=K_{\mathrm{set}}=4a\ln(4n)$. At a layer with record
budget $m$ maintain coefficient mass at most $e^{m\ln(4n)}$.
This holds initially with $m=1$. At most $(4n)^{(a-1)m}$ union
guesses per term and a conjunction error of $e^{-\lambda}$, with
$\lambda=v+\ln(2L)+am\ln(4n)$, add at most $e^{-v}/(2L)$ error.
Use a conjunction of degree $d\le c\sqrt{n\lambda}$
(\cref{lem:nor-small}). Its Boolean Fourier mass is at most
$2(n+1)^{d/2}\le(4n)^d$ for $d\ge1$; a constant approximant
can be counted with degree budget one. Substituting the tests
gives at most $am+d$ records and maintains the invariant.

For an envelope $v/K\le M\le n/K$ with $m\le M$ we have
$\lambda\le3KM$ and $am\le\sqrt{nKM}$, so the new envelope can be
$3c\sqrt{nKM}$. Starting at $M_0=v/K$, after $j$ steps this is at most
$9c^2nK(v/(nK^2))^{2^{-j}}$. Since $v/(nK^2)=(R_0K^2)^{-Q}$,
for $j<L$ this is at most $9c^2n/(R_0K)$, within the envelope range.
The recurrence is nondecreasing while $M\le n/K$.
After the final approximate pullback the degree is at most
$9c^2n/\sqrt{R_0}\le\vartheta n$. This last approximation matters:
fixing the winning records alone leaves a test at each losing position.

On the formal $q$-ary cube the entire approximant is bounded by two,
so Parseval bounds its character mass by twice the square root of the
number of characters of degree at most $\vartheta n$. This number is at most
$\exp[n(\Phi(\vartheta)+\vartheta\ln(q-1))]$ by the subset entropy bound.
The spectral hypothesis gives
correlation at most $e^{-\gamma n}$ for each of these characters
against any nontrivial target character.
Their total contribution is at most $2e^{-\gamma n/2}$.
Since $v\le\gamma n/4$, for sufficiently large $n$ the total correlation
is at most $2e^{-v}$. Apply this to the conjugate character of the
predicted digit, using zero for a nondigit, and expand digit equality.
This gives the stated success bound, including at nondigit readouts.

For multiplication the extra target digit adds at most one to the
degree of the arithmetic mask. Choose $\rho$ below the sparse-mask
cutoff, using $n=\Theta(N)$.

\emph{Dependence on $H,k$.} All the preceding estimates hold for the parameters at a single length. The condition is
equivalent to
\[
 n\ge R_0^QK_{\mathrm{set}}^{2Q-1}\bigl(1+\ln(2L)/K_{\mathrm{set}}\bigr).
\]
For fixed $L$ and sufficiently large $n$, $\ln(2L)\le K_{\mathrm{set}}$, so the right side is at most
$2R_0^Q4^{2Q-1}a^{2^L-1}(\ln(4n))^{2Q-1}$, which is $o(n)$ in the stated range with slack $\varepsilon$.
\emph{Growing depth.} Let $b=4+2\log_2a$, so $2^{h(a)}\ge2b$. For $L\le J$ and
$u\ge\max(4,8\sqrt{R_0})$ we have $L\le\log_2u$ and $K_{\mathrm{set}}\le8u^{1+\log_2a}$.
Because $64R_0\le u^2$, we have $\ln(R_0K_{\mathrm{set}}^2)\le b\ln u$ uniformly in $a$.
Also $Q\le u/(2^{h(a)+1}\ln u)\le u/(4b\ln u)$. Thus $(R_0K_{\mathrm{set}}^2)^Q\le n^{1/4}$ and
$t\ge n^{3/4}K_{\mathrm{set}}^2$. Since $K_{\mathrm{set}}\ge2$ and $K_{\mathrm{set}}\ge\ln(2L)$, this implies $t\ge K_{\mathrm{set}}+\ln(2L)$ for varying $H,k$ as well.
\end{proof}

\begin{proof}[Proof of \cref{cor:topk-loglog}]
The advantage is at most $\exp(-n^{3/4})$ for $L\le J$, by the digit-accuracy bound and $t\ge n^{3/4}K_{\mathrm{set}}^2$.
This is strictly below $n^{-p}$ once $n^{3/4}>p\ln n$.
\end{proof}

\fi

\begin{lemma}[Marked-position code]\label[lemma]{lem:marked-position-code}
Fix a causal total-length model on at most $n$ positions and $r$ specified
generated slots. Consider runs in which every other raw slot is either a
fixed symbol or a binary symbol. Let $a=1+Hk$. The number of distinct strings
at the specified slots is at most
\[
 \exp\!\left(nL\Phi(a^Lr/n)+a^Lr\ln2\right).
\]
\end{lemma}

\begin{proof}
Mark the $r$ top readouts for the designated slots. From every marked
position at layer $j$, mark its own position and every selected position
in layer $j-1$. Then $|S_j|\le a^{L-j}r$. Encode $S_0,\ldots,S_{L-1}$
and the marked raw bits at nondesignated slots. For a given marked-set
tuple, the positions of these bits are fixed. The subset entropy
bound counts the marked-set tuples and there are at most $2^{a^Lr}$
bit strings per tuple.

Decode the designated symbols in chronological order. At each marked
readout, restrict replay to positions at or before that readout.
By causality all raw marks used there are known from the code, fixed symbols,
or earlier decoded tokens. Computing on the marked eligible positions
preserves every ordered winner list because it includes all the actual
winners. The same scores and tie order apply when restricted to these
positions. Layer induction recovers the readout, completing the decoding.
\end{proof}

A two-layer replay is illustrated in \cref{sec:icml-cot}.

For the set $\mathcal A_t$ of realized length-$t$ prefix traces, $t\le\sigma n$,
\cref{lem:marked-position-code} gives
\begin{equation}\label{eq:icml-linear-traces}
 \ln |\mathcal A_t|\le n\bigl[L\Phi(a^L\sigma)+a^L\sigma\ln2\bigr].
\end{equation}

\begin{lemma}[Last-character moment]\label[lemma]{lem:icml-character-moment}
Let the task input and target symbols lie in a fixed finite abelian group.
In a literal causal context ending before target $u$, fix any inserted
symbols and supply a subsequence of the true earlier symbols. Let $F$
be a bounded complex probe of specified top records there.
If an approximant has uniform error $d$ and character coefficient
mass $D_0$, then for every nontrivial character $\nu$ at $u$ and
any character $\chi$ of the preceding target symbols,
\[
 \left|\E\big[F\,\nu(Z_u)\chi(Z_{<u})\big]\right|
 \le d+\epsilon_u D_0 ,
\]
where $\epsilon_u$ bounds the input and strict-prefix correlations at $u$
through degree $\deg P+\deg\chi$.
\end{lemma}
\begin{proof}
Extend the approximant to the full formal input and strict prefix by
ignoring omitted coordinates. Multiplication by $\chi$ permutes its
character coefficients. Expand against $\nu(Z_u)$ and add the
uniform approximation error.
\end{proof}

For a binary block, write $g_i$ for the signed readouts on the
literal prefixes. If $|W|\le1$ and $W\prod_{i\in J}g_i$ is a real probe of at most
$\eta n$ top records no later than $u=\max J$, then
\cref{lem:icml-character-moment} and \eqref{eq:icml-linear-approximant} give
\begin{equation}\label{eq:icml-common-moment}
 \left|\E\left[W\prod_{i\in J}\widehat M_i g_i\right]\right|
 \le e^{-\eta n}/2+\epsilon_u e^{\delta n/4}.
\end{equation}
Here $\widehat M_i$ is the true sign at position $i$. This is the
moment estimate used for the teacher-forced block and exact generation.

\paragraph{Prediction in both digit orders.}
For the fixed-resource guarantee in \cref{thm:topk-depth}, part 1 of
\cref{thm:icml-transfer} uses the two sparse-mask estimates in
\cref{app:icml-interior}. The proof uses no dense-mask estimate.
For binary MSB-first contexts, the same character argument against
the signed readout bounds its absolute correlation, allowing
non-digit outputs with sign zero.

\paragraph{Joint moments for the teacher-forced block.}
Take $b=\lfloor\eta n/4\rfloor$, $I=\{N-b+1,\ldots,N\}$, and
$s=\lfloor\sigma n\rfloor$. Choose $0<\sigma\le\eta/4$ so small that
the bracket in \eqref{eq:icml-linear-traces} is at most $\eta/4$.
For each $t\le s$, fix its total-length model. For $\tau\in\mathcal A_t$
let $W_\tau$ indicate that each of its symbols is the readout on the
corresponding prefix of $x\tau$. Exactly one $W_\tau$ equals one per input.
For $i\in I$, let $g_i$ be $+1$ if the readout on $x\tau z_1\cdots z_{i-1}$
is the bit $1$, and $-1$ otherwise. For nonempty $J\subseteq I$, set
$u=\max J$ and $F=W_\tau\prod_{i\in J}g_i$. By causality this reads at
most $t+|J|\le\eta n$ top records of the same model on the literal
context $x\tau z_1\cdots z_{u-1}$, of length at most $n$ for large $N$.
Apply \eqref{eq:icml-common-moment} with $W=W_\tau$ and
$\epsilon_u\le e^{-\delta n}$. The two terms are
$\tfrac12 e^{-\eta n}$ and at most $e^{-3\delta n/4}$.
The choice $\eta<\delta/4$ in \cref{lem:position-union}
bounds their sum by $2e^{-\eta n}$.
These are the nonempty Fourier moments used in
\cref{app:icml-cot-partial}. The edit-distance extension is proved in \cref{app:icml-cot-edit}.

\paragraph{Direct proof of the exact-generation lower bound.}
Fix $t\le s$ and use the block and literal readouts above. On a
successful free run, supplying the correct prefix gives the same
history at every tested position. Thus exact success implies
agreement of all $b$ signed readouts with the true bits, using the
realized trace in each readout. Expand its agreement indicator as
$2^{-b}\prod_{i\in I}(1+\widehat M_i g_i)$. The empty term contributes
$2^{-b}$. For each nonempty $J\subseteq I$, split the expectation
over realized traces by the indicators $W_\tau$. Each fixed-trace
term costs at most $2e^{-\eta n}$ by \eqref{eq:icml-common-moment},
as above. Since the weights of all nonempty subsets sum to less
than one, the probability of exact success is at most
\[
 2^{-b}+2|\mathcal A_t|e^{-\eta n}
 \le 2^{-b}+2e^{-3\eta n/4}.
\]
Sum over the $s+1$ candidate lengths. The sum is exponentially small
in $N$, even when the operands determine which candidate is used.
This proves the lower bound of \cref{thm:cot-avg} for an initial
scratchpad. The edit corollary additionally takes a union over
alignment masks (\cref{app:icml-cot-edit}).

\paragraph{Quantitative losses in one layer.}
In the exact proof in \cref{sec:icml-cot}, the winner-union and
assignment count uses $5N/256$ bits of the $1023N/40960$
correlation exponent. The larger contribution is the empty term
of the agreement expansion, set by the tested block size and the
union over candidate lengths.

\ificmlextended
\subsection{Finite one-layer scratchpad bounds}\label{sec:icml-finite-one-trace}

\begin{proposition}[Joint enumeration of one-layer readouts]\label[proposition]{prop:icml-finite-scratchpad}
Let $X,Y$ be independent uniform $N$-bit operands, $N\ge951$.
For each $0\le t\le s$, take one causal total-length model of
\cref{def:topk} with one layer, at most $H$ heads, and top-$k$ selection.
It generates $t$ auxiliary tokens from its own finite vocabulary before
writing the $2N$ product digits MSB first. Width and precision are arbitrary.
Let $n$ bound the context length through the readout at $M_N$ for all these
models, including fixed separators and auxiliary slots. For each $t$ choose
$1\le b_t\le N-143$ and put $I_t=\{N-b_t+1,\ldots,N\}$. Write
\[
 C_{t,j}=2^{j+\mathbf 1_{t>0}}
 \sum_{r=0}^{\min\{n,Hk(t+j)\}}\binom nr 2^r .
\]
Even if the length is chosen using the operands, full-product success has probability at most
\begin{equation}\label{eq:icml-finite-scratchpad}
 \sum_{t=0}^s 2^{-b_t}\!
 \left(1+\sum_{\emptyset\ne J\subseteq I_t}
 C_{t,|J|}\,2^{-\max J/6}\right).
\end{equation}
\end{proposition}

\begin{proof}
Fix $t$ and generate the actual auxiliary trace. For each $i\in I_t$ let
$g_i$ equal $+1$ when the model reads out token $1$ on this trace and
the true prefix through $M_{i-1}$, and $-1$ otherwise. Exact generation
implies all these signs agree with $\widehat M_i=2M_i-1$. Expand
$2^{-b_t}\prod_{i\in I_t}(1+\widehat M_i g_i)$. The empty subset contributes
$2^{-b_t}$ even when the trace depends on the operands.

For a nonempty subset $J$, put $j=|J|$ and $u=\max J$. On the cube of
operand bits and formal prefix bits $z_1,\ldots,z_{u-1}$, compute the
$t$ auxiliary readouts and the $j$ designated digit readouts.
Guess the union $S$ of all their selected positions. It has at most
$Hk(t+j)$ elements. Fix the binary raw symbols in $S$ and at the query
positions. In addition to $S$ these cost at most $j+\mathbf 1_{t>0}$ bits, since
after the first auxiliary readout each auxiliary query uses a token
already generated. Separators are fixed.

Replay each readout on $S$ restricted to eligible positions, in chronological
order, using the original scores and tie priority. This determines the
auxiliary symbols inductively and hence every list, cutoff, and readout
on the branch. Discard guesses with fewer eligible keys than the list size $\min(k,\text{eligible count})$
or whose retained lists have union different from $S$. For every remaining
eligible key outside $S$, check that it loses to the cutoff of each
applicable head and query. Each check sees one free raw bit or a known
symbol, since all auxiliary symbols have been replayed. The checks and
guessed bit assignments together form a conjunction of literals or a
constant. This cylinder indicator has Walsh norm at most one. On any
input the true union and its bit assignment pass. Every passing branch
reproduces the true lists by causal induction, so it is unique.
Thus $\prod_{i\in J}g_i$ is a signed sum of at most $C_{t,j}$ such
indicators. This counts positions and bits regardless of the vocabulary.

Multiplying by $\prod_{i\in J\setminus\{u\}}(2z_i-1)$ preserves Walsh
norm. Set $z_i=M_i$ and apply \eqref{eq:walsh} at $u$. The certified
range $N\ge951$, $144\le u\le N$ gives correlation at most
$C_{t,j}2^{-u/6}$ in absolute value (\cref{app:explicit-d,app:rates}).
Summing the expansion and then taking the union over lengths proves
\eqref{eq:icml-finite-scratchpad}.
\end{proof}

\begin{table}[ht]
\centering
\caption{Upper bounds on exact full-product success for one top-$1$ head
in one layer, $N\ge1024$ and $n\le4N$. Each row allows any input-selected
length $0\le t\le s$. The tested block has $b_t=15-t$ digits for $t\le10$
and $b_t=16-t$ for $11\le t\le14$. Percentages are rounded upward.}
\label{tab:icml-finite-scratchpad}
\small
\begin{tabular}{@{}lrrrrrrrrr@{}}
\toprule
Budget $s$ & 0 & 1 & 2 & 3 & 5 & 8 & 10 & 12 & 14 \\
Digits $b_s$ & 15 & 14 & 13 & 12 & 10 & 7 & 5 & 4 & 2 \\
Success (\%) & .004653 & .013954 & .029356 & .056962 & .209822 & 1.586551 & 6.280397 & 18.891128 & 59.620526 \\
\bottomrule
\end{tabular}
\end{table}

To evaluate \cref{tab:icml-finite-scratchpad}, group subsets with
$\max J=N-b_t+1+\ell$ and $|J|=j$. This gives
$\binom{\ell}{j-1}$ such subsets. Evaluate at $N=1024$, $n=4096$.
The companion script \texttt{check\_finite\_scratchpad.py} encloses
the six residues of $2^{-u/6}$ by rational sixth-power comparisons.
For $N\ge1024$ the terms can only decrease. Increasing $N$ by one
multiplies each $\binom{4N}{r}$ by at most
$(1+4/(4096-15))^{16}<2^{1/6}$, since $r\le16$, while $2^{-u/6}$
shrinks by $2^{-1/6}$.

\begin{proof}[Proof of \cref{cor:icml-one-layer-linear}]
Take $b=\lfloor N/(4096h)\rfloor$ at every length, so
$d=h(t+j)\le65N/4096$. Weighting terms by $512^{-r}\ge512^{-d}$ gives
\[
 \sum_{r\le d}\binom nr 2^r\le 512^d(257/256)^{4N}.
\]
Since $(257/256)^{512}<8$, the base-two logarithm of each nonempty
moment bound $C_{t,j}2^{-u/6}$ is at most
\[
 1+\frac{N}{4096}+\frac{9\cdot65N}{4096}
 +\frac{12N}{512}-\frac{4095N}{6\cdot4096}
 =1-\frac{N}{8192}.
\]
The empty term costs $2^{-b}\le2^{1-N/(4096h)}$ per length.
Substitution into \eqref{eq:icml-finite-scratchpad} proves the claim.
\end{proof}

In particular, one top-$1$ head with up to $\lfloor N/64\rfloor$
auxiliary tokens succeeds with probability at most $4(s+1)2^{-N/8192}$.

\fi

\icmlcotpartialproof
\subsection{Free-running edit distance}\label[appendix]{app:icml-cot-edit}

Let the task alphabet be a fixed abelian group of order $q\ge2$.
Suppose $\epsilon_u\le \epsilon$ at a set $I$ of $b$ target
indices, and that any bounded probe of at most $b+r$ top records
has uniform error $d$, character mass $D_0$ and degree at most $D_1$
after approximation. Correlations are only needed at degree at most $D_1+b-1$
in the context coordinates.
Let $n$ bound the longest context considered and $a=1+Hk$. Define
\[
 S(n,z)=\sum_{j=0}^{\min(n,\lfloor z\rfloor)}\binom nj,\qquad
 C_r=q^{\min(n,a^Lr)}\prod_{j=1}^L S(n,a^jr).
\]
A radius-$r$ alignment deletes at most $r$ generated tokens and
at most $r$ target symbols. The probability that any length model
admits such an alignment is at most
\begin{equation}\label{eq:icml-general-alignment}
 (2r+1)S(n,r)^2
 \bigl[q^{-(b-r)_+}+C_r(d+\epsilon D_0)\bigr].
\end{equation}

Fix a raw length $m$ and deletion masks $U$ on raw slots and $V$ on
target slots, pairing their complements in order. In a causal replay,
generate at raw slots in $U$ and supply the paired true symbol
elsewhere. If the free run realizes the alignment, its history agrees
with this replay by induction. Put $I'=I\setminus V$ and expand the
prediction equalities at $I'$ in the $q$ characters, decoding a
nongroup prediction to a fixed group symbol for an upper bound.
The all-trivial term contributes $q^{-|I'|}$ on the replay.

In any other term let $u$ be the last index with a nontrivial
character. Split the expectation only over the generated deleted
symbols before the raw slot for $u$. There are at most $C_r$ such
lists by the marked-position code of \cref{lem:marked-position-code}
with $q$ possibilities at each varying nondesignated raw mark.
The same code applies with externally supplied retained symbols:
the replay needs only their values on the marked positions.
For a fixed list, its realization indicator tests the readouts at
its deleted slots. Multiply this indicator by the readout character
factors. This is a bounded probe of at most $r+b$ top records on
one literal context, by causality. The other target factors form
a strict-prefix character of degree at most $b-1$ at $u$, so
\cref{lem:icml-character-moment} bounds this list's contribution by
$d+\epsilon D_0$. The nontrivial terms have total coefficient
mass at most one. Summing over lists charges $C_r$ only to this
error term. Finally $|I'|\ge (b-r)_+$, and there are at most
$2r+1$ lengths and $S(n,r)^2$ mask pairs per length. This gives
\eqref{eq:icml-general-alignment}.

For fixed resources and $\epsilon_t\le e^{-\gamma N}$ at
$\beta N$ indices, choose $n=\Theta(N)$ with slack for all lengths
under consideration. The pullback of \cref{lem:position-union} gives exponentially small $d$,
$D_0\le e^{\gamma N/2}$ and $D_1\le\rho N/2$ simultaneously.
Take $b$ small enough that $b\le\rho N/2$ and the probe budget holds, then $r=\lfloor cN\rfloor$ sufficiently smaller
than $b$. Subset entropy makes the exponential rates of the mask
and list counts arbitrarily small as $c$ decreases. Both terms
in \eqref{eq:icml-general-alignment} are then exponentially small.
This proves \cref{thm:icml-scratchpad-transfer}, with exact
generation after $t$ auxiliary tokens obtained from $U=[t]$ and
$V=\emptyset$. Deleting both sides of every substitution gives
the Levenshtein formulation. For \cref{thm:cot-avg,cor:icml-generation-edits}, use the
binary discrepancy estimate at a linear block ending at the
midpoint. The same framework applies even when the hard indices
are scattered.

\paragraph{Endogenous stopping.}
An admissible causal generator that stops from its current history is
covered if its output before stopping is a prefix of an admissible
length-model run, with the same decoding tape. In particular, for a
single causal token generator with local STOP decisions, use the
same generation rule up to each candidate length, ignoring STOP at
or after that length and defining arbitrary continuations on histories
that have already stopped. On any input that stops at that candidate
length, the two runs agree up to it by causality. Its success is thus
one event in the length union already bounded. A stopping mechanism
that also passes additional state to the token generator needs an
admissible simulation of that state.

\paragraph{Randomized decoding and external advice.}
An independent random tape may select the symbol from the current readout
distribution at each step, including temperature sampling. For each
fixed tape these choices are deterministic functions of the current
records and position. The estimates are uniform in that tape, so averaging
retains the bound. The same argument permits input-independent randomness
in the model under the stated routing and layer bounds after fixing the
tape. An external $r$-bit message may select among at most $2^r$ model
families, even using the operands. Sum the near-product success probabilities over all
messages to bound near-product success by $2^r e^{-c_1N}$. This remains
exponentially small for $r\le c_1N/(2\ln2)$. In the single-digit problem,
one bit of advice can specify the target.

\paragraph{An analytic certificate in one layer.}
For binary MSB-first generation at $L=H=k=1$ the master bound gives success at most
$2^{-N/32768}$ with deletion budget $R=\lfloor N/32768\rfloor$ per
side for $N\ge32768$, provided every candidate fits within $n=5N$
context rows including separators.
Set $v=N/32768$, $r=1/32768$, $b_*=1/736$, and test the last
$b=\lfloor b_*N\rfloor$ targets through the midpoint. For a
nontrivial term, guess the winner union for at most $b+R$
readouts. Enumerate the bits in the union and at the queries,
then keep the exact conjunction of raw losing tests.
Each assignment contributes Walsh norm at most one, so the
total is at most $2^{2(b+R)}S(5N,b+R)\le2^{16(b_*+r)N}$.
The last selected target has index at least $(1-b_*)N$.
The bound $46\cdot2^{-u/40}$ in \cref{prop:finite-fourier}
therefore bounds each signed moment by
$2^{6+16(b_*+r)N-(1-b_*)N/40}$.

For each mask, $S(5N,R)\le2^{19v}$ and
the marked-list count is $C_R\le 2^{2R}S(5N,2R)\le2^{38v}$,
using $S(n,j)\le(en/j)^j$ and
$\log_2(5e/(2r))<18$. There are at most $2R+1\le2^{2v}$ lengths.
In the trivial term the marked-list count is absent.
Substituting into \eqref{eq:icml-general-alignment} gives
\[
  \Pr[\text{success}]
     \le 2^{1-(81/23)v}+2^{6-(270/23)v}
     \le 2^{-v},\qquad v\ge1 .
\]
For these parameter choices, concatenating the length, mask, and replay
counts charges at most $78v$ bits to the nonconstant term, in addition
to the $16(b_*+r)N$ bits from winner verification. The first term of
the displayed bound is larger for $v\ge1$. At fixed $b$ it still
pays for the length and mask counts if the discrepancy term improves.

The closed form first gives a nonzero budget at $N=32768$, where
one deletion per string gives success at most $1/2$. At $N=327680$ it
gives ten deletions per string with success at most $2^{-10}$.
The specialized decision-tree calculation certifies particular
smaller instances by verified integer estimates
(\cref{app:decision-tree-edits}).

\ificmlextended
The growing-resource argument permits $k$ to grow when
$2^L\ln((1+Hk)\ln(2n))=o(\ln n)$. Its sufficient record fractions
can be of order $(C(1+Hk)\ln^2(2n))^{-4\cdot2^L}$.
\fi

% Drop-in appendix section. Uses amsmath, amssymb, and booktabs.
\subsection{Finite edit bounds by sharing decision-tree probes}
\label{app:decision-tree-edits}
Consider independent uniform $N$-bit operands supplied one bit per slot,
fixed separators, an MSB-first binary product, and one causal top-$k$
layer with $H$ heads. Fix the model for each candidate total length. An
alignment deletes at most $r$ generated tokens and $r$ product bits.
Write $B(p,h)=\sum_{0\le j\le\min(p,h)}\binom pj$, taking an empty
sum as zero. Let $\delta(u)$ bound the absolute correlation of the
$u$th target sign against every operand and genuine-prefix parity.
For $p\ge1$ define
\[
 U(p)=\max_{a+b=p}\min\{2^{-a},(2+0.7b)2^{-b}\},\qquad
 Q(u,m)=\min\left\{1,\inf_{p\ge1}2^m
   \big[B(p,Hkm)\delta(u)+B(p,Hkm-1)U(p)\big]\right\},
\]
where the maxima and infima here are over nonnegative or positive integers.

\paragraph{The replay moment.}
Fix a length and the two deletion masks. Supply the paired genuine bits at
retained raw slots and generate the other slots chronologically. On any
witness this replay equals the free run. For a nonempty set $J$ of
retained targets put $u=\max J\le N$. Let $Z_j$ be the product of the true
sign and the predicted sign at $j$, mapping a nonbit to a fixed bit for
the test. If at most $i$ deleted raw slots precede $u$'s readout, then
\[
               \left|\mathbb E\prod_{j\in J}Z_j\right|
                      \le Q(u,i+|J|).
\]
Indeed, simulate the earlier deleted-slot readouts and those in $J$ in
chronological order. These are at most $m=i+|J|$ prescribed rows.
Guess all distinct ordinary query bits first, in at most $2^m$ ways.
Previous generated tokens are available in full once their rows have been
processed. At each head sort every eligible (position, possible token)
pair by score and fixed priority and scan until the prescribed number of
records has been selected. Fixed tokens and known raw bits cost no probes.
For a fresh binary coordinate, probe whether its candidate token occurs
and retain the coordinate's value for all subsequent rows. A hit fills one
head record immediately; at most $Hkm$ fresh probes hit. Deterministic
operations are exhausted before another probe is requested.

Cut the binary probe tree at depth $p$. For each fixed query assignment,
completed branches inject into length-$p$ words with at most $Hkm$ ones
by padding with zeros; terminal words are prefix-free. An unfinished
branch has exactly $p$ fresh probes and at most $Hkm-1$ hits, since a
record is still missing. Each branch is a conjunction of ordinary
operand or strict-prefix literals. At a completed leaf the predicted
signs are fixed. The cylinder indicator has Walsh coefficient mass at
most one, and multiplication by the earlier true signs preserves this
mass. Its signed contribution is therefore at most $\delta(u)$.

An unfinished leaf specifies $a$ operand and $b$ prefix bits among its
$p$ fresh probes. Its genuine probability is at most $2^{-a}$. To bound
the prefix probability, write $m_0=2^N$, $K=2^{u-1}$ and $z=2^{-b}$.
There are $Kz$ allowed prefix words. Conditional on $X=x>0$, each word
restricts $Y$ to an interval of length $m_0^2/(Kx)$, giving total
conditional probability at most $\min(1,zm_0/x)+Kz/m_0$.
At $x=0$ use one. The left Riemann sum of $\min(1,z/t)$ exceeds its
integral $z(1+b\ln2)$ by at most $1/m_0$. As $K/m_0\le1/2$ and
$1/m_0\le z/2$, the resulting bound is $z(2+b\ln2)\le z(2+0.7b)$.
Hence each unfinished leaf costs at most $U(p)$. Summing both types of
leaves proves the moment estimate. The number of fixed separator slots
and the vocabulary do not enter this calculation.

\paragraph{From signed moments to whole-string edits.}
Take $r<b\le w\le N$ and $1\le t\le b-r$. For $0\le i,j\le r$ put
$d=b-j$, $D=B(d,t)$, and
\[
 A_\ell(d,t)=\sum_{a=0}^{\ell}\binom{\ell}{a}
          B(d-\ell,t-\max(a,\ell-a)).
\]
The success probability at any input-selected permitted length is at most
\begin{align*}
 \sum_{i,j=0}^r &(2r-i-j+1)\binom{w+i-j}{i}\binom wj\left[\frac1D+
  \frac1{D^2}\sum_{\ell=1}^{\min(2t,d)}A_\ell(d,t)
  \sum_{v=\ell}^d\binom{v-1}{\ell-1}
     \max_{0\le s\le j}Q(w-b+v+s,i+\ell)\right].
\end{align*}
To see this, clip a witnessing alignment at its last retained raw slot
paired to a target among the first $w$. Include all $j$ target deletions
up to $w$. The clipped raw prefix has $w+i-j$ slots when it contains $i$
raw deletions. The remaining budgets allow raw length minus $2N$ from
$i-r$ through $r-j$. The factors outside the brackets thus count masks
and predetermined models. Among the last $b$ targets through $w$, test
the earliest $d$ survivors. The $v$th tested coordinate lies between
$w-b+v$ and $w-b+v+j$. Use the square
\[
 \left(D^{-1}\sum_{S\subseteq[d],\ |S|\le t}\prod_{v\in S}Z_v\right)^2
\]
as a pointwise upper bound for the all-correct indicator. Its constant
coefficient is $1/D$. A support of size $\ell$ has coefficient
$A_\ell(d,t)/D^2$: splitting the symmetric difference between two
subsets leaves their common part free outside the support. There are
$\binom{v-1}{\ell-1}$ supports with largest ordinal $v$. The stated
bound follows from the replay moment and the union over clipped masks.

\begin{center}
\begin{tabular}{rrrrrll}
\toprule
$N$ & $r$ & $w$ & $b$ & $t$ & arithmetic bound & success at most\\
\midrule
\ificmlextended 384  & 1 & 361  & 134 & 3  & analytic transfer & $0.444911$\\\fi
\ificmlextended 512  & 1 & 471  & 130 & 5  & analytic transfer & $0.002564$\\\fi
490  & 1 & 423  & 144 & 3  & elementary cylinders & $0.495552$\\
512  & 1 & 444  & 163 & 3  & elementary cylinders & $0.334588$\\
640  & 1 & 537  & 183 & 4  & elementary cylinders & $0.008231$\\
\ificmlextended 768  & 2 & 706  & 244 & 7  & analytic transfer & $0.009456$\\\fi
\ificmlextended 1000 & 1 & 1000 & 275 & 8  & $2^{-u/6}$ & $2.662\cdot10^{-9}$\\\fi
\ificmlextended 1024 & 2 & 942  & 323 & 10 & analytic transfer & $1.279\cdot10^{-7}$\\\fi
1024 & 2 & 874  & 253 & 7  & elementary cylinders & $0.016752$\\
4096 & 1 & 4096 & 817 & 4  & $46\,2^{-u/40}$ & $0.001940$\\
\bottomrule
\end{tabular}
\end{center}
All rows use $H=k=1$ and allow arbitrary length-specific models, width,
precision and finite vocabulary.
\ificmlextended
For the analytic rows use, at $48\le
u\le N$, the smaller of the separate grid and profiled transfer bounds
\begin{align*}
 &3982\,2^{-u/4}+3\,2^{u-N}+(8+1.39N)2^{-N},\\
 &758.72(37.64+24.66N)2^{-(3N-2u)/5}
       +38101\,2^{-u/4}+(8+3.8N)2^{-N}.
\end{align*}
The first combines the Mellin symbol estimate with the
continuum-to-integers grid transfer. The second uses the profiled
transfer. See \cref{thm:symbol,thm:transfer,thm:rates} and the
proof of the sharper exponents for the $2^{-u/6}$ row.
\fi
The last row uses the finite Fourier bound in Appendix~A.1.

For the elementary cylinder rows, a completed leaf is a signed prefix
cylinder times separate bounded operand tests. Let $s\le m-1$ be its
number of earlier parity signs and $\theta=37/(30\sqrt2)$. The elementary
signed-cylinder proof in \cref{app:midpoint-diagnostic} gives
continuum operator norm $\mu\le50\theta^u2^{s/2}$ and discrete
correlation at most $40\theta^{u/2}2^{s/4}$. A second discrete bound is
$\mu+8\,2^{u-N}$. Indeed extend each operand constantly on its full
integer grid cell, put $M=2^N$ and $Q=M^2/2^u$, and call a cell bad if
its open product interval contains a positive multiple $hQ$. Row $x=0$
has no bad cells since $M\le Q$. For $1\le x<M$ there are at most
$(x+1)M/Q$ potentially crossed multiples. For each, the eligible $y$
lie between $hQ/(x+1)-1$ and $hQ/x$, hence number at most $M/x+2$.
Their product is at most $4M^2/Q$ since $(x+1)/x\le2$ and $x+1\le M$.
The bad fraction is thus at most $4\,2^{u-N}$ and each cell costs at
most two. The checker substitutes the smaller discrete bound with $s=m-1$
for $\delta(u)$ in $Q(u,m)$ on these rows; only completed leaves incur it.
The checker
\texttt{check\_decision\_tree\_edits.py} uses integer upward rounding
at precision $2^{-320}$ and exact integer-power comparisons for every
fractional exponent. Each evaluation uses finitely many choices of $p$.

\subsection{Linear-length multiplication by a packed FFT}
\label[appendix]{app:icml-cot-packed}

\begin{proof}
We implement the packed complex FFT of \citet[\S4.1]{Furer2014}
using the radix-two schedule and causal routing of \citet[App.~C.2.3]{AHRH25}. For $N\ge2$ set
$b=\lfloor\log_2 N\rfloor$, $B=2^b$, $m=\lceil N/b\rceil$, and let $M$ be the least power of
two at least $2m$. Write $X=\sum_{j<m}x_jB^j$ and $Y=\sum_{j<m}y_jB^j$. For each digit,
starting from zero, append its operand bits from high to low by the update $v\gets2v+u$ on a
previous word $v$ and a raw bit $u$. Fixed addresses handle any prescribed input order. This
costs $2N$ gates in all, including the shorter leading digits.

Use normalized radix-two transforms, halving at every butterfly.
Write $h=\log_2 M$. Let $\Lambda$ be the least power of two at least
$32M^2(h+1)(B+1)^2$, and $\Delta=1/\Lambda$.
Let $R_\Delta$ round the two complex coordinates to the grid of spacing
$\Delta$, with fixed ties. For the exact twiddle $w$, use a dyadic
$\tau$ with $|\tau|\le1$ and $|\tau-w|\le2\Delta$, constructed below.
Each butterfly emits $R_\Delta((u\pm\tau v)/2)$. Start from exact inputs
of modulus at most $A$. The exact values have modulus at most $A$
at every level. A level with input error at most $e$ has output error
at most $e+\Delta(A+1)$ by the triangle inequality. Thus each forward
transform on packed digits has error $F=h\Delta(B+1)\le1$.
Pointwise multiplication followed by grid rounding costs error at
most $(2B+1)F+\Delta$, and the exact products have modulus at most $B^2$.
Apply the normalized opposite-sign transform to those products. Its
error is at most
\[
 (2B+1)F+\Delta+h\Delta(B^2+1)
 \le4(h+1)\Delta(B+1)^2\le1/(8M^2).
\]
The exact result is $c_j/M^2$ at position $j$: the two normalized
forward transforms supply $1/M^2$, and the normalized opposite-sign
transform is the inverse of the unnormalized forward transform.
Root orthogonality gives cyclic convolution, which is ordinary convolution
because $2m-2<M$. Multiply the real part by $M^2$ and round to the nearest
integer to recover $c_j$ exactly.

To construct the twiddles, let $P=16M^2\Lambda$ and round toward zero
at spacing $1/P$ throughout. Starting from $i$ for length four, repeatedly
normalize $1+z$ to unit modulus and truncate to obtain the root for the
next doubled length. Normalization here is $2$-Lipschitz in the first
quadrant since $|1+z|\ge1$. Truncation preserves modulus at most one
and contributes at most $2/P$ error per step. Thus the computed primitive
root is within $2M/P$ of the true root. If $z=(a+ib)/P$, each numerator
of the next root is the largest integer $j\le P$ with
$j^2((P+a)^2+b^2)\le P^2t^2$, for $t=P+a$ or $t=b$. Binary search finds
it with integer comparisons. Form powers by repeated grid multiplication
and truncation. Each step adds at most $(2M+2)/P$ error, so all powers
are within $4M^2/P$. Truncating to spacing $\Delta$ adds at most
$\sqrt2\Delta$ and keeps modulus at most one. Conjugation gives the
other sign. This proves the twiddle claims.

The radix-two split into even and odd indices uses $M$ instructions
per level, one for each output of a butterfly. The three transforms
and rounding steps take $O(M\log M)=O(N)$ word instructions. Each grid
word has two log-bit numerators, because all computed moduli are at
most $B^2+2$ and $\Lambda$ is polynomial in $N$. Complex multiplications
and grid rounding use integer intermediates of $O(\log N)$ bits.

Starting with $h_0=0$, compute one pair $(d_j,h_{j+1})$ per $j=0,\ldots,2m-1$ by dividing
$c_j+h_j$ by $B$, with remainder $d_j$ and quotient $h_{j+1}$ (here $c_{2m-1}=0$).
Induction gives $0\le h_j\le m(B-1)$. Multiplying the division identities by $B^j$ and
summing gives $XY=\sum_{j<2m}d_jB^j+h_{2m}B^{2m}$, hence $h_{2m}=0$ since $XY<B^{2m}$.
For $r=2N-1,\ldots,0$ read the stored digit $d_{\lfloor r/b\rfloor}$ and emit its bit at
offset $r\bmod b$. This omits the extra leading zeros from the digit padding.

It remains to implement the word gates autoregressively. The readout at the final input
position first emits a zero word. Append a fixed spacer after this word and after each
gate result. To compute $f(u,v)$, the readout at the spacer attends to $u$ and emits a
staged copy of $u$. The next readout has this copy in its residual, attends to $v$, and emits
$f(u,v)$. All operands occupy positions strictly before their requesting readouts. A unary
gate ignores one argument. At a bit readout the head reads the scheduled digit record.
Zero and spacer instructions may attend to the first input position and ignore its value.
The fixed schedule uses at most three auxiliary symbols per gate, and every word, including
a carry pair or a staged word, has $O(\log(N+1))$ bits. Fixed type tags preserve this bound.

For the request at position $i$ with target $a_i$, use positional query $(2a_i,1)$ and key
$(j,-j^2)$, whose dot product uniquely selects $j=a_i$. Positional and symbol coordinates
occupy disjoint banks in the additive embedding. The value map copies the selected symbol
to a fresh bank, so the shared pointwise map sees $i$, the residual symbol and the selected
symbol. Position $i$ determines the operation and its constants. The map can write the
one-hot vector of the next symbol for linear readout. This implements every instruction
with one layer and one head. Padding the auxiliary phase with fixed spacers can make the
total length $K N$ for a sufficiently large fixed integer $K$, so each total-length model
has a single schedule. The case $N=1$ is a fixed binary gate.
\end{proof}

\paragraph{Linear vocabulary and constant hidden dimension.}
Refine the same fixed-address program into limb instructions before
using the tagged affine implementation in \cref{cor:icml-vocab}.
For $N\ge2$, take $w=\max(1,\lfloor(\log_2 N)/3\rfloor)$ and
$Q=2^w$. Choose an absolute constant $K$ so that all signed integer
registers, including intermediate products and rounding offsets,
fit in $Kw$ bits. This is possible since the scale $\Lambda$, the
integer numerators and their products are polynomially bounded in
$N$, and $w=\Theta(\log N)$. Represent a register by $K$ base-$Q$
limbs in two's complement. Each addition uses a fixed carry ripple:
the low limb and carry of $x+y+c$ are separate ternary instructions.
Negation complements the limbs and adds one. Schoolbook multiplication
forms each shifted row by splitting $xy+c$ into low and high limbs,
adds the rows, and drops overflow above the reserved word. This computes
modulo $Q^K$ and agrees with signed arithmetic whenever the result
fits the reserved width. All these expansions have constant length
because $K$ is fixed.

Each fixed shift uses copies for whole limbs, and combines two
adjacent limbs for the remaining shift by a positional literal of
at most $w-1$. A right shift uses sign extension. For division by
$2^s$ and nearest rounding with ties toward $+\infty$, add $2^{s-1}$
before the arithmetic right shift when $s>0$. Rounding and
rescaling therefore use only fixed shifts and additions. The
error in nearest-$\Delta$ rounding, including at a tie, is at most $\Delta/2$. Twiddle
construction is offline, and its signed dyadic numerators are
emitted as positional limb literals. Operand packing, both signs
of the butterflies, pointwise complex multiplication, coefficient
recovery and the binary carry pass now have constant-size
expansions per word instruction. A final bit extraction is a
lookup of one limb and its prescribed offset. For any fixed radix
$q$, use packed base $q^b$ with $b=\lfloor\log_q N\rfloor$ as in
\cref{app:icml-radix-edit}. During the final carry pass, divide each
nonnegative coefficient-plus-carry register $b$ times by $q$.
A high-to-low ripple returns the quotient limb and new remainder
from $cQ+x$, where $c<q$ and $x<Q$. Each division uses a constant
number of limb instructions for $Q\ge q$. The remainders give the
output digits, and the last quotient is the carry. There are
$O_q(N/b)$ registers, so these divisions also cost $O_q(N)$ in total.
Finitely many small sizes have fixed lookup programs.

The limb library has a fixed number of gates of arity at most
three, each with input and output of one limb. The positions and
literal constants are constructible in polynomial time. Applying
the tagged implementation gives $O(Q^3)=O(N)$ symbols and
$O(N)$ auxiliary tokens, scalar parameters and positional entries.
The hidden dimension is constant and all scores and parameters
are exact $O(\log(N+1))$-bit integers. Any fixed answer order can
be implemented by permuting the final digit reads because the
limb program has already computed the digits. Its size follows directly from the FFT word schedule.
The tagged implementation encodes individual limbs and their gate
arguments, so it does not store a distinct symbol for every
possible full FFT register.

\paragraph{An explicit ReLU implementation.}
One may encode positions and symbols in separate one-hot banks, with a
fresh bank for the selected value. For each instruction and each pair of
symbol arguments, one ReLU row tests the position and the required
symbol coordinates. For $r$ such tests use the sum of their indicators
minus $r-1$. Exactly one row is active on valid records and its output is
one. Its output column writes the prescribed next-symbol coordinate.
This gives a shared one-hidden-layer ReLU map with polynomially many
rows. Enumerating positions and symbol argument tuples, and evaluating
their integer gate tables, constructs all weights in deterministic
polynomial time in $N$ and a supplied input layout. For a polynomial-time
specified layout this yields a polynomial-time uniform family on unary
$N$. Both positional scores and grid registers have $O(\log N)$ bits.

\paragraph{Executable construction.}
The supplement's dyadic FFT verifier interprets the integer weights of this
construction. It materializes the full dense model at widths one and two.
Evaluating the needed weight columns on demand gives the same predictions
and covers all pairs through width six and selected pairs through width
$1024$. A signed-limb replay checks the refined arithmetic in binary,
ternary and decimal. The tagged-head replay checks the shared affine
readout and causal routing separately. A modular NTT checker also
tests the serialization schedule.

\ificmlextended
\paragraph{Constant hidden width with scalar arithmetic.}
The same prime-free schedule also has a one-head realization with
constant hidden width and exact $O(\log(N+1))$-bit integers. Use
signed integer symbols $b\in[-R,R]$ with finitely many types and
numerical embedding $(b,b^2)$. A staged symbol holds the left
operand $x$ in the residual, and the scheduled head reads the
right operand $y$. The affine decoder sees $x,x^2,y,y^2$ together
with a positional offset $c$, denominator $J>0$, and type flags.
The positional queries and keys are those of the preceding
construction, and the pointwise map can be the identity.

For each of a fixed finite collection of linear forms $L$ on
$(x,x^2,y,y^2)$, let $Z=L+c$. Distinct type flags select floor or
nearest rounding with the same form $L$. Candidate $b$ has the
corresponding unflagged logit
\[
 4bZ-2Jb^2-2Jb+2b \quad\text{or}\quad 4bZ-2Jb^2+b.
\]
These are affine readouts since the candidate $b$ is fixed per row.
The floor score has vertex $(2Z-J+1)/(2J)$, whose unique nearest
integer is $\lfloor Z/J\rfloor$. The second vertex is
$Z/J+1/(4J)$, whose unique nearest integer is
$\lfloor Z/J+1/2\rfloor$. Both assertions follow by writing
$Z$ as a quotient and integer remainder modulo $J$. Choose $R$
to contain all intended results. A positional type bonus exceeding
twice the maximum absolute unflagged score suppresses inactive
types. The scores are integers, so every correct argmax has gap
at least one. A copy type stages a source, and a literal type uses
$L=0$, $J=1$ and a positional offset to emit a constant.

In addition to affine operations, three instructions give
\[
 t=x+y,\qquad u=t^2-x^2,\qquad (u-y^2)/2=xy.
\]
The middle and final instructions revisit the original sources
by fixed causal addresses. Each instruction uses the spacer and
stage protocol above. A complex word occupies two scalar symbols.
For instance a butterfly with twiddle numerators $s,t$ emits the
rounded coordinates with integer numerators
$\Lambda u_1\pm(sv_1-tv_2)$ and $\Lambda u_2\pm(sv_2+tv_1)$ over
$2\Lambda$. The position table emits twiddles and other multiplication
constants as literal words. Pointwise multiplication, inverse
rescaling, and carries each use a constant number of the same
instructions. Prepare the bit at offset $j$ of a result digit $d$
as $v-2\lfloor v/2\rfloor$, where $v=\lfloor d/2^j\rfloor$.
The answer phase copies the prepared bits with output rows for
only the binary symbols active.

All interfaces of the packed program are polynomially bounded in
$N+1$. The constant-size instruction expansions therefore have
polynomially bounded intermediates as well. The positions, flags,
numeric weights and score gaps fit exact logarithmic-bit arithmetic.
The position table has $O(N)$ rows and the polynomial vocabulary
uses scalar-square embeddings and affine score rows. There is no
table indexed by instruction and operands. A polynomial-time printer
can enumerate the rows given the layout. This bound concerns
hidden dimension and emitted length. Scoring the full vocabulary
can still require polynomial work per token.
\fi

\begin{corollary}[Vocabulary-sensitive upper bound]\label[corollary]{cor:icml-vocab}
For $1\le b\le\lceil\log_2(N+1)\rceil$, a one-layer, one-head model
with input-independent routing and vocabulary size $O(2^b)$ writes the
exact product of two $N$-digit operands in any fixed radix using
\[
 T=O\!\left(N+\frac Nb\log(2+N/b)\right)
\]
auxiliary tokens. The hidden dimension is constant. Including the
position table, the model has $O(T+2^b)=O(T)$ scalar parameters of
$O(\log(N+1))$ bits each, constructible in polynomial time given the
input layout. Either output order is allowed. Constants may depend on the radix.
\end{corollary}

For binary multiplication the construction below takes a supplied
bounded-fan-in multiplier of size $S(m)$ for $m=\Theta(N/b)$.
Its token bound is $O(N+S(m))$ with the same vocabulary
$O(2^b)$ and constant hidden dimension. The size of the supplied
schedule and token table determines its $O(N+S(m)+2^b)$ parameter
count. A polynomial-time constructible circuit gives a
polynomial-time constructible transformer family. This is one
family member per length and layout. The positional scores need
$O(\log(N+S(m)+1))$ bits in this general form.

\begin{proof}
First consider binary inputs. Use logical words of
$w=\max(1,\lfloor b/3\rfloor)$ bits and put
$Q=2^w$. Each instruction applies a function from a fixed library to
at most three earlier logical words and returns one such word. We first
construct this program. We then implement each instruction with a
constant number of tokens and a vocabulary of size $O(Q^3)$.

\emph{Split and evaluate.}
Put $d=\lfloor(w+1)/2\rfloor$, $t=2d-1$, and $K=\lceil N/d\rceil$.
Thus $t\le w$ and $d=\Theta(b)$. Write the operands as $A(2^K)$ and
$B(2^K)$ with polynomials of degree less than $d$ and nonnegative
$K$-bit coefficients. Form their base-$Q$ limbs from the raw bits by
successive shift-and-add instructions in $O(N)$ steps, padding with
zeros as necessary. Evaluate $A$ and $B$ at each of $0,\ldots,t-1$ by
Horner's rule. Every value has at most
\[
 m=K+\lceil d\log_2(2d)\rceil+1=O(1+N/b)
\]
bits. Here $d=O(\log(N+1))$ gives $d^2\log(2d)=O(N)$.
Multiplying an $m$-bit value by a given evaluation point and adding a
coefficient takes $O(1+m/w)$ limb instructions. A ripple pass for a
small multiplier $j$ returns the low limb and the carry of $jx+c$ in
two instructions with the same three arguments; the carry is at most
$j$ and fits a limb. Another ripple pass adds the coefficient. All
$O(d^2)$ such operations therefore cost $O(N)$ instructions.

\emph{Multiply the evaluations in parallel.}
Pack the $i$th bit of every $A(j)$ into the single logical word
$\sum_{0\le j<t}2^j\operatorname{bit}_i A(j)$, and similarly for $B$.
Extracting individual bits from known limbs and packing them by
shift-and-add takes $O(tm)=O(N)$ instructions. Take a bounded-fan-in
Boolean circuit for multiplying two $m$-bit integers. Replace every
gate by its coordinatewise action on two $w$-bit words, ignoring
unused lanes. Thus each gate of one multiplier simultaneously evaluates
that gate in all $t$ products. Unpack the $2m$ output planes into
base-$Q$ limbs by individual bit extraction and assembly, again in
$O(N)$ instructions. This gives all $P(j)=A(j)B(j)$ for
$0\le j<t$ using the size of a \emph{single} $m$-bit multiplier plus
$O(N)$ instructions.

\emph{Interpolate at the requested argument.}
Set $C_{i,0}=P(i)$ and fill successive rows by
\[
 C_{i,r}=\frac{C_{i+1,r-1}-C_{i,r-1}}{r},\qquad 0\le i<t-r.
\]
These are integers. Indeed, expand $P(i+x)$ in the falling factorials
$(x)_j=x(x-1)\cdots(x-j+1)$, with $(x)_0=1$. This monic triangular
basis has integer coefficients in both directions. Since
$\Delta(x)_j=j(x)_{j-1}$, the coefficient of $(x)_r$ equals
$\Delta^rP(i)/r!=C_{i,r}$. Each numerator in the table has
$O(m+t)=O(m)$ bits by the binomial formula for differences.
It is nonnegative because $P$ has nonnegative monomial coefficients,
and taking a forward difference preserves that property. To divide
by a positive one-limb integer $r$, scan high to low, replacing
remainder $u$ and digit $z$ by the quotient digit and remainder of
$(Qu+z)/r$. Both outputs fit a limb since $u<r$, and two ternary
instructions compute them. Thus the table costs $O(d^2(1+m/w))=O(N)$.

Newton's formula now gives $P(x)=\sum_{r<t}C_{0,r}(x)_r$.
Starting at $V=C_{0,t-1}$, for $r=t-2,\ldots,0$ replace
$V$ by $(2^K-r)V+C_{0,r}$. The factor is nonnegative for all
sufficiently large $N$. Each step uses a fixed shift, a multiplication
by one limb, and additions or subtractions. A shift by a prescribed
number of bits takes constant work per output limb by combining
adjacent source limbs. Every intermediate has $O(tK+m+\log(t+1))=O(N)$
bits by the recurrence and the bounds on $C_{0,r}$. These $O(d)$
steps cost $O(d(1+N/w))=O(N)$. The result is $XY$; prepare its
$2N$ individual bits in the requested order in another $O(N)$ steps.
Finitely many small widths are covered by constant-size circuits.

\emph{Instruction count and uniformity.}
For any supplied $m$-bit multiplier of size $S(m)$, the binary program
has $O(N+S(m))$ instructions. We use the $O(m\log(2+m))$ Boolean
circuit size of the explicit algorithm of \citet{HarveyHoeven21}.
The published bit-time bound uses a multitape machine. For the circuit
bound we use its fixed array wiring and Boolean arithmetic, whose
gate count is accounted for by \citet[\S5]{Harvey26MultiplicationNote}.
Fixed array rearrangements are wires. The nonrecursive
cost in the circuit model comes from the fixed-point butterfly
arithmetic and the preprocessing and postprocessing of the
algorithm. Harvey's recurrence has subproblems of length
$m'=m^{1/d+o(1)}$ with total length at most $Km$ and
$O(m\log m)$ nonrecursive gate work. At fixed $d>K$, division by
$m\log m$ makes the recursive term a contraction.
The schedule and precision depend only on $m$. Generating the recursive
wiring and the rounded, fixed-precision constants takes polynomial time.
With $b=\lceil\log_2(N+1)\rceil$ this yields a linear trace and vocabulary.
The separate packed FFT above gives its linear trace through a direct
word schedule with a linear vocabulary.
At $b=\Theta(\log N)$, a supplied size bound
$O(m\log m\log\log m)$ would give $O(N\log\log N)$ tokens, and
$O(m\log m\,2^{O(\log^*m)})$ would give
$O(N\,2^{O(\log^*N)})$. This substitution uses only the
circuit-to-word reduction.
The remaining instruction library has a fixed number of functions
independent of $N,b$. For example, shift amounts within a limb and
small divisors are operand words. Low and high outputs of an addition,
subtraction, small multiplication or division are separate functions.
The tables of these functions on $w$-bit arguments are computable in
polynomial time, with arbitrary fixed values on invalid arguments.

\emph{Other fixed radices.}
For radix $q\ge2$, increase $w$ to at least
$\lceil\log_2(q+1)\rceil$, which changes the vocabulary bound by a
constant depending on $q$. Split each operand into
$h=\lceil N/\ell\rceil$ chunks of $\ell=\lceil\log_2(N+1)\rceil$
digits, padding the leading chunk. Each chunk is an integer less than
$D=q^\ell$. Encode the chunks in increasing order of significance in
separate binary slots of width
$s=\lceil\log_2(h+1)\rceil+2\lceil\ell\log_2q\rceil+1$.
The binary operands have $hs=O_q(N)$ bits. Their product has one
integer convolution coefficient $c_j$ in each $s$-bit slot, because
$0\le c_j<hD^2<2^s$. To construct each binary operand, append the
raw radix-$q$ digits to its chunk by repeated multiplication by $q$
and addition, then extract the chunk's bits and pad its slot with zeros.
The $N$ updates act on $O_q(\log(N+1))$ bits each and cost
$O_q(N(1+\log(N+1)/b))$ instructions altogether. Apply the binary
program to the resulting bit positions. Regroup its prepared bits into
the limbs for all $c_j$ at cost $O_q(N)$.

Starting with $H_0=0$, divide $c_j+H_j$ successively by $q$, taking
the $\ell$ remainders as output digits and the last quotient as
$H_{j+1}$. Include the final zero coefficient at $j=2h-1$.
The carry satisfies $0\le H_j\le h(D-1)$ by induction using
$c_j\le h(D-1)^2$. There are $O(N)$ divisions, each on
$O_q(\log(N+1))$ bits with one-limb divisor $q$. Both preprocessing
and carry propagation therefore fit the stated $T$, since
$\log(N+1)=O(\log(2+N/b))$ throughout the range of $b$.
Read the prepared $2N$ digits in either order, omitting leading
padding. All addresses are fixed in advance.

\emph{Constant-width causal implementation.}
A token carries a scalar embedding value. The value of a staging token
with index $p$ is $p$. A result token tagged by a gate $g$ of arity $a$
and an index $p<Q^a$ has as its value $g$ applied to the $a$ base-$Q$
digits of $p$, including leading zeros. The raw digit tokens have their
usual values. The tag set is fixed, so this table has $O(Q^3)=O(2^b)$
entries. Begin each gate at a spacer of value zero. At each of its
$a$ readouts, attend to the scheduled argument of value $y$. With
the previous partial index $x$ in the residual, emit the token of
index $Qx+y$ under the scheduled tag. Use a staging tag until the
last readout, then the result tag of $g$. The result is available to
every later gate by its fixed address. Emit a fresh spacer before
the next gate. All arguments precede their requesting positions.

Use the positional query $(2a_i,1)$ and key $(j,-j^2)$ to select the
prescribed address $a_i$ at position $i$. The value map writes $y$
in a fresh coordinate. Within a scheduled tag, give candidate $p$ the
affine logit $2pZ-p^2$ with $Z=Qx+y$. Its unique maximizer is the
integer $Z$. For literals use the same score with $Z$ equal to a
positional constant. For final outputs use $Z=y$ with the
output digit tag. A spacer has one candidate. Fixed positional flags and
a bonus exceeding twice the largest absolute unflagged score suppress
inactive tags. Thus the shared pointwise map can be the identity.
To make the size accounting explicit, the residual uses two scalar
word coordinates, two positional key coordinates, one address
coordinate, a constant coordinate, one literal coordinate, and one
flag per tag. Query and key dimension is two and value dimension is
one. The tag set is fixed. The initial position table schedules the
first spacer from the last input position. It uses a fixed legal
dummy address whenever the readout ignores the head. On the scheduled
reads $0\le x<Q^2$ and $0\le y<Q$, so $0\le Z<Q^3$.
Even the maximum-magnitude inactive logit is polynomial in $Q$:
the index is less than $Q^3$ and each affine target uses only a
fixed combination of $x,y$ and a literal. Positional scores are
$O((N+T)^2)$ in modulus. All weights, activations, and scores therefore
fit exact signed $O(\log(N+1))$-bit integers, and every active
integer argmax has a gap of at least one. The input and output
embeddings cost $O(2^b)$ scalars, the position table $O(N+T)$, and
the shared projections $O(1)$. We can ensure $T\ge N$ with spacers.

A direct evaluator that scores every visible key and vocabulary
entry uses $O(N+T+2^b)=O(T)$ log-bit operations per emitted symbol,
or $O(T^2)$ for the complete trace including output. A random-access
evaluator can instead use the scheduled address and the exact
argmax index $Z$ directly. Given the precomputed tables, this
takes $O(T)$ log-bit word operations for the same trace. These
evaluation counts are separate from the emitted length and
the polynomial time to construct the weights.

If models are indexed only by total length, let $\ell$ be the
scheduled total length before padding. Insert spacers to reach
$2^{\lceil\log_2(\ell+1)\rceil}+N$ total positions. This is $O(\ell)$
and its leading binary bit can be removed to recover $N$ for any
prescribed choice $b=b(N)$.
\end{proof}

\icmlsecondarytransfer
\section{Transfer over fixed finite abelian alphabets}\label[appendix]{app:icml-transfer-task}\label[appendix]{app:icml-transfer-alphabet}

A character is taken in the dual group $\widehat G$ of the fixed
alphabet $G$. Its degree counts positions carrying a nontrivial
character, regardless of its order.

\paragraph{Proof of the single-prediction part of \cref{thm:icml-transfer}.}
Use \cref{lem:position-union} with $n=\Theta(N)$, one record,
a degree cutoff of $\rho N$, and a coefficient mass budget
at most $e^{\gamma N/2}$. For each nontrivial target character
$\nu$, compose its complex conjugate with the predicted symbol
and apply \cref{lem:icml-character-moment} with trivial additional
prefix character. The error and the product of coefficient mass
and $\epsilon_t$ are both exponentially small. Character
orthogonality expands the equality indicator into one trivial
term $1/q$ and the bounded nontrivial terms. Map predictions outside
$G$ to a fixed symbol for an upper bound.
Part 2 uses the same moment bound in the alignment argument
(\cref{app:icml-cot-edit}).

\begin{corollary}[Carryless multiplication]\label[corollary]{cor:icml-carryless}
For two independent uniform $d$-bit polynomial operands, output their
product coefficients in descending degree. At fixed depth, head count,
and fan-in, the next-bit advantage with the true prefix is exponentially
small throughout any fixed-margin interior. For some $c>0$, the
probability of matching the product after at most $cd$ deletions from
each string is exponentially small.
\end{corollary}
\begin{proof}
Take two independent uniform binary polynomials of degree less than $d$
and emit the $2d-1$ coefficients of their product in descending order.
A mask with last position $T$ has character $(-1)^{x^\top B y}$, where
$B_{ij}=a_{i+j}$ for $0\le i,j<d$. The mask's lowest selected degree is
$e=2d-1-T$, so $a_e=1$ and $a_j=0$ for $j<e$. The anti-diagonal
$i+j=e$ has $r=\min(T,2d-T)$ entries. Restrict to the rows and columns it
meets and reverse the columns. This gives a triangular minor with ones
on the diagonal, hence $\operatorname{rank}_{\mathbb F_2}(B)\ge r$.
Averaging $(-1)^{x^\top B y+u^\top x+v^\top y}$ over $y$ leaves only
$x$ satisfying $B^\top x=v$, a fraction at most $2^{-r}$.
For a linear block around $T=d$, the transfer criterion gives exponential
next-bit hardness and an $\Omega(d)$ auxiliary-token lower bound for
constant exact-success probability. This is a true-prefix top-$k$ application
of the criterion. \citet[Thm.~2]{RemscrimSipser13} treat the middle output block
against $\mathrm{AC}^0$ tests.
\end{proof}

\begin{proof}[Proof of \cref{cor:icml-dual-code}]
Write $AU$ for the parity bits and let $d_\perp$ be the dual distance.
The mean of any character on the full word $(U,AU)$ is zero unless
the character mask is in $\{(A^\top v,v):v\in\mathbb F_2^p\}$.
Fix $n=\Theta(N)$ with slack for the alignment lengths and choose
$\rho>0$ so that $\rho n+1\le d_\perp/3$ for large $N$.
Expand a single approximated readout using
\cref{lem:position-union-degree}. Multiplying by the target sign
introduces a new position, so every character has nonempty support
of size below $d_\perp$. Only the uniform approximation error remains.

For free runs, repeat the alignment argument of
\cref{app:icml-cot-edit} with $b$ tested positions and $r$ deletions on
each side. Choose $b=\Theta(N)$ sufficiently small that
$b\le d_\perp/3$ and $b\le\eta n/2$. For $r\le\eta n/2$, the
probe on any fixed replay uses at most $b+r$ records.
After applying \cref{lem:position-union-degree}, each term with last
nontrivial target $u$ gains at most $b$ character positions from the
target factors. The coordinate $u$ remains in its support because the
probe only uses the strict prefix. Its degree is below $d_\perp$,
so its mean is zero. Thus \eqref{eq:icml-general-alignment} applies
with the nontrivial moment term $C_r e^{-\eta n}/2$. Take $r=\lfloor
cN\rfloor$ with $c$ small relative to $b/N$ and $\eta$. The mask and
list counts then preserve exponential decay as in the alignment proof.

\end{proof}

Such systematic codes exist. For a uniform binary matrix $A$,
each nonzero $v$ gives uniform $A^\top v$, so a union bound over
low-weight pairs $(a,v)$ costs at most
$2^{-N}\sum_{j\le\rho(N+p)}\binom{N+p}{j}$, exponentially
small for sufficiently small positive $\rho$.

\ificmlextended
\paragraph{Finite-field multiplication.}
Let $X,Y$ be independent uniform elements of $\mathbb F_{2^d}$, in a fixed
binary basis, and output the $d$ coordinates of $XY$ in any fixed order.
Here $N=2d$. Every output parity containing its current coordinate has the
form $(-1)^{\ell(XY)}$ for a nonzero binary linear functional $\ell$.
For $B_{xy}=(-1)^{\ell(xy)}$, distinct rows have inner product
$\sum_y(-1)^{\ell((x+x')y)}=0$ by character orthogonality, so
$BB^\top=2^d I$. Thus $|\E f(X)g(Y)(-1)^{\ell(XY)}|\le 2^{-d/2}$ for all
$|f|,|g|\le1$ by Cauchy--Schwarz. This gives the criterion at every
coordinate and hence both bounds above for finite-field multiplication.

\paragraph{A width-four predictor under the same correlation bound.}
Take $d=2m$ with $m$ even. Output the first $m$ coordinates $V$ of the field
product, followed by $q(V)=\sum_{j=1}^{m/2}V_{2j-1}V_{2j}\pmod 2$.
The first $m$ steps have the rectangle bound $2^{-m}$ just proved.
At the last step, for each $C\subseteq[m]$ the function
$h(v)=(-1)^{q(v)}\chi_C(v)$ has all $2^m$ normalized Walsh coefficients
of modulus $2^{-m/2}$. Indeed, each two-bit factor $(-1)^{ab}$ has four
coefficients of modulus $1/2$, and twisting by $\chi_C$ permutes the
coefficients. In the expansion of $h(V)$ the constant term costs at most
$2^{-m/2}$, and each nonconstant character has rectangle bound $2^{-m}$.
Consequently $|\E f(X)g(Y)h(V)|\le 2^{1-m/2}$ for every $|f|,|g|\le1$.
A scan predicts the last bit exactly by ignoring the operands and storing
the accumulated parity and the first bit of the current pair while reading
$V$. The scan lower bound for integer multiplication
uses additional control of finite-state prefix tests in \cref{app:icml-recurrent}.

\fi

\ificmlextended
\subsection{Further field-arithmetic examples}
For example, take $G=(\mathbb F_q,+)$ and let $K=\mathbb F_{q^d}$ have order
$Q=q^d$. Encode field elements in any fixed basis over $\mathbb F_q$ and
emit the coordinates in any fixed order. Input and output characters
take the form $\psi(ax)$ for a fixed nontrivial additive character $\psi$
of $K$ and suitable $a\in K$. A character mask containing the current
output coordinate has nonzero coefficient $b$ on the field output.
For independent uniform $X,Y\in K$, the field multiplication sum
$\E\psi(aX+cY+bXY)$ has magnitude $Q^{-1}$ by summing over $Y$.
For uniform $X$ in odd characteristic, the field squaring sum
$\E\psi(aX+bX^2)$ has magnitude $Q^{-1/2}$: after squaring the magnitude,
the shift $h$ contributes an inner average of $\psi(2bhX)$, which vanishes
except at $h=0$. For field inversion in any characteristic with $0$ mapped
to $0$, the sum vanishes if $a=0$ by bijectivity. For $a,b\ne0$, set
$S(a,b)=\sum_{x\ne0}\psi(ax+b/x)$. Orthogonality gives
$\sum_{a,b}|S(a,b)|^4=Q^2 E$, where $E$ counts pairs of ordered pairs of
nonzero elements with the same sum and reciprocal sum. A nonzero sum
and its reciprocal sum determine the product, allowing at most two
orderings. Sum zero contributes $(Q-1)^2$ quadruples. Thus
$E\le 3(Q-1)^2$. Since $S(a,b)=S(1,ab)$ for nonzero $a,b$, each such value
occurs at least $Q-1$ times in the fourth moment. Including the term at
zero, the normalized inversion sum is $O(Q^{-1/4})$.
These estimates apply to every coordinate, so each task has exponentially
small advantage and joint success under the stated short scratchpad budget.
\fi

\icmlselectionappendix
\icmldenseappendix
\begingroup
\def\pr{\mathbb P}
\def\hard{P_{\mathrm{hard}}}
\def\rob{P_{\mathrm{rob}}}
\def\eps{\varepsilon}
\def\ceil#1{\left\lceil #1\right\rceil}
\def\floor#1{\left\lfloor #1\right\rfloor}
\subsection{Scratchpad transfer for robust dense attention}\label[appendix]{app:dense-scratchpad}

A causal model has $L\ge1$ layers and at most $H\ge1$ heads per layer.
Initial rows are token- and position-local. The mask is fixed at the
specified total length and each head has a nonempty set of eligible
sources. Scores depend on the current query and eligible source records
and their positions. Each value can be reconstructed from those two
records (source-only values suffice). The update sees only the incoming
row and its head outputs. A next-token rule depends on its final row,
position and input-independent coins. Fixing its random tape must give
a deterministic local readout at each position. Public architecture
randomness is permitted subject to the same bounds on every realization.
A head's exact output at its current incoming state is its softmax mean.
The dense adversary can replace each such mean by any point of its
closed $\eps$-ball, independently subject to this constraint at every
row, head and layer, and may use its own randomness. Its choices can
depend on the input, previous emissions and the present head computation,
without access to future private decoder coins. It can choose a fresh error at each newly visited head and may
retain earlier choices as the context grows. In a recomputing
implementation, the exhibited strategy repeats a saved error if
that head and row are evaluated again. An adversary allowed to
choose arbitrarily at each recomputation can make the same choices.
A prescribed law of independent errors need not do so.

Fix $\eps>0$. For a finite family with one model for each of $F$
candidate total lengths, each having at most $n$ rows, let $E$ be the event that some candidate's free output aligns
to the true target under the stipulated tolerance. Shared decoder coins
between lengths are allowed. One strong version of robust success is
\[
 \rob(E)=\E_X\inf_{\mathcal A}
       \pr_{\mathrm{decoder},\mathcal A}(E\mid X).
\]
Here $\mathcal A$ specifies admissible perturbations in all candidate
runs. We construct one online strategy for each length, so the upper
bounds also apply to a length selected by the input, to a maximum
over per-length robust probabilities, and to deterministic pathwise
certificates. A priori fixing a different admissible strategy for every
input is unnecessary. Randomness in the exhibited strategy can also
be removed per input by averaging when the infimum allows only
deterministic strategies. If the architecture evaluates its earlier rows
again on every prefix, the strategy repeats its recorded old
perturbations. Causality and the fixed total-length model reproduce
exactly the same old rows. For a causal stopping rule in a single
fixed-horizon model, use a truncation of that model at each
candidate output length, retaining its positional encodings and
causal masks. Reuse the same clocks and decoder tape along common
prefixes. Every successful stopped output is one of these
truncations. A separately emitted stop symbol costs at most one
deletion. If the terminal length changes the earlier computation,
the argument requires the explicit family of length models.
Different error radii per layer or head can
be used, with the largest ratio of each bound to its own radius below.

\begin{lemma}[Sampling and domination]\label[lemma]{lem:dense-sampling-domination}
Suppose all head values have norm at most $B$ at every state reachable
under admissible perturbations on an actual run of each length model.
Given $0<p<1$, set
\[
 k=\max\left\{1,\ceil{(B/\eps)^2
               \max\{4,8\ln(1/p)\}}\right\},\qquad a=1+Hk.
\]
Let $U$ be a uniform upper bound on the probability of $E$ for every
hard top-one family of the same depth with $Hk$ heads and deterministic
local updates and readouts. Then
\[
 \rob(E)\le (1-p)^{-FnLH} U,\qquad
 \rob(E)\le F(1-p)^{-nLH} U.                         \tag{D.1}\label{eq:dense-domination}
\]
In particular choose $p=(2FnLH)^{-1}$ for $\rob(E)\le2U$.
The same conclusions hold under a conditional attention-weighted
variance bound $B^2$ in place of the value bound, with
\[
 k=\max(1,\ceil{144B^2/\eps^2})\ceil{\log_2(1/p)}.
\]
\end{lemma}

\begin{proof}
For each possible length, row, layer, head, replicate, and eligible
source $j$, draw an independent unit exponential $D_j$. Choosing the
largest $s_j-\ln D_j$ has law $e^{s_j}/\sum_d e^{s_d}$, since
$D_j e^{-s_j}$ are independent exponential waiting times of rates
$e^{s_j}$. Use a fixed tie priority throughout. For fixed tables each
replicate is a valid top-one selection with pairwise position-dependent
scores. Each family thus becomes an ordinary causal hard model when
both the tables and the independent decoder tape are fixed. Its local
update computes the empirical mean of the $k$ selected values and
passes it to the dense update. Even a query-dependent value can be
reconstructed locally from the winning source record and incoming query.
No precision or implementation-size bound on this hard simulator is
needed. On formal or failed histories extend partially specified
operations by arbitrary local total maps where necessary.

For a bounded value law $V$ of mean $\mu$, independence gives
$\E\|k^{-1}\sum_i V_i-\mu\|^2\le B^2/k$.
Changing one draw changes the norm error by at most $2B/k$.
The bounded-differences inequality gives
\[
 \pr\{\|k^{-1}{\textstyle\sum_i}V_i-\mu\|>\eps\}
 \le \exp[-k\eps^2/(8B^2)]
\]
for $k\ge 4B^2/\eps^2$ and $B>0$.
Indeed, center the norm error by its expectation, at most $B/\sqrt{k}$.
The successive reveal martingale has increments of conditional range
length at most $2B/k$. The conditional exponential-moment bound
$\exp(\lambda^2(2B/k)^2/8)$ for a centered variable in such an
interval and optimization in $\lambda$ give the displayed inequality.
If $B=0$ the samples are exact.
With variance alone, group the samples into independent batches of
$\max(1,\ceil{144 B^2/\eps^2})$. Each batch mean lies within
$\eps/3$ with probability at least $15/16$ by the second moment.
Return the batch mean whose closed ball containing a strict majority
of batch means has minimum radius, with fixed ties. When a strict
majority is good, the minimum radius is at most $2\eps/3$ and the
chosen ball contains a good mean. With $d$ batches the remaining
failure probability is at most $2^d(1/16)^{d/2}=2^{-d}$.
Both estimates hold conditionally at the current admissible state.

Fix the input and expose the simulated free run in row order, computing
all its layers bottom-up before exposing the next row. A current layer
reads only the preceding layer at that row and earlier rows.
When a row's symbol is generated, only earlier clock entries and
appropriate decoder coins have been used. Each newly sampled head
has a fresh block of clocks. Caching makes the number of head sites
at most $nLH$ per length, even though a literal attention implementation
could recompute its past. Randomness for a head or decoder step may
be addressed by position. No sampling table is indexed by the realized
transcript. After fixing the clocks, causal replay is therefore
consistent on all alternative literal histories, as required by the
hard lower bound.

Write $P$ for the ordinary hard simulator law before fixing its tapes.
Define $Q$ by using the same decoder kernels and, at every head,
sampling the local block conditional on its estimate being within
$\eps$ of the exact mean at its \emph{current} state. Equivalently the
adversary draws fresh blocks until acceptance, using the estimate as
the perturbed dense head output. Acceptance has conditional probability
$p_i\ge1-p$, so this terminates almost surely and uses no future coins.
Under $Q$ every queried state is admissibly reachable. The variance or
norm assumption is needed only there, even when $P$ can reach
uncontrolled states. Earlier sampled errors may change the next
emission and all its future logits. No comparison with a fixed clean
trajectory enters the proof.

On the paths with all heads accepted, the unconditioned kernel at
site $i$ restricts to $p_i$ times the conditioned kernel. Successive
integration, leaving the decoder kernels unchanged, gives for any event
$A$ of a sequence of at most $M$ sites
\[
 P(A)\ \ge\ \E_Q\!\left[\mathbf 1_A\prod_i p_i\right]
       \ \ge\ (1-p)^M Q(A).                         \tag{D.2}\label{eq:dense-path-domination}
\]
This also holds conditional on the input and on an entire fixed private
decoder tape independent of the clocks. The strategy itself only sees
the dense computation to date. Expose all length models with independent
clock tables and their prescribed joint decoder tape. Taking
$M=FnLH$ proves the first inequality of \eqref{eq:dense-domination} by averaging the fixed-table
hard bound over the original independent clocks and decoder tape.
Alternatively apply \eqref{eq:dense-path-domination} to each individual length with $M=nLH$ and use
$Q(\bigcup_t E_t)\le\sum_t Q(E_t)$, followed by the fixed-table hard
bound on each marginal, to prove the second. This marginal argument
also bounds $\E_X\max_t \inf_{\mathcal A_t}\pr(E_t\mid X)$ by the same
quantity. Finally $(1-p)^M\ge1-Mp$ proves the factor two. A union
over all possible literal scratch strings or generated histories is
unnecessary for \eqref{eq:dense-path-domination}.
\end{proof}

\paragraph{Growing attention resources.}

We use the position-union pullback (\cref{lem:position-union}) and the
marked replay bound (\cref{lem:marked-position-code} and
\eqref{eq:icml-general-alignment}). Write $q$ for the size ($q\ge2$) of
the fixed finite abelian input/target alphabet. Assume a linear number
of target positions $u$ have
$|\E[\nu(Y_u)\chi(X,Y_{<u})]|\le e^{-\gamma n}$ for all nontrivial
target characters $\nu$ and all input and strict-prefix characters
$\chi$ of degree at most $\rho n$, for fixed $\gamma,\rho>0$.
The number of rows and target positions are of the same order.
The sparse-mask estimates supporting
\cref{thm:topk-depth,thm:icml-radix-edit} give this property for
multiplication in either monotone digit order and every fixed radix. Scratch tokens may use any finite alphabet.

Write $S(n,t)=\sum_{j=0}^{\min(n,\floor{t})}\binom{n}{j}$.
For $b$ qualifying target indices and $r$ deletions per side, marked
replay bounds the success of any hard family by
\[
 D_r\left[q^{-(b-r)_+}+C_r(d+e^{-\gamma n}A)\right],\quad
 D_r=(2r+1) S(n,r)^2,\quad
 C_r=q^{\min(n,a^Lr)}\prod_{j=1}^L S(n,a^jr).      \tag{D.3}\label{eq:dense-hard-alignment}
\]
Here every bounded probe on at most $b+r$ specified top records on a
literal context must have a character polynomial of uniform error
$d$, coefficient mass $\le A$, and degree $d_1$ with $d_1+b-1\le\rho n$.
The positions deleted on each side specify an alignment.
A literal replay generates just the deleted output slots and supplies
the matched true symbols. Expanding the remaining target equalities
and stopping at the last nontrivial character leaves a probe at the
tested and deleted slots and a strict-prefix character of degree at
most $b-1$. The $C_r$ possible deleted symbol lists follow by marking
the queried winners down the layers. Conditional on the sets of
marked positions and their raw alphabet symbols, chronologically
replaying the marked winners reconstructs the list. These position
sets have respective sizes at most $a^jr$. This explains why \eqref{eq:dense-hard-alignment}
applies to every frozen clock table even for a large scratch alphabet.

\begin{theorem}[Quantitative dense scratchpad bound]\label{thm:dense-scratchpad-rates}
Set $p=1/(2nLH(2n+1))$ and let $k,a$ be as in
Lemma~\ref{lem:dense-sampling-domination}. Put
\[
 g=\ln(4n),\quad K=4ag,\quad Q_0=2^{L-1},\qquad
 v=\frac{n K^2}{(R K^2)^{Q_0}},\qquad
 b=\floor{\frac{v}{8K}} .
\]
Here $R$ is a sufficiently large constant depending only on the
task constants and the absolute conjunction-approximation constant.
For a sufficiently small $c_q>0$, set
\[
 r=\left\lfloor c_q\min\left\{\frac{b}{g},\,
       \frac{v}{(L+2)a^L(g+\ln q)}\right\}\right\rfloor.             \tag{D.4}\label{eq:dense-edit-budget}
\]
As soon as $b\to\infty$, $v\ge \ln(2L)$ and $b$ qualifying indices
exist, the robust probability that any candidate's free transcript
aligns after at most $r$ deletions from each side is
$\exp[-\Omega_q(b)]$. At most $2r+1$ lengths can align. The result
covers exact multiplication after an input-selected auxiliary length
$t\le r$, and $s$ interleaved scratch tokens together with $e$
Levenshtein edits whenever $s+e\le r$.
\end{theorem}

\begin{proof}
For completeness optimize the hard resource budget in \eqref{eq:dense-hard-alignment} by keeping
the log inverse approximation error $v$ separate from $b$.
For each bounded probe on $m_0\le b+r$ records start with coefficient
mass one. Keep mass at most $\exp(gm)$ at a pullback with record budget
$m$. Guess the union of the winning source positions of the current
queries. There are at most $(4n)^{(a-1)m}$ guesses. The records of the
queries and guessed positions determine the relative winners there.
One conjunction on at most $n$ Boolean tests (one per outside position)
verifies that the guessed winners beat all outsiders. The unique
correct union gives an exact sum. Replace its conjunction by the
standard small-error AND polynomial of degree at most
$c\sqrt{n\lambda}$ and uniform error $e^{-\lambda}$, where
$\lambda=v+\ln(2L)+amg$. If that degree exceeds $n$, use the exact
conjunction. The resulting sum adds at most $e^{-v}/(2L)$ uniform
error. On Boolean inputs the bounded polynomial has coefficient mass
at most $2(n+1)^{d/2}\le(4n)^d$ at degree budget $d\ge1$, by Parseval.
Substitution yields at most $am+d$ records and preserves the mass
invariant. Inflate $c\ge1$ to absorb integer ceilings.

Let $M_0=v/K$, which bounds $b+r$. As long as $v/K\le M_j\le n/K$,
$\lambda\le 3KM_j$ and $aM_j\le\sqrt{nKM_j}$. Hence
$M_{j+1}=3c\sqrt{nKM_j}$ is a valid envelope.
It increases in this range, and iteration gives
\[
 \frac{M_j}{n}\le 9c^2K
       \left(\frac{v}{nK^2}\right)^{2^{-j}} .
\]
For $j<L$ this is at most $9c^2/(RK)\le 1/K$.
At $j=L$ it is at most $9c^2/\sqrt R$. Choose
$\vartheta<\min(\rho/2,1/2)$ so small that
$h(\vartheta)+\vartheta\ln(q-1)<\gamma$, where $h$ is binary entropy
in nats. Choose $R$ so $9c^2/\sqrt R\le\vartheta$,
$b\le\rho n/2$ and $v\le \gamma n/4$. The final approximant has
uniform error $d\le e^{-v}/2$ and degree $\le\vartheta n$. Its
supremum on the entire formal alphabet cube is at most two.
Cauchy--Schwarz and Parseval therefore give
$A\le2\exp\{n[h(\vartheta)+\vartheta\ln(q-1)]/2\}$.
The strict-prefix factor costs up to $b-1$ extra coordinates. Thus
$d+e^{-\gamma n}A\le 2e^{-v}$ for large $n$.

For $r\ge1$, $S(n,j)\le (4n)^j$ for each nonnegative integer $j$
and $\ln(2r+1)\le 2rg$. Consequently
$\ln D_r\le4rg$ and $\ln C_r\le La^Lrg+a^Lr\ln q$.
Using both terms in \eqref{eq:dense-edit-budget} gives
$\ln D_r\le (b-r)\ln(q)/2$ and
$\ln(D_r C_r)\le v/2$ for a sufficiently small $c_q$,
uniformly in $L,a,n$. The same estimates hold when $r=0$.
Substitute in \eqref{eq:dense-hard-alignment} to obtain $U\le e^{-\Omega_q(b)}$.
There are at most $2r+1\le2n+1$ possible raw lengths since an
alignment changes the target length by at most $r$. Lemma~\ref{lem:dense-sampling-domination}
now proves the dense bound without changing its exponent.
\end{proof}

\begin{corollary}[Asymptotic regimes]\label[corollary]{cor:asymp}
Let $T=1+(B/\eps)^2$. If
\[
 2^L\ln\!\left((1+HT\ln(2n^2LH))\ln(4n)\right)=o(\ln n),
\]
then \eqref{eq:dense-edit-budget} is $r=n^{1-o(1)}$ and robust alignment success is at most
$\exp[-n^{1-o(1)}]$. For fixed $L,H$ this applies whenever
$1+B/\eps=n^{o(1)}$. For fixed $L,H$ and
$1+B/\eps\le n^{\alpha+o(1)}$, where
$0\le \alpha<1/[2(2^L+L-2)]$, \eqref{eq:dense-edit-budget} gives at least
\[
 \begin{split}
 r&\ge n^{1-2\alpha(2^L+L-2)-o(1)},\\
 b&\ge n^{1-2\alpha(2^L-1)-o(1)} .
 \end{split}
\]
In depth one this includes every $\alpha<1/2$. The exponent tradeoffs
for $\alpha>0$ are sufficient lower bounds, with no claim of sharpness.
\end{corollary}
\begin{proof}
Here $k=O(T\ln(2n^2LH))$. Under the first condition,
$\ln(v/n)=-o(\ln n)$, $\ln K=o(\ln n)$ and
$\ln((L+2)a^L)=o(\ln n)$. Consequently $v,b,r=n^{1-o(1)}$,
and $b=o(n)$ as required for a linear supply of qualifying targets.
With $L,H$ fixed, $a,K\le n^{2\alpha+o(1)}$.
The denominators defining $b$ and the second term of \eqref{eq:dense-edit-budget} contain,
up to powers of logarithms and fixed constants, respectively
$a^{2^L-1}$ and $a^{2^L+L-2}$. The first term of \eqref{eq:dense-edit-budget} costs just
one more logarithm relative to $b$. This proves the stated powers.
\end{proof}

\paragraph{Fixed relative error.}

\begin{corollary}\label[corollary]{cor:dense-linear-scratchpad}
For fixed $L,H$ and bounded $B/\eps$, some $c,c'>0$ give
$\rob(E)\le e^{-c'n}$ at deletion radius $\floor{cn}$.
\end{corollary}
\begin{proof}
We make explicit why fixed $k$ can offset the probability cost of an
entire transcript. Let $h_0(t)$ be the binary entropy in nats of
$\min(t,1/2)$, with $h_0(0)=0$.
For hard effective fan-in $a$, the position-union proof permits
initial probe size and inverse-error exponent $\eta n$ with
\[
 \eta=c_L\,a^{-(2^L-1)}
                     [\ln(ea)]^{-(2^{L-1}-1)}.
                                                               \tag{D.5}\label{eq:dense-fixed-fanin}
\]
Here $c_L>0$ can depend on the task. At a pullback write $x=m/n$
for record fraction and $y=(\ln W)/n$ for log coefficient mass.
For $n\eta\ge\ln(2L)$, start with $x_0=\eta,y_0=0$ and use
\[
 z=C\sqrt{2\eta+y+h_0(ax)},\qquad
 x'=ax+z,\qquad y'=y+h_0(ax)+\eta+\tfrac12h_0(z).
\]
The union count is at most $\exp[n h_0(ax)]$, and $nz$ bounds
the AND degree at layer error $e^{-\eta n}/(2L)$. Parseval
bounds its Boolean coefficient mass by $2\exp[n h_0(z)/2]$.
The constant $C$ absorbs the degree ceiling. Since $x'$ dominates
$x,ax,z$ and $x\ge\eta$, induction gives $y_j\le 3j h_0(x_j)$
for $\eta\le1/2$. Thus $z_j\le C_L\sqrt{h_0(ax_j)}$.

For $j<L$, put $u_j=2^{L-j}-1$, $v_j=2^{L-1-j}-1$, and
$\ell_a=\ln(ea)$. Use envelopes
$x_j\le \delta_j a^{-u_j}\ell_a^{-v_j}$ with positive constants
$\delta_j\le 1/2$. For $j<L-1$, monotonicity and
$h_0(t)\le t\ln(e/t)$ for $0<t\le1/2$ give
\[
  x_{j+1}\le
  C_L\sqrt{\delta_j(1+\ln(1/\delta_j))}
     \,a^{-u_{j+1}}\ell_a^{-v_{j+1}} .
\]
In fact $a x_j\le\delta_j a^{-(u_j-1)}\ell_a^{-v_j}$ and the
logarithm of the reciprocal of this envelope is at most
$C_L(1+\ln(1/\delta_j))\ell_a$. The powers follow from
$u_j-1=2u_{j+1}$ and $v_j-1=2v_{j+1}$. At $j=L-1$ we have
$a x_j\le\delta_j$, so $x_L$ is at most
$\delta_j+C_L\sqrt{h_0(\delta_j)}$.
Choose the $\delta_j$ backwards to ensure
$x_L\le\rho/2$ and $3Lh_0(x_L)+x_L\ln(q)/2\le\gamma/2$.
This gives \eqref{eq:dense-fixed-fanin} with $c_L=\delta_0$.
The dependence on $a$ in this proof is $1/a$ at one layer and
$1/[a^3\ln(ea)]$ at two layers. The resulting numerical threshold
in $n$ is still conservative. Expanding each bottom term on its
raw records now gives total character mass at most $e^{\gamma n/2}$,
degree at most $\rho n/2$, and uniform error $e^{-\eta n}/2$.

Choose $c_L$ small enough that the required number of qualifying target indices exists. Take $b=\floor{\eta n/4}$ and
$r=\floor{c_{L,q}\eta^2 n/a^L}$ with $c_{L,q}$ small, so that
$b+r\le\eta n$, $r\le b/4$ eventually and the strict-prefix
degree budget holds. Formula \eqref{eq:dense-hard-alignment} with
$S(n,j)\le\exp[n h_0(j/n)]$ gives
$\ln D_r+\ln C_r\le n\{(L+2)h_0(c_{L,q}\eta^2)
+c_{L,q}\eta^2\ln q\}+o(n)$, where $\ln(2r+1)=o(n)$ for each
fixed $a$. Using $h_0(t)\le C\sqrt t$ again and reducing
$c_{L,q}$ makes this at most
$\min(1,\ln q)\eta n/32$ for large $n$. Thus \eqref{eq:dense-hard-alignment} gives
$U\le e^{-d_q\eta n}$ with $d_q>0$ independent of $a$, for all
large enough $n$ at each fixed $a$. The task and $L$ are fixed.

Take a fixed number $k$ of replicas. Under the value assumption we
can choose $p=e^{-c k}$ for all sufficiently large $k$, where $c>0$
depends only on the uniform bound on $B/\eps$. For bounded variance
the batches give the same form after changing $c$. Since
$a=1+Hk$, the polynomial \eqref{eq:dense-fixed-fanin} satisfies
$2LHp<d_q\eta/2$ for sufficiently large fixed $k$. For $B=0$
use exact samples. Apply the second inequality of \eqref{eq:dense-domination} with $p\le1/2$,
so $(1-p)^{-nLH}\le e^{2p nLH}$. The polynomial factor $F\le2n+1$
costs $o(n)$ in the logarithm. The resulting exponential bound holds
with the positive linear fraction $c_{L,q}\eta^2/a^L$ obtained
at this fixed $k$.
\end{proof}

\paragraph{Scope.}
Admissible errors occur at all rows and layers, measured against the
exact head mean at the actual perturbed incoming state. A guarantee
only on the clean trajectory, perturbation of the final readout alone,
or a noise law imposed in advance does not supply this adversary.
Sampling distributions may be concentrated or diffuse. Neither smooth
updates, a positive logit margin, nor reproduction of the clean
scratch string is required. The result concerns simultaneous
per-head Euclidean radii, including dimensions that grow with $n$.
It gives no bound on hardware mantissa width.

\endgroup

\begin{proof}[Proof of \cref{prop:icml-effective-robust}]
An online adversary first checks the inequality
at the current dense incoming array. If it holds it returns
the truncated head mean. This is an allowed perturbation of
the dense mean; otherwise it sets all remaining errors to
zero. Row by row the adversary agrees with the truncated
network up to its first failed certificate. Reuse the same
decoder tape in the coupling, and repeat previously chosen
perturbations if rows are recomputed. Neither step inspects
future decoder coins. Thus the attack fails to reproduce any
candidate tape with probability at most $\beta_k$. For each
fixed decoder tape the truncated model is a deterministic
hard top-$k$ model, so averaging preserves its success bound.
If the top-$k$ tail condition holds on every perturbation-reachable
array, induction gives $\beta_k=0$. The relevant tails are along
the truncated trajectory, which can be generated in a separate
rollout. Clean dense-run tails alone do not certify this attack.
\end{proof}

\paragraph{Learned attention at small operand lengths.}
We trained two-layer, four-head multipliers on four- and five-bit
operands with two seeds per length. On 514 held-out model-input
cases, dense greedy generation got 239 products right. At relative
radius $1/4$, the exact displacement certificate failed on 395
cases for $k=4$ and on two for $k=8$. The sufficient $2Bt$ test
failed on all 514 at both $k$. Here $B$ for each head is the
maximum value norm observed across the tested rollouts. The
supplement provides the training script and input-level counts.
This is a diagnostic of truncating learned networks with modest
accuracy. The lengths are below the certified generation regime.

\subsection{Independent Gaussian noise in fixed-depth generation}
\label[appendix]{app:gaussian-transcript}

\begin{proposition}[Gaussian scratchpad lower bound]
\label[proposition]{prop:gaussian-scratchpad}
Fix depth $L$, head count $H$, and let $R=B/\sigma$ with $\sigma>0$.
Use the length models and local readouts of
\cref{app:dense-scratchpad}, with Euclidean head values of norm at most
$B$ on states reached under the noise. Perturb each new head output by an independent
$\mathcal N(0,\sigma^2 I)$ vector. Either cache previously computed
states or use fresh independent noise for every row of each
next-token forward pass. In the latter convention each row is
computed once within a pass, and the same causal length model is
used on every prefix at that candidate length.
On independent uniform $N$-bit operands let $E_r$ be alignment of any
free-running output to the binary product in either specified digit
order after at most $r$ deletions per side. This includes exact generation
after at most $r$ auxiliary tokens. Scratch vocabularies and value
dimensions may be arbitrary and finite.
\begin{itemize}
\item At bounded $R$, some $c,c'>0$ give
$\Pr(E_{\lfloor cN\rfloor})\le e^{-c'N}$ for large $N$.
\item If $1+R=N^{o(1)}$, there is a budget $r=N^{1-o(1)}$ with
$\Pr(E_r)\le\exp[-N^{1-o(1)}]$.
\item Write $h=2^L-1$, $s=h+\min\{1,L-1\}$. If $1+R\le N^{\alpha+o(1)}$
and $0\le\alpha<1/(2s)$, then for every fixed $\xi>0$
sufficiently small, one may take
$r=\lfloor N^{1-2\alpha s-\xi}\rfloor$ with
$\Pr(E_r)\le\exp[-N^{1-2\alpha h-\xi}]$ for large $N$.
\end{itemize}
\end{proposition}

\begin{lemma}[Gaussian simulation from bounded samples]
\label[lemma]{lem:gaussian-sample-extrapolation}
Let $V$ be a random vector in a Euclidean ball of radius $R$ about the
origin, with mean $\mu$, and let $p$ be the density of
$\mathcal N(\mu,I)$. For each integer $D\ge1$ there is a constant
$C_D$ such that, whenever $k\ge C_D(1+R)^2$, a rule using $Dk$
independent samples of $V$ and independent auxiliary randomness has
output density $q$ satisfying
\[
   \chi^2(p\Vert q)+\chi^2(q\Vert p)
      \le C_D\big((1+R)^2/k\big)^{2D}.
\]
The rule uses the sampled vectors alone. The bound is uniform in the
dimension, $R$, and the law of $V$.
\end{lemma}

\begin{proof}
Put $u=(1+R)/\sqrt k$. Write $M_j$ for the mean of the first $jk$
samples, $v_j=M_j-\mu$, and $\phi_j$ for the unit-covariance Gaussian
density centered at $M_j$. We use $C_D$ for constants depending only
on $D$, increasing its value in the sampling threshold as needed.

\emph{Moments of a random center.}
For the centered mean $v$ of $l\ge k$ samples,
$\mathbb E\|v\|^2\le R^2/l$, and changing one sample changes its
norm by at most $2R/l$. Bounded differences give
$\Pr(\|v\|>u(1+t))\le e^{-t^2/2}$ for $t\ge0$ (the case $R=0$ is immediate).
For fixed $m,c\ge0$, if $u$ is sufficiently small in terms of $c$, then
\begin{equation}\label{eq:gaussian-center-moments}
  \mathbb E\big[\|v\|^m e^{c\|v\|^2}\big]\le C_{m,c}u^m .
\end{equation}
Indeed $Y=\|v\|/u$ has $\Pr(Y>y)\le e^{-y^2/8}$ for $y\ge2$.
Integrate this tail against the derivative of $y^m e^{cu^2y^2}$,
taking $cu^2\le1/16$. All bounds are dimension independent.

\emph{Signed approximation.}
Vandermonde interpolation gives coefficients $c_j$, $1\le j\le D$,
with $\sum_jc_j=1$ and $\sum_jc_j j^{-d}=0$ for $1\le d<D$.
Set $S=\sum_j c_j\phi_j$. The order-$m$ Hermite coefficient tensor
of the ratio of a Gaussian shifted by $v$ to $p$, centered at $\mu$,
is $v^{\otimes m}$. This follows from the generating function
$\exp(\langle t,v\rangle)$. The squared $L^2(p)$ norm of the
order-$m$ component is $\|v\|^{2m}/m!$, by product Hermite
orthogonality (each multi-index has $m!/\alpha!$ ordered entries).

For $1\le m<2D$, the centered sample mean of $l$ draws satisfies
\[
  \mathbb E v^{\otimes m}
  =l^{-m}\sum_{\pi}(l)_{|\pi|} T_\pi .
\]
Here $\pi$ ranges over partitions with no singleton, $(l)_d$ is the
falling factorial, and each $T_\pi$ uses one independent centered draw
per block. Thus $\|T_\pi\|_{\rm HS}\le(2R)^m$.
The scalar factors are polynomials in $l^{-1}$ whose powers lie
between $m-|\pi|\ge\lceil m/2\rceil$ and $m-1$.
Substituting $l=jk$ and combining with $c_j$ cancels all powers
below $D$. The surviving tensor norm is at most
$C_D(1+R)^{2D}k^{-D}=C_Du^{2D}$ at each of these orders.
For a fixed $v$ the norm of the entire Hermite tail from order $2D$
is at most $\|v\|^{2D} e^{\|v\|^2/2}/\sqrt{(2D)!}$, by the
exponential series. Convexity of the norm and
\eqref{eq:gaussian-center-moments} give the same $O_D(u^{2D})$
order for the averaged tail. The constant term is one, so
$\|\mathbb E S/p-1\|_{L^2(p)}\le C_Du^{2D}$.

\emph{Positive simulation.}
Fix $\beta=1/2$, write $W=S-\beta\phi_1=\sum_j b_j\phi_j$, and use
the conditional output density
\[
 r=\beta\phi_1+(1-\beta)W_+/\!\int W_+ .
\]
Since $\int W_+\ge1-\beta$, one can generate the second component
by proposing from $G/\int G$ with $G=\sum_j(b_j)_+\phi_j$ and
accepting with probability $W_+(z)/G(z)$. The acceptance rate is at
least $(1-\beta)/\sum_j(b_j)_+>0$ and all quantities are functions of
the sampled centers. On a null tape that never accepts use a fixed
local default.

Write $\|\cdot\|_p$ for the $L^2(p)$ norm in the rest of the proof.
For a unit Gaussian $Z$ and a fixed shift $v$ set
$L_v(Z)=\exp(\langle Z,v\rangle-\|v\|^2/2)$.
For each fixed $d\ge1$,
\[
 \|L_v-1\|_{L^d}\le C_d\|v\|e^{C_d\|v\|^2}.
\]
To see this, integrate $L_{tv}(\langle Z,v\rangle-t\|v\|^2)$
over $0\le t\le1$ and use H\"older with exponents $2d,2d$.
The two norms are at most $e^{d\|v\|^2}$ and
$C_d\|v\|+\|v\|^2$. Now $W/p=1-\beta+T$ with
$T=\sum_jb_j(\phi_j/p-1)$. For real $T$,
$(1-\beta+T)_-\le(1-\beta)^{1-2D}|T|^{2D}$.
By the preceding shift estimate at $d=4D$, the triangle inequality
and \eqref{eq:gaussian-center-moments},
\[
  \mathbb E\|W_-/p\|_p^2
  \le C_D\,\mathbb E\|T\|_{L^{4D}(p)}^{4D}
  \le C_D u^{4D}.
\]
Also $\mathbb E\|W_+/p\|_p^2\le C_D$, since
$\|\phi_j/p\|_p=e^{\|v_j\|^2/2}$. With $a=\int W_-\le\|W_-/p\|_p$,
normalizing $W_+$ yields
\[
 \|(r-S)/p\|_p
 \le \big(1+\|W_+/p\|_p/(1-\beta)\big)\|W_-/p\|_p .
\]
Cauchy--Schwarz over the sampled centers bounds the expectation by
$C_Du^{2D}$. In particular $q=\mathbb E r$ has
$\chi^2(q\Vert p)=\|q/p-1\|_p^2\le C_Du^{4D}$.
Finally Jensen's inequality and $\mathbb E v_1=0$ give pointwise
$q(z)/p(z)\ge\beta\mathbb E L_{v_1}(z-\mu)
\ge\beta e^{-R^2/(2k)}$.
Integrating $(q-p)^2/q$ proves the reverse bound as well.
\end{proof}

\begin{proof}[Proof of \cref{prop:gaussian-scratchpad}]
Write $n=\Theta(N)$ for the common row bound. Project values onto the
radius-$B$ ball on every formal history. The reachable-value hypothesis
preserves the free-run law. Fix a candidate length, its input-independent
architecture and decoder tapes, and the sets $U,V$ of deleted emissions
and target digits in one alignment. Replay the run to the fixed horizon,
generating the slots in $U$ and supplying the paired true digit at every
other slot. Couple by row of birth in the cached model, and by pass and
row in the fresh model. On the event that the free run realizes this
alignment, every retained replay prediction is correct. This inclusion
uses the stipulated continuation past a local stop.

\emph{Removing the generated slots.}
Let $P$ be the replay law, and let $Q$ replace every head output in
each layer of the prediction row for a slot in $U$ by an independent
$\mathcal N(0,\sigma^2 I)$ vector. Make this replacement when the
row is born in the cached model, and only in the pass that emits
that slot in the fresh model. Include the operands and all noisy
vectors in the finite transcript, exposing new sites in causal order.
At an altered site the conditional squared likelihood integral is
$\exp(\|\mu\|^2/\sigma^2)\le e^{R^2}$, where $\mu$ is the true
conditional head mean. Elsewhere the two kernels coincide. Thus
backward integration and Cauchy--Schwarz give, for any event $A$,
\[
 P(A)\le e^{|U|LHR^2/2}Q(A)^{1/2}.
\]
Under $Q$ freeze the centered replacement outputs, which are
independent of the operands and of the other innovation tapes.
At an altered prediction row every local update now sees its own
incoming state and fixed vectors. Its emission is a unary function
of the previous literal token. Composing consecutive such functions
makes each supplied slot a position-dependent encoding of a constant,
an operand digit, or a retained strict-prefix digit. The literal tape
is consequently determined by the operands and frozen tapes.

\emph{One coherent test.}
Choose $b$ qualifying target positions as below, retain those outside
$V$ in $I$, and put $m=|I|\ge b-r$. Let $A$ require agreement on $I$.
Evaluate the supplied tape once in causal order, using Gaussian heads.
For cached generation this evaluation must preserve the frozen head
outputs at the altered rows in all later uses of their records.
For fresh generation use ordinary Gaussian heads at those rows:
a retained pass recomputes them. These prescriptions specify two
coherent evaluators with position-local updates. Let $S$ count errors
on $I$ in the prescribed evaluator. In the cached case it reproduces
the joint law under $Q$, so agreement on all tests is bounded by the
mean of $2^{-S}$. In the fresh case, conditional on the operands and
frozen tapes, retained passes are independent. Their success
probabilities $p_i$ equal the single-row success marginals of the
coherent evaluator by causality and the common model across prefixes.
Consequently
\[
 \prod_{i\in I}p_i\le 2^{-\sum_{i\in I}(1-p_i)}
                =2^{-\mathbb E S}\le\mathbb E 2^{-S},
\]
with the same conditioning. In both cases this last expectation is the
probability of passing the coherent test on $J$, where $J$ includes each
element of $I$ independently with probability $1/2$. Average over the
operands before bounding that test.

At every ordinary site of the coherent evaluator apply
\cref{lem:gaussian-sample-extrapolation} in units of $\sigma$. The
$Dk$ samples use independent exponential clocks as in
\cref{lem:dense-sampling-domination}. At a frozen site both
evaluations use the identical deterministic vector. Let $G$ and
$T$ denote the coherent Gaussian and simulated laws. Integrating
out the new clocks and simulator tape at each ordinary site gives
conditional densities with $1+\chi^2(G_{\rm site}\Vert
T_{\rm site})\le\exp[C_D((1+R)^2/k)^{2D}]$ on each common history.
There are at most $nLH$ such sites. Applying the preceding
finite-transcript calculation to the coherent test $F$ gives
\begin{equation}\label{eq:gaussian-transcript-rate}
 G(F)\le\big[T(F)\exp\{C_DnLH((1+R)^2/k)^{2D}\}\big]^{1/2}.
\end{equation}
Fixing the clocks and local simulator tapes by length, row, layer,
head and replica gives a hard top-one model with effective fan-in
$a=1+HDk$. A frozen row ignores its winners and uses its prescribed
vectors; its resulting records remain available to later queries.

For each fixed $J\subseteq I$ the hard bound is
$2^{-|J|}+e^{-c b}$. Indeed map a nonbinary prediction to a fixed bit
for this upper bound and expand simultaneous correctness in binary
characters. The constant term is $2^{-|J|}$. For each nonempty term
the probe reads at most $b$ top records through its latest target.
Apply the position-union pullback on the literal tape. At its bottom,
substitution of the fixed unary encodings preserves the number of
variables per term, allowing repeated copies. At fixed fan-in the
Walsh mass of a bounded term on $t$ binary variables is at most
$2^{t/2}$. At growing fan-in the substituted full approximant
retains its degree bound and supremum bound on the genuine formal
cube. The other selected truths cost at most $b-1$ strict-prefix
coordinates. The sparse correlation and approximation errors are
therefore those in the proofs of \cref{cor:dense-linear-scratchpad}
and \cref{thm:dense-scratchpad-rates}. Specifically, at fixed fan-in
take $b=\lfloor\eta n/4\rfloor$ with $\eta$ from
\eqref{eq:dense-fixed-fanin}, giving error at most
$e^{-\eta n}/2+e^{-\gamma n/2}$ after reducing the task constant in
$\eta$ if needed. At growing fan-in use $K=4a\ln(4n)$ and
$b=\lfloor v/(8K)\rfloor$ from \cref{thm:dense-scratchpad-rates},
giving error at most $2e^{-v}$. Hence $c>0$ can be independent of
the fan-in. Averaging over $J$ gives
$T(F)\le(3/4)^m+e^{-c b}$. This is uniform in the frozen vectors
and permits averaging over them after applying
\eqref{eq:gaussian-transcript-rate}.

\emph{Budgets.}
If $r\le b/4$ and the exponent in
\eqref{eq:gaussian-transcript-rate} is a sufficiently small multiple
of $b$, then $Q(A)\le e^{-c_1 b}$ for a fixed $c_1>0$.
The replay comparison costs $O(rLHR^2)$ in the exponent. The union
over the two deletion sets and candidate lengths costs
$O(r\ln(en/r)+\ln n)$ for $1\le r\le n/2$. For bounded $R$,
choose a fixed integer $D$ with $2D>h=2^L-1$ and then a sufficiently
large fixed $k$, using the uniform upper bound on $R$. Indeed
$\eta\asymp a^{-h}[\ln(ea)]^{-(2^{L-1}-1)}$, so the coherent
likelihood cost is an arbitrarily small multiple of $b$ for large
enough fixed $k$. A sufficiently small fixed $r/n>0$ now absorbs
both remaining costs.

For growing $R$ fix $\tau>0$, take $D$ with $2D>2h$ and
$(2+\tau)(2D-h)>4D$, and use
$k=\lceil K_0(1+R)^{2+\tau}g\rceil$, where $g=\ln(4n)$.
Then $b\asymp n/(kg)^h$ whenever this last ratio diverges, and all
hard thresholds hold since $v\ge\ln(2L)$ eventually and a fixed
fraction of targets qualify. Taking $K_0$ sufficiently large gives
$k^{2D-h}\ge C_{D,L,H}(1+R)^{4D}g^h$ and pays the coherent
likelihood cost. Take
\[
 r=\left\lfloor\frac{c_2 b}{g+LHR^2}\right\rfloor
\]
with $c_2>0$ small. In the stated growing-ratio regimes $b\gg\ln n$,
so the mask union and the replay comparison leave success
$e^{-\Omega(b)}$. This proves the subpolynomial conclusion. For
$1+R\le N^{\alpha+o(1)}$, $L\ge2$ and $\alpha<1/[2(h+1)]$,
choose $\tau$ with $\alpha\tau h<\xi/2$. The resulting radius and
negative log success dominate the claimed powers with $s=h+1$.

At $L=1$ the old head outputs are unused by every later query, so
both free-run output laws agree. For the sharper polynomial radius,
apply \eqref{eq:gaussian-transcript-rate} directly to the cached
free-run alignment event with ordinary heads. Here $T$ is a coherent
hard free run, bounded per length by $e^{-\Omega(b)}$ at the radius
in \eqref{eq:dense-edit-budget}. With $L=1$ this radius is of order
$b/g$ before flooring. The same $D,k$ pay the likelihood cost, and
the length union costs $O(\ln n)$. This gives the claimed powers
with $s=h=1$.
\end{proof}

\subsection{Sparse digit masks and output orders}\label[appendix]{app:icml-interior}

\ificmlextended
\begin{theorem}\label{thm:icml-interior}
For each fixed $0<\eta\le1$, there exist $c_\eta>0$ and $N_\eta$ such that
for $N\ge N_\eta$ and every mask $S$ with $\eta N\le T=\max S\le(2-\eta)N$,
\[
                    \disc_N(S)\le 2^{-c_\eta N}.
\]
\end{theorem}

\begin{lemma}[Bourgain's Type II estimate, dyadic form]\label[lemma]{lem:icml-interior-typeii}
Take integer exponents $1\le\mu<\nu$ and
$\lambda=\mu+\nu+2$. Let $w_A$ multiply the signs of the
least-significant-first binary digits indexed by
$A\subseteq\{0,\ldots,\lambda-1\}$. On
$[2^\mu,2^{\mu+1})\times[2^\nu,2^{\nu+1})$, with each
coefficient array bounded by one in modulus,
$\big|\sum_{x,v} f(x)g(v) w_A(xv)\big|$ is at most
\[
 C2^\lambda\big(2^{-c_1\epsilon\rho}
       +2^{2\rho-c_2\mu}+2^{C_2\rho-c_3(\mu/\lambda)|A|}\big).
\]
Here $\epsilon>0$ is sufficiently small and fixed,
$\rho$ and $\epsilon\rho$ are positive integers,
$\rho<\mu/100$, and $\epsilon\rho\ge C_0(\log\lambda)^2$
for a sufficiently large absolute $C_0$. The constants in
the bound are independent of the coefficient arrays.
\ifdefined\ICMLEXTENDED\else
Fix also $\sigma>0$ and assume $\rho\ge\sigma\lambda$.
The constants may depend on $\sigma$.
\fi
\end{lemma}
The following correspondence applies to \citet[\S2, (2.1), (2.14), and
(2.31)]{Bourgain11}. The last column specializes it to the
dense-mask boxes below.
\begin{center}
\begin{tabular}{@{}lll@{}}
\toprule
Bourgain & Dyadic form & Box with exponents $a\le b$ \\
\midrule
$m,n$ & $x,v$ & $x,2y$ \\
$M,N$ & $2^\mu,2^\nu$ & $2^a,2^{b+1}$ \\
$X=2^\lambda$ & $4\cdot 2^\mu 2^\nu$ & $2^{a+b+3}$ \\
$L$ & $2^\rho$ & $2^\rho$ \\
$S$ & $A$ & retained bit positions shifted by one \\
Weights in (2.1) & $f(x),g(v)$ & $f(x),\mathbf 1_{\{v\ {\rm even}\}}g(v/2)$ \\
\bottomrule
\end{tabular}
\end{center}
The active set $S'$ in Bourgain's (2.29) retains the selected mask bits
in the effective window $[K,K+\mu+(1+\epsilon)\rho)$.
Use $K=0$, all $K=j\mu-\rho$ with $j\ge1$ and $j\mu\le\nu-1$,
and $K=\nu-\rho-1$. Nonzero shifts satisfy
$\mu-\rho\le K\le\nu-\rho-1$, so $2^\mu\le 2^{K+\rho}<2^\nu$,
as required in (2.23). They also satisfy
$t=\epsilon\rho<K/2$ since $\rho<\mu/100$. This is the upper
cutoff in his Lemma~5, whose other requirement is the integer
lower cutoff $t>C(\log\lambda)^2$. It follows from
$(\log\lambda)^2\ll_\epsilon\rho$.
Consecutive window starts differ by at most $\mu$.
The last window reaches $\lambda-3+\epsilon\rho\ge\lambda$,
so $O(\lambda/\mu)$ windows cover the bit positions.
By averaging, one window retains $\gg(\mu/\lambda)|A|$ bits.
This is the weight bound needed in (2.29) to obtain (2.31).
For the diagonal term (2.14), set $D=2^{\mu+\rho+t}$.
Bourgain smooths the short operand at scale
$M_1=M^{1-\epsilon_1}$ in (2.11), with $M=2^\mu$. When
$\rho\ge\sigma\lambda$, take a fixed $\epsilon_1\le\epsilon\sigma/2$,
so $\epsilon_1\lambda\le\epsilon\rho$.
The diagonal $k=k'$ at nonzero shift $|\ell|<L=2^\rho$
has $|k|\le DL^\epsilon$ and
$\|k\ell/D\|<1/M_1$, where $\|\cdot\|$ denotes distance to an integer.
Writing $g=\gcd(D,|\ell|)$, this condition is periodic in $k$
with period $D/g$. Within a period it holds for
$O(1+D/(g M_1))$ frequencies. There are $O(g L^\epsilon)$ periods in
the frequency range, so the count is $O(L^\epsilon(g+D/M_1))$.
When $\epsilon_1\lambda\le t$, we have
$D/M_1\le L^{1+2\epsilon}$ and $g\le L$, giving $O(L^2)$ terms
for $\epsilon<1/3$. The factor $1/L$ averages over the shifts.
The Fourier sup-norm bound (2.6) or (2.9) then gives
$O(2^{2\lambda}L^2 2^{-c|S'|})$ in the squared sum, as in (2.14).
The off-diagonal counts starting from (2.12) use the same bound
$2^{\mu+\rho+t}/M_1\le L^{1+2\epsilon}$, explicitly in (2.17)
and in the interval count preceding (2.25). After that substitution,
the estimates (2.22) and (2.28) use the Fourier bounds (2.6),
(2.9) and the counting bounds with $\epsilon$ fixed. Their exponents
are independent of $\epsilon_1$; the implicit constant for the
smooth summation may depend on the chosen cutoff.

\ifdefined\ICMLEXTENDED
\paragraph{Uniform cutoff used for the lower-end extension.}
For uniformity in $\rho$, one can instead take $M_1=M$ directly.
In the Cauchy--Schwarz step leading to (2.2), first discard the
carry and Fourier-truncation errors using (2.3)--(2.4) on the short
dyadic box, retaining the coefficients in the shifted quadratic
form. Before Fourier expansion, its summand at $m$ is proportional to
\[
 \sum_{n\in\mathbb Z}
 \left|\sum_{j=1}^{L}\beta_{n+j2^K}
    W_{S'}\bigl(m(n+j2^K)\bigr)\right|^2 .
\]
Here $\beta$ is extended by zero off the long operand range, and
$W_{S'}=w_{S'}$ for $K=0$. This is nonnegative for every $m$. Majorize
$\mathbf 1_{[M,2M)}(m)$ by a fixed constant times
$\bigl(\sin(\pi(m-3M/2)/(4M))/(\pi(m-3M/2)/(4M))\bigr)^2$,
with the value at zero argument defined by continuity. This weight
has sum $O(M)$ and discrete Fourier transform supported on
$\|\theta\|\le 1/(4M)$, by periodizing the triangular Fourier transform
of the squared sinc. Expansion therefore gives (2.11) with $M_1=M$
and no smoothing remainder. All subsequent resonance bounds count
frequencies in a set that only shrinks as $M_1$ increases.
This gives the same estimate for sublinear $\rho$ as for $\rho$
a fixed fraction of $\lambda$.
\fi

The three contributions in (2.31) are the carry and Fourier
truncation error, the short-side term, and the retained-mask term.
They give the three terms in \cref{lem:icml-interior-typeii} after
substituting $L=2^\rho$ and the window bound on $|S'|$. The bilinear
estimate applies directly to the two operand tests.

\fi
\paragraph{Fourier mass of sparse digit masks.}
Let the radix be $b$. A product of digit characters on $T$ slots
with $k$ nontrivial choices has normalized discrete Fourier mass
at most $b^k[C_b(1+T/(k+1))]^{k+1}$.
For the character at digit $j$ with index $a_j$, its Fourier
factor in modulus is $H_{a_j}(h/b^j)$, where
$H_a(z)=|b^{-1}\sum_{u=0}^{b-1}e^{2\pi i u(a/b-z)}|$.
This follows by independently summing each digit of the bin index.
Summing over preimages of multiplication by $b$ gives positive
operators with weight $H_a$ and sup norm at most $b$ per selected
step. Across $\ell$ unselected digits the geometric sums multiply
to a normalized Dirichlet kernel of length $b^\ell$.
The sum of its modulus over the $b^\ell$ preimages is $O_b(1+\ell)$:
the nearest terms are bounded by one and the others sum a harmonic
series by distance to the nearest endpoint. Multiplying these
bounds and applying AM--GM to the $k+1$ zero runs proves the claim.

\ificmlextended
\paragraph{Dense masks in any digit positions.}
If the mask selects $k\ge\delta N$ digits for fixed $\delta>0$,
the discrepancy is $O(2^{-c k})$ for an absolute $c>0$ and
sufficiently large $N$. No condition on its extreme indices is needed.
Let $d=\lfloor k/16\rfloor$. Discarding operands
below $2^{N-d}$ costs at most $2^{1-d}$. Partition the rest into dyadic boxes
with lower endpoints $2^a,2^b$, swapping operands so $N-d\le a\le b<N$.
In least-significant-bit indexing, bits at positions $\ge a+b+2$ vanish there,
so at most $2d$ selected digits are lost. Replace $y$ by $v=2y$, put its
coefficient to zero on odd $v$, and shift the retained positions up by one to
form $A$. This preserves the sum. Put $\mu=a$, $\nu=b+1$, $\lambda=a+b+3$.
Apply \cref{lem:icml-interior-typeii} with coefficient arrays $f(x)$
and $\mathbf1_{\{v\ {\rm even}\}}g(v/2)$. Their moduli are at most
one. The change of variables shifts every retained bit by one
and preserves its sign on this support. The dyadic side lengths
are $2^\mu$ and $2^\nu$, and $2^\lambda$ is four times their
product.
Fix a sufficiently small reciprocal-integer $\epsilon>0$. Choose $\rho$
the largest multiple of $1/\epsilon$ below $\kappa k$ for a sufficiently small
absolute $\kappa>0$. On these boxes $k\le 2N$, $\mu\ge7N/8$,
$\mu/\lambda\ge1/3$ for large $N$, and $|A|\ge7k/8$.
In particular $\mu/k\ge7/16$. Choose $\kappa$ so that
\[
 \kappa<7/1600,\quad 2\kappa\le7c_2/32,\quad
 C_2\kappa\le7c_3/48 .
\]
The first inequality ensures $\rho<\mu/100$. Since $k\ge\delta N$,
rounding by a fixed multiple still gives $\rho\ge\kappa k/2$ and
$(\log\lambda)^2\ll_\epsilon\rho$ eventually.
Choose $\epsilon$ first. The exponent constants in
\citet[(2.31)]{Bourgain11} come from the Fourier and resonance
bounds there and are fixed before choosing the smoothing exponent.
Choose $\kappa$ by the preceding inequalities, then take
$\sigma=\kappa\delta/6$ and $\epsilon_1=\epsilon\sigma/2$.
Indeed $\lambda\le3N$ and $\rho\ge\kappa\delta N/2$ give
$\rho\ge\sigma\lambda$ and $\epsilon_1\lambda\le\epsilon\rho/2$.
Any prefactor depending on smoothing is fixed once $\delta$ is
fixed and is uniform over the boxes.
The last two inequalities give
$2\rho-c_2\mu\le-7c_2 k/32$ and
$C_2\rho-c_3(\mu/\lambda)|A|\le-7c_3 k/48$.
Thus all three terms save $2^{-\Omega(k)}$ uniformly in the box.
Its factor $2^\lambda$ is eight times the original box area.
The disjoint original boxes have total area at most $4^N$; dividing
their summed bound by $4^N$ and restoring the discarded strips
gives $O(2^{-c k})$.

\fi
\paragraph{Sparse MSB-first masks in every fixed radix.}
Fix a base $b$ and a margin $\eta$. There is $\delta>0$ such that
every product of at most $\delta N$ digit characters with last
nontrivial index $T\in[\eta N,(2-\eta)N]$ has exponentially small
correlation against bounded separate operand tests.

Write $b$ for the radix, $H=b^N$ for the operand grid length,
$P=H^2$ for the product grid length, and $Q_T=b^T$ for the
number of target-scale cells. Put $m=2N-T$ and
$r=\lfloor\min(T,m)/8\rfloor$. Write $e(t)=e^{2\pi i t}$.
Let $W_S$ denote the digit phase and let $A(h)$ be its discrete
Fourier coefficient modulus, as above. Integration within each
cell shows that its continuous coefficient $c_h$ satisfies
$c_0=0$ and $|c_h|\le A(h)\min(1,Q_T/|h|)$.
The last character is nontrivial, so its factor vanishes at zero
and $A(h)\le C_b|h|/Q_T$. By periodicity of $A$ and harmonic summation,
$\sum_{0<|h|<Q_T2^r}|c_h|\le C(1+r)M$, where
$M=\sum_{0\le h<Q_T}A(h)$. Choose $\delta$ sufficiently small that
$M\le2^{\eta N/64}$ by the sparse mass bound.

\emph{Truncation on integer products.}
Let $V=W_S*F_K$ with the positive Fej\'er kernel of order $K=Q_T2^r$ and put
$v=Q_T^{-1}2^{-r/2}$. Its multipliers are $(1-|h|/K)_+$.
At circle distance more than $v$ from every grid endpoint $j/Q_T$, the
pointwise error is $O(1/(Kv))=O(2^{-r/2})$, since
$F_K(t)\le (4K\|t\|^2)^{-1}$. As $Q_T\mid P$, the exceptional set contains
at most $O(P(2^{-r/2}+2^{-m}))$ grid points $n/P$, $0<n<P$.
Each positive product has at most $d(n)$ representations and
$\sum_{n\le P}d(n)^2\le\sum_{n\le P}d_4(n)\le P(1+\log P)^3$
by the prime-power inequality and three harmonic sums. Thus Cauchy--Schwarz,
$|V|,|W_S|\le1$, and $\Pr[XY=0]\le2/H$ give
\[
  \mathbb E|V(XY/P)-W_S(XY/P)|\ll(1+N)^{3/2}2^{-r/4}+H^{-1}.
\]

\emph{Bilinear frequencies.}
The frequencies $0<|h|\le2^r$ contribute $O_b(2^{2r-T})$ by the last-bit bound.
For $2^r<|h|<K$, put $\theta=|h|/P$, so $P\theta>2^r$ and
$\theta<2^{-m+r}\le2^{-7r}$. Cauchy--Schwarz followed by geometric
summation gives, for $|f|,|g|\le1$ and independent uniform $X,Y\in\{0,\ldots,H-1\}$,
\[
 |\mathbb E f(X)g(Y)e(hXY/P)|^2\le
 H^{-1}+2H^{-2}\sum_{d=1}^{H-1}\min\{H,(2\|d\theta\|)^{-1}\}.
\]
Group the $d$'s in consecutive blocks of length $\lfloor1/(2\theta)\rfloor$,
with the final block possibly shorter. The circle points $d\theta$ in one block
are $\theta$-separated. Summing in distance intervals of width $\theta$ gives
$O(H+\theta^{-1}\log(2/\theta))$ per block, using $H$ in the first interval.
There are $O(1+H\theta)$ blocks. Since $1/P\le\theta$, the squared form is
$O(\theta+(1+N)/H+(1+N)/(P\theta))=O((1+N)2^{-r})$.
After summing all frequencies, the discrepancy for $k\le\delta_\eta N$ is
\[
 \ll (1+N)^{3/2}2^{-r/4}+H^{-1}+2^{2r-T}
      +(1+r)M\sqrt{1+N}\,2^{-r/2}.
\]
Here $r\ge\eta N/8-1$, $T\ge8r$, and $M\le2^{\eta N/64}$,
so the right side is at most $2^{-\eta N/64}$ for large $N$.

\ificmlextended
The all-mask binary interior result follows by using the dense
estimate above for the remaining masks.
\fi

\paragraph{Sparse LSB-first masks in every fixed radix.}
Use $r$ for the most significant selected position, indexed from
the least significant digit, and set $Q=b^r$.
Translation by $Q/b$ multiplies the product phase by $e(a/b)$,
where $a$ is the nonzero character index at position $r$.
Thus only frequencies $h\equiv a\pmod b$ are present. In particular
$b\nmid h$, so some prime $p\mid b$ has $v_p(h)<v_p(b)$.
The reduced denominator $Q'=Q/\gcd(h,Q)$ then satisfies
$v_p(Q')=rv_p(b)-v_p(h)\ge r$ and $Q'\ge2^r$.
At that frequency the complete bilinear matrix is a column
permutation of the $Q'$-point Fourier matrix. If $Q'\ge H=b^N$,
restriction to the operand grid bounds the normalized correlation
by $\sqrt{Q'}/H$. If $Q'<H$, aggregate each operand test by
residues modulo $Q'$. Each aggregate has modulus at most
$\lceil H/Q'\rceil\le 2H/Q'$, giving a bound $4/\sqrt{Q'}$.
Both bounds are exponentially small for
$\eta N\le r\le(2-\eta)N$, since $Q'\le b^r$.
The sparse Fourier mass can absorb this decay by taking a small
enough degree fraction.

\begin{corollary}[Least-significant-first scratchpads]\label[corollary]{cor:icml-lsb}
The linear lower bound in \cref{thm:cot-avg}, including exponential
failure with a small linear deletion budget, holds for
least-significant-first product generation in every fixed integer
radix $q\ge2$. The upper bounds of \cref{thm:cot-avg} also hold
in this order.
\end{corollary}
\begin{proof}
Apply the sparse LSB-first estimate at a linear block of interior
positions and use the low-degree transfer theorem. The packed and
bit-parallel radix constructions both store the result before
emitting it (\cref{app:icml-radix-edit,cor:icml-vocab}), so their
final addresses may follow any prescribed permutation.
\end{proof}

\begin{corollary}[Fixed output orders with an interior block]
\label[corollary]{cor:icml-permutation}
Index digits from the least significant end. Fix $\eta,\xi>0$
and an input-independent output permutation. Suppose it has at least
$\xi N$ interior targets in $[\eta N,(2-\eta)N]$ such that, at each of
those targets, the indices generated earlier avoid at least one of the two
extreme bands outside that interval. The band avoided may depend on
the target. Each such position has exponentially small next-digit
advantage under fixed depth, heads, and fan-in and a genuine prefix.
The edit-tolerant linear lower bound and the upper bounds of
\cref{thm:cot-avg} hold for this order.
\end{corollary}
\begin{proof}
For each tested target the associated character moment contains that
digit and a sparse set of digits generated earlier. If those indices
avoid the high extreme band, the most significant selected position
lies in the interior, so the residue estimate applies. If they
avoid the low extreme band, the least significant selected position
lies in the interior, so the sparse MSB-first Fej\'er estimate
applies. The single-readout part of \cref{thm:icml-transfer} gives
the prediction bound. For generation, the transfer
proof uses the last tested digit in the \emph{generation} order of
each moment. Its earlier factors, including those from the sparse
approximant, fall under the corresponding estimate for that target.
The two fixed-address constructions store the product before
emitting any digit, so their final addresses may be permuted.
\end{proof}

\ificmlextended\subsection{Independent operands above the collision-entropy threshold}
\label[appendix]{app:icml-source-threshold}

For a probability mass function $p$, write $H_{2,q}(p)=-\log_q\sum_z p(z)^2$.
Index product digits by their place $q^j$, starting at $j=0$.
A width-$m$ operand block at offset $a$ has the form
$X=x_0+q^a U$, with $0\le U<q^m$ and the corresponding $m$ digits
of $x_0$ zero. All other digits of $x_0$ may be prescribed.

\begin{theorem}[Hardness from two independent operand blocks]
\label[theorem]{thm:icml-source-threshold}
Fix a radix $q\ge2$ and constants $\tau,\delta>0$. In two $N$-digit
operands fix any digits outside one width-$m$ block per operand, where
$m\ge\tau N$. Write $X=x_0+q^aU$, $Y=y_0+q^bV$. Suppose $U,V$
are independent and
\[
              H_{2,q}(U)+H_{2,q}(V)\ge (1+\delta)m.
\]
There are $\kappa,\rho,c>0$ depending only on $q,\tau,\delta$
such that the following holds for all sufficiently large $N$.
For every $|j-a-b-m|\le \kappa m$, multiply a nontrivial character of
the product digit at $q^j$ by characters of at most $\rho N$ earlier
product digits in MSB-first order. The resulting phase $W$ satisfies
\[
               |\mathbb E f(X)g(Y) W(XY)|\le e^{-cN}
               \qquad (|f|,|g|\le1).
\]
Write $B=\min_{p\mid q}p^{v_p(q)}$, where $p$ ranges over primes. In
LSB-first order the same moment estimate holds for target character
index $\alpha$ whenever $p^{v_p(q)}\nmid\alpha$ for every $p\mid q$.
At fixed depth, heads, and fan-in the true-prefix accuracy bound in the
window is $1/q+e^{-\Omega(N)}$ MSB first and $1/B+e^{-\Omega(N)}$
LSB first. Here $B=q$ for every prime power. In both orders and every
fixed radix the chance of a free run within $c'N$ deletions on each
side of the true product is $e^{-\Omega(N)}$ for sufficiently small
$c'>0$, with any finite scratch vocabulary.
\end{theorem}

\begin{proof}
Put $H=q^m$, $P=q^{2N}$, $D=q^j$, $D_0=q^{j-a-b}$ and $Q=P/D$.
We can assume $0<\delta\le1$. Take $\kappa=\delta/32$, so
$H^{1-\kappa}\le D_0\le H^{1+\kappa}$. The product of the Euclidean
norms of the two probability vectors is at most $H^{-(1+\delta)/2}$.
For the phase $e(tXY)$, where $e(z)=\exp(2\pi i z)$, the constant
and the terms linear in $U,V$ can be absorbed into separate bounded tests.
Thus any such weighted expectation is at most $H^{-(1+\delta)/2}$
times the operator norm of $(e(tq^{a+b}uv))_{0\le u,v<H}$.

We use two elementary norm bounds. At a rational frequency of reduced
denominator $s$, the $H$-row matrix is obtained from a permuted
$s$-point Fourier matrix by restricting or repeating residue classes
in each coordinate. Its norm is at most
\[
             \lceil H/s\rceil\sqrt{s}\le 2\max(\sqrt{s},H/\sqrt{s}).
\]
At frequency $0<\theta<1/2$, the squared norm is at most
\[
 C(1+H\theta)\bigl(H+\theta^{-1}\log(2/\theta)\bigr).
\]
Indeed a Gram entry at row difference $d$ is bounded by
$\min(H,(2\|d\theta\|)^{-1})$, with the value $H$ at an integer.
In each consecutive block of $\lfloor1/(2\theta)\rfloor$ differences
the phases are $\theta$-separated on the circle. Summing by distance
to zero bounds the row sum by the displayed expression (enlarging
$C$ if necessary); the same bound holds on every column. The same
estimate applies for negative frequency by conjugation.

For MSB-first order, regard $W$ as a step function on the unit circle
evaluated at $XY/P$, constant on the $Q$ cells. Let $A(h)$ be the modulus
of its normalized discrete Fourier coefficient on the cell indices, and
let $A_0(h)$ omit the character at the last selected digit.
The sparse Fourier-mass estimate in \cref{app:icml-interior} bounds both
corresponding sums over $h\bmod Q$ by $H^{\kappa/4}$ for small enough
fixed $\rho$ and large $N$. This follows since $m\ge\tau N$ and the
logarithm of that mass is $O_q(\rho N\log(2/\rho)+\log N)$.
Writing $c_h$ for the circle Fourier coefficients, integration within a
cell gives $|c_h|\le A(h)\min(1,Q/|h|)$ for $h\ne0$, and $c_0=0$.
For $|h|\le QH^{-\kappa}$ the digit factorization gives
\[
        A(h)\le C_q H^{-\kappa} A_0(h).
\]
In fact the factor for the last digit is
$|q^{-1}\sum_{s=0}^{q-1}e(s(\alpha/q-h/Q))|$ with $\alpha\ne0$,
which is $O_q(|h|/Q)$ by cancellation at zero. The corresponding factor
with $\alpha=0$ is bounded below by $1/2$ throughout this range for
large $N$, and all other factors agree. The total Fourier mass of these
low frequencies is therefore $O_q(H^{-3\kappa/4})$.

Let $F_K$ be the unit-integral Fej\'er kernel of order
$K=\lceil QH^{2\kappa}\rceil$. We next show that replacing $W$ by
$F_K*W$ costs $O(H^{-\kappa})$ in expected absolute error.
Put $l=\lfloor H^\kappa\rfloor$. For $1\le t<l$,
$t/D_0$ has reduced denominator between $H^{1-2\kappa}$ and
$H^{1+\kappa}$. The rational norm estimate then gives
$|\mathbb E e(tXY/D)|\le 2H^{-(\delta-2\kappa)/2}$.
Since $F_l(z)\gg l$ for $\|z\|\le1/(2l)$ and its coefficients have
modulus at most one,
\[
  \Pr[\|XY/D\|\le1/(2l)]\ll l^{-1}+ H^{-(\delta-2\kappa)/2}.
\]
Outside this event, $XY/P$ has distance at least $1/(2lQ)$ from
every step boundary, so $F_K(z)\le1/(4K\|z\|^2)$ bounds the
pointwise smoothing error by $O(lQ/K)$. On the event the error is
at most two, proving the claim.

Only $QH^{-\kappa}<|h|<K$ remain in the smoothed Fourier expansion.
Their bilinear frequencies $\theta=|h|q^{a+b}/P$ satisfy
$H^{-1-2\kappa}<\theta\le 2H^{-1+3\kappa}$. Expanding the second
norm bound gives $O((1+m)H^{1+3\kappa})$ for its square. The
weighted expectation at each frequency is thus
$O(\sqrt{1+m}\,H^{-(\delta-3\kappa)/2})$.
Periodicity of $A$ and the within-cell coefficient bound give total
mass $O((1+\log(2K/Q))H^{\kappa/4})$ throughout this range.
All contributions are exponentially small.

For LSB-first order the phase is a function on residues modulo $q^{j+1}$.
Translation by $q^j$ multiplies it by $e(\alpha/q)$, so only
$h\equiv\alpha\pmod q$ occur. For every $p\mid q$ the hypothesis
gives $v_p(h)=v_p(\alpha)<v_p(q)$. Consequently the reduced denominator
of $h/q^{j+1-a-b}$ lies between $D_0$ and $qD_0$. The rational norm
estimate bounds each weighted harmonic by $O_q(H^{-(\delta-\kappa)/2})$.
The sparse Fourier mass bound gives the stated LSB-first moment estimate.

The tested window contains $\Theta(m)=\Theta(N)$ valid product places
because $0\le a,b\le N-m$. In MSB-first order apply
\cref{thm:icml-transfer}, absorbing operand characters into $f,g$.
For LSB-first order use the nonnegative trigonometric polynomial
\[
        K_B(z)=\left|B^{-1}\sum_{t=0}^{B-1}e(tz/q)\right|^2,
        \qquad z\in\mathbb Z/q.
\]
It majorizes $\mathbf1_{z=0}$, its constant coefficient is $1/B$,
and all coefficients are nonnegative with sum $K_B(0)=1$. Its nonzero
frequencies have representatives $d$ with $0<|d|<B$, hence
$p^{v_p(q)}\nmid d$ for every $p\mid q$. For one digit expand
$K_B(\text{digit}-\text{prediction})$ in place of its equality
indicator in the single-readout proof of \cref{thm:icml-transfer}.
For generation majorize the surviving digit equalities by the product
of these kernels in the replay of \cref{app:icml-cot-edit}. The
all-constant term becomes $B^{-|I'|}$ for surviving positions $I'$.
Every other term has an allowed character on its last tested digit.
The earlier factors may be arbitrary digit characters, and all coefficients
together have mass one. Thus the master bound
\eqref{eq:icml-general-alignment} holds with $B$ in place of $q$ in
its trivial term; the marked-position count still uses the actual alphabet
size $q$. Since $B\ge2$, a sufficiently small positive linear radius
still gives exponential decay.

\end{proof}

In particular take $m=N$, $a=b=0$. Combined collision entropy
$(1+\delta)N$ suffices even if one operand supplies only $\delta N$
units and the other supplies $N$. At the endpoint $N$, a uniform
operand and an identically zero operand have identically zero product.
Even with both entropies $N/2$ for even $N$, taking both operands
uniform below $q^{N/2}$ makes the digit at $q^N$ identically zero.
These examples give the sharp threshold for a guarantee based only
on the combined entropy rate.

For independent operands with i.i.d. digits inside the two blocks,
the sum of the two per-digit base-$q$ Shannon entropies may replace
the collision-entropy rate in the strict inequality above. Indeed,
after omitting zero-probability letters, self-information per digit is
bounded for each fixed digit law. Its lower tail below the mean by
a fixed constant is exponentially small by Hoeffding. Conditioning
on the typical part separately for each operand leaves independence
and gives min-entropy, hence collision entropy, at least the original
Shannon entropy minus an arbitrarily small linear slack. Returning the
discarded mass changes each estimate by at most $e^{-\Omega(m)}$.
For example, independent binary digits with probability $1/8$ of one
have Shannon entropy greater than one half per bit. The theorem permits
any mixture of the stated fixed-outside-digit laws for the generation
conclusion by conditioning and averaging, provided $\tau,\delta$ are
uniform across the mixture.

\paragraph{Composite radix and stronger entropy.}
Use the independent block setting and notation of the collision-entropy
theorem, and write $H=q^m$ and
$\sigma=(H_{2,q}(U)+H_{2,q}(V))/m$.
For $\alpha\not\equiv0\pmod q$ define
\[
 G_\alpha=\prod_{p^e\parallel q,\ p^e\nmid\alpha}p^e,
 \qquad B=\min_{p^e\parallel q}p^e .
\]
The LSB-first sparse-mask moment is exponentially small for target
character $\alpha$ if $\sigma$ exceeds $2-\log_q G_\alpha$ by any
fixed positive margin. In particular, the LSB-first next-digit
accuracy bound improves to $1/q+e^{-\Omega(N)}$ when $\sigma$ exceeds
$2-\log_q B$ by a fixed positive margin. Uniform inputs have
$\sigma=2$. This result in the block window uses only the elementary
rational norm estimate from the collision-entropy proof.

Indeed, translation by $q^j$ restricts every contributing harmonic
to $h\equiv\alpha\pmod q$. Put $d=j+1-a-b$. For each prime counted
in $G_\alpha$, $v_p(h)=v_p(\alpha)<e$. The reduced denominator $s$
of $h/q^d$ thus satisfies $q^{-1}G_\alpha^d\le s\le q^d$.
With $|j-a-b-m|\le\kappa m$, the upper bound is
$qH^{1+\kappa}$ and the lower is
$q^{-1} H^{(1-\kappa)\log_q G_\alpha}$. The normalized rectangle
bound per harmonic is
$2H^{-\sigma/2}\max(\sqrt{s},H/\sqrt{s})$.
Both terms are exponentially small when
$\sigma>\max(1+\kappa,2-(1-\kappa)\log_q G_\alpha)$ with strict
uniform slack. Choose $\kappa$ then the maximum mask density
sufficiently small to absorb the sparse Fourier mass. A union over
the finitely many permitted target characters gives common constants.
Since $G_\alpha\ge B$ for every nonzero character, ordinary Fourier
expansion of equality proves the full-baseline corollary. More generally,
if $A\subseteq\mathbb Z/q$ and every nonzero difference of elements
of $A$ meets the moment condition, the nonnegative majorant
$|\sum_{t\in A} e(tz/q)/|A||^2$ yields baseline $1/|A|$ by the same
transfer argument. The earlier $1/B$ bound uses $A=\{0,\ldots,B-1\}$.

\emph{A full-prefix obstruction.}
For any nontrivial proper unitary divisor $R\mid q$, put $S=q/R$,
so $\gcd(R,S)=1$. Take $a=b=0$, $m=N$, and independent
$X,Y$ with $Y$ uniform below $q^m$ and $X$ uniform on the multiples
of $R^{m+1}$ below $q^m$, including zero. The combined base-$q$
collision entropy is $(2-\log_q R)m-O_q(1)$. For $j=m$ let
$Z=XY\bmod q^m$ be the full genuine LSB prefix and
$t=\lfloor XY/q^m\rfloor\bmod q$ the next digit.
As $R^{m+1}$ divides $XY$ and $q^{m+1}$,
$Z+q^m t\equiv 0\pmod {R^{m+1}}$. Thus $R^m\mid Z$ and
\[
             t\equiv - (S^m)^{-1}(Z/R^m)\pmod R.
\]
The observed prefix leaves at most $S$ possible digits, so its Bayes
success is at least $1/S=R/q$. With $R=B$, the example identifies a primary-factor obstruction at
rate approaching $2-\log_q B$. With $R=q/B$, the success is at least
$1/B$ at combined
rate approaching $1+\log_q B>1$. The calculation does not implement
the modular reduction of the full prefix by a fixed-depth bounded-
selection reader. Whether its $1/B$ guarantee under merely superunit
entropy is best possible for that architectural class remains open.
\fi
\subsection{Scratchpad transfer with growing resources}
\label[appendix]{app:icml-radix-edit}

The alignment proof also allows growing depth and fan-in. The key point is
that its approximant has small degree even after adding the characters of
earlier tested targets. The proof therefore needs only the sparse arithmetic
estimates.

\icmlgrowingstatement

\begin{proof}[Proof of \cref{thm:icml-growing-scratchpad}]
Apply the master alignment bound \eqref{eq:icml-general-alignment}.
Choose $b$ of the qualifying targets, where $b$ will be sublinear and
$r\le b$. Use a coefficient budget exponential in the record count. A probe on
at most $b+r$ top
records starts with record budget $m_0=2b$ and coefficient mass one.
Let $g=\ln(4n)$, $K=4ag$ and $v=b$. At a layer with record budget $m$
and coefficient mass at most $e^{gm}$, choose the uniform error of
each conjunction to be $e^{-\lambda}$, where
$\lambda=v+\ln(2L)+amg$. At most $e^{(a-1)mg}$ union guesses per
term give error at most $e^{-v}/(2L)$ in this layer. By
\cref{lem:nor-small}, the replacement uses at most
$c\sqrt{n\lambda}$ additional records per term for an absolute
constant $c\ge1$. A degree budget $d$ multiplies the coefficient
mass of each guessed term
by at most $(4n)^d$. Thus the new budget $am+d$ preserves the
coefficient invariant.

Set $b=\lfloor n K^{-2\cdot 2^L}\rfloor$. The growth condition yields
$b=n^{1-o(1)}$, $b=o(n)$, and $b\ge\ln(2L)$ eventually, so there
are enough qualifying targets. Start an envelope at $M_0=2b$.
For $v/K\le M_j\le n/K$, both $\lambda\le3KM_j$ and
$am_j\le\sqrt{nKM_j}$, so use $M_{j+1}=3c\sqrt{nKM_j}$. It follows
that
\[
 \frac{M_j}{n}\le 9c^2 K
     \left(\frac{2b}{nK}\right)^{2^{-j}},\qquad 0\le j\le L.
\]
For $j<L$ the right side is $O(K^{-3})$ and hence respects the upper
envelope condition. The recurrence is nondecreasing on this range,
so the lower condition persists as well. At $j=L$ the bound is
$O(1/K)=o(1)$. The approximant therefore has degree at most
$m_L=o(n)$ and uniform error $e^{-b}/2$. Its supremum is at most two.
Parseval bounds its character mass by twice the square root of the
number of characters of degree at most $m_L$, which is $\exp(o(n))$
by subset entropy and the fixed alphabet size. The two degrees together
are at most $\rho N$ for all sufficiently large $N$, even
after the last-target argument multiplies the approximant by the
earlier tested characters. With $\epsilon=e^{-\gamma N}$, the total
nontrivial-moment bound in \eqref{eq:icml-general-alignment} is therefore
$e^{-b}/2+\exp(-\gamma N+o(n))$. This uses only the stated low-degree
hypothesis.

Take
\[
 r=\left\lfloor\frac{\min(1,\ln q)b}
 {64(L+2)a^L(g+\ln q)}\right\rfloor .
\]
The growth condition implies $\ln(L+2)+L\ln a=o(\ln n)$,
so $r=n^{1-o(1)}$. For
$A=(2r+1)S(n,r)^2 C_r$, the definition of $C_r$ charges at most
$a^Lr\ln q$ for raw symbol values and at most $La^Lrg$ for marked
position sets. Adding the mask and length counts gives
\[
 \ln A\le \ln(2r+1)+(L+2)a^Lr(g+\ln q)
          \le o(b)+\min(1,\ln q)b/64.
\]
Now $r\le b/64$ and the first term in \eqref{eq:icml-general-alignment}
is at most $A q^{-(b-r)}$. Hence every term in that bound is
$\exp(-\Omega_q(b))$ or smaller.
\end{proof}

\begin{theorem}[Radix and depth tradeoff]\label[theorem]{thm:icml-radix-edit}
Let two independent uniform $N$-digit operands in a fixed radix $q\ge2$
be given with one token per digit. The padded $2N$-digit product may be
generated in either digit order, or in any prescribed order satisfying
the interior-target condition of \cref{cor:icml-permutation}.
For fixed $L,H,k$ there are $c,c'>0$ such that the probability that some
free run admits a radius-$\lfloor cN\rfloor$ alignment is at most
$e^{-c'N}$ for all sufficiently large $N$. Under the growth condition
of \cref{thm:icml-growing-scratchpad} the radius is $n^{1-o(1)}$ and the
success bound is $\exp(-n^{1-o(1)})$, where $n=\Theta(N)$. Both bounds
allow a growing scratch vocabulary and cover $s$ interleaved scratch
tokens and $e$ Levenshtein edits whenever $s+e$ fits within the radius.
\end{theorem}

\begin{proof}
The strict-prefix character and a nontrivial target character multiply
to a sparse product-digit mask containing that target. On a qualifying
target, either its most significant selected position or its least
significant selected position lies in the fixed-margin interior, by
the prescribed-order condition. The sparse residue or Fej\'er estimate
of \cref{app:icml-interior}, respectively, bounds its correlation with
the two operand characters by $e^{-\Omega(N)}$ uniformly up to a small
linear degree. Reducing the degree cutoff by a constant factor leaves
room for the extra target character. Either monotone digit order has a
linear number of qualifying interior targets. Apply
\cref{thm:icml-transfer,thm:icml-growing-scratchpad}. For the
Levenshtein formulation, delete the scratch tokens from the raw output
and delete both sides of each substitution to realize the alignment
with at most $s+e$ deletions per side.
\end{proof}

\paragraph{Matching construction in every fixed radix.}
For $N\ge q$, replace the binary packing in \cref{app:icml-cot-packed}
by blocks of $b=\max(1,\lfloor\log_q N\rfloor)$ digits and set
$B=q^b$. Appending a digit $u$ to a partial word now uses
$v\gets qv+u$. The FFT still uses dyadic butterflies, independent
of $q$. Its error proof uses the bound $B\le N$ on the input
words, so it recovers the ordinary integer convolution. Propagate
carries in base $B$, then emit radix-$q$ digit $j$ by reading the
word at $\lfloor j/b\rfloor$ and taking its digit at offset
$j\bmod b$. The transform uses $O_q(N/\log N)$ entries, giving
$O_q(N)$ instructions and $O_q(\log N)$ bits per auxiliary token.
The same causal, fixed-routing one-head implementation executes
the instructions, using the limb implementation of \cref{app:icml-cot-packed} with constant width and $O_q(N)$ vocabulary. For the finitely
many $N<q$, use a fixed product lookup.

\subsection{Growing input and output word sizes}\label[appendix]{app:icml-packed-replay}
\begin{theorem}[Growing binary words]\label{thm:icml-packed-replay}
Fix a depth $L$, a head bound $H$, and a per-head selection bound $k$
in the causal position-pair top-$k$ model with fixed tie priority.
Let two independent uniform $B$-bit integers be given in a fixed layout
of width-$w$ words, where $1\le w\le B$ and $p=\lceil B/w\rceil$.
One may use up to $2p+O(1)$ input positions, each a predetermined function
of at most $w$ designated operand bits. Positions may overlap.
The target is the padded $2B$-bit product in consecutive words of
width $w$ in either digit order, with a possible shorter last word. Format-dependent
word embeddings and scratch symbols may be arbitrary. For some constants
$c,C>0$ depending on the architecture, and all sufficiently large $p$,
the probability that any generated string admits an alignment deleting
at most $cp$ generated tokens and $cp$ target words is at most
$\exp(-Cp)$. The constants are uniform in $B,w$. The target words
have a fixed injective serialization at each position and retained words
feed back by that serialization. There is one predetermined model per
candidate total length, which the operands may select. An independent
random tape may be fixed before the argument and averaged afterward.
\end{theorem}

For example $w=\lceil\log_2 B\rceil$ requires $\Omega(B/\log B)$
scratch tokens for constant exact-product success, even when the scratch
vocabulary grows without restriction. Length, substitutions, insertions
and deletions here are in tokens. Taking $s$ interleaved scratch tokens
and $e$ ordinary word edits gives the stated alignment whenever $s+e\le cp$.
For a fixed predictor and context layout independent of the operands,
a true-prefix word of width $w$ contained in $[B/2,B]$ in the chosen
generation order has
exact accuracy at most $2^{-w}+\exp(-\Omega(p))$.

\paragraph{Arithmetic in the low half.}
For an LSB-first parity with last selected index $u\le B$, use indices
$0,\ldots,u-1$ for the bits in the residue ring modulo $q_0=2^u$.
Writing $W$ for the resulting sign table, translation by $q_0/2$
negates $W$, since it toggles just the highest selected bit.
Thus its normalized discrete Fourier coefficients vanish at all even
frequencies. The binary factorization and two-digit contraction in
\cref{prop:finite-fourier} give
$\sum_h|\widehat W(h)|\le 2(2\kappa)^u$. The factorization here is the
discrete sum over all $q_0$ cells, before integration within a cell.
For odd $h$, condition each bounded operand test on its residue
modulo $q_0$. Both residues are uniform and the matrix
$(e^{2\pi i hxy/q_0})_{x,y}$ is a row permutation of the Fourier
matrix, of operator norm $\sqrt{q_0}$. Hence each normalized bilinear
term is at most $q_0^{-1/2}$. The discrepancy is at most
$2^{1-\gamma u}$ for $\gamma=-\tfrac12\log_2\cos(\pi/8)>0$.

\paragraph{An alphabet-independent consequence of position-union approximation.}
Write $a=1+Hk$. At token length at most $n$, for each fixed $\xi>0$
there is $\eta>0$ depending only on $\xi,L,H,k$, such that a bounded
probe of at most $\eta n$ top rows has, on arbitrary formal bottom
token inputs, a uniform approximant of error $e^{-\eta n}/2$ that is a
finite sum of functions each depending on at most $\xi n$ bottom rows.
The approximant has modulus at most two. This is the pre-Fourier
portion of the position-union lemma in \cref{lem:position-union}:
guess at most $Hkm$ winning positions for a term on $m$ rows,
verify with one AND over outside positions, and approximate this AND on
its Boolean test inputs. The sums of absolute coefficients and counts
of shared rows in the first $L$ pullbacks depend exclusively on $n,m,L,H,k$
and the error. Specifically, if $x=m/n$ and $y=n^{-1}\log$
coefficient mass with mass floored at one, their next upper envelopes are
$ax+z$ and $y+\Phi(Hkx)+\eta+\tfrac12\Phi(z)$, with
$z=C_0\sqrt{2\eta+y+\Phi(Hkx)}$ for the absolute AND constant $C_0$
and large $n$. Here $\Phi(t)$ is binary entropy in nats at $\min(t,1/2)$.
Start at $(\eta,0)$. At fixed depth the resulting envelopes vanish as
$\eta\downarrow0$. This argument requires no alphabet bound at its bottom.
All upper and lower positions keep their original indices and masks.

\paragraph{Symbolic marks.}
Fix output length $m$ and deletion sets $U$ on raw slots, $V$ on
target words, pairing the complements in order. The teacher-forced
replay uses the matched target words outside $U$ and generates the tokens
in $U$ using the model. On a witnessing input it agrees with the free
run. From the readout rows for $U$, mark their own rows and the members
of their ordered winner lists at every preceding layer. Guess just the
resulting sets of positions $S_0,\ldots,S_{L-1}$, with
$|S_{L-j}|\le a^j|U|$ for $1\le j\le L$. The set at level $L$ is prescribed.
For each tuple, define a total symbolic completion of the slots in $U$.
In chronological order, compute its designated readout using just the
marked rows at each lower level, allowing only rows at or before the
query. Missing residuals or too few eligible keys for the original
required list size can return a fixed fallback token. Equivalently,
propagate a failure flag to that readout; all scores, original positions
and tie priorities are retained on valid branches. At the raw level
use ordinary words on the marks and the previously completed token at
any marked slot in $U$. In a consistent tuple, every actual winner is in
the set. Layer induction recovers the correct state on each mark through
the current readout, and chronological induction recovers every symbol
in $U$. In an inconsistent tuple the fallback still defines causal
functions. No test of consistency is needed for the upper bound.

For every tuple, each completed token is a function of the ordinary
bottom tokens on $S_0$ that precede its emission. This same set of
at most $a^L|U|$ ordinary token positions can support all the completions.
In particular no future target value is read. Now evaluate ordinary
full-model retained readouts on this completed tape. The marks restrict
only the computation of the replaced raw symbols; the retained readouts
still see the full context. If a real run witnesses the alignment,
at least one tuple's completed tape passes all retained equality tests.

\paragraph{Proof of the theorem.}
Take $n=6p$, sufficient for the length including separators and the edit
budget, eventually. Then $nw\le12B$. Choose $\theta>0$ as specified below, then $\xi>0$ small enough,
and then $\eta$ from the position-union estimate,
decreasing it to at most $1/40$. Test $b=\lfloor\eta n\rfloor$ full
target words within bit positions $[B/2,B]$ in generation order. For large $p$ the interval
contains at least $p/2-3$ such words. Fix an integer radius $r\le s n$
with $s>0$ selected later. For fixed masks and marks, expand equality
on the $b'\ge b-r$ surviving words using the $2^w$ binary Walsh
characters per word. A readout outside the valid word tokens is mapped
to a fixed word for the equality test only. The normalized expansion
has total coefficient mass one, with constant term $2^{-wb'}$.

For a nontrivial summand, let $u$ be the latest index in generation order of any
selected true bit in its last nontrivial word. Product signs in that word
are lifted to the full bit cube. All selected predicted-word characters
are one bounded probe of at most $b$ retained top rows in the same
completed causal context through that word's readout. It omits all true
bits in its current word. Use the alphabet-independent approximation
on formal raw slots of that context, then substitute the symbolic
completions at the slots in $U$. Each bottom term now depends on at most
$(\xi n+a^Lr)w$ genuine operand or strict-prefix bits: take the
ordinary positions of the term and add the common marked set $S_0$
once. Restrict this set to the causal prefix at hand. The approximant
is bounded by two and retains its uniform error pointwise after
substitution, regardless of overlap. Multilinear degree is at most
$(\xi n+a^Lr)w$ because every term depends on that many bits.
On the full cube of the $2B$ operand bits and $u-1$ strict-prefix bits
unseen earlier bits are independent formal coordinates. If
$12(\xi+a^L s)\le 3\theta$ with $0<\theta<1/2$, Parseval on a cube
padded to dimension $3B$ bounds its Walsh mass by
$2\exp(3B\Phi(\theta)/2)$. Multiplying by the true selected bits
strictly before $u$ leaves this mass invariant. The coefficient at
the target couples to the actual $u$th bit using the all-mask
first-half estimate in \cref{thm:main} for MSB first or the
residue estimate above for LSB first. Either is at most
$e^{-c_{\rm d} B}$ for $u\in[B/2,B]$ and a fixed $c_{\rm d}>0$.
Choose $3\Phi(\theta)/2<c_{\rm d}/2$, then sufficiently small $\xi$
and $s$. The contribution is bounded in absolute value by
$e^{-\eta n}/2+2e^{-c_{\rm d} B/2}$.

Write $S(d,t)=\sum_{i=0}^{\min(d,\lfloor t\rfloor)}\binom di$.
The total union factor for lengths, masks and marked position tuples
is at most
\[
 P=(2r+1)S(n,r)^2\prod_{j=1}^L S(n,a^j r).
\]
There is no raw value factor. Thus the probability is bounded by
$P[2^{-w(b-r)}+e^{-\eta n}/2+2e^{-c_{\rm d} B/2}]$. Since $B\ge n/12$,
the bracket is exponentially small in $n$ for $s<\eta/4$. Its rate
is bounded below uniformly when $s$ decreases further.
The exponential rate of $P$ is at most
$2\Phi(s)+L\Phi(a^Ls)+o(1)$, which tends to zero with $s$.
This proves the theorem. For a single word omit the masks and marks
and expand equality at that word alone.

\paragraph{What the marks are buying.}
Enumerating $2^{wa^Lr}$ possible values of the raw marks before
a uniform approximate pullback could erase the $e^{-\eta n}$ gain.
Conversely, ignoring the raw values entirely inside the predictor
is invalid. For example a deleted token can copy an arbitrary operand
word at a fixed winning row, then a retained output can copy the deleted
token. The fixed marked positions carry no such value as constants.
In the symbolic substitution that operand word is counted once among
the common ordinary raw variables.

\paragraph{Matching word count with unrestricted vocabulary.}
For ordinary injective packed input words, one layer and one head
suffice with $2p+O(1)$ scratch tokens. In the accumulation stage the
current raw query is a scratch symbol containing all operand words
seen so far. The head reads the next operand word at a fixed address
and a row-local update emits the enlarged accumulator. A fixed prompt
symbol initializes this procedure. After the last word, the head at
each output step selects the final accumulator at its fixed address.
The local update now multiplies and emits the requested word in the
chosen order. This uses at most $2^{O(B)}$ distinct scratch symbols
and unrestricted local computation. Hence the unrestricted-vocabulary
scratchpad complexity for these layouts is $\Theta(B/w)$ in either
order when $B/w\to\infty$.

\section{Scratchpads with deterministically rounded dense attention}
\label{app:icml-rounded-dense}

We give upper and lower bounds for the following implemented-arithmetic
model. The input presents contiguous blocks for $X$ and $Y$, fixed
separators, and a fixed delimiter. All auxiliary tokens precede the product,
which is generated most significant bit first as literal binary tokens.
The receiver for the next emission is the last available position and
attention is inclusive causal. In a single layer, a source record depends
on its token and position alone. The receiver and its raw token may determine
the score rule. The readout receives that local record and the head outputs.
Parameters may depend on $N$. In this section arithmetic is deterministic
and fixed-point. The robustness results in
\cref{app:dense-scratchpad} instead quantify over adversarial head errors.

\begin{theorem}[Rounded dense scratchpad tradeoff]
\label{thm:icml-rounded-dense}
Suppose each row with its receiver position and raw token fixed can be
evaluated in a constant number $d$ of left-to-right passes with $c\ge1$
persistent bits, positions and fixed parameters being public. If at most
$S$ auxiliary tokens are used and the probability of exact multiplication
on independent uniform $N$-bit operands is at least a fixed $a>0$, then
$(S+1)c=\Omega_a(N)$ for fixed $d$.

With a constant number of constant-width heads and $O(\log N)$-bit
fixed-point words, a polynomial-time constructible family computes every
product exactly with $O(N/\log N)$ auxiliary tokens when the score has
a learned relative-position bias. The same resources permit
$O(N/\log\log N)$ tokens with ordinary dot-product scores and only
absolute position tables. Both constructions use length-dependent
parameters and a polynomial-size untied vocabulary.
\end{theorem}

The usual stable row calculation first finds the maximum logit, then
accumulates rounded exponential masses, then accumulates values weighted
by rounded normalized masses. A sequential left fold needs
$c=O(H(w+1)p)$ bits for $H$ heads of value dimension $w$ and $p$ bits
per word, including exponent or integer range. Thus the relative-bias
bound is tight at constant hidden size and logarithmic word precision
for this order of evaluation. Other accumulation orders require
their own live-state bound.

\subsection{Lower bound from scan memory}

We use the balanced-prefix-table norm bound from
\citet{ArithmeticRobustnessCompanion}. Specifically, if a real table $w$
on the $2^t$ dyadic cells in $[0,1)$ is bounded by one, sums to zero on
each sibling pair, and has matrix rank at most $R$ over every cut of its
$t$-bit address, its product kernel $K_w(x,y)=w(xy)$ has
$L^2[0,1]$ operator norm at most $C\sqrt R\,2^{-t/4}$.
Here is the reduction from that bound to the row-memory statement.

With all random coins fixed, simulate the scratch generation between two
parties who hold $X$ and $Y$. The receiver and its raw token are initially
public. For each pass the first party scans $X$ and transmits its $c$-bit
exit state. The other scans $Y$ and the public scratch symbols and
transmits the state for the next pass. Fixed separators can be treated
by either party at the corresponding position. The readout makes the
new scratch token public from this transcript; its name is never
transmitted. A public stop decision costs at most one further row.
There are at most $2^{O(d(S+1)c)}$ transcript rectangles, each fixing
the entire scratch string and its emission schedule.

Fix a transcript rectangle and test the last $b$ positions through bit
$N$, for a constant $b$ to be chosen. Run each selected row on its
\emph{genuine} preceding product prefix. A scan over $X,Y$ and a formal
prefix can evaluate these rows in $d$ passes and $m=O(b(c+1))$ bits:
the literal preceding bit of each receiver takes only two values.
Run both possible row scans in parallel until that preceding bit is
available, then use the actual one. Stop each row calculation at the
specified inclusive receiver. Reinsert the fixed scratch string between
$Y$ and the prefix for each row. Predictions outside the bit alphabet
can be mapped to arbitrary bits when upper bounding the success event.

For a nonempty subset of the $b$ positions let $t$ be its last position.
The signed agreement moment is the expected product of $(-1)^{M_t}$
and a width-$W=2^m$ scan of the operands and the prefix before $t$.
The scan computes all selected prediction signs and incorporates earlier
target signs from this prefix. Each of its $d$ passes reads $X,Y$, then
that prefix. Record the exits of all but the final block. There are at
most $K=W^{3d-1}$ formal exit transcripts. Their operand checks specify
rectangles, and any operand pair belongs to at most $J=W^{d-1}$ of
them, since the first $d-1$ prefix exits determine all operand exits.
Intersect each with the fixed scratch rectangle. For a fixed exit
transcript, a one-pass width-$W^d$ machine on the formal prefix checks
its claimed exits and returns a sign if they all agree, and zero
otherwise. Its product with the final bit sign is a balanced cell table
$w$ with every address-cut rank at most $W^d$ (up to an absolute factor
at the final bit), since its row is determined by the parallel state at
the cut. The smoothed contribution from a rectangle of mass $q_i$ is
bounded by $C\sqrt{q_i}W^{d/2}2^{-t/4}/\rho$ when each operand is
extended to its positive interval of relative width $\rho\le1$.
By Cauchy--Schwarz, $\sum_i\sqrt{q_i}\le\sqrt{KJ}$.

For each operand pair and formal prefix there is exactly one consistent
exit transcript. The summed table has modulus at most one and is
constant on the common level-$t$ cells. Consequently the positive
cell smoothing incurs an $O(\rho)$ boundary error once for the sum,
since $t\le N$ by the test choice. For completeness, write
$Q=2^{2N-t}$. For each $x$, any crossing needs
$xy\equiv -j\bmod Q$ with $1\le j\le 2^{N+1}\rho$.
With $g=\gcd(x,Q)$, only at most $2^{N+1}\rho/g$ values of $j$
qualify; each allows at most $1+2^Ng/Q$ values of $y$.
The implied probability is at most $2\rho(1+2^{t-N})$.
Optimizing $\rho$ now bounds the \emph{unconditional} moment on the
scratch rectangle by $O(W^{(5d-2)/4}2^{-t/8})$; if the minimizing
choice exceeds one the bound follows from the trivial estimate.

Exact free-running success implies all $b$ genuine-prefix agreements.
The empty term of their expanded joint moment is one, by the partition.
The nonempty terms and $t\ge N-b+1$ give probability at most
\[
  2^{-b}+ C\,2^{O(d(S+1)c+db(c+1))-(N-b+1)/8}.
\]
Choose $b=\lceil\log_2(2/a)\rceil$ and then a sufficiently small
constant in the contrary assumption $(S+1)c<\delta_a N$.
The bound is eventually below $a$. The proof holds on each fixed
independent random tape, if such a tape is allowed. Endogenous stopping
computed from the public simulation is covered. If an external rule
selects any of $S+1$ fixed-length models on the input, union-bound the
last display with $b=\lceil\log_2(2(S+1)/a)\rceil$; this gives the
same order since $\log_2(S+1)=O(S+1)$.

\subsection{Constructive upper bounds}

\paragraph{Recovering a positive integer sum.}
For a scheduled row, give each designated source an integer mass
$u_i\le N^{O(1)}$ via logit $\log u_i$ and bounded integer values $v_i$.
The sum of designated masses $Z$ is public. Suppress unrelated
positions by a sufficiently large multiple of $\log N$. On a common
mesh $\Delta=N^{-C}$ (rounded down to a binary power), score error
$N^{O(1)}\Delta$, leakage $N^{-\Omega(1)}$, and stable coefficient
rounding each perturb the computed mean by inverse-polynomial
amounts. More explicitly, rounding each max-shifted exponential
costs $O(\Delta)$ and the true denominator is at least one, so the
$O(N)$-term denominator error is $O(N\Delta)$. Rounded normalized
coefficients have error $O(N\Delta)$ apiece, and the values and $Z$
are polynomially bounded. Summing grid values incurs no extra roundoff
in range. Making the grid finer and the penalty larger by constant
factors in $\log N$ ensures error below $1/8$ after multiplication
by $Z$. Store all words in polynomial range.
Quadratic mismatch penalties $-A(i-j)^2$ are realized by a constant
number of dot-product coordinates, using public stage-and-group IDs;
different stages and irrelevant positions have different IDs.

\paragraph{Packing and convolution.}
Pack both operands into $m=\lceil N/b_0\rceil$ pairs of
base-$U$ limbs, with $b_0=\max\{1,\lfloor\log_2(N+1)/8\rfloor\}$
and $U=2^{b_0}$. A row for each limb pair weights raw bits in that
limb by their powers of two. Use a transform length $K$, the least
power of two at least $2m$. Choose a prime $q>m(U-1)^2$ with
$2K\mid(q-1)$ and $q=N^{O(1)}$: take a power of two $D$ divisible by
$2K$ and greater than $m(U-1)^2$, and use the least prime in the
progression $1\bmod D$ by Linnik's theorem~\citep{Linnik1944}. Searching to the first
prime and finding an element $\zeta$ of exact order $2K$ by testing
$\zeta^K=-1$ costs polynomial time. Put $\omega=\zeta^2$.

The forward kernel is
$\omega^{jk}=\zeta^{j^2}\zeta^{k^2}\zeta^{-(j-k)^2}$.
Put the nonnegative representative of $\zeta^{-(j-k)^2}$ in the
integer mass at relative displacement $j-k$. A dedicated head's bias
stores its logarithm; the actual displacement includes the known
offset between the source block and receiver block. Tag the source
block to suppress all other positions. Store $\zeta^{k^2}$ times
the two source residues in the token's values. A scheduled row
recovers integer sums in both coordinates, reduces them modulo $q$,
applies the output chirp, and stores the product of the two
transformed residues. Run the inverse with $\zeta^{-1}$ and factor
$K^{-1}$ to recover the ordinary convolution exactly, since its
coefficients are below $q$. Emit carried product limbs by a
least-significant-first pass, selecting the coefficient and the
preceding carry. The extra zero coefficient flushes the carry.
These stages use $O(m+K)=O(N/\log N)$ rows. The known
off-by-one between the position of a receiver and its emitted
token is included in the position schedule.

\paragraph{Removing the relative bias.}
Factor each length-$K$ transform in mixed radix by decimating into
residue classes, with power-of-two radices at most
$R\asymp \log N/\log\log N$, all but possibly the last equal to $R$.
For a transform of length $rn'$ group input positions by their
index modulo $r$, and write $A_k(i)$ for the length-$n'$ inner
transform on residue class $k$. Combining at a fixed $i$ gives a
radix-$r$ transform of the pre-multiplied entries
$x_k=\omega_{rn'}^{ik}A_k(i)$ under the root $\omega_r$ of order $r$,
with $\omega_\ell=\omega^{K/\ell}$. The resulting coefficient at
frequency $i+n'j$ is $\sum_k x_k\omega_r^{jk}$.
Each input token's pre-multiplication and group assignment is public.

For a radix-$r$ block choose a root $\eta$ with
$\eta^2=\omega_r$ and set $h_a=\eta^{a^2}$ for $0\le a\le2r-2$.
Solve
\[
    h_a=\sum_{u=1}^{2r-1} d_u u^a\pmod q,\qquad 0\le a\le2r-2 .
\]
The distinct positive nodes have an invertible Vandermonde matrix
over the field, so this is a polynomial-time preparation step.
With input values $v_k=\eta^{-k^2}x_k\bmod q$, one sweep at receiver
$u$ uses the ordinary dot-product score $k\log u$ and stores
$d_u\sum_k u^k v_k$ reduced modulo $q$. A second sweep at receiver
$j$ uses score $j\log u$ and recovers its sum, then reduces and
multiplies by $\eta^{-j^2}$. This produces the radix-$r$ transform
because $\eta^{-j^2-k^2}h_{j+k}=\omega_r^{jk}$. Each mass
$u^k$ or $u^j$ is at most $(2R)^R=N^{O(1)}$. The log-node grid
induces at most $R$ times its mesh in score error, so the same
positive-sum argument applies. After every sweep, the stored
integer is recovered before reuse. Field elimination imposes
no real conditioning requirement. The group gate and the
bilinear power score use a constant number of coordinates.
At most $3K$ rows per mixed-radix stage suffice, so the total is
$O(K\lceil\log_R K\rceil+m)=O(N/\log\log N)$.

\paragraph{Realizing the vocabulary and bit readout.}
Capture the tuple $y$ of rounded head words and the public instruction
in fresh residual coordinates. On the binary grid there are only
polynomially many such tuples in this constant dimension and polynomial
range. For a vocabulary entry $v$ representing a tuple, set its
affine logit to $\Delta^{-1}(2v\cdot y-\|v\|^2)$; the captured tuple
uniquely maximizes it. The coefficients and logit arithmetic are
representable in the same grid and polynomial range. The untied
input embedding of that entry contains the scheduled nearest-integer
sum and its required bounded field or limb arithmetic values.
Off-path entries have bounded values. Position biases distinguish
scratch-token emissions from the two digit candidates.

To read out digits in MSB order, use a head to select the final
product limb, and a second to sum prior emitted bits at higher
offsets $s$ in that same limb, weighted by masses $2^s$.
A fixed zero-valued sentinel supplies the remaining mass to make
the total $2U$. Leading unused bit positions are known zeros.
An affine threshold compares the recovered limb minus the
prior-bit contribution to $2^r-1/2$ for current offset $r$.
The reconstruction error is below $1/4$ in total. Induction gives
all $2N$ literal digits. Constant hidden size suffices throughout;
positional, bias and vocabulary tables are generated separately
for each $N$ in polynomial time. The supplement's
\texttt{check\_rounded\_dense.py} checks the two transform schedules,
carries and prefix extraction on complete small products.

\ificmlextended
\begin{lemma}[Mellin--Hankel bound]\label[lemma]{lem:hankel}
Let $w:[0,1)\to\C$ be bounded and measurable, let $K_w$ have kernel $w(st)$ on $L^2[0,1]$ and let
$\Wt(\xi)=\int_0^1w(u)u^{-1/2+i\xi}du$. Then $\|K_w\|\le\sup_{\xi\in\R}|\Wt(\xi)|$.
\end{lemma}

\begin{proof}
The map $(Vf)(a)=e^{-a/2}f(e^{-a})$ is unitary from $L^2[0,1]$ onto $L^2[0,\infty)$. With $s=e^{-a}$, $t=e^{-b}$,
\vsw{\begin{multline*}
\int_0^1\!\!\int_0^1f(s)\,w(st)\,g(t)\,ds\,dt\\=\int_0^\infty\!\!\int_0^\infty(Vf)(a)\,\Psi(a+b)\,(Vg)(b)\,da\,db,
\end{multline*}}{\[
\int_0^1\!\!\int_0^1f(s)\,w(st)\,g(t)\,ds\,dt=\int_0^\infty\!\!\int_0^\infty(Vf)(a)\,\Psi(a+b)\,(Vg)(b)\,da\,db,
\]}
where $\Psi(c)=e^{-c/2}w(e^{-c})$ for $c\ge0$ and $\Psi=0$ on $(-\infty,0)$. Extend $Vf,Vg$ by zero to $\R$ and put $\check G(b)=(Vg)(-b)$. The right side is
$\int(Vf)(a)(\Psi*\check G)(a)\,da$, so by Cauchy--Schwarz and Plancherel its modulus is at most
$\|f\|_2\|g\|_2\sup_\xi|\hat\Psi(\xi)|$; here $\Psi\in L^1$ because $|\Psi(c)|\le\|w\|_\infty e^{-c/2}$. Finally
$\hat\Psi(\xi)=\int_0^\infty e^{-c/2}w(e^{-c})e^{-ic\xi}dc=\int_0^1w(u)u^{-1/2+i\xi}du=\Wt(\xi)$ by $u=e^{-c}$.
\end{proof}

% Shared main-body piece (section: warm-up); input by main-body.tex (FOCS and named preprint) and by icml.tex (ICML 2027).
\section{Warm-up: multiplication becomes a convolution}\label{sec:warmup}

\vsw{We reduce the arithmetic problem to a one-dimensional Fourier transform. Full shell estimates appear in \cref{app:icml-shell}.}{This section proves the Hankel bound, the shell identity and the off-resonance bound.} (The smallest integer case, $N=2$, where the largest correlation is
$\NtwoDisc$, is also checkable by hand: \cref{ex:small}.)\vsw{ Write $z=\frac12+i\xi$, $r_k=(-1)^{b_k}$ (so that
$W_U=\prod_{k\in U}r_k$), $U^{(j)}=\{k-j:\ k\in U,\ k>j\}$ for the set $U$ shifted down by $j$, and
$J_U(\xi)=\int_{1/2}^1W_U(u)u^{z-1}du$ for the \emph{shell integral} of $W_U$ over $[\frac12,1)$.}{}

\ificml\else\begin{lemma}[Mellin--Hankel bound]\label[lemma]{lem:hankel}
Let $w:[0,1)\to\C$ be bounded and measurable, let $K_w$ have kernel $w(st)$ on $L^2[0,1]$ and let
$\Wt(\xi)=\int_0^1w(u)u^{-1/2+i\xi}du$. Then $\|K_w\|\le\sup_{\xi\in\R}|\Wt(\xi)|$.
\end{lemma}

\begin{proof}
The map $(Vf)(a)=e^{-a/2}f(e^{-a})$ is unitary from $L^2[0,1]$ onto $L^2[0,\infty)$. With $s=e^{-a}$, $t=e^{-b}$,
\vsw{\begin{multline*}
\int_0^1\!\!\int_0^1f(s)\,w(st)\,g(t)\,ds\,dt\\=\int_0^\infty\!\!\int_0^\infty(Vf)(a)\,\Psi(a+b)\,(Vg)(b)\,da\,db,
\end{multline*}}{\[
\int_0^1\!\!\int_0^1f(s)\,w(st)\,g(t)\,ds\,dt=\int_0^\infty\!\!\int_0^\infty(Vf)(a)\,\Psi(a+b)\,(Vg)(b)\,da\,db,
\]}
where $\Psi(c)=e^{-c/2}w(e^{-c})$ for $c\ge0$ and $\Psi=0$ on $(-\infty,0)$. Extend $Vf,Vg$ by zero to $\R$ and put $\check G(b)=(Vg)(-b)$. The right side is
$\int(Vf)(a)(\Psi*\check G)(a)\,da$, so by Cauchy--Schwarz and Plancherel its modulus is at most
$\|f\|_2\|g\|_2\sup_\xi|\hat\Psi(\xi)|$; here $\Psi\in L^1$ because $|\Psi(c)|\le\|w\|_\infty e^{-c/2}$. Finally
$\hat\Psi(\xi)=\int_0^\infty e^{-c/2}w(e^{-c})e^{-ic\xi}dc=\int_0^1w(u)u^{-1/2+i\xi}du=\Wt(\xi)$ by $u=e^{-c}$.
\end{proof}
\fi

\newcommand{\multiplicationshellproof}{%
\begin{icmltight}\begin{lemma}[shell identity]\label[lemma]{lem:shell}
For $\max S=T$ and $z=\frac12+i\xi$,
\[
\Wt_S(\xi)=\sum_{j=0}^{T-1}2^{-jz}J_{S^{(j)}}(\xi)+\frac{2^{-Tz}}{z},\qquad \max S^{(j)}=T-j .
\]
\end{lemma}\end{icmltight}

\begin{proof}
For $u\in[0,\frac12)$ we have $b_1(u)=0$ and $b_k(u)=b_{k-1}(2u)$, so $W_S(u)=W_{S^{(1)}}(2u)$. Substituting $u=v/2$,
$\int_0^{1/2}W_S(u)u^{z-1}du=2^{-z}\Wt_{S^{(1)}}(\xi)$, so $\Wt_S=J_S+2^{-z}\Wt_{S^{(1)}}$. Iterate $T$ times and use
$S^{(T)}=\emptyset$ and $\Wt_\emptyset(\xi)=\int_0^1u^{z-1}du=1/z$.
\end{proof}

Since $|2^{-jz}|=2^{-j/2}$, shell $j$ is a Walsh function of top bit $T'=T-j$ on $[\frac12,1)$, damped by $2^{-j/2}$.
Three elementary bounds control one shell. Fix $U$ with $\max U=T'\ge1$ and $\xi\ge0$ (the case $\xi<0$ follows by
complex conjugation).

\begin{lemma}[one shell]\label[lemma]{lem:oneshell}
\textup{(i)} $|J_U(\xi)|\le\int_{1/2}^1u^{-1/2}du=2-\sqrt2$.
\textup{(ii)} If $T'\ge2$ then $|J_U(\xi)|\le2(\sqrt2-1)\,|z-1|\,2^{-T'}$.
\textup{(iii)} $|J_U(\xi)|\le(2^{T'}-0.29)/|z|$.
\end{lemma}

\begin{proof}
(i) is immediate. (ii) Let $F(u)=\int_{1/2}^uW_U$. The interval $[\frac12,1)$ is a union of aligned blocks of length
$2^{1-T'}$; on each block $r_{T'}$ changes sign at the midpoint and the other factors of $W_U$ are constant, so $W_U$ has
mean zero on each block. Hence $F(\frac12)=F(1)=0$ and $|F|\le2^{-T'}$. Integrating by parts,
$J_U=-(z-1)\int_{1/2}^1F(u)u^{z-2}du$, and $\int_{1/2}^1u^{-3/2}du=2(\sqrt2-1)$. (iii) On each of the $2^{T'-1}$ cells
$[a_i,a_{i+1})$ of $[\frac12,1)$ where $W_U$ is constant, $\int u^{z-1}du=(a_{i+1}^z-a_i^z)/z$. Summation by parts
leaves two endpoint terms of modulus at most $1$ and $2^{-1/2}$ and at most $2^{T'-1}-1$ sign changes, each contributing
at most $2$ because $|a^z|\le1$.
\end{proof}

\begin{proposition}[off-resonance bound]\label[proposition]{prop:warmup}
For every $S$ with $\max S=T$ and every $\xi\ge0$,
\[
|\Wt_S(\xi)|\le\min\Bigl\{\bigl(5(\xi+\tfrac12)^{1/2}+3\bigr)2^{-T/2},\ \frac{2^{T+1}}{\xi}\Bigr\}.
\]
Consequently $|\Wt_S(\xi)|\le12\cdot2^{-T/4}$ for all real $\xi$ unless $2^{T/2}<|\xi|<2^{5T/4}$.
\end{proposition}

\begin{proof}
By \cref{lem:shell}, $|\Wt_S(\xi)|\le\sum_{T'=1}^{T}2^{-(T-T')/2}|J_{S^{(T-T')}}(\xi)|+2^{-T/2}/|z|$, where the
shell $S^{(T-T')}$ has top bit $T'$, and $2^{-T/2}/|z|\le2\cdot2^{-T/2}$.

\emph{First bound.} Let $a=\xi+\frac12\ge|z-1|$ and define $t_0$ by $2^{t_0}=a$. Use \cref{lem:oneshell}(i) for the shells
with $T'<t_0$ and for $T'=1$, and \cref{lem:oneshell}(ii) for $T'\ge\max(t_0,2)$. With $q=(1-2^{-1/2})^{-1}<3.42$,
\begin{gather*}
\sum_{T'<t_0}0.586\cdot2^{-(T-T')/2}\le0.586\,q\,a^{1/2}2^{-T/2},\\
\sum_{T'\ge t_0}0.8285\,a\,2^{-T'}2^{-(T-T')/2}\le0.8285\,q\,a^{1/2}2^{-T/2},
\end{gather*}
and the shell $T'=1$, if not already counted, adds at most $0.586\cdot2^{-(T-1)/2}<0.83\cdot2^{-T/2}$. The total is at most
$(4.83a^{1/2}+0.83+2)2^{-T/2}$.

\emph{Second bound.} By \cref{lem:oneshell}(iii), $|J_{S^{(T-T')}}|\le2^{T'}/\xi$, so the shells contribute at most
$\xi^{-1}2^{-T/2}\sum_{T'\le T}2^{3T'/2}\le1.547\cdot2^T/\xi$; the tail adds $2^{-T/2}/\xi$, and
$1.547\cdot2^T+2^{-T/2}\le2^{T+1}$.

\emph{Consequence.} If $\xi\le2^{T/2}$, then $(5(\xi+\frac12)^{1/2}+3)2^{-T/2}\le(5\cdot2^{T/4}+6.6)2^{-T/2}\le12\cdot2^{-T/4}$.
If $\xi\ge2^{5T/4}$, then $2^{T+1}/\xi\le2\cdot2^{-T/4}$.
\end{proof}

\noindent The elementary bounds never used that $W_S$ is a parity: they hold for any function that is constant on cells
of length $2^{-T}$ and has mean zero on pairs of cells. \vsw{The resonance band requires the particular digit factorization of $W_S$ (\cref{sec:symbol}).}{\Cref{prop:necessity} shows that inside the band this
generality fails.}

}
\ificml\else\multiplicationshellproof\fi

\section{Mellin estimates}\label[appendix]{app:icml-shell}
For $w=W_S$ write $\widetilde W_S$ for its Mellin symbol and
$\mu_S=\sup_{\xi\in\mathbb R}|\widetilde W_S(\xi)|$.
\mellinsymbolstatement
\multiplicationshellproof
% Shared main-body piece (section: decay at resonance); input by main-body.tex (FOCS and named preprint) and by icml.tex (ICML 2027).
\section{Decay at resonance: parity and a transfer operator}\label{sec:symbol}

Inside the band we bound each shell $J_U$ by $\CJ\cdot2^{-T'/4}$ (\cref{cor:shell}); summing the shells with weights $2^{-j/2}$ then gives
\cref{thm:symbol}. Two exact facts about parities do the work.

\begin{lemma}[Fourier transform of a Walsh function]\label[lemma]{lem:walshft}
For finite $V\subseteq\{1,2,\dots\}$ and $\omega\in\R$,
$\bigl|\int_0^1W_V(x)e^{i\omega x}dx\bigr|=\prod_{i\ge1}\bigl|\cos\bigl(\omega2^{-i-1}-[i\in V]\tfrac\pi2\bigr)\bigr|$.
\end{lemma}

\begin{proof}
If $x$ is uniform on $[0,1)$, its digits $b_i$ are independent fair bits, $W_V(x)=\prod_{i\in V}(-1)^{b_i}$ and
$e^{i\omega x}=\prod_ie^{i\omega b_i2^{-i}}$. By independence the integral is $\prod_i\frac12(1\pm e^{i\omega2^{-i}})$, with the minus
sign for $i\in V$; and $|1+e^{i\alpha}|/2=|\cos\frac\alpha2|$, $|1-e^{i\alpha}|/2=|\sin\frac\alpha2|$.
\end{proof}

The factors with $i>\max V$ multiply to $|\sin(\omega2^{-\max V-1})/(\omega2^{-\max V-1})|$, so only the factors
$i\le\max V$ matter. A product of $|\cos|$ and $|\sin|$ at dyadic frequencies is small \emph{on average}:

\begin{lemma}[transfer bound]\label[lemma]{lem:transfer}
Let $h_0(y)=|\cos\pi y|$, $h_1(y)=|\sin\pi y|$ and $h_*\equiv1$. For a word $d=(d_0,\dots,d_{n-1})\in\{0,1,*\}^n$,
$\int_0^1\prod_{i=0}^{n-1}h_{d_i}(2^i\theta)\,d\theta\le2^{-n_d/2}$, where $n_d$ is the number of letters different from $*$.
\end{lemma}

\begin{proof}
Let $(\mathcal Pf)(x)=\frac12[f(\frac x2)+f(\frac{x+1}2)]$ act on $1$-periodic functions. For $1$-periodic $G$,
\[\int_0^1\varphi(\theta)G(2\theta)\,d\theta=\int_0^1G(x)(\mathcal P\varphi)(x)\,dx.\]
Applying this $n$ times, the integral equals
$\int_0^1(\mathcal T_{d_{n-1}}\cdots\mathcal T_{d_0}1)(x)dx$ with $\mathcal T_ef=\mathcal P(h_ef)$ for a letter $e$. Each
$\mathcal T_e$ is positive and $\mathcal T_*1=1$. For $e\in\{0,1\}$ we have
$\mathcal T_e1(x)=\frac12(|\cos\frac{\pi x}2|+|\sin\frac{\pi x}2|)\le2^{-1/2}$, hence
$\|\mathcal T_ef\|_\infty\le\|\mathcal T_e1\|_\infty\|f\|_\infty$, and the claim follows.
\end{proof}

\paragraph{Plan of the proof.} Split $[\frac12,1)$ into $2^{\lambda-1}$ windows of length $2^{-\lambda}$,
$\lambda=\lceil\log_4\xi\rceil$. Parity is used twice: on each window $W_U$ is a sign times the \emph{same} inner Walsh
function (step~(i) of the proof of \cref{prop:window}), whose partial integrals factor into one cosine or sine per digit
(step~(iii)). After linearising the phase (step~(ii)) the window frequencies are spaced about $1$ apart, so the sum over
windows becomes an average (step~(v)), bounded by \cref{lem:transfer} (step~(vi)); step~(iv) handles the edges of the
band (\cref{app:windowproof}; parameters in \cref{app:symbol}).

Throughout, $\rho=2^{-1/2}$, $\Gamma_0=1$ and $\Gamma_1=\sqrt2$. The window bound (\cref{prop:window}), with its
parameters and the two auxiliary \cref{lem:avg,lem:riemann}, is stated in \cref{app:windowstmt}; its consequence,
proved in \cref{app:symbol}, is:

\begin{corollary}[shell bound]\label[corollary]{cor:shell}
For every $U$ with $\max U=T'\ge1$ and every $\xi\in\R$, $|J_U(\xi)|\le\CJ\cdot2^{-T'/4}$.
\end{corollary}

\begin{proof}[Proof of the bound on $\mu_S$ in \cref{thm:symbol}(a)]
By \cref{lem:shell} and \cref{cor:shell} (all printed constants are rounded up),
\[\mu_S\le\sum_{j\ge0}2^{-j/2}\CJ\,2^{-(T-j)/4}+2\cdot2^{-T/2}\le\bigl(\CJ/(1-2^{-1/4})+2\bigr)2^{-T/4}\le\CAexact\cdot2^{-T/4},\]
and \cref{lem:hankel} gives $\|K_S\|\le\mu_S$.
\end{proof}

% FOCS-MOVED3: remark -> app:rates
\noindent The frequency profile in (a) and part (b) are proved in \cref{app:rates}.

% Shared main-body piece (section: continuum to integers); input by main-body.tex (FOCS and named preprint) and by icml.tex (ICML 2027).
\section{From the continuum to integers}\label{sec:discrete}

The integer problem is the bilinear form $4^{-N}\sum_{x,y}f(x)g(y)W_S(xy/4^N)$. \Cref{thm:transfer}
(\cref{app:transferstmt}) bounds it in three ways, for $|f|,|g|\le1$ and $\max S=T\le N$: (a) by
$\|K_S\|+O(2^{T-N}+N2^{-N})$, which is useful only for $T\le cN$, $c<1$; (b) if $N\ge8$ and $\mu_S\le1$, by
$O(N)\mu_S^{1/2}+O(N2^{-N})$, which loses a square root but reaches $T=N$; and (c) by a bound that also uses the decay
of $\Wt_S$ above $2^T$ and loses only the power $\frac1{1+\nu}$ of the bound $\mu$ on $|\Wt_S|$, which gives the sharper exponents of
\cref{thm:rates}. Parts (a) and (b) compare the integer form with the continuum in two different ways.

% Shared main-body piece (section: proof of the main theorem); input by main-body.tex (FOCS and named preprint) and by icml.tex (ICML 2027).
\section{Proof of the main theorem}\label{sec:assembly}

\noindent\emph{Proof outline for \cref{thm:main}.} The continuum and Dirichlet transfers give the two
explicit ranges beginning at $T=64$ (\cref{app:mainproof}). Pairing columns $y$ and $2y$ yields a uniform
discrepancy gap for small $T$. The analytic Mellin estimate covers $T\ge106$ for $D=1/512$ and $T\ge157$
for $D=1/16$ (\cref{app:explicit-d}). The exponents $1/8$ and $1/6$ are proved in \cref{app:rates}. Past the midpoint, complete residue systems control the smoothing error (\cref{app:midpoint}).

\let\appendix\relax % appendix.tex opens with \appendix; keep one continuous lettering A, B, C, ...
\appendix
\sloppy
\ificml\else\clearpage\fi

\ificml\else
% FOCS-MOVED from main-body.tex (verbatim)
\section{Further results}\label{sec:further}

This section summarises the remaining results. Their statements and proofs are in the appendices.

\paragraph{How much depth does multiplication need?}\label{sec:depthgap} Depth $\lceil\log_2(2N+1)\rceil$, with $\Theta(N)$-bit
states, suffices for every function of the operands; for fixed $H$ and $k$, inverse-polynomial advantage on
multiplication needs at least $\log_2\ln n-\log_2\ln\ln n-O(1)$ layers (\cref{cor:topk-loglog}). One unique-hard-attention head per layer with a fixed doubling
routing gathers all operand bits at one position in $\lceil\log_2(2N+1)\rceil$ layers, after which the positionwise map
computes any function of $(X,Y)$; this is standard~\cite{LiuAGKZ23,SHT24,Feng24}. Fixed routing needs
$\log_{1+Hk}n-O(1)$ layers for inverse-polynomial advantage, and exact prediction needs $\Omega(\log n/\log\log n)$ layers
by Beame and H\aa{}stad~\cite{BeameHastad89} (\cref{rem:depth-upper}). We sketch how input-dependent routing reaches error $N^{-a}$
with $O(\log n/\log\log n)$ layers, the classical CRCW-PRAM time of multiplication~\cite{CSV84}, with
$2^{O(\sqrt{\log n})}$-bit states and only for astronomically large $N$ (\cref{rem:depth-gap}). \emph{Open question:}
what depth between $\log_2\ln n-\log_2\ln\ln n-O(1)$ and the $O(\log n/\log\log n)$ of the sketched construction does inverse-polynomial advantage need,
and with $O(\log n)$-bit states?

\paragraph{Softmax attention with margins.} With a score gap and inverse temperature of order $\log n$, softmax tracks
hard attention layer by layer~\cite[Lemma~25, Thm.~26]{YSCA24},~\cite[Lemma~B.7]{EGKZ22}, so under the strong
hypotheses listed below our bounds transfer; we add
that the hypotheses are needed only on true contexts, at a top-$k$ cutoff (\cref{cor:softmax} in \cref{app:softmax}).
The hypotheses are strong: a unit gap between the $k$-th and $(k+1)$-th raw scores at every layer, head and query
position $i>k$ of every true context, temperature at most $1/(A\ln n)$ with $A>1+d$, maps that are $B$-Lipschitz and
values bounded by $B$ for one constant $B$ at every layer, and an output margin $n^{-d}$. Under them, for large $n$, greedy
decoding agrees with a top-$k$ model of the same depth and head counts, uniformly for $L\le\log_2\ln n$, so the transfer covers
the full depth range of \cref{cor:topk-loglog}. The corollary is a conditional transfer: it says nothing about softmax
transformers without these margins, the regime used in practice. Without the output margin, or without any gap,
softmax can predict $M_T$ exactly (\cref{rem:softmax-sharp,rem:averaging}).

\paragraph{Averaging attention escapes the bound.} Fixed-depth softmax attention, and averaging hard attention, with
$O(\log n)$-bit precision can predict $M_T$ exactly (\cref{rem:averaging} in \cref{app:averaging}). This is Feng et
al.'s construction for multiplication~\cite[Thm.~5.3 and App.~E.3]{Feng24}, taken in base $2$. Its only use of
attention is to gather all $2N$ operand bits at one position, which a fixed number of top-$k$ heads cannot do at fixed
depth; so the separation is about averaging rather than about arithmetic.

\paragraph{Constant-depth circuits} (\cref{sec:ac0}). Depth-$d$ circuits of size $N^u$, $u<2^{d-1}/(2^{d-1}-1)$, that
read $X$, $Y$ and the whole true prefix have advantage $2^{-N^{\Omega(1)}}$, and DNFs and CNFs of bounded width and any
size have advantage $2^{-\Omega(N)}$ (\cref{cor:ac0-full}). These bounds rest on an arithmetic estimate, not on
\cref{thm:main}. Polynomial-size DNFs and CNFs whose terms mix bits of $X$, $Y$ and the prefix have advantage at most
$(\log N)^{O(1)}/N$, shown by hiding a polylogarithmic block of one operand (\cref{thm:ac0-mixeddnf,thm:ac0-hiddenblock}).
Mixed circuits of depth three or more are open, and we explain why:
three unrestricted features can predict $M_T$ (\cref{prop:ac0-barrier}), we characterise, up to a factor $2$, what \cref{thm:main}
can prove (\cref{prop:ac0-duality}), and the route we use for transformers would need new circuit lower bounds
(\cref{prop:ac0-uniform}). \Cref{app:ac0-results} adds circuits that read one operand and the entire prefix, DNFs with
few terms, decision trees and predictors with few operand queries.

\paragraph{Other models.} \Cref{tab:transformers} (\cref{app:transfull}) lists the main transformer results. Besides the
one-layer bound (\cref{thm:onelayer}), these are three-layer bounds by explicit polynomials and with $k$ allowed to grow
(\cref{thm:three,cor:topk-three}), worst-case and windowed models, and a single middle bit
(\cref{thm:worstcase,thm:window,prop:singlebit}). \Cref{thm:main} also bounds the communication complexity of a parity
of top bits (\cref{rem:cc}).

\paragraph{Chains of thought} (\cref{sec:cot}). A chain of thought that writes the whole product, computed by unique hard
attention with $L$ layers and $H$ heads ($L$, $H$ fixed), needs $N^{\Omega(1/L)}$ auxiliary tokens on some input, even
if its length depends on the input (\cref{thm:cot-worst}; worst case, does not use \cref{thm:main}). On uniform operands, $L$-layer top-$k$ models write the product after at most $N^{2^{1-L}-o(1)}$
auxiliary tokens only with probability $\exp(-N^{2^{1-L}-o(1)})$ (\cref{thm:cot-avg}; asymptotic).

\section{Related work}\label{sec:related}

\paragraph{Bits of products.} Remscrim and Sipser~\cite{RemscrimSipser13} do not cover the top window or parities of
many bits, so with~\cite{HAF22} their result does not bound the prediction of $M_T$ from a context containing $M_{<T}$.
Mohammadkhani, Krishna, Sarrof and Hahn~\cite[Thm.~B.17]{MKSH26} bound the operator norm of the bit of weight
$2^{N-1}$, our $M_{N+1}$, by additive characters modulo $2^N$; this bit lies just outside the range $T\le N$, and their
method uses that it depends only on $xy\bmod2^N$, which fails for the top bits.

\paragraph{Communication and branching programs.} The middle bit is the standard hard bit for branching programs
(\cref{app:overview}); we found there no discrepancy bound under $X\,|\,Y$ for a parity of top bits, which
\cref{thm:main} gives (\cref{rem:cc}).

\paragraph{Digital functions, bilinear sums and Mellin transforms.} The truncation idea of Mauduit and
Rivat~\cite{MauduitRivat10} is the classical form of our passage to the continuum, and \cref{lem:walshft,lem:transfer}
are Walsh versions of standard tools of~\cite{MauduitRivat10,DMR11} (see also~\cite{Bourgain15}). \Cref{lem:mv} is a weak
form of the mean-value theorem of Montgomery and Vaughan~\cite{MontgomeryVaughan74}, and the Mellin bound for kernels
$w(st)$ is classical~\cite[Ch.~IX]{HLP52}; what is new is its use for bits of products, where it isolates the arithmetic
(\cref{thm:transfer}) from the harmonic analysis of one function of one variable (\cref{thm:symbol}).

\paragraph{Polynomials and constant-depth circuits.} Our transformer proofs are approximate-degree arguments on the whole
formal cube, with Sherstov's small-error polynomials for OR~\cite{Sherstov18}, the shared-input composition
of~\cite{BKT19}. For circuits we use Tal's
Fourier tails~\cite{Tal17}, \cite{BKT19} and polylogarithmic independence~\cite{Bazzi09,Braverman10,HarshaSrinivasan19}.

\paragraph{Lower bounds for transformers.} Besides~\cite{Hahn20,HAF22,CPW24} (\cref{sec:intro}; see
also~\cite{BKLP24}), \cite{AHRH25} and~\cite{MKSH26} proved chain-of-thought bounds (\cref{sec:cot}). For multiplication,
Buss~\cite{Buss92} showed that checking a product is as hard as counting, and Feng et al.~\cite[Thm.~4.3]{Feng24} embed
majority into the first output token of a truncated product, with no earlier product bits in context (\cref{app:overview}).

\fi
\ificml
\else
\section{Notation}\label[appendix]{app:notation}

\Cref{tab:notation} lists the notation, in the order in which it is introduced.

\begin{table}[!h]
\centering\footnotesize
\begin{tabular}{@{}>{\raggedright}p{0.17\linewidth}>{\raggedright}p{0.62\linewidth}>{\raggedright\arraybackslash}p{0.15\linewidth}@{}}
\toprule
Symbol & Meaning & Where\\
\midrule
\multicolumn{3}{@{}l}{\emph{Discrepancy part}}\\
$N$; $X,Y$; $[N]$ & bit length; independent uniform integers in $\{0,\dots,2^N-1\}$; $\{1,\dots,N\}$ & throughout\\
$M_k(x,y)$ & $k$-th most significant bit of the $2N$-bit product $xy$, $1\le k\le2N$ & throughout\\
$\hat Z$ & $2Z-1$ for a bit $Z$ (so $1\mapsto+1$); note $(-1)^Z$ maps $1\mapsto-1$, so $\hat M_k=-(-1)^{M_k}$ & throughout\\
$T$, $S$, $C$ & $S\subseteq[T]$ nonempty with $\max S=T$; $S=C\cup\{T\}$ in \cref{conj:C5} & throughout\\
$\chi_S(x,y)$ & $\prod_{k\in S}(-1)^{M_k(x,y)}=(-1)^{|S|}\hat M_T\prod_{k\in C}\hat M_k$ & throughout\\
$\disc_N(S)$ & $\max\{|\E f(X)g(Y)\chi_S(X,Y)|:\ |f|,|g|\le1\}$, $f,g$ complex-valued & \cref{sec:setting}\\
$b_k(u)$, $r_k$, $W_U$ & $k$-th binary digit of $u\in[0,1)$; $r_k=(-1)^{b_k}$; Walsh function $\prod_{k\in U}r_k$; $\chi_S(x,y)=W_S(xy/4^N)$ & \cref{sec:setting}\\
$U^{(j)}$ & shifted set $\{k-j:\ k\in U,\ k>j\}$ & \cref{sec:warmup}\\
$K_S$; $\|\cdot\|$ & operator on $L^2[0,1]$ with kernel $W_S(st)$; operator norm, for a matrix the spectral norm & \cref{sec:warmup}\\
$e(\theta)$; $z$ & $e^{2\pi i\theta}$; $z=\frac12+i\xi$, $\xi\in\R$ & \cref{sec:warmup,sec:discrete}\\
$\Wt_S(\xi)$, $\mu_S$ & Mellin symbol $\int_0^1W_S(u)u^{z-1}du$; $\mu_S=\sup_\xi|\Wt_S(\xi)|$ & \cref{sec:warmup}\\
$J_U(\xi)$ & shell integral $\int_{1/2}^1W_U(u)u^{z-1}du$ & \cref{sec:warmup,sec:symbol}\\
$h_0,h_1$; $\rho$, $\Gamma_0$, $\Gamma_1$ & $|\cos\pi y|$, $|\sin\pi y|$; $2^{-1/2}$, $1$, $\sqrt2$ (other values of $\Gamma_0,\Gamma_1$ in \cref{app:rates}) & \cref{sec:symbol}\\
$\kappa$, $\nu$ & transfer rate per digit; $\nu=-\frac12\log_2\kappa$, the Mellin exponent ($\nu=\frac14$ for $\kappa=\rho$) & \cref{thm:symbol}, \cref{app:rates}\\
$\lambda$, $T_{\mathrm{in}}$, $n$, $\delta$, \dots & parameters of the window bound (\cref{tab:params}); $n$ there is a digit count & \cref{sec:symbol}\\
$Q$ & $2^{2N-T}$: as a function of the integer $xy$, $\chi_S$ jumps only at multiples of $Q$ & \cref{app:cells,app:dirichlet}\\
$\mathcal H$ & harmonic sum $\sum_{1\le x<2^N}1/x\le1+N\ln2$ & \cref{app:cells,app:dirichlet}\\
$\log$, $\log_2$ & natural and binary logarithm & throughout\\
\midrule
\multicolumn{3}{@{}l}{\emph{Transformer part}}\\
$n$; $L$, $H$ & context length; numbers of layers and heads ($H_1,H_2$ per layer in \cref{thm:full-two}) & \cref{sec:transformers}\\
$\mathbf z$, $m$ & formal bits of the context, treated as independent; $m=2N+T-1$ & \cref{sec:overview}\\
$A(f)$; $A_0$; $S_2$ & Walsh norm $\sum_U|\widehat f(U)|$; number of first-layer transcripts; number of tuples & \cref{sec:overview,sec:transformers}\\
$\hat O_T$ & signed output: $1$, $-1$ or $0$ for output token $1$, $0$ or other & \cref{thm:onelayer}\\
$\eta$; $\vartheta$ & an advantage; a uniform approximation error & \cref{sec:transformers}\\
$\widetilde{\mathrm{AND}}$; $t$ & low-degree approximation of AND; its degree & \cref{thm:full-two}\\
selector & family of polynomials approximating the indicator vector of a value in $\ell^1$ & \cref{app:selectors}\\
$N_{m,\varepsilon}$, $D(m,\varepsilon)$ & approximant of $\mathrm{NOR}_m$ with error $\varepsilon$; its degree bound & \cref{app:selectors}\\
$R$; $\Lambda$; $\deg p$ & number of middle-head orders; $\log(1/\vartheta)$; degree of the approximating polynomial & \cref{thm:threeorders}\\
$\Sigma_j$; $A_1$ & set of values of the first-layer state at position $j$; a bound on $|\Sigma_j|$ & \cref{thm:boundedstates}\\
$d$; $\mathcal P^{(j,b,h)}$, $\mathrm{rk}_{j,h}$ & depth threshold; list of a first-layer head; rank of its winner in the list & \cref{app:three}\\
$s'_j$, $\Sigma'_j$; $\Psi_j$, $\mathrm{dec}_j$, $\bot$ & coarse state and its set of values; state from token and ranks, decoding, dummy state & \cref{app:three}\\
$w$; $\mathcal W$ & attention-window width; set of earlier output positions in the window & \cref{app:window-stmt}\\
\bottomrule
\end{tabular}
\caption{Notation. A few letters are reused locally with other meanings and are defined where used: $m$ (a window index
in \cref{sec:symbol}), $n$ (a digit count in \cref{sec:symbol}, an integer argument in \cref{app:dirichlet}),
$\varepsilon$, $\delta$, $\ell$, $r$, $b$, the shifts $\nu_x,\nu_y$ in \cref{lem:sidon}, and Bourgain's notation in
\cref{app:classical}. The appendices on top-$k$ attention, constant-depth circuits and chains of thought reuse further letters
(among them $Q$, $K$, $a$, $A$, $s$, $w$, $U$, $V$, $\nu$) with local meanings, defined where used. In
\cref{thm:topk-depth} and \cref{app:topk}, $S$ denotes the correlation $|\E[\hat M_T\hat O_T]|$, not a
mask.}\label{tab:notation}
\end{table}

\fi
\section{Proofs for \vsw{Appendix}{Section}~\ref{sec:symbol}}\label[appendix]{app:symbol}

\subsection{The window bound}\label[appendix]{app:windowstmt}

Recall $\rho=2^{-1/2}$, $\Gamma_0=1$ and $\Gamma_1=\sqrt2$. By \cref{lem:transfer}, a word of length $n$ with no $*$ has integral at most
$\Gamma_0\rho^n$, and a word of length $n$ with exactly one $*$ at most $\Gamma_1\rho^n$.

\paragraph{Two auxiliary lemmas} (proofs in \cref{app:auxproofs}).

\begin{lemma}[averages over an interval]\label[lemma]{lem:avg}
Let $0\le k<n$, let $d_{k+1},\dots,d_n\in\{0,1\}$ and $g(\gamma)=\prod_{k'=k+1}^{n}|\cos(\gamma2^{-k'-1}-d_{k'}\frac\pi2)|$. Let
$I$ be an interval of length $\ell\ge\pi_n:=2^{n+1}\pi$ and $c_{\rm cov}=1+2\pi_n/\ell$. Then $\mathrm{avg}_Ig\le c_{\rm cov}\Gamma_0\rho^{n-k}$
and $\mathrm{avg}_I|g'|\le c_{\rm cov}\Gamma_1\rho^{n-k}2^{-k-1}$.
\end{lemma}

\begin{lemma}[Riemann sums]\label[lemma]{lem:riemann}
Let $f\ge0$ be absolutely continuous on $[a,b+\delta_0]$ and let $\gamma_1,\dots,\gamma_r\in[a,b]$ have pairwise gaps at
least $\delta_0$. Then $\sum_mf(\gamma_m)\le\delta_0^{-1}\int_a^{b+\delta_0}f+\int_a^{b+\delta_0}|f'|$.
\end{lemma}

\paragraph{Parameters.} Fix $U$ with $\max U=T'$, $\xi>0$ and an integer $\lambda$ with $1\le\lambda\le T'-1$. Put
$T_{\mathrm{in}}=T'-\lambda$, $\delta=\xi4^{-\lambda}$, $\beta=1+2^{-\lambda}+2\delta$, $\ell'=\xi2^{-\lambda}+\delta$, assume
$\ell'\ge2\pi$, let $n_0=\max\{n:2^{n+1}\pi\le\ell'\}\ge0$ and fix an integer $0\le n\le\min(T_{\mathrm{in}},n_0)$. Let
$\bar\alpha=\min(1,2^{T'+1}/\xi)$, let $\bar p=\min(1,\xi2^{-T'})$ if $n<T_{\mathrm{in}}$ and $\bar p=1$ if $n=T_{\mathrm{in}}$,
and let $c_{\rm cov}=1+2^{n+2}\pi/\ell'$.

\begin{proposition}[window bound]\label[proposition]{prop:window}
\[
|J_U(\xi)|\le\sqrt2\,\beta\,\bar\alpha\Bigl[(1+2^{-\lambda})c_{\rm cov}\bar p\sum_{k=0}^{n-1}2^{-k}\rho^{n-k}\bigl(\Gamma_0+\delta2^{-k-1}\Gamma_1\bigr)
+\tfrac12\Bigl(\bar p\sum_{k=n}^{T_{\mathrm{in}}-1}2^{-k}+2\cdot2^{-T_{\mathrm{in}}}\Bigr)\Bigr].
\]
\end{proposition}

\subsection{Proof of Proposition~\ref{prop:window}}\label[appendix]{app:windowproof}

\begin{proof}
\emph{(i) Windows.} Split $[\frac12,1)$ into $I_m=[m2^{-\lambda},(m+1)2^{-\lambda})$, $2^{\lambda-1}\le m<2^\lambda$. On
$I_m$ the digits $1,\dots,\lambda$ of $u$ are those of $m$, so $W_U(2^{-\lambda}(m+v))=\sigma_mW_{\mathrm{in}}(v)$ with $\sigma_m=\pm1$ and
$W_{\mathrm{in}}=W_{U^{(\lambda)}}$, whose top digit is $T_{\mathrm{in}}$. Write $(m+v)^{z-1}=m^{i\xi}e^{i\gamma_mv}\psi_m(v)$ with $\gamma_m=\xi/m$ and
$\psi_m(v)=(m+v)^{-1/2}\exp(i\xi[\ln(1+v/m)-v/m])$. Then
$J_U=\sum_m\sigma_m2^{-\lambda z}m^{i\xi}\int_0^1W_{\mathrm{in}}(v)e^{i\gamma_mv}\psi_m(v)dv$.

\emph{(ii) Summation by parts.} Let $\Phi_\gamma(v)=\int_0^vW_{\mathrm{in}}(t)e^{i\gamma t}dt$ and $\Phi^*(\gamma)=\sup_{v\in[0,1]}|\Phi_\gamma(v)|$.
Then $|\int_0^1W_{\mathrm{in}}e^{i\gamma v}\psi_m|\le(|\psi_m(1)|+\|\psi_m'\|_1)\Phi^*(\gamma)$. From
$\psi_m'=\psi_m[-\frac1{2(m+v)}+i\xi(\frac1{m+v}-\frac1m)]$ and $|\frac1{m+v}-\frac1m|\le v/m^2$ we get
$|\psi_m(1)|+\|\psi_m'\|_1\le m^{-1/2}(1+\frac1{2m}+\frac\xi{2m^2})\le2^{-(\lambda-1)/2}\beta$, using $m\ge2^{\lambda-1}$. With
$|2^{-\lambda z}|=2^{-\lambda/2}$: $|J_U|\le\sqrt2\beta2^{-\lambda}\sum_m\Phi^*(\gamma_m)$.

\emph{(iii) Dyadic decomposition.} For $v<1$, $[0,v)$ is the disjoint union of one dyadic interval of length $2^{-k}$ for
each $k\le T_{\mathrm{in}}$ with $b_k(v)=1$ and a remainder of length below $2^{-T_{\mathrm{in}}}$ on which $W_{\mathrm{in}}$ is constant. On an aligned dyadic interval of
length $2^{-k}$, $W_{\mathrm{in}}$ is a sign times a rescaled copy of $W_{(U^{(\lambda)})^{(k)}}$ (the second use of parity), so rescaling and \cref{lem:walshft} give
$|\int W_{\mathrm{in}}e^{i\gamma t}|=2^{-k}\Pi_k(\gamma)\,\mathrm{sinc}(\gamma)$ with $\Pi_k=\prod_{k'=k+1}^{T_{\mathrm{in}}}\phi_{k'}$,
$\phi_{k'}(\gamma)=|\cos(\gamma2^{-k'-1}-[k'\in U^{(\lambda)}]\frac\pi2)|$, $\Pi_{T_{\mathrm{in}}}=1$, and
$\mathrm{sinc}(\gamma)=|\sin(\gamma2^{-T_{\mathrm{in}}-1})/(\gamma2^{-T_{\mathrm{in}}-1})|\le\min(1,2^{T_{\mathrm{in}}+1}/\gamma)$. The remainder contributes at most
$\min(2^{-T_{\mathrm{in}}},2/\gamma)$. (The term $k=0$, an interval of length $1$, covers $v=1$, since $\Phi^*$ is a supremum over $[0,1]$.) All $\gamma_m$ lie in $[\xi2^{-\lambda},\xi2^{1-\lambda}]$, so
$\Phi^*(\gamma_m)\le\bar\alpha\bigl(\sum_{k=0}^{T_{\mathrm{in}}}2^{-k}\Pi_k(\gamma_m)+2^{-T_{\mathrm{in}}}\bigr)$.

\emph{(iv) The top digit.} Since $T_{\mathrm{in}}\in U^{(\lambda)}$, $\phi_{T_{\mathrm{in}}}(\gamma)=|\sin(\gamma2^{-T_{\mathrm{in}}-1})|\le\xi2^{-T'}$ at $\gamma=\gamma_m$.
If $n<T_{\mathrm{in}}$, then $\Pi_k\le\bar p\,g_k$ for $k<n$, with $g_k=\prod_{k'=k+1}^n\phi_{k'}$, and $\Pi_k\le\bar p$ for $n\le k<T_{\mathrm{in}}$. If
$n=T_{\mathrm{in}}$, then $\Pi_k=g_k$ and $\bar p=1$.

\emph{(v) Riemann step.} The gaps $\gamma_m-\gamma_{m+1}=\xi/(m(m+1))$ are at least $\xi4^{-\lambda}=\delta$. Each $g_k$ is
Lipschitz, hence absolutely continuous, and \cref{lem:riemann} with $\delta_0=\delta$
on $I'=[\xi2^{-\lambda},\xi2^{1-\lambda}+\delta]$, of length $\ell'$, gives
$2^{-\lambda}\sum_mg_k(\gamma_m)\le(1+2^{-\lambda})[\mathrm{avg}_{I'}g_k+\delta\,\mathrm{avg}_{I'}|g_k'|]$, since
$2^{-\lambda}\ell'/\delta=1+2^{-\lambda}$. Constant terms $c$ contribute $2^{-\lambda}2^{\lambda-1}c=c/2$.

\emph{(vi) Transfer step.} Since $n\le n_0$, \cref{lem:avg} applies to each $g_k$ with $\ell=\ell'$. Collecting (ii)--(vi)
gives the claim.
\end{proof}

\subsection{Proofs of Lemmas~\ref{lem:avg} and~\ref{lem:riemann}}\label[appendix]{app:auxproofs}

\begin{proof}[Proof of \cref{lem:avg}]
With $\gamma=2^{n+1}\pi\theta$, factor $k'$ becomes $h_{d_{k'}}(2^{n-k'}\theta)$, so $g$ is a $1$-periodic product of
$n-k$ factors as in \cref{lem:transfer} (the substitution reverses the order of the digits). The interval $I$ meets at most
$\ell/\pi_n+2$ periods, which gives the first bound. For the second, $|g'|\le\sum_{k''=k+1}^n2^{-k''-1}\prod_{k'\ne k''}h_{d_{k'}}$
almost everywhere; each product is a word with one $*$, and $\sum_{k''>k}2^{-k''-1}\le2^{-k-1}$.
\end{proof}

\begin{proof}[Proof of \cref{lem:riemann}]
For $\gamma\in[\gamma_m,\gamma_m+\delta_0)$, $f(\gamma_m)\le f(\gamma)+\int_{[\gamma_m,\gamma_m+\delta_0)}|f'|$. Average over $\gamma$
and sum over the disjoint intervals.
\end{proof}

\subsection{Proof of Corollary~\ref{cor:shell}}

\emph{Proof idea.} Outside the band \cref{lem:oneshell} suffices. Inside it,
write $\xi=2^{T'+\tau}$ and take $\lambda=\lceil\log_4\xi\rceil\approx\frac12(T'+\tau)$ and $n=\min(T_{\mathrm{in}},\lambda-5)$. The
first part of \cref{prop:window} is then $O(\bar\alpha\bar p\rho^n)$ with $\rho^n=O(2^{-(T'-|\tau|)/4})$, and the top
digit ($\bar p\le2^\tau$ if $\tau<0$) or the sinc factor ($\bar\alpha\le2^{1-\tau}$ if $\tau\ge0$) supplies a factor
$2^{-|\tau|}$; so this part is $O(2^{-T'/4}2^{-3|\tau|/4})$. The second part is $O(\bar\alpha2^{-T_{\mathrm{in}}})=O(2^{-(T'+|\tau|)/2})$.

\Cref{tab:params} lists the parameters of \cref{prop:window} and the values that the proof below gives them inside the
resonance band.

\begin{table}[h]
\centering\footnotesize
\begin{tabular}{@{}>{\raggedright}p{0.1\linewidth}>{\raggedright}p{0.36\linewidth}>{\raggedright\arraybackslash}p{0.48\linewidth}@{}}
\toprule
Parameter & Definition & Value or bound in the band ($4^5<\xi\le4^{T'-1}$, $T'\ge14$)\\
\midrule
$\lambda$ & window level & $\lceil\log_4\xi\rceil$, so $6\le\lambda\le T'-1$ and $\lambda\in[\frac{T'+\tau}2,\frac{T'+\tau}2+1)$, $\xi=2^{T'+\tau}$\\
$T_{\mathrm{in}}$ & top digit of the inner Walsh function & $T'-\lambda>\frac{T'-\tau}2-1$\\
$n$ & number of averaged digits & $\min(T_{\mathrm{in}},\lambda-5)\ge\frac{T'-|\tau|}2-5$; $n\le n_0$\\
$\delta$ & $\xi4^{-\lambda}$, minimal gap of the $\gamma_m$ & $(\frac14,1]$\\
$\beta$ & $1+2^{-\lambda}+2\delta$ (summation by parts) & $\le3+2^{-6}$\\
$\ell'$ & $\xi2^{-\lambda}+\delta$, length of the frequency interval & $>2^{\lambda-2}\ge2^{n+3}$, so $\ell'\ge2\pi$\\
$c_{\rm cov}$ & $1+2^{n+2}\pi/\ell'$ (periods met) & $\le1+\pi/2$\\
$\bar\alpha$ & $\min(1,2^{T'+1}/\xi)$ (sinc factor) & $\le2^{1-\tau}$ if $\tau\ge0$; $\le1$\\
$\bar p$ & $\min(1,\xi2^{-T'})$ if $n<T_{\mathrm{in}}$, else $1$ (top digit) & $\le2^{\tau}$ if $\tau<0$ (then $n<T_{\mathrm{in}}$); $\le1$\\
$\Sigma'$, $\Upsilon$ & constants below & $\Sigma'=4.50805\ldots$, $\Upsilon=12.77036\ldots$\\
\bottomrule
\end{tabular}
\caption{Parameters of the window bound in the proof of \cref{cor:shell}.}\label{tab:params}
\end{table}

\begin{proof}[Proof of \cref{cor:shell}]
Take $\xi\ge0$ and write $\xi=2^{T'+\tau}$. If $T'<14$, use \cref{lem:oneshell}(i):
$2-\sqrt2\le0.586\cdot2^{13/4}2^{-T'/4}<5.6\cdot2^{-T'/4}$. Let $T'\ge14$. If $\xi\le2^{3T'/4}$, \cref{lem:oneshell}(ii) gives
$|J_U|\le0.8285(\xi+\frac12)2^{-T'}\le2^{-T'/4}$, because $2^{-3T'/4}\le2^{-10}$. If $\xi\ge2^{5T'/4}$,
\cref{lem:oneshell}(iii) gives $|J_U|\le2^{-T'/4}$. Otherwise $4^5<\xi\le4^{T'-1}$; take $\lambda=\lceil\log_4\xi\rceil$, so that
$6\le\lambda\le T'-1$, $\delta\in(\frac14,1]$ and $\beta\le3+2^{-6}$, and take $n=\min(T_{\mathrm{in}},\lambda-5)$. Then $n\le n_0$ because
$\ell'>\xi2^{-\lambda}>2^{\lambda-2}>2^{\lambda-4}\pi$, and $c_{\rm cov}\le1+\pi/2$.

Using $2^{-k}\rho^{n-k}=\rho^n(2\rho)^{-k}$, $\delta\le1$ and $2^{-n}\le\rho^n$, \cref{prop:window} becomes
$|J_U|\le\sqrt2\beta\bar\alpha[\Upsilon\bar p\rho^n+2^{-T_{\mathrm{in}}}]$ (the factor $\frac12$ there turns $2\cdot2^{-T_{\mathrm{in}}}$ into $2^{-T_{\mathrm{in}}}$ and
$\bar p\sum_{k\ge n}2^{-k}\le2\bar p\rho^n$ into the summand $1$ of $\Upsilon$), where $\Sigma'=\Gamma_0/(1-\frac1{2\rho})+\Gamma_1/(2(1-\frac1{4\rho}))\le\Sigcon$ and $\Upsilon=(1+2^{-6})(1+\frac\pi2)\Sigma'+1\le\Acon$. From
$\lambda\in[\frac{T'+\tau}2,\frac{T'+\tau}2+1)$ we get $T_{\mathrm{in}}>\frac{T'-\tau}2-1$ and $n\ge\frac{T'-|\tau|}2-5$.
If $\tau\ge0$: $\bar\alpha\le2^{1-\tau}$, $\bar p\le1$, so $\bar\alpha\bar p\rho^n\le2\rho^{-5}2^{-T'/4}2^{-3\tau/4}$ and
$\bar\alpha2^{-T_{\mathrm{in}}}\le4\cdot2^{-(T'+\tau)/2}$. If $\tau<0$: then $n<T_{\mathrm{in}}$ (equality $n=T_{\mathrm{in}}$ would force $\tau>3$), so
$\bar p\le2^\tau$, $\bar\alpha\le1$, $\bar\alpha\bar p\rho^n\le\rho^{-5}2^{-T'/4}2^{-3|\tau|/4}$ and $2^{-T_{\mathrm{in}}}<2\cdot2^{-(T'+|\tau|)/2}$.
In both cases $|J_U|\le\sqrt2(3+2^{-6})(2\Upsilon\rho^{-5}+4)2^{-T'/4}$, and this constant, evaluated with the exact values of
$\Sigma'$ and $\Upsilon$, is $633.2288\ldots\le\CJ$. The margin is about $0.0012$: with the rounded values $\Sigcon$ and
$\Acon$ printed above the constant would be $633.69\ldots$, so the exact values are needed.
\texttt{checks/constants\_interval.py} verifies $\Sigma'\le\Sigcon$, $\Upsilon\le\Acon$, the value $633.2288\ldots$, the
failure of the rounded version and $\CJ/(1-2^{-1/4})+2\le\CAexact$ in interval arithmetic.
\end{proof}

\emph{How to read the window bound.} The prefactor $\sqrt2\beta\bar\alpha$ comes from the summation by parts in step (ii)
and, when $\xi>2^{T'}$, from the decay $\bar\alpha$ of the sinc factor. Inside the bracket, the first sum collects the
dyadic pieces $k=0,\dots,n-1$, whose digits are averaged: the dyadic piece of length $2^{-k}$ carries weight $2^{-k}$ and gains $\rho=2^{-1/2}$
for each of the $n-k$ averaged digits below it, and $\bar p\le\xi2^{-T'}$ is the factor of the top digit when
$\xi<2^{T'}$. The second part collects the digits that are not averaged; they are bounded trivially, but their weights
$2^{-k}$, $k\ge n$, are already small. \Cref{cor:shell} then chooses $\lambda=\lceil\log_4\xi\rceil\approx\frac12\log_2\xi$ and
$n=\min(T_{\mathrm{in}},\lambda-5)$, which makes both parts of order $2^{-T'/4}$.

\emph{Intuition for steps (v)--(vi) of the proof of \cref{prop:window}.}
Because $4^{\lambda-1}<\xi\le4^\lambda$ and $2^{\lambda-1}\le m<2^\lambda$, consecutive
window frequencies $\gamma_m=\xi/m$ differ by $\xi/(m(m+1))$, which lies between $\xi4^{-\lambda}\in(\frac14,1]$ and about
four times that; so the $\gamma_m$ form a grid of spacing of order $1$ on an interval of length between $2^{\lambda-2}$
and $2^\lambda$. The factor of digit $k'$ is periodic in $\gamma$ with period $2^{k'+1}\pi$. For the digits
whose period is shorter than the interval ($2^{k'+1}\pi\le2^{\lambda-2}$ holds for $k'\le\lambda-5$, since $\pi<4$), the grid sweeps through many full periods, so the
sum over windows behaves like an average over a uniformly distributed phase, which is exactly what \cref{lem:transfer}
bounds. The remaining digits have periods longer than the interval and are bounded trivially.

\emph{Where the exponent $1/4$ comes from.} Each active digit contributes a factor $2^{-1/2}$, and only the about
$T'/2$ digits below the window level are active. Summing the windows with the triangle inequality discards the signs
$\sigma_m$, which carry the information in the top $\lambda$ digits; this is why the method stops at $1/4$\vsw{}{, while the computed decay is faster (\cref{app:numerics})}.

\section{The transfer theorem}\label[appendix]{app:transferstmt}

% FOCS-MOVED from main-body.tex (verbatim)
\paragraph{Cell by cell (part (a)).} Extend $f,g$ to step functions $F,G$ on $[0,1]$. The integer form then equals
$\int\!\!\int F(s)W_S(st)G(t)\,ds\,dt$, of modulus at most $\|K_S\|$, up to the $O(2^{N+T}+N2^N)$ cells on which $st$
crosses a multiple of $2^{-T}$; this gives the error in (a) (\cref{app:cells}).

\paragraph{Through Dirichlet polynomials (part (b)).} Mellin inversion writes the double sum as an integral of the
Mellin transform of $v\mapsto W_S(v/4^N)$, which is $2^N\Wt_S$ on the critical line, against two Dirichlet polynomials
in $f$ and $g$. Smoothing in $\log v$ over a relative width $\varepsilon=2^{-N}\mu_S^{1/2}$ makes the integral converge
and cuts off frequencies beyond $1/\varepsilon$, where a mean-value theorem (\cref{lem:mv}) is affordable. The
smoothing error is counted by residues of $xy$ modulo $2^{2N-T}$ (\cref{lem:residues}), which needs $T\le N$. Balancing
gives $O(N)\mu_S^{1/2}$, with no $2^{T-N}$ term (\cref{app:dirichlet}).
Part~(c) runs the same argument but uses that above $2^T$ the symbol is smaller by $(2^T/|t|)^{1-\nu}$: a much
narrower smoothing then costs little, and the balance gives $\mu^{1/(1+\nu)}$ (\cref{app:rates}).

\Cref{sec:discrete} describes how the integer problem is compared with the continuum. The precise statement is the
following.

\begin{theorem}[from the continuum to integers]\label{thm:transfer}
Let $1\le T\le N$, let $\max S=T$, and let $f,g:\{0,\dots,2^N-1\}\to\C$ with $|f|,|g|\le1$.
\begin{enumerate}
\item[(a)] $|\E f(X)g(Y)\chi_S(X,Y)|\le\|K_S\|+3\cdot2^{T-N}+(8+1.39N)2^{-N}$.
\item[(b)] If $N\ge8$ and $\mu_S\le1$, then $|\E f(X)g(Y)\chi_S(X,Y)|\le(29+15N)\mu_S^{1/2}+(8+4N)2^{-N}$.
\item[(c)] If $T\ge9$, $\frac14\le\nu<\frac12$, $0<\mu\le1$ and $|\Wt_S(t)|\le\mu\min\{1,(2^T/|t|)^{1-\nu}\}$ for all
real $t$, then
\[
|\E f(X)g(Y)\chi_S(X,Y)|\le\bigl(c_1\mathcal H+8.02(3+1.9N)\bigr)\mu^{\frac1{1+\nu}}2^{-\frac{1-\nu}{1+\nu}(N-T)}+c_2\mu+(8+3.8N)2^{-N},
\]
where $\mathcal H=\sum_{1\le x<2^N}1/x\le1+0.694N$ and $c_1,c_2$ depend only on $\nu$ (\cref{app:rates}); for
$\nu=\frac14$, $c_1=13.5737\ldots$, $c_2=9.5682\ldots$, and the bracket is at most $37.64+24.66N$.
\end{enumerate}
\end{theorem}

\noindent Part~(a) loses $2^{T-N}$, so it is useful only for $T\le cN$, $c<1$. Part~(b) loses a square root but
reaches $T=N$; its hypothesis $\mu_S\le1$ holds for $T\ge\TAone$ by \cref{thm:symbol}. Part~(c) uses the decay of
$\Wt_S$ above $2^T$ and loses only the power $\frac1{1+\nu}$, which is $\frac45$ by \cref{thm:symbol}(a). The resulting sharper rates, an exponent $\frac15-o(1)$ for all
$T\le N$ and $\frac14-o(1)$ for $T\le12N/13$, are \cref{thm:rates} in \cref{app:rates}.

\section{Proof of Theorem~\ref{thm:transfer}(a)}\label[appendix]{app:cells}

Let $F(s)=f(\lfloor2^Ns\rfloor)$ and $G(t)=g(\lfloor2^Nt\rfloor)$; then $\|F\|_2,\|G\|_2\le1$ and
\begin{multline*}
\E f(X)g(Y)W_S(XY/4^N)=\int\!\!\int F(s)W_S(st)G(t)\,ds\,dt\\
-\sum_{x,y}f(x)g(y)\int\!\!\int_{\mathrm{cell}(x,y)}\bigl[W_S(st)-W_S(xy/4^N)\bigr]\,ds\,dt.
\end{multline*}
The first term is at most $\|K_S\|$ in modulus. On $\mathrm{cell}(x,y)=[x,x+1)\times[y,y+1)\cdot2^{-N}$, $st$ ranges
over $[xy,(x+1)(y+1))/4^N$, and $W_S$ is right-continuous with jumps only at multiples of $2^{-T}$. So the bracket vanishes
unless the open interval $(xy,(x+1)(y+1))$ contains a multiple of $Q=2^{2N-T}$; call such cells bad and let $B$ be
their set. Each bad cell contributes at most $2\cdot4^{-N}$.

For $x=0$ there are at most $2^N$ bad cells. For $x\ge1$ and $q\ge1$, the cell is bad for the multiple $qQ$ only if
$qQ/(x+1)-1<y<qQ/x$, which allows at most $qQ/(x(x+1))+2$ values of $y$, and only if $q<(x+1)2^{T-N}$. Summing over $q$,
row $x$ has at most $2^T\frac{x+1}{2x}+\frac{2^{N-1}}x+2(x+1)2^{T-N}$ bad cells. Summing over $1\le x<2^N$ with
$\mathcal H=\sum_{x<2^N}1/x\le1+N\ln2$,
\[
|B|\le1.5\cdot2^{N+T}+2^N+2^{T+1}+(2^{T-1}+2^{N-1})\mathcal H .
\]
With $T\le N$, $2\cdot4^{-N}|B|\le3\cdot2^{T-N}+(8+1.39N)2^{-N}$.

\section{Proof of Theorem~\ref{thm:transfer}(b)}\label[appendix]{app:dirichlet}

\paragraph{Outline.} Write $\Phi(v)=W_S(v/4^N)$ for $0<v<4^N$ and $\Phi(v)=0$ for
$v\ge4^N$. This is a step function of $v>0$ whose jumps lie at multiples of $Q$ and at $4^N$. Mellin inversion expresses $\Phi(xy)$ through $x^{-1/2-it}y^{-1/2-it}$, so the double sum becomes
\[
\frac1{2\pi}\int_\R\widetilde\Phi(\tfrac12+it)\,\mathcal F(t)\,\mathcal G(t)\,dt,\qquad
\mathcal F(t)=\sum_{1\le x<2^N}f(x)x^{-1/2-it},
\]
with $\mathcal G$ defined from $g$ in the same way, and $|\widetilde\Phi(\frac12+it)|=2^N|\Wt_S(t)|\le2^N\mu_S$. Two
problems remain. First, the integral does not converge absolutely. Second, the mean square of $\mathcal F$ over $|t|\le\Omega$ grows like
$\Omega\log2^N+2^N$ (a mean-value theorem for Dirichlet polynomials, \cref{lem:mv}). Both are solved by smoothing $\Phi$ in
$\log v$ over a window of relative width $\varepsilon=2^{-N}\mu_S^{1/2}$. By the convolution theorem, this smoothing
multiplies the Mellin transform by the transform of a small bump, which decays like $t^{-2}$. So the integral converges
absolutely, and frequencies beyond $|t|\approx1/\varepsilon$ are cut off, where the mean-value theorem is still
affordable. The price is that smoothing changes $\Phi(n)$ only at integers $n$ whose residue $n\bmod Q$ is below
$\varepsilon n$. The number of pairs $(x,y)$ with $xy\bmod Q<\ell$ is at most $\ell\,2^N(3+1.9N)$ (\cref{lem:residues});
this needs $Q\ge2^N$, that is $T\le N$. Balancing the two errors gives $O(N)\mu_S^{1/2}$. No step
compares sums with integrals cell by cell, so there is no $2^{T-N}$ term, and nothing in this argument uses the parity
structure: it enters only through $\mu_S$.

\begin{lemma}[mean value of a Dirichlet polynomial]\label[lemma]{lem:mv}
For $\Omega\ge2$ and finitely many complex numbers $c_n$ ($n\ge1$), $\int_{-\Omega}^\Omega|\sum_nc_nn^{-it}|^2dt\le\sum_n|c_n|^2(4.935\Omega+17.17n)$.
\end{lemma}

\begin{proof}
Let $\alpha=\pi/(4\Omega)\le\pi/8$ and $w(t)=\frac{\pi^2}8(\frac{\sin\alpha t}{\alpha t})^2$. For $|t|\le\Omega$, $\alpha|t|\le\pi/4$, so $w(t)\ge1$;
and $\hat w(\zeta)=\frac{\pi^3}{8\alpha}(1-\frac{|\zeta|}{2\alpha})_+\ge0$. Hence the integral is at most
$\int w|\sum c_nn^{-it}|^2=\sum_{n,n'}c_n\bar c_{n'}\hat w(\ln n-\ln n')\le\frac{\pi^3}{8\alpha}\sum_{|\ln(n/n')|<2\alpha}|c_n||c_{n'}|$.
By $|c_nc_{n'}|\le\frac12(|c_n|^2+|c_{n'}|^2)$ this is at most $\frac{\pi^3}{8\alpha}\sum_n|c_n|^2\#\{n':|\ln(n'/n)|<2\alpha\}$, and the
count is at most $1+2n\sinh2\alpha\le1+4.427\alpha n$. Finally $\frac{\pi^3}{8\alpha}\le4.935\Omega$ and $4.935\Omega\cdot4.427\alpha<17.17$.
\end{proof}

\begin{lemma}[residues of products]\label[lemma]{lem:residues}
Let $T\le N$ and $Q=2^{2N-T}$. For every $j$, $\#\{0\le x,y<2^N:\ xy\equiv j\bmod Q\}\le2^N(3+1.9N)$. Consequently, for an
integer $\ell\ge1$, $\#\{(x,y):\ xy\bmod Q<\ell\}\le\ell\,2^N(3+1.9N)$.
\end{lemma}

\begin{proof}
Write $Q=2^{m_0}$ with $m_0=2N-T\ge N$ and $e(\theta)=e^{2\pi i\theta}$. The count is
$Q^{-1}\sum_{h\bmod Q}e(-hj/Q)\Sigma(h)$ with $\Sigma(h)=\sum_{x,y}e(hxy/Q)$, so it is at most $Q^{-1}\sum_h|\Sigma(h)|$. Here
$\Sigma(0)=4^N$ and, for $h\ne0$, $|\Sigma(h)|\le\sum_x\min(2^N,1/(2\|hx/Q\|))$. The term $x=0$ gives $2^N$ for each $h$. For
$x\ge1$ let $g=2^{v_2(x)}$, where $2^{v_2(x)}$ is the largest power of $2$ dividing $x$; since $v_2(x)<N\le m_0$, $g$ divides $Q$, and as $h$ runs over $1,\dots,Q-1$ the residue $hx\bmod Q$
equals $0$ exactly $g-1$ times and each nonzero multiple of $g$ exactly $g$ times. So
$\sum_{h\ne0}\min(2^N,\frac1{2\|hx/Q\|})\le g2^N+Q(1+m_0\ln2)$. Using $\sum_{1\le x<2^N}2^{v_2(x)}=N2^{N-1}$ and dividing by $Q$
gives at most $2^T(1+\frac N2)+2^N(2+m_0\ln2)\le2^N(3+1.9N)$. The second claim sums the first over $j<\ell$.
\end{proof}

\begin{lemma}[smoothing in $\log v$]\label[lemma]{lem:smooth}
Let $\Phi(v)=W_S(v/4^N)$ for $0<v<4^N$ and $\Phi(v)=0$ for $v\ge4^N$. Let $p_\triangle$ be the triangle density on $[0,1]$
(the convolution of two uniform densities on $[0,\frac12]$), let $0<\varepsilon\le0.01$ and
$\Phi_\varepsilon(v)=\int_0^1\Phi(ve^{-\varepsilon\sigma})p_\triangle(\sigma)d\sigma$.
\textup{(a)} $|\Phi_\varepsilon|\le1$, and $\Phi_\varepsilon(n)=\Phi(n)$ for every integer $1\le n<4^N$ with $n\bmod Q\ge\varepsilon n$.
\textup{(b)} The Mellin transforms satisfy $\widetilde\Phi_\varepsilon(z)=\widetilde\Phi(z)k(\varepsilon z)$ with
$k(\zeta)=((e^{\zeta/2}-1)/(\zeta/2))^2$; on $z=\frac12+it$, $|\widetilde\Phi(z)|=2^N|\Wt_S(t)|$ and $|k(\varepsilon z)|\le\min(1.01,(4.01/(\varepsilon|t|))^2)$.
\textup{(c)} $\Phi_\varepsilon(n)=\frac1{2\pi}\int_\R\widetilde\Phi_\varepsilon(\frac12+it)n^{-1/2-it}dt$ for every $n\ge1$.
\end{lemma}

\begin{proof}
(a) $\Phi(ne^{-\varepsilon\sigma})$, $\sigma\in[0,1]$, is constant unless a jump of $\Phi$ lies in $(ne^{-\varepsilon},n]$. The jumps are at
multiples of $Q$ below $4^N$ and at $4^N>n$. If $p=jQ\in(ne^{-\varepsilon},n]$, the largest multiple of $Q$ not exceeding $n$ is at
least $p$, so $n\bmod Q\le n-p<n(1-e^{-\varepsilon})<\varepsilon n$.
(b) The Mellin transform of $v\mapsto\Phi(ve^{-\varepsilon\sigma})$ is $e^{\varepsilon\sigma z}\widetilde\Phi(z)$, and
$\int_0^1p_\triangle(\sigma)e^{\zeta\sigma}d\sigma=k(\zeta)$. With $v=4^Nu$, $\widetilde\Phi(z)=4^{Nz}\Wt_S(t)$. Further
$|(e^{\zeta/2}-1)/(\zeta/2)|=|\int_0^1e^{\zeta\tau/2}d\tau|\le e^{\varepsilon/4}$ for $\zeta=\varepsilon z$, and $|e^{\zeta/2}-1|\le1+e^{\varepsilon/4}$ with $|\zeta/2|\ge\varepsilon|t|/2$.
(c) In the variable $y=\ln v$, $\Phi_\varepsilon(e^y)=\int\Phi(e^{y'})\,\varepsilon^{-1}p_\triangle((y-y')/\varepsilon)\,dy'$ is the
convolution of the bounded step function $y'\mapsto\Phi(e^{y'})$ (jumps at $\ln(jQ)$ and at $\ln4^N$, by (a)) with a
continuous compactly supported density, so it is continuous. It vanishes for $y\ge\ln4^N+\varepsilon$ and is bounded by
$1$, so $y\mapsto\Phi_\varepsilon(e^y)e^{y/2}$ is continuous and integrable. Its Fourier transform
$\widetilde\Phi_\varepsilon(\frac12+it)$ is $O(t^{-2})$ by (b), so Fourier inversion holds at every point.
\end{proof}

\begin{proof}[Proof of \cref{thm:transfer}(b)]
Let $\mu=\mu_S\le1$ and $\varepsilon=2^{-N}\mu^{1/2}\le0.01$ (as $N\ge8$). Pairs with $x=0$ or $y=0$ contribute at most $2^{1-N}$.
For $x,y\ge1$ replace $\Phi(xy)$ by $\Phi_\varepsilon(xy)$. By \cref{lem:smooth}(a) they differ, by at most $2$, only when
$xy\bmod Q<\varepsilon xy\le\varepsilon4^N=2^N\mu^{1/2}$; by \cref{lem:residues} with $\ell=\lceil2^N\mu^{1/2}\rceil$ this costs at most
$2(3+1.9N)(\mu^{1/2}+2^{-N})$ after dividing by $4^N$. By \cref{lem:smooth}(c),
\[
\sum_{x,y\ge1}f(x)g(y)\Phi_\varepsilon(xy)=\frac1{2\pi}\int_\R\widetilde\Phi(\tfrac12+it)\,k(\varepsilon(\tfrac12+it))\,\mathcal F(t)\mathcal G(t)\,dt .
\]
\Cref{lem:mv} with $c_x=f(x)x^{-1/2}$ and Cauchy--Schwarz give $\int_{-\Omega}^\Omega|\mathcal F\mathcal G|\le4.935\Omega \mathcal H+17.17\cdot2^N$,
where, as in \cref{app:cells}, $\mathcal H=\sum_{1\le x<2^N}1/x\le1+N\ln2$. Let
$\Omega_0=4.01/\varepsilon$. Using $|k|\le1.01$ on $|t|\le\Omega_0$ and $|k|\le4^{-i}$ on $2^i\Omega_0\le|t|\le2^{i+1}\Omega_0$,
$\int|k||\mathcal F\mathcal G|\le24.73\Omega_0\mathcal H+40.25\cdot2^N$. With $|\widetilde\Phi|\le2^N\mu$ and division by $4^N$, the main term is at
most $\frac\mu{2\pi}(99.2\mathcal H2^{-N}/\varepsilon+40.25)=15.8\mathcal H\mu^{1/2}+6.41\mu$. With $\mathcal H\le1+0.694N$ and $\mu\le\mu^{1/2}$, the total
is at most $(\TRa+\TRb N)\mu^{1/2}+(8+3.8N)2^{-N}$.
\end{proof}

\section{Proof of Theorem~\ref{thm:main}}\label[appendix]{app:mainproof}

\begin{proof}[Proof of \cref{thm:main}]
\emph{Explicit range 1 ($64\le T\le4N/5$).} Then $2^{T-N}\le2^{-T/4}$ and $(8+1.39N)2^{-N}\le5.6\cdot2^{-T/4}$, so
\cref{thm:transfer}(a) and \cref{thm:symbol} give $\disc_N(S)\le3991\cdot2^{-T/4}$, which is at most $2^{-T/16}$ for
$T\ge\frac{16}3\log_23991$, and $\frac{16}3\log_23991\le\thrtwo<64$.

\emph{Explicit range 2 ($N\ge8$, $16\log_2N+180\le T\le N$).} Here $\mu_S\le3982\cdot2^{-45}<1$, and
\cref{thm:transfer}(b) with $\mu_S^{1/2}\le63.11\cdot2^{-T/8}$ gives
$\disc_N(S)\le63.11(29+15N)2^{-T/8}+(8+4N)2^{-N}$. The first term is at most $\frac12 2^{-T/16}$ as soon as
$2^{T/16}\ge126.22\cdot18.7N\ge126.22(29+15N)$, which holds for $T\ge16\log_2N+\throne$. The second is at most
$\frac122^{-N/16}\le\frac122^{-T/16}$ because $(16+8N)\le2^{15N/16}$ for $N\ge8$.

\emph{Coverage for $N\ge400$.} For $N\ge400$ we have $4N/5\ge16\log_2N+180$ (at $N=400$: $\NzeroLHS\ge\NzeroRHS$, and
the left side grows faster). So each $T\in[64,N]$ lies in range~1 or range~2.

\emph{A uniform explicit constant.}
The column-pair estimate and two comparisons in \cref{app:explicit-d} give
$\disc_N(S)\le 2^{-T/512}$ for all $N\ge T\ge1$ with $N\ge2$.
The same comparisons give the $1/16$ exponent uniformly for $T\ge157$.
\end{proof}

\noindent The sharper $1/8$ and $1/6$ rates are stated in \cref{app:explicit-d} and proved in \cref{app:rates}.

\section{Explicit discrepancy at every size}\label[appendix]{app:explicit-d}

Put $m=2^N$ and $q=2^T$. For two columns $y,y'$ of the sign matrix
$W_S(xy/m^2)$, suppose the fraction of rows on which their signs agree,
and the fraction on which they disagree, are both at least $p\le1/2$.
For any complex column weights $g(y),g(y')$ of modulus at most one,
let $a$ be the agreement fraction. After taking absolute values in the row
sum, the contribution divided by $2m$ is at most
\[
 \tfrac12\bigl(a|g(y)+g(y')|+(1-a)|g(y)-g(y')|\bigr)
 \le \sqrt{(1-p)^2+p^2}=:B(p).
\]
Here Cauchy--Schwarz uses $|g(y)+g(y')|^2+|g(y)-g(y')|^2\le4$.
The estimate includes all admissible complex row weights.

For $T\ge2$, pair each integer $y\in[m/4,m/2)$ with $y'=2y$.
These disjoint pairs cover half the columns. Write $\lambda=qy/m$, so
the two column signs on row $x$ are $W_S(z/q)$ and $W_S(2z/q)$, with
$z=\lambda x/m$. On each full unit interval $[j,j+1)$ in $[0,\lambda)$,
the first sign is constant. The second takes opposite signs on the two halves:
the index of its length-$1/q$ cell goes from $2j$ to $2j+1$, which flips only
digit $T$, and $T\in S$. This uses half-open cells, including at dyadic
boundaries. Each half contains at least $\lfloor m/(2\lambda)\rfloor$ sampled
rows. Thus each of agreement and disagreement has fraction at least
\[
 p_y=\frac{\lfloor\lambda\rfloor\lfloor m/(2\lambda)\rfloor}{m}.
\]
For $2\le T\le6$ both floored arguments are at least one, so $p_y\ge1/8$.
When $T\ge7$ the first argument is at least $32$, hence $p_y\ge8/33$,
using $\lfloor u\rfloor/u\ge a/(a+1)$ for $u\ge a$ and integer $a\ge1$.
A second family pairs $y\in[m/16,m/8)$ with $2y$. It is disjoint from the
first, covers $1/8$ of the columns, and the two floored arguments are at least
$8$ and $4$, so $p_y\ge16/45$. Unpaired columns contribute at most their fraction.
Consequently for $7\le T\le105$ the discrepancy is bounded by
$3/8+B(8/33)/2+B(16/45)/8<866/1000<2^{-105/512}$.
At $T=1$, pair $y<m/4$ with $y'=y+3m/4$. The low column is positive throughout,
while the high column is positive on $x<m/2$ and negative on $x\ge3m/4$.
Thus $p\ge1/4$ on half the columns. The bounds
$1/2+B(1/4)/2$ at $T=1$ and $1/2+B(1/8)/2$ at $2\le T\le6$
are both below $63/64<2^{-6/512}$.

For large $T$, use the analytic bound of \cref{thm:rates}. For fixed
$T\ge48$ its first and last terms decrease in $N\ge T$, since the ratios
of their positive linear factors under $N\mapsto N+1$ are at most $49/48$,
and the exponential factors shrink by $2^{-3/5}$ or $1/2$.
At $N=T$, multiply the bound by $2^{DT}$. Each of the three terms is
decreasing in $T\ge106$ for $D\le1/16$, since the linear factors have
ratios at most $107/106<2^{11/80}$ and the slowest exponential shrinkage
is at most $2^{-11/80}$. The normalized sums at $(D,T)=(1/512,106)$
and $(1/16,157)$ are both strictly below one. The elementary comparisons
are recorded in \texttt{checks/explicit\_discrepancy.py}, using integer and
rational arithmetic. This uses the analytic Mellin rate.

\ificml
\paragraph{Discrepancy ranges.}
For every mask $S$ with $\max S=T$, the following exponents $D$ bound
$\disc_N(S)$ by $2^{-DT}$.
\begin{center}
\small
\begin{tabular}{@{}lll@{}}
\toprule
$D$ & Range & Proof \\
\midrule
$1/2560$ & $N\ge2$, $1\le T\le N$ & additive Fourier, \cref{app:finite-fourier} \\
$1/512$ & $N\ge2$, $1\le T\le N$ & analytic Mellin, this appendix \\
$1/16$ & $157\le T\le N$ & analytic Mellin, this appendix \\
$1/8$ & $N\ge402$, $96\le T\le N$ & explicit transfer, \cref{app:rates} \\
$1/6$ & $N\ge951$, $144\le T\le N$ & explicit transfer, \cref{app:rates} \\
$D_\eta$ & $N<T\le(4/3-\eta)N$, large $N$ & analytic midpoint transfer, \cref{app:midpoint} \\
\bottomrule
\end{tabular}
\end{center}
The $1/16$ rate also holds for $64\le T\le4N/5$, or for $N\ge8$
and $16\log_2 N+180\le T\le N$. The sharper first-half rates use
the certified transfer estimates in \cref{app:rates}.
\fi

\section{Sharper exponents}\label[appendix]{app:rates}

% FOCS-MOVED3 from main-body.tex (verbatim)
\noindent The profile in (a) comes from running the window bound on the whole range $4^5<\xi\le4^{T'-1}$. Part (b)
replaces the factor $2^{-1/2}$ per digit of \cref{lem:transfer} by $0.6646$, averaged over blocks of $12$ digits and
certified in outward-rounded interval arithmetic (\cref{lem:cert}). Both proofs are in \cref{app:rates}.

This appendix proves the frequency profile in \cref{thm:symbol}(a), part~(b) of \cref{thm:symbol},
\cref{thm:transfer}(c), \cref{thm:rates}, and the sharper rates in \cref{app:explicit-d}. Everything that uses
the pair $(\kappa,r)=(\rho,1)$ below is a hand proof; the pair $(0.6646,12)$ rests on the computer certificate of
\cref{lem:cert}. The printed constants are evaluated from their formulas in interval arithmetic by
\texttt{checks/exponent\_constants.py}; upper bounds are rounded up and lower bounds down.

\begin{theorem}[sharper rates]\label{thm:rates}
Let $48\le T\le N$ and $\max S=T$. Then
\[
\disc_N(S)\le758.72\,(37.64+24.66N)\,2^{-T/5-3(N-T)/5}+38101\cdot2^{-T/4}+(8+3.8N)2^{-N}.
\]
So $\disc_N(S)\le C(1+N)2^{-T/5}$, and $\disc_N(S)\le C(1+N)2^{-T/4}$ if $T\le12N/13$, for an absolute $C$. With
\cref{thm:symbol}(b) instead, the exponent is at least $0.2276-o(1)$ for all $T\le N$ and $0.2947-o(1)$ for
$T\le0.77N$, as $T/\log N\to\infty$.
\end{theorem}

\subsection{A parametrised transfer bound and a frequency profile for one shell}

Let $\mathcal T_e$ ($e\in\{0,1,*\}$) be the operators of the proof of \cref{lem:transfer}. For a word
$d=(d_0,\dots,d_{m-1})$ over $\{0,1,*\}$ put $F_d=\mathcal T_{d_{m-1}}\cdots\mathcal T_{d_0}1$; explicitly
$F_d(x)=2^{-m}\sum_{j<2^m}\prod_{i<m}h_{d_i}(2^i(x+j)/2^m)$. That proof shows that the integral in \cref{lem:transfer}
equals $\int_0^1F_d\le\sup F_d$. For $r\ge1$ let $R_r=\max_{w\in\{0,1\}^r}\sup_xF_w(x)$; so $R_1=\rho$.

\begin{lemma}[parametrised transfer bound]\label[lemma]{lem:ptransfer}
Let $r\ge1$ and $\frac12<\kappa\le\rho$ satisfy $R_r\le\kappa^r$, and put $\Gamma_0=(\rho/\kappa)^{r-1}$ and
$\Gamma_1=\kappa^{-1}(\rho/\kappa)^{2r-2}$. Then the integral of \cref{lem:transfer} is at most $\Gamma_0\kappa^m$ for
every word of length $m$ with no $*$, and at most $\Gamma_1\kappa^m$ for every word of length $m$ with exactly one $*$.
\end{lemma}

\begin{proof}
Each $\mathcal T_e$ is positive, $\mathcal T_*1=1$ and $\mathcal T_e1\le\rho$ for $e\in\{0,1\}$, so
$\|\mathcal T_wf\|_\infty\le\|\mathcal T_w1\|_\infty\|f\|_\infty$ for every product $\mathcal T_w$ of these operators
and every $f\ge0$. Cut the word, in the order in which the operators are applied, into $q$ blocks of length $r$ and a
remainder of length $s<r$, so $m=qr+s$. Without $*$, each block contributes a factor at most $R_r\le\kappa^r$ and each
remaining letter a factor $\rho$, so $\sup F_d\le\kappa^{qr}\rho^s=\kappa^m(\rho/\kappa)^s\le\Gamma_0\kappa^m$. If
the $*$ lies in a block, that block contributes at most $\rho^{r-1}$, and the product is
$\kappa^{(q-1)r}\rho^{r-1+s}=\kappa^m\kappa^{-1}(\rho/\kappa)^{r-1+s}\le\Gamma_1\kappa^m$; if it lies in the
remainder, the product is $\kappa^{qr}\rho^{s-1}=\kappa^m\kappa^{-1}(\rho/\kappa)^{s-1}\le\Gamma_1\kappa^m$.
\end{proof}

For $(\kappa,r)=(\rho,1)$ this is \cref{lem:transfer}, with $\Gamma_0=1$ and $\Gamma_1=\sqrt2$. Throughout this appendix
$\nu=-\frac12\log_2\kappa$, so that $\kappa=2^{-2\nu}$ and $\frac14\le\nu<\frac12$; the hand pair has $\nu=\frac14$.

\begin{proposition}[shell bound with a frequency profile]\label[proposition]{prop:profile}
Let $(\kappa,r)$ be as in \cref{lem:ptransfer}, let
$\Sigma'=\Gamma_0/(1-\frac1{2\kappa})+\Gamma_1/(2(1-\frac1{4\kappa}))$, $\Upsilon=(1+2^{-6})(1+\frac\pi2)\Sigma'+1$ and
$C_J=\sqrt2(3+2^{-6})(2\Upsilon\kappa^{-5}+4)$, and suppose that $C_J\ge2^{13\nu}$ and $C_J\ge849\cdot2^{-14(1-\nu)}$.
Then for every $U$ with $\max U=T'\ge1$ and every real $\xi$,
\[
|J_U(\xi)|\le C_J\,2^{-\nu T'}\min\bigl\{1,(2^{T'}/|\xi|)^{1-\nu}\bigr\}.
\]
\end{proposition}

\noindent For the hand pair, $\Sigma'$, $\Upsilon$ and $C_J=633.2288\ldots$ are the constants of the proof of
\cref{cor:shell}; both pairs used below satisfy the two hypotheses. On the part $|\tau|<T'/4$ of the resonance band
(with $\xi=2^{T'+\tau}$) the factor $2^{-3|\tau|/4}$ already appears in that proof; what is new is that the window
argument is run on the whole range $4^5<\xi\le4^{T'-1}$, which makes the bound decay in $\xi$ above $2^{T'}$.

\begin{proof}
\emph{The window bound with $\kappa$.} The proof of \cref{prop:window} (\cref{app:windowproof}) uses
\cref{lem:transfer} only through \cref{lem:avg}, and \cref{lem:avg} uses it only for words with no $*$ and words with
one $*$. By \cref{lem:ptransfer}, \cref{lem:avg} and \cref{prop:window} therefore hold with $\rho^{n-k}$ replaced by
$\kappa^{n-k}$ and with the present $\Gamma_0,\Gamma_1$.

By conjugation let $\xi\ge0$, and write $\xi=2^{T'+\tau}$ when $\xi>0$. \emph{Case $T'\le13$.} If $\xi\le2^{T'}$,
\cref{lem:oneshell}(i) gives $|J_U|\le0.586\le2^{13\nu}2^{-\nu T'}$. If $\xi>2^{T'}$, \cref{lem:oneshell}(iii) gives
$|J_U|\le2^{T'}/\xi\le(2^{T'}/\xi)^{1-\nu}\le2^{13\nu}2^{-\nu T'}(2^{T'}/\xi)^{1-\nu}$.

\emph{Case $T'\ge14$, $\xi\le4^5$.} Then $\xi<2^{T'}$, and \cref{lem:oneshell}(ii) gives
$|J_U|\le0.8285\cdot1024.5\cdot2^{-T'}\le849\cdot2^{-14(1-\nu)}2^{-\nu T'}$.

\emph{Case $T'\ge14$, $4^5<\xi\le4^{T'-1}$.} Choose $\lambda=\lceil\log_4\xi\rceil$ and $n=\min(T_{\rm in},\lambda-5)$
as in the proof of \cref{cor:shell}; the parameters satisfy the bounds of \cref{tab:params}. Using
$2^{-k}\kappa^{n-k}=\kappa^n(2\kappa)^{-k}$, $\delta\le1$, $2^{-\lambda}\le2^{-6}$, $c_{\rm cov}\le1+\frac\pi2$ and
$\bar p\sum_{k\ge n}2^{-k}\le2\bar p\kappa^n$ (as $\kappa\ge\frac12$), the window bound becomes
$|J_U|\le\sqrt2\beta\bar\alpha[\Upsilon\bar p\kappa^n+2^{-T_{\rm in}}]$. From $n\ge\frac{T'-|\tau|}2-5$ we get
$\kappa^n\le\kappa^{-5}2^{-\nu(T'-|\tau|)}$, and $|\tau|\le T'$ in this range. If $\tau\ge0$, then $\bar\alpha\le2^{1-\tau}$
and $\bar p\le1$, so $\bar\alpha\bar p\kappa^n\le2\kappa^{-5}2^{-\nu T'-(1-\nu)\tau}$, and
$\bar\alpha2^{-T_{\rm in}}\le4\cdot2^{-(T'+\tau)/2}\le4\cdot2^{-\nu T'-(1-\nu)\tau}$, because
$\frac{T'+\tau}2=\nu T'+(1-\nu)\tau+(\frac12-\nu)(T'-\tau)$. If $\tau<0$, then $n<T_{\rm in}$, $\bar p\le2^\tau$ and
$\bar\alpha\le1$, so $\bar\alpha\bar p\kappa^n\le\kappa^{-5}2^{-\nu T'-(1-\nu)|\tau|}$, and
$2^{-T_{\rm in}}<2\cdot2^{-(T'+|\tau|)/2}\le2\cdot2^{-\nu T'-(1-\nu)|\tau|}$. With $\beta\le3+2^{-6}$, in both cases
$|J_U|\le C_J2^{-\nu T'}2^{-(1-\nu)|\tau|}$, which is at most the claimed bound.

\emph{Case $T'\ge14$, $\xi>4^{T'-1}$.} Then $\tau>T'-2$, and \cref{lem:oneshell}(iii) gives
$|J_U|\le2^{-\tau}=2^{-\nu\tau}2^{-(1-\nu)\tau}\le2^{2\nu}2^{-\nu T'}2^{-(1-\nu)\tau}$, and $2^{2\nu}\le2^{13\nu}\le C_J$.
\end{proof}

\begin{proof}[Proof of the profile in \cref{thm:symbol}(a), and of \cref{thm:symbol}(b)]
Let $(\kappa,r)$ satisfy the hypotheses of \cref{prop:profile}, and let $C_A=C_J/(1-2^{-(1/2-\nu)})+2$ and
$C_P=C_J/(1-2^{-(3/2-2\nu)})+1$. We show $|\Wt_S(\xi)|\le C_A2^{-\nu T}\min\{1,(2^T/|\xi|)^{1-\nu}\}$ for all $T\ge1$,
all $S$ with $\max S=T$ and all $\xi\ge0$ (negative $\xi$ by conjugation). By \cref{lem:shell},
$\Wt_S=\sum_{j<T}2^{-jz}J_{S^{(j)}}+2^{-Tz}/z$ with $\max S^{(j)}=T-j$, and $|2^{-Tz}/z|\le2\cdot2^{-T/2}$. By
\cref{prop:profile}, $|\Wt_S(\xi)|\le\sum_j2^{-j/2}C_J2^{-\nu(T-j)}+2\cdot2^{-T/2}\le C_A2^{-\nu T}$. If
$\xi=2^{T+\tau}$ with $\tau\ge0$, then $2^{T-j}/\xi=2^{-(\tau+j)}$, so shell $j$ contributes at most
$2^{-j/2}C_J2^{-\nu(T-j)-(1-\nu)(\tau+j)}=C_J2^{-\nu T-(1-\nu)\tau}2^{-j(3/2-2\nu)}$, and the last term is at most
$2^{-T/2}/\xi=2^{-3T/2-\tau}\le2^{-\nu T-(1-\nu)\tau}$. Hence $|\Wt_S(\xi)|\le C_P2^{-\nu T}(2^T/\xi)^{1-\nu}$, and
$C_P\le C_A$.

For the hand pair, $C_A\le\CAexact\le3982$, which is \cref{thm:symbol}(a). The pair $(\kappa,r)=(0.6646,12)$ satisfies
the hypothesis of \cref{lem:ptransfer} by \cref{lem:cert}; for it $\nu=0.29472\ldots$, $\Gamma_0\le1.9778$,
$\Gamma_1\le5.8855$, $\Sigma'\le12.7027$, $\Upsilon\le34.1663$, $C_J\le2264.66$ and $C_A\le17077.1$, which is
\cref{thm:symbol}(b).
\end{proof}

\subsection{The certificate}

\begin{lemma}[certificate for $\kappa=0.6646$]\label[lemma]{lem:cert}
$R_{12}\le0.0074151<0.6646^{12}$. Hence $(\kappa,r)=(0.6646,12)$ satisfies the hypothesis of \cref{lem:ptransfer}.
\end{lemma}

\begin{proof}[Proof (computer-assisted)]
Let $k=12$, $G=10$ and $M=2^{G+k}$. The program \texttt{checks/kappa\_cert\_interval.py} works in binary64 arithmetic
and moves the result of every product and sum up by one unit in the last place; since all quantities are nonnegative
and each $\mathcal T_e$ is monotone, it computes upper bounds. It starts from upper bounds for $\cos(\pi j/M)$,
$0\le j\le M/2$, obtained in interval arithmetic (\texttt{mpmath.iv}, $90$ bits), which give upper bounds for $h_0$ and
$h_1$ at the points $j/M$.

\emph{(i) Cells.} By the recursion $F_{(d,e)}=\mathcal T_eF_d$ it computes, for each word, an upper bound for the
supremum of $F_d$ on each cell $[c2^{-G},(c+1)2^{-G}]$: for $x$ in a cell of the coarser level, $x/2$ and $(x+1)/2$
lie in two known cells of the finer level, and on a dyadic cell $h_0$ and $h_1$ are monotone (a dyadic cell never
contains $\frac12$ in its interior), so their suprema there are endpoint values. Over all $4096$ words
$w\in\{0,1\}^{12}$ the largest cell bound is at most $0.0074151$, and $0.6646^{12}\ge0.0074254$ in interval
arithmetic. This proves the lemma.

\emph{(ii) Cross-checks.} The same program bounds $F_w$ at the grid points $j2^{-G}$ (largest value at most
$0.0074040$) and adds $\pi L_12^{-G-1}$, where $L_1\le0.011213$ is the largest cell bound over the $24576$ words with
exactly one $*$. This is a second valid bound: by the explicit form of $F_d$ and $|h_e'|\le\pi$,
$|F_w'|\le\pi\sum_{i<k}2^{i-k}F_{w^{(i)}}\le\pi L_1$ almost everywhere, where $w^{(i)}$ is $w$ with letter $i$ replaced
by $*$; it gives $R_{12}\le0.0074212$. A second, separately written program, \texttt{checks/kappa\_cert.py}, evaluates
the grid values in ordinary floating point from a table of cosines computed at $80$-bit precision, with an a priori
bound on the accumulated rounding error ($6\cdot10^{-15}$, padded to $10^{-13}$), and obtains $R_{12}\le0.0074236$;
its grid maximum agrees with that of the first program to within $10^{-16}$. With $\kappa=0.6645$ in place of $0.6646$
the first program reports, as it should, that the inequality is not certified.
\end{proof}

\subsection{Proof of \cref{thm:transfer}(c)}

Let $\Delta=N-T$, $b=1-\nu$, and let $\mu$, $\nu$ be as in \cref{thm:transfer}(c). Let
$J=\lceil(2\Delta+\log_2(1/\mu))/(1+\nu)\rceil\ge0$ and $\varepsilon=4.01\cdot2^{-(T+J)}$; as $T\ge9$,
$\varepsilon\le4.01/512<0.01$. Let $\Phi$, $\Phi_\varepsilon$, $k$, $Q=2^{2N-T}$, $\mathcal F$, $\mathcal G$ and
$\mathcal H$ be as in \cref{app:dirichlet}.

\emph{Smoothing error.} Pairs with $x=0$ or $y=0$ contribute at most $2^{1-N}$. For $x,y\ge1$, by \cref{lem:smooth}(a)
and \cref{lem:residues} (which needs $T\le N$) with $\ell=\lceil\varepsilon4^N\rceil$, replacing $\Phi(xy)$ by
$\Phi_\varepsilon(xy)$ changes the expectation by at most $2(3+1.9N)(\varepsilon2^N+2^{-N})=
2(3+1.9N)(4.01\cdot2^{\Delta-J}+2^{-N})$.

\emph{Main term.} By \cref{lem:smooth}(b),(c) the main term is
$4^{-N}(2\pi)^{-1}\int\widetilde\Phi(\frac12+it)k(\varepsilon(\frac12+it))\mathcal F(t)\mathcal G(t)\,dt$, with
$|\widetilde\Phi(\frac12+it)|=2^N|\Wt_S(t)|$ and $|k|\le\min(1.01,(4.01/(\varepsilon|t|))^2)$. By \cref{lem:mv} and
Cauchy--Schwarz, $I(\Omega)=\int_{-\Omega}^\Omega|\mathcal F\mathcal G|\le4.935\Omega\mathcal H+17.17\cdot2^N$ for
$\Omega\ge2$. Let $\Omega_j=2^{T+j}$ and split the line into $B_0=[-\Omega_0,\Omega_0]$ and
$B_j=\{\Omega_{j-1}<|t|\le\Omega_j\}$, $j\ge1$. On $B_j$ the hypothesis gives $|\Wt_S|\le\mu w_j$ with $w_0=1$ and
$w_j=2^{-b(j-1)}$, and $|k|\le k_j$ with $k_j=1.01$ for $j\le J$ and $k_j=4^{-(j-1-J)}$ for $j>J$ (there
$|t|>2^{T+j-1}$ and $4.01/\varepsilon=2^{T+J}$). Hence
\[
\int|\Wt_S||k||\mathcal F\mathcal G|\le\mu\sum_{j\ge0}w_jk_jI(\Omega_j)
\le\mu\Bigl[4.935\mathcal H2^T\sum_jw_jk_j2^j+17.17\cdot2^N\sum_jw_jk_j\Bigr].
\]
In the first sum the term $j=0$ is $1.01$; the terms $1\le j\le J$ add
$1.01\cdot2^b\sum_{j=1}^J2^{\nu j}\le2.02\cdot2^{\nu J}/(2^\nu-1)$; and with $j=J+1+i$ the terms $j>J$ add
$\sum_{i\ge0}2^{J+1+i-b(J+i)-2i}=2\cdot2^{\nu J}/(1-2^{-(1+b)})$. So the first sum is at most $c_A2^{\nu J}$; the second
is at most $1.01(1+\sum_{j\ge1}2^{-b(j-1)})=1.01c_B$. Multiplying by $2^N/(2\pi)$ and dividing by $4^N$, the main term is
at most $(\mu/2\pi)[4.935c_A\mathcal H2^{-\Delta+\nu J}+17.17\cdot1.01c_B]$.

\emph{Balancing.} By the choice of $J$, $2^J\ge(4^\Delta/\mu)^{1/(1+\nu)}$ and
$2^{\nu J}\le2^\nu(4^\Delta/\mu)^{\nu/(1+\nu)}$, so
$2^{\Delta-J}\le\mu^{1/(1+\nu)}2^{-\Delta(1-\nu)/(1+\nu)}$ and
$\mu2^{-\Delta+\nu J}\le2^\nu\mu^{1/(1+\nu)}2^{-\Delta(1-\nu)/(1+\nu)}$. Adding the three contributions, with
$2(3+1.9N)2^{-N}+2^{1-N}=(8+3.8N)2^{-N}$, gives the claim. For $\nu=\frac14$, $\mathcal H\le1+0.694N$ gives the bracket
$c_1\mathcal H+8.02(3+1.9N)\le37.64+24.66N$.

\emph{What changed.} \Cref{thm:transfer}(b) bounds $|\Wt_S|\le\mu$ on the whole range $|t|\le4.01/\varepsilon$ and
balances $\mu/(\varepsilon2^N)$ against $\varepsilon2^N$, which gives $\mu^{1/2}$. Here the range
$2^T<|t|\le4.01/\varepsilon$ costs $\mu2^{\nu J}$ instead of $\mu2^J$, because the symbol has already decayed there.
The same computation with a profile $(2^T/|t|)^{b}$ gives $\mu^{1/(2-b)}$; so the profile exponent is what limits
the rate at $T=N$, and $b=1$ would remove the loss.

\subsection{Proof of \cref{thm:rates} and of the sharper exponents}

\begin{proof}[Proof of \cref{thm:rates}]
By \cref{thm:symbol}(a) the hypothesis of \cref{thm:transfer}(c) holds with $\nu=\frac14$ and $\mu=3982\cdot2^{-T/4}$,
and $\mu\le1$ for $T\ge48$, since $4\log_23982<47.84$. Then $\mu^{4/5}\le758.72\cdot2^{-T/5}$ and
$c_2\mu\le38101\cdot2^{-T/4}$ (with the unrounded $c_2=9.56821\ldots$, $c_2\cdot3982<38100.7$), which gives the
displayed bound. Its first term is at most $C(1+N)2^{-T/5}$, and at most $C(1+N)2^{-T/4}$ when
$\frac T5+\frac35(N-T)\ge\frac T4$, that is when $T\le12N/13$.

\emph{Computer-assisted rates.} By \cref{thm:symbol}(b) the hypothesis of \cref{thm:transfer}(c) holds with
$\nu=-\frac12\log_20.6646$ (exactly this value; the profile exponent $1-\nu$ may not be rounded up) and
$\mu=17077.1\cdot2^{-\nu T}$, and $\mu\le1$ for $T\ge48$. With $1/(1+\nu)\ge0.77236$, $\nu/(1+\nu)\ge0.22763$,
$(1-\nu)/(1+\nu)\ge0.54473$, $17077.1^{1/(1+\nu)}\le1857.8$ and $c_2\cdot17077.1\le169030$,
\begin{multline*}
\disc_N(S)\le1857.8\,(36.40+23.81N)\,2^{-0.22763T-0.54473(N-T)}\\+169030\cdot2^{-0.29472T}+(8+3.8N)2^{-N}
\end{multline*}
for $48\le T\le N$. So the exponent is at least $0.2276-o(1)$ for all $T\le N$ as $T/\log N\to\infty$, and
$\disc_N(S)\le2^{-T/5}$ whenever $T\ge36.19\log_2N+603.9$. By \cref{thm:transfer}(a),
$\disc_N(S)\le17077.1\cdot2^{-0.29472T}+3\cdot2^{T-N}+(8+1.39N)2^{-N}$, and $2^{T-N}\le2^{-0.2987T}$ for $T\le0.77N$, so
the exponent is $0.2947-o(1)$ there.
\end{proof}

\begin{proof}[Proof of the sharper exponents]
\emph{$D=1/8$.} If $96\le T\le4N/5$, explicit range~1 of \cref{app:mainproof} gives $\disc_N(S)\le3991\cdot2^{-T/4}$,
which is at most $2^{-T/8}$ for $T\ge8\log_23991=95.70\ldots$. Let now $T\ge\frac{40}3\log_2N+205.92$; then
$N\ge T\ge246$. In \cref{thm:rates} drop the factor $2^{-3(N-T)/5}$. As $N\ge8$, the first term is at most
$758.72\cdot29.365N\,2^{-T/5}$, which is at most $\frac122^{-T/8}$ because
$2^{3T/40}\ge2\cdot758.72\cdot29.365N$ for $T\ge\frac{40}3\log_2N+205.92$ (the exact threshold is
$205.9109\ldots$). The second term is at most $\frac142^{-T/8}$ for $T\ge8\log_2(4\cdot38101)=137.74\ldots$, and the
third is at most $\frac142^{-N/8}\le\frac142^{-T/8}$ for $N\ge9$. Finally $\frac45N\ge\frac{40}3\log_2N+205.92$ for
$N\ge402$ (at $N=402$: $321.6\ge321.3$, and the left side grows faster), so for $N\ge402$ every $T\in[96,N]$ lies in
one of the two ranges.

\emph{$D=1/6$.} The same computation with $\frac15-\frac16=\frac1{30}$ and $\frac14-\frac16=\frac1{12}$: range~1 needs
$T\ge12\log_23991=143.55\ldots$; the other range is $T\ge30\log_2N+463.31$ (exact threshold $463.2996\ldots$), where
the second term needs $T\ge12\log_2(4\cdot38101)=206.61\ldots$; and the two ranges cover $[144,N]$ for every $N\ge951$.
\end{proof}

\section{The discrepancy bound beyond the midpoint}\label[appendix]{app:midpoint}

\begin{proof}[Proof of the post-midpoint discrepancy estimate]
Let $m=2N-T$ and $Q=2^m$. Both input ranges are unions of complete residue systems modulo $Q$.
For a fixed residue $a$, the number of solutions of $xy=a\bmod Q$ in $y$ for a given $x$ is at most
$\gcd(x,Q)$. Hence $\Pr[XY\equiv a\bmod Q]\le(1+m/2)/Q$, since for a uniform residue $x$,
$\mathbb E\gcd(x,Q)=1+m/2$ by summing over binary valuations, including $x=0$.
In the smoothing proof of \cref{thm:transfer}(c) use any $J\ge0$ and $\varepsilon=4.01\cdot2^{-(T+J)}$.
The number of boundary residues is at most $\lceil\varepsilon4^N\rceil$, so the smoothing error is
$O((1+m)(2^{-J}+2^{-m})+2^{-N})$, including zero operands.
The main-term frequency bounds in the same proof use $T$ only through the cutoff $2^T$ and apply here
without change. With $\nu=1/4$ and $\mu=3982\cdot2^{-T/4}$ they give
$O(\mu[(1+N)2^{T-N+J/4}+1])$.
Take $J=\lfloor(4N-3T)/5\rfloor$. The combined result is
\[
 O\!\left((1+N)
  \left[2^{-(4N-3T)/5}+2^{-T/4}+2^{-(2N-T)}+2^{-N}\right]\right),
\]
which decays exponentially throughout the asserted range. The estimate becomes uninformative at $T=4N/3$.
Since $4N-3T\ge3\eta N$, all four terms decay exponentially in $N$ and $T\le4N/3$.
Absorb the polynomial prefactor by increasing $N_\eta$.
\end{proof}

\ificml
The $4N/3$ cutoff comes from the two competing error terms in this proof.
Exhaustive real-sign optima for widths two through four extend past this
point (code supplement). They measure real bounded tests at finite widths.
\fi

\ificml\else
\section{Cell structure alone is not enough}\label[appendix]{app:necsec}

\begin{figure}[t]
\centering
\includegraphics[width=0.8\linewidth]{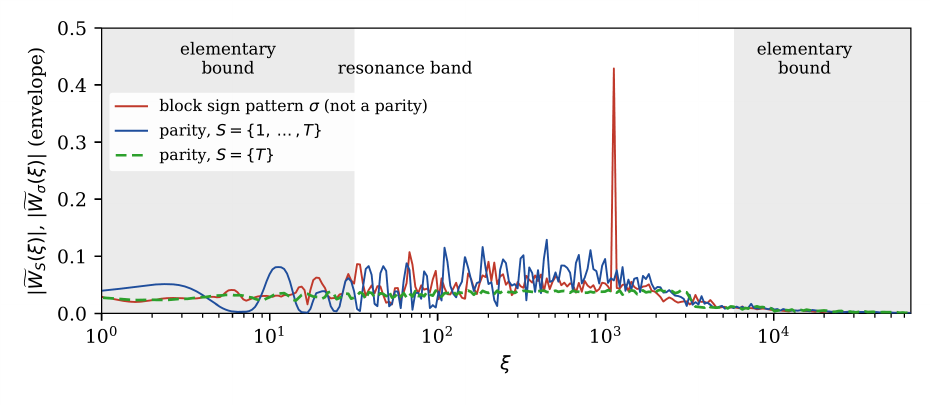}
\caption{Where parity enters: upper envelope of $|\Wt(\xi)|$ for $T=10$ (maximum over bins of width $0.04$ in
$\ln\xi$). Shaded: the ranges covered by the elementary bounds of \cref{prop:warmup}. Blue (solid): the parity $S=[T]$;
green (dashed): the parity $S=\{T\}$. Red: a sign
pattern of the kind used in \cref{prop:necessity}, chosen to align the block integrals at $\xi=\neckappa\pi\cdot2^{T}$; it
produces a large peak inside the resonance band, which neither plotted parity does; for large $T$, \cref{thm:symbol} rules
such a peak out for every parity.}
\label{fig:mellin}
\end{figure}

\begin{proposition}[cell structure is not enough]\label[proposition]{prop:necessity}
Let $T\ge3$. For $\sigma\in\{\pm1\}^{2^{T-1}}$ let $W_\sigma(u)=\sigma_{\lfloor2^{T-1}u\rfloor}\,r_T(u)$, a function that,
like $W_S$, is constant on cells of length $2^{-T}$ and has mean zero on each pair of cells, and let $K_\sigma$ be the
operator with kernel $W_\sigma(st)$. Then some $\sigma$ satisfies $\|K_\sigma\|\ge\necc-\necerr\cdot2^{-T}$; for
$T\ge\necTthr$ this is at least $\necLarge$.
\end{proposition}

\Cref{prop:warmup} and \cref{thm:transfer} never use that $W_S$ is a parity. \Cref{prop:necessity} shows that the
Mellin bound itself needs more: among functions that are constant on cells of length $2^{-T}$ and have mean zero on each
pair of cells, some have operator norm bounded away from zero, so the parities are special.

\paragraph{Idea of the proof.} Test $K_\sigma$ against $\varphi(s)=s^{-1/2+i\xi}\mathbf 1_{[1/2,1]}(s)$ with
$\xi=\kappa\pi2^T$. The substitution $u=st$ turns $\langle\bar\varphi,K_\sigma\varphi\rangle$ into
$\int_{1/4}^1W_\sigma(u)u^{z-1}\Delta(u)du$ with $\Delta(u)=\ln2-|\ln2u|$, a sum of one integral $a_b$ per pair of
cells. On a pair of cells the phase is nearly linear, so $a_b$ is close to an explicit number whose size does not
decay with $T$. Choosing $\sigma_b$ as the sign of $\mathrm{Re}(e^{-i\theta}a_b)$ for a good angle $\theta$ makes the
terms add up coherently, which is impossible for a parity. With $\kappa=\neckappa$ this gives
$\|K_\sigma\|\ge\necc-\necerr\cdot2^{-T}$ (\cref{app:necessity}).

\begin{figure}[t]
\centering
\includegraphics[width=0.85\linewidth]{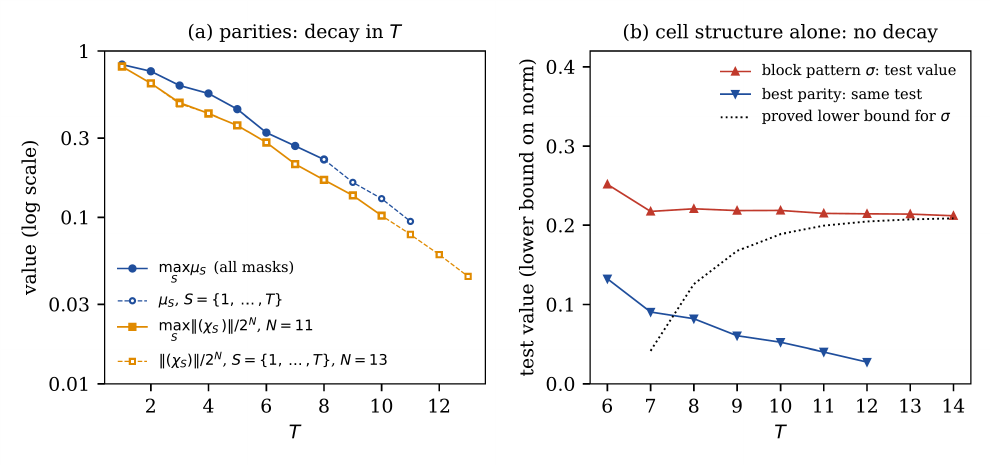}
\caption{(a, log scale) Parities decay. Blue: computed $\mu_S$; orange: $\|(\chi_S(x,y))\|/2^N$. Solid: maximum over
all masks (as far as all masks were computed); dashed: $S=[T]$, which is the maximiser wherever all masks were computed and
$T\ge4$. The dashed curves extend to larger $T$; there they are the values marked $\dagger$ in \cref{tab:num}.
(b) Cell structure alone gives no decay. Red: the test value $|\langle\bar\varphi,K_\sigma\varphi\rangle|/\|\varphi\|^2$,
a lower bound for $\|K_\sigma\|$, for the explicit pattern of \cref{prop:necessity}; blue: the largest value of the same
test over all parities; dotted: the proved lower bound $\necc-\necerr\cdot2^{-T}$, drawn from $T=7$ because it is
negative for $T\le6$.}
\label{fig:decay}
\end{figure}

Numerically (\cref{fig:decay}), the test value for the explicit pattern lies between $\necSigMin$ and $\necSigMax$
for $\necTmin\le T\le\necTmax$, while the largest test value over all parities at the same frequency decreases from
$\necWfirst$ at $T=\necWTfirst$ to $\necWlast$ at $T=\necWTlast$. The proof of \cref{thm:symbol} uses parity in exactly
two places: the same inner pattern in every window (step~(i) of the proof of \cref{prop:window}) and the product formula (step~(iii)). Any bound on the
norm of the continuum operator that uses only the cell structure (constancy on cells of length $2^{-T}$ and mean zero on
pairs of cells) cannot decay in $T$; a proof of \cref{conj:C5} through the continuum operator must use some property of
parities beyond these two.

\begin{remark}[a discrete version; sketch]\label[remark]{rem:discrete}
The pattern also defeats the integer problem. The pattern $\sigma$ depends on $T$ (it is a sign pattern on the top $T$
bits), so this is a family of integer problems, one for each $T$. Define $\disc_N(\sigma)=\max\{|\E
f(X)g(Y)W_\sigma(XY/4^N)|:\ |f|,|g|\le1\}$, as for $\disc_N(S)$. Take $f(x)=g(x)=2^{-1/2}\varphi(x2^{-N})$, which
satisfies $|f|\le1$. The cell-by-cell comparison of \cref{sec:discrete} applies to $W_\sigma$ unchanged, and
$\varphi$ varies by $O(2^{T-N})$ on a cell, so the integer form is $\frac12\langle\bar\varphi,K_\sigma\varphi\rangle
+O(2^{T-N}+N2^{-N})$. Hence $\disc_N(\sigma)\ge\frac{\ln2}2(\necc-\necerr\cdot2^{-T})-O(2^{T-N}+N2^{-N})$, which does
not decay in $T$ for $T\le N-O(1)$. We have not tracked the implied constants, so this remark is a sketch and is not
used elsewhere.
\end{remark}

\subsection{Proof of Proposition~\ref{prop:necessity}}\label[appendix]{app:necessity}

The pairs of cells are $B_b=[b\,2^{1-T},(b+1)2^{1-T})$. Let $\kappa>0$, $\xi=\kappa\pi2^T$,
$z=\frac12+i\xi$ and $\varphi(s)=s^{z-1}\mathbf 1_{[1/2,1]}(s)$, so $\|\varphi\|_2^2=\ln2$.

\emph{Step 1.} With $u=st$, $\int\!\!\int\varphi(s)\varphi(t)W_\sigma(st)\,ds\,dt=\int_{1/4}^1W_\sigma(u)u^{z-1}\Delta(u)du$, where
$\Delta(u)=\int\mathbf 1[s,u/s\in[\frac12,1]]\,ds/s=\ln2-|\ln2u|$. Its modulus is at most $\|K_\sigma\|\ln2$.

\emph{Step 2.} The integral equals $\sum_b\sigma_ba_b$ with $a_b=\int_{B_b}r_Tu^{z-1}\Delta$. Since the average of
$|\mathrm{Re}(e^{-i\theta}a)|$ over $\theta$ is $\frac2\pi|a|$, some $\theta$ has $\sum_b|\mathrm{Re}(e^{-i\theta}a_b)|\ge\frac2\pi\sum_b|a_b|$;
with $\sigma_b=\sgn\mathrm{Re}(e^{-i\theta}a_b)$, $|\sum_b\sigma_ba_b|\ge\frac2\pi\sum_b|a_b|$.

\emph{Step 3.} Let $c$ be the centre of $B_b\subseteq[\frac14,1]$ and $\omega=\xi/c$. On $B_b$,
$|u^{z-1}\Delta(u)-c^{z-1}\Delta(c)e^{i\omega(u-c)}|\le(11+35\kappa)2^{-T}$, because $u^{-1/2}\Delta(u)$ is $10.8$-Lipschitz on
$[\frac14,1]$, $c^{-1/2}\Delta(c)\le1.39$, and $|\ln u-\ln c-(u-c)/c|\le8\cdot4^{-T}$. The model integral is
$\int_{B_b}r_T(u)e^{i\omega(u-c)}du=-2i(1-\cos(\omega2^{-T}))/\omega$. Hence
$|a_b|\ge\frac{2^{-T}}{\kappa\pi}\,2g(c)-2(11+35\kappa)4^{-T}$, where $g(c)=c^{1/2}\Delta(c)(1-\cos(\kappa\pi/c))$.

\emph{Step 4.} There are $3\cdot2^T/8$ pairs of cells in $[\frac14,1]$. The midpoint rule for $g$, which is
$(5.4+35\kappa)$-Lipschitz, gives
$\sum_b|a_b|\ge I(\kappa)-\bigl(\frac38\frac{5.4+35\kappa}{\kappa\pi}+\frac34(11+35\kappa)\bigr)2^{-T}$ with
$I(\kappa)=\frac1{\kappa\pi}\int_{1/4}^1g$. Combining with Steps 1--2,
$\|K_\sigma\|\ge\frac2{\pi\ln2}\sum_b|a_b|$. At $\kappa=\neckappa$, $I(\kappa)\ge\necI$, and the bound becomes
$\|K_\sigma\|\ge\necc-\necerr\cdot2^{-T}$. \jmlrQED

\emph{Rigorous evaluation of the constants.} $I(\kappa)$ is computed in interval arithmetic by the composite midpoint
rule on $[\frac14,\frac12]$ and $[\frac12,1]$, where $g$ is smooth ($\Delta(c)=\ln4c$ and $\Delta(c)=-\ln c$
respectively), with $3000$ subintervals on each piece and the error term $h^3\sup|g''|/24$ per subinterval, the supremum
taken by interval evaluation of $g''$ over the subinterval. This gives $I(\neckappa)\in[0.2285982,0.2285984]$, hence
$I(\neckappa)\ge\necI$. The same script (\texttt{checks/constants\_interval.py}) certifies $2I/(\pi\ln2)\ge\necc$, the
error coefficient $21.544\ldots\le\necerr$, the value $0.1993\ldots\ge\necLarge$ at $T=\necTthr$, and, by interval
bounds on the derivatives over the same subintervals, the Lipschitz constants $8.0\le10.8$ for $u^{-1/2}\Delta(u)$ and
$4.36\le5.4+35\kappa$ for $g$, and $\max u^{-1/2}\Delta(u)\le0.981\le1.39$.

\section{How sharp are the bounds?}\label[appendix]{app:numerics}

\paragraph{Why the exponents are $1/8$ and $1/6$, and not $\expSpec$.} The Mellin bound has exponent $1/4$ because the proof sums
the resonance windows with the triangle inequality; a computer-assisted transfer bound over blocks of $12$ digits raises
it to $0.2947$. Near $T=N$ the passage to integers loses the power $\frac45$ (\cref{thm:transfer}(c)), and absorbing the
factor $O(N)$ and the constants yields $1/8$ and $1/6$. Least-squares fits for $S=[T]$ give $2^{-\expSpec\,T}$ for the
spectral norm ($\expSpecRange$; \cref{app:numerics}); this is numerical evidence, not a bound.

\fi
\begin{example}[small cases]\label[example]{ex:small}
For $N=1$ and $S=\{1\}$, $\chi_S\equiv1$, so $\disc_1(\{1\})=1$. For $N=2$, an exhaustive computation over all $S$ and
all $\pm1$-valued $f,g$ gives the largest value $\NtwoDisc$, attained at $S=\NtwoS$, so
$\disc_2(\NtwoS)\ge\NtwoDisc$; the largest correlation of a single $\chi_S$ (with $f=g=1$) is also $\NtwoChar$. By \cref{thm:main} the value is below $1$ for every $N\ge2$.
\end{example}
\ificml\else

\begin{table}[t]
\centering\small
\begin{tabular}{@{}cccc@{}}
\toprule
$T$ & $\max_S\mu_S$ & $\max_S\|(\chi_S)\|/2^N$, $N=11$ & $\max_{S\in\mathcal S_T}\|(\chi_S)\|/2^N$, $N=13$\\
\midrule
\input{tab_num.tex} 
\bottomrule
\end{tabular}
\caption{Computed values. Column 2: $\mu_S$ maximised over all masks for $T\le\melTmaxAll$ and, marked $\dagger$, the value
for $S=[T]$ beyond (supremum over a fine $\xi$-grid with local refinement). Column 3: spectral
norm of the $2^N\times2^N$ matrix $(\chi_S(x,y))$, divided by $2^N$, over all masks.
Column 4: the same over a family $\mathcal S_T$ of six masks: $[T]$, $\{T\}$, $\{T-1,T\}$, $\{1,T\}$ and the two
alternating masks $\{T,T-2,T-4,\dots\}$ and $\{T,T-1,T-3,T-5,\dots\}$ (all within $[T]$). The spectral norm bounds $\disc_N(S)$. Floating-point evidence, not part of any proof.}
\label{tab:num}
\end{table}

\Cref{tab:num} lists computed values. Wherever all masks were computed and $T\ge4$, the largest values are attained at
$S=[T]$, both for $\mu_S$ and for the spectral norm; within the six-mask family at $N=13$, $S=[T]$ is the largest for
every $T\ge4$ up to $\famTmax$. The normalised spectral norms are essentially independent of $N$: for the values of $S=[T]$
with $T\ge4$, the relative spread over $N=11,\dots,14$ is at most $\specSpreadPct\%$. A least-squares fit gives decay
$2^{-\expSpec\,T}$ for the spectral norm ($\expSpecRange$, $N=13$) and $2^{-\expMel\,T}$ for $\mu_{[T]}$ ($\expMelRange$).
The proved exponents $1/4$ and, computer-assisted, $0.2947$ are smaller, and the proved constant $3982$ makes
\cref{thm:symbol} non-trivial only for $T\ge\TAone$; the sharper rates in \cref{app:explicit-d} reflect the
loss $\mu\mapsto\mu^{4/5}$ of \cref{thm:transfer}(c) near $T=N$ and a generous treatment of the factor $O(N)$.

\section{Classical Fourier and bilinear estimates}\label[appendix]{app:classical}

For $C=\emptyset$ the bit $M_T$ is a single square
wave in $xy/4^N$; its Fourier coefficients live at frequencies $|h|\ge2^{T-1}$, and the type~II bound gives decay in $T$.
\paragraph{Fourier expansion.} A natural attempt for general $C$ expands $W_S(u)=\sum_hc_S(h)e(hu)$ in Fourier series and applies to each term the per-frequency
bilinear bound $|\E\,\alpha(X)\beta(Y)e(hXY/4^N)|\lesssim((1+\ln|h|)/|h|)^{1/2}$, valid for $|\alpha|,|\beta|\le1$ and
$0<|h|\le2^{N-1}$. Estimating the low-frequency portion costs the weighted absolute mass
$\Lambda_{\mathrm{cl}}(S)=\sum_{h\ne0}|c_S(h)|\,((1+\ln|h|)/|h|)^{1/2}$.
In the computed range its maximum is \clSeq{} for $T=1,\dots,\clTmax$
(rounded to two decimals). A two-digit Fourier coefficient contraction
gives asymptotic decay of this mass. To obtain a uniform lattice bound,
one must also control the higher-frequency phases and the error of
smoothing the jumps.

\paragraph{Bourgain's bound.} Bourgain~\cite[\S2]{Bourgain11} bounds bilinear sums $\sum_{m\sim M,\,n\sim N'}\alpha_m\beta_nw_S(mn)$ with
$|\alpha_m|,|\beta_n|\le1$, $MN'\sim X=2^\lambda$, $M\sim2^\mu\le N'$, where $w_S$ is a Walsh function with digits counted
from the bottom. His estimate~(2.29) reads
\[
\Bigl|\sum_{m,n}\alpha_m\beta_nw_S(mn)\Bigr|\lesssim X\bigl(L^{-c\varepsilon}+L^2M^{-c}+L^C2^{-c|S'|}\bigr),\qquad
S'=S\cap[K,K+\mu+(1+\varepsilon)\rho],
\]
with $L=2^\rho$ a parameter satisfying $\mu/100>\rho\gg(\log\lambda)^2$, $\varepsilon>0$ fixed, absolute constants
$c,C>0$, a window position $K$ that may be chosen, and $\lesssim$ hiding constant factors (Bourgain writes $<$). The bound is non-trivial only when $|S'|$ exceeds a constant
multiple of $\rho$. So if every window of $\mu+(1+\varepsilon)\rho$ digits contains only boundedly many bits of the mask,
for example $S=\{T\}\cup C$ with $|C|$ bounded, the right-hand side of (2.29) is not below the trivial bound and gives no
decay in $T$. This is a statement about that bound, not about all methods: for $C=\emptyset$ the type~II argument above
does give decay. A finite Fourier proof can handle sparse and dense masks together by
using a two-digit coefficient bound and smoothing on the integer grid.

\section{Consequences for transformers: full discussion}\label[appendix]{app:transfull}

\begin{table}[ht]
\centering\footnotesize
\begin{tabular}{@{}>{\raggedright}p{0.40\linewidth}>{\raggedright}p{0.07\linewidth}>{\raggedright\arraybackslash}p{0.47\linewidth}@{}}
\toprule
Model (result) & Case & Statement\\
\midrule
\multicolumn{3}{@{}l}{\emph{New; uses \cref{thm:main} uniformly over masks}}\\
one layer (\cref{thm:onelayer,cor:onelayer}) & average & $\le2(2n)^H2^{-DT}$; $\Omega(N/\log N)$ heads for advantage $2^{-DN/2}$ at $T=N$\\
two layers, fixed heads (\cref{thm:full-two}) & average & $\exp(-\Omega(N/\log^2N))$\\
three layers, unique, explicit polynomials only (\cref{thm:three}) & average & $\le\exp(-n^\gamma)$ for $\gamma<\frac1{13+4H_1}$\\
three layers, top-$k$, fixed $H$ (\cref{thm:topk-depth,cor:topk-three}) & average & $\exp(-\Omega_{\alpha,H}(n/(\ln n)^{12}))$ for $1\le k\le\ln n$\\
$L$ layers, top-$k$ (\cref{thm:topk-depth,cor:topk-loglog}) & average & $\exp(-\Omega(n/(\ln n)^{2^L-2}))$ at fixed $L,H,k$. Inverse-polynomial advantage requires $L>J$\\
greedy softmax with score and output margins (\cref{cor:softmax}) & average & same fixed-depth bounds as top-$k$. Inverse-polynomial advantage requires $L>J$\\
last $w$ output bits (\cref{cor:avgwindow}) & average & $\le2^{Lw+2-\Omega(T/(\log N)^d)}$ for $L$ layers, $d=d(L)$\\
full product, $L\ge2$ layers, top-$k$ (\cref{thm:cot-avg}) & average &
for $q=2^{1-L}$ and $b=\lfloor c_L N^q/(\ln N)^{2-q}\rfloor$, probability $\le e^{-c^\prime b}$ of a correct trace with at most $c_L b/\ln N$ auxiliary tokens\\
\midrule
\multicolumn{3}{@{}l}{\emph{New; worst case, does not use \cref{thm:main}}}\\
full product, $L$ layers (\cref{thm:cot-worst}) & worst & $N^{\Omega(1/L)}$ auxiliary tokens on some input\\
\midrule
\multicolumn{3}{@{}l}{\emph{Known method, or only the case $C=\emptyset$}}\\
full attention, $O(1)$ layers (\cref{thm:worstcase}) & worst & fails if $\min(T,2N-T)\ge N^\varepsilon$ (Buss's embedding)\\
last $w$ output bits, $w$ fixed (\cref{thm:window}) & worst & fails at every $T\in[\alpha N,N]$\\
chain of thought ending in $M_T$ (\cref{prop:singlebit}) & worst & length $\Omega(N)$; uses only $C=\emptyset$\\
\midrule
\multicolumn{3}{@{}l}{\emph{Upper bounds (constructions)}}\\
$\lceil\log_2(2N+1)\rceil$ layers, unique, fixed routing (\cref{rem:depth-upper}) & worst & exact; generic (any function of the operands)\\
$O(\log n/\log\log n)$ layers, top-$1$ (\cref{rem:depth-gap}, sketch) & average & error $N^{-a}$; $2^{O(\sqrt{\log n})}$-bit states\\
nine layers, averaging or softmax (\cref{rem:averaging}, after~\cite{Feng24}) & worst & exact, $O(\log n)$-bit activations\\
\bottomrule
\end{tabular}
\caption{Consequences for transformers that output $M_1M_2\cdots$ under teacher forcing, and upper-bound constructions.
Here $\alpha\in(0,1]$ is fixed, $\alpha N\le T\le N$, $n$ is the context length, $H_1$ counts first-layer heads and $H$
bounds the heads per layer. For $u=\ln n$, the hard and greedy softmax depth bounds use
$J=\log_2u-\log_2\ln u-\lceil\log_2(8+4\log_2(1+Hk))\rceil$, also for varying $H,k$.
The softmax row assumes the temperature, uniform Lipschitz and value bounds, and margins of \cref{cor:softmax}.
Advantage is the probability of a correct prediction on uniform operands minus $1/2$. The last row of the first block
and the row of the second block concern chains of thought that write the whole product.}
\label{tab:transformers}
\end{table}

\subsection{Overview: windowed models and full attention}\label[appendix]{app:overview}

\begin{remark}[communication]\label[remark]{rem:cc}
Let $N\ge2$ and $\max S=T\le N$, and give $X$ to Alice and $Y$ to Bob. For sets $\mathcal A$, $\mathcal B$ of values of $X$ and $Y$, the bias of
$\mathrm{par}_S=\bigoplus_{k\in S}M_k$ on the rectangle $\mathcal A\times\mathcal B$ is
$|\E\mathbf 1_{\mathcal A}(X)\mathbf 1_{\mathcal B}(Y)\chi_S(X,Y)|\le\disc_N(S)$. So by the discrepancy
method~\cite{KN97} every public-coin protocol computing $\mathrm{par}_S$ with error at most $\varepsilon<\frac12$ on
every input exchanges at least $DT-\log_2\frac1{1-2\varepsilon}$ bits, with $D$ as in \cref{thm:main} ($D=1/8$ if $N\ge402$ and $96\le T\le N$). This is a routine consequence of
\cref{thm:main}; for $S=\{T\}$ it follows from the classical case $C=\emptyset$ above, and \cref{sec:related} compares it
with the literature on bits of multiplication. Given the prefix $M_{<T}$, by contrast, one step is a GREATER-THAN instance
of randomised cost $O(\log N)$ (\cref{prop:gt}).
\end{remark}

\paragraph{Branching programs.} The middle bit (our $M_{N+1}$) is the standard hard bit for branching
programs~\cite{Bryant91,Woelfel05,AK03,SW03,WW07}, and the most significant bit also needs exponential
OBDDs~\cite{Bollig08}; these are worst-case bounds for partitions induced by a variable order. OBDDs for the middle bit need
exponential size~\cite{Bryant91,Woelfel05}, randomized OBDD bounds go through one-way communication
complexity~\cite{AK03}, randomized read-$k$ programs obey time--space lower bounds~\cite{SW03} while read-$O(\log n)$ ones
work in log space~\cite{WW07}; on uniform inputs, log-space read-once programs approximate the middle bit with
polynomially small error (\cite{SW03}, as reported in~\cite{WW07}).

\paragraph{Further related work.} Communication complexity gives lower bounds for one layer:
Sanford, Hsu and Telgarsky~\cite{SHT23} reduce set disjointness to a single attention unit, and Peng, Narayanan and
Papadimitriou~\cite{PNP24} simulate one softmax layer by a short protocol; Chen, Peng and Wu~\cite{CPW24} reach
multi-layer decoder-only transformers with an autoregressive communication model. Head-count lower bounds for one layer
are known for parity: Rajaraman, Sundaram and Tesfaye~\cite{RST26} and Hsu~\cite{Hsu26}; \cref{thm:onelayer} gives one
for multiplication.

\paragraph{Attention to recent output bits, and full attention.}\label[appendix]{sec:window} A \emph{$w$-windowed} model is a
UHAT in which a head at the position of $M_q$ ranges only over the input positions and the positions of
$M_{q-w},\dots,M_q$ (\cref{def:window}); $w=0$ is a Markovian chain of thought. With $L$ layers, the output at the
position of $M_{T-1}$ is then a function $F(X,Y,M_{\mathcal W})$ with $|\mathcal W|\le Lw+1$ (\cref{lem:cone}).
\Cref{thm:avgwindow} bounds the advantage of any such $F$ by $2^{|\mathcal W|-1}((2N+1)^{t/2}2^{-DT}+\delta_t^{1/2})$,
where $\delta_t$ is the Fourier weight above degree $t$ of the functions $F(\cdot,\cdot,\sigma)$; Tal's Fourier tail
bound for $\mathrm{AC}^0$~\cite{Tal17} and the circuits of Hao, Angluin and Frank~\cite[Thm.~1]{HAF22} then give
advantage $2^{Lw+2-\Omega(T/(\log N)^d)}$, with $d$ depending only on $L$ (\cref{cor:avgwindow}). The Fourier analysis is
over $(X,Y)$ only, for each value of the window; this is what the original transfer, which Fourier-expanded over all $3N-1$ bits, could not do. For full attention the worst case is classical: Buss~\cite{Buss92} chose $X$
and $Y$ so that one block of the product holds the count $\sum_iz_i$ of some input bits while every other product bit
is $0$ or a single $z_i$, and with the $\mathrm{AC}^0$ bound for UHATs this shows that no UHAT with $O(1)$ layers and
heads outputs $M_T$ correctly for all inputs, at $T=N$ and at every $T$ with $\min(T,2N-T)\ge N^\varepsilon$
(\cref{thm:worstcase}; we use a denser variant of the embedding built on a Sidon set, \cref{lem:sidon}). This does not
use \cref{conj:C5} and says nothing about the average case.

\paragraph{What remains open.} For unique and top-$k$ hard attention with fixed head counts the advantage at
step $T\in[\alpha N,N]$ is below $\exp(-n^\gamma)$ for every $\gamma<1$ at every fixed depth, for fixed $k$ and a fixed
tie priority (\cref{thm:topk-depth}), and the same holds for softmax attention with score and output margins
(\cref{cor:softmax}). Averaging hard attention and softmax without margins can predict $M_T$ exactly at
fixed depth (\cref{rem:averaging}). Open are the depth needed for inverse-polynomial advantage, between
$\log_2\ln n-\log_2\ln\ln n-O(1)$ and the $O(\log n/\log\log n)$ of the sketched construction of \cref{rem:depth-gap}
(\cref{sec:depthgap}), and the optimal exponents.
\Cref{app:barriers-disc} shows why communication (\cref{prop:disj,prop:gt}), restriction
(\cref{prop:restriction}) and uniform-measure Fourier arguments each fail here, for precise reasons.

\fi
\ificml\else
\subsection{The model in detail}\label[appendix]{app:model}

We use the unique-hard-attention transformers (UHATs) of~\cite[\S2]{AHRH25}. Position $i$ holds a token; layer $0$ is
$y_i^{(0)}=\mathrm{emb}(\text{token}_i)+p_i$ with arbitrary token embeddings $\mathrm{emb}$ and positional encodings
$p_i$. In layer $\ell$, each of $H$ heads at position $i$ computes bilinear scores
$a_{i,j}=(y_j^{(\ell-1)})^\top\mathbf K_h^\top\mathbf Q_hy_i^{(\ell-1)}$ for $j\le i$ (causal masking) and attends to
the single position $j$ with the largest score; then $y_i^{(\ell)}=f^{\rm MLP}(y_i^{(\ell-1)}+\sum_hV_hy_{j_h}^{(\ell-1)})$
with an arbitrary map $f^{\rm MLP}$, and the prediction at position $i$ is read from $Oy_i^{(L)}$. Width and precision
are unrestricted. In~\cite{AHRH25} ties between equal scores are broken by a fixed rule, with two named examples: the
left-most and the right-most maximiser. We allow any fixed priority order on positions, which covers both. As
in~\cite[Def.~2.2]{AHRH25}, a model is a family with one transformer per total length and a bounded number $L$ of
layers and $H$ of heads. This non-uniformity can only help the model, and it costs our lower bounds nothing, because
each of them is proved at one length at a time.

\emph{Symbols in this section and in \cref{app:transformers}.} The model writes $M_T$ at step $T$; $n$ is the context
length, $L$ and $H$ are the numbers of layers and heads ($H_1,H_2$ per layer in \cref{thm:full-two}), $w$ is the width
of an attention window (\cref{sec:window}), and $\mathcal W\subseteq[T-1]$ is a set of earlier output positions. The
\emph{advantage} of a prediction of $M_T$ is the probability over uniform $(X,Y)$ that it is correct, minus $\frac12$; it
is at most $\frac12|\E\hat M_T\hat O_T|$ for the signed output $\hat O_T$ of \cref{thm:onelayer}, and $\eta$ always denotes
an advantage. \Cref{tab:notation} lists the letters that are also used, with other meanings, in
\cref{sec:symbol,sec:discrete}.

\emph{Input and output.} The input $x$ lists the $2N$ bits of $X$ and $Y$, one per position, in any fixed order, with
possibly $O(1)$ separator tokens (such as `$=$') at fixed positions. The model writes $M_1M_2\cdots M_{2N}$. Following
\cite[Def.~2.2]{AHRH25} it is judged on the \emph{true} prefix (teacher forcing): step $T$ is correct on $(X,Y)$ if
the output at the last position of $xM_1\cdots M_{T-1}$ is $M_T$. By causal masking this output depends only on that
prefix, so a model that is correct at every step on every input generates the product by itself. Throughout,
$\hat Z=2Z-1$ as in \cref{sec:intro}.

\emph{Why ties, masking and the true prefix matter.} Three hypotheses are used. (1) The tie rule is a priority order
on positions. A rule that depends on the whole set of tied positions (say, the median one) splits the attention event
of \cref{thm:onelayer} further, and its count of terms is lost. (2) Causal masking: under~\cite[Def.~2.2]{AHRH25} the
model runs on the whole string $xM_1\cdots M_{2N}$, and without the mask the position that predicts $M_T$ could attend
to $M_T$ itself. (3) The model is judged on the true prefix; if it were run on its own possibly wrong earlier outputs,
those tokens would not be bits of the product.

\subsection{Explicit polynomials: selectors}\label[appendix]{app:selectors}

The proofs for two and three UHAT layers below construct, for a function $G$ on a Boolean cube, a real polynomial that
is uniformly close to $G$ on the \emph{whole} cube and has small degree. This subsection collects the tools. They are
elementary: two facts about Chebyshev polynomials and the triangle inequality in $\ell^1$. Throughout, $\mathcal X$ is
a Boolean cube $\{0,1\}^M$, a \emph{polynomial} is a real polynomial in the $M$ coordinates, and $\deg$ is total degree.
All degree statements concern the specific polynomials constructed, and all error statements are pointwise on
$\mathcal X$. For a finite set $\Omega$ and $w\in\R^\Omega$ we write $\|w\|_1=\sum_\omega|w_\omega|$, and $\mathbf
1_\omega$ is the indicator vector of $\omega\in\Omega$.

\begin{definition}[selectors]\label[definition]{def:selector}
Let $f:\mathcal X\to\Omega$ with $\Omega$ finite. A \emph{selector for $f$ with error $\delta$} is a family
$w=(w_\omega)_{\omega\in\Omega}$ of polynomials on $\mathcal X$ with $\|w(x)-\mathbf 1_{f(x)}\|_1\le\delta$ for every
$x\in\mathcal X$. Its degree is $\max_\omega\deg w_\omega$. It is \emph{exact} if $\delta=0$.
\end{definition}

\noindent Note that $\|w(x)\|_1\le1+\delta$. A selector is the polynomial counterpart of a procedure that outputs $f(x)$
except with probability $\delta$; the $w_\omega$ play the role of outcome probabilities, but they may be negative and
need not sum to one.

\begin{lemma}[post-processing and values]\label[lemma]{lem:sel-fibre}
Let $w$ be a selector for $f$ with error $\delta$. \textup{(a)} For $g:\Omega\to\Omega'$, the family
$u_\nu=\sum_{\omega:g(\omega)=\nu}w_\omega$ is a selector for $g\circ f$ with error at most $\delta$ and degree at most
$\deg w$. \textup{(b)} For $c:\Omega\to\R$ with $|c|\le B$, the polynomial $p=\sum_\omega c(\omega)w_\omega$ satisfies
$|p(x)-c(f(x))|\le B\delta$ on $\mathcal X$.
\end{lemma}

\begin{proof}
(a) $u(x)-\mathbf 1_{g(f(x))}$ is the image of $w(x)-\mathbf 1_{f(x)}$ under summation over the fibres of $g$, which
does not increase the $\ell^1$ norm. (b) $p(x)-c(f(x))=\sum_\omega c(\omega)(w_\omega(x)-[\omega=f(x)])$.
\end{proof}

\begin{lemma}[sequential composition]\label[lemma]{lem:sel-compose}
Let $w$ be a selector for $f:\mathcal X\to\Omega$ with error $\delta_1$. For every $\omega\in\Omega$ let
$g_\omega:\mathcal X\to\Omega'$ have a selector $v^\omega$ with error at most $\delta_2$, at every $x$, also when
$\omega\ne f(x)$. Then $u_\nu=\sum_\omega w_\omega v^\omega_\nu$ is a selector for $h(x)=g_{f(x)}(x)$ with error at
most $(1+\delta_1)(1+\delta_2)-1$ and degree at most $\max_\omega(\deg w_\omega+\deg v^\omega)$.
\end{lemma}

\begin{proof}
$u(x)-\mathbf 1_{h(x)}=\sum_\omega(w_\omega(x)-[\omega=f(x)])v^\omega(x)+(v^{f(x)}(x)-\mathbf 1_{h(x)})$. The second
term has norm at most $\delta_2$, and the first at most $\delta_1\max_\omega\|v^\omega(x)\|_1\le\delta_1(1+\delta_2)$.
\end{proof}

\noindent The cross term $\delta_1\delta_2$ cannot be dropped: $w=(1+\delta_1)\mathbf 1_f$ and
$v^f=(1+\delta_2)\mathbf 1_h$ attain the bound. Three consequences are used. (S2a) \emph{Chains:} $t$ stages with
errors $\delta_1,\dots,\delta_t$ give error at most $\prod_i(1+\delta_i)-1$ and degree at most the sum of the stage
degrees; an exact constant stage costs nothing. (S2b) \emph{Products:} if $w^{(i)}$ is a selector for
$f_i:\mathcal X\to\Omega_i$ with error $\delta_i$, then $(\prod_iw^{(i)}_{\omega_i})_{(\omega_1,\dots,\omega_t)}$ is a
selector for $(f_1,\dots,f_t)$ with error at most $\prod_i(1+\delta_i)-1$ and degree at most $\sum_i\deg w^{(i)}$.
(S2c) \emph{Values:} if $|p_\omega-\Gamma_\omega|\le\theta'$ on $\mathcal X$ for every $\omega$, where
$|\Gamma_\omega|\le1$, then $P=\sum_\omega w_\omega p_\omega$ satisfies $|P(x)-\Gamma_{f(x)}(x)|\le\theta'+
\delta_1(1+\theta')$, and $\deg P\le\max_\omega(\deg w_\omega+\deg p_\omega)$. In words: a selector lets us ``read
$f$ and then act on the outcome''; the sum runs over all outcomes, right or wrong, so the hypothesis on the second stage
must hold for every outcome.

\begin{lemma}[a multilinear polynomial off the cube]\label[lemma]{lem:sel-offcube}
Let $q$ be a multilinear polynomial in $m$ variables with $|q|\le B$ on $\{0,1\}^m$. For $z\in\{0,1\}^m$ and arbitrary
$Z\in\R^m$, $|q(Z)-q(z)|\le B\bigl(\prod_{i=1}^m(1+2|Z_i-z_i|)-1\bigr)$.
\end{lemma}

\begin{proof}
Both sides of $q(Z)=\sum_{b\in\{0,1\}^m}q(b)\prod_iu^{(i)}_{b_i}$, with $u^{(i)}=(1-Z_i,Z_i)$, are multilinear in $Z$
and agree on the cube, so the identity holds. Since $\|u^{(i)}-\mathbf 1_{z_i}\|_1=2|Z_i-z_i|$, the pointwise
computation of (S2b) shows that the tensor $(\prod_iu^{(i)}_{b_i})_b$ is within $\prod_i(1+2|Z_i-z_i|)-1$ of
$\mathbf 1_z$ in $\ell^1$; apply \cref{lem:sel-fibre}(b).
\end{proof}

\paragraph{Two approximants of NOR.} Let $\mathrm{NOR}_m(z)=\prod_{i\le m}(1-z_i)$. Only two facts about the
\fi
\ificml\else\fi
\ificml\else
\paragraph{The first-present selector.} This replaces maximum finding. Let $\mathcal S$ be a finite set of $m$
\emph{slots}, $\mathcal T_k$ a finite alphabet for each slot $k$, $\Pi=\{(k,\tau):k\in\mathcal S,\tau\in\mathcal T_k\}$
the set of \emph{pairs}, and $>$ a strict total order on $\Pi$. Let $c_k:\mathcal X\to\mathcal T_k$; the pair $(k,\tau)$
is \emph{present} at $x$ if $c_k(x)=\tau$, so exactly one pair per slot is present, and the \emph{winner} $F(x)$ is the
$>$-largest present pair. We are given, for every slot $k$, a selector $y_k=(y_{k,\tau})_{\tau}$ for $c_k$ with error
$\eta_k$ and degree $d_k$; a partition $\mathcal S=\mathcal S_1\cup\dots\cup\mathcal S_g$ into groups; and for every
group $a$ and slot $k$ a multilinear polynomial $N^{(k)}_a$ in the variables indexed by $\mathcal S_a\setminus\{k\}$ with
$|N^{(k)}_a-\mathrm{NOR}|\le\varepsilon_a$ and $|N^{(k)}_a|\le1$ on the cube. For a pair $p=(k,\tau)$ and a slot
$k'\ne k$ put $Z^p_{k'}=\sum_{\tau':(k',\tau')>p}y_{k',\tau'}$, and define
\[
w_p=y_{k,\tau}\prod_{a=1}^gN^{(k)}_a\bigl((Z^p_{k'})_{k'\in\mathcal S_a\setminus\{k\}}\bigr).
\]

\begin{lemma}[first-present selector]\label[lemma]{lem:sel-first}
$(w_p)_{p\in\Pi}$ is a selector for the winner $F$ with error at most $mE+(1+E)\sum_k\eta_k$, where
$E=\prod_a\bigl(\varepsilon_a+\prod_{k\in\mathcal S_a}(1+2\eta_k)\bigr)-1$, and degree at most
$\max_kd_k+\sum_a\max_k\deg N^{(k)}_a\cdot\max_{k\in\mathcal S_a}d_k$. In particular, with exact inner selectors of
degree at most $d$ and one group, the error is at most $m\varepsilon_1$ and the degree at most $d(1+\max_k\deg N^{(k)}_1)$.
\end{lemma}

\begin{proof}
Fix $x$ and a pair $p=(k,\tau)$. For $k'\ne k$ let $z^p_{k'}=[(k',c_{k'}(x))>p]\in\{0,1\}$. Then $|Z^p_{k'}-z^p_{k'}|=
|\sum_{\tau':(k',\tau')>p}(y_{k',\tau'}-[c_{k'}(x)=\tau'])|\le\eta_{k'}$. By \cref{lem:sel-offcube} with $B=1$ and the
error of $N^{(k)}_a$, $|N^{(k)}_a(Z^p)-\mathrm{NOR}(z^p)|\le E_a:=\varepsilon_a+\prod_{k'\in\mathcal S_a}(1+2\eta_{k'})-1$
on each group. The numbers $t_a=\mathrm{NOR}(z^p|_{\mathcal S_a\setminus\{k\}})$ lie in $\{0,1\}$, so telescoping gives
$|\prod_aN^{(k)}_a(Z^p)-\prod_at_a|\le\prod_a(1+E_a)-1=E$, and in particular $|\prod_aN^{(k)}_a(Z^p)|\le1+E$ for every
pair, present or not. If $p$ is present, $\prod_at_a=[\text{no other present pair exceeds }p]=[p=F(x)]$. Hence for a
slot $k$ with present value $\tau_0=c_k(x)$,
$|w_{(k,\tau_0)}-[(k,\tau_0)=F(x)]|\le|y_{k,\tau_0}-1|(1+E)+E$ and
$\sum_{\tau\ne\tau_0}|w_{(k,\tau)}|\le(1+E)\sum_{\tau\ne\tau_0}|y_{k,\tau}|$; the two add up to at most
$(1+E)\eta_k+E$. Sum over the $m$ slots. The degree is read off the definition.
\end{proof}

\noindent A slot with one value and $y_k\equiv1$ is a \emph{default}: a pair that is always present. Groups matter
when slots have very different degrees $d_k$. When $N^{(k)}_a=N_{|\mathcal S_a\setminus\{k\}|,\varepsilon_a}$, its degree
is at most $D(|\mathcal S_a|,\varepsilon_a)$, and we always make this choice.

\paragraph{From the record cube back to the true prefix.} All constructions below end the same way. The model's
first-layer states are encoded at each position $j$ by $\ell$ \emph{record bits} $e_j(\mathbf z)$, each of which, as a
sign, is a function of $\mathbf z$ of Walsh norm at most $A_0=2(2n)^{H_1}$ (\cref{app:twospectral}); the output
is $G(e(\mathbf z))$ for a function $G$ on all of $\{0,1\}^{n\ell}$; and we construct a polynomial $p$ on
$\{0,1\}^{n\ell}$ with $|p-G|\le2\vartheta$ on the whole cube. Reduce $p$ modulo $e^2=e$ (the degree does not grow).
Then $|p|\le1+2\vartheta\le3$ on the cube, so in $\pm1$ variables Parseval and Cauchy--Schwarz give
$A(p)\le3(n\ell+1)^{\deg p/2}$, and substituting the true encoded sign bits multiplies this by at most
$A_0^{\deg p}$. The function $P=p\circ e$ of $\mathbf z$ satisfies $|P-\hat O_T|\le2\vartheta$ on the whole formal
cube and $\log A(P)=O_H(\deg p\cdot\log n)$, and by~\eqref{eq:walsh}
\begin{equation}\label{eq:return}
|\E[\hat M_T\hat O_T]|\le2\vartheta+2^{-DT}\exp\bigl(O_H(\deg p\cdot\log n)\bigr).
\end{equation}
For the record bits we use \emph{exact record selectors}: for a position $j$ and a value $\tau$,
$y_{j,\tau}(e)=\sum_{c\in\{0,1\}^\ell:\,\mathrm{dec}_j(c)=\tau}\prod_ke_{j,k}^{c_k}(1-e_{j,k})^{1-c_k}$ has degree at
most $\ell$ in the bits of $e_j$ and equals $[\mathrm{dec}_j(e_j)=\tau]$ on Boolean $e$.

\subsection{Two layers with full attention}\label[appendix]{app:twospectral}
% FOCS-MOVED2 from main-body.tex (verbatim)
\begin{theorem}[two layers with the entire true prefix]\label{thm:full-two}
Fix $\alpha\in(0,1]$ and let $\alpha N\le T\le N$. Consider two layers with $H_1$ and $H_2$ heads and the model of
\cref{sec:model}, and let $\hat O_T$ be as in \cref{thm:onelayer}. For fixed $H_1,H_2$ and all sufficiently large $N$,
$|\E[\hat M_T\hat O_T]|\le\exp(-\Omega_{H_1,H_2,\alpha}(N/(\log N)^2))$. A fixed positive advantage $\eta$ at step $T$
requires $(H_1+1)^3(H_2+1)=\Omega_{\alpha,\eta}(N/(\log N)^3)$.
\end{theorem}

\paragraph{Proof idea.} In four steps (\ificml\cref{lem:position-union}\else\cref{fig:idea}(b)\fi):

\emph{(1) Layer-one winners are conjunctions.} A first-layer \emph{transcript} at a position records the query
token and, for each head, the index and token of its winner; there are at most $A_0=2(2n)^{H_1}$. A transcript occurs
exactly when the stated tokens occur and every other key loses to the stated winner, which are one-token conditions;
so every $\pm1$-valued predicate of one first-layer state has Walsh norm at most $A_0$.

\emph{(2) Layer two is a conjunction of $n$ checks.} A final-layer \emph{tuple} specifies the query state and, for each
head, the index and state of its winner; there are at most $S_2=A_0(nA_0)^{H_2}$. With the tuple fixed, ``the stated
states occur'' and ``no key outranks the stated winner'' are conditions on one first-layer state at a time; grouped by
position they give $n$ checks $v_j$ with $A(v_j)\le A_0$. Exactly one tuple passes, and its readout $o\in\{-1,0,1\}$
determines the output, so $\hat O_T=\sum_{\text{tuples}}o\cdot\mathrm{AND}(v_1,\dots,v_n)$.

\emph{(3) One explicit polynomial.} An explicit polynomial built from Chebyshev polynomials (Sherstov's
construction~\cite[Thm.~4.5]{Sherstov18}, with explicit constants in \cref{lem:nor-small}) approximates AND on $n$
signs with error at most $\varepsilon$ on \emph{all} of $\{\pm1\}^n$ and has degree $t\le3.34\sqrt{n\ln(2/\varepsilon)}+1$.
By Parseval and Cauchy--Schwarz its Walsh norm is at most $(n+1)^{t/2}$, and substituting $v_j$ for the $j$-th sign
multiplies this by at most $A_0^t$: the exact AND has exponential norm, but degree $t$ costs only $2^{O(t\log n)}$.
Summing over tuples with $\varepsilon=\vartheta/S_2$ gives $P$ with $|P-\hat O_T|\le\vartheta$ on the whole cube and
$\log A(P)=O_H(\log n+\sqrt{n\log(n/\vartheta)}\,\log n)$.

\emph{(4) Parameters.} With $\vartheta=\exp(-aN/(\log N)^2)$ and $n\le3N+O(1)$, $\log A(P)=O_H(\sqrt a\,N)+o(N)$, at most
$DT\ln2/2$ for large $N$ once $a=a(H_1,H_2,\alpha)>0$ is small; the head count follows with $\vartheta=\min(\eta,1/4)$.
After substitution $P$ contains characters with arbitrary sets of earlier output bits, so uniformity over masks is
essential.

\paragraph{Proof of \cref{thm:full-two}.}
Use the $m=2N+T-1$ formal raw bits $\mathbf z$ of the operands and the prefix, and $\hat O_T$ as in \cref{thm:onelayer}.
For a function on this cube write $A(f)=\sum_U|\widehat f(U)|$ for its Walsh coefficient norm. The product rule for
characters gives $A(fg)\le A(f)A(g)$ by convolution. Each conjunction of raw literals has norm at most one, by
multiplying factors $(1\pm\chi_i)/2$. For the actual context $Z=(X,Y,M_{<T})$, the discrepancy theorem gives
\[
 |\mathbb E[\hat M_T f(Z)]|\le 2^{-DT}A(f).
\]
Indeed every character factors as an operand character for $X$, an operand character for $Y$, and a parity of earlier
product bits.

At a fixed position a first-layer transcript records its raw query token and the index and raw token of the winner for
every head. There are at most $A_0=2(2n)^{H_1}$ transcripts. Each transcript event is a conjunction of literals: fix the
query and winner tokens, and at each visible key exclude every token that would outrank the specified winner under the
fixed priority. Conditions at a binary coordinate combine to a literal, a tautology or a contradiction. Separators are
constants. The events are disjoint and exhaust the formal cube, because the model can be run on every string. Hence any
$\{-1,+1\}$-valued predicate of the state at a single first-layer position has norm at most $A_0$: write it as the
signed sum of these transcript indicators. This bound allows arbitrary state values, including states that no
consistent input produces.

A final-layer tuple specifies its query state and each head's winning index and state. There are at most
$S_2=A_0(nA_0)^{H_2}$ tuples. A particular tuple occurs exactly when the $H_2+1$ stated states match and no competing
index outranks the purported winner of any head. Each check is a Boolean predicate of a single first-layer state because
the query state and winning scores in the tuple are fixed. Use signs for the checks, with $+1$ denoting success.
Combine all checks referring to the same position into a single signed predicate there, using the constant $+1$ when
there is no check. This gives $n$ checks, regardless of the number of heads, each defined and $\pm1$-valued on the whole
formal cube. Each check $v_j$ satisfies $A(v_j)\le A_0$. On every formal input exactly one tuple passes all checks; its
readout $o\in\{-1,0,1\}$ is determined by the tuple.

Put $\varepsilon=\vartheta/S_2$ with $0<\vartheta\le1$. Write a check as $v_j=1-2u_j$ with $u_j\in\{0,1\}$ the
indicator that it fails; the zero-one conjunction of the checks is $\mathrm{NOR}_n(u)$. Let $N=N_{n,\varepsilon}$ be the
polynomial of \cref{lem:nor-small}: $|N-\mathrm{NOR}_n|\le\varepsilon$ and $|N|\le1$ on $\{0,1\}^n$, and
$t=\deg N\le3.34\sqrt{n\ln(2/\varepsilon)}+1$, with an explicit constant and for every $\varepsilon\in(0,1]$. Let
$\widetilde{\mathrm{AND}}(w)=N((1-w)/2)$ for signs $w\in\{\pm1\}^n$: a multilinear polynomial of degree at most $t$ with
$|\widetilde{\mathrm{AND}}|\le1$ on $\{\pm1\}^n$, within $\varepsilon$ of the zero-one conjunction. In Walsh form, by
Cauchy--Schwarz and Parseval, $A(\widetilde{\mathrm{AND}})\le(\sum_{j\le t}\binom{n}{j})^{1/2}$. Substitute the $v_j$ for
the sign variables. Every Walsh monomial has at most $t$ factors, so
$A(\widetilde{\mathrm{AND}}(v_1,\ldots,v_n))\le A(\widetilde{\mathrm{AND}})A_0^t$. Summing over all tuples with
coefficient $o$ gives a function $P$ that uniformly approximates the signed two-layer output within
$S_2\varepsilon=\vartheta$. Thus for $n$ sufficiently large
\[
 |\mathbb E[\hat M_T\hat O_T]|\le
 \vartheta+2^{-DT}\exp\!\left(C_H\bigl(\log n+
 \sqrt{n(\log n+\log(1/\vartheta))}\,\log n\bigr)\right),
\]
with $C_H$ depending only on $H_1$, $H_2$. Here $S_2$ is polynomial in $n$ for fixed heads and
$\sum_{j\le t}\binom{n}{j}\le (n+1)^t$ for integer $t\ge1$, by encoding a subset in its padded, sorted list. Choose
$\vartheta=\exp(-aN/(\log N)^2)$. Since $n\le3N+O(1)$, $\log(S_2/\vartheta)=o(n)$ and the second exponent above is
$O_H(\sqrt a\,N)+o(N)$. Choose fixed $a>0$ small enough in terms of $H_1$, $H_2$, $D$ and $\alpha$ to make it at most
$DT\ln2/2$ for all large $N$. This proves the correlation bound. With a nonbinary readout,
$2\Pr[\mathrm{correct}]-1\le\mathbb E[\hat M_T\hat O_T]$, giving the claimed bound on success.

\paragraph{Head count.}
Fix a desired positive advantage $\eta$. In the same proof $A_0=2(2n)^{H_1}$ and $S_2=A_0(nA_0)^{H_2}$. Set
$\vartheta=\min(\eta,1/4)$. In the regime $(H_1+1)^3(H_2+1)\le c N/(\log N)^3$ one has
$\log(S_2/\vartheta)=O_\eta((H_1+1)(H_2+1)\log n)=o(n)$. Writing $t=O_\eta(\sqrt{n(H_1+1)(H_2+1)\log n})$ gives
\begin{align*}
 \log A(P)&\le \log S_2+O(t(\log n+\log A_0))\\
 &=O_\eta\!\left((H_1+1)(H_2+1)\log n+
 \sqrt{n(H_1+1)^3(H_2+1)(\log n)^3}\right).
\end{align*}
For $T\ge\alpha N$ and small enough $c=c(\alpha,\eta)>0$ this is at most $DT\ln2/2$. The resulting correlation is
strictly less than $2\eta$ for large $N$, while advantage $\eta$ needs correlation at least $2\eta$. Hence advantage
$\eta$ requires $(H_1+1)^3(H_2+1)=\Omega_{\alpha,\eta}(N/(\log N)^3)$.

\subsection{Three layers and shared middle-head orders}\label[appendix]{app:orders}

\begin{theorem}[shared middle-head orders]\label{thm:threeorders}
Take three full causal layers, fixed head counts and $\alpha N\le T\le N$ for fixed $\alpha>0$.
Suppose each middle head selects, from its query, one of at most $R$ shared orders on
(position, first-layer state) pairs, including ties. If $R=o(\sqrt n/(\log n)^{11/2})$,
then $|\E[\hat M_T\hat O_T]|=o(n^{-a})$ for every fixed $a>0$.
For fixed $R$ it is at most $\exp(-\Omega(n^{1/3}/(\log n)^{8/3}))$.
A rank-one bilinear middle score with a shared fixed tie priority has $R\le3$.
\end{theorem}

\noindent The proof constructs an explicit polynomial in the first-layer records that is uniformly close to the
output of the top two layers on \emph{all} record strings, with the tools of \cref{app:selectors}, and then returns
to the true prefix by~\eqref{eq:return}.

\begin{proof}[Proof of \cref{thm:threeorders}]
Set $n=2N+T-1+O(1)$, fix the head counts, let $\vartheta=e^{-\Lambda}$ with $\Lambda\ge\log n$, and let $\hat O_T$ be the
signed output of the three-layer model at step $T$, a function of the formal raw bits $\mathbf z$. If $\Lambda>n$ the
bounds below exceed $1$, so assume $\Lambda\le n$; assume also $R\le\sqrt n$.

\emph{Middle-head orders.} For a middle head $h$, an order compares pairs $(j,s_j)$, where $s_j$ is a possible
first-layer state at position $j$; refined by the head's fixed tie priority, it is a strict order on pairs with distinct
positions, and the head at query position $i$ attends to the maximal pair among $\{(j,s_j):j\le i\}$. The hypothesis is
that the order is one of at most $R$ fixed orders $\succ_{h,1},\dots,\succ_{h,R}$, selected by the first-layer state at
$i$: the head uses $\succ_{h,r_h(s_i)}$. Different heads may have different order families.

\emph{The record cube.} At each position $j$ encode the first-layer transcript (raw token, and the index and token of
each first-layer winner) by $\ell=O(\log n)$ bits $e_j$; all fields take polynomially many values. Each bit of $e_j$, as
a function of $\mathbf z$, is a function of the transcript at $j$, hence a signed sum of the disjoint transcript
indicators, so as a sign it has Walsh norm at most $A_0=2(2n)^{H_1}$ (\cref{app:twospectral}). Decode an invalid
string $e_j$ as a fixed admissible transcript. The first-layer states are functions of the transcripts, and the middle
and final layers depend on the input only through the first-layer states. So there is a function $G$ on all of
$\{0,1\}^{n\ell}$, the three-layer output computed from the decoded transcripts $s_1,\dots,s_n$, with
$G(e(\mathbf z))=\hat O_T(\mathbf z)$ for every formal $\mathbf z$. The exact record selectors $y_{j,\tau}$ of
\cref{app:selectors} have degree $\ell$.

\emph{The polynomial.} Let $1\le b\le n$, $n_e=H_2R\lceil n/b\rceil$, $\varepsilon_1=\vartheta/(2bn_e)$,
$\varepsilon_2=\vartheta/(120H_2H_3n^3)$ and $\varepsilon_3=\vartheta/(16H_3n)$. We construct $p$ with
$|p-G|\le2\vartheta$ on $\{0,1\}^{n\ell}$ and
\begin{equation}\label{eq:degorders}
\deg p\le n_e\ell(1+D(b,\varepsilon_1))+\ell\bigl(1+H_2(1+D(b,\varepsilon_2))\bigr)\bigl(1+H_3(1+D(n,\varepsilon_3))\bigr).
\end{equation}
Partition the positions into consecutive blocks of size $b$, the last possibly shorter.

\emph{Step 1: the table.} An \emph{entry} is $(h,r,B)$: a middle head, one of its orders and a block. Its true value
$\tau^*_{h,r,B}(e)$ is the $\succ_{h,r}$-largest pair $(j,s_j)$, $j\in B$. \Cref{lem:sel-first}, with slots $B$, the
order $\succ_{h,r}$ (refined arbitrarily between two pairs of one position), the exact record selectors and one group
with $N_{|B|-1,\varepsilon_1}$, gives a selector for $\tau^*_{h,r,B}$ with error at most $b\varepsilon_1$ and degree at
most $\ell(1+D(b,\varepsilon_1))$. By (S2b) the products over the $n_e$ entries form a selector $W$ for the whole table
$\tau^*(e)$ with error $\delta_1\le(1+b\varepsilon_1)^{n_e}-1\le e^{\vartheta/2}-1\le0.65\vartheta$ and degree at most
$n_e\ell(1+D(b,\varepsilon_1))$. Its outcomes $\tau$ are all assignments of a pair of the block to each entry.

\emph{Step 2: middle states from a table.} Fix an outcome $\tau$. For a position $i$ let $f_\tau(i)$ be the following
function of $e$: with $\sigma=s_i$, for each middle head $h$ and $r=r_h(\sigma)$ let $w_h$ be the $\succ_{h,r}$-largest
among the table pairs $\tau_{h,r,B'}$ of the blocks $B'$ before the block of $i$ and the present pairs $(j,s_j)$, $j\le
i$, of the block of $i$; and $f_\tau(i)=\phi_i(\sigma,(w_h)_h)$, where $\phi_i$ is the residual connection and the MLP.
It is defined for every $\tau$, and $f_{\tau^*(e)}(i)$ is the true middle state $m_i$. A selector for $f_\tau(i)$, by
(S2a): the exact selector $y_{i,\cdot}$ for $s_i$; then, given $\sigma$, for each middle head \cref{lem:sel-first} with
slots the positions $j\le i$ of the block of $i$ and, if there are earlier blocks, a default slot holding the best table
pair (at most $b+1$ slots, so every $N^{(k)}$ has at most $b$ inputs), exact record selectors and one group with
$\varepsilon_2$; then \cref{lem:sel-fibre}(a) for $\phi_i$. Its error is $\eta\le(1+(b+1)\varepsilon_2)^{H_2}-1$ and its
degree $d_{\rm mid}\le\ell(1+H_2(1+D(b,\varepsilon_2)))$.

\emph{Step 3: the final layer from a table.} Let $\Gamma_\tau(e)\in\{-1,0,1\}$ be the output of the final layer run on
the states $f_\tau(1),\dots,f_\tau(n)$; so $\Gamma_{\tau^*(e)}(e)=G(e)$. The output is $o(\mu,(i_h,f_\tau(i_h))_h)$, where
$\mu=f_\tau(n)$ is the query state and $(i_h,f_\tau(i_h))$ is the largest present pair among $(i,f_\tau(i))$, $i\le n$,
in the order given by the score of final head $h$ for the query $\mu$, then its priority (and arbitrarily between two
pairs of one position). A selector for the tuple $(\mu,\text{winners})$, by (S2a): the selector of Step~2 for
$f_\tau(n)$; then, given $\mu$, for each final head \cref{lem:sel-first} with slots $1,\dots,n$, as alphabets the finite
sets of values $\phi_i(\sigma,(w_h)_h)$ over all $\sigma$ and all $w_h$, the selectors of Step~2 as inner selectors and
one group with $\varepsilon_3$. Each head has error at most $\delta_F=nE+(1+E)n\eta$ with
$E=\varepsilon_3+(1+2\eta)^n-1$ and degree at most $d_{\rm mid}(1+D(n,\varepsilon_3))$. So the tuple has a selector with
error $\vartheta_2=(1+\eta)(1+\delta_F)^{H_3}-1$ and degree at most $d_{\rm mid}+H_3d_{\rm mid}(1+D(n,\varepsilon_3))$, and
by \cref{lem:sel-fibre}(b) the polynomial $p_\tau=\sum o(\mu,\text{winners})\cdot(\text{selector})$ satisfies
$|p_\tau-\Gamma_\tau|\le\vartheta_2$ on the cube, for every outcome $\tau$.

\emph{Step 4: assembly and error.} Let $p=\sum_\tau W_\tau p_\tau$. By (S2c), $|p-G|\le\vartheta_2+\delta_1(1+\vartheta_2)$,
and the degree is~\eqref{eq:degorders}. With $x=H_2(b+1)\varepsilon_2\le\vartheta/(60H_3n^2)$ we get
$\eta\le e^x-1\le\vartheta/(58H_3n^2)$; then $(1+2\eta)^n-1\le e^{2n\eta}-1\le\vartheta/(27.9H_3n)$, so
$E\le0.099\vartheta/(H_3n)$, $\delta_F\le0.11\vartheta/H_3$ (using $n\ge2$), $(1+\delta_F)^{H_3}\le1+0.13\vartheta$ and
$\vartheta_2\le0.14\vartheta$. Hence $|p-G|\le0.14\vartheta+0.65\cdot1.14\,\vartheta\le\vartheta$ on the whole cube.

\emph{Parameters.} Since $R\le\sqrt n$, $\ln(2/\varepsilon_i)=O_H(\Lambda)$ for $i=1,2,3$, so
$D(m,\varepsilon_i)=O_H(\sqrt{m\Lambda})+1$; with $\lceil n/b\rceil\le2n/b$,~\eqref{eq:degorders} is
$O_H(\ell[Rn\sqrt{\Lambda/b}+\Lambda\sqrt{nb}])$. Set $b=\lceil R\sqrt{n/\Lambda}\rceil$, which lies in $[1,n]$. Then
$\deg p=O_H(\ell\sqrt R\,n^{3/4}\Lambda^{3/4})$, and~\eqref{eq:return} gives
\[
|\E[\hat M_T\hat O_T]|\le2e^{-\Lambda}+2^{-DT}\exp\bigl(C_H\sqrt R\,n^{3/4}\Lambda^{3/4}(\log n)^2\bigr).
\]
Since $T\ge\alpha N$ and $n\le3N+O(1)$, $DT\ln2/2\ge c_0n$ with $c_0=c_0(\alpha,D)>0$. For $R=o(\sqrt n/(\log n)^{11/2})$
take $\Lambda=a\log n$ for any fixed $a\ge1$: the exponent is $o(n)$, at most $DT\ln2/2$ for large $N$, and the bound is
$2n^{-a}+2^{-DT/2}$. For fixed $R$ take $\Lambda=c\,n^{1/3}/(\log n)^{8/3}$ with $c$ small enough that the exponent is at
most $DT\ln2/2$; this gives the stated rate.

\paragraph{Score rank.}
For a rank-one middle score $u(q)v(j,s_j)$, the sign of $u(q)$ chooses among
increasing value, decreasing value, and the tie priority alone. With a common
fixed priority within each head, $R\le3$. Rank two alone cannot give a
comparable order bound. Take $m$ old keys at distinct unit vectors on a
circle, followed by $m$ positions whose keys are zero and whose queries are
the respective unit vectors. Each later query uniquely selects a different
old key. One order always ranks the same old key highest, so these queries
require at least $m$ orders. The example has fixed keys and illustrates a
limitation of the order hypothesis, not an average-case predictor.
\end{proof}

\subsection{Three layers with unrestricted middle and final heads}\label[appendix]{app:boundedstates}

In the proof of \cref{thm:threeorders} the hypothesis on the middle heads is used in one place only: Step~1, which
tabulates block maxima for each of the $R$ orders. With unrestricted middle heads the order used at position $i$
depends on $i$ and on the first-layer state there, so $R$ is at least the number of positions and the table is useless.
Everything else (the record cube, the selector calculus and the return~\eqref{eq:return} to the true prefix) does not
use the hypothesis. \Cref{thm:boundedstates} replaces the table by an adaptively built \emph{knowledge state}; its cost is
polynomial in the number of values of a first-layer state. For a position $j$ let $\Sigma_j$ be the set of values that
the first-layer state $s_j$ at $j$ takes over the formal cube.

\begin{theorem}[unrestricted middle and final heads]\label{thm:boundedstates}
Take three full causal layers with $H_1,H_2,H_3$ heads and $\alpha N\le T\le N$ for fixed $\alpha>0$. Suppose
$|\Sigma_j|\le A_1$ for every position $j$, where $2\le A_1\le n$. No hypothesis is made on the middle and final heads.
Then for all large $n$ and every $\Lambda\ge\log n$,
\[
|\E[\hat M_T\hat O_T]|\le2e^{-\Lambda}+2^{-DT}\exp\bigl(C_H(n\Lambda)^{13/14}A_1^{2/7}(\log n)^2\bigr),
\]
with $D$ the constant of \cref{thm:main} and $C_H$ depending only on the head counts and $\alpha$. Consequently
$|\E[\hat M_T\hat O_T]|=o(n^{-a})$ for every fixed $a>0$ if $A_1=o(n^{1/4}/(\log n)^{41/4})$, and
$|\E[\hat M_T\hat O_T]|\le\exp(-\Omega(n^{1/13}/(\log n)^{28/13}))$ for fixed $A_1$.
\end{theorem}

\noindent\emph{Example.} If the value matrices of the first layer vanish on the positional encodings (the heads may
attend by position and content in any way, but copy only tokens), then the first-layer state at $j$ is a function of
the token at $j$ and the $H_1$ attended tokens, since the positional encoding of $j$ itself is fixed; so
$A_1\le|\Xi|^{H_1+1}$ for the vocabulary $\Xi$ (two bits and $O(1)$ separators). Two layers are the case of a first
layer that copies nothing, so \cref{thm:boundedstates} contains a form of \cref{thm:full-two} with a weaker rate. What
it does not cover directly is a first layer that writes \emph{pointers} (the position of a content-dependent winner)
into its state: then $A_1$ is of order $n$ and the bound is vacuous. \Cref{app:three} covers it by truncating the
pointers first, using the form of the theorem in \cref{lem:bstar}.

\begin{proof}[Proof of \cref{thm:boundedstates}]
\emph{Records and pairs.} Encode $s_j$ by $\ell=\lceil\log_2A_1\rceil$ bits $e_j$, its index in a fixed enumeration of
$\Sigma_j$ (unused indices decode to the first element). Each bit of $e_j$, as a sign, is a function of the first-layer
state at $j$, hence of Walsh norm at most $A_0=2(2n)^{H_1}$ (\cref{app:twospectral}). As in the proof of
\cref{thm:threeorders} there is a function $G$ on all of $\{0,1\}^{n\ell}$, the three-layer output computed from the
decoded states, with $G(e(\mathbf z))=\hat O_T(\mathbf z)$ on the formal cube, and the exact record selectors $y_{j,\tau}$ of \cref{app:selectors} have degree $\ell$. From now
on $s=(s_1,\dots,s_n)$ is an arbitrary element of $\prod_j\Sigma_j$ (the decoded record string). A
\emph{pair} is $(j,\tau)$ with $\tau\in\Sigma_j$; it is \emph{present} if $s_j=\tau$, and exactly one pair is present at
each position.

\emph{Lists.} A \emph{list} is $\mathsf L=(i,\sigma,h)$: a position $i$, a state $\sigma\in\Sigma_i$ and a middle
head $h$. It carries the strict order $>_{\mathsf L}$ on $\Pi_{\mathsf L}=\{(j,\tau):j<i\}\cup\{(i,\sigma)\}$ given by
the score of head $h$ for the query state $\sigma$ at position $i$, ties broken by the head's priority on positions
(and arbitrarily between two pairs of one position, which are never both present). There are at most $nA_1H_2$ lists.
The middle state at $i$ is $m_i=\phi_i(s_i,(\mathrm{win}_h(i))_h)$ for a fixed map $\phi_i$ (residual connection and
MLP), where $\mathrm{win}_h(i)$ is the $>_{\mathsf L}$-largest present pair of $\Pi_{\mathsf L}$, $\mathsf L=(i,s_i,h)$.

\emph{Knowledge and potential.} A knowledge state $\mathcal K$ consists of a set $K$ of positions with their records
$(s_j)_{j\in K}$ and a set $\mathrm{Dead}$ of pairs at positions outside $K$. A pair $(j,\tau)$ is \emph{live} if
$j\notin K$ and $(j,\tau)\notin\mathrm{Dead}$; $\mathcal K$ is \emph{correct} if the stored records are the true ones
and no present pair is dead. For a list $\mathsf L=(i,\sigma,h)$ let $b_{\mathsf L}$ be the $>_{\mathsf L}$-largest
among $(i,\sigma)$ and the known pairs $(j,s_j)$, $j\in K$, $j<i$; let $C_{\mathsf L}$ be the set of live pairs
$(j,\tau)$ with $j<i$ and $(j,\tau)>_{\mathsf L}b_{\mathsf L}$; let $P_{\mathsf L}$ be the set of positions with a pair
in $C_{\mathsf L}$ and $r_{\mathsf L}=|P_{\mathsf L}|$; for $j\in P_{\mathsf L}$ let $\mathrm{top}_{\mathsf L}(j)$ be the
largest pair of $j$ in $C_{\mathsf L}$; and for $p=(j,\tau)\in C_{\mathsf L}$ let
$\mathrm{red}_{\mathsf L}(p)=1+\#\{j'\in P_{\mathsf L}\setminus\{j\}:\mathrm{top}_{\mathsf L}(j')<_{\mathsf L}p\}$. Put
$\mathrm{gain}(p)=\sum_{\mathsf L:\,p\in C_{\mathsf L}}\mathrm{red}_{\mathsf L}(p)$ and
$\mathrm{pot}(\mathcal K)=\sum_{\mathsf L}r_{\mathsf L}\le n^2A_1H_2$. Fix a threshold $g_0\ge1$; a live pair is
\emph{heavy} if its gain is at least $g_0$.

\emph{(1) Reading a middle state.} If $\mathcal K$ is correct, then for every list $\mathsf L=(i,s_i,h)$ the pair
$\mathrm{win}_h(i)$ is the $>_{\mathsf L}$-largest of $b_{\mathsf L}$ and the present pairs of positions in
$P_{\mathsf L}$ that lie in $C_{\mathsf L}$. Indeed the pair of $i$ and the known pairs are at most $b_{\mathsf L}$, and
for an unknown $j<i$ outside $P_{\mathsf L}$ no live pair of $j$ exceeds $b_{\mathsf L}$, while the present pair of $j$
is live because $\mathcal K$ is correct.

\emph{(2) Potential drop.} Let $p=(j,\tau)$ be live, and let $\mathcal K'$ be $\mathcal K$ with $j$ added to $K$ with
record $\tau$ (the pairs of $j$ leave $\mathrm{Dead}$). Then $\mathrm{pot}(\mathcal K')\le\mathrm{pot}(\mathcal K)-
\mathrm{gain}(p)$, and declaring pairs dead never increases $\mathrm{pot}$. Indeed live pairs only disappear and the
thresholds $b_{\mathsf L}$ only rise, so $P'_{\mathsf L}\subseteq P_{\mathsf L}$ for every list. If $p\in C_{\mathsf L}$
then $b'_{\mathsf L}=p$; so $j$ leaves $P_{\mathsf L}$, and so does every $j'$ with
$\mathrm{top}_{\mathsf L}(j')<_{\mathsf L}p$. Thus $r_{\mathsf L}$ drops by at least $\mathrm{red}_{\mathsf L}(p)$;
sum over the lists.

\emph{(3) No heavy pair means short lists.} If no live pair is heavy, then
$\sum_{\mathsf L}r_{\mathsf L}(r_{\mathsf L}+1)/2<nA_1g_0$, so for every $r_{\max}\ge1$ fewer than
$2nA_1g_0/r_{\max}^2$ lists have $r_{\mathsf L}>r_{\max}$. Indeed, order $P_{\mathsf L}=\{j_1,\dots,j_r\}$ with
$\mathrm{top}_{\mathsf L}(j_1)>_{\mathsf L}\dots>_{\mathsf L}\mathrm{top}_{\mathsf L}(j_r)$; then
$\mathrm{red}_{\mathsf L}(\mathrm{top}_{\mathsf L}(j_k))=r-k+1$, so
$\sum_p\mathrm{gain}(p)\ge\sum_{\mathsf L}r_{\mathsf L}(r_{\mathsf L}+1)/2$; and there are at most $nA_1$ live pairs,
each of gain below $g_0$.

\emph{Parameters.} Thresholds $g_0\ge1$ and $r_{\max}\ge1$ are fixed at the end. Let $\vartheta=e^{-\Lambda}$,
$R_{\max}=\lfloor n^2A_1H_2/g_0\rfloor+1$, $\varepsilon_1=\vartheta/(2nR_{\max})$,
$\varepsilon_2=\vartheta/(120H_2H_3n^3)$ and $\varepsilon_3=\vartheta/(16H_3n)$.

\emph{Phase 1: the knowledge state, as a selector.} A knowledge state now also carries a flag \emph{stopped}. For a
state $\mathcal K$ that is not stopped define \emph{levels}: $\mathcal K_0=\mathcal K$; while the set $\mathcal D_k$ of
heavy pairs of $\mathcal K_k$ is nonempty, $\mathcal K_{k+1}$ is $\mathcal K_k$ with $\mathcal D_k$ declared dead. The
level of a pair is the $k$ with $p\in\mathcal D_k$, and $\infty$ if there is none (in particular for dead pairs and for
pairs at known positions). Let $>_{\mathcal K}$ be the strict order on all pairs in which a pair is larger if its level is
smaller, then if its position is smaller, then arbitrarily. The \emph{update} $U(\mathcal K,p)$ for an arbitrary pair
$p=(j,\tau)$ is: $\mathcal K$ itself if $\mathcal K$ is stopped; the last $\mathcal K_k$, stopped, if $p$ has level
$\infty$; and otherwise, with $k$ the level of $p$, $\mathcal K_k$ with $j$ added to $K$ with record $\tau$. The
\emph{exact process} on a record string $s$ is $\mathcal K^0=\emptyset$ and $\mathcal K^{t+1}=U(\mathcal K^t,
F_{\mathcal K^t}(s))$ for $t<R_{\max}$, where $F_{\mathcal K}(s)$ is the $>_{\mathcal K}$-largest present pair; let
$\mathcal K^*(s)=\mathcal K^{R_{\max}}$. Let $\mathcal C$ be the finite set of all states obtained from the empty state by
$R_{\max}$ updates with \emph{arbitrary} pairs.

\emph{(P1) Every $\mathcal K\in\mathcal C$ is stopped and has no heavy pair.} An update with a pair of finite level $k$
uses a pair that is heavy, hence live, in $\mathcal K_k$, so by (2) the potential drops by at least $g_0$ relative to
$\mathrm{pot}(\mathcal K_k)\le\mathrm{pot}(\mathcal K)$. The potential starts at most $n^2A_1H_2$ and is never negative,
so there are at most $R_{\max}-1$ such updates. If the first $R_{\max}-1$ updates are all of this kind, then
$\mathrm{pot}<g_0$, no pair is heavy (as $\mathrm{gain}(p)\le\mathrm{pot}$, because
$\mathrm{red}_{\mathsf L}(p)\le r_{\mathsf L}$), every level is $\infty$ and the last update stops. An update with a pair
of level $\infty$ stops in the last $\mathcal K_k$, which has no heavy pair by construction.

\emph{(P2) $\mathcal K^*(s)$ is correct for $s$.} By induction over the rounds: if the minimal level of a present pair
is $k<\infty$, no present pair lies in $\mathcal D_0,\dots,\mathcal D_{k-1}$, and the added record is the true one; if
it is $\infty$, no present pair lies in any $\mathcal D_k$.

For a state $\mathcal K$ that is not stopped, \cref{lem:sel-first} with slots $1,\dots,n$, alphabets $\Sigma_j$, the
order $>_{\mathcal K}$, the exact record selectors and one group with $\varepsilon_1$ is a selector for $F_{\mathcal K}$
with error at most $n\varepsilon_1$ and degree at most $\ell(1+D(n,\varepsilon_1))$; by \cref{lem:sel-fibre}(a) it gives
a selector for $s\mapsto U(\mathcal K,F_{\mathcal K}(s))$. For a stopped $\mathcal K$ use the exact constant selector. By
(S2a) over the $R_{\max}$ rounds there is a selector $W=(W_{\mathcal K})_{\mathcal K\in\mathcal C}$ for $\mathcal K^*$
with error $\delta_1\le(1+n\varepsilon_1)^{R_{\max}}-1\le e^{\vartheta/2}-1\le0.65\vartheta$ and degree at most
$R_{\max}\ell(1+D(n,\varepsilon_1))$.

\emph{Phase 2: the final layer, for every outcome.} Fix $\mathcal K\in\mathcal C$, and compute $b_{\mathsf L}$,
$C_{\mathsf L}$, $P_{\mathsf L}$, $r_{\mathsf L}$ from it. Let $I_{\rm long}$ be the set of positions $i$ with
$r_{\mathsf L}>r_{\max}$ for some list $\mathsf L=(i,\sigma,h)$; by (P1) and (3), $|I_{\rm long}|<2nA_1g_0/r_{\max}^2$,
for \emph{every} $\mathcal K\in\mathcal C$. For every $i$ let $f_{\mathcal K}(i)=\phi_i(s_i,(w_h)_h)$, where $w_h$ is the
$>_{\mathsf L}$-largest of $b_{\mathsf L}$ and the present pairs of positions in $P_{\mathsf L}$ that lie in
$C_{\mathsf L}$, $\mathsf L=(i,s_i,h)$. This is a function of the record string for every $\mathcal K\in\mathcal C$;
let $\Gamma_{\mathcal K}\in\{-1,0,1\}$ be the output of the final layer run on the states
$f_{\mathcal K}(1),\dots,f_{\mathcal K}(n)$. By (1) and (P2), $\Gamma_{\mathcal K^*(s)}(s)=G(s)$.

\emph{Middle selectors.} By (S2a): the exact selector for $s_i$; then, given $\sigma$, for each middle head
\cref{lem:sel-first} for the list $\mathsf L=(i,\sigma,h)$ with slots the positions of $P_{\mathsf L}$ and a default
slot holding $b_{\mathsf L}$, the order in which the pairs of $C_{\mathsf L}$ are ordered by $>_{\mathsf L}$ above the
default and all other pairs lie below it, exact record selectors and one group with $\varepsilon_2$ (each $N^{(k)}$ has
at most $r_{\mathsf L}$ inputs); then \cref{lem:sel-fibre}(a) twice (winners outside $C_{\mathsf L}$ are mapped to
$b_{\mathsf L}$; then $\phi_i$). With $r_i=\max_{\sigma,h}r_{(i,\sigma,h)}\le n-1$ this is a selector for
$f_{\mathcal K}(i)$ with error $\eta\le(1+n\varepsilon_2)^{H_2}-1$ and degree $d_i\le\ell(1+H_2(1+D(r_i,\varepsilon_2)))$.
Put $d_s=\ell(1+H_2(1+D(r_{\max},\varepsilon_2)))$ and $d_l=\ell(1+H_2(1+D(n,\varepsilon_2)))$; then $d_i\le d_s$ for
$i\notin I_{\rm long}$ and $d_i\le d_l$ for all $i$.

\emph{Final selector.} As in Step~3 of the proof of \cref{thm:threeorders}, with \cref{lem:sel-first} applied to the
slots $1,\dots,n$ in two groups, $\mathcal S_1=[n]\setminus I_{\rm long}$ and $\mathcal S_2=I_{\rm long}$, each with
$\varepsilon_3$: each final head has error at most $\delta_F=nE+(1+E)n\eta$ with
$E\le(\varepsilon_3+(1+2\eta)^n)^2-1$, and degree at most $d_l+D(n,\varepsilon_3)d_s+D(|I_{\rm long}|,\varepsilon_3)d_l$.
So there is a polynomial $p_{\mathcal K}$ with $|p_{\mathcal K}-\Gamma_{\mathcal K}|\le\vartheta_2:=(1+\eta)(1+\delta_F)^{H_3}-1$
on the cube, for every $\mathcal K\in\mathcal C$, and
$\deg p_{\mathcal K}\le d_l+H_3(d_l+D(n,\varepsilon_3)d_s+D(|I_{\rm long}|,\varepsilon_3)d_l)$.

\emph{Total.} Let $p=\sum_{\mathcal K\in\mathcal C}W_{\mathcal K}p_{\mathcal K}$. By (S2c),
$|p-G|\le\vartheta_2+\delta_1(1+\vartheta_2)$. As in the proof of \cref{thm:threeorders}, now with the square in $E$,
$\eta\le\vartheta/(58H_3n^2)$, $\delta_F\le0.213\vartheta/H_3$, $\vartheta_2\le0.27\vartheta$, so
$|p-G|\le0.27\vartheta+0.65\cdot1.27\,\vartheta\le1.1\vartheta\le2\vartheta$ on the whole cube, and
\begin{equation}\label{eq:degbounded}
\deg p\le R_{\max}\ell(1+D(n,\varepsilon_1))+(1+H_3)d_l+H_3D(n,\varepsilon_3)d_s+H_3D(|I_{\rm long}|,\varepsilon_3)d_l.
\end{equation}

\emph{Choice of thresholds.} If $\Lambda>n$ the claimed bound exceeds $1$ for large $n$, so let $\Lambda\le n$. Since
$A_1\le n$ and $\Lambda\ge\log n$, $\ln(2/\varepsilon_i)=O_H(\Lambda)$, so $D(m,\varepsilon_i)=O_H(\sqrt{m\Lambda})+1$,
$d_l=O_H(\ell\sqrt{n\Lambda})$ and $d_s=O_H(\ell\sqrt{r_{\max}\Lambda})$. Take
$g_0=\lceil n^{11/7}A_1^{5/7}\Lambda^{-3/7}\rceil$ and $r_{\max}=\lceil n^{6/7}A_1^{4/7}\Lambda^{-1/7}\rceil$ (both
unrounded quantities are at least $1$). Then the three main terms of~\eqref{eq:degbounded} are
$O_H(\ell(n^2A_1/g_0+1)\sqrt{n\Lambda})$, $O_H(\ell\Lambda\sqrt{nr_{\max}})$ and
$O_H(\ell\Lambda n\sqrt{A_1g_0}/r_{\max}+\ell\sqrt{n\Lambda})$, each $O_H(\ell(n\Lambda)^{13/14}A_1^{2/7})$. So
$\deg p=O_H((n\Lambda)^{13/14}A_1^{2/7}\log n)$, and~\eqref{eq:return} gives the displayed bound.

\emph{Consequences.} $T\ge\alpha N$ and $n\le3N+O(1)$ give $DT\ln2/2\ge c_0n$ with $c_0=c_0(\alpha,D)>0$. If
$A_1=o(n^{1/4}/(\log n)^{41/4})$, then with $\Lambda=a\log n$ the exponent is $o(n)$, and the bound is $2n^{-a}+2^{-DT/2}$
for large $n$. For fixed $A_1$ take $\Lambda=c\,n^{1/13}/(\log n)^{28/13}$ with $c$ small.
\end{proof}

\noindent A classical simulation of the combinatorial core, with exact searches, arbitrary orders for every list (so
unrestricted middle heads) and arbitrary record strings, checks (1)--(3), the round bound and the correctness of the
final knowledge state on $108$ random and adversarial instances (\texttt{checks/phase1\_check.py}); three
deliberately broken variants are each caught. The polynomial construction itself is analytic.

\paragraph{The form of \cref{thm:boundedstates} that is used.} \Cref{thm:boundedstates} is stated for the true
first-layer states $s_j$ and their exact ranges $\Sigma_j$. Its proof uses less, and \cref{thm:three} needs the
weaker hypotheses, because there the states are coarsened.

\begin{lemma}[bounded states, abstract form]\label[lemma]{lem:bstar}
Let $H_1,H_2,H_3$, $\alpha$, $T$ be as in \cref{thm:boundedstates} and $2\le A_1\le n$. For every position $j$ let
$\Sigma'_j$ be a finite set with $|\Sigma'_j|\le A_1$ and $s'_j:\{0,1\}^m\to\Sigma'_j$ a map such that:
\begin{enumerate}
\item[(i)] $s'_j(\mathbf z)$ is a function of the first-layer transcript at $j$ (\cref{app:twospectral}) of a UHAT first
layer with $H_1$ heads;
\item[(ii)] $F(\mathbf z)=G(s'_1(\mathbf z),\dots,s'_n(\mathbf z))$, where $G:\prod_j\Sigma'_j\to\{-1,0,1\}$ has the
following form. For each position $i$, each $\sigma\in\Sigma'_i$ and each of $H_2$ middle heads $h$ there is a strict
total order $>_{(i,\sigma,h)}$ on $\{(j,\tau):j<i,\ \tau\in\Sigma'_j\}\cup\{(i,\sigma)\}$; $\mathrm{win}_h(i)$ is the
largest pair $(j,s_j)$, $j\le i$, under $>_{(i,s_i,h)}$; the middle state is $m_i=\phi_i(s_i,(\mathrm{win}_h(i))_h)$ for
arbitrary maps $\phi_i$ into a vector space; and $G(s)$ is the output at position $n$ of a UHAT layer with $H_3$ heads
applied to the states $(m_i)_i$.
\end{enumerate}
Then for all $n\ge n_0$ and every $\Lambda\ge\log n$,
$|\E[\hat M_TF(Z)]|\le2e^{-\Lambda}+2^{-DT}\exp(C_H(n\Lambda)^{13/14}A_1^{2/7}(\log n)^2)$, where $n_0$ and $C_H$ depend
only on $H_1,H_2,H_3$ and $\alpha$, and not on $A_1$, the sets $\Sigma'_j$ or the maps $s'_j,\phi_i$.
\end{lemma}

\begin{proof}
Read the proof of \cref{thm:boundedstates} with $s'_j$, $\Sigma'_j$ and $F$ in place of $s_j$, $\Sigma_j$ and $\hat
O_T$. The first layer enters that proof only in the paragraph \emph{Records and pairs}, in two ways. (1) Each sign bit
of the encoding $e_j$ of the state at $j$ has Walsh norm at most $A_0$, because it is a function of the first-layer
transcript at $j$; by (i) this holds for $s'_j$. (2) There is a function $G$ on all strings of records with
$G(e(\mathbf z))=\hat O_T(\mathbf z)$ on the formal cube; by (ii) this holds for $F$. From the sentence ``From now on
$s=(s_1,\dots,s_n)$ is an arbitrary element of $\prod_j\Sigma_j$'' on, the proof uses about the middle and final layers
only what (ii) states: a strict order for every list $(i,\sigma,h)$ (its use of scores and priorities is only to define
such an order), the maps $\phi_i$, and the final layer as a UHAT layer on the middle states. It never uses that
$\Sigma_j$ is the exact range of a first-layer state. Every constant in it depends only on the head counts, and the
estimates use $A_1$ only through $2\le A_1\le n$. The construction gives $p$ with $|p-G|\le2\vartheta$ on the whole
record cube, hence $P=p\circ e$ with $|P-F|\le2\vartheta$ on the whole formal cube and $\log A(P)=O_H(\deg p\cdot\log n)$,
and~\eqref{eq:walsh} applied to $P$ gives the bound. \Cref{thm:boundedstates} is the case $s'_j=s_j$,
$\Sigma'_j=\Sigma_j$, $F=\hat O_T$.
\end{proof}

\paragraph{A limit of exact simulation by query algorithms.} The constructions in the proofs of
\cref{thm:threeorders,thm:boundedstates} approximate $G$ on \emph{all} strings of first-layer records, and the argument
bounds $\log A(P)$ by $O(\deg p\cdot\log n)$, which must stay below $DT\ln2/2$; so it needs approximating polynomials of
degree $o(n/\log n)$ on the record cube. Simulating $G$ instead by a query algorithm that is correct on all record
strings would likewise need $o(n/\log n)$ record lookups. \Cref{prop:surj} shows that for three
unrestricted layers no such query algorithm exists on the true first-layer records. It is a statement about query
algorithms only, not about polynomials: the approximate degree of surjectivity is $O(n^{3/4})$~\cite{Sherstov18} in the
one-hot encoding of the records, and $O(n^{3/4}\log n)$ on binary records, so \cref{prop:surj} does not exclude an
approximating polynomial of degree $o(n/\log n)$, and the backward argument of \cref{app:topkdepth} provides one for
every fixed number of top-$k$ layers on all strings of its Cartesian alphabets: at inverse-polynomial error, the last
two of three layers get record degree $\widetilde O(n^{3/4})$, and the last $L-1$ of $L$ layers record degree
$\widetilde O(n^{1-2^{-(L-1)}})$ (\cref{lem:onelayerback}, \cref{rem:topk-escape}).
\Cref{thm:three} does not need one: there the construction runs on coarse records with
$2d^{H_1}+1\le n^{1/4}$ values, and the true and the coarse model differ only on rare true contexts.

\begin{proposition}[a limit of exact simulation by query algorithms]\label[proposition]{prop:surj}
For every $R\ge2$ there is a three-layer UHAT with one head per layer and context length $n=4R+1$ whose record function
$G$ (the output as a function of arbitrary strings of first-layer records, as in the proof of
\cref{thm:boundedstates}) computes, on suitable record strings, whether a function $f:[2R]\to[R]$ is onto. So every
bounded-error quantum query algorithm that computes $G$ on all record strings makes $\Omega(n)$ record lookups.
\end{proposition}

\begin{proof}[Proof of \cref{prop:surj}]
Positions $1,\dots,R$ are \emph{sources}, $R+1,\dots,3R$ \emph{data}, $3R+1,\dots,4R$ \emph{checkers}, and $4R+1$ is
the output position. First layer: a data position attends to the left-most source holding the token $1$ (a fixed
source if there is none), and its state records the index $v\in[R]$ of the attended source; so each data position
has $R$ possible states, each realised by some input. The other positions keep their own token and position. Middle
layer: checker $3R+c$ gives score $1$ to data positions whose state is $v=c$ and score $0$ to all other positions; its
state becomes \emph{found} if its winner is a data position with state $c$, and \emph{missing} otherwise. Final
layer: the output position gives score $1$ to checkers in state \emph{missing} and $0$ to all other positions, and
outputs $1$ iff its winner is not a checker in state \emph{missing}. Scores of this form are bilinear in one-hot
encodings of (position, state), which unrestricted width allows. Fix the records of all non-data positions. On a string
of records in which data position $R+x$ has state $f(x)$, the output is $1$ iff every $c\in[R]$ is found, that is iff
$f:[2R]\to[R]$ is onto. So a query algorithm that computes $G$ on all record strings, with each record lookup counted as
one query, decides surjectivity of $f$ with the same number of queries to $f$. Beame and
Machmouchi~\cite{BeameMachmouchi} prove that deciding whether a function $f:[2m-2]\to[m]$ is surjective needs
$\Theta(m)$ quantum queries. Surjectivity for $[2R-2]\to[R]$ reduces to $[2R]\to[R]$ by setting $f(2R-1)=f(2R)=f(1)$, at
the cost of one query each, so the lower bound $\Omega(R)=\Omega(n)$ holds for record lookups, and for bit queries,
each of which is answered by one record lookup.
\end{proof}

\noindent Nor does \cref{prop:surj} limit an argument that uses only the record strings that are typical on true
contexts; \cref{app:three} takes that route.

\subsection{Three unrestricted layers}\label[appendix]{app:three}

Throughout this subsection $2\le T\le N$, the context $xM_1\cdots M_{T-1}$ has length $n$, and the formal cube
$\{0,1\}^m$, $m=2N+T-1$, has one coordinate for each bit position of the context (separators are constants). The true
context is $Z=(X,Y,M_{<T}(X,Y))$ with $X,Y$ uniform, and a \emph{literal} fixes one coordinate. We first show that
conjunctions of many literals are rare on true contexts, then that a deep first-layer pointer forces such a
conjunction, and then combine this with \cref{lem:bstar}.

\begin{theorem}[three UHAT layers, by explicit polynomials]\label{thm:three}
Fix $\alpha\in(0,1]$ and let $\alpha N\le T\le N$. Consider three full causal layers with $H_1,H_2,H_3$ heads and the
model of \cref{sec:model}, with no other restriction, and let $\hat O_T$ be as in \cref{thm:onelayer}. For fixed
$H_1,H_2,H_3$ and every fixed $\gamma<1/(13+4H_1)$, $|\E[\hat M_T\hat O_T]|\le\exp(-n^\gamma)$ for all sufficiently
large $N$, where $n\ge2N$ is the context length; in particular $|\E[\hat M_T\hat O_T]|=o(N^{-a})$ for every $a>0$.
\end{theorem}

\noindent The proof below uses only the explicit polynomials of \cref{app:selectors}, and no query algorithm.
\Cref{thm:topk-depth} gives the stronger bound $\exp(-\Omega(n/(\ln n)^6))$ for the same model (the case $k=1$), also by
explicit polynomials.

\fi
\newcommand{\icmlraredirect}{%
\ificml
\section{Rare deep ranks on true prefixes}
The statements below use $2\le T\le N$, the true context $Z=(X,Y,M_{<T})$ for uniform independent operands,
and $n$ context positions including any fixed separators.
\fi
\begin{lemma}[literals on prefix bits]\label[lemma]{lem:prefixlit}
Let $2\le T\le N$, $1\le k_1<\dots<k_a\le T-1$ and $b_1,\dots,b_a\in\{0,1\}$. Then
$\Pr[M_{k_1}=b_1,\dots,M_{k_a}=b_a]\le(2+N\ln2)2^{-a}\le(N+2)2^{-a}$.
\end{lemma}

\begin{proof}
Put $L'=2^{2N-T+1}$. The integer with binary digits $M_1\cdots M_{T-1}$ is $w=\lfloor XY/L'\rfloor\in\{0,\dots,2^{T-1}-1\}$.
The event fixes $a$ digits of $w$, so it is $\{w\in W\}$ for a set $W$ of $2^{T-1-a}$ values. Fix $w_0$. The pairs with
$X=0$ number $2^N$ and all have $w=0$. For $1\le X<2^N$ the $Y$ with $w_0L'\le XY<(w_0+1)L'$ lie in a half-open interval of
length $L'/X$, so there are at most $L'/X+1$ of them. Since $\sum_{X=1}^{2^N-1}1/X\le1+N\ln2$,
$\#\{(X,Y):w=w_0\}\le L'(1+N\ln2)+2^N+[w_0=0]2^N$. Summing over $w_0\in W$ and using $|W|L'=2^{2N-a}$, the number of pairs in
the event is at most $2^{2N-a}(1+N\ln2)+2^{T-1-a}2^N+2^N$. Divide by $4^N$: the second term becomes
$2^{-a}2^{T-1-N}\le2^{-a}/2$ as $T\le N$, and the third $2^{-N}\le2^{-a}/2$ as $a\le T-1\le N-1$.
\end{proof}

\begin{lemma}[conjunctions on true contexts]\label[lemma]{lem:conj}
Let $2\le T\le N$ and let $C$ be a conjunction of $d$ literals on $d$ distinct coordinates of the formal cube. Then
$\Pr[C(Z)=1]\le(N+2)2^{-\lceil d/3\rceil}$.
\end{lemma}

\begin{proof}
The coordinates are bits of $X$, bits of $Y$ and prefix bits. One of the three kinds carries $a\ge\lceil d/3\rceil$ of the
literals, and $C$ implies their conjunction. For bits of $X$ (or of $Y$) its probability is $2^{-a}$; for prefix bits it
is at most $(N+2)2^{-a}$ by \cref{lem:prefixlit}.
\end{proof}

\emph{Lists and ranks.} Fix a position $j$, a first-layer head $h$ and a token $b$ that position $j$ can hold. The
\emph{candidates} are the pairs $(k,c)$ with $k<j$ and $c$ a token that position $k$ can hold (two for a bit position,
one for a separator), and the pair $(j,b)$. The score of head $h$ for the query $(j,b)$ and the key $(k,c)$ is
$(\mathrm{emb}(c)+p_k)^\top\mathbf K_h^\top\mathbf Q_h(\mathrm{emb}(b)+p_j)$, a function of $(j,b,h)$ and $(k,c)$ only.
Order the candidates by decreasing score, then by the head's priority on positions, and arbitrarily between the two
pairs of one position; this is the \emph{list} $\mathcal P^{(j,b,h)}_1,\mathcal P^{(j,b,h)}_2,\dots$. A pair $(k,c)$ is
\emph{present} in $\mathbf z$ if $z_k=c$ (the pair of a separator is always present), so exactly one pair per position
is present. On an input with $z_j=b$ the head attends to the position of the first present entry of the list, whose
index is the \emph{rank} $\mathrm{rk}_{j,h}(\mathbf z)\ge1$: among present pairs the list order is the attention rule of
\cref{sec:model}. The first-layer state at $j$ is a function $\Psi_j(z_j,\mathrm{rk}_{j,1},\dots,\mathrm{rk}_{j,H_1})$, since the
token and the ranks determine the winners. Fix an integer $d\ge1$; head $h$ at $j$ is \emph{deep} on $\mathbf z$ if
$\mathrm{rk}_{j,h}(\mathbf z)>d$.

\begin{lemma}[truncating the first layer]\label[lemma]{lem:truncate}
Consider a UHAT with any number of layers and $H_1$ first-layer heads.
\textup{(a)} If head $h$ at $j$ is deep on $\mathbf z$ and $z_j=b$, then the first $d$ entries $(k_1,c_1),\dots,(k_d,c_d)$
of $\mathcal P^{(j,b,h)}$ lie at $d$ distinct bit positions and $z_{k_i}=1-c_i$ for all $i$.
\textup{(b)} $\Pr[\text{some first-layer head at some position is deep on }Z]\le2nH_1(N+2)2^{-\lceil d/3\rceil}$.
\textup{(c)} Let the \emph{coarse state} be $s'_j(\mathbf z)=(z_j,\mathrm{rk}_{j,1},\dots,\mathrm{rk}_{j,H_1})$ if no head at $j$ is
deep and $s'_j(\mathbf z)=\bot$ otherwise. It takes at most $2d^{H_1}+1$ values and is a function of the first-layer
transcript at $j$. Let $\mathrm{dec}_j(s')=\Psi_j(s')$ for $s'\ne\bot$ and let $\mathrm{dec}_j(\bot)$ be a fixed
first-layer state of position $j$, and let $\hat O'_T(\mathbf z)$ be the output of the layers above the first, run on
the states $\mathrm{dec}_j(s'_j(\mathbf z))$. Then $\hat O'_T=\hat O_T$ wherever no head is deep, and
$\E|\hat O_T(Z)-\hat O'_T(Z)|\le4nH_1(N+2)2^{-\lceil d/3\rceil}$.
\end{lemma}

\begin{proof}
(a) The first $d$ entries are absent. So none is $(j,b)$ or the pair of a separator, which are present, and no two are
the two pairs of one bit position, one of which is present. Hence they lie at $d$ distinct bit positions, and absent
means $z_{k_i}\ne c_i$. (b) Fix $(j,h)$. For each of the at most two tokens $b$ of $j$, by (a) the event ``deep and
$z_j=b$'' lies in a conjunction of $d$ literals on distinct coordinates (it is empty if the list is too short), of
probability at most $(N+2)2^{-\lceil d/3\rceil}$ by \cref{lem:conj}. Sum over $b$, the $n$ positions and the $H_1$ heads.
(c) There are at most two tokens and $d^{H_1}$ rank vectors, plus $\bot$. Given the token, a head's rank and its winner's
pair determine each other, so $s'_j$ is a function of the transcript. If no head is deep anywhere, every
$\mathrm{dec}_j(s'_j(\mathbf z))$ is the true first-layer state, so $\hat O'_T(\mathbf z)=\hat O_T(\mathbf z)$. Both
take values in $\{-1,0,1\}$, so $\E|\hat O_T(Z)-\hat O'_T(Z)|\le2\Pr[\text{some head deep on }Z]$; apply (b).
\end{proof}

}
\ificml\ificmlextended\icmlraredirect\fi\else\icmlraredirect\fi
\ificml\else
\noindent No property of the scores is used: the order of each list is arbitrary. \Cref{lem:truncate} is a statement
about true contexts; it fails for arbitrary record strings and for worst-case inputs. In the instance of
\cref{prop:surj} a data position whose pointer is source $v$ has rank $v$ in its list, so the hard record strings use
pointers of rank above $d$, which are deep, and \cref{lem:truncate} says that these are rare on true contexts.

\begin{proposition}[explicit form of \cref{thm:three}]\label[proposition]{prop:threeexplicit}
In the setting of \cref{thm:three}, for all $n\ge n_0(H_1,H_2,H_3,\alpha)$, every integer $d\ge1$ with $2d^{H_1}+1\le n$
and every $\Lambda\ge\log n$,
\[
|\E[\hat M_T\hat O_T]|\le2e^{-\Lambda}+2^{-DT}\exp\bigl(C_H(n\Lambda)^{13/14}(2d^{H_1}+1)^{2/7}(\log n)^2\bigr)
+4nH_1(N+2)2^{-\lceil d/3\rceil},
\]
with $D$ the constant of \cref{thm:main} and $C_H$ depending only on the head counts and $\alpha$.
\end{proposition}

\begin{proof}
Take $s'_j$, $\mathrm{dec}_j$ and $\hat O'_T$ from \cref{lem:truncate}, $\Sigma'_j$ the set of coarse states at $j$ and
$A_1=2d^{H_1}+1$. Then $F=\hat O'_T$ satisfies the hypotheses of \cref{lem:bstar}: (i) is \cref{lem:truncate}(c). For
(ii), for a middle head $h$, a position $i$ and $\sigma\in\Sigma'_i$, order the pairs $(j,\tau)$ by the score of $h$ for
the query state $\mathrm{dec}_i(\sigma)$ and the key state $\mathrm{dec}_j(\tau)$, then by the head's priority, then
arbitrarily between pairs of one position. The middle layer run on the states $\mathrm{dec}_j(s'_j)$ attends to the
largest present pair in this order, its output state is a function $\phi_i$ of $s'_i$ and the winning pairs, and the
final layer is a UHAT layer on the middle states. By \cref{lem:bstar} and $|\hat M_T|=1$,
$|\E\hat M_T\hat O_T|\le|\E\hat M_T\hat O'_T|+\E|\hat O_T-\hat O'_T|$, and \cref{lem:truncate}(c) bounds the last term
($T\ge\alpha N\ge2$ for large $N$).
\end{proof}

\begin{proof}[Proof of \cref{thm:three}]
Note $2N\le n\le3N+O(1)$ and $DT\ln2/2\ge c_0n$ for some $c_0=c_0(\alpha,D)>0$. \emph{Polynomial rate.} Fix $a>0$, and
take $d=\lceil3(a+3)\log_2n\rceil$ and $\Lambda=(a+1)\ln n$ in \cref{prop:threeexplicit}. Then
$2d^{H_1}+1=O((\log n)^{H_1})$, so the exponent in the middle term is $o(n)$, at most $DT\ln2/2$ for large $n$, and the
middle term is at most $2^{-DT/2}$. The first term is $2n^{-(a+1)}$, and the last is at most
$4nH_1(N+2)n^{-(a+3)}\le4H_1n^{-(a+1)}$. \emph{Stretched-exponential rate.} Fix $\gamma<\gamma'<1/(13+4H_1)$ and take
$d=\lceil n^{\gamma'}\rceil$ and $\Lambda=n^{\gamma'}$. Then $2d^{H_1}+1\le3\cdot2^{H_1}n^{\gamma'H_1}\le n$ for large $n$,
since $\gamma'H_1<\frac14$, and the exponent in the middle term is
$O(n^{13/14+\gamma'(13+4H_1)/14}(\log n)^2)$, because $\frac{13}{14}(1+\gamma')+\frac27\gamma'H_1=\frac{13}{14}+
\frac{\gamma'(13+4H_1)}{14}$, which is less than $1$. So the middle term is again at most $2^{-DT/2}$, the first is
$2e^{-n^{\gamma'}}$ and the last at most $4n^2H_12^{-n^{\gamma'}/3}$; their sum is at most $\exp(-n^\gamma)$ for large
$n$, and $\exp(-n^\gamma)\le\exp(-N^\gamma)$.
\end{proof}

\noindent\emph{Checks.} \Cref{lem:prefixlit} was verified by exact counting for all $2\le T\le N\le12$ and all
patterns (largest ratio of probability to bound below $0.889$), and \cref{lem:conj} for all conjunctions with
$m\le15$, $N\le7$ (largest ratio $0.469$); variants with the factor $2+N\ln2$ removed, or with the exponent $d$ in place
of $\lceil d/3\rceil$, are violated, as they must be (\texttt{checks/prefix\_conj\_check.py}). A second script simulates
$14$ random three-layer UHATs built from the definition in \cref{app:model} (small integer matrices, so ties are
frequent; a separator; random priorities), exhaustively over the formal cube, and checks the list description of the
first layer, \cref{lem:truncate}(a) and (c), the probability bound behind (b), the Walsh norm of the encoded coarse
states, and the form (ii) of \cref{lem:bstar} on random record strings; three broken variants (lists that ignore the
query token, a median tie rule, no truncation by depth) are each caught (\texttt{checks/uhat\_sim\_check.py}). A
third, separately written script checks \cref{lem:prefixlit,lem:conj} for smaller sizes and \cref{lem:truncate}(a) on
random list models (\texttt{checks/conj\_check.py}). These are consistency checks, not part of the proofs.

\begin{remark}[why the truncation does not reach four layers]\label[remark]{rem:four}
For four layers the natural plan is to truncate the second layer in the same way and apply \cref{lem:bstar} to the top
two layers. This fails: second-layer pointers are spread out on typical inputs. Let $N\ge64$, let the bits
$X_1,\dots,X_N$ of $X$ occupy positions $1,\dots,N$, and let $r=\lfloor\log_2(N/(4\log_2N))\rfloor\ge1$. In the first
layer every position $j$ attends to the nearest position $k\le j$ holding the token $1$ (score $1$ for keys with token
$1$, $0$ otherwise, ties to the right) and records $j$, its token and the winner's position and token. For $r\le j\le N$
the flag $f_j=[X_{j-r+1}=\dots=X_j=0]$ is a function of this state. In the second layer the last position gives score
$1$ to the positions $j\in[r,N]$ with $f_j=1$ and $0$ to all others, ties to the left, and copies the position it
attends to, $J$. For every set $S$ of positions, $\{J\in S\}$ lies in the union of the events $f_j=1$, $j\in S\cap[r,N]$,
and the event that no $f_j$ equals $1$. The first have probability $2^{-r}$ each. The flags $f_r,f_{2r},\dots$ depend on
disjoint blocks of bits, so the last has probability at most $\exp(-\lfloor N/r\rfloor2^{-r})\le e^{-3}$, because
$2^r\le N/(4\log_2N)$ and $r\le\log_2N$. Hence $\Pr[J\in S]\le|S|2^{-r}+e^{-3}$. So if the second-layer state at the
last position is mapped to $A$ values from which $J$ can be recovered with probability $1-\theta$, then
$A\ge(1-\theta-e^{-3})2^r>(1-\theta-e^{-3})N/(8\log_2N)$, far more than the $o(n^{1/4})$ values that \cref{lem:bstar}
can afford. This is an obstacle to one method, not a lower bound: later layers need not use $J$, and an approximation
need only be right on most true contexts. A second obstacle is that second-layer states are not functions of a first-layer
transcript, so hypothesis (i) of \cref{lem:bstar} fails for them as well. So no input-independent coarsening of the
second-layer states reaches four layers. For top-$k$ (in particular unique) hard attention, the argument of
\cref{app:topkdepth} does reach every fixed depth (\cref{thm:topk-depth}): it never coarsens, but keeps the full
Cartesian alphabets and pays for their size only logarithmically, through Parseval (\cref{rem:topk-escape}).
\end{remark}

\subsection{Why the exact Fourier norm can grow exponentially}\label[appendix]{app:onelayer-rem}

\begin{remark}[exact Fourier norm at two layers]\label[remark]{rem:twolayer}
Two layers can compute, at an output position, the AND of two earlier output bits, and then an OR of $(T-1)/2$ such
ANDs over disjoint pairs. The function $1-ab$ of two bits has Fourier $\ell^1$ norm $3/2$, so the product over the
pairs has $\ell^1$ norm $(3/2)^{(T-1)/2}\ge2^{\gadgetExp(T-1)}$, which outgrows the gain $2^{-DT}$ of
\cref{thm:main} for every $D<\gadgetExp$, in particular for every rate proved here at $T=N$. The uniform approximation in \cref{thm:full-two} bypasses this exact-norm obstruction. A more basic obstruction is that the first layer can
already combine bits of $X$ and $Y$ (\cref{prop:disj}).
\end{remark}

\begin{remark}[one softmax layer with a chain of thought]\label[remark]{rem:pnp}
Peng, Narayanan and Papadimitriou~\cite{PNP24} simulate one softmax attention layer with $H$ heads, width
$d_{\rm model}$ and $p$-bit arithmetic by a protocol in which one party sends $H(d_{\rm model}+1)p$ bits per query. Let the input end with a separator `$=$', give Alice the positions of $X$ and
Bob all other positions, and let Bob also send each generated token ($\lceil\log_2|\Xi|\rceil$ bits, for a vocabulary
$\Xi$). A one-layer model that writes an arbitrary chain of thought $g$ of $|g|$ tokens and then outputs $M_T$,
correctly with probability at least $\frac12+\eta$, becomes a deterministic protocol of
$c=|g|(H(d_{\rm model}+1)p+\lceil\log_2|\Xi|\rceil)+H(d_{\rm model}+1)p$ bits. Such a protocol splits the inputs into
at most $2^c$ rectangles $\mathcal A\times\mathcal B$ with constant output, and $|\E\hat M_T\mathbf 1_{\mathcal A}(X)\mathbf 1_{\mathcal B}(Y)|\le2^{-DT}$, so
$2\eta\le2^{c-DT}$, that is, $c\ge DT+1-\log_2(1/\eta)$. This uses only the case $C=\emptyset$ of \cref{thm:main}, which is classical
(\cref{sec:intro}), and the simulation is that of~\cite{PNP24}; we record it only for comparison.
\end{remark}

\subsection{A single middle bit, and windowed models}\label[appendix]{app:window-stmt}

The simplest consequence needs no parities. A subcube that fixes $k$ bits of $X$ and $Y$ is a product $\mathcal A\times\mathcal B$,
where $\mathcal A$ is a set of values of $X$ and $\mathcal B$ a set of values of $Y$, and it has probability $2^{-k}$. If $M_T$ is
constant on it, then $f=\mathbf 1_{\mathcal A}$ and $g=\mathbf 1_{\mathcal B}$ give $2^{-k}=|\E f(X)g(Y)(-1)^{M_T}|\le2^{-DT}$, so $k\ge DT$.
This is the property that the restriction method of Hahn~\cite{Hahn20} and~\cite[Thm.~3.3]{AHRH25} needs. The proof
of~\cite[Thm.~4.5]{AHRH25} does not spell out this step (``Now if one fixes $CN$ input bits, for $C$ small,
the result cannot be fixed.''); the bound $k\ge DT$, which needs only the classical case $C=\emptyset$, supplies it.

\begin{proposition}[a single middle bit]\label[proposition]{prop:singlebit}
Fix $\alpha\in(0,1]$ and $T=T(N)$ with $\alpha N\le T\le N$. A UHAT chain of thought $g$ in the sense
of~\cite[Def.~2.2]{AHRH25} (any vocabulary), with fixed numbers of layers and heads, that ends with $M_T$ has length
$|g|=\Omega(N)$, the constant depending on the numbers of layers and heads.
\end{proposition}

To go beyond one bit we restrict how far back a step can look.

\begin{definition}[windowed attention]\label[definition]{def:window}
Fix $w\ge0$. A \emph{$w$-windowed} model is a UHAT as in \cref{sec:model} with one extra mask: a head at the position
of $M_q$ ($q\ge0$; $M_0$ means the last input position) ranges only over the input positions and the positions of
$M_{q-w},\dots,M_q$. Heads at input positions attend causally as before.
\end{definition}

\noindent So each step sees the whole input, its own token and the last $w$ output bits; $w=0$ is a Markovian chain of
thought. Information cannot travel further back than $w$ positions per layer, because input positions never see output
positions.

\begin{lemma}[dependency cone]\label[lemma]{lem:cone}
In a $w$-windowed model with $L$ layers, the output at the position of $M_{T-1}$ is a function
$F(X,Y,M_{\mathcal W})$ of the input and of the bits $M_k$, $k\in\mathcal W$, where
$\mathcal W=[T-1-Lw,T-1]\cap[1,T-1]$; so $|\mathcal W|\le Lw+1$.
\end{lemma}

\begin{theorem}[windowed models, worst case]\label{thm:window}
Fix $w$, $L$, $H$ and $\alpha\in(0,1]$. For all large $N$ and every $T$ with $\alpha N\le T\le N$, no $w$-windowed
model with $L$ layers and $H$ heads outputs $M_T$ correctly for all $X,Y$.
\end{theorem}

\noindent The proof (\cref{app:window}) applies Hahn's depth reduction once for each of the $2^{|\mathcal W|}$ values
of the window, so that after fixing few input bits the output depends on $O(1)$ input bits and on the window; it then
expands in characters and applies \cref{thm:main} with $S=U\cup\{T\}$ for every $U\subseteq\mathcal W$. Here the parities of earlier
bits are needed. A windowed model is a special case of a full-attention model, so the conclusion of
\cref{thm:window} also follows from \cref{thm:worstcase} below, which does not use \cref{conj:C5}; the point of
\cref{thm:window} is that the restriction method of~\cite{AHRH25}, combined with \cref{thm:main}, does reach every step
$T\in[\alpha N,N]$ once the window is bounded. Its average-case companion below has no such classical counterpart.

\begin{theorem}[windowed predictors, average case]\label{thm:avgwindow}
Let $N\ge2$, $1\le T\le N$, $\mathcal W\subseteq[T-1]$ and $t\ge0$. Let
$F:\{0,1\}^{2N}\times\{0,1\}^{\mathcal W}\to\{0,1\}$ be any function, write $F_\sigma(x,y)=F(x,y,\sigma)$, and let
$\delta_t\ge\max_\sigma\mathrm{tail}_{>t}[F_\sigma]$, where $\mathrm{tail}_{>t}[F_\sigma]$ is the Fourier weight of
$2F_\sigma-1$ above degree $t$ in the $2N$ input bits. Then
\[
\Bigl|\Pr_{X,Y}\bigl[F(X,Y,M_{\mathcal W}(X,Y))=M_T(X,Y)\bigr]-\tfrac12\Bigr|
\le2^{|\mathcal W|-1}\bigl((2N+1)^{t/2}2^{-DT}+\delta_t^{1/2}\bigr).
\]
\end{theorem}

\begin{corollary}\label[corollary]{cor:avgwindow}
\textup{(a)} If every $F_\sigma$ is computed by a circuit of depth $d$ and size $s\ge2$ (unbounded fan-in AND and OR gates over the literals
$x_i,\neg x_i$), then the advantage in \cref{thm:avgwindow} is at most
\[2^{\,|\mathcal W|+1-DT/(2c_d(\log_2s)^{d-1}\log_2(2N+1))},\]
where $c_d=60d\cdot192^{d-1}\le2^{16d}$ is the depth-dependent
constant of Tal's Fourier tail bound~\cite[Thm.~1, explicit as Thm.~18]{Tal17} (circuits over literals; a circuit that also has NOT gates can be converted at no cost in depth and at most doubling its size, so then $s$ is replaced by $2s$).
\textup{(b)} A $w$-windowed UHAT with $L$ layers and $H$ heads is correct at step $T\le N$ on at most a
$\frac12+2^{Lw+2-\Omega(T/(\log N)^{d})}$ fraction of inputs, where $d$ depends only on $L$ and the implied constant
only on $L$ and $H$. For $T\ge\alpha N$ the advantage is $o(1)$ as long as $w\le\gamma N/(\log N)^d$ for a small enough
$\gamma>0$ depending on $L$, $H$ and $\alpha$.
\end{corollary}

\noindent For each value $\sigma$ of the window, the part of $F_\sigma$ of degree at most $t$ is a combination of few
input parities, and \cref{thm:main} makes each of them nearly useless; $\delta_t$ is the part of higher degree, which for
small circuits is tiny. For (b), each $F_\sigma$ is a fixed-length hard-attention computation, which the construction of
Hao, Angluin and Frank~\cite[Thm.~1]{HAF22} turns into a circuit of depth $O(L)$ and size $n^{O((H+1)^L)}$.

\subsection{Full attention}\label[appendix]{app:full-stmt}

For unrestricted models the worst case is classical. Buss~\cite{Buss92}
chose $X$ and $Y$ so that one block of the product holds the count $\sum_iz_i$ of some input bits, while every other
product bit is $0$ or a single $z_i$. Combined with the $\mathrm{AC}^0$ bound for UHATs~\cite[Thm.~1]{HAF22}, this shows
that no UHAT outputs every bit of the product correctly. We state the per-step form explicitly and prove it with a
denser variant of the embedding built on a Sidon set (\cref{lem:sidon}), with about $\sqrt{N/\log N}$ counted variables
instead of about $N^{1/3}$ in~\cite{Buss92}.

\begin{theorem}[full attention, worst case]\label{thm:worstcase}
Fix $L$, $H$ and $\varepsilon\in(0,1]$. For all large $N$ and every $T$ with $\min(T,2N-T)\ge N^\varepsilon$, in
particular $T=N$, no UHAT with at most $L$ layers and $H$ heads outputs $M_T$ correctly on the true prefix for all $X,Y$.
The same holds if the chain of thought contains $O(1)$ extra tokens from a fixed finite vocabulary at fixed positions
before $M_T$.
\end{theorem}

\noindent \Cref{thm:worstcase} does not use \cref{thm:main} or \cref{conj:C5}, and it says nothing about the average
case: the embedded inputs have probability $2^{r-2N}$ for $r$ counted variables. It holds for direct output and $O(1)$ extra tokens at fixed positions from a finite
vocabulary, but not for extra tokens at input-dependent positions, a vocabulary that grows with $N$, or $\omega(1)$
extra tokens.

\noindent The proof (\cref{app:worstcase}) substitutes the embedding into the $\mathrm{AC}^0$ circuit that computes
$M_T$ from $(X,Y,M_{<T})$ and obtains a small constant-depth circuit for MAJORITY.

\subsection{Deeper models and three barriers}\label[appendix]{app:barriers-disc}

\emph{The question, answered.} Is a transformer of fixed depth that writes $M_1M_2\cdots$ directly correct at step
$N$ on at most a $\frac12+o(1)$ fraction of uniform inputs? For unique and top-$k$ hard attention at every fixed depth,
and for softmax with margins, the answer is yes (\cref{thm:topk-depth,cor:softmax}); for averaging hard attention and
softmax without margins it is no (\cref{rem:averaging}). The truncation behind \cref{thm:three}
does not extend to the second layer (\cref{rem:four}), and three further natural methods fail, each for a precise
reason.

\emph{(i) Communication.} Communication arguments do reach several layers: Chen, Peng and Wu~\cite{CPW24} prove
unconditional lower bounds for multi-layer decoder-only transformers, for sequential function composition, through a
multi-party autoregressive communication model. We do not know how to apply such a model to multiplication. For the
natural two-party split $X\,|\,Y$, two layers cannot be simulated cheaply, and given the true prefix a single step is
easy to communicate.

\begin{proposition}[two layers compute set intersection]\label[proposition]{prop:disj}
For every $N$ there is a two-layer, one-head UHAT with width $O(1)$ and integer scores of $O(\log N)$ bits that outputs
$\bigvee_i(x_i\wedge y_i)$ at the last position of $x_1\cdots x_Ny_1\cdots y_N{=}$. Hence a deterministic two-party
protocol that reproduces such models across $X\,|\,Y$ needs at least $N$ bits for some of them~\cite[Fact~5]{SHT23}
(see also~\cite{KN97}).
\end{proposition}

\begin{proposition}[one step given the prefix]\label[proposition]{prop:gt}
Let $1\le T\le 2N$, $Y\ge1$ and $a=\lfloor XY/2^{2N-T+1}\rfloor$, the integer formed by $M_1\cdots M_{T-1}$. Then
$M_T=\mathbf 1\bigl[X\ge\lceil(2a+1)2^{2N-T}/Y\rceil\bigr]$.
\end{proposition}

\noindent So a party who holds $Y$ and the prefix computes a threshold, and step $T$ becomes the GREATER-THAN function
of $X$ and that threshold, whose randomised communication complexity is $O(\log N)$~\cite{KN97}. So no lower bound of
the form ``bits per step times number of steps'' can be proved this way: the difficulty of one step given its prefix is
computational, not informational.

\emph{(ii) Restriction.} The method of~\cite[Thm.~3.3]{AHRH25} fixes input bits to make chain-of-thought tokens constant
and charges a constant number of fixed bits per token. For multiplication the next bit is cheap to fix.

\begin{proposition}[two fixed bits decide the next bit]\label[proposition]{prop:restriction}
Let $\mathcal Q$ be a subcube of inputs (some bits of $X$ and $Y$ fixed), $1\le T\le2N$, and let $V\subseteq\mathcal Q$ be any set on
which $M_1,\dots,M_{T-1}$ are constant. Then fixing the most significant free bit of $X$ and that of $Y$, both to a
suitable common value $v$, makes $M_T$ constant on what remains of $V$.
\end{proposition}

\noindent Making $M_1,\dots,M_T$ all constant costs at most $T+1$ fixed bits (set the top $\lceil T/2\rceil$ bits of
$X$ and of $Y$ to $0$), so the per-token accounting never produces a contradiction. \Cref{prop:restriction} also shows
that no subcube restriction that keeps the earlier bits constant can embed PARITY or MAJORITY into $M_T$; an embedding
must let the earlier bits vary, as Buss's does (there they are single variables). Projections that identify input
variables, or substitute negations, are not covered by this remark.

\emph{(iii) Fourier analysis and density.} A step with full attention is a function $F$ of all $2N+T-1$ bits, but only
its values on consistent inputs $(X,Y,M_{<T}(X,Y))$ matter. The uniform distribution on consistent inputs has density
$2^{T-1}$ with respect to the uniform measure on $\{0,1\}^{2N+T-1}$. So a Fourier tail bound for $\mathrm{AC}^0$ under
the uniform measure controls the high-degree part of $F$ on consistent inputs only up to a factor $2^{(T-1)/2}$. This
loss exceeds the gain $2^{-DT}$ of \cref{thm:main}, because $D\le\log_2(8/7)<\frac12$ (\cref{ex:small}); and no gain of
the form $2^{-DT}$ can have $D>1+1/T$, because setting the top $\lceil T/2\rceil$ bits of $X$ and of $Y$ to $0$ forces $M_T=0$, so
$\disc_N(\{T\})\ge2^{-T-1}$. Making the uniform tail small enough to absorb the factor needs degree
$t=\Omega(T(\log s)^{d-1})$, and then the $(3N)^{t/2}$ low-degree terms swamp $2^{-DT}$. \Cref{thm:avgwindow} avoids this by conditioning on the window, at cost $2^{|\mathcal W|}$ with $|\mathcal W|\le Lw+1$, which is affordable only for
$w=o(T/\mathrm{polylog}\,N)$.

A solution would need a new ingredient: a restriction lemma that never fixes output positions, Fourier concentration
of $\mathrm{AC}^0$ under the consistent distribution, or a structural argument like Buss's that also works on average.

\subsection{Lemma C.6 of~\cite{AHRH25} and average sensitivity}\label[appendix]{app:c6}

In arXiv v3, \citet[App.~C.2.2]{AHRH25} give the sensitivity claim
in Lemma~C.6 as follows.
\begin{quote}
Assume Conjecture C.5. Then any Boolean function implementing $F_N$
needs to satisfy $\operatorname{as}(F_N)=\Omega(N)$.
\end{quote}
Their Eq.~(25) defines implementation by equality on genuine
operand and prefix triples. The sensitivity calculation in the
proof uses the uniform measure on the full Boolean cube.

Let \(G=\{(x,y,M_{<N}(x,y)):x,y\in\{0,1\}^N\}\) inside the
formal cube \(\{0,1\}^{3N-1}\). Eq.~(29) computes the moments
\(\lambda_{A,B,C}\) under the uniform law on \(G\), which has
density \(2^{-(N-1)}\) in that cube.
For fixed \(C\), they are the operand Fourier coefficients of the
Boolean function \(\hat M_N(x,y)\chi_C(M_{<N}(x,y))\), so Parseval
gives \(\sum_{A,B}|\lambda_{A,B,C}|^2=1\). Summing over \(C\) gives
\(\sum_{A,B,C}|\lambda_{A,B,C}|^2=2^{N-1}\). Multiplying each moment
by \(2^{-(N-1)}\) would give the uniform-cube coefficients of the
real-valued zero extension.

To compute the effect of extending the predictor outside its support,
let \(F(z)=\hat M_N(x,y)\) when \(z=(x,y,M_{<N}(x,y))\in G\), and set
\(F(z)=+1\) elsewhere. This is a Boolean function implementing the
midpoint predictor on every genuine context, as required by Eq.~(25).
Every edge of the cube across which \(F\) changes touches \(G\).
There are at most \((3N-1)|G|\) such undirected edges, each counted
twice in the sum of the pointwise sensitivities. Thus
\[
  \operatorname{as}(F)\le 2(3N-1)2^{-(N-1)}.
\]
The universal sensitivity claim in Eq.~(27) therefore fails even
when the genuine-context correlations obey Conjecture~C.5. For
the Fourier argument to give linear sensitivity for a chosen
Boolean extension, it would suffice to have constants
$\theta,\delta>0$ such that
$\sum_{|U|\le\theta N}|\widehat F(U)|^2\le1-\delta$
under the uniform cube measure. Parseval and the Fourier formula for
average sensitivity would then give $\operatorname{as}(F)\ge
\theta\delta N$. The genuine-context moments concern the
restriction to $G$. The asserted UHAT consequence was deduced
from Eq.~(27), so that deduction loses its premise. The full-product
lower bound here follows from a separate argument on genuine
contexts (\cref{thm:cot-avg}). The proof of the single-bit worst-case
lower bound of \citet[Thm.~4.5]{AHRH25} is in their Appendix~C.2.1
and is independent of Lemma~C.6.

\section{Proofs for Section~\ref{sec:transformers}}\label[appendix]{app:transformers}

Throughout, $\hat M_k=2M_k-1=-(-1)^{M_k}$, and \cref{thm:main} is used in the form
$|\E f(X)g(Y)\prod_{k\in S}(-1)^{M_k}|\le2^{-DT}$ for $|f|,|g|\le1$, $\max S=T\le N$, $N\ge2$.

\subsection{One attention layer}\label[appendix]{app:onelayer}

\begin{theorem}[one layer]\label{thm:onelayer}
Let $N\ge2$ and $1\le T\le N$. Consider a one-layer UHAT with $H$ heads as in \cref{sec:model}, run on
$xM_1\cdots M_{T-1}$, a context of length $n=|x|+T-1$. Let $O_T$ be its output token and $\hat O_T=1$ if $O_T=1$,
$\hat O_T=-1$ if $O_T=0$, and $\hat O_T=0$ otherwise. Then
\[
\bigl|\E[\hat M_T\hat O_T]\bigr|\le2(2n)^H\,2^{-DT},
\]
with $D$ the constant of \cref{thm:main}.
\end{theorem}

\begin{corollary}\label[corollary]{cor:onelayer}
If such a model is correct at step $T$ with probability at least $\frac12+\eta$ over uniform $X,Y$, then
$H\ge(DT-\log_2(1/\eta))/\log_2(2n)$. At $T=N$ the context has length $n\le3N+O(1)$, so predicting $M_N$ with
advantage $\eta\ge2^{-DN/2}$ needs $H\ge DN/(2\log_2(2n))=\Omega(N/\log N)$ heads.
\end{corollary}

\begin{proof}[Proof of \cref{cor:onelayer}]
$\Pr[O_T=M_T]-\Pr[O_T=1-M_T]=\E[\hat M_T\hat O_T]$, and the left side is at least $2\Pr[O_T=M_T]-1\ge2\eta$. So
$\eta\le(2n)^H2^{-DT}$ by \cref{thm:onelayer}; take logarithms.
\end{proof}

\begin{proof}[Proof of \cref{thm:onelayer}]
The proof uses only that every token of $x$ is a fixed symbol or a symbol determined by a single bit of $X$ or $Y$, so
it also allows repeated bits and any order. Let the query position be $n$ (the last one) and let $\Sigma_j$ be the set
of values that the token $\mathrm{tok}_j$ at position $j\le n$ can take; $|\Sigma_j|\le2$, and the token at a
chain-of-thought position $|x|+k$ is the bit $M_k$. The query token $q=\mathrm{tok}_n$ is $M_{T-1}$ or, for $T=1$, the
last input token. Fix $q$. With one layer, $y_j^{(0)}=\mathrm{emb}(\mathrm{tok}_j)+p_j$, so the score of head $h$ at key $j$ is a
number $\mathrm{sc}_h(j,\mathrm{tok}_j)$ that depends on position $j$ and its token only. Let $\pi$ be the priority order of the
tie rule, and write $(j,\sigma)\succ_h(j',\sigma')$ if $\mathrm{sc}_h(j,\sigma)>\mathrm{sc}_h(j',\sigma')$, or the scores are equal and $j$
precedes $j'$ in $\pi$. Causal masking means that only positions $j\le n$ compete, and these are exactly the positions
of the prefix. Head $h$ attends to $j^*$ if and only if $(j^*,\mathrm{tok}_{j^*})\succ_h(j,\mathrm{tok}_j)$ for every
$j\ne j^*$. Hence, for $\sigma^*\in\Sigma_{j^*}$,
\[
I_h(j^*,\sigma^*):=\mathbf 1[\text{$h$ attends to $j^*$ and }\mathrm{tok}_{j^*}=\sigma^*]
=\mathbf 1[\mathrm{tok}_{j^*}=\sigma^*]\prod_{j\ne j^*}\mathbf 1[\mathrm{tok}_j\in \mathcal L_{h,j}],
\]
where $\mathcal L_{h,j}=\{\sigma\in\Sigma_j:(j^*,\sigma^*)\succ_h(j,\sigma)\}$.
Each factor depends on one token, so it is the constant $0$, the constant $1$, or a literal in one bit of $X$, of $Y$
or of $M_{<T}$. The output vector is $f^{\rm MLP}(\mathrm{emb}(q)+p_n+\sum_hV_h(\mathrm{emb}(\sigma_h)+p_{j_h}))$ when head $h$ attends to
$j_h$ holding $\sigma_h$, so $O_T=\mathrm{Out}_q((j_h,\sigma_h)_{h\le H})$ for some function $\mathrm{Out}_q$. Therefore
\[
\hat O_T=\sum_{q}\ \sum_{(j_h,\sigma_h)_h}\mathbf 1[\mathrm{tok}_n=q]\prod_{h=1}^HI_h(j_h,\sigma_h)\,
\varepsilon_q\bigl((j_h,\sigma_h)_h\bigr),
\]
with $\varepsilon_q=1$ where $\mathrm{Out}_q=1$, $-1$ where $\mathrm{Out}_q=0$, and $0$ otherwise (the $I_h$ inside the sum are those for
the query value $q$). For each input exactly one term is nonzero. There are at most $2$ values of $q$ and at most $2n$
pairs $(j,\sigma)$ per head, so at most $2(2n)^H$ terms. Each term is $\pm1$ or $0$ times a product of literals, which
we group as $f(X)\,g(Y)\,\mathrm{Lit}(M_{<T})$ with $f,g\in\{0,1\}$ and $\mathrm{Lit}$ a product of literals in the bits $M_k$, $k\in R\subseteq
[T-1]$. Two literals in the same bit multiply to that literal or to $0$, so $\mathrm{Lit}=\prod_{k\in R}\frac12(1+\varsigma_k(-1)^{M_k})$
with signs $\varsigma_k$, or $\mathrm{Lit}\equiv0$. Expanding,
$\mathrm{Lit}=\sum_{C\subseteq R}2^{-|R|}\varsigma_C\prod_{k\in C}(-1)^{M_k}$ with $\varsigma_C=\prod_{k\in C}\varsigma_k$, whose coefficients have absolute values summing to $1$. Hence
\begin{align*}
|\E\hat M_T f(X)g(Y)\mathrm{Lit}(M_{<T})|&\le\sum_{C\subseteq R}2^{-|R|}\Bigl|\E f(X)g(Y)\prod_{k\in C\cup\{T\}}(-1)^{M_k}\Bigr|\\
&\le\max_{C\subseteq[T-1]}\disc_N(C\cup\{T\})\le2^{-DT}
\end{align*}
by \cref{thm:main}, since $\max(C\cup\{T\})=T\le N$. Summing over the at most $2(2n)^H$ terms proves the theorem.
\end{proof}

\subsection{A single bit, and windowed models}\label[appendix]{app:window}

\begin{proof}[Proof of \cref{prop:singlebit}]
Let the input have $n_0=2N+s_0$ positions, of which $s_0=O(1)$ are separators, and let $|g|$ be the length of the
chain of thought. The proof uses two statements of~\cite{AHRH25}, quoted in \cref{app:quotes}. The first is their
depth-reduction lemma (Lemma~B.3, from Hahn's Lemma~4~\cite{Hahn20}): if a restriction leaves at least $\vartheta n_0$
positions free, then for any $\vartheta'<\vartheta$ a further restriction leaving at least $\vartheta'n_0$ positions free
removes one layer, at the price of letting each layer-$0$ activation depend on boundedly many positions. Applied $L$
times, it makes every activation depend on at most $c$ free positions. The second is the inductive step of the proof
of~\cite[Thm.~3.3]{AHRH25}: each chain-of-thought token is then made constant by fixing at most $c\,H\,L$ further
positions, so all of them by fixing at most $c\,H\,L\,|g|$ positions.

We run this proof for the function $f=M_T$, starting from the restriction $\varrho^*$ that fixes the separator
positions to their symbols. (This is needed: otherwise the restriction produced by Lemma~B.3 could set a separator
to a wrong symbol and leave no well-formed input.) Lemma~B.3 accepts any initial restriction that leaves linearly many
positions free. For any $\vartheta\in(0,1)$ the proof produces a restriction under which $M_T$ is constant and which
fixes at most $(1-\vartheta)n_0+s_0$ positions in the $L$ applications of Lemma~B.3, plus at most $c\,H\,L\,|g|$ further
positions, where $c$ depends only on $\vartheta$, $L$ and $H$. All fixed positions other than separators are bits of $X$
and $Y$. Choose $\vartheta$ with $(1-\vartheta)(2N+s_0)+s_0<D\alpha N/2$ for large $N$. If the chain of thought
did not have length $\Omega(N)$, then for every $\varepsilon>0$ there would be infinitely many $N$ at which every chain
of thought has length at most $\varepsilon N$; the proof works one length at a time, so it applies along these $N$. With
$cHL\varepsilon<D\alpha/2$, $M_T$ would be constant on a subcube fixing fewer than $D\alpha N\le DT$ bits of $X$ and
$Y$. This contradicts the bound $k\ge DT$ proved in \cref{app:window-stmt}.
\end{proof}

\begin{proof}[Proof of \cref{lem:cone}]
By induction on the layer $\ell$: the activation at the position $q$ of a generated bit depends on the input and on the
generated tokens at positions $[q-\ell w,q]$. For $\ell=0$ it is $\mathrm{emb}(\mathrm{tok}_q)+p_q$. For $\ell\ge1$, a head at $q$
compares positions that are input positions, whose activations depend on input tokens only (they attend causally to
input positions), or generated positions $j\in[q-w,q]$, whose activations depend, by induction, on the input and on
the tokens at $[j-(\ell-1)w,j]\subseteq[q-\ell w,q]$. The residual term and the MLP add nothing new. After $L$ layers,
the output at the position $p$ of $M_{T-1}$ depends on the input and on the tokens at $[p-Lw,p]$, which hold the bits
$M_k$, $k\in\mathcal W$.
\end{proof}

\begin{proof}[Proof of \cref{thm:window}]
Let $p$ be the position of $M_{T-1}$. By \cref{lem:cone} the output at $p$ is $F(X,Y,M_{\mathcal W})$ with
$|\mathcal W|\le Lw+1$, and $\mathcal W\subseteq[T-1]$ once $T>Lw+1$. Generated positions outside the window do not affect $p$; fix them to arbitrary
symbols. We use Lemma~B.3 of~\cite{AHRH25}, which is Hahn's depth-reduction lemma~\cite[Lemma~4]{Hahn20} (both quoted
in \cref{app:quotes}): given a transformer with $L$ layers and a restriction leaving at least $\vartheta n$ positions
free, for any $\vartheta'<\vartheta$ there is a further restriction leaving at least $\vartheta'n$ positions free under
which the function is computed by a transformer with $L-1$ layers in which each layer-$0$ activation depends on
boundedly many positions. Hahn proves it for an arbitrary attention score
function $f^{\rm att}(y_i,y_j)$ (and the proof in~\cite{AHRH25} extends it to causal masking). The window mask is such a
score function once positions are carried in the activations: give every masked pair a score below all unmasked
ones, which is consistent because the position $q$ itself is never masked. Every head's allowed set contains all input
positions, among them all free ones, so the lemma applies unchanged.

For each $\sigma\in\{0,1\}^{\mathcal W}$, let $\varrho^*_\sigma$ fix the window positions to $\sigma$. Go through the
$2^{|\mathcal W|}$ values of $\sigma$ in turn; for each, apply Lemma~B.3 $L$ times, starting from the current restriction
of the positions of $X$ and $Y$ together with $\varrho^*_\sigma$ and the fixed separators. The free positions are always positions of $X$ and
$Y$, initially about half of the $n\le4N+O(1)$ positions, and each application loses an arbitrarily small constant
fraction. After this constant number of rounds we have one restriction $\varrho$ of $X$ and $Y$ that fixes $k\le\varepsilon N$
bits, for any $\varepsilon>0$ and all large $N$, and such that for every $\sigma$ the output at $p$ (with window $\sigma$)
depends on at most $c$ free positions; a $0$-layer transformer in which each activation depends on at most $c$ positions
has this property, and further restriction preserves it. Let $R$ be the union of these positions over all $\sigma$; then
$|R|\le2^{Lw+1}c=:c'$, a constant. So on every input with $(X,Y)$ in $\varrho$, correctness gives
$(-1)^{M_T}=\Theta(M_{\mathcal W},(X,Y)_R)$ for a $\pm1$-valued function $\Theta$ of $|\mathcal W|+|R|$ bits. Write
$\Theta=\sum_{U\subseteq\mathcal W,V\subseteq R}c_{UV}\prod_{j\in U}(-1)^{M_j}\chi_V$, with $|c_{UV}|\le1$ and
$\chi_V=\prod_{v\in V}(-1)^{(X,Y)_v}$. Let $\mathbf 1_\varrho=\mathbf 1_{\varrho_X}(X)\mathbf 1_{\varrho_Y}(Y)$. Then
\[
2^{-k}=\E\bigl[\mathbf 1_\varrho\Theta\,(-1)^{M_T}\bigr]\le\sum_{U,V}\Bigl|\E\Bigl[f_V(X)g_V(Y)\prod_{j\in U\cup\{T\}}(-1)^{M_j}\Bigr]\Bigr|
\le2^{|\mathcal W|+|R|}2^{-DT},
\]
where $f_V$ is $\mathbf 1_{\varrho_X}$ times the part of $\chi_V$ in the bits of $X$, and $g_V$ likewise; $|f_V|,|g_V|\le1$,
and \cref{thm:main} applies with $S=U\cup\{T\}$. Hence $k\ge DT-(Lw+1)-c'\ge D\alpha N-O(1)$, which contradicts
$k\le\varepsilon N$ for $\varepsilon<D\alpha$ and $N$ large.
\end{proof}

\begin{proof}[Proof of \cref{thm:avgwindow}]
Write $\hat F=2F-1$ and $w'=|\mathcal W|$. The advantage is $\frac12|\E\hat F(X,Y,M_{\mathcal W})\hat M_T|$. Expand the
indicator of the window value:
$\mathbf 1[M_{\mathcal W}=\sigma]=2^{-w'}\prod_{j\in\mathcal W}(1+\varsigma_j(-1)^{M_j})
=2^{-w'}\sum_{U\subseteq\mathcal W}\varsigma_U\prod_{j\in U}(-1)^{M_j}$ with signs $\varsigma_j$ depending on
$\sigma_j$ and $\varsigma_U=\prod_{j\in U}\varsigma_j$. Then
\[
\E\hat F(X,Y,M_{\mathcal W})\hat M_T=2^{-w'}\sum_\sigma\sum_{U\subseteq\mathcal W}\varsigma_U\,\E\Bigl[\hat F_\sigma(X,Y)\hat M_T\prod_{j\in U}(-1)^{M_j}\Bigr],
\]
so $|\E\hat F\hat M_T|\le\max_\sigma\sum_U|\E[\hat F_\sigma\prod_{j\in U\cup\{T\}}(-1)^{M_j}]|$. Fix $\sigma$ and $U$ and split
$\hat F_\sigma=\mathrm{Lo}+\mathrm{Hi}$ at degree $t$ in the $2N$ input bits. The low part is
$\sum_{|A|+|B|\le t}c_{AB}\chi_A(X)\chi_B(Y)$, and \cref{thm:main} with $f=\chi_A$, $g=\chi_B$ bounds each term by
$|c_{AB}|2^{-DT}$. By Cauchy--Schwarz and Parseval ($\hat F_\sigma=\pm1$), $\sum|c_{AB}|\le(\#\{(A,B):|A|+|B|\le t\})^{1/2}
\le(2N+1)^{t/2}$. For the high part, $|\E[\mathrm{Hi}\cdot h(X,Y)]|\le\|\mathrm{Hi}\|_2\le\delta_t^{1/2}$ for any $|h|\le1$; this
is exact because every factor is a function of $(X,Y)$ under the uniform measure. Summing over the $2^{w'}$ sets $U$ and
multiplying by $\frac12$ gives the claim.
\end{proof}

\begin{proof}[Proof of \cref{cor:avgwindow}]
(a) Tal~\cite[Thm.~18]{Tal17} gives $\mathrm{tail}_{\ge k}[f]\le2\cdot2^{-k/(c_d(\log m)^{d-1})}$ for a circuit $f$ of depth $d$ and size
$m>1$, with $c_d=60d\cdot192^{d-1}$ and $\log=\log_2$ (\cite[\S2]{Tal17}; see \cref{app:quotes}). Put
$\kappa_s=c_d(\log_2s)^{d-1}\ge1$ and $t=\lfloor DT/\log_2(2N+1)\rfloor$. Then $(2N+1)^{t/2}2^{-DT}\le2^{-DT/2}$, and
$\delta_t\le\mathrm{tail}_{\ge t+1}\le2\cdot2^{-(t+1)/\kappa_s}$ with $t+1\ge DT/\log_2(2N+1)$. With
$\Delta_T=DT/(2\kappa_s\log_2(2N+1))\le DT/2$, \cref{thm:avgwindow} gives advantage at most
$2^{|\mathcal W|-1}(2^{-\Delta_T}+\sqrt2\,2^{-\Delta_T})\le2^{|\mathcal W|+1-\Delta_T}$.
(b) By \cref{lem:cone} the step computes $F(X,Y,M_{\mathcal W})$ with $|\mathcal W|\le Lw+1$. For fixed $\sigma$, $F_\sigma$ is the output of a
fixed-length UHAT whose window positions are fixed to $\sigma$ and whose other positions vary only at the bits of $X$
and $Y$. Hao, Angluin and Frank define generalised unique-hard-attention transformers with arbitrary activation values
and arbitrary functions computing activations and scores; causal and window masks and a priority tie rule are realised
as in \cref{app:worstcase}. Their proof of~\cite[Thm.~1]{HAF22} is per input length: activation values at layer $\ell$
take at most $(H+1)^\ell(2\lceil\log_2(n+1)\rceil+O(1))$ bits~\cite[Lemma~2]{HAF22} (quoted in \cref{app:quotes}), which
gives a circuit of depth $d=d(L)=O(L)$ and size $s=n^{O((H+1)^L)}$ with $n\le4N+O(1)$. Fixed tokens become constants.
Part (a) with $Lw+1$ in place of $|\mathcal W|$ and
$\log_2s=O_{L,H}(\log N)$ gives the bound. The last sentence follows by taking $\gamma$ small.
\end{proof}

\subsection{The embedding and the worst case}\label[appendix]{app:worstcase}

A set $A\subset\mathbb Z_{\ge0}$ is a \emph{Sidon set} if all differences $a-a'$ with $a\ne a'$ in $A$ are distinct.

\begin{lemma}[a Sidon variant of Buss's embedding]\label[lemma]{lem:sidon}
Let $b\ge2$, $r=2^b-1$, and let $A=\{a_0,\dots,a_{r-1}\}$ be a Sidon set with $0\in A$ and $\tau=\max A$. Let $N,T$
satisfy $\min(T,2N-T)\ge(\tau+1)b+1$. Then there are shifts $\nu_x,\nu_y\ge0$ such that, for every $z\in\{0,1\}^r$, the
numbers $X(z)=2^{\nu_x}\sum_iz_i2^{a_ib}$ and $Y=2^{\nu_y}\sum_j2^{(\tau-a_j)b}$ are below $2^N$ and
\textup{(1)} $M_T(X(z),Y)=\mathrm{MAJ}_r(z)$; \textup{(2)} for each $k<T$, $M_k(X(z),Y)$ is either $0$ for all $z$
or equal to $z_i$ for one $i$ depending only on $k$; \textup{(3)} every bit of $X(z)$ is $0$ or a single $z_i$, and
$Y$ does not depend on $z$.
\end{lemma}

\begin{proof}
Let $\nu=\nu_x+\nu_y$. Then $X(z)Y=2^\nu\sum_{i,j}z_i2^{(\tau+a_i-a_j)b}$. Group the terms by $\delta=a_i-a_j\in[-\tau,\tau]$.
For $\delta=0$ the coefficient of $2^{\nu+\tau b}$ is $\sum_iz_i\le r<2^b$; for $\delta\ne0$ the Sidon property leaves at
most one pair $(i,j)$, so the coefficient of $2^{\nu+(\tau+\delta)b}$ is $0$ or a single $z_i$. All coefficients are below
$2^b$ and the blocks are $b$ bits apart, so there are no carries, and the binary expansion of $XY$ is the concatenation
of the blocks. Since $M_T$ is the bit of weight $2^{2N-T}$, choose $\nu=(2N-T)-((\tau+1)b-1)$: then $M_T$ is the top bit of
the block holding $\sum_iz_i$, which is $\mathbf 1[\sum_iz_i\ge2^{b-1}]=\mathrm{MAJ}_r(z)$. The higher bits belong to
blocks with $\delta\ge1$ or lie above all blocks, which gives (2). For the sizes, $X<2^N$ iff $\nu_x+\tau b+1\le N$ and
$Y<2^N$ iff $\nu_y+\tau b+1\le N$ (using $0\in A$). With $\nu_{\max}=N-1-\tau b$, the hypothesis gives $0\le\nu\le2\nu_{\max}$,
so $\nu$ splits as $\nu_x+\nu_y$ with $0\le\nu_x,\nu_y\le\nu_{\max}$. (3) is immediate.
\end{proof}

\noindent Buss's integers $x,y$~\cite{Buss92} are of the same shape, with the spacings $n^2+j$ in place of a Sidon set,
and span about $n^3$ bits for $n$ counted bits. A Sidon set gives a quadratic span. For an odd prime $p\ge r$, the set
$a_i=2pi+(i^2\bmod p)$, $0\le i<r$, is a Sidon set (Erd\H{o}s and Tur\'an): $a_i+a_j$ determines $i+j$ (the quotient by
$2p$) and $i^2+j^2\bmod p$, hence $\{i,j\}$, and equal differences $a_i-a_j=a_{i'}-a_{j'}$ give equal sums
$a_i+a_{j'}=a_{i'}+a_j$. With $p$ the least prime $\ge r$ (so $p<2r$), $a_0=0$ and $\tau=a_{r-1}<2pr<4r^2$. Given
$K\ge256$ (so that $b\ge2$, as \cref{lem:sidon} requires), let $b$ be the largest integer with $8\cdot4^bb\le K$. Then $(\tau+1)b+1\le4r^2b+1\le K$, and maximality of
$b$ gives $r+1=2^b>\sqrt{K/(\sidonConst\log_2K)}$.

\begin{proof}[Proof of \cref{thm:worstcase}]
\emph{(a) From the transformer to a circuit.} Fix $N$, $T$ and the transformer $\mathcal T_n$ for the total length. By causal
masking, its output at the position before $M_T$ is a function of the prefix $(x,M_1,\dots,M_{T-1})$, of fixed length
$m=|x|+T-1$. This computation is a generalised unique-hard-attention computation in the sense of Hao, Angluin and
Frank~\cite{HAF22}, whose definition places no restriction on activation values or on the functions that compute
activations and attention scores: carry the position in the activation; use the score $\arctan(a_{ij})$ for $j\le i$
and $-2$ for $j>i$ (causal masking); realise the priority tie rule by adding $\delta\cdot(\text{rank of }j)$ with $\delta$
below the smallest gap between the finitely many distinct scores that occur at this length; absorb residual connections,
value maps and $f^{\rm MLP}$ into the activation function; and, since their output is read at an end marker, add one
layer in which the marker attends to position $m$ and copies it. Their proof of~\cite[Thm.~1]{HAF22} is per input
length (see the proof of \cref{cor:avgwindow}(b)), so it gives a circuit $\mathcal C$ of depth $d=d(L)$ and size at most
$m^{a}$, $a=a(L,H)$, over literals of the input bits, with $\mathcal C(x,M_{<T}(x))=M_T(x)$ for every input. Applying it to each
length separately avoids having to define one language over all lengths.

\emph{(b) Substitution.} Let $K=\min(T,2N-T)\ge N^\varepsilon$ and choose $b$, $r$ and $A$ as above. Substitute the
inputs of \cref{lem:sidon}. Every input wire of $\mathcal C$ becomes a constant or a literal $z_i$ or $\neg z_i$ (the symbol
encoding of an input bit may negate it, which is harmless for circuits over literals); every earlier product bit becomes
a constant or some $z_i$. So $\mathcal C$ becomes a circuit of depth $d$ and size at most $m^a\le(5N)^a$ that computes
$\mathrm{MAJ}_r$, where $r\ge N^{\varepsilon/3}$ for large $N$; its size is polynomial in $r$. Majority has no such
circuits: with $u\in\{0,1\}^{m'}$, $r=2m'-1$ and $\mathrm{TH}_j(u)=\mathrm{MAJ}_r(u,1^{m'-j},0^{j-1})=\mathbf 1[|u|\ge j]$,
we get $\mathrm{PARITY}_{m'}=\bigvee_{j\text{ odd}}(\mathrm{TH}_j\wedge\neg\mathrm{TH}_{j+1})$ ($\mathrm{TH}_{m'+1}\equiv0$),
a circuit of depth $d+3$ and size polynomial in $m'$. This contradicts the theorem of Furst, Saxe and
Sipser~\cite{FSS84} that parity has no polynomial-size circuits of constant depth.

\emph{(c) Extra tokens.} Suppose $c=O(1)$ extra tokens from a finite vocabulary $\Xi$ sit at fixed positions before
$M_T$. Each is the output of the transformer at a fixed position, so by (a) it is given by $|\Xi|$ circuits (one per
symbol) of the earlier prefix. Under the substitution of (b) that prefix consists of constants, literals and earlier
extra tokens, so by induction the $c$ extra tokens and then $M_T$ are computed by a circuit in $z$ of depth at most
$(c+1)(d+1)$ and size at most $(c+1)|\Xi|m^a$, and (b) applies. Extra tokens at input-dependent positions, a vocabulary
that grows with $N$, or $\omega(1)$ extra tokens are not covered.
\end{proof}

\subsection{The barriers}\label[appendix]{app:barriers}

\begin{proof}[Proof of \cref{prop:disj}]
Layer 1: the $X$-position $j$ has key $(2j,-j^2,1)$, and the $Y$-position $N+i$ has query $(i,1,N^2+1)$; all other keys
are $0$. The score $2ij-j^2+N^2+1=N^2+1+i^2-(i-j)^2$ is positive and uniquely maximal at $j=i$, while every other position
scores $0$. The value copies $x_j$, and the MLP writes $z=x_iy_i$ at $Y$-position $N+i$ and $z=0$ elsewhere. Layer 2: the
last position scores each position by its $z$ and breaks ties to the left; if some $z_i=1$ it attends to such a position,
and otherwise to position $1$, where $z=0$. It outputs the attended $z$. All scores are integers of absolute value at most
$3N^2+1$. A protocol that reproduces this model across $X\,|\,Y$ decides whether two $N$-bit sets intersect, and every
deterministic protocol for that needs $N$ bits~\cite[Fact~5]{SHT23}.
\end{proof}

\begin{proof}[Proof of \cref{prop:gt}]
By definition of $a$, $XY\in[a2^{2N-T+1},(a+1)2^{2N-T+1})$, and $M_T=1$ if and only if $XY\ge(2a+1)2^{2N-T}$. Since $Y>0$
and $X$ is an integer, this holds if and only if $X\ge\lceil(2a+1)2^{2N-T}/Y\rceil$.
\end{proof}

\begin{proof}[Proof of \cref{prop:restriction}]
Let $a_X$ and $a_Y$ be the weights of the most significant free bits of $X$ and $Y$ in $\mathcal Q$ (take $a_X=0$ if $X$
has no free bit, and fix only the bit of $Y$; similarly for $Y$), and let $X_0$, $Y_0$ be the values with all free bits
$0$. On $V$, $\lfloor XY/2^{2N-T+1}\rfloor=q$ is constant, so $M_T=\mathbf 1[XY\ge\theta]$ with $\theta=(2q+1)2^{2N-T}$.
The lower free bits contribute between $0$ and $a_X-1$ to $X$ (and $a_Y-1$ to $Y$). So if both chosen bits are $0$ then
$X\le X_0+\max(a_X-1,0)$ and $Y\le Y_0+\max(a_Y-1,0)$, which gives $XY<\Pi:=(X_0+a_X)(Y_0+a_Y)$ unless $XY=0$; if both
are $1$ then $XY\ge\Pi$. (If neither $X$ nor $Y$ has a free bit, $V$ has at most one point.) Let
$v=\mathbf 1[\Pi\ge\theta]$ and fix both bits to $v$: if $v=1$ then $XY\ge\Pi\ge\theta$, and if $v=0$ then
$XY<\max(\Pi,1)\le\theta$. Either way $M_T=v$ on what remains of $V$.
\end{proof}

\subsection{Statements used from prior work}\label[appendix]{app:quotes}

For the reader's convenience we quote the external statements used in \cref{sec:transformers}, verbatim except for
typesetting.

\emph{Tie rule~\cite[\S2.2]{AHRH25}.} ``If more than one $s$ attains the maximal attention score, ties are broken
according to some fixed rule (e.g., choosing the left- or right-most match, \dots).''

\emph{$c$-Transformer and depth reduction~\cite[Def.~B.2 and Lemma~B.3]{AHRH25}.} ``A $c$-Transformer conforms to the
definition of transformers from Section~2.2, except that each $y_i^{(0)}$ is a function of $\le c$ input positions.''
``Lemma~B.3 (Depth Reduction Lemma). Given a UHAT $c$-Transformer $T$ with $L$ layers, a constant $C\in(0,1]$, and
restrictions $\rho_N=\rho_1,\rho_2,\dots$ such that $|\{i\le N:\rho_N(i)=*\}|\ge CN$ for all sufficiently large $N$.
Choose any $C'\in(0,C)$. Then there are restrictions $\rho'_1\succ\rho_1,\rho'_2\succ\rho_2,\dots$ such that
$|\{i\le N:\rho'_N(i)=*\}|\ge C'N$ for all sufficiently large $N$, and such that there is a
$c\cdot(|\Sigma|^ckH+1)$-transformer $T'$ with $L-1$ layers, for some integer $k$ (depending on $C'$), such that, for
each $i\in[1,N]$ and each $l\in[1,L]$, $y_i^{(l)}$ in $T$ equals $y_i^{(l-1)}$ in $T'$ whenever the input is in
$\rho'\Sigma^*$.'' The same appendix adds: ``Causal masking can be easily simulated in UHAT, \dots\ Hence, the proof of
the Depth Reduction Lemma continues to work in the presence of hard-coded causal masking.''

\emph{Hahn's original~\cite[Lemma~4]{Hahn20}.} ``Lemma~4 (Depth Reduction Lemma). Given a $c$-transformer with $L$
layers, and some restriction $\rho$ such that $|\{i\le n:\rho_n(i)=*\}|\ge Cn$ ($C\in(0,1]$) for all sufficiently large
$n$. Choose any $C'<C$. Then there is a restriction $\rho'$ such that $|\{i\le n:\rho'_n(i)=*\}|\ge C'n$ for all
sufficiently large $n$, and such that the resulting function is computed by a $(c\cdot(2^ckH+1))$-transformer with
$L-1$ layers, for some integer $k$ (depending on $C'$), where $H\ge1$ is the number of attention heads at each layer and
position.''

\emph{Cost of the chain-of-thought tokens (proof of~\cite[Thm.~3.3]{AHRH25}).} ``Overall, by taking
$\rho:=\rho^{(M)}$, we have fixed an additional $\le c\cdot H\cdot L\cdot|g(x)|=o(N)$ input positions to fix all CoT
tokens.'' Here $|g(x)|$ is the length of the chain of thought; the bound $c\cdot H\cdot L$ per token holds for any
length, and we use it with $|g|$ in place of $o(N)$.

\emph{Bit size of activations~\cite[Lemma~2]{HAF22}.} ``Let $T$ be a GUHAT in informative normal form with $K$ layers
and $H$ heads, and alphabet $\Sigma$. Let $s=\ell(|\Sigma|+1)$. Then for any input of length $n$ and any $k\in[0..K]$,
the activation values at layer $k$ can be represented by $(H+1)^k(2\ell(n)+s)$ bits, and for $k\ge1$, the attention
scores at layer $k$ can be represented by $2(H+1)^{k-1}(2\ell(n)+s)$ bits.'' Here $\ell(n)=\lceil\log_2(n+1)\rceil$,
and by their Lemma~1 every GUHAT has an equivalent one in informative normal form with the same numbers of layers and
heads. In a GUHAT the attention scores are arbitrary functions of the two activations, and ties go to the least
position; we encode causal masking, window masks and a priority tie rule in the scores as in the proof of
\cref{thm:worstcase}.

\emph{Fourier tail of $\mathrm{AC}^0$~\cite[Thm.~18]{Tal17}.} ``Theorem~18 (Theorem~1, restated). Let $f$ be an Boolean
circuit of depth $d$ and size $m>1$. Then, $W^{\ge k}[f]\le2\cdot2^{-k/(c_d\log^{d-1}(m))}$ where
$c_d=60d\cdot192^{d-1}\le2^{16d}$.'' Here $W^{\ge k}[f]$ is what we write $\mathrm{tail}_{\ge k}[f]$, and Tal's \S2
fixes the base: ``We denote by $\log$ and $\ln$ the logarithms in bases $2$ and $e$''.

\fi
% Appendix: top-k attention at every fixed depth, the depth picture, softmax with margins, the averaging separation,
% and three layers with growing k.
% Proves lem:topk-poly, thm:topk-depth, cor:topk-loglog, cor:softmax and cor:topk-three.
\ifoldpullback
\ificml
\section{Proofs for top-$k$ attention}\label[appendix]{app:topk}
\ificml\multiplicationtopktechnical\fi

Throughout, $n$ is the number of positions, $\log$ is natural, and $A(\cdot)$ is the Walsh norm.
The formal input is a binary cube, with fixed tokens allowed.
\else
\section{Top-$k$ attention: every fixed depth, and what lies beyond}\label[appendix]{app:topk}

This appendix proves \cref{lem:topk-poly}, \cref{thm:topk-depth} and \cref{cor:topk-loglog}
(\cref{app:topkdepth}), gives the depth constructions behind \cref{sec:depthgap} (\cref{app:depthpicture}), proves
\cref{cor:softmax} (\cref{app:softmax}), describes the exact construction of \cref{rem:averaging}
(\cref{app:averaging}), and derives the three-layer \cref{cor:topk-three} (\cref{app:topkk}). Throughout, $n$ is the
context length, $\alpha N\le T\le N$, and $\hat O_T\in\{-1,0,1\}$ is the signed output of \cref{thm:onelayer}. We write
$A(f)$ for the Walsh norm of a function on the formal cube, as in \cref{sec:overview}, so that \eqref{eq:walsh} reads
$|\E\hat M_Tf(Z)|\le2^{-DT}A(f)$. Since $T\ge\alpha N$ and $n\le3N+O(1)$, we have $T\ge\alpha n/4$ for large $n$.

\subsection{The model and its scope}\label[appendix]{app:topkmodel}

We use \cref{def:topk}. Fix a layer and a head. For a query at position $i$ the head orders the positions $j\le i$
by its score, which may be any function of the two states and the two positions, and then by a fixed priority order on
positions. It selects the first $\min(k,i)$ positions in this order. The new state at $i$ is an arbitrary function of the
old state at $i$ and, for every head, the ordered list of selected (position, state) pairs. This contains the residual
connection, an MLP and an average of the selected values, with any width, precision and positional encodings. A head may
also have a smaller fixed capacity $k'\le k$; it can use the first $k$ entries in the arguments below and keep $k'$ of
them at the end. The model may differ at every length. With $k=1$ and bilinear scores this is the UHAT of
\cref{app:model}.

\paragraph{What is not covered.} The tie rule is a fixed priority on positions. We do not treat averaging hard attention,
in which a head averages all positions tied at the maximum score however many there are, nor softmax attention. The
method cannot reach averaging hard attention (\cref{rem:averaging}). Every bound below is obtained by approximating the model's output, on the
whole formal cube, by a polynomial of sublinear degree. One averaging layer computes MAJORITY, whose bounded-error quantum
query complexity and approximate degree are linear~\cite{BBCMW01}. No exponent below, and no depth threshold, is claimed
to be tight.

\fi
\subsection{Every fixed depth: pulling polynomials back through the layers}\label[appendix]{app:topkdepth}

The generic pullback bounds the degree and Walsh norm in \cref{lem:topk-poly}.
For \cref{thm:topk-depth} we sharpen the Walsh-norm argument by guessing position sets
and keeping their states as shared records. The final correlation estimate then uses
\eqref{eq:walsh} with \cref{thm:main}.

Fix an $L$-layer top-$k$ model with at most $H$ heads per layer on $n$ positions, each holding a raw token with at most
one variable bit (a bit, or a fixed separator). Let $a=1+Hk$ and $K_{\mathrm{full}}=4a^L\ln(4n)$.

\paragraph{Cartesian alphabets.} Let $\Sigma_{0,i}$ be the set of tokens position $i$ can hold ($|\Sigma_{0,i}|\le2$).
Given $\Sigma_{j-1,1},\dots,\Sigma_{j-1,n}$, let $\Sigma_{j,i}$ be the set of all values of the layer-$j$ update at $i$ on
the whole product $\prod_{i'}\Sigma_{j-1,i'}$, including combinations of states that no input produces. So each layer maps
every lower string to an allowed upper string. Let $A_j=\max_i|\Sigma_{j,i}|$. A \emph{local transcript} at layer $j$ and
position $i$ is an old query state in $\Sigma_{j-1,i}$ and, for each head, an ordered list of selected (position, state)
pairs, of the length $\min(k,i)$ that the head returns at $i$. Let $J_j$ be the largest number of local transcripts at a
position. Counting gives
\begin{equation}\label{eq:depth-alph}
A_j\le J_j\le A_{j-1}(nA_{j-1})^{Hk},\qquad nA_j\le(2n)^{a^j},\qquad nJ_j\le(2n)^{a^j},
\end{equation}
the last two by induction from $nA_0\le2n$. Encode each state of $\Sigma_{j,i}$ by a word of
$\ell_j=\max(1,\lceil\log_2A_j\rceil)$ bits and decode unused words to a fixed state. Every suffix of the network is then a
function on the \emph{whole} cube of words of its lowest layer. A character on this cube has \emph{record degree} equal
to the number of positions whose words it reads.

\begin{lemma}[Walsh norm and record degree]\label[lemma]{lem:recordnorm}
If $|p|\le B$ on $(\{0,1\}^\ell)^n$ and $p$ has record degree at most $d$, then its Walsh norm is at most
$B(1+n2^\ell)^{d/2}$.
\end{lemma}

\begin{proof}
At most $\sum_{j\le d}\binom nj(2^\ell-1)^j\le(1+n2^\ell)^d$ characters involve at most $d$ words. By Parseval the sum of
the squared coefficients is at most $B^2$, and Cauchy--Schwarz gives the bound.
\end{proof}

\begin{lemma}[joint transcripts]\label[lemma]{lem:jointtr}
Fix a layer $j$, a set of $m$ query positions and, at each of them, a local transcript. On the cube of layer-$(j-1)$ words,
the indicator that all $m$ actual transcripts equal the guessed ones is $0$ or a product $\prod_{i=1}^nb_i(v_i)$ of
Boolean functions $b_i$, each of the word $v_i$ at a single position. Exactly one combination of guesses holds on each
string.
\end{lemma}

\begin{proof}
Discard guesses with invalid lengths, a position repeated within one head, two different states guessed for one
position, or a selected list that is not in the head's order for the guessed query state. For the others, require at
every query position and every selected position that its state equals the guessed one, and intersect the requirements
made at one position. For every head and every eligible position not selected by it, require that its key, scored against
the guessed query state, comes after the last selected key in the head's order (score, then priority). A head that
returns no position imposes nothing. Every requirement is a test of one position's state, so the conjunction has the
stated form, with at most $n$ factors however many queries read a position. The tests are necessary by the definition of
selection. They are sufficient, because the guessed lists have the right length and order and every other key loses to
the cutoff. The actual transcripts form one guess, so exactly one guess holds.
\end{proof}

\noindent For example, if a query selects positions $3$ and $1$ in that order, the guess fixes the query state and the
states at $3$ and $1$, checks their order, and requires every other eligible key to rank below position $1$; a second
query that also uses position $3$ adds a second test at $3$, which is intersected with the first.

\begin{lemma}[one layer]\label[lemma]{lem:onelayerback}
Let $1\le j\le L$ and let $P$ be a function on the cube of layer-$j$ words with record degree at most $d\ge1$ and
$|P|\le2$. Let $F_j$ be the layer-$j$ map from layer-$(j-1)$ words to layer-$j$ words.
\textup{(a)} $P\circ F_j$ is a linear combination of indicators of \cref{lem:jointtr} with total absolute coefficient at
most $e^{K_{\mathrm{full}}d}$.
\textup{(b)} For every $s\ge1$ there is a function $P'$ on the layer-$(j-1)$ cube of record degree at most
$c\sqrt{ns}$, with $c$ as in \cref{lem:topk-poly}, and $|P'-P\circ F_j|\le e^{K_{\mathrm{full}}d-s}$ everywhere.
\textup{(c)} If $j=1$, $P\circ F_1$ has Walsh norm at most $e^{K_{\mathrm{full}}d}$ on the raw cube.
\end{lemma}

\begin{proof}
(a) A character $\chi$ of $P$ reads at most $d$ positions, and the upper word at a position is determined by its local
transcript. Summing over the at most $J_j^d$ guesses at those positions writes $\chi\circ F_j$ as a sum of indicators with
signs $\pm1$; the constant character uses the empty guess. By \cref{lem:recordnorm} with $B=2$ and $2^{\ell_j}\le2A_j$,
the Walsh norm of $P$ is at most $2(1+2nA_j)^{d/2}$, so the total coefficient is at most
$2(1+2nA_j)^{d/2}J_j^d\le2\cdot3^{d/2}(2n)^{3a^jd/2}\le e^{K_{\mathrm{full}}d}$ by \eqref{eq:depth-alph}. (b) By \cref{lem:nor-small}
applied to the negated inputs, there is a polynomial in $n$ Boolean variables, of degree at most
$\min(n,\lfloor3.34\sqrt{n(s+\ln2)}\rfloor+1)\le c\sqrt{ns}$, within $e^{-s}$ of $\mathrm{AND}_n$ on $\{0,1\}^n$.
Substitute the exact Boolean tests $b_i(v_i)$ for its variables, in every indicator of (a). A monomial then reads at most
$c\sqrt{ns}$ positions, whatever the number of bits in a word, and each indicator moves by at most $e^{-s}$ at every point.
(c) At the bottom each $b_i$ is a function of the raw token at $i$, which has at most one variable bit; so $b_i$ is a
literal, a constant $1$ or $0$, of Walsh norm at most $1$, and so is the product. Use (a).
\end{proof}

\noindent The degree bound in (b) is where the joint conjunction matters. Approximating each queried output separately
and substituting would multiply the degrees of consecutive layers; here all tests at one position are intersected into one
test before the single AND is approximated. This is the shared-input composition of Bun, Kothari and
Thaler~\cite[Thm.~2]{BKT19}.

\paragraph{The recurrence.} Fix $v$ with $K_{\mathrm{full}}+\ln(2L)\le v\le n$ and put $\omega=4c^2$. Start from the readout, a function of
the top word at the last position, with record degree $d_0=1$ and values in $[-1,1]$. Remove layers $L,L-1,\dots$ one at a
time; at the step that removes the $(r+1)$-th layer apply \cref{lem:onelayerback}(b) with
\begin{equation}\label{eq:depth-rec}
s=v+\ln(2L)+K_{\mathrm{full}}d_r,\qquad d_{r+1}\le c\sqrt{n(v+\ln(2L)+K_{\mathrm{full}}d_r)} .
\end{equation}
Each step adds uniform error at most $e^{-v}/(2L)$. Composing an existing uniform error with a map into its domain keeps
its bound, so after $r$ steps the error against the exact suffix is at most $re^{-v}/(2L)\le\frac12$, everywhere on the
current word cube, and the hypothesis $|P|\le2$ of the lemma holds at every step. For $r\ge1$ put
\begin{equation}\label{eq:record-envelope}
E_r=\omega nK_{\mathrm{full}}\Bigl(\frac{v}{nK_{\mathrm{full}}^2}\Bigr)^{2^{-r}} .
\end{equation}
Then $d_r\le E_r$. Indeed $d_1\le c\sqrt{2nv}\le\omega\sqrt{nv}=E_1$, because $v+\ln(2L)+K_{\mathrm{full}}\le2v$. Since $v\le n$ and
$K_{\mathrm{full}}\ge1$, the $E_r$ increase with $r$ and $E_1\ge v$. If $d_r\le E_r$, the argument of the square root in
\eqref{eq:depth-rec} is at most $n(2v+K_{\mathrm{full}}E_r)\le3nK_{\mathrm{full}}E_r$, so $d_{r+1}\le\sqrt{3c^2nK_{\mathrm{full}}E_r}\le\sqrt{\omega nK_{\mathrm{full}}E_r}=E_{r+1}$.

\begin{proof}[Proof of \cref{lem:topk-poly}]
For the degree bound, run all $L$ steps, the last one at the bottom layer too. The final function is a polynomial on the
raw cube with error at most $Le^{-v}/(2L)=e^{-v}/2$, and its record degree $d_L\le E_L$ is its ordinary degree, because a
raw word has at most one variable bit. For the Walsh-norm bound with $L\ge2$, run $L-1$ steps and expand the bottom layer
exactly by \cref{lem:onelayerback}(c): the error is at most $(L-1)e^{-v}/(2L)\le e^{-v}/2$ and
$\ln A(P)\le K_{\mathrm{full}}d_{L-1}\le K_{\mathrm{full}}E_{L-1}$.
\end{proof}

\paragraph{Grouping by selected positions.}
For the attention theorem we improve the dependence on $a$ in \cref{lem:topk-poly}
by keeping position guesses and their selected states separate.
Represent an approximant after $r$ pullbacks by a sum of bounded functions, each depending on
at most $m_r$ records, with total absolute coefficient $B_r$ and $\ln B_r\le m_r\ln(4n)$.
The readout starts with $m_0=1$ and $B_0=1$.

For a term reading $m$ upper records, guess only the unordered sets of selected positions
at these queries, at most $n^{Hkm}$ choices. Their union with the query positions has at most
$am$ lower records. On these records, recover the ordered lists by sorting each guessed set
with the actual score and tie order. The term evaluated on the resulting states is a bounded
function of these records. Extend it by zero if a guessed state lies outside the upper alphabet.
To check the guess, each eligible nonselected key must rank below its head's last selected key,
including keys used by other heads. These conditions form one conjunction of $n$ Boolean tests,
each on its own record and the shared records. Exactly one guess passes for each lower input.

Set $K_{\mathrm{set}}=4a\ln(4n)$ and $s=v+\ln(2L)+a m\ln(4n)$. Replacing each conjunction by the
small-error polynomial of \cref{lem:nor-small} (after negating the test bits) gives error
at most $B_r n^{Hkm}e^{-s}\le e^{-v}/(2L)$ at this step. The polynomial is bounded by one
on its Boolean test inputs and has degree $d\le c\sqrt{ns}$; its Walsh norm on those
inputs is at most $(1+n)^{d/2}$ by Parseval. After substitution each character uses at most
$d$ records besides the shared ones. Thus the new number of records is at most $am+d$,
and the logarithm of the coefficient mass is at most $(am+d)\ln(4n)$.
If $am+d\ge n$, instead represent the exact suffix directly as one bounded function
on all $n$ records, with error zero. For $m'=\min(n,am+d)$ we have
$(m')^2\le d^2+2nam$, whether $m'\le d$ or $m'>d$.
Therefore $m'\le c\sqrt{n(v+\ln(2L)+K_{\mathrm{set}}m)}$, precisely the recurrence
\eqref{eq:depth-rec} with record counts and the smaller $K_{\mathrm{set}}$. Here and below $E_r$ is defined by \eqref{eq:record-envelope} with $K_{\mathrm{set}}$ in place of $K_{\mathrm{full}}$. The estimate $m_r\le E_r$ follows by the same
induction. For the exact final pullback through the first layer, expand each bounded term
over full transcripts. There are at most $(2n)^{am}$ such transcripts for $m$ queried
records, and each is a conjunction of raw literals, of Walsh norm at most one.
Thus the final log Walsh norm is at most $(a+1)m\ln(4n)\le K_{\mathrm{set}}m$.

\begin{proof}[Proof of \cref{thm:topk-depth,rem:topk-explicit}]
\ificml
For fixed $L,H,k$, apply \eqref{eq:icml-linear-approximant} to the one signed top record
with any constant $\delta_\star>0$ satisfying $\delta_\star\le D\alpha\ln 2/4$.
Indeed $T\ge\alpha n/4$, and \eqref{eq:walsh} then contributes a factor at most
$e^{-\delta_\star n}$. The approximation and correlation errors are exponentially small in $n$.
The following explicit bound controls growing parameters.
\fi

Use $\delta,R_0,Q$ from \cref{rem:topk-explicit} and write
\begin{equation}\label{eq:depth-t}
t=\frac{nK_{\mathrm{set}}^2}{(R_0K_{\mathrm{set}}^2)^Q}.
\end{equation}
If $t\ge K_{\mathrm{set}}+\ln(2L)$, take $v=t\le\delta n/16$. For $L\ge2$ the position-set
pullbacks followed by the exact expansion give a uniform error at most $e^{-t}/2$ and
$\ln A(P)\le K_{\mathrm{set}} E_{L-1}=4c^2n/R_0=\delta n/4$. For $L=1$ the exact expansion alone
has $\ln A(P)\le K_{\mathrm{set}}\le t$. Since $T\ge\alpha n/4$ and $D$ is applicable in
\cref{thm:main}, $DT\ln 2\ge\delta n$. It follows from \eqref{eq:walsh} that
$|\E\hat M_T\hat O_T|\le e^{-t}/2+e^{-3\delta n/4}\le2e^{-t}$.

\emph{Dependence on $H,k$.} All the preceding estimates hold for the parameters at a single length. The condition is
equivalent to
\[
 n\ge R_0^QK_{\mathrm{set}}^{2Q-1}\bigl(1+\ln(2L)/K_{\mathrm{set}}\bigr).
\]
For fixed $L$ and sufficiently large $n$, $\ln(2L)\le K_{\mathrm{set}}$, so the right side is at most
$2R_0^Q4^{2Q-1}a^{2^L-1}(\ln(4n))^{2Q-1}$, which is $o(n)$ in the stated range with slack $\varepsilon$.
For fixed $L,H,k$, $K_{\mathrm{set}}\le8a\ln n$ once $n\ge4$. Since $t\to\infty$, one may take
$c_0=1/(2R_0^Q(8a)^{2Q-2})$ by absorbing the factor $2$ into $e^{t/2}$.

\emph{Growing depth.} Let $b=4+2\log_2a$, so $2^{h(a)}\ge2b$. For $L\le J$ and
$u\ge\max(4,8\sqrt{R_0})$ we have $L\le\log_2u$ and $K_{\mathrm{set}}\le8u^{1+\log_2a}$.
Because $64R_0\le u^2$, we have $\ln(R_0K_{\mathrm{set}}^2)\le b\ln u$ uniformly in $a$.
Also $Q\le u/(2^{h(a)+1}\ln u)\le u/(4b\ln u)$. Thus $(R_0K_{\mathrm{set}}^2)^Q\le n^{1/4}$ and
$t\ge n^{3/4}K_{\mathrm{set}}^2$. Since $K_{\mathrm{set}}\ge2$ and $K_{\mathrm{set}}\ge\ln(2L)$, this implies $t\ge K_{\mathrm{set}}+\ln(2L)$ for varying $H,k$ as well.
\end{proof}

\begin{proof}[Proof of \cref{cor:topk-loglog}]
The advantage is at most $\frac12|\E\hat M_T\hat O_T|\le\exp(-n^{3/4})$ for $L\le J$.
This is strictly below $n^{-p}$ once $n^{3/4}>p\ln n$.
\end{proof}

\noindent For fixed $H,k$, the boundary of the finite condition is
$L=\log_2\ln n-\log_2\ln\ln n+o(1)$ before rounding to an integer depth:
$\ln(R_0K_{\mathrm{set}}^2)=2\ln\ln n+O(1)$ and $\ln(K_{\mathrm{set}}+\ln(2L))=O(\ln\ln n)$ at this boundary.
A bound of the form $\exp(-n^{C^{-L}})$ also yields a depth lower bound of order $\log\log n$.
Reaching order $\log n/\log\log n$ remains open.

\paragraph{Checks.} We ran the construction on literal top-$k$ networks with $4$ to $6$ positions, depths $2$ to $4$,
$(H,k)\in\{(1,1),(1,2),(2,1),(2,2)\}$ and alphabets with unused codewords, in exact rational arithmetic: the pulled-back
functions equal the network on the whole word cube, including unused codewords; record degrees and errors stay within
\cref{lem:onelayerback}; and dropping the cutoff test, the query-state test or the Cartesian closure, or approximating
each query's conjunction separately, produces violations. The parameter algebra of \eqref{eq:depth-alph}--\eqref{eq:depth-t}
was checked in high precision over a grid of constants. These are consistency checks, not part of the proofs.

\ificml\else
\begin{remark}[how the proof avoids two obstructions]\label[remark]{rem:topk-escape}
\Cref{rem:four} shows that a second-layer pointer can be spread over many positions on typical inputs, so any
\emph{input-independent} coarsening of the second-layer state that recovers the pointer needs many values. The proof
above never coarsens, truncates or samples. It keeps the full Cartesian alphabets, of size up to $(2n)^{a^j}$, and pays
only $\log(nA_j)$ per unit of record degree, through Parseval; so the obstruction of \cref{rem:four} does not arise. It is
also consistent with \cref{prop:surj}, which concerns query algorithms that are correct on all record strings: no query
algorithm is used here, and the first layer is expanded exactly in literals. On records, the last two of three layers
get record degree $\widetilde O(n^{3/4})$ at inverse-polynomial error, which is compatible with the $O(n^{3/4})$
approximate degree of surjectivity~\cite{Sherstov18}.
\end{remark}

\fi
\else
\section{Attention constructions}\label[appendix]{app:topk}
\fi
\ificml\subsection{An adaptive upper bound for the upper half}\label[appendix]{app:icml-depth-sketch}

\begin{proposition}\label[proposition]{prop:icml-adaptive-depth-upper}
Fix $a>0$ and $\alpha\in(0,1]$. For all sufficiently large $N$ and each
$\alpha N\le T\le N$, a causal transformer with one top-$1$ head per layer,
unrestricted states, and depth $O_{a,\alpha}(\log n/\log\log n)$ predicts
$M_T$ with error at most $N^{-a}$ on independent uniform operands.
It uses only the given context positions and ignores the supplied prefix values.
\end{proposition}

\begin{proof}
All logarithms in the construction are base two. Put $k=2N-T$, $q=2^k$,
$D=2^N$, $w=\lceil(2a+3)\log N\rceil$, $d=k-w$ and $W=w+1$.
Eventually $d\ge0$. Number operand bits from the units place and retain
$S=\sum_{d\le i+j\le k}x_i y_j2^{i+j-d}$ modulo $2^W$.
Higher terms are divisible by $2q$, and the discarded lower terms have total
$E\le N(2^d-1)<\epsilon q$, where $\epsilon=N2^{-w}<1$.
If bit $w$ of $S$ is wrong, adding $E$ crossed a positive multiple of $q$,
so $sq\le XY<sq+\epsilon q$ for some integer $s\ge1$.

Write $R=q/D\ge1$. For fixed $X=x>0$ there are at most $xD/q$ eligible
positive multiples and at most $\epsilon q/x+1$ integers $y$ per interval.
This bounds the conditional probability by $\epsilon+x/q\le\epsilon+1/R$.
Alternatively, enlarge the range of $y$ to $0\le y<q$.
Modulo $q$, multiplication by $x$ hits each multiple of $g=\gcd(x,q)$
exactly $g$ times, giving at most $\epsilon q+g$ values with residue below
$\epsilon q$. Since $\sum_{0<x<D}\gcd(x,q)=ND/2$ and $x=0$ gives no error,
the second probability bound is $\epsilon R+N/(2D)$.
Splitting at $R=\epsilon^{-1/2}$ bounds the error by
$\epsilon+\sqrt{\epsilon}+N/(2D)\le N^{-a}$ for large $N$.

For the network put $f=\lfloor\sqrt{\log N}\rfloor$, $b=2^{2f}$ and
$m=\lceil N/b\rceil$. Partition each operand by significance into blocks of
$b$ bits, clipping the last. In $2f$ fixed-address layers pack each block
at its last physical position. At stride $s=1,2,\ldots,b/2$ (powers of two),
each block position prepends the string held $s$ places earlier in that
block's physical order, when such a position exists. This doubles its
available suffix, capped by its block prefix. The known ordering identifies
every bit. For an $X$-block starting at index $p$, the possible $Y$ indices
lie in $[d-p-b+1,k-p]$, of length $b+w\le2b$. At most three aligned
$Y$-blocks intersect it. One prefix position per $X$-block reads that block
and its at most three partner blocks in four layers. Its local update computes
the contribution to $S$ modulo $2^W$. This gives $m$ rows of $W$ bits.

Use parallel pattern tests and carry-save reduction as in~\cite{CSV84}.
While $r>2f$ rows remain, split them into $\lceil r/f\rceil$ groups and
pad the last with zeros. For each group and column place $2^f$ pattern
positions after its sources, then a count position after the patterns.
In one layer a pattern position scores self $1$, a source witnessing any
disagreement $2$, and all other keys $0$. It marks itself valid exactly
when self wins and all padded bits agree. The predicate on a source tests
all wires assigned to that source. In the next layer the count position
selects the unique marked pattern for its group and column and stores its
Hamming weight. Both layers use one head.

Put $h=\lceil\log(f+1)\rceil$. A group produces $h$ logical rows.
Bit $t$ of the count for column $c$ supplies column $c+t$ of row $t$,
discarding overflow and filling unwired cells with zero. These are fixed
addresses of bits in count states. The new rows have the same sum modulo
$2^W$, and their number is $r'=h\lceil r/f\rceil\le 2hr/f$.
Eventually $2h\le\sqrt f$, so there are $O(\log m/\log f)$ stages and
their input row counts sum to at most $2m$. Place successive stages in
increasing physical order. At termination copy the at most $2fW$
remaining bits into distinct positions and gather them by suffix doubling
in $\lceil\log(2fW)\rceil$ layers. The last context position reads the
result, sums the rows locally and outputs bit $w$.

Including a reserved readout, the total number of prefix positions used is
at most
\[
 m+\frac{4m}{f}W(2^f+1)+2fW+1=o(N),
\]
since $m\le2N/2^{2f}$ and $W=O_a(f^2)$. Hence all of them fit in
$T-1$ for sufficiently large $N$. The depth is
$2f+O(\log m/\log f+\log(2fW)+1)=O(\log n/\log\log n)$.
For bilinear scores, one initial positionwise update can encode each
(position, token) as a basis vector. Encode subsequent (position, state)
pairs likewise. Each finite score table is then bilinear, and a value map
can copy the winner to coordinates disjoint from the residual for the
positionwise update. This also implements all fixed-address reads.
\end{proof}
\fi

\ificml\else
\subsection{The depth picture}\label[appendix]{app:depthpicture}

The two remarks below are stated in the model of \cref{app:model}, but allowing a family with $L(n)$ layers. Neither
construction is new in technique; we record what they show about the depth of multiplication.

\begin{remark}[the depth bound is the right order for fixed routing]\label[remark]{rem:depth-upper}
In the model of \cref{app:model}, allowing $L(n)$ layers, there is a causal UHAT with one head per layer and
$\lceil\log_2(2N+1)\rceil$ layers that outputs $M_T$ exactly at every step $1\le T\le 2N$, for all operands and every
supplied prefix. Its attention pattern is fixed and does not depend on the input. Each layer doubles the block of
operand bits held at the readout position, and the last positionwise map multiplies. The construction is generic: the
same pattern computes any function of $(X,Y)$. It needs $\Theta(N)$ bits of state at one position, either one integer
coordinate of $2N+1$ bits or $4N+O(1)$ coordinates of $O(\log n)$ bits, and the last map must compute one bit of the
product of two $N$-bit integers. We do not know whether $O(\log n)$ bits of state per position suffice. Conversely, for any
fixed $H,k$, a top-$k$ model whose selected positions do not depend on the input has advantage at most
$2^{(1+Hk)^L/2-DT-1}$ at step $T\le N$. For $\alpha N\le T\le N$, inverse-polynomial advantage therefore needs $L\ge\log_{1+Hk}n-O(1)$, which
matches the construction for $H=k=1$ up to an additive constant. With input-dependent routing, exact correctness at one
step $\alpha N\le T\le N$ needs $\Omega(\log n/\log\log n)$ layers. This follows from \cref{lem:sidon}, a constant-time
CRCW-PRAM simulation of each layer, and Beame and H\aa{}stad~\cite[Cor.~4.1(a), 4.2]{BeameHastad89}. Whether
$O(\log\log n)$ layers can reach inverse-polynomial advantage is open.
\end{remark}

\noindent\emph{Sketches.} \emph{Construction.} Let $r_i=\min(i,2N+1)$. In layer $\ell$ the head at $i$ attends to the
unique position $\max(1,r_i-2^{\ell-1})$ (bilinear scores $(r_i-s)^2-(j-(r_i-s))^2$ built from the coordinates $i$, $i^2$
and $r_i$), and the update appends the word held there to the word held at $i$. After $\lceil\log_2(2N+1)\rceil$ layers
position $i$ holds all $2N$ operand bits (and, beyond position $2N$, the current token, which is discarded), and the last
map outputs the requested bit. This is the standard log-depth aggregation~\cite{LiuAGKZ23,SHT24}; Sanford, Hsu and
Telgarsky note that a single embedding that aggregates the whole input can compute any function of
it~\cite[footnote~2]{SHT24}. \emph{Fixed routing.} If the selected positions do not depend on the input, the output depends
on a fixed set of at most $(1+Hk)^L$ input positions; expanding it in characters on the formal cube, Parseval and
Cauchy--Schwarz give Walsh norm at most $2^{(1+Hk)^L/2}$, and \eqref{eq:walsh} with \cref{thm:main} gives the bound.
\emph{Exact prediction.} \Cref{lem:sidon} gives a projection under which every context token is a constant or one input
bit $z_i$ and $M_T$ is a majority of $r>n^{1/3}$ of them. A top-$k$ layer with fixed $H$ and $k$ is simulated in $O_{H,k}(1)$
steps of a PRIORITY CRCW PRAM with polynomially many processors and unbounded cells, which needs
$\Omega(\log n/\log\log n)$ steps for majority~\cite[Cor.~4.1(a), 4.2]{BeameHastad89}, even with infinitely many memory
cells. Beame and H\aa{}stad already list multiplication among the problems with this bound.

\begin{remark}[sublogarithmic depth for one product bit, with inverse-polynomial error]\label[remark]{rem:depth-gap}
Fix $a>0$ and $\alpha\in(0,1]$. For all sufficiently large $N$ and each $\alpha N\le T\le N$, there is a causal top-$1$
attention network in the model of \cref{app:model}, allowing $L(n)$ layers, with one head per layer and
$L(n)=O_{a,\alpha}(\log n/\log\log n)$ layers. On independent uniform $N$-bit operands it outputs $M_T$ with error at
most $N^{-a}$. It reads only the operand tokens. The $T-1$ prefix positions are used as scratch positions, and their tokens
are ignored.\footnote{Using $M_{T-1}$ to correct carries across even boundaries halves the error but is not needed.} The
routing is input-dependent, as it must be, since fixed routing needs $\log_2n-O(1)$ layers (\cref{rem:depth-upper}).
The network keeps $O(\log N)$ columns of the multiplication array near bit $T$. These determine $M_T$ except with
probability $N^{-a}$. It adds the per-block column sums with carry-save counters, each of which reads the unique matching
pattern of $f=\lfloor\sqrt{\log_2N}\rfloor$ bits by hard attention. Each position stores at most $2^{O(\sqrt{\log n})}$
bits. In the bilinear form this means width $\exp(2^{O(\sqrt{\log n})})$ with $0/1$ coordinates. The order
$\log n/\log\log n$ matches the classical CRCW-PRAM time of multiplication~\cite[Cor.~4.2]{BeameHastad89},~\cite{CSV84}.
All thresholds are asymptotic, and the network is shallower than the generic $\lceil\log_2(2N+1)\rceil$-layer
construction only for $\log_2N\ge484$. With $O(\log n)$ bits per position, and for the optimal depth between the
$\Omega(\log\log n)$ lower bound and this upper bound, the question is open.
\end{remark}

\noindent The threshold $484$ refers to the explicit layer count of the construction sketched next.

\noindent\emph{Sketch.} Let $q=2^{2N-T}$ and take $w=\lceil(2a+3)\log_2N\rceil$ guard columns. Write
$XY=4qK'+2^dS+E$, where $S$ collects the columns $d\le i+l\le 2N-T+1$ ($d=2N-T-w$) and $E$ the lower columns, so that
$0\le E<N2^{-w}q$. Then $\lfloor XY/q\rfloor\bmod 2$ is read off $S\bmod2^{w+2}$ unless $XY$ lies within $N2^{-w}q$ above
a multiple of $q$, which has probability $O(N2^{-w}+\sqrt{N2^{-w}}+N2^{-N})\le N^{-a}$ by counting residues. To add the
column sums in sublogarithmic depth, the network first packs blocks of nearby operand bits into single positions by
fixed doubling layers, and then runs stages of carry-save counting: groups of positions test all $2^f$ patterns of $f$
bits in parallel, and one input-dependent head returns the unique matching count. The packing takes $O(f)$ layers, each
stage takes $O(1)$ layers and shrinks the number of summands by a factor of order $f/\log f$, and so
$O(\log n/\log\log n)$ layers suffice in all; the scratch positions are the prefix
positions, which number $T-1$. This is the classical parallel technique: test all patterns, then carry-save
reduce~\cite{CSV84}.

\subsection{Softmax attention with margins}\label[appendix]{app:softmax}

\begin{corollary}[softmax with margins]\label[corollary]{cor:softmax}
Fix $\alpha\in(0,1]$, $B\ge1$, $d\ge0$ and $A>1+d$, independent of length and depth. Consider an $L$-layer causal
softmax transformer with at most $H$ heads per layer and common temperature $0<\tau\le1/(A\ln n)$. Its raw score maps
(in the two states), value maps, layer updates (including residual, normalisation and MLP, as maps of the old state and
the tuple of head outputs) and individual output logits are $B$-Lipschitz in the max norm, and its values have norm at
most $B$. On every true context of step $\alpha N\le T\le N$, at each soft layer, head and query $i>k$, suppose the
$k$-th largest raw score exceeds the $(k+1)$-th by at least $1$. Suppose the two highest soft output logits differ by at least $n^{-d}$.

For all sufficiently large $n$, uniformly for $1\le L\le\log_2\ln n$, its greedy prediction agrees on every true
context with that of a top-$k$ model of the same depth and head counts. Thus the fixed-depth bounds of
\cref{thm:topk-depth,cor:topk-three} transfer. With $J$ from \cref{rem:topk-explicit}, $|\E[\hat M_T\hat O_T]|$ is at most
$2\exp(-n^{3/4})$ for $1\le L\le J$, also for varying $H,k$. For each fixed $p>0$, advantage at least $n^{-p}$
requires $L>J$ for all sufficiently large $n$. How large $n$ must be depends only on $A,B,d$ for the agreement, and only
on $p,\alpha,D,A,B,d$ for the last two claims.
\end{corollary}

\noindent The corollary is a conditional transfer: it says nothing about softmax transformers without these margins,
the regime used in practice. Removing the output margin or the score gap allows exact prediction of $M_T$
(\cref{rem:softmax-sharp,rem:averaging}).

The mechanism is that of Yang, Strobl, Chiang and Angluin~\cite[Lemma~25, Thm.~26]{YSCA24}: with a score gap and
inverse temperature of order $\log n$, each softmax layer stays close to its hardened version, and the errors compose.
We give the version needed for \cref{cor:softmax}. Its differences from~\cite{YSCA24} are that the gap is required only on
true contexts, at a top-$k$ cutoff rather than among tied maxima, with Lipschitz constants that may grow like $n^b$, and
that greedy and sampled outputs are treated separately.

\paragraph{Truncation.} Fix a layer, a head and a query $i$, and let $p_j\propto e^{s_j/\tau}$ be the softmax weights on
the keys $j\le i$. Let $r=\min(k,i)$, let $S$ be the first $r$ keys in the hard order (score, then the fixed priority), and
let $q$ be $p$ conditioned on $S$. Then $\mathrm{TV}(p,q)=p(S^c)=:t(p)$, and if the $r$-th score exceeds the $(r+1)$-th
by $\Delta$, then $t(p)\le\frac{i-r}re^{-\Delta/\tau}\le ne^{-\Delta/\tau}$, the truncation bound
of~\cite[Lemma~B.7]{EGKZ22}. The \emph{hard network} replaces every softmax by its truncation $q$, computed on the hard
network's own incoming states. It is a top-$k$ model (\cref{def:topk}): the update sees the old query state and the
ordered selected pairs, so it can recompute their scores and the conditional weights.

\paragraph{The recurrence.} Measure states in the max norm. At layer $\ell$ let $U_\ell$ be a Lipschitz bound of the update
(including residual, normalisation and MLP), let values differ by at most $V_\ell$ times the distance of their arguments
and have norm at most $B_\ell$ on the hard run, let scaled scores $s_j/\tau$ differ by at most $D_\ell$ times that
distance, and let every soft head have $t(p)\le\eta_\ell$ on true contexts. These bounds are needed only along the soft
and the hard run on true contexts. Put $a_\ell=U_\ell\max\{1,V_\ell+2B_\ell D_\ell\}$, $b_\ell=2U_\ell B_\ell\eta_\ell$,
$E_0=0$ and $E_\ell=a_\ell E_{\ell-1}+b_\ell$. Then on every true context the soft and the hard states after layer
$\ell$ differ by at most $E_\ell$. Indeed, softmax is $\frac12$-Lipschitz from the max norm on logits to total
variation, so the full softmax laws $p$ (soft run) and $p'$ (hard run) satisfy $\mathrm{TV}(p,p')\le D_\ell d/2$ when the
incoming states differ by $d$. Since $t(\cdot)$ is the least mass outside $r$ keys, $t(p')\le t(p)+\mathrm{TV}(p,p')$,
and so $\mathrm{TV}(p,q')\le\eta_\ell+D_\ell d$ for the truncation $q'$ of $p'$; the two top sets may differ. Swapping the
values costs $V_\ell d$ and swapping the laws costs $2B_\ell\,\mathrm{TV}(p,q')$, and the update multiplies by $U_\ell$.
If the logits are $R$-Lipschitz, each moves by at most $RE_L$, so the greedy decisions agree wherever the top two soft
logits differ by more than $2RE_L$; for a sampled output with $R$-Lipschitz signed mean, the correlations differ by at
most $RE_L$.

\begin{proof}[Proof of \cref{cor:softmax}]
Write $u=\ln n$ and $x=1/\tau$. With all bounds at most $B$, $D_\ell\le Bx$, $\eta_\ell\le ne^{-x}$,
$a_\ell\le B^3(1+2x)$ and $b_\ell\le2B^2ne^{-x}$. Unrolling from $E_0=0$ gives
$2RE_L\le4LB^{3L}n(1+2x)^{L-1}e^{-x}$.
For $L\le\log_2 u$ and all large $u$ we have $Au>L$. The logarithmic derivative
$2(L-1)/(1+2x)-1$ is negative for $x\ge Au$, so $2RE_L\le F n^{1-A}$ with
$F=4LB^{3L}(1+2Au)^{L-1}$. Since $4L\le4^L$ and $1+2Au\le(1+2A)u$,
\[
 \ln F\le\frac{\ln u}{\ln2}\bigl(\ln u+\ln(4B^3(1+2A))\bigr)
        =O_{A,B}((\ln\ln n)^2)=o(\ln n).
\]
In particular $\ln F<(A-1-d)u$ eventually, uniformly here. Then $2RE_L<n^{-d}$, so the greedy outputs agree on
every true context. Every fixed depth and the entire range $1\le L\le J$ are contained in $L\le\log_2u$ for large $n$.
Apply \cref{thm:topk-depth,cor:topk-loglog} to the hard model. The transfer uses the exact maps of \cref{def:topk}.
At fixed depth the same calculation allows $B=n^b$ if $A>1+3bL+d$. With $B$ fixed, a sampled readout with
$R$-Lipschitz signed mean incurs additive correlation error at most $RE_L=O_{L,B,A}(n^{1-A}(\ln n)^{L-1})$.
On a given query the unit gap can also occur at any cutoff $r\le k$ with an excluded key. The first $k$ keys
contain the first $r$, hence the same tail bound holds.

\emph{The counterexamples.} (i) Let $Q\ge2$, give one head the scores $0$ at position $1$ and
$-1-(j-2)\log_Q2$ at $j\ge2$, temperature $1/\ln Q$, and values equal to the token bits. The weights are proportional to
$1$ and $Q^{-1}2^{-(j-2)}$, so the raw gap is at least $1$ and the tail is below $2/Q$; at the last position the
denominator does not depend on the input, and the head output $b_1+Q^{-1}\sum_{j\ge2}2^{-(j-2)}b_j$ encodes the whole
context injectively. A piecewise-linear readout recovers $M_T$ exactly. Rescaled to be $1$-Lipschitz, it has an output
margin far below $n^{-d}$. (ii) Give the first $k$ keys score $1$ and the others $0$, put $e^{1/\tau}=n^A$, and give
key $j$ the value $b_je_j$ (a first layer of $2$-Lipschitz maps produces these values from additive embeddings). The
head output is then an injective code of the context with codes separated by $r_0=1/(kn^A+n-k)$, and a $1$-Lipschitz
readout $g$ equal to $\pm r_0/2$ on the codes, with the sign of $\hat M_T$, and logits $(g,0)$ predict $M_T$
perfectly with output margin $r_0/2\approx n^{-A}/(2k)$. For $A<d$ the output gap is at least $n^{-d}$ for large $n$, so the hypothesis $A>1+d$ cannot be
replaced by $A<d$; unit-temperature sampling with logits $(g,0)$ gives advantage
$(1+e^{-r_0/2})^{-1}-1/2=\Theta(r_0)$, so polynomially small tails of this form
can leave a polynomial sampling advantage.
\end{proof}

\begin{remark}[sharpness of \cref{cor:softmax}]\label[remark]{rem:softmax-sharp}
For sampled output the advantage exceeds the top-$k$ bound by at most $O_{L,B,A}(n^{1-A}(\ln n)^{L-1})$. At fixed depth,
bounds $B=n^b$ are allowed if $A>1+3bL+d$; the depth clause needs $B$ independent of $n$ and $L$. The gap may also sit
at any cutoff $r\le k$ instead of the $k$-th. The condition cannot be weakened to $A<d$: there is such a model, with all
maps $2$-Lipschitz, that predicts $M_T$ perfectly. Without the output-margin or Lipschitz-readout hypothesis, one softmax
head with a unit score gap and arbitrarily small attention tail predicts $M_T$ exactly. These are the counterexamples
(i) and (ii) in the proof above.
\end{remark}

\noindent That one soft-attention head can encode a whole context, and that unbounded sensitivity defeats the classical
soft-attention bounds, has antecedents: Hahn's Lipschitz bound for soft attention~\cite[Lemma~5]{Hahn20},
Chiang and Cholak on layer normalisation~\cite{ChiangCholak22}, and the one-head memorisation of Kajitsuka and
Sato~\cite{KajitsukaSato24}.

\fi
\subsection{The exact averaging construction}\label[appendix]{app:averaging}

\begin{remark}[averaging and softmax attention escape the bound]\label[remark]{rem:averaging}
At fixed $L,H,k$ the advantage of top-$k$ hard attention is \ificml$\exp(-\Omega(n))$\else$\exp(-n^{\Omega(1)})$\fi (\cref{thm:topk-depth}, with its
hypotheses: fixed $L$, $H$ and $k$, a fixed positional tie priority, $\alpha N\le T\le N$). Fixed-depth softmax attention
with $O(\log n)$-bit precision can predict it exactly. This is Feng et al.'s construction for
multiplication~\cite[Thm.~5.3 and App.~E.3]{Feng24}, taken in base $2$ with output length $2N$. That construction already
emits the product most significant bit first. It reads each digit from the operands and the output position alone, so it
ignores the earlier product bits in the context. A routine exact form of it (\cref{app:averaging}): with zero attention scores, one head per layer computes the uniform mean $1/i$ over the visible
positions. That mean is exact for averaging hard attention and for softmax, and it survives rounding to
$\lceil\log_28N\rceil$ fractional bits. On the same context $X\,Y\,{=}\,M_{<T}$, a causal transformer then computes every
$M_T$ exactly with nine layers, one such head per layer, ReLU feed-forward maps of width $O(N^2)$, and every activation a
rational of $O(\log n)$ bits. Its advantage is therefore $1/2$. The only use of attention is to gather all $2N$ operand
bits at one position. This is exactly what a fixed number of top-$k$ heads cannot do at fixed depth, and it is why the separation is
about averaging rather than about arithmetic.
\end{remark}

The construction of \cref{rem:averaging} is Feng et al.'s multiplier~\cite[App.~E.3]{Feng24} in base $2$, written out
exactly. Their Theorem~5.3 covers every base $p\ge2$ and output length $l\le2n$, and their task (\S3) lists the product
digits from the most significant one. We use the transformer of~\cite[\S2.2]{YSCA24}: word plus positional embedding, one head per layer with the
inclusive causal mask, residual connections, and feed-forward maps $c+W_2R(W_1c+b_1)+b_2$ with $R(s)=\max(0,s)$. With
$Q=K=0$ every score is $0$, so averaging hard attention and softmax both give weight exactly $1/i$ to every key $j\le i$.
The context is $X\,Y\,{=}\,M_{<T}$ with each operand most significant bit first; one network per $N$ serves every
$1\le T\le2N$. Let $m=\lfloor\log_2N\rfloor+1$, $B=2^m$ (so $N<B\le2N$), $G=\lceil2N/m\rceil$ and
$H_t(v)=R(v-t+1)-R(v-t)=[v\ge t]$ for integers $v$.
\begin{enumerate}
\item \emph{Tag.} Positional flags mark which operand slot a position holds; a ReLU writes the operand bit into that slot.
\item \emph{Gather.} The only nonzero value map averages the tags: slot $h$ becomes $0$ or $1/i$, and
$b_h=1-R(1-8Na_h)$ recovers the bit. From here on attention is not used.
\item \emph{Block sums.} With $d_{ab}=R(x_a+y_b-1)$, write $S_j=\sum_{\lfloor(a+b)/m\rfloor=j}2^{(a+b)\bmod m}d_{ab}$ for
$0\le j<G$, so that $XY=\sum_jS_jB^j$ and $0\le S_j\le N(B-1)$.
\item \emph{Split.} $Q_j=\sum_{t\le N}H_{tB}(S_j)=\lfloor S_j/B\rfloor$ and $u_j=S_j-BQ_j+Q_{j-1}$, so
$0\le u_j\le B+N-2$.
\item \emph{Generate and propagate.} $g_j=H_B(u_j)$ and $p_j=H_{B-1}(u_j)$.
\item \emph{Carry witnesses.} $w_{sj}=R(g_s+\sum_{s<h\le j}p_h-(j-s))$, which is $1$ exactly when block $s$ generates a
carry that every block up to $j$ propagates.
\item \emph{Carries.} $c_j=\bigvee_{s\le j}w_{sj}$ and $z_j=u_j+c_{j-1}<2B$; the base-$B$ digits are $z_j\bmod B$.
\item \emph{Digit bits.} Bit $r$ of $z_j\bmod B$ is the alternating hinge sum $\sum_{\ell}(-1)^{\ell+1}H_{\ell2^r}(z_j)$.
\item \emph{Read.} A positional flag selects the bit of weight $2^{2N-T}$; every old coordinate is cancelled, so the output
is exactly an output-symbol embedding.
\end{enumerate}
Every activation is a rational with denominator dividing $i\le4N$ and magnitude below $(32(N+1))^6$, so $O(\log n)$
bits suffice. Rounding $1/i$ to $\lceil\log_28N\rceil$ fractional bits keeps the decoding in step~2 exact, since the
rounded weight is at least $3/(16N)>1/(8N)$; all later steps are exact on that grid. The width is $O(N^2)$. For example,
with $N=3$ and $X=Y=7$: $B=4$, $S=(5,7,1)$, $u=(1,4,2)$, the middle block generates a carry, and the digits $(1,0,3)$
give $110001_2$. We checked the construction by a literal simulation with explicit integer matrices, exhaustively for
$N\le5$ and on sampled operands up to $N=30$, under both attention rules and both arithmetic modes; deleting a carry or
reading the digits least significant first is detected.

\ificml\else
\subsection{Three layers, with $k$ allowed to grow}\label[appendix]{app:topkk}

At three layers \cref{thm:topk-depth} gives a bound that stays useful when $k$ grows with $n$.

\begin{corollary}[three layers, $k$ allowed to grow]\label[corollary]{cor:topk-three}
Fix $\alpha\in(0,1]$ and $H\ge1$, and let $\alpha N\le T\le N$. Consider three top-$k$ layers (\cref{def:topk}) with at
most $H$ heads per layer, run on the true context of length $n$, where $k=k(n)$ may vary with $1\le k\le\ln n$. There are
$c_1>0$ and $n_0$, depending only on $\alpha$ and $H$, such that $|\E[\hat M_T\hat O_T]|\le\exp(-c_1n/(\ln n)^{12})$ for
$n\ge n_0$. For fixed $k$ the bound is $\exp(-\Omega(n/(\ln n)^6))$.
\end{corollary}

\begin{proof}
Both conclusions follow from \cref{rem:topk-explicit}. With $L=3$ its exponent is
$t=n/(R_0^4K_{\mathrm{set}}^6)$, and the feasibility condition is $n\ge R_0^4K_{\mathrm{set}}^7(1+\ln6/K_{\mathrm{set}})$, with $K_{\mathrm{set}}=4(1+Hk)\ln(4n)$.
At fixed $H,k$ this gives $\exp(-\Omega(n/(\ln n)^6))$.
At fixed $H$ and $1\le k\le\ln n$, we have $K_{\mathrm{set}}\le8(H+1)(\ln n)^2$ for large $n$, so the condition holds and
$t=\Omega_{\alpha,H}(n/(\ln n)^{12})$. Since $t\to\infty$, the factor $2$ in $2e^{-t}$ is absorbed into $c_1$.
\end{proof}

\noindent Head counts may differ between layers: take $H$ to be the largest. With $k=1$ and bilinear scores the
model is the UHAT of \cref{app:model}, so the corollary also improves the exponent $1/(13+4H_1)$ of \cref{thm:three}.
\fi

\ificml\else
% Appendix section (moved from the main body in the structural refocus): constant-depth circuits that read the true
% prefix. Input from appendix.tex. Proofs and further results: app-ac0.tex.
\section{Constant-depth circuits, and the limits of the method}\label[appendix]{sec:ac0}

Throughout this section $Q=2^N$, $X,Y$ are independent and uniform on $\{0,\dots,Q-1\}$, $1\le T\le N$,
$P=(M_1,\dots,M_{T-1})$ is the true prefix and $B=(-1)^{M_T}$. A predictor is a $\{0,1\}$-valued function $C$ of the
formal context $V=(X,Y,P)\in\{0,1\}^n$, $n=2N+T-1$, and its \emph{advantage} is
$\Pr[C(V)=M_T]-\frac12=\frac12\E[B(-1)^{C(V)}]$; we bound its absolute value. Circuits have unbounded fan-in AND and
OR gates, NOT gates only at the inputs, depth $d$ and size $s$ (the number of AND and OR gates). In asymptotic
statements $\alpha\in(0,1]$ is fixed and $\alpha N\le T\le N$.

\paragraph{What we prove.} \Cref{thm:main} controls exactly the correlations of $M_T$ with the functions
$f(X)g(Y)\chi_S(P)$. The usual route from such bounds to circuit lower bounds, Fourier concentration of
$\mathrm{AC}^0$~\cite{Tal17}, concerns the uniform measure, while the true context is supported on a $2^{-(T-1)}$
fraction of the formal cube. Any two of the three blocks are harmless: $(X,Y)$ is uniform and $(X,P)$ has density at
most $1.5\cdot2^N/X$ (\cref{lem:ac0-density}). We are not aware of earlier bounds for predicting top product bits from
the true prefix by constant-depth circuits; the tools are standard. Circuits of size $N^u$ for a range of $u>1$ are
handled by an arithmetic estimate that does not use \cref{thm:main} (\cref{cor:ac0-full}). \Cref{app:ac0-results}
adds circuits that read one operand and the entire prefix, or both operands and all but a short top window of the
prefix~\cite{Braverman10,Tal17,HarshaSrinivasan19}, DNFs with $o(T^2)$ terms of any width, decision trees with fewer
than $2^{\Omega(T)}$ leaves and predictors with $c_\alpha N$ operand queries
(\cref{thm:ac0-twoblock,thm:ac0-window,thm:ac0-dnf,thm:ac0-query,prop:ac0-small}). Polynomial-size DNFs and CNFs
whose terms mix bits of $X$, $Y$ and the prefix have advantage at most $(\log N)^{O(1)}/N$ (\cref{thm:ac0-mixeddnf}
in \cref{app:ac0-mixed}). Mixed circuits of polynomial size and depth three or more are open, and the rest of this
section explains why: three unrestricted
features can predict $M_T$ (\cref{prop:ac0-barrier}), we characterise, up to a factor $2$, what \cref{thm:main} can prove
(\cref{prop:ac0-duality}), and the route we use for transformers would need new circuit lower bounds
(\cref{prop:ac0-uniform}).

\paragraph{Circuits of size $N^u$.} By \cref{thm:ac0-operand} (\cref{app:ac0-results}), the advantage is
$2^{-N^{\Omega(1)}}$ even if an arbitrary function of the prefix selects a depth-$d$ circuit of size $O(N^u)$ on the
operand bits. Fixing $P$ in a depth-$d$ circuit of size $m$ on $(X,Y,P)$ leaves a circuit on $(X,Y)$ of depth at most $d$ and size at
most $m$, and fixing $P$ in a width-$w$ DNF leaves a width-$w$ DNF. Hence:

\begin{corollary}[circuits on the whole context]\label[corollary]{cor:ac0-full}
Let $d\ge2$, $b=2^{d-1}$ and $1\le u<b/(b-1)$. Every depth-$d$ circuit of size $O(N^u)$ that reads $X$, $Y$ and the
whole true prefix has advantage at most $2^{-N^\gamma}$ for every $\gamma<b-(b-1)u$ and large $N$, and
$2^{-\Omega(N/(\log N)^{A_d})}$ at linear size, uniformly over $\alpha N\le T\le N$. Every DNF or CNF of width $w$
on $(X,Y,P)$, of any size, has advantage $2^{-\Omega_{\alpha,w}(N)}$. In particular depth-$3$ circuits of size
$N^{4/3-\eta}$, $0<\eta\le1/3$, have advantage $2^{-N^{\Omega(1)}}$.
\end{corollary}

\subsection{A barrier, and what \cref{thm:main} can prove}

\begin{proposition}[three features predict]\label[proposition]{prop:ac0-barrier}
For all $6\le T\le N$ there are functions $f(X),g(Y),h(P)$ with values in $\{-1,1\}$ such that
\[
\bigl|\E\,B\,f(X)g(Y)h(P)\bigr|\ge\tfrac3{8192}\sin\tfrac18.
\]
\end{proposition}

\noindent \emph{Idea.} With $t=2^T/32$, the phases $(X/Q)^{it}$ and $(Y/Q)^{it}$ multiply to $(XY/Q^2)^{it}$ because
$\ln xy=\ln x+\ln y$, and the prefix supplies the phase at the midpoint of its cell; the remaining phase changes sign
exactly where $M_T$ does. So \cref{thm:main} (an arbitrary prefix feature is not allowed) and \cref{thm:ac0-arith} (an
arbitrary $Y$-feature is not allowed) are both compatible with \cref{prop:ac0-barrier}, and a lower bound for general
$\mathrm{AC}^0$ must use that the features are computed by small circuits.

The next two statements are about the method. Let $\Pi$ be the set of functions $f(X)g(Y)\chi_S(P)$ with real
$|f|,|g|\le1$ and $S\subseteq[T-1]$, and for a real function $p$ of $(X,Y)$ let $\|p\|_\Pi$ be the least total weight
$\sum_i|c_i|$ of a representation $p=\sum_ic_i\varphi_i$ with $\varphi_i\in\Pi$. \Cref{thm:main} says exactly that
$|\E[B\varphi]|\le2^{-DT}$ for every $\varphi\in\Pi$.

\begin{proposition}[what \cref{thm:main} can prove]\label[proposition]{prop:ac0-duality}
Fix $N,T$, a function $\widetilde C$ of $(X,Y)$ and $\delta>0$. Let $\Sigma_\delta$ be the set of functions
$\sigma'$ of $(X,Y)$ with values in $[-1,1]$ and $|\E[\varphi\sigma']|\le\delta$ for every $\varphi\in\Pi$. Then
\[
\max_{\sigma'\in\Sigma_\delta}\E[\widetilde C\sigma']=\min_p\bigl\{\E|\widetilde C-p|+\delta\|p\|_\Pi\bigr\},
\]
the minimum over all real functions $p$ of $(X,Y)$.
\end{proposition}

\noindent With $\widetilde C=(-1)^{C(V)}$ and $\delta=2^{-DT}$: an argument that uses about $M_T$ only the inequalities
of \cref{thm:main} and $|B|\le1$ proves for $C$ exactly the bound on the right, and nothing better, since a
randomised bit $\sigma'\in\Sigma_\delta$ satisfies all those inequalities and is predicted by $C$ with that
correlation. If complex $f,g$ are allowed in $\Pi$, as in \cref{thm:main}, the constraint set lies between
$\Sigma_{\delta/2}$ and $\Sigma_\delta$, so the characterisation holds up to a factor $2$ in $\delta$.

\begin{proposition}[the uniform route needs new circuit lower bounds]\label[proposition]{prop:ac0-uniform}
Fix $d$, $c$ and $\varepsilon<\frac12$. Suppose every depth-$d$ circuit $f$ of size $n^c$ on $n$ variables has a real
$p$ with $|f-p|\le\varepsilon$ on $\{0,1\}^n$ and Walsh norm at most $2^{w(n)}$. Then for every $m$, every such $f$ and
every affine map $\Lambda:\mathbb F_2^{2m}\to\mathbb F_2^n$,
$\Pr_{x,y}[f(\Lambda(x,y))=\mathrm{IP}_m(x,y)]\le\frac12+\varepsilon+(\frac12+2^{w(n)})2^{-m}$. In particular, if
$w(n)\le\gamma n$ then depth-$d$ circuits of size $n^c$ over at most $(1-o(1))m/\gamma$ parities of the inputs do not
compute $\mathrm{IP}_m$ with advantage $\varepsilon+o(1)$.
\end{proposition}

\noindent The argument is that of Bun, Kothari and Thaler~\cite[Prop.~26 and Thm.~8]{BKT19}, with Walsh norm in place
of the number of monomials. The transformer method of \cref{sec:transformers} needs such approximants with $\gamma$ of
order $\alpha D$, so for polynomial-size $\mathrm{AC}^0$ it would yield inner-product lower bounds against
polynomial-size $\mathrm{AC}^0$ over linearly many parity gates, which as far as we know goes beyond the published
bounds for $\mathrm{AC}^0\circ\oplus$ (\cref{app:ac0-method}).

% Appendix: further results and proofs for the constant-depth section (sec-ac0.tex).
\section{Further results for constant-depth circuits}\label[appendix]{app:ac0-results}

This appendix states the circuit results that are not stated in \cref{sec:ac0}, in the notation of
\cref{sec:ac0}; proofs are in \cref{app:ac0}. The first group rests on an arithmetic estimate and not on
\cref{thm:main}; the second uses \cref{thm:main}.

\subsection{Operand circuits chosen by the prefix}

The first theorem needs neither \cref{thm:main} nor any structure of the prefix map. Its constants depend only on
$\alpha$, $d$, $u$ and $\gamma$ (or $w$).

\begin{theorem}[operand circuits chosen by the prefix]\label{thm:ac0-operand}
Fix $d\ge2$, put $b=2^{d-1}$, and let $1\le u<b/(b-1)$. Suppose that for each value $p$ of the prefix an arbitrary
depth-$d$ circuit $F_p$ of size $O(N^u)$ on the $2N$ operand bits is used, the choice of $F_p$ being an arbitrary
function of $p$. Then the predictor $F_P(X,Y)$ has advantage at most $2^{-N^\gamma}$ for every $0<\gamma<b-(b-1)u$
and all large $N$, and at most $2^{-\Omega(N/(\log N)^{A_d})}$ at linear size, for a constant $A_d$ depending only on
$d$; both bounds hold uniformly over $\alpha N\le T\le N$. The same holds for DNFs and CNFs of width $w$ with any
number of terms, with rate $2^{-\Omega_{\alpha,w}(N)}$, and rate $2^{-N^{1-o(1)}}$ if $w=o(\sqrt{\log N})$. The same
bounds hold for the average over $p$, weighted by the prefix probabilities, of the best conditional advantage of a
circuit of the class.
\end{theorem}

\noindent The proof (\cref{app:ac0-operand}) combines pointwise approximate-degree bounds on the operand cube (Bun,
Kothari and Thaler~\cite[Thm.~15]{BKT19}, and Sherstov~\cite[Thm.~5.1]{Sherstov18} for bounded width) with an
estimate in which the prefix test is arbitrary and the operand $Y$ enters through one additive character:

\begin{theorem}[an arithmetic estimate]\label{thm:ac0-arith}
There is an absolute constant $C$ such that for all $N\ge2$, $1\le T\le N$, all integers $b$ and all functions
$f,h$ with $|f|,|h|\le1$,
\[
\bigl|\E\,B\,h(P)f(X)\,\mathrm e(bY/Q)\bigr|\le C(N+1)2^{-T/16},\qquad\mathrm e(v)=e^{2\pi iv}.
\]
\end{theorem}

\noindent \emph{Idea.} $Bh(P)$ is a function $H$ of $XY/Q^2$ that takes opposite values on the two halves of each
prefix cell; pairing them kills the low Fourier coefficients of $H$, $|\widehat H(k)|\le\pi|k|/2^T$. The high
coefficients meet the bilinear sums $\E f(X)\mathrm e(bY/Q+kXY/Q^2)$, which are bounded by summing the geometric series
in $Y$. A Fej\'er smoothing and a separate count of products near cell boundaries finish the proof. To pass from one
character to a monomial in the bits of $Y$ we use that $j$ marked binary digits have additive Fourier norm at most
$[C(1+N/j)]^j$ on $\mathbb Z/Q$ (\cref{lem:ac0-digits}). \Cref{thm:ac0-arith} and \cref{thm:main} are incomparable:
here the prefix test is arbitrary but one operand enters through a character, there both operand tests are
arbitrary but the prefix enters through a parity.

\subsection{Further consequences of the arithmetic estimate and of \cref{thm:main}}

\begin{theorem}[DNFs of any width]\label{thm:ac0-dnf}
There is an absolute constant $c>0$ such that every DNF or CNF on $(X,Y,P)$ with $s\ge1$ terms or clauses, of
arbitrary width, has advantage $O(2^{-c\min(T,T^2/s)})$, for all $N\ge2$ and $1\le T\le N$.
\end{theorem}

\noindent So $s=o(T^2)$ terms give vanishing advantage and $s=O(T)$ terms exponentially small advantage. The constant
$c$ depends only on the constant $D$ of \cref{thm:main} and on the absolute constant of the small-error approximation
of OR; it is not explicit. In the range $N\ge400$, $64\le T\le N$ the proof may use $D=1/16$. The proof composes a
univariate approximation of OR with the average of the term indicators and bounds the Walsh norm of the result by
Lagrange interpolation, without a $\log s$ loss (\cref{app:ac0-dnf}).

\begin{theorem}[few operand queries]\label{thm:ac0-query}
There is $c_\alpha>0$ such that every predictor that receives the full prefix, performs arbitrary computation on it,
and makes at most $c_\alpha N$ adaptive queries to individual operand bits, has advantage $2^{-\Omega(T)}$ for
$\alpha N\le T\le N$. The query positions and the output may depend arbitrarily on the prefix and on earlier answers,
and the predictor may be randomised.
\end{theorem}

\noindent We do not know the largest $c$ for which $cN$ operand queries still give vanishing advantage; reading all
$2N$ operand bits determines $M_T$. Exact computation at $T=N$ shows that the unrestricted
predictor that reads all of $X$ and the prefix, $\mathrm{sign}\,\E[B\mid X,P]$, has advantage
$\frac12\E|\E[B\mid X,P]|$ between $0.0823$ ($N=6$) and $0.0575$ ($N=12$), decreasing slowly
(\texttt{checks/ac0\_bayes\_TN.py}). This is an observation at small $N$, not a theorem; it indicates that the product
structure $f(X)h(P)$ in \cref{thm:ac0-arith} cannot be replaced by an arbitrary function of $(X,P)$.

\begin{theorem}[one operand and a short message about the other]\label{thm:ac0-message}
Let $d\ge2$, let $s\ge2$ be polynomial in $N$, and let $c'>0$ be small enough in terms of $d$ and $D$.
\textup{(a)} Let $j(Y)$ be an arbitrary message with $L$ values, and suppose that after seeing $X$ and $j(Y)$ the
predictor chooses, arbitrarily, a depth-$d$ circuit of size at most $s$ on the $T-1$ prefix bits. If
$\log_2L\le c'T/(\log_2(2s))^{d-1}$, the advantage is $2^{-\Omega_d(T/(\log_2(2s))^{d-1})}$.
\textup{(b)} Suppose an arbitrary function of $P$ selects one of $L$ depth-$d$ circuits of size at most $s$ on all
$2N$ operand bits. If $\log_2L\le c'_{\alpha}T/(\log_2(2s))^{d-1}$ for $c'_\alpha>0$ small in terms of $\alpha,d$,
the advantage is $2^{-\Omega_{\alpha,d}(T/(\log_2(2s))^{d-1})}$.
\end{theorem}

\subsection{Two of the three blocks}

The last results use \cref{thm:main} together with Tal's Fourier tail bound~\cite[Thm.~18]{Tal17} (quoted in
\cref{app:quotes}). All their constants are asymptotic: with Tal's $c_d=60d\cdot192^{d-1}$ they bite only for very
large $T$. Let $\kappa=c_d(\log_2s)^{d-1}$.

\begin{lemma}[two blocks are nearly uniform]\label[lemma]{lem:ac0-density}
For $2\le T\le N$, every $x\ge1$ and every prefix value $a$, $\Pr[X=x,P=a]\le1.5\,(2^N/x)\,2^{-(N+T-1)}$. The same
holds with $X$ and $Y$ exchanged.
\end{lemma}

\begin{theorem}[one whole operand and the entire prefix]\label{thm:ac0-twoblock}
Let $2\le T\le N$, $n_1=N+T-1$, $s>1$, $k=\lfloor DT/(4\log_2(n_1+1))\rfloor$ and $\beta=2^{-k/(4\kappa)}$.
\begin{enumerate}
\item[(1)] Every depth-$d$, size-$s$ circuit that reads all of $X$ and the entire true prefix (but not $Y$) predicts
$M_T$ with advantage at most $4\beta+2^{-DT/2}=\exp(-\Omega(DT/(c_d(\log s)^{d-1}\log N)))$.
\item[(2)] The same holds, with bound $(2+2^{u+1})\beta+2^{-DT/2}$, if the predictor also receives an arbitrary
$u$-bit function $\psi(Y)$ and uses, for each value of $\psi$, its own depth-$d$, size-$s$ circuit on $(X,P)$; this is
nontrivial for $u=O(T/\mathrm{polylog}\,N)$.
\item[(3)] If $C=\phi(A(X,P),B'(Y,P))$, where $A$ and $B'$ have $r_1$ and $r_2$ output bits, every event $\{A=v\}$ and
$\{B'=w\}$ is decided by a depth-$d$, size-$s$ circuit, and $\phi$ is arbitrary, the advantage is at most
$(2+2^{r_1+r_2})\beta+2^{-DT/2}$.
\end{enumerate}
The same holds with $X$ and $Y$ exchanged.
\end{theorem}

\noindent \emph{Idea.} By \cref{lem:ac0-density}, after removing the event $X<2^{N-j}$ the true law of $(X,P)$ has
density at most $1.5\cdot2^j$ with respect to the uniform law, so Tal's uniform-measure tail bound transfers to the
true context. The low-degree part is a combination of functions in $\Pi$, which \cref{thm:main} handles. The proof
fails for a single gate that touches bits of $X$, of $Y$ and of the prefix.

Let $r_d(s,\varepsilon)$ be the least $r$ such that every $r$-wise independent distribution $\varepsilon$-fools every
depth-$d$ circuit of size $s$; by Harsha and Srinivasan~\cite{HarshaSrinivasan19},
$r_d(s,\varepsilon)\le(\log s)^{O(d)}\log(1/\varepsilon)$ (earlier bounds are due to Braverman~\cite{Braverman10} and
Tal~\cite{Tal17}).

\begin{theorem}[all but a top window of the prefix]\label{thm:ac0-window}
Let $1\le K\le T\le N$, let $W\subseteq[K-1]$ with $|W|=u$, and let $C$ be a $\{0,1\}$-valued function of $X$, $Y$,
the prefix bits $M_W$ and $M_K,\dots,M_{T-1}$ such that for every fixed value $v$ of $M_W$ the function $C_v$ of the
remaining $n_K=2N+T-K$ bits is computed by a depth-$d$ circuit of size $s$ (the dependence on $M_W$ is arbitrary).
Let $\varepsilon_0>0$ and $r=r_d(s,\varepsilon_0)$. Then
\[
\bigl|\E[(-1)^{C}B]\bigr|\le2^{-DT}+2^{u+1}\bigl[2\varepsilon_0+3(1+\varepsilon_0)(n_K+1)^r2^{-DK}\bigr].
\]
\end{theorem}

\noindent With $u=0$ and $s$ polynomial in $N$ this lets the circuit read $X$, $Y$ and every prefix bit except the top
$O((\log N)^{O(d)}\ell/D)$ ones, with correlation at most $2^{-DT}+2^{3-\ell}$ (\cref{app:ac0-window}). The gap cannot
be closed this way: with $W=[K-1]$ the bound needs $(1-D)K+r\log_2n<0$, which is impossible since $D<1$.

\begin{proposition}[small circuits on the full context]\label[proposition]{prop:ac0-small}
Let $1\le T\le N$. If $C$ is a sum of $s$ functions with values in $\{0,1\}$ and pairwise disjoint supports, each of
the form $f(X)g(Y)c(P)$ with $f,g$ $\{0,1\}$-valued and $c$ a conjunction of prefix literals or of conditions
$\chi_R(P)=\pm1$ on prefix parities, then
$|\E[(-1)^CB]|\le(2s+1)2^{-DT}$. This covers unambiguous DNFs over the formal bits (with arbitrary tests of $X$ and of
$Y$ allowed as literals) and decision trees with $s$ leaves whose queries are arbitrary tests of $X$, arbitrary tests
of $Y$, or parities of prefix bits.
\end{proposition}

\subsection{Mixed DNFs and CNFs on the full prefix}\label[appendix]{app:ac0-mixed}

For DNFs whose terms mix bits of $X$, $Y$ and the prefix, the results above need $o(N^2)$ terms
(\cref{thm:ac0-dnf,cor:ac0-full}) or bounded width (\cref{cor:ac0-full}). We now allow any polynomial number of terms
of any width, at a rate that is inverse-linear up to a polylogarithmic factor; for $o(N^2)$ terms the earlier bounds
are much stronger. The proofs are in
\cref{app:ac0-mixed-proofs}.

\begin{theorem}[mixed DNFs on the full prefix]\label{thm:ac0-mixeddnf}
Fix $0<\alpha\le1$ and $b>0$. There are an absolute constant $c$ and constants $C_{\alpha,b}$ and $N_0(\alpha,b)$
such that for all $N\ge N_0(\alpha,b)$ and all $\alpha N\le T\le N$, every DNF with at most $N^b$ terms, and every
CNF with at most $N^b$ clauses, over the literals of $X$, $Y$ and $P$ has advantage at most
$C_{\alpha,b}(\log N)^c/N$ in absolute value. Terms may mix literals of all three blocks and may have any width. The
bound also holds if the formula is chosen as a function of $Y$.
\end{theorem}

\noindent The proof hides a block of $\ell$ consecutive bits of $X$. Given the other operand bits, the product then runs
along an arithmetic progression whose slope is $Y$. On average over $Y$, sparse tests of the local product digits cannot
tell this progression from a continuous interval, and the interval has bounded density on the local digit cube. Bounded
independence~\cite{Bazzi09,Braverman10} turns this into a statement about circuits. In the next theorem, and in its use in
the proof of \cref{thm:ac0-mixeddnf}, the bits of $X$ (and of $Y$) are numbered from the most significant end: bit $i$ has
weight $2^{N-i}$, so bits $a+1,\dots,a+\ell$ have weights $2^{N-a-1},\dots,2^{N-a-\ell}$. The block length $\ell$ is
the one chosen in the proof; for much shorter blocks the conclusion fails (see the paragraph after the proof).

\begin{theorem}[a hidden operand block]\label{thm:ac0-hiddenblock}
Fix an integer $d\ge1$ and $b,A>0$. There are $c_d$, depending only on $d$, $C_{d,b,A}$, and for each $N$ a block
length $\ell=\ell(N)\le C_{d,b,A}(\log N)^{c_d}$ such that the following holds for all $N\ge N_0(d,b,A)$, all $T\le N$
and all $a\ge1$ with $a+\ell\le T$. Hide bits $a+1,\dots,a+\ell$ of $X$, and let an arbitrary function of $Y$ and of the
other bits of $X$ select a depth-$d$ circuit of size $N^b$ on $P$. Then the mean, over the exposed bits, of the
absolute conditional advantage of the selected circuit for $M_T$ is at most $N^{-A}$. The same holds with $X$ and $Y$
exchanged.
\end{theorem}

\noindent The rate in \cref{thm:ac0-mixeddnf} is set by the term-deletion step of the proof, which loses
$O(\ell\log N/N)$; it is not exponential, and we do not know whether it is tight. \Cref{thm:ac0-mixeddnf} is an
average-case statement about depth two; its proof uses \cref{thm:ac0-hiddenblock} with $d=2$. It gives no consequence
for transformers: UHATs lie in $\mathrm{AC}^0$~\cite{HAF22}, and that inclusion does not place them in
polynomial-size depth two.
Polynomial-size mixed circuits of depth three or more remain open (\cref{app:ac0-mixed-proofs} explains where the
argument stops). Together with \cref{prop:ac0-barrier}, \cref{thm:ac0-mixeddnf} gives a separation: for large $N$ and $\alpha N\le T\le N$, no DNF or CNF with $N^b$ terms
agrees with the product $f(X)g(Y)h(P)$ of the three features there except on a set of probability $o(1)$, since such a
formula would predict $M_T$ with constant advantage.

\section{Proofs for Appendices~\ref{sec:ac0} and~\ref{app:ac0-results}}\label[appendix]{app:ac0}

Notation is that of \cref{sec:ac0}: $Q=2^N$, $P=M_{<T}$, $B=(-1)^{M_T}$, $V=(X,Y,P)\in\{0,1\}^n$ with $n=2N+T-1$, and
$Z=XY/Q^2\in[0,1)$. For a function $p$ on a Boolean cube, $A(p)=\sum_U|\widehat p(U)|$ is its Walsh norm, as in
\cref{sec:overview}. On the cyclic group $\mathbb Z/Q$ we use normalised Fourier coefficients
$\widehat v(\xi)=Q^{-1}\sum_{y}v(y)\mathrm e(-\xi y/Q)$, so that $v(y)=\sum_\xi\widehat v(\xi)\mathrm e(\xi y/Q)$, and
write $\|v\|_{A_Q}=\sum_\xi|\widehat v(\xi)|$. For a real $\theta$, $\|\theta\|$ is its distance to the nearest
integer. $C$ denotes an absolute constant, not always the same.

\subsection{A transfer lemma}\label[appendix]{app:ac0-transfer}

\begin{lemma}[pointwise approximants]\label[lemma]{lem:ac0-transfer}
Let $F$ be a $\{-1,1\}$-valued function on the formal cube $\{0,1\}^n$, and let $p$ be a real polynomial of Walsh
degree at most $r$ with $|p-F|\le e$ on the whole cube. Then
\[
|\E BF(V)|\le e+(1+e)2^{-DT}\Bigl(\sum_{i=0}^r\binom ni\Bigr)^{1/2}.
\]
If $r\le\rho n$ with $0<\rho<\frac12$, the sum is at most $2^{nH_2(\rho)}$; if $\alpha N\le T\le N$ and
$3H_2(\rho)<D\alpha$, the second term is at most $(1+e)2^{-DT/2}$.
\end{lemma}

\begin{proof}
$|\E B(F-p)(V)|\le e$. By \eqref{eq:walsh}, $|\E Bp(V)|\le2^{-DT}A(p)$, and by Cauchy--Schwarz and Parseval on the
uniform cube $A(p)\le(\#\{U:|U|\le r\})^{1/2}\|p\|_2\le(\sum_{i\le r}\binom ni)^{1/2}(1+e)$. The binomial sum is at most
$\rho^{-r}(1-\rho)^{-(n-r)}\sum_i\binom ni\rho^i(1-\rho)^{n-i}\le2^{nH_2(\rho)}$. Finally $n\le3N\le3T/\alpha$.
\end{proof}

\begin{remark}[a necessary condition]\label[remark]{rem:ac0-necessary}
If a family of predictors has advantage at least $\eta>0$ infinitely often, then along that subsequence its formal-cube
approximate degree at error $\frac13$ (for $\{0,1\}$ output) is at least $c\,n$ with $c=c(\alpha,D,\eta)>0$. Indeed, if
$p$ approximates $C$ within $\frac13$, then $(3p+1)/5$ lies in $[0,\frac25]$ on zeros of $C$ and in $[\frac35,1]$ on
ones; composing with the acceptance polynomial of a majority vote over a fixed odd number of independent trials reduces
the error below $\eta/4$ at a constant factor in degree, and \cref{lem:ac0-transfer} then excludes degree below $cn$.
This is only an implication from a counterexample to an approximate-degree lower bound.
\end{remark}

\subsection{The arithmetic estimate}\label[appendix]{app:ac0-arith}

\begin{lemma}[bilinear sums]\label[lemma]{lem:ac0-bilinear}
For every nonzero integer $k$, every integer $b$ and every $f$ with $|f|\le1$, the sum
$A_k=\E f(X)\mathrm e(bY/Q+kXY/Q^2)$ satisfies $|A_k|\le C(1+\log Q)(|k|^{-1}+Q^{-1}+|k|Q^{-2})$, and $|A_k|\le1$.
\end{lemma}

\begin{proof}
Put $\theta_x=b/Q+kx/Q^2$. Summing the geometric series in $Y$,
\[
\Bigl|\frac1Q\sum_y\mathrm e(y\theta_x)\Bigr|\le\min\Bigl\{1,\frac1{2Q\|\theta_x\|}\Bigr\},\qquad\text{so}\qquad
|A_k|\le\E_X\min\Bigl\{1,\frac1{2Q\|\theta_X\|}\Bigr\}.
\]
The $Q$ numbers $\theta_x$ form an arithmetic progression with step $|k|/Q^2$ and range of length at most $|k|/Q$. For
$0<u\le\frac12$, at most $3+|k|/Q$ integers have a $u$-neighbourhood meeting that range, and each neighbourhood contains
at most $1+2uQ^2/|k|$ terms of the progression. Hence
\[
\Pr_X[\|\theta_X\|\le u]\le Q^{-1}(3+|k|/Q)(1+2uQ^2/|k|)\le C\bigl(Q^{-1}+|k|Q^{-2}+u(1+Q/|k|)\bigr).
\]
Split according to $\|\theta_X\|\le1/Q$ and the shells $\|\theta_X\|\in(2^{i-1}/Q,2^i/Q]$, $1\le i\le\log_2Q$, on which
the integrand is at most $2^{-i}$. Summing $2^{-i}\Pr[\|\theta_X\|\le2^i/Q]$ over the $O(1+\log Q)$ shells gives the
bound. The argument is uniform in $b$, $f$ and the sign of $k$.
\end{proof}

\begin{proof}[Proof of \cref{thm:ac0-arith}]
Let $q=2^T\le Q$ and define the $1$-periodic function $H(z)=(-1)^{\lfloor qz\rfloor}h(\lfloor qz/2\rfloor)$, where a
prefix is identified with its integer value. Then $\lfloor qZ\rfloor=2P+M_T$, so $H(Z)=Bh(P)$ and the quantity to
bound is $E:=\E H(Z)f(X)\mathrm e(bY/Q)$. $H$ is constant on the cells $[c/q,(c+1)/q)$, and the cells $2a$ and $2a+1$
carry opposite values $\pm h(a)$. Pairing each even cell with its translate by $1/q$,
\[
|\widehat H(k)|=\Bigl|\sum_a h(a)\int_{2a/q}^{(2a+1)/q}\bigl(\mathrm e(-kz)-\mathrm e(-k(z+1/q))\bigr)dz\Bigr|\le
\frac12\cdot\frac{2\pi|k|}q=\frac{\pi|k|}q ,
\]
and $\widehat H(0)=0$; by Parseval $\sum_k|\widehat H(k)|^2\le1$.

\emph{Smoothing.} Let $K=\lceil q^{9/8}\rceil$ and $F_K(v)=K^{-1}|\sum_{j=0}^{K-1}\mathrm e(jv)|^2$, the Fej\'er
kernel, with $\int F_K=1$ and $\widehat F_K(k)=(1-|k|/K)_+$. For $H_K=H*F_K$,
\[
\E H_K(Z)f(X)\mathrm e(bY/Q)=\sum_{0<|k|<K}(1-|k|/K)\widehat H(k)A_k,
\]
with $A_k$ as in \cref{lem:ac0-bilinear}. The
terms with $|k|\le q^{1/4}$ contribute at most $\sum_{0<|k|\le q^{1/4}}\pi|k|/q=O(q^{-1/2})$. By Cauchy--Schwarz and
\cref{lem:ac0-bilinear} the others contribute at most
\[
C(1+\log Q)\Bigl(\sum_{|k|>q^{1/4}}k^{-2}+\frac K{Q^2}+\frac{K^3}{Q^4}\Bigr)^{1/2}=O\bigl((1+\log Q)q^{-1/8}\bigr),
\]
using $q\le Q$. No smoothness of $h$ is used.

\emph{Jumps.} Let $\eta=q^{-1/16}$ and $a=\eta/q$. $H$ changes value only at multiples of $1/q$. If $\|qz\|>\eta$, every
shift of circular length below $a$ preserves $H(z)$, and $F_K(v)\le1/(4K\|v\|^2)$ gives
$|H(z)-H_K(z)|\le2\int_{\|v\|\ge a}F_K\le1/(Ka)\le q^{-1/16}$. Everywhere $|H-H_K|\le2$. It remains to show
$\Pr[\|qZ\|\le\eta]=O((1+\log Q)q^{-1/16})$; we may assume $\eta\le\frac14$. Put $J=\lfloor1/(2\eta)\rfloor$. If
$\|qZ\|\le\eta$, then $J\|qZ\|\le\frac12$ and $F_J(qZ)=\sin^2(\pi Jv)/(J\sin^2\pi v)|_{v=qZ}\ge4J/\pi^2$. Expanding
$F_J$ in its Fourier series,
\[
\Pr[\|qZ\|\le\eta]\le\frac{\pi^2}{4J}\E F_J(qZ)\le\frac C J+C\max_{1\le r<J}|\E\,\mathrm e(rqZ)|.
\]
$\E\,\mathrm e(rqZ)$ is $A_k$ with $f=1$, $b=0$ and $k=rq$, where $q\le k\le q^{17/16}$, so by
\cref{lem:ac0-bilinear} it is $O((1+\log Q)q^{-15/16})$; and $1/J=O(\eta)$. Collecting the three parts,
$|E|=O((1+\log Q)q^{-1/16})=O((N+1)2^{-T/16})$.
\end{proof}

\begin{lemma}[additive Fourier norm of marked digits]\label[lemma]{lem:ac0-digits}
Let $v:\{0,\dots,Q-1\}\to\C$ be a product of the signs $(-1)^{y_i}$ of $j\ge1$ distinct binary digits of $y$, or of
literal indicators on $j$ distinct digits. Then $\|v\|_{A_Q}\le[C(1+N/j)]^j$; for $j=0$ the norm is $1$.
\end{lemma}

\begin{proof}
Order the marked digits from the most significant end. The unmarked leading block is left alone; each marked digit
starts a block that extends up to, but excludes, the next marked digit or the end of the word. Let the blocks have
lengths $\ell_1,\dots,\ell_j$, so $\sum_i\ell_i\le N$. Appending a block of $\ell$ digits to a word $u$ with $H$
possible values gives the function $a(u)b(w)$ of the integer $uL+w$, $L=2^\ell$, $0\le w<L$, where $b$ depends only on
the most significant digit of $w$. At frequency $\xi=j_0+mH$ ($0\le j_0<H$, $0\le m<L$) its normalised coefficient on
$\mathbb Z/(HL)$ is
\[
\widehat a(j_0)\cdot\frac1L\sum_{w=0}^{L-1}b(w)\,\mathrm e\bigl(-(m+j_0/H)w/L\bigr).
\]
The function $b$ is a signed sum of at most two indicators of intervals. For one interval and a shift
$\theta\in[0,1)$, geometric summation bounds the second factor at $m$ by $\min\{1,1/(2\,\mathrm{dist}(m+\theta,L\mathbb
Z))\}$; at most four of the points $m+\theta$ have distance to $L\mathbb Z$ in each interval $[i,i+1)$, so the sum over
$m$ is at most $C(1+\ell)$, uniformly in $j_0/H$. So appending the block multiplies the Fourier norm by at most
$C(1+\ell)$. The unmarked leading word has norm $1$. Hence $\|v\|_{A_Q}\le\prod_iC(1+\ell_i)\le[C(1+N/j)]^j$ by the
inequality of arithmetic and geometric means.
\end{proof}

\noindent Since $v\mapsto v\log(C(1+N/v))$ is increasing for $v>0$ and $C\ge1$, the bound $[C(1+N/i)]^i$ also covers
products of at most $i$ marked digits.

\subsection{Proof of \cref{thm:ac0-operand} and \cref{cor:ac0-full}}\label[appendix]{app:ac0-operand}

\begin{lemma}[prefix-selected approximants]\label[lemma]{lem:ac0-selected}
Suppose that for each prefix value $p$ the sign function $F_p$ of the operand bits has a real polynomial $p_p$ on the
operand cube $\{0,1\}^{2N}$ of Walsh degree at most $r$ with $|p_p-F_p|\le e\le1$ everywhere. Then
\[
|\E BF_P(X,Y)|\le e+C(1+e)(N+1)2^{-T/16}\sum_{i=0}^r\binom{2N}i[C(1+N/i)]^i ,
\]
with the summand $1$ at $i=0$. If $\alpha N\le T\le N$, there is $c_\alpha>0$ such that for $r\le c_\alpha N$ the
second term is at most $2^{-\Omega(T)}$.
\end{lemma}

\begin{proof}
Write $p_p(x,y)=\sum_Uc_U(p)\chi_{U_X}(x)\chi_{U_Y}(y)$ over $|U|\le r$. By Parseval on the uniform operand cube,
$|c_U(p)|\le\E|p_p|\le1+e$. Expand $\chi_{U_Y}(Y)$ in additive characters of $\mathbb Z/Q$; by \cref{lem:ac0-digits}
its norm is at most $[C(1+N/i)]^i$ with $i=|U|$. Apply \cref{thm:ac0-arith} with $h=c_U/(1+e)$ and
$f=\chi_{U_X}$ to each character and sum. The term with $B(F_P-p_P)$ is at most $e$. For the last claim,
$\binom{2N}i\le(2eN/i)^i$, so for $1\le r\le N$ the sum is at most $\exp(O(r\log(eN/r)))$, which is at most $2^{T/32}$
once $r\le c_\alpha N$ with $c_\alpha$ small.
\end{proof}

\begin{proof}[Proof of \cref{thm:ac0-operand}]
We use the theorem of Bun, Kothari and Thaler~\cite[Thm.~15, eq.~(18)--(19)]{BKT19}: a function $h:\{0,1\}^{n'}\to\{0,1\}$
computed by a depth-$d$ circuit of size $m\ge1$ has
\[
\deg_\varepsilon(h)=\widetilde O\bigl(\sqrt{n'}\,m^{1/2-2^{-d}}(\log(1/\varepsilon))^{2^{-d}}\bigr)\qquad\text{if }
\varepsilon>2^{-m},
\]
where size counts AND and OR gates and the hidden factor is polylogarithmic in $m$, uniformly in $\varepsilon$. Here
$n'=2N$. A $\{0,1\}$ approximant with error $\varepsilon$ gives a sign approximant with error $e=2\varepsilon$ of the
same degree, and multilinearisation and the change between $\{0,1\}$ and $\pm1$ variables keep the degree. We may pad
each circuit with dummy gates so that $m=\Theta(N^u)$.

\emph{Polynomial size.} Let $\gamma<\gamma'<b-(b-1)u$ and $\log_2(1/\varepsilon)=N^{\gamma'}$; then
$\varepsilon>2^{-m}$ because $\gamma'<u$. The degree is $\widetilde O(N^{1/2+u(1/2-2^{-d})+\gamma'2^{-d}})=o(N)$, since
the exponent is below $1$ exactly when $\gamma'<2^{d-1}-u(2^{d-1}-1)=b-(b-1)u$. So for large $N$ the degree is at most
$c_\alpha N$, and \cref{lem:ac0-selected} bounds $|\E BF_P|$ by $2^{1-N^{\gamma'}}+2^{-\Omega(T)}\le2\cdot2^{-N^\gamma}$
for large $N$. The advantage is half of this.

\emph{Linear size.} Take $\log_2(1/\varepsilon)=N/(\log N)^{A_d}$ with $A_d$ larger than $2^d$ times the exponent of
the polylogarithmic factor; then the degree is $o(N)$.

\emph{Bounded width.} Sherstov~\cite[Thm.~5.1]{Sherstov18} (on the full cube) gives, for a $w$-DNF or $w$-CNF on
$n'$ variables and $0<\varepsilon<\frac12$, $\deg_\varepsilon\le c(\sqrt2)^wn'^{w/(w+1)}(\log(1/\varepsilon))^{1/(w+1)}$
with an absolute $c$. Put $\log_2(1/\varepsilon)=\Delta=n'(c_\alpha/(2c\,2^{w/2}))^{w+1}$; then the degree is at most
$c_\alpha n'/2=c_\alpha N$, and \cref{lem:ac0-selected} gives the bound $2^{1-\Delta}+2^{-\Omega(T)}$. Since
$\Delta=N2^{-O_\alpha(w^2+1)}$, this is $2^{-\Omega_{\alpha,w}(N)}$, and $2^{-N^{1-o(1)}}$ for $w=o(\sqrt{\log N})$.

\emph{Prefix selection and conditional optima.} The approximants $p_p$ are chosen separately for each $p$, which
\cref{lem:ac0-selected} allows. For each prefix value of positive probability the class of circuits is finite;
choose the circuit with the largest conditional absolute correlation and complement its output if needed (the
classes are closed under negation by De Morgan's laws, at the same depth and size). The resulting $F_P$ has
correlation equal to the average over $p$ of the best conditional correlation, and the bounds above apply to it.
\end{proof}

\begin{proof}[Proof of \cref{cor:ac0-full}]
For a circuit $F$ on $(X,Y,P)$ let $F_p$ be $F$ with $P=p$ fixed; it has depth at most $d$ and size at most that of
$F$, and $F(V)=F_P(X,Y)$ on true contexts. The same holds for width-$w$ DNFs and CNFs. Apply \cref{thm:ac0-operand}.
For depth $3$, $b=4$ and $u<4/3$.
\end{proof}

\noindent The proof of \cref{thm:ac0-operand} uses only \cref{thm:ac0-arith} and \cref{lem:ac0-digits}, not
\cref{thm:main}; its constants are absolute apart from the dependence on $\alpha$, $d$, $u$, $\gamma$ and $w$.

\subsection{Proof of \cref{thm:ac0-dnf}}\label[appendix]{app:ac0-dnf}

Let $C=t_1\vee\dots\vee t_s$ be a DNF on the formal bits, and let $C_0$ be the absolute constant below. Every literal indicator has Walsh norm $1$, and $A$ is
submultiplicative, so each term indicator has $A(t_i)\le1$, and the average $a=s^{-1}\sum_it_i$ has $A(a)\le1$; $sa$
counts the satisfied terms, so $a$ takes values in the grid $\{0,1/s,\dots,1\}$.

\emph{A univariate approximant.} For $\log_23\le\Delta\le s$, Sherstov's explicit construction for symmetric
functions~\cite[Thm.~1.4]{Sherstov18} (valid for error $2^{-s}\le\varepsilon\le\frac13$) gives a polynomial on
$\{0,1\}^s$ of degree $O(\sqrt{s\Delta})$ approximating OR within $2^{-\Delta}$; averaging it over permutations of the
variables gives a univariate polynomial in the Hamming weight of the same degree (a monomial of degree $i$ averages
to $\binom ji/\binom si$ on the layer of weight $j$). Converting to signs, there is a univariate $R$ of degree
$k\le C_0\sqrt{s\Delta}$, $k\le s$, with $|R(j/s)-\mathrm{sgn}(j)|\le2^{1-\Delta}$ for all $0\le j\le s$, where
$\mathrm{sgn}(0)=1$ and $\mathrm{sgn}(j)=-1$ for $j\ge1$. For $\Delta>s$ take exact interpolation, of degree $k\le s$.
In every case $|R|\le3$ on the grid.

\emph{Norm after substitution.} For $1\le k\le s$ let $u_j=j\lfloor s/k\rfloor/s$, $0\le j\le k$; these are grid points
with spacing at least $1/(2k)$. By Lagrange interpolation $R(a)=\sum_jR(u_j)\prod_{i\ne j}(a-u_i)/(u_j-u_i)$. Each
factor $a-u_i$ has Walsh norm at most $2$, and $\prod_{i\ne j}|u_j-u_i|\ge(2k)^{-k}j!(k-j)!$. Hence
\[
A(R(a))\le3(4k)^k\sum_{j=0}^k\frac1{j!(k-j)!}=\frac{3(8k)^k}{k!}\le3(8e)^k;
\]
the constant case $k=0$ obeys the same bound.

\emph{Choice of parameters.} Let $\Delta=\lambda\min(T,T^2/s)$ for a small absolute $\lambda>0$. If $\Delta<2$ the claim
holds after enlarging the absolute prefactor. If $2\le\Delta\le s$ then $k\le C_0\sqrt{s\Delta}\le C_0\sqrt\lambda T$, and if
$\Delta>s$ then $k\le s<\lambda T$. Choose $\lambda$ so small (in terms of $D$ and $C_0$) that $3(8e)^k2^{-DT}\le3\cdot2^{-DT/2}$.
Since $|R(a(V))-(-1)^{C(V)}|\le2^{1-\Delta}$ everywhere, \eqref{eq:walsh} gives
$|\E B(-1)^{C(V)}|\le2^{1-\Delta}+3\cdot2^{-DT/2}=O(2^{-c\min(T,T^2/s)})$ with $c=\min(\lambda,D/2)$. For a CNF apply
this to its negation.

\subsection{Proof of \cref{thm:ac0-query}}\label[appendix]{app:ac0-query}

Fix a deterministic strategy. For each prefix value unfold its decision tree, pruning repeated queries. Distinct
leaves of one tree have disjoint nonempty subcubes of the operand cube, so for a fixed prefix each consistent set of
signed operand literals is the literal set of at most one reachable leaf. Grouping leaves by their \emph{unordered}
literal sets $\mathcal L$, the predictor's sign output equals
$\sum_{\mathcal L}c_{\mathcal L}(P)\mathbf 1_{\mathcal L_X}(X)\mathbf 1_{\mathcal L_Y}(Y)$, where
$c_{\mathcal L}(p)\in\{-1,0,1\}$ is the output at that leaf (zero if absent) and $\mathcal L_X,\mathcal L_Y$ are the
literals on $X$ and on $Y$; this holds even though the order of queries may depend on the prefix. There are at most
$2^i\binom{2N}i$ sets of size $i\le k$. For each, expand $\mathbf 1_{\mathcal L_Y}$ in additive characters
(\cref{lem:ac0-digits}, norm at most $[C(1+N/i)]^i$) and apply \cref{thm:ac0-arith} with $h=c_{\mathcal L}$ and
$f=\mathbf 1_{\mathcal L_X}$. The correlation is at most
$C(N+1)2^{-T/16}\sum_{i=0}^k2^i\binom{2N}i[C(1+N/i)]^i\le C(N+1)2^{-T/16}e^{O(k\log(eN/k))}$, and $c_\alpha$ small makes
the last factor at most $2^{T/32}$. A randomised strategy is an average of deterministic ones.

\subsection{Proof of \cref{thm:ac0-message}}\label[appendix]{app:ac0-message}

\emph{(a)} Let $m=T-1$, $V_t=\sum_{i\le t}\binom mi$, and let $\delta_t$ bound the uniform Fourier weight above degree $t$
of every circuit $F_{x,j}$ (on the cube of the $m$ prefix bits). We show
\[
|\E BF_{X,j(Y)}(P)|\le LV_t2^{-DT}+4\sqrt{L\delta_t}.
\]
Truncate each $F_{x,j}$ to Walsh degree $t$; for fixed $j$ and $S$ the coefficient is a function $f_{j,S}(x)$ with
$|f_{j,S}|\le1$ by Parseval. The truncated part contributes $\sum_j\sum_{|S|\le t}\E f_{j,S}(X)\mathbf
1_{j(Y)=j}\chi_S(P)B$, and each term is at most $2^{-DT}$ by \cref{thm:main} applied to $S\cup\{T\}$, with
$g=\mathbf 1_{j(Y)=j}$. For the residual, condition on $X=x$. For $x>0$ a fixed value of $P$ confines $Y$ to an
interval of length $Q^2/(2^mx)$, so the law of $P$ given $X=x$ has density at most $b_x=Q/x+2^m/Q$ with respect to the
uniform law on $\{0,1\}^m$; for $x=0$ put $b_0=2^m$. The squared residual of the selected circuit is at most the sum
of the squared residuals of all $L$ circuits $F_{x,j}$, whose conditional expectation is at most $b_xL\delta_t$. By
Cauchy--Schwarz at fixed $x$ and averaging, the residual contributes at most $\E_X\sqrt{b_XL\delta_t}$, and
$\E_X\sqrt{b_X}\le Q^{-1}(\sqrt Q+\sum_{x\ge1}(\sqrt{Q/x}+1))\le4$, using $2^m\le Q$ and
$\sum_{x=1}^{Q-1}x^{-1/2}\le2\sqrt Q$.

Now take $t=\lfloor\lambda m\rfloor$ with $0<\lambda<\frac12$ and $H_2(\lambda)<D/4$, so that $V_t\le2^{DT/4+O(1)}$, and
Tal's bound $\delta_t\le2\cdot2^{-(t+1)/(c_d(\log_2(2s))^{d-1})}$~\cite[Thm.~18]{Tal17}. With
$\log_2L\le c'T/(\log_2(2s))^{d-1}$ and $c'$ small, both terms are $2^{-\Omega_d(T/(\log_2(2s))^{d-1})}$.

\emph{(b)} Let $\ell(P)$ be the selected index. Truncate each of the $L$ operand circuits to Walsh degree
$r=\lfloor c_\alpha N\rfloor$ on the operand cube, $c_\alpha$ as in \cref{lem:ac0-selected}. The truncated part of the
selected circuit has coefficients that are functions of $P$ bounded by $1$, so the argument of \cref{lem:ac0-selected}
(with $e=0$, applied to the exact low-degree part) bounds its correlation by $2^{-\Omega(T)}$, whatever $L$ is. The
residual of the selected circuit is pointwise at most the square root of the sum of the $L$ squared residuals; the
operands are uniform, so by Jensen its expectation is at most $\sqrt{L\delta_r}$ with
$\delta_r\le2\cdot2^{-(r+1)/(c_d(\log_2(2s))^{d-1})}$, and $r\ge c_\alpha T-1$. The stated condition on $L$ gives the
rate.

\subsection{Proof of \cref{prop:ac0-barrier}}\label[appendix]{app:ac0-barrier}

First allow complex factors of modulus at most $1$. Let $q=2^T$ and $t=q/32$. Let $f(x)=\mathbf 1[Q/4\le x\le Q/3]
(x/Q)^{it}$ and $g(y)=(y/Q)^{it}$ for $y>0$, $g(0)=0$; on the support of $f$ their product is $Z^{it}$. Keep the prefix
values $a$ whose cells $[2a/q,(2a+2)/q)$ lie in $I=[\frac1{16},\frac18]$; there are exactly $q/32$ of them. Let
$h(a)=z_a^{-it}$ with $z_a=(2a+1)/q$ the midpoint for these $a$, and $h(a)=0$ otherwise. On the support of $fgh$ the
phase $\phi=t\ln(Z/z_a)$ satisfies $|\phi|\le t\cdot\frac{1/q}{1/16}=\frac12$, by the mean value theorem and $Z\ge\frac1{16}$;
it is negative on the left half of the cell, where $M_T=0$ and $B=1$, and nonnegative on the right half, where $B=-1$.
So $-B\sin\phi\ge0$ throughout. If $2a/q\le Z<(2a+\frac12)/q$, then $|Z-z_a|\ge1/(2q)$ and $Z,z_a\le\frac18$, so
$|\phi|\ge t\cdot\frac{1/(2q)}{1/8}=\frac18$.

For every retained $x$, the event $2a/q\le Z<(2a+\frac12)/q$ is an interval of $y$ inside $[0,Q)$ of length
$Q^2/(2qx)\ge3Q/(2q)\ge1$, which contains at least half its length in integers; summing over the $q/32$ retained cells,
its conditional probability is at least $3/128$. The retained $x$ have probability at least $\frac1{16}$ when $Q\ge64$.
Hence $-\mathrm{Im}\,\E[Bfgh]\ge\frac3{2048}\sin\frac18$. The imaginary part of a product of three complex numbers is a
signed sum of four products of their real and imaginary parts, so for some choice of real functions
$f_1,g_1,h_1$ with values in $[-1,1]$, $|\E Bf_1(X)g_1(Y)h_1(P)|\ge\frac3{8192}\sin\frac18$. A function $u$ with values
in $[-1,1]$ is the expectation of $\mathrm{sgn}(u-\theta)$ over $\theta$ uniform in $(-1,1)$; with independent
thresholds for the three functions the correlation is the average of the correlations of sign functions, so some
choice of thresholds gives sign functions $f,g,h$ with at least the same absolute correlation.

\noindent The mechanism needs only the logarithmic phase: recentring at the left endpoint of each cell instead of the
midpoint changes $h$ by a cell-dependent constant phase.

\subsection{Proofs of \cref{prop:ac0-duality,prop:ac0-uniform,prop:ac0-small}}\label[appendix]{app:ac0-method}

\begin{proof}[Proof of \cref{prop:ac0-duality}]
For fixed $g$ and $S$ the expression $\E[f(X)g(Y)\chi_S(P)\sigma']$ is linear in $f$, so its maximum over
$|f|\le1$ is attained at a $\{-1,1\}$-valued $f$, and likewise for $g$. Hence $\Sigma_\delta$ is cut out by the
finitely many constraints $|\E[\varphi_i\sigma']|\le\delta$, $\varphi_i$ ranging over the finite set $\Pi_\pm$ of
functions in $\Pi$ with $\{-1,1\}$-valued $f,g$, together with $-1\le\sigma'\le1$. Moreover $\|p\|_\Pi$ equals the least
weight of a representation by elements of $\Pi_\pm$: a $[-1,1]$-valued function is a convex combination of
$\{-1,1\}$-valued ones, so every element of $\Pi$ is a convex combination of elements of $\Pi_\pm$.

The left side is the value of the finite linear programme: maximise $\E[\widetilde C\sigma']$ over $\sigma'$ subject
to $-1\le\sigma'\le1$ and $-\delta\le\E[\varphi_i\sigma']\le\delta$. It is feasible ($\sigma'=0$) and bounded. Its dual
is: minimise $\E|\widetilde C-p|+\delta\sum_i|c_i|$ over $p=\sum_ic_i\varphi_i$. (Writing the box constraint with
multipliers $\mu^\pm\ge0$ and the other constraints with $c_i^\pm\ge0$, dual feasibility reads
$\mu^+-\mu^-=\widetilde C-\sum_i(c_i^+-c_i^-)\varphi_i$ pointwise, and at the optimum
$\E(\mu^++\mu^-)=\E|\widetilde C-p|$ and $\sum_i(c_i^++c_i^-)=\sum_i|c_i|$.) By strong duality for finite linear
programmes the two optima are equal and attained, and minimising over representations of $p$ gives $\|p\|_\Pi$.

For complex $f,g$: the complex constraint set is contained in $\Sigma_\delta$, since real atoms are among the complex
ones; and it contains $\Sigma_{\delta/2}$, since $\mathrm{Re}(e^{i\theta}fg)=\mathrm{Re}(e^{i\theta}f)\mathrm{Re}\,g-
\mathrm{Im}(e^{i\theta}f)\mathrm{Im}\,g$ is a sum of two real atoms.
\end{proof}

\begin{proof}[Proof of \cref{prop:ac0-uniform}]
Let $p$ approximate $f$. Each Walsh character of $\{0,1\}^n$ composed with the affine map $\Lambda$ is $\pm$ a character
of $\mathbb F_2^{2m}$, possibly the trivial one, so $p\circ\Lambda$ is a combination of characters with total weight
at most $A(p)$. Every character of $\mathbb F_2^{2m}$ has correlation exactly $2^{-m}$ in absolute value with the bent
function $(-1)^{\mathrm{IP}_m}$. Hence
$|\E(1-2f\circ\Lambda)(-1)^{\mathrm{IP}_m}|\le2\varepsilon+(1+2A(p))2^{-m}$, and the probability of agreement is half of
one plus this. If $w(n)\le\gamma n$ and $n<(1-o(1))m/\gamma$, then $2^{w(n)-m}=o(1)$.
\end{proof}

\noindent The published bounds for $\mathrm{AC}^0\circ\oplus$ bound the total size: Cheraghchi et al.~\cite{CGJ16} prove
superlinear worst-case size lower bounds for inner product at every fixed depth and $\widetilde\Omega(n^2)$ at depth $4$,
Bun, Kothari and Thaler~\cite{BKT19} prove average-case size lower bounds $\widetilde\Omega(n^{1/(1-2^{-d})})$ at depth
$d+1$, and Servedio and Viola posed as a challenge to show that inner product has no polynomial-size
$\mathrm{AC}^0\circ\oplus$ circuits~\cite{ServedioViola12}.

\begin{proof}[Proof of \cref{prop:ac0-small}]
A conjunction of $w$ prefix literals, or of $w$ conditions $\chi_{R}(P)=\pm1$, is $2^{-w}\prod(1\pm\chi_{R_i}(P))$, an
average of $2^w$ signed parities of prefix bits, so each summand $f(X)g(Y)c(P)$ has $\|\cdot\|_\Pi\le1$. Hence
$\|1-2C\|_\Pi\le2s+1$, and \cref{thm:main}, applied to each atom with $S\cup\{T\}$, gives the bound. In an unambiguous
DNF the terms have disjoint supports, and the literals on $X$ (on $Y$) of one term multiply to a $\{0,1\}$-valued
function of $X$ (of $Y$). In a decision tree the leaves with output $1$ have disjoint supports, and the indicator of a
leaf is the product of the indicators of the outcomes along its path.
\end{proof}

\noindent For DNFs with arbitrary terms, composing the small-error approximation of OR~\cite{deWolf08,Sherstov18}
with the term indicators gives in the same way, for DNFs with at most
$c(DT)^2/(\log^2T\log(1/\varepsilon))$ terms, the bound $\varepsilon+2^{1-DT/2}$ on the advantage; \cref{thm:ac0-dnf} is stronger.

\subsection{Proofs of \cref{lem:ac0-density} and \cref{thm:ac0-twoblock}}\label[appendix]{app:ac0-twoblock}

Here $2\le T\le N$, $L'=2^{2N-T+1}$, and $\mu$ is the law of the true context; $U$ is the uniform law on the cube in
question.

\begin{proof}[Proof of \cref{lem:ac0-density}]
$P=a$ if and only if $aL'\le xY<(a+1)L'$, so $Y$ lies in a real interval of length $L'/x$, which contains at most
$L'/x+1$ integers. As $x<2^N$ and $T\le N$, $L'/x>2^{N-T+1}\ge2$, so $L'/x+1<1.5L'/x$, and
$\Pr[X=x,P=a]\le1.5L'/(x4^N)=1.5(2^N/x)2^{-(N+T-1)}$.
\end{proof}

Consequently, for $0\le j\le N$ and $G_X=\{X\ge2^{N-j}\}$ (so $\Pr[G_X^c]=2^{-j}$), every $h\ge0$ on
$\{0,1\}^N\times\{0,1\}^{T-1}$ satisfies
\begin{equation}\label{eq:ac0-dn}
\E[\mathbf 1_{G_X}h(X,P)]\le1.5\cdot2^j\,\E_U[h],
\end{equation}
and the same with $Y$ in place of $X$.

A \emph{partition of unity} is a family of $\{0,1\}$-valued functions summing to $1$. For a circuit of depth $d$ and size
$s>1$ computing $a$, Tal's theorem~\cite[Thm.~18]{Tal17} applied to $1-2a$ gives
$\E_U[(a-a^{<k})^2]\le\tau/4$ with $\tau=\tau_k=2\cdot2^{-k/\kappa}$, where $a^{<k}$ is the truncation to Walsh degree
below $k$ and $\kappa=c_d(\log_2s)^{d-1}$.

\begin{lemma}[two-block inequalities]\label[lemma]{lem:ac0-twoblock-ineq}
Let $n_1=N+T-1$, let $\{a_v\}_{v\in\mathcal V}$ be a partition of unity of functions of $(x,m)\in\{0,1\}^N\times\{0,1\}^{T-1}$,
each computed by a depth-$d$, size-$s$ circuit, let $\{b_w\}_{w\in\mathcal W}$ be a partition of unity of functions of
$(y,m)$, let $\phi:\mathcal V\times\mathcal W\to\{-1,1\}$, $F=\sum_{v,w}\phi(v,w)a_vb_w$, and let $k\ge1$, $0\le j\le N$.
\textup{(a)} If every $b_w$ is a function of $y$ alone (arbitrary), and even if for each $w$ a different partition
$\{a^w_v\}_v$ is used, $F=\sum_wb_w(y)\sum_v\phi(v,w)a^w_v(x,m)$, then
\[
|\E F(V)B|\le2^{-j}+|\mathcal V|\sqrt{|\mathcal W|}\sqrt{0.375\cdot2^j\tau}+|\mathcal W|(n_1+1)^k2^{-DT}.
\]
\textup{(b)} If every $b_w$ is computed by a depth-$d$, size-$s$ circuit in $(y,m)$, then
\[
|\E F(V)B|\le2^{1-j}+1.4\cdot2^j|\mathcal V||\mathcal W|\sqrt\tau+(n_1+1)^{2k}2^{-DT}.
\]
\end{lemma}

\begin{proof}
Let $p_v=a_v^{<k}$ on the cube $\{0,1\}^{n_1}$. As $a_v\ge0$, $|\widehat a_v(S)|\le\E_Ua_v$, so
$\sum_v|\widehat a_v(S)|\le1$ for every $S$ and
\begin{equation}\label{eq:ac0-n1}
\textstyle\sum_vA(p_v)\le\#\{S:|S|<k\}\le(n_1+1)^k .
\end{equation}
(a) Let $G=G_X$. Since $|F|\le1$, we have $|\E\mathbf 1_{G^c}FB|\le2^{-j}$. For each $(v,w)$ write
$\E[\mathbf 1_Ga_vb_wB]=\E[\mathbf 1_Gp_vb_wB]+\E[\mathbf 1_G(a_v-p_v)b_wB]$. By Cauchy--Schwarz with $b_w^2=b_w$ and
\eqref{eq:ac0-dn}, the second term is at most
\[
\sqrt{\Pr[b_w(Y)=1]}\,\sqrt{1.5\cdot2^j\tau/4},\qquad\text{and}\qquad\sum_w\sqrt{\Pr[b_w(Y)=1]}\le\sqrt{|\mathcal W|}.
\]
{\sloppy The function $\mathbf 1_G(X)p_v(X,P)b_w(Y)$ is a combination, of total weight $A(p_v)$, of the
functions $[\mathbf 1_G\chi_A](X)\,b_w(Y)\,\chi_S(P)$, which lie in $\Pi$; so by \cref{thm:main} (with
$S\cup\{T\}$) the first term is at most $A(p_v)2^{-DT}$; summing with \eqref{eq:ac0-n1}, applied for each $w$
separately, gives $|\mathcal W|(n_1+1)^k2^{-DT}$. Nothing used that the partition is the same for different $w$.\par}

(b) Let $q_w=b_w^{<k}$ on the cube of $(y,m)$, $G=G_X\cap G_Y$, $\Pr[G^c]\le2^{1-j}$. Pointwise,
$\mathbf 1_{G_X}a_v\mathbf 1_{G_Y}b_w-\mathbf 1_{G_X}p_v\mathbf 1_{G_Y}q_w=\mathbf 1_{G_X}a_v\cdot\mathbf
1_{G_Y}(b_w-q_w)+\mathbf 1_{G_Y}q_w\cdot\mathbf 1_{G_X}(a_v-p_v)$. The expectation of the absolute value of the first
product is at most $\sqrt{\Pr[a_v=1]}\sqrt{0.375\cdot2^j\tau}$ by \eqref{eq:ac0-dn} for $(Y,P)$, which sums to at most
$\sqrt{|\mathcal V|}|\mathcal W|\sqrt{0.375\cdot2^j\tau}$. The second is at most
$\sqrt{1.5\cdot2^j\E_U[q_w^2]}\sqrt{0.375\cdot2^j\tau}$, and $\E_Uq_w^2\le\E_Ub_w$, so it sums to at most
$0.75\cdot2^j|\mathcal V|\sqrt{|\mathcal W|}\sqrt\tau$. Together these are at most $1.4\cdot2^j|\mathcal V||\mathcal
W|\sqrt\tau$. The main term $\mathbf 1_{G_X}p_v\cdot\mathbf 1_{G_Y}q_w$ is a combination of weight $A(p_v)A(q_w)$ of
functions $[\mathbf 1_{G_X}\chi_A](X)[\mathbf 1_{G_Y}\chi_{B'}](Y)\chi_{S\triangle S'}(P)\in\Pi$; by \cref{thm:main} and
\eqref{eq:ac0-n1} for both families it is at most $(n_1+1)^{2k}2^{-DT}$.
\end{proof}

\begin{proof}[Proof of \cref{thm:ac0-twoblock}]
Let $k$ be as stated and $j=\lfloor k/(4\kappa)\rfloor$; if $k=0$ there is nothing to prove. Then $j\le N$, because
$k/(4\kappa)\le DT/16$, and $2^{-j}\le2\beta$, $2^j\sqrt\tau\le\sqrt2\beta$, $(n_1+1)^k\le2^{DT/4}$. In both parts of
\cref{lem:ac0-twoblock-ineq} this gives $|\E FB|\le(4+2|\mathcal V||\mathcal W|)\beta+2^{-DT/2}$ (in (a) the last term
$|\mathcal W|2^{-3DT/4}$ is absorbed, since $2^{-3DT/4}\le\beta$). The advantage is half of $|\E FB|$. (1) is (a) with the
partition $\{C,1-C\}$ ($|\mathcal V|=2$, $|\mathcal W|=1$; $1-C$ has the same depth and size by De Morgan's laws). (2) is
(a) with $b_w=\mathbf 1[\psi(Y)=w]$, $|\mathcal W|=2^u$, and for each $w$ the partition $\{C_w,1-C_w\}$. (3) is (b) with
$a_v=\mathbf 1[A=v]$, $b_w=\mathbf 1[B'=w]$, $|\mathcal V|=2^{r_1}$, $|\mathcal W|=2^{r_2}$. Exchanging $X$ and $Y$ uses
the symmetric form of \cref{lem:ac0-density}.
\end{proof}

\subsection{Proof of \cref{thm:ac0-window}}\label[appendix]{app:ac0-window}

\begin{lemma}[the tail of the prefix has small bias]\label[lemma]{lem:ac0-tail}
For $1\le K\le T\le N$, every nonempty parity of the bits of $(X,Y,M_K,\dots,M_T)$ has expectation at most $2^{-DK}$
in absolute value.
\end{lemma}

\begin{proof}
A parity is $\chi_A(X)\chi_B(Y)\chi_S(M)$ with $S\subseteq[K,T]$. If $S=\emptyset$ then $A\cup B\ne\emptyset$ and the
expectation is $0$, since $(X,Y)$ is uniform. Otherwise \cref{thm:main} at level $t=\max S\in[K,N]$ gives at most
$2^{-Dt}\le2^{-DK}$.
\end{proof}

\begin{lemma}[sandwiching polynomials~\cite{Bazzi09}]\label[lemma]{lem:ac0-sandwich}
Let $f:\{0,1\}^m\to\{0,1\}$ be $\varepsilon$-fooled by every $r$-wise independent distribution. Then there are real
polynomials $q_l\le f\le q_u$ (pointwise) of degree at most $r$ with $\E_U[f-q_l]\le\varepsilon$,
$\E_U[q_u-f]\le\varepsilon$ and $A(q_l),A(q_u)\le(1+\varepsilon)(m+1)^r$.
\end{lemma}

\begin{proof}
Maximise $\E_U[q]=\widehat q(\emptyset)$ over polynomials $q$ of degree at most $r$ with $q\le f$ pointwise. This finite
linear programme is feasible ($q=0$) and bounded by $1$. Its dual is to minimise $\sum_z\lambda_zf(z)$ over
$\lambda\ge0$ with $\sum_z\lambda_z\chi_S(z)=[S=\emptyset]$ for all $|S|\le r$; the feasible $\lambda$ are exactly the
$r$-wise independent distributions. By strong duality the optimum is $\min_\lambda\E_\lambda f\ge\E_Uf-\varepsilon$, which
gives $q_l$; $q_u=1-(\text{the lower polynomial for }1-f)$, and $1-f$ is fooled as well as $f$. Finally
$|\widehat q_l(S)|\le\E_U|q_l|\le\E_U|f|+\E_U(f-q_l)\le1+\varepsilon$, there are at most $(m+1)^r$ sets of size at most
$r$, and likewise for $q_u$.
\end{proof}

\begin{proof}[Proof of \cref{thm:ac0-window}]
$(-1)^C=1-2C$ and $|\E B|\le2^{-DT}$ by \cref{thm:main}, so it suffices to show
$|\E CB|\le2^u[2\varepsilon_0+3(1+\varepsilon_0)(n_K+1)^r2^{-DK}]$. Write $C=\sum_v\mathbf 1[M_W=v]C_v$ and
$\mathbf 1[M_W=v]=2^{-u}\sum_{R\subseteq W}\chi_R(v)\chi_R(M_W)$; then
$|\E CB|\le2^u\max_{v,R}|\E\chi_R(M_W)C_vB|$. Fix $v,R$, take $q_l\le C_v\le q_u$ from \cref{lem:ac0-sandwich} on the
$n_K$ bits of $(X,Y,M_K,\dots,M_{T-1})$, and write $\E\chi_RC_vB=\E\chi_Rq_lB+\E\chi_R(C_v-q_l)B$. Since $C_v-q_l\ge0$,
the second term is at most $\E_\mu[C_v-q_l]\le\E_\mu[q_u-q_l]=\E_U[q_u-q_l]+\sum_{S\ne\emptyset}\widehat{(q_u-q_l)}(S)
\E_\mu\chi_S$, which by \cref{lem:ac0-tail} is at most $2\varepsilon_0+2(1+\varepsilon_0)(n_K+1)^r2^{-DK}$. The first
term is a combination of weight $A(q_l)$ of functions $\chi_A(X)\chi_B(Y)\chi_{S'}(M)B$ whose top prefix position is
$T$, each at most $2^{-DT}\le2^{-DK}$ by \cref{thm:main}.
\end{proof}

\noindent\emph{The corollary in \cref{sec:ac0}.} Insert $r_d(s,\varepsilon_0)\le(\log_2s)^{\beta d}\log_2(1/\varepsilon_0)$
(Harsha and Srinivasan~\cite{HarshaSrinivasan19}, $\beta$ absolute) and $\varepsilon_0=2^{-u-\ell}$. The bound of
\cref{thm:ac0-window} is then at most $2^{-DT}+2^{3-\ell}$ provided
$DK\ge(u+\ell)[(\log_2s)^{\beta d}\log_2(n+1)+1]+3$. The bound is on $|\E(-1)^CB|$, twice the advantage. For
$u=0$ and $s$ polynomial in $N$ the circuit may read $X$, $Y$ and every prefix bit except the top
$K-1=O((\log N)^{\beta d+1}\ell/D)$ ones. For $u>0$, $W$ may be any set of top positions; the unread positions are
$[K-1]\setminus W$. With $W=[K-1]$ the condition would need $2^u(n+1)^r2^{-DK}<1$ with $u=K-1$, that is
$(1-D)K+r\log_2(n+1)<0$, which fails since $D<1$.

\subsection{Proofs of \cref{thm:ac0-mixeddnf,thm:ac0-hiddenblock}}\label[appendix]{app:ac0-mixed-proofs}

\paragraph{Bounded independence.} We use Braverman's theorem in the parameterised form of his Corollary~2: for every
$d$, $s\ge2$ and $0<\varepsilon<1$, $r_d(s,\varepsilon)\le(\log(s/\varepsilon))^{O(d^2)}$, with an absolute
constant in the exponent~\cite[Cor.~2]{Braverman10}. The bound of Harsha and Srinivasan quoted before
\cref{thm:ac0-window} would also do. Polynomials that sandwich a fooled function come from Bazzi's linear-programming
duality (\cref{lem:ac0-sandwich}). The polylogarithmic exponents below are inherited from these bounds.

\paragraph{Setting.} Fix $a\ge1$ and a block length $\ell\ge16$ divisible by $8$, with $a+\ell\le T\le N$. Put
\[
m=T-a+1,\qquad\nu=2^m,\qquad R=2^\ell=\lambda^8,\qquad S=2^{2N-a+1}.
\]
Then $m\ge\ell+1$, so $\nu\ge2R$; $\lambda\ge4$; and $\nu\le Q$, since $m\le T\le N$. Only $\nu\le2Q$ is used. The
\emph{exposure} is $\xi=(x_0,y)$, where $x_0$ is $X$ with bits $a+1,\dots,a+\ell$ set to $0$, and $y=Y$. Given $\xi$,
$X=x_0+2^{N-a-\ell}U$ with $U$ uniform on $\{0,\dots,R-1\}$, and $XY/S\equiv Z(U)\pmod 1$, where
\[
Z(u)=x_0y/S+yu/(2QR).
\]
As $XY<Q^2$, the integer part of $XY/S$ is the word of the top $a-1$ product bits, and the first $m$ binary digits of
the fractional part are $M_a,\dots,M_T$. So the local word $j=\lfloor\nu Z(U)\rfloor\in\mathbb Z/\nu$ has the digits
$M_a,\dots,M_T$. The word of the top $a$ bits, the integer part of $2XY/S$, is $c$ or $c+1$ with
$c=\lfloor2x_0y/S\rfloor$, because $0\le yU/(QR)<1$. Its parity is $M_a$, the first digit of $j$. Hence, given $\xi$,
each of $M_1,\dots,M_{a-1}$ is a constant, $M_a$ or $\neg M_a$, and $P=\pi_\xi(j_{<m})$ for a map $\pi_\xi$ that uses
only constants and literals of the first $m-1$ digits $j_{<m}$ of $j$.

\paragraph{The reference interval.} Replace $U$ by $V$, uniform on the real interval $[0,R)$ given $\xi$, and put
$j'=\lfloor\nu Z(V)\rfloor$. For $y>0$, $Z(V)$ is uniform on a circular arc of length $y/(2Q)$, so the law of $j'$ on
the digit cube $\{0,1\}^m$ has density at most $2Q/y$ with respect to the uniform law. Moreover, for every function $g$
of the first $m-1$ digits,
\begin{equation}\label{eq:ac0-pairs}
\bigl|\Pr[g(j'_{<m})=j'_m\mid\xi]-\tfrac12\bigr|\le2Q/(y\nu).
\end{equation}
Indeed the cells $2i$ and $2i+1$ of $\mathbb Z/\nu$ agree in their first $m-1$ digits ($m\ge2$) and differ in the last;
a pair covered completely by the arc contributes equally to both outcomes, and each of the at most two incomplete pairs
changes the success probability by at most half the mass of one cell, $2Q/(y\nu)$. No restriction on $g$ is needed.

\paragraph{Digit moments.} For $I\subseteq[m]$ let $\omega_I(z)$ be the product of the signs $(-1)^{\text{digit}}$ of
$\lfloor\nu z\rfloor$ at the positions in $I$, with $\omega_\emptyset=1$. We show
\begin{equation}\label{eq:ac0-moments}
\E_\xi\bigl|\E[\omega_I(Z(U))\mid\xi]-\E[\omega_I(Z(V))\mid\xi]\bigr|\le C\lambda^{-1}+C\lambda^{-2}[C(N+1)]^{|I|+1}.
\end{equation}
Write $G_y(h)=R^{-1}\sum_{u<R}\mathrm e(hyu/(2QR))$ and $\Phi_y(h)=\int_0^1\mathrm e(hyv/(2QR))\,dv$. Writing
$V=U+V_0$ with $V_0$ uniform on $[0,1)$, the conditional expectations of $\mathrm e(hZ(U))$ and $\mathrm e(hZ(V))$ are
$\mathrm e(hx_0y/S)G_y(h)$ and $\mathrm e(hx_0y/S)G_y(h)\Phi_y(h)$. Let $K=\nu\lambda^2$.

\emph{(i) Every frequency.} For $0<|h|\le K$, $\E_y|G_y(h)(1-\Phi_y(h))|\le C\lambda^{-2}$. \emph{(ii) High
frequencies.} For $\lambda^6<|h|\le K$, $\E_y|G_y(h)|\le C\lambda^{-2}$. For $|h|\le\lambda^6$, (i) follows from
$|1-\Phi_y(h)|\le\pi|h|y/(2QR)\le\pi|h|/(2R)\le\frac\pi2\lambda^{-2}$. For (ii), which gives (i) on the remaining range
because $|1-\Phi_y|\le2$, let $\delta=\lambda^{-4}$. The $Q$ numbers $hy/(2QR)$ have spacing $|h|/(2QR)$ and lie in an
interval of length at most $|h|/(2R)$. At most $|h|/(2R)+2$ integers have a closed $\delta$-neighbourhood meeting this
interval, and each neighbourhood contains at most $1+2\delta\cdot2QR/|h|$ of the numbers. So the fraction of $y$ with
$\|hy/(2QR)\|\le\delta$ is at most
\[
\frac{|h|}{2QR}+\frac2Q+2\delta+\frac{8\delta R}{|h|}\le\lambda^{-6}+\frac2R+2\lambda^{-4}+8\lambda^{-2},
\]
using $|h|/(2QR)\le K/(2QR)=\nu\lambda^{-6}/(2Q)\le\lambda^{-6}$ (here $\nu\le2Q$ is used), $Q\ge R$ and
$|h|>\lambda^6$. For the other $y$, summing the geometric series and $|\sin\pi t|\ge2\|t\|$ give
$|G_y(h)|\le(2R\delta)^{-1}=\lambda^{-4}/2$.

\emph{(iii) Mass near digit boundaries.} For $W=Z(U)$ or $W=Z(V)$, $\Pr[\|\nu W\|\le1/\lambda]\le C/\lambda$. Let
$l=\lambda/2$ and $F_l(t)=l^{-1}|\sum_{i=0}^{l-1}\mathrm e(it)|^2=\sum_{|i|<l}(1-|i|/l)\mathrm e(it)$, the Fej\'er
kernel. On the event, $F_l(\nu W)\ge4l/\pi^2$. Every nonzero frequency of $F_l(\nu\,\cdot)$ is $i\nu$ with
$1\le|i|<l$, and $\lambda^6<R\le\nu\le|i\nu|<K$; here $m\ge\ell$ is used. By (ii) each
$|\E\,\mathrm e(i\nu W)|\le\E_y|G_y(i\nu)|\le C\lambda^{-2}$, under either law. So
$\Pr[\|\nu W\|\le1/\lambda]\le\frac{\pi^2}{4l}(1+2lC\lambda^{-2})\le C/\lambda$.

\emph{(iv) Assembly.} The function $z\mapsto\omega_I(z)$ on the circle is a step function of $\lfloor\nu z\rfloor$, and
its Fourier coefficient at $h\in\mathbb Z$ is $\widehat\omega(h\bmod\nu)\cdot\nu\int_0^{1/\nu}\mathrm e(-ht)\,dt$,
where $\widehat\omega$ is the normalised coefficient on $\mathbb Z/\nu$. By \cref{lem:ac0-digits}, applied with $m$
digits, $\sum|\widehat\omega|\le[C(1+m/|I|)]^{|I|}\le[C(N+1)]^{|I|}$. The second factor has modulus at most
$\min\{1,\nu/(\pi|h|)\}$, and its sum over the $|h|<K$ in one residue class is at most $C(1+\log\lambda)\le C(N+1)$.
Let $F_K$ be the Fej\'er kernel of order $K$, with $\widehat F_K(h)=(1-|h|/K)_+$ and $F_K(t)\le1/(4K\|t\|^2)$. Then
$F_K*\omega_I$ has at most $[C(N+1)]^{|I|+1}$ total coefficient weight, all at $|h|<K$. The difference of its two
conditional expectations is $\sum_{0<|h|<K}\widehat{F_K*\omega_I}(h)\,\mathrm e(hx_0y/S)G_y(h)(1-\Phi_y(h))$. Take
absolute values termwise, average over $\xi$, and apply (i). This gives the second term of \eqref{eq:ac0-moments}.
The phase $\mathrm e(hx_0y/S)$ has modulus $1$ and drops out. If $\|\nu z\|>1/\lambda$, every shift of circular
length below $1/(\nu\lambda)$ preserves all digits of $\lfloor\nu z\rfloor$, so
$|\omega_I(z)-F_K*\omega_I(z)|\le2\int_{\|t\|\ge1/(\nu\lambda)}F_K\le\nu\lambda/K=1/\lambda$; and always the error is
at most $2$. By (iii) the mean smoothing error is at most $C/\lambda$ under each law. This proves
\eqref{eq:ac0-moments}.

\begin{proof}[Proof of \cref{thm:ac0-hiddenblock}]
Let $C_\xi$ be the selected circuit, and let $f_\xi(j)=[C_\xi(\pi_\xi(j_{<m}))=j_m]$ on $\{0,1\}^m$. Substituting
constants and literals for the inputs of $C_\xi$ adds no gates, and $\neg C_\xi$ has the same depth and size by De
Morgan's laws. So $f_\xi=(C_\xi\wedge j_m)\vee(\neg C_\xi\wedge\neg j_m)$ has depth $d+2$ and size at most
$s'=2N^b+3$. So bounded independence is invoked at depth $d+2$ (at depth $4$ when the theorem is used for DNFs). The
conditional success probability of $C_\xi$ is $\E[f_\xi(j)\mid\xi]$.

Let $\varepsilon=N^{-A-3}$ and $k=r_{d+2}(s',\varepsilon)$. By Braverman's bound, $k\le C_{d,b,A}(\log N)^{O(d^2)}$,
and $k$ does not depend on $\ell$, $a$ or $\xi$. \Cref{lem:ac0-sandwich} gives polynomials $q_l\le f_\xi\le q_u$ of
degree at most $k$ on $\{0,1\}^m$ with uniform gaps at most $\varepsilon$ and every coefficient at most $1+\varepsilon$
in modulus. They may depend on $\xi$. For $y>0$, by the density bound,
\[
\E[f_\xi(j)\mid\xi]\le\E[q_u(j)\mid\xi]=\E[q_u(j')\mid\xi]+\Delta_u\le\E[f_\xi(j')\mid\xi]+2Q\varepsilon/y+\Delta_u,
\]
with $\Delta_u=|\E[q_u(j)-q_u(j')\mid\xi]|\le(1+\varepsilon)\sum_{|I|\le k}|\E[\omega_I(Z(U))-\omega_I(Z(V))\mid\xi]|$,
and symmetrically from below with $q_l$. With \eqref{eq:ac0-pairs} for $g=C_\xi\circ\pi_\xi$, the absolute
conditional advantage is at most $2Q/(y\nu)+2Q\varepsilon/y+\Delta_u+\Delta_l$. Average over $\xi$. Each coefficient
is bounded by $1+\varepsilon$ before averaging, and \eqref{eq:ac0-moments} is then applied to each fixed $I$; there are
at most $(m+1)^k$ sets $I$. With $\Pr[y=0]=1/Q$ and $Q^{-1}\sum_{y=1}^{Q-1}Q/y\le C(N+1)$, the mean absolute
conditional advantage is at most
\[
Q^{-1}+C(N+1)(\varepsilon+\nu^{-1})+2\lambda^{-1}[C'(N+2)]^{2k+2}
\]
for an absolute $C'$. Let $\ell$ be the least multiple of $8$ with
$\ell/8=\log_2\lambda\ge(2k+2)\log_2(C'(N+2))+(A+3)\log_2N+2$. Then $\ell\ge16$, the last term is at most
$N^{-A-3}/2$, $\nu^{-1}\le\lambda^{-8}\le N^{-A-3}$, and the total is at most $N^{-A}$ for $N\ge N_0(d,b,A)$. Also
$\ell\le C_{d,b,A}(\log N)^{c_d}$ with $c_d=O(d^2)$. Exchanging $X$ and $Y$ changes nothing, since $XY=YX$.
\end{proof}

\noindent\emph{The size bound is needed, and $\ell$ must be at least $b\log_2N$.} If $m\ge\ell+3$ and $y\ge Q/2$,
consecutive points of the progression $Z(u)$ are at least $2/\nu$ apart, on an arc shorter than $\frac12$, so distinct values of $U$ give distinct
$j_{<m}$, which $P$ contains. The exposure can then select a DNF on $P$ with at most $R$ terms, one for each value of
$U$ with $M_T=1$, that computes $M_T$ whenever $y\ge Q/2$. Its mean absolute conditional advantage is at least
$\frac14$. If $\ell\le b\log_2N-1$, this DNF has size at most $N^b$.

\begin{proof}[Proof of \cref{thm:ac0-mixeddnf}]
A CNF is the negation of a DNF with as many terms, and has the same absolute advantage, so let
$F=t_1\vee\dots\vee t_s$ be a DNF, $s\le N^b$, possibly chosen as a function of $Y$. Remove contradictory terms. Let
$w=\lceil\log_2(2sN^3)\rceil$, and let $F'$ be $F$ without the terms that contain more than $w$ distinct variables of
$X$. Since $X$ is uniform given $Y$, $\Pr[F\ne F']\le s2^{-w-1}\le N^{-3}/4$, whatever the other literals of the
deleted terms are.

Let $\ell$ be the block length of \cref{thm:ac0-hiddenblock} for $d=2$, size exponent $b+1$ (a DNF with at most $s$
terms has at most $s+1\le N^{b+1}$ gates) and $A=3$. Place $J=\lfloor(\lfloor T/2\rfloor-1)/\ell\rfloor$ disjoint
blocks of $\ell$ consecutive bits of $X$ in positions $2,\dots,\lfloor T/2\rfloor$. Each block is bits
$a+1,\dots,a+\ell$ with $a\ge1$ and $a+\ell\le T$, and $J\ge\alpha N/(4\ell)\ge1$ for $N\ge N_0(\alpha,b)$. Pick a
block $\beta$ uniformly at random, independently of $(X,Y)$, and let $F''$ be $F'$ without every term that contains a
variable of $\beta$. Then $F''\le F'$ pointwise. If $F'$ accepts an input, its first satisfied term (in a fixed order)
contains at most $w$ variables of $X$, so it meets at most $w$ blocks and survives unless $\beta$ is one of them.
Hence $\Pr_\beta[F''\ne F']\le w/J$ on every input. For each fixed $\beta$, $F''$ does not read the block $\beta$.
Given $Y$ and the other bits of $X$, it is a DNF on $P$ with at most $s$ terms. So \cref{thm:ac0-hiddenblock} bounds
its advantage by $N^{-3}$. Therefore the advantage of $F$ is at most
$N^{-3}+N^{-3}/4+w/J\le C_{\alpha,b}\,\ell\log N/N$, since $w=O(b\log N)$. With $\ell\le C_b(\log N)^{c_2}$ this is the
claim with $c=c_2+1$.
\end{proof}

\noindent\emph{Where the argument stops.} At depth three an AND of DNFs is accepted only if every child has a satisfied
term. The union of these witnesses can meet every block, so the loss $w/J$, which controls one DNF, does not survive a
union bound over polynomially many gates. \Cref{thm:ac0-hiddenblock} alone says nothing about a residual circuit that
still reads the hidden block.

% Appendix section (moved from the main body): chain of thought for the full product. Input from appendix.tex.
% Proofs in app-cot.tex.
\section{Chain of thought for the full product}\label[appendix]{sec:cot}

\paragraph{The setting.} In the definition of~\cite[Defs.~2.1--2.2]{AHRH25}, a \emph{chain of thought} (CoT) for a
function $f$ maps each input $x$ to a string $g(x)$ over a vocabulary $\Xi$ that ends with $f(x)$; it is
\emph{expressible} in UHAT if a family $(\mathcal T_n)$, one transformer per total length $n=|x\,g(x)|$, with at most $L$
layers and $H$ heads, outputs at every position $|x|+i-1$ of $x\,g(x)$ the next symbol $g(x)_i$. Being correct on $x$
means that $\mathcal T_{|x g(x)|}$, run autoregressively on $x$, writes $g(x)$. For multiplication we take $f(x)$ to be
the zero-padded $2N$-bit product of the operands, and $x=x(X,Y)$ is the input format of \cref{sec:transformers}: the
$2N$ operand bits in a fixed order and $O(1)$ separators at fixed positions. So $g(x)$ consists of $t(x)\ge0$ \emph{auxiliary} symbols followed by
$M_1\cdots M_{2N}$; the length $t(x)$ may depend on $x$, and the vocabulary may grow with $N$. Amiri Bavandpour et
al.\ proved that a CoT ending with a single middle bit needs length $\Omega(N)$~\cite[Thm.~4.5]{AHRH25}, and their
Lemma~C.6 asserted, under \cref{conj:C5}, a lower bound for CoTs that write the whole product. Our \cref{thm:main}
removes the conjecture from the average-case route at every fixed depth, and a separate argument handles every fixed
depth in the worst case. The average-case theorem allows the stronger family in which $\mathcal T_n$ is a top-$k$ model
of \cref{def:topk} (one model per total length); UHAT is the case $k=1$.

\ificml\noindent The theorem for uniform operands, \cref{thm:cot-avg}, is stated in \cref{sec:icml-cot}.\else
% Statement of thm:cot-avg, shared: input by sec-cot.tex (FOCS and named preprint, appendix) and by icml.tex (ICML 2027
% main body). Do not edit the statement here without the statement gate.
\begin{theorem}[full product, every fixed depth, uniform operands]\label{thm:cot-avg}
Fix integers $H,k\ge1$ and the input format of the setting above. Let $D$ be the constant of \cref{thm:main},
$c=3.34\sqrt{1+\ln2}+1$, $\rho=\min\{1,\,D\ln2/(40960\,c^2),\,\ln2/12\}$ and $c'=\ln2/4$. For $L\ge2$ put
$q=2^{1-L}$ and $c_L=\rho\,(1+Hk)^{-2L}$. Consider any family $(\mathcal T_d)$ of $L$-layer top-$k$ models
(\cref{def:topk}; at most $H$ heads per layer, fixed positional tie priority, arbitrary width, precision and
positional encodings, arbitrary finite vocabularies, one model per total length $d$). For $N\ge N_0(L,H,k)$ put
\[
 b=\Bigl\lfloor\frac{c_LN^q}{(\ln N)^{2-q}}\Bigr\rfloor,\qquad s=\Bigl\lfloor\frac{c_Lb}{\ln N}\Bigr\rfloor .
\]
For independent uniform $N$-bit operands $X,Y$, the probability that for some integer $0\le t\le s$ the model
$\mathcal T_{|x|+t+2N}$, run autoregressively on $x$, emits $t$ symbols followed by the zero-padded $2N$-bit product
(most significant bit first) is at most $e^{-c'b}$. Hence a chain of thought computed by such a family on a set of
inputs of probability $\delta$ has more than $s$ auxiliary symbols on a set of probability at least $\delta-e^{-c'b}$.
For fixed $L,H,k$, $s=\Theta_{L,H,k}(N^{q}/(\ln N)^{3-q})=N^{2^{1-L}-o(1)}$.

Moreover, for fixed $H,k$ a single threshold $N_0(H,k)$ suffices for all integers $L$ with
$2\le L\le\log_2\ln N-\log_2\ln\ln N-\lceil\log_2(4+2\log_2(1+Hk))\rceil$, with the same $\rho$, $c'$ and the
displayed $c_L$; in this range $s\ge(\rho^2/4)(\ln N)^5$.
\end{theorem}
\fi

\noindent The power of $N$ in $s$ is $2^{1-L}$: $1/2$, $1/4$, $1/8$ at $L=2,3,4$. At depth three the power is $1/4$,
independently of $H$ and $k$. The threshold
$N_0$ is asymptotic: with $H=k=1$ and $D=1/16$ the proof gives $s\ge1$ only from $N\approx10^{45}$, $10^{100}$ and
$10^{220}$ at $L=2$, $3$ and $4$, and with $D=1/6$ from $N\approx10^{43}$, $10^{96}$ and $10^{213}$
(\texttt{checks/cot\_depth\_thresholds.py}).

\paragraph{Proof idea} (\cref{app:cot}). Fix an auxiliary string; its symbols become constant tokens. Correct
generation forces the teacher-forced predictions of the $b$ digits $M_{N-b+1},\dots,M_N$ to be right simultaneously.
Each prediction sign is uniformly close, on the whole formal cube, to a bounded function of Walsh norm
$e^{O(N^{1-q}(\ln N)^{2-q})}$ (\cref{lem:topk-poly}, applied to the readout ``the next symbol is $1$''). For every
$J\subseteq\{N-b+1,\dots,N\}$ one polynomial truncated in total degree amplifies the product of the $|J|$ prediction
signs (\cref{lem:joint}), and \cref{thm:main} at the top index of $J$ bounds every term of the expansion of the joint
success indicator, so the whole block is right with probability $e^{-\Omega(b)}$. The block length $b$ is set by
balancing the Walsh norm of the joint polynomial against $2^{-DN/2}$. The model can emit only $(10N)^{(1+Hk)^L}$
different next symbols at each step, whatever its vocabulary (\cref{lem:nexttok}), so a union bound over the
auxiliary strings costs $e^{O(s\ln N)}$.

\begin{theorem}[full product, every fixed depth, worst case]\label{thm:cot-worst}
Fix $L$ and $H$. If a CoT for $N$-bit multiplication, with the product as its $2N$-bit suffix, is expressible in UHAT
with at most $L$ layers and $H$ heads~\cite[Def.~2.2]{AHRH25}, then its worst-case number $s$ of auxiliary symbols
satisfies
\[
s+1=\Omega_{L,H}\Bigl(\Bigl(\frac{\sqrt{N/\log N}}{(\log N)^{2L+2}}\Bigr)^{1/(2L+3)}\Bigr).
\]
In particular, the UHAT sentence of Lemma~C.6 of~\cite{AHRH25} holds in this worst-case form without \cref{conj:C5}.
\end{theorem}

\paragraph{Proof idea.} On the Sidon embedding of \cref{lem:sidon} one block of $b=\Theta(\log N)$ product digits holds
the full binary count $|z|$ of $r=\Theta(\sqrt{N/\log N})$ free bits, and every earlier product digit is a constant or
a single $z_i$. For each candidate auxiliary length $t\le s$, guessing the auxiliary symbols and the count digits and
verifying all next-token tests in parallel gives a circuit of depth $2L+2$ and size $e^{O((s+1)\log N)}$ for the count
that the model would write at length $t$. These $s+1$ candidates form a list that contains $|z|$ for every $z$. Along a
random monotone chain through the middle of the cube the list must change often, which forces large total influence
(\cref{lem:listcover}); Boppana's bound on the influence of small constant-depth circuits~\cite{Boppana97} then forces
$s$ to be polynomial in $N$.

\paragraph{Relation to Lemma C.6 of~\cite{AHRH25}.} The route of~\cite{AHRH25} from \cref{conj:C5} to a
chain-of-thought lower bound is an outline. Its Lemma~C.6 (arXiv v3) asserts, under the conjecture, average
sensitivity $\Omega(N)$ for every implementation of $(X,Y,M_{<N})\mapsto M_N$, with the uniform measure on all $3N-1$
bits; this fails for the implementation that is $0$ off the consistent inputs (\cref{app:c6}). Our applications therefore
work on consistent inputs. We do not claim that the lower bound of~\cite{AHRH25} is false, and \cref{thm:cot-worst}
proves its transformer conclusion in a worst-case form.

\paragraph{What is new, and what is not.} For $s=0$, and for $O(1)$ auxiliary tokens at fixed positions from a fixed
vocabulary, the worst-case statement follows from Buss's embedding~\cite{Buss92} and the inclusion of UHATs in
$\mathrm{AC}^0$~\cite{HAF22}; this is \cref{thm:worstcase}. At a \emph{fixed} auxiliary length a routine combination of
these with the device of fixing the whole CoT string, which makes the event ``the model emits this string'' an
$\mathrm{AC}^0$ indicator~\cite[Lemma~B.5]{MKSH26}, plausibly gives a polynomial bound as well; we did not find it in print
and do not claim it. What \cref{thm:cot-worst} adds is the case of the definition itself: an auxiliary length that
depends on the input, so that the positions of the product digits are unknown, and a vocabulary that may grow with $N$;
the list-cover argument handles both. For average-case CoT length, Mohammadkhani, Krishna, Sarrof and
Hahn~\cite[Thm.~3.1, Eq.~(3), Cor.~B.6]{MKSH26} prove a near-linear lower bound for fixed-precision transformers that
end with the single middle bit of a product, with a fixed finite alphabet and no product digits in the trace; their
per-string indicator device is related to our guess-and-verify step. Their bound does not apply when the trace contains
the earlier product digits, which carry about $T$ bits of information and are exactly the teacher-forced context that
\cref{thm:main} is built to handle. \Cref{thm:cot-avg} covers that case on average, at every fixed depth. From $L=6$
on, the worst-case exponent $1/(4L+6)$ of \cref{thm:cot-worst} exceeds the average-case exponent $2^{1-L}$.

\fi
% Appendix: proofs for the chain-of-thought section (sec-cot.tex). Input from appendix.tex.
\section{Proofs for \vsw{Section~\ref{sec:icml-cot}}{Appendix~\ref{sec:cot}}}\label[appendix]{app:cot}

\ifoldpullback
Throughout, $\hat M_i=2M_i-1$, $\log$ is natural, $L$ and $H$ are fixed, and constants may depend on them. A context
is \emph{literal} if every position holds either a fixed token or one of two tokens chosen by one formal bit; the
formal cube $\{0,1\}^m$ consists of the free bits. For a function $P$ on it, $A(P)$ is the Walsh norm of
\cref{sec:overview}. Contexts in which the auxiliary symbols are fixed tokens are literal: the operand bits and the
supplied earlier product bits are the formal bits. A fixed position carries no formal bit, so \cref{lem:topk-poly},
which allows at most one variable bit per position, applies to literal contexts, and inequality~\eqref{eq:walsh} holds
for every $f$ on the formal cube, since fixed tokens contribute no character.

\subsection{Three tools}

\begin{lemma}[next tokens]\label[lemma]{lem:nexttok}
Fix a UHAT with $L\ge1$ layers and $H$ heads and a literal context of length $n$. The next output takes at most
$n^{C_{L,H}}$ values, whatever the vocabulary; for an $L$-layer top-$k$ model (\cref{def:topk}) with at most $H$ heads
per layer it takes at most $(2n)^{(1+Hk)^L}$ values. For a UHAT, every Boolean predicate of the next output is computed, as a function of
the formal bits, by an unbounded fan-in AND/OR circuit over literals of depth at most $2L$ and size at most
$n^{C_{L,H}}$.
\end{lemma}

\begin{proof}
At layer $0$ the number $V_0$ of pairs (position, possible state) is at most $2n$. A state at layer $\ell+1$ is a
function of the query pair and the at most $H$ winning pairs of layer $\ell$, so $V_{\ell+1}\le V_\ell^{H+1}$ and
$V_L\le(2n)^{(H+1)^L}$; the output is a function of the last state, which proves the first claim. For a top-$k$
model the query position fixes the length of each head's list, and the new state is a function of the query pair and
the at most $Hk$ selected pairs, in order; so $V_{\ell+1}\le V_\ell^{1+Hk}$, which proves the second claim. For each pair
$(j,\sigma)$ of layer $\ell$ let $I_{j,\sigma}$ indicate that the state at $j$ is $\sigma$. At layer $0$ these are
constants or literals. Given layer $\ell$, enumerate the tuples consisting of a query pair $(i,\sigma)$ and one proposed
winner $(j_h,\tau_h)$, $j_h\le i$, for each head. All scores are constants for a fixed tuple, so the tuple occurs exactly
when $I_{i,\sigma}$ and every $I_{j_h,\tau_h}$ hold and $\neg I_{k,\rho}$ holds for every visible pair $(k,\rho)$ that
outranks $(j_h,\tau_h)$ for head $h$ (score, then the fixed priority). Then $I_{i,\sigma'}$ at layer $\ell+1$ is the OR,
over the tuples that produce $\sigma'$, of these ANDs, and $\neg I_{i,\sigma'}$ is its dual. Inductively $I$ is
OR-topped and $\neg I$ is AND-topped of depth $2\ell$; an AND gate absorbs the AND-topped inputs, so each layer adds two
levels, and the readout predicate is an OR of final indicators at the last position. The numbers of tuples and gates
are polynomial in $n$ for fixed $L,H$.
\end{proof}

\fi
\ificml\icmlcotpartialproof\icmlrecordproof\subsection{Constant vocabulary and Boolean circuits}\label[appendix]{app:icml-cot-circuits}

Use bounded-fan-in Boolean circuits with arbitrary fan-out and let
$C_{\rm mult}(N)$ be the minimum gate count for the full $2N$-bit product.
Fix positive architecture bounds $L,H,k$ and a constant token alphabet
containing $0,1$. With input-independent routing and an input-independent
auxiliary length, exact multiplication has minimum auxiliary length
$\Theta(C_{\rm mult}(N))$.
Indeed, for $N\ge2$ the $2N$ output functions are distinct and none
is an input wire or constant. Products of powers of two isolate
each bit below the top one, and $(2^N-1)^2$ attains the top bit.
Each output vanishes when one operand is zero, so it cannot copy
an input. Thus $C_{\rm mult}(N)\ge 2N$.

Indeed, the record at each readout has at most $D=(1+Hk)^L$ fixed
ancestor positions at layer zero. For a fixed total length, the next
symbol is therefore a fixed function of at most $D$ symbols from a
constant alphabet. After a fixed binary encoding this function uses a
constant number of bounded-fan-in gates, by its truth table. Unrolling
all $s+2N$ generated tokens gives $C_{\rm mult}(N)\le O(s+N)$.
\Cref{thm:cot-avg} gives $s=\Omega(N)$ for exact multiplication.

Conversely, topologically order a size-$C_{\rm mult}(N)$ circuit.
Use the staged-copy schedule in \cref{app:icml-cot-packed} with one
binary symbol per gate result and a constant number of binary staging
symbols per gate. The final $2N$ token positions read the output wires in
most-significant-first order. This uses $O(C_{\rm mult}(N)+1)$ auxiliary
symbols with one layer and one head. The minimum gate count is nondecreasing in $N$
by restriction, so fixed padding can give distinct total lengths across widths,
including for layouts with varying bounded separator counts.
A linear scratchpad in this fixed-length model is equivalent to a
linear-size Boolean multiplier.

\paragraph{Circuit size of fast multiplication.}
Theorem~1.1 of \citet{HarveyHoeven21} is stated for multitape Turing
time. The same algorithm gives $C_{\rm mult}(N)=O(N\log N)$ by
the direct circuit accounting in \citet[\S5]{Harvey26MultiplicationNote}.
For fixed input length, its fixed rearrangements, shifts and zero
padding can be implemented by wiring and constants. The butterfly
additions and subtractions on $p$-bit fixed-point numbers use $O(p)$
bounded-fan-in gates via ripple carry. The nonrecursive arithmetic
cost is $O(N\log N)$ and the recursive circuit cost obeys the same
size recurrence, with at most $KN/N'$ calls of size
$N'=N^{1/d+o(1)}$ for an absolute $K$ and a fixed $d>K$.
Then $(KN/N')N'\log N'=(K/d+o(1))N\log N$, and the recurrence
gives the claimed gate bound. The staged-copy simulation therefore
uses $O(N\log N)$ fixed-routing constant-alphabet auxiliary tokens.

The conditional $\Omega(N\log N)$ circuit lower bound of
\citet[Thm.~1]{AfshaniEtAl19} assumes the Li--Li undirected
multiple-unicast conjecture. \citet{BravermanHe26} and
\citet{ZhangLiLi26} report counterexamples in recent arXiv preprints.
An unconditional superlinear bound for the fixed-routing
constant-vocabulary problem would also give one for Boolean
multiplication circuits.

\paragraph{Input-dependent selection.}
With an externally chosen total length there may be one circuit per
candidate length. A per-length simulation alone gives no circuit for
selecting among those candidates. Adaptive attention changes the
simulation. For example, fix an arbitrary Boolean matrix $A$ of size
$N\times N$ and consider the $N$ outputs $\bigvee_{j:A_{ij}=1}x_j$.
One adaptive head uses keys $(e_j,x_j,0)$ on raw inputs and
$(0,z_j,1)$ on generated positions with bit $z_j$. At output $i$ its
query $(2A_{i,*},1,-2)$ gives input score $2A_{ij}+x_j$ and suppresses
every generated position. For an empty row emit a fixed zero.
This family uses zero auxiliary symbols. Its $2^{N^2}$ distinct functions
are distinguished by single-one inputs. Circuits with $S$ bounded-fan-in
gates, $N$ inputs and $N$ outputs number at most
$2^{O((S+N)\log(S+N))}$ by encoding gates, predecessors, and output wires.
Some matrices therefore require $\Omega(N^2/\log N)$ Boolean gates.
The fixed-routing circuit simulation does not yield a linear
constant-vocabulary bound for adaptive multiplication.

\paragraph{One adaptive head.}
Let $D_{\rm mult}(N)$ be the minimum gate count for the full binary
product using literal decision-list gates \citep{Rivest1987DecisionLists}. A gate tests its input wires
or their negations in a fixed order and returns the bit attached to
the first true test, with a default. The test order and branch labels
are nonuniform data. For a fixed finite alphabet and fixed auxiliary
length, one adaptive top-$1$ layer with one head needs
$\Theta(D_{\rm mult}(N))$ auxiliary tokens. This holds for arbitrary
pair scores and for bilinear scoring with additive position and token
embeddings, unrestricted width and unrestricted local updates.

To unroll a generator, condition on the symbol at the readout slot.
This fixes its raw query, so all eligible (position, symbol) records
have a fixed score order. The first present record is the winner.
For each possible output symbol a literal list tests record presence
and returns the indicator that the winner emits that symbol.
Generated records use their stored one-hot indicators; raw bits use
literals. A constant-size multiplexer combines the hypothetical
queries. The alphabet is fixed, so unrolling costs $O(s+N)$ gates.

For the reverse simulation, emit both signs of each raw and
previously computed wire, using a fixed-position read to complement
it. An arbitrary literal list now tests only whether its selected
signed wires equal one. Give each tested position a distinct priority
$1,\ldots,h$, subtract $h+1$ from that position's score, then add
$h+1$ on token one. Put a fixed default record at score zero and
unlisted records below it. Position and token one-hot embeddings
implement the sum by a shared bilinear form with the query position
selecting the coefficients. A shared local update reads the query
position and selected record to emit the corresponding branch bit.
Writing the gates in topological order and then copying the designated
answer wires uses $O(N+D_{\rm mult}(N))$ scratch tokens.
As with the bounded-fan-in basis, the distinct, nontrivial output
functions imply $D_{\rm mult}(N)\ge 2N$, and the generation lower bound
gives $s=\Omega(N)$. These remove the additive $N$ in both directions.
Padding can ensure distinct total lengths across operand widths:
$D_{\rm mult}$ is nondecreasing by setting the highest input bits to
zero and dropping the corresponding leading output zeros. Whether these decision-list circuits have linear gate count for
multiplication remains open. Each bounded-fan-in Boolean gate uses
a constant number of literal lists, so $D_{\rm mult}=O(C_{\rm mult})$.
A superlinear lower bound for $D_{\rm mult}$ would therefore also
be a superlinear bounded-fan-in Boolean circuit lower bound.

\paragraph{An unrestricted accumulator.}
Any deterministic task on $n$ raw symbols from a finite alphabet,
with prescribed output length, has a one-layer, one-head top-one
implementation with at most $n$ auxiliary tokens. At step $i$, the
query row contains the accumulated word of length $i-1$, initially
empty. A fixed positional query reads input $i$, and the update
emits the old word with that symbol appended. The scratch vocabulary
contains all words up to length $n$. Each answer row reads the last
accumulator by fixed position and its update emits the required
coordinate of the task output. The score can be bilinear in
position one-hots. This proves a universal linear-token upper bound
with unrestricted local computation and finite length-dependent
vocabulary. Superlinear token bounds require additional restrictions.

Copying is simpler: each output row points directly to its input
coordinate and uses no scratch. Binary addition in increasing order
of significance uses no scratch with four fixed top-one heads.
At bit $i>0$, the literal readout row holds the previously emitted
digit $z_{i-1}$. Read the two operand bits at positions $i-1$ and
the two at $i$. With previous operand sum $d_{i-1}$, the carry
into position $i$ is zero for $d_{i-1}=0$, one for $d_{i-1}=2$,
and $1-z_{i-1}$ for $d_{i-1}=1$. Add the carry to the current
operand bits and emit parity; the initial carry is zero. A terminal
row uses zero for both nonexistent next operand bits. The genuine
prefix makes this predictor exact from five binary coordinates,
so its Fourier expansion has a nonnegligible signed correlation
with a degree-five character.

\paragraph{Adaptive routing saves a linear scratchpad on powers of two.}
Suppose $X=2^i,Y=2^j$ for unknown $0\le i,j<N$.
In either digit order one adaptive bilinear top-$1$ head generates the
full product with zero auxiliary tokens. With input-independent routing,
fixed auxiliary length and fixed $L,H,k$, the minimum is $\Theta(N)$.

For MSB-first output at exponent $t$, give the active $X$ record
priority $i$ and the active $Y$ record priority $t-j-1/2$.
Return one if $X$ wins. Any earlier output one takes precedence
over both operands and returns zero. Thus the test is $i+j\ge t$
until the first one, which occurs exactly at $t=i+j$.
For LSB-first use priorities $-i$ and $j-t-1/2$ to test $i+j\le t$.
These are bilinear scores with additive token and position embeddings:
the query position selects the row of position biases, and a common
token-one bonus larger than their range places every active record
above all zero records. The local map distinguishes the winning
position's role. No auxiliary symbols store the exponents.

For fixed routing let $D=(1+Hk)^L$. At exponent $t=N-1$ take the
union of the raw $X$ positions in the fixed layer-zero footprints
of the $s$ auxiliary readouts and of this particular answer readout.
Its size is at most $D(s+1)$. In MSB order compare
$(2^i,2^{N-1-i})$ with $(2^{i-1},2^{N-1-i})$ for $1\le i<N$.
All earlier output symbols are zero on both inputs. If the union
misses $\{i-1,i\}$, induction through the auxiliary symbols and then
the target readout makes their target predictions identical, although
their target digits differ. The union must hit every adjacent pair
and therefore has size at least $\lfloor N/2\rfloor$. In LSB order
replace $i-1$ by $i+1$ for $0\le i<N-1$. Conversely a bounded-fan-in
circuit encodes a one-hot word's index in $O(N)$ gates: at each node
of a balanced binary tree retain occupancy and the local binary index.
Nodes spanning $2^h$ positions cost $O(h)$ gates each, summing to
$O(N)$ (pad to a power of two). Add the two indices and expand the
sum into one-hot indicators by a binary prefix tree in $O(N)$ gates.
The fixed-routing staged simulation gives the upper bound.
\else
\noindent Outputs that are not a valid symbol are decoded by a fixed deterministic rule; this changes no valid trace.

\begin{lemma}[joint amplification]\label[lemma]{lem:joint}
Let $a_1,\dots,a_b\in\{\pm1\}$ be functions on a common cube and $p_1,\dots,p_b$ real functions there with $|p_j|\le1$,
$|p_j-a_j|\le\eta$ everywhere, $2b\eta\le\frac12$ and $\log A(p_j)\le B$ for some $B\ge1$. For every integer
$m\ge0$ there is $P$ with $|P|\le1$, $|P-\prod_ja_j|\le2(2b\eta)^{m+1}$ everywhere and $\log A(P)\le4(b+m)B$.
\end{lemma}

\begin{proof}
Put $x_j=1-p_j^2$. Since $\eta<1$, $\sgn p_j=a_j$ and $|p_j|\ge1-\eta$, so $x_j\in[0,2\eta]$ and
$a_j=p_j(1-x_j)^{-1/2}$. With $c_i=4^{-i}\binom{2i}i\in[0,1]$, the coefficients of $(1-x)^{-1/2}$, let
\[
P=\Bigl(\prod_jp_j\Bigr)\sum_{\nu\in\mathbb Z_{\ge0}^b,\ |\nu|\le m}\ \prod_jc_{\nu_j}x_j^{\nu_j}.
\]
The full series equals $\prod_jp_j\prod_j(1-x_j)^{-1/2}=\prod_ja_j$. All its terms are nonnegative multiples of
$\prod_jp_j$, so every partial sum has modulus at most $|\prod_ja_j|=1$. The terms with $|\nu|=k$ sum to at most
$(\sum_jx_j)^k\le(2b\eta)^k$ (multinomial expansion, $c_i\le1$), and the tail $k>m$ is at most $2(2b\eta)^{m+1}$. For
the norm, $A(x_j)\le1+A(p_j)^2\le2e^{2B}$, $A(\prod_jp_j)\le e^{bB}$ and there are $\binom{b+m}b\le2^{b+m}$ terms, so
$A(P)\le2^{b+m}e^{bB}(2e^{2B})^m\le e^{4(b+m)B}$.
\end{proof}

\ificml\else
\begin{lemma}[a list that covers the weight]\label[lemma]{lem:listcover}
Let $F_1,\dots,F_k:\{0,1\}^r\to\{0,1\}^b$, read as integers, be such that $|z|\in\{F_1(z),\dots,F_k(z)\}$ for every
$z$. Let $I(f)=2^{-r}\sum_z\sum_{i\le r}\mathbf 1[f(z)\ne f(z^{\oplus i})]$ be the total influence of a bit $f$, and
$F_{j,\ell}$ the $\ell$-th bit of $F_j$. There is an absolute $c>0$ such that for all large $r$ either $k\ge c\sqrt r$ or
$\sum_{j\le k}\sum_{\ell\le b}I(F_{j,\ell})\ge cr/k$.
\end{lemma}

\begin{proof}
Let $w_0=\lfloor r/2\rfloor$ and $W=\lfloor\sqrt r/4\rfloor$. Take a uniformly random ordering of the coordinates and the
monotone chain that sets them to $1$ in that order; let $v_w$ be its vertex of weight $w$. On a run of consecutive
vertices $v_w$ on which the list $(F_1,\dots,F_k)$ is constant, at most $k$ values $w$ can be covered, so the run has at
most $k$ vertices. Hence among $v_{w_0},\dots,v_{w_0+W}$ the list changes at least $(W+1)/k-1\ge\sqrt r/(8k)$ times
when $k\le\sqrt r/8$, and every change changes some bit $F_{j,\ell}$. The step of the chain from weight $w$ to $w+1$ is a
uniform edge between the slices $w$ and $w+1$. Let $p_w(f)$ be the fraction of those edges on which $f$ changes, and
$\pi_w=\binom rw(r-w)/(r2^{r-1})=2^{-(r-1)}\binom{r-1}w$ the fraction of all edges of the cube that join the two slices.
Counting bichromatic edges gives $\sum_w\pi_wp_w(f)=I(f)/r$. By Stirling's formula $\pi_w\ge c_2/\sqrt r$ for
$w_0\le w\le w_0+W$ and large $r$, with an absolute $c_2>0$. So the expected number of changes of $f$ on the segment is at
most $\sum_w p_w(f)\le(\sqrt r/c_2)\sum_w\pi_wp_w(f)=I(f)/(c_2\sqrt r)$. Summing over the $kb$ bits and comparing with
the deterministic lower bound gives $\sum I(F_{j,\ell})\ge c_2r/(8k)$.
\end{proof}

\noindent We use Boppana's theorem~\cite{Boppana97}: an unbounded fan-in AND/OR circuit over literals of depth $d$ and
size $S$ has total influence at most $C_d(\log(2S))^{d-1}$, with $C_d$ depending only on $d$.

\fi
\subsection{Proof of \cref{thm:cot-avg}}

Write $w=\ln N$, $a=1+Hk$, $A=a^L$ (a number; $A(f)$ with an argument is the Walsh norm) and $F=1280c^2$, so that
$\rho=\min\{1,D\ln2/(32F),\ln2/12\}$ and $c_LA^2=\rho$. Only indices $N/2\le T\le N$ enter, so any value of $D$ valid
there may be used. Each size condition below holds for $N\ge N_0(L,H,k)$, which depends also on the number of separators
of the input format. Since $c_L\le1$ and $q\le\frac12$, $s\le b\le\sqrt N$.

\emph{One auxiliary string, one approximant per digit.} Fix $0\le t\le s$, the model $\mathcal T=\mathcal
T_{|x|+t+2N}$ and a string $\tau$ of length $t$, and let $I=\{N-b+1,\dots,N\}$. For $i\in I$ let $g_i$ be $+1$ if the
next symbol of $\mathcal T$ on the literal context $x\,\tau\,z_1\cdots z_{i-1}$ is $1$, and $-1$ otherwise (an invalid
symbol counts as $-1$); it is a function of the $2N+i-1$ formal bits. If $\mathcal T$, run autoregressively on $x$, emits
$\tau$ and then the product, then $g_i=\hat M_i$ for every $i\in I$ on the true context. Since every head at a position
selects among earlier or equal positions, the first $n_i=|x|+t+i-1$ positions of $\mathcal T$ form an $n_i$-position
top-$k$ model with the same scores, encodings and readout, and $g_i$ is a $\{\pm1\}$-valued function of its top state at
the last position. For large $N$, $2N\le n_i\le5N$ and $i\ge N/2$.

Put $n=5N$, $K=4A\ln(4n)$, $v=2K$ and $K_i=4A\ln(4n_i)$. As $A\ge2^L\ge\ln(2L)$, for large $N$ we have
$K_i+\ln(2L)\le K+\ln(2L)\le v\le2N\le n_i$ and $v\ge4w+\ln2$. The last sentence of \cref{lem:topk-poly}, at $n_i$,
$K_i$ and $v$, gives $P_i$ on the formal cube with $|P_i-g_i|\le\epsilon:=e^{-v}/2$ everywhere and
$\ln A(P_i)\le4c^2n_i^{1-q}K_i^{2-2q}v^q$. This bound increases in $n_i$ and $K_i$, so
\[
\ln A(P_i)\le B:=4c^2nK^2\Bigl(\frac v{nK^2}\Bigr)^q\le F A^2N^{1-q}w^{2-q},
\]
the last step by $K\le8Aw$ (for $N\ge20$), $5^{1-q}2^q\le5$ and $(8A)^{2-q}\le64A^2$. Let $p_i=P_i/(1+\epsilon)$. Then
$|p_i|\le1$, $|p_i-g_i|\le\eta:=e^{-v}$ and $\ln A(p_i)\le B$; also $B\ge1$, and $2b\eta\le N^{-3}\le\frac12$ since
$b\le N$ and $v\ge4w+\ln2$.

\emph{Every parity of the block.} Let $\emptyset\ne J\subseteq I$ and $u=\max J$, and regard $p_i,g_i$, $i\in J$, as
functions on the formal cube of $x\,\tau\,z_1\cdots z_{u-1}$, ignoring the extra coordinates (which leaves Walsh norms
unchanged). \Cref{lem:joint} with $m=\lceil b/w\rceil$ gives $P_J$ with $|P_J|\le1$,
$|P_J-\prod_{i\in J}g_i|\le2(2b\eta)^{m+1}\le2e^{-3b}$ everywhere (as $w(m+1)\ge b$), and, since $m\le b$ and
$b\le c_LN^q/w^{2-q}$,
\[
\ln A(P_J)\le4(b+m)B\le8FA^2bN^{1-q}w^{2-q}\le8F\rho N\le\tfrac{D\ln2}4N .
\]
Let $\chi=\prod_{i\in J\setminus\{u\}}(2z_i-1)$; it reads only coordinates before $z_u$, and $A(\chi P_J)=A(P_J)$. Every
$g_i$, $i\in J$, reads only $X$, $Y$ and $z_{<u}$, so on the true context $Z=(X,Y,M_{<u})$ we have
$\prod_{i\in J}\hat M_ig_i=\hat M_u\,\chi\prod_{i\in J}g_i$. By~\eqref{eq:walsh} at $T=u$, $N/2\le u\le N$,
\[
\Bigl|\E\prod_{i\in J}\hat M_ig_i\Bigr|\le2e^{-3b}+|\E\,\hat M_u\,\chi P_J(Z)|\le2e^{-3b}+2^{-DN/2}A(P_J)\le
2e^{-3b}+2^{-DN/4}.
\]

\emph{The whole block.} The indicator that all predictions in $I$ are right is
$2^{-b}\prod_{i\in I}(1+\hat M_ig_i)=2^{-b}\sum_{J\subseteq I}\prod_{i\in J}\hat M_ig_i$, so its probability is at most
$2^{-b}+2e^{-3b}+2^{-DN/4}\le3\cdot2^{-b}\le e^{-(\ln2/2)b}$, using $b\ge4$ and, for $N\ge(4/D)^2$,
$b\le\sqrt N\le DN/4$. No independence between the predictions is used.

\emph{All auxiliary strings.} Fix $t\le s$. Every string that $\mathcal T_{|x|+t+2N}$ emits as its first $t$ symbols on
some input is a leaf of the tree whose children at a node $\tau'$ of depth $j<t$ are the possible next symbols on the
literal context $x\,\tau'$ ($\tau'$ fixed, $x$ formal). By \cref{lem:nexttok} each node has at most $(10N)^A$ children,
so there are at most $(10N)^{At}$ leaves, independently of the vocabulary. If generation with $t$ auxiliary symbols is
correct on $(X,Y)$, its auxiliary string is one of these leaves and the block event holds for it. A union bound over
$t\le s$ and the leaves gives probability at most $(s+1)(10N)^{As}e^{-(\ln2/2)b}$, and for $N\ge20$
\[
\ln(s+1)+sA\ln(10N)\le3Aws\le3Ac_Lb=3\rho b/A\le\tfrac{\ln2}4b,
\]
by $s\le c_Lb/w$, $A\ge4$ and $\rho\le\ln2/12$. This gives the bound $e^{-c'b}$. It covers an input-dependent choice of
$t$, hence of the model. For the second sentence, on the inputs where the CoT is computed with at most $s$ auxiliary
symbols, generation succeeds with some $t\le s$. At fixed $L,H,k$ both floors are asymptotically negligible, so
$s=\Theta(N^q/w^{3-q})$.

\emph{One threshold for all depths in the range.} Fix $H,k$, let $r=\log_2a$ and $h=\lceil\log_2(2r+4)\rceil$, and
let $2\le L\le\log_2w-\log_2\ln w-h$. Then $2^L\le w$, so $A=2^{rL}\le w^r$, and
$qw\ge2^{h+1}\ln w$, so $N^q\ge w^{2^{h+1}}\ge w^{4r+8}$. Hence $c_LN^q/w^{2-q}\ge\rho w^{2r+6}$, so
$b\ge(\rho/2)w^{2r+6}$ once this is at least $2$, and then $c_Lb/w\ge(\rho^2/2)w^5$ and $s\ge(\rho^2/4)w^5$. Also
$v\le16Aw\le16w^{r+1}\le2N$. So every size condition above ($b\ge4$, $v\le2N$, $2N\le n_i\le5N$, $N\ge20$,
$N\ge(4/D)^2$) holds beyond one threshold that depends only on $H$, $k$ and the input format, and no other estimate
depends on $L$. \qed
\fi

\ificml\else
\subsection{Proof of \cref{thm:cot-worst}}

\emph{The embedding.} Apply \cref{lem:sidon} with $T=N$, and $b$, $r=2^b-1$ and the Erd\H{o}s--Tur\'an set chosen as
after it with $K=N$; then $b=\Theta(\log N)$ and $r+1>\sqrt{N/(\sidonConst\log_2N)}$. The proof of \cref{lem:sidon} shows
more than its statement: the binary expansion of $X(z)Y$ is a concatenation of $b$-bit blocks without carries, the
block of $\delta=0$ is the binary representation of $|z|$, and every product digit above it is $0$ or a single $z_i$. Let
this block occupy the digits $M_{k_0},\dots,M_{k_0+b-1}$.

\emph{A list of candidates.} We may assume $s\le N$. For $t=0,\dots,s$ let $\mathcal T^{(t)}=\mathcal T_{|x|+t+2N}$ and
define $F_t(z)\in\{0,1\}^b$ by running $\mathcal T^{(t)}$ on the embedded operands: generate $t$ symbols
autoregressively; append the known digits $M_1,\dots,M_{k_0-1}$ (constants or single $z_i$); generate $b$ more symbols
autoregressively, decoding an output that is not a bit as $0$. On every $z$, at the true auxiliary length $t(z)$ the
generated auxiliary string and the appended digits are the true trace, so $F_{t(z)}(z)=|z|$. The other $F_t$ may be
arbitrary on $z$. So the list $F_0,\dots,F_s$ covers the weight; no procedure needs to know $t(z)$.

\emph{Circuits.} Fix $t$ and a bit of $F_t$. Enumerate the paths consisting of a leaf $\tau$ of the tree of
\cref{lem:nexttok} at depth $t$ and a word $w\in\{0,1\}^b$; there are at most $N^{Ct}2^b$. On a fixed path every
transition reads a literal context: constants (guessed symbols and guessed count digits), operand literals and
appended digits that are constants or literals. By \cref{lem:nexttok} each test ``the next output decodes to the guessed
symbol'' has depth $2L$ and size $N^{O(1)}$. The AND of the $t+b$ tests of a path has depth $2L+1$, and by induction from
the first transition exactly one path passes on each $z$. The OR over the paths whose word $w$ has the requested bit
equal to $1$ computes that bit, in depth $d=2L+2$ and size $S\le\exp(C(s+1)\log N)$.

\emph{Influence.} By Boppana's theorem each of the $(s+1)b$ list bits has influence at most $C_d(C(s+1)\log N)^{d-1}$. If
$s+1\ge c\sqrt r$ we are done. Otherwise \cref{lem:listcover} with $k=s+1$ gives
\[
\frac{cr}{s+1}\le(s+1)b\,C_d\bigl(C(s+1)\log N\bigr)^{d-1},\qquad\text{so}\qquad cr\le C'(s+1)^{d+1}(\log N)^{d},
\]
using $b=O(\log N)$. With $r=\Omega(\sqrt{N/\log N})$ and $d=2L+2$ this is the claimed bound. \qed

\noindent The influence bound is applied on the whole cube of the free bits $z$ to explicit circuits for the list
bits; no sensitivity property is needed of extensions of a multiplication predictor to inconsistent inputs, which is
where the uniform-measure Parseval step in the proof of~\cite[Lemma~C.6]{AHRH25} does not apply as stated (\cref{app:c6}).
\fi

\ificml\else
\section{What is computed, and how it is checked}\label[appendix]{app:provenance}

The proofs of \cref{thm:main,thm:symbol,thm:transfer,thm:rates} and \cref{prop:necessity} are analytic; apart from
evaluating explicit constants (for \cref{prop:necessity} this includes a certified interval quadrature of $I(\kappa)$),
they use no computer calculation, with one exception: \cref{thm:symbol}(b) and the
computer-assisted rates in \cref{thm:rates} rest on the certificate of \cref{lem:cert}. That certificate is computed in
binary64 arithmetic with every operation rounded upward, from cosine values bounded in interval arithmetic, by
\texttt{checks/kappa\_cert\_interval.py} (about a minute with four processes; recorded output
\texttt{kappa\_cert\_interval.out}), and re-verified by the separately written \texttt{checks/kappa\_cert.py}, which
uses an a priori rounding-error bound instead (about $15$ seconds). The constants of \cref{app:rates} are evaluated in
interval arithmetic by \texttt{checks/exponent\_constants.py}. The constants that are delicate are certified in interval
arithmetic by \texttt{checks/constants\_interval.py} (about $5$ seconds): the shell constant of \cref{cor:shell} with its
margin, and the integral $I(\kappa)$ and the Lipschitz constants of \cref{app:necessity}. Every printed constant that is an upper bound is rounded up, and every lower bound is rounded down. The script \texttt{paper/make\_numbers.py} re-evaluates their numerical constants from the formulas (for
example $\CAexact\le3982$, the thresholds $\thrtwo$ and $\throne$, and the coverage inequality for $N\ge400$) and exits
with an error if one fails. It also computes the cases $N=1,2$ of \cref{ex:small} exactly in rational arithmetic.

\Cref{tab:num} and the figures are floating-point computations and serve only as evidence and illustration. The spectral
norms use single-precision matrices and an iterative eigensolver; the values of $\mu_S$ are maxima over a fine grid of
$\xi$ with local refinement, hence (up to grid effects) lower bounds for the supremum. Every sentence in the paper about
these data (which masks attain the maximum, the spread over $N$, the fitted exponents, the ranges in
\cref{app:necsec}, and the values of $\Lambda_{\mathrm{cl}}(S)$ in \cref{app:classical}) is computed from the stored files by \texttt{make\_numbers.py}, which asserts it.

In addition, the lemmas behind \cref{thm:symbol,thm:transfer} were tested numerically by separately written programs:
the steps of \cref{prop:window} on random instances and exhaustively for small top digits, \cref{lem:transfer} for all
words of length at most $16$, the bad-cell count of \cref{app:cells} exhaustively for $N\le11$, \cref{lem:residues}
exactly for $N\le11$, and the identities of \cref{lem:shell,lem:smooth} against high-precision quadrature. These tests found
no violation; they are consistency checks, not part of the proofs.

The results of \cref{sec:transformers} are proved analytically: \cref{thm:full-two,thm:threeorders,thm:boundedstates,thm:three}
by the explicit polynomials of \cref{app:selectors}, \cref{thm:topk-depth} by the backward argument of \cref{app:topkdepth}, \cref{cor:topk-three} as a consequence of it, and \cref{prop:surj} from a cited query lower bound. \texttt{make\_numbers.py} checks their
combinatorial statements exactly, by enumeration in integer arithmetic, and exits with an error if one fails:
\cref{prop:gt} for all $N\le\BtwoNmax$ and $T\le2N$; \cref{prop:restriction} on every subcube and every set
$\{M_{<T}=q\}$ for $N\le\propBNmax$ and $T\le2N$; \cref{lem:sidon} with the Erd\H{o}s--Tur\'an set for $b\le\sidonBmax$, all
$z$ and every admissible $k$ at two values of $N$ each; the parameter inequalities after \cref{lem:sidon} on a geometric
grid of $N$ from $\sidonNlo$ to $\sidonNhi$; and the norm $3/2$ in \cref{rem:twolayer}. Separately written programs also
compared the decomposition in the proof of \cref{thm:onelayer} with direct simulation of random one-layer models
(exhaustively over $X$ and $Y$, for small $N$), and checked \cref{prop:disj} exhaustively for small $N$; they found no
violation. The script \texttt{checks/phase1\_check.py} simulates the combinatorial part of the proof of
\cref{thm:boundedstates} classically (\cref{app:boundedstates}). The scripts \texttt{checks/prefix\_conj\_check.py},
\texttt{checks/uhat\_sim\_check.py} and \texttt{checks/conj\_check.py} check the lemmas behind \cref{thm:three} in exact
arithmetic, exhaustively at small sizes (\cref{app:three}), and \texttt{checks/four\_layer\_check.py} evaluates the
probability bound of \cref{rem:four} exactly for $64\le N\le2048$. The script \texttt{checks/c6\_sensitivity.py} checks
the two facts about Lemma~C.6 in \cref{app:c6} in exact arithmetic. The script \texttt{checks/nonvacuous.py} computes, in exact integer arithmetic, the sizes quoted in
\cref{sec:limits} at which the explicit transformer bounds stop being vacuous. The script \texttt{checks/ac0\_bayes\_TN.py}
computes, in exact integer arithmetic, the Bayes advantages at $T=N$ quoted after \cref{thm:ac0-query}.

\paragraph{Code and data.}
{\sloppy The scripts \texttt{scripts/spec.py} and \texttt{scripts/mellin.py} produce the data in \texttt{results/r0/out/};
\texttt{scripts/mellin\_curve.py}, \texttt{scripts/necessity\_curve.py} and \texttt{scripts/classical\_l1.py} produce
\texttt{results/r1/};
\texttt{scripts/make\_figs.py} draws the figures; and \texttt{paper/make\_numbers.py} generates the arithmetic tables and constants from these files, re-evaluates the constants of the proofs, computes the cases $N\le2$ exactly, checks the
statements made about the data, and checks the combinatorial statements of \cref{sec:transformers} by enumeration.
The finite two-layer table is generated by \texttt{checks/finite\_two\_layer.py}. All of these files, and the scripts in \texttt{checks/}, are in the supplementary material\vsw{}{ (code under the MIT
license)}\ifdefined\ZENODODOI, \texttt{https://doi.org/\ZENODODOI}\fi.\par}

\fi

\section{Finite-size attention bounds}\label[appendix]{app:finite-two}

A direct estimate gives numerical upper bounds on success for two layers. Its main observation is that, on any input, the $n$ proposed winners of a head have distinct ranks. We approximate the winner by a nonnegative polynomial in its rank and charge the total weight of the wrong ranks.

\begin{theorem}[finite two-layer bound]\label{thm:finite-two}
Consider the top-$1$ model of \cref{def:topk} on the true context of length $n$, with $H$ heads in the first layer and $J$ in the second. Let $2\le T\le N$, and let $\Delta$ bound $\disc_N(S)$ for every $S$ with $\max S=T$. For integers $\ell\ge1$ and $0\le d<m=n-1$, put
\[
 B=2(\ell+1)^H,\quad
 u_\ell=\max_{0\le a\le\ell,\,a\in\mathbb Z}
 \min\{2^{-a},(2+(\ell-a)\ln2)2^{-(\ell-a)}\},\quad
 R=\prod_{i=0}^{d}\frac{m-i}{m+1+i}.
\]
Then the absolute signed correlation is at most
\begin{equation}\label{eq:finite-two}
 4nH u_\ell+(1-R)^{-J}-1+\Delta n^J B^{J+1}(2B+4)^{2dJ}.
\end{equation}
For $\ell\ge n$ the first term may be set to zero.
\end{theorem}

\noindent At $T=N$ with $n=3N-1$ and $H=J=1$, the following are upper bounds on success. The hand discrepancy is \cref{thm:rates}; the sharper discrepancy in its proof uses \cref{lem:cert}. The displayed decimals are rounded up for the indicated operand sizes.
\begin{center}
\begin{tabular}{r@{\quad}rr@{\qquad}rr}
\toprule
&\multicolumn{2}{c}{Hand discrepancy}&\multicolumn{2}{c}{Certified discrepancy}\\
$N$ & $(\ell,d)$ & Success & $(\ell,d)$ & Success\\
\midrule
$16{,}384$  & $(49,209)$  & $0.859597$ & $(52,236)$  & $0.740416$\\
$32{,}768$  & $(54,415)$  & $0.609743$ & $(56,470)$  & $0.561678$\\
$65{,}536$  & $(60,819)$  & $0.518380$ & $(64,922)$  & $0.507017$\\
$131{,}072$ & $(70,1599)$ & $0.500837$ & $(76,1795)$ & $0.500149$\\
\bottomrule
\end{tabular}
\end{center}
\noindent With the certified discrepancy, \eqref{eq:finite-two} also gives success at most $0.986669$ at $N=10{,}240$ for $H=J=1$ and $(\ell,d)=(48,148)$, and $0.567760$ at $N=524{,}288$ for $H=J=2$ and $(\ell,d)=(66,2108)$.

\begin{proof}
\emph{Truncate the first layer.} A prescribed pattern on $b$ prefix bits has probability at most $(2+b\ln2)2^{-b}$. Indeed, writing $Q=2^N$, $K=2^{T-1}$ and $z=2^{-b}$, the pattern permits $Kz$ prefix words. Given $X=x>0$, each word restricts $Y$ to an interval of length $Q^2/(Kx)$, so the conditional probability is at most $\min(1,zQ/x)+Kz/Q$. At $x=0$ take the minimum to be $1$. The left-endpoint average of the decreasing function $\min(1,z/t)$ on the $Q$-grid exceeds its integral $z(1+b\ln2)$ by at most $1/Q$; both $1/Q$ and $Kz/Q$ are at most $z/2$. Any $\ell$ fixed literals, $a$ of them operand bits, therefore have joint probability at most $u_\ell$, since the operand bits are independent.

Given the query token, each first-layer head selects the first present entry of a fixed list. A rank exceeding $\ell$ means the first $\ell$ candidates are absent. These lie at distinct bit positions, since fixed tokens are present and the two values at one bit position cannot both be absent. There are at most $n-1$ absent entries. Keep each of the $H$ ranks up to $\ell$ and give each head a sentinel for a larger rank. Together with the query token these ranks form at most $B$ records at a position. Each record event is a conjunction of literals, including the absences for each sentinel. Decode any record containing a sentinel to a fixed state. Changing the output this way costs at most $4nH u_\ell$ in mean absolute error by a union over positions, heads and the two possible query tokens. Each record indicator has Walsh norm at most $1$. An indicator of a subset of records has norm at most $U=(B+1)/2$: for a subset of size $s$ use the smaller of $s$ and $1+B-s$ by expanding the subset or its complement.

\emph{Filter the final winners.} Below we construct a polynomial $q$ of degree at most $d$ with $q(0)=1$, $\sum_{r=0}^m q(r)^2=(1-R)^{-1}$, and $A(q(r))\le(2B+4)^d$ whenever $r$ counts $m$ Boolean predicates of Walsh norm at most $U$. Guess a query record and, for each final head, a winner index and record. Multiply their exact indicators, of joint norm at most one. For a head $h$, let $r_h$ count the other positions whose keys outrank its proposed winner under the guessed query, including the fixed tie-break. Each comparison is a predicate of one record. Multiply the anchor indicators by $\prod_h q(r_h)^2$ and the tuple's signed readout, and sum over tuples. On a fixed input only the true records at the proposed positions survive. For each head the possible indices run through all ranks $0,\ldots,m$, so the weight of the true winner tuple is one and the total weight of all other tuples is $(1-R)^{-J}-1$. This bounds the pointwise error. The Walsh norm of the sum is at most $n^J B^{J+1}(2B+4)^{2dJ}$ by the tuple count and submultiplicativity. Applying \eqref{eq:walsh} gives \eqref{eq:finite-two}.

\emph{The rank polynomial.} We use the classical Hahn (Gram) polynomials $G_j(x)=Q_j(x;0,0,m)$ and their reproducing kernel at zero (NIST DLMF, \href{https://dlmf.nist.gov/18.20.E5}{Eq.~18.20.5} and \href{https://dlmf.nist.gov/18.2.E12}{Eq.~18.2.12}). For $0\le j\le d$ let $D$ be the forward-difference operator and set
\[
 h_j(x)=\binom{x}{j}\binom{x-m-1}{j},\qquad
 G_j(x)=\frac{D^j h_j(x)}{(-1)^j\binom mj}.
\]
The $G_j$ have degree $j$ and $G_j(0)=1$. They are orthogonal for the counting measure on $0,\ldots,m$: summation by parts $j$ times shifts the differences onto the other factor, since $h_j$ vanishes on $0,\ldots,j-1$ and $m+1,\ldots,m+j$. Against $G_j$ itself the $j$th backward difference is $\binom{2j}{j}/[(-1)^j\binom mj]$ by its leading coefficient. Thus Vandermonde's identity gives
\[
 \sum_{x=0}^m G_j(x)^2
 =\frac{\binom{2j}j}{\binom mj^2}\sum_{x=0}^m\binom{x}{j}\binom{m+j-x}{j}
 =\frac{(m+j+1)!\,(m-j)!}{(2j+1)(m!)^2}.
\]
The reciprocal $w_j$ of this squared norm is $R_{j-1}-R_j$, where $R_j$ is the product defining $R$ stopped at $j$ and $R_{-1}=1$. Hence $q=(\sum_j w_jG_j)/(1-R)$ has the asserted endpoint and squared norm.

Finally, Newton expansion gives
\[
 G_j(r)=\sum_{s=0}^j(-1)^s\binom js\binom{j+s}s\frac{\binom rs}{\binom ms}.
\]
Indeed $D^{j+s}h_j(0)$ has terms only at $j+t$, $0\le t\le s$; taking out $(-1)^j\binom{j+s}s$ leaves the $s$th difference of $\binom{m-t}{j}$ at zero, namely $(-1)^s\binom{m-s}{j-s}$. After substitution the binomial ratio in the display averages products of $s$ distinct predicates, of norm at most $U^s$. Thus
\[
 A(G_j(r))\le\sum_s\binom js\binom{j+s}s U^s
 =\sum_t\binom jt^2 U^t(1+U)^{j-t}
 \le(\sqrt U+\sqrt{1+U})^{2j}\le(2B+4)^j.
\]
The equality follows by matching powers with Vandermonde's identity, and the first inequality after it by squaring the positive binomial expansion. Taking the convex combination defining $q$ finishes the proof.
\end{proof}

The outward-rounded integer witness calculation is reproduced by \texttt{checks/finite\_two\_layer.py}.

\begin{corollary}[all operand lengths from $65{,}536$]\label[corollary]{cor:finite-two-uniform}
For every $N\ge65{,}536$, at step $T=N$ with context length $n=3N-1$, a two-layer top-$1$ model with one head per layer predicts the next bit with probability less than $0.507967$.
\end{corollary}

\begin{proof}
Let $N_0=65{,}536$, $x=N/N_0$, $\ell=64+\lceil4\log_2x\rceil$ and $d=\lfloor917\sqrt x\rfloor$. We bound each term of \eqref{eq:finite-two}. First,
$u_\ell\le\sqrt{2+\ell}\,2^{-\ell/2}$ by taking the geometric mean in its definition. Since $\sqrt{2+\ell}\le\sqrt{67+4\log_2x}\le9x$ and $2^{-\ell/2}\le 2^{-32}x^{-2}$, clipping costs at most $12N_0\cdot9\cdot2^{-32}$.

Next, applying $\ln((1-y)/(1+y))\le-2y$ with $y=(2i+1)/(2m+1)$ gives
\[
 R\le\exp\!\left(-\frac{(d+1)^2}{m+1/2}\right)
 <\exp\!\left(-\frac{917^2}{3N_0}\right)<\frac1{71}.
\]
The first eleven terms of the exponential series at $917^2/(3N_0)$ sum to more than $71$. Thus the wrong ranks have total weight less than $1/70$.

Finally, the estimate from \cref{lem:cert} in the proof of \cref{thm:rates} gives $\Delta\le50{,}000N\,2^{-0.22763N}$ at these lengths, by using the slowest rate in its three terms. Write $M=2B+4$. From $\ell\le65+4\log_2 x$ and $e^t\ge1+t$,
\[
 B\le132+8\log_2x\le132\sqrt x,\qquad
 M\le268+16\log_2x\le268 x^{1/8}.
\]
Hence $\log_2 M\le8.07+(\log_2 x)/8\le8.07\sqrt x$, using $268^{100}<2^{807}$. Also $50{,}000N nB^2\le150{,}000N_0^2\,132^2 x^3\le2^{64}x^3\le2^{64x}$. Since $0.22763N_0-2\cdot917\cdot8.07-64>53$, the discrepancy term is at most $2^{-53x}$. The success probability is therefore less than $(1+12N_0\cdot9\cdot2^{-32}+1/70+2^{-53})/2<0.507967$.
\end{proof}

The closing numerical comparisons are reproduced by \texttt{checks/finite\_two\_uniform.py}.

\ificml
\paragraph{A one-head bound at 128 bits.}
At $N=128,T=106$ and with two separators, one-layer, one-head UHAT
success is at most $0.505406347$. Guess the raw query bit, winning
position and winning raw bit. There are at most $4n$ guesses with
$n=363$. For each guess the winning comparisons and fixed bits give
a conjunction of literals or a constant, of Walsh norm at most one.
The signed readout thus has Walsh norm at most $4n$, and
\eqref{eq:walsh} gives success at most
$1/2+2n\max_{\max S=T}\disc_N(S)$.
\Cref{thm:symbol}(b) and \cref{thm:transfer}(a) bound the discrepancy
by $17077.1(0.6646)^{53}+3\cdot2^{-22}+(8+1.39\cdot128)2^{-128}$.
The supplement evaluates this certificate with exact rational arithmetic
and rounds the result upward.
\fi

\newif\ifshowroundedsoftmax
\ificml
  \ificmlextended\showroundedsoftmaxtrue\fi
\else
  \showroundedsoftmaxtrue
\fi
\ifshowroundedsoftmax
\section{Rounded softmax at logarithmic width}\label[appendix]{app:precision}

Consider the inclusive-causal softmax model with iterative rounding of
Li et al.~\cite[Defs.~3.2--3.3]{LiEtAl24}. Fix the scalar format with alphabet size $F$, and the depth and head count.
Allow all vector dimensions at most $c_0\log_2 N$ for any fixed $c_0<1/\log_2 F$.
Serialize the two operands as contiguous blocks $X,Y$ followed by the true prefix $P=M_{<T}$, with any
fixed separators. Absolute positional embeddings are allowed, encoded in the finite scalar format. Scores
are computed from the resulting queries and keys. At $\alpha N\le T\le N$ for fixed $\alpha>0$, the
next-bit advantage is at most
\ificml $\exp(-\Omega(N))$ by \cref{thm:icml-recurrent}.\else
$\exp[-\Omega(N/(\ln N)^{d-1})]$, where $d$ is the fixed depth of its Boolean
circuit simulation.\fi The constant can depend on the scalar format, architecture and $\alpha$.

\paragraph{Finite information after the first operand.}
Write $[\cdot]$ for scalar rounding and $\oplus$ for addition followed by rounding. For each layer,
head and possible query vector $q$, keep the left-folded denominator $D_q$ of the rounded
exponentiated scores. For each candidate nonzero representable denominator $a$, also keep
the vector left fold $V_{q,a}$ of the value summands that would be used with final
denominator $a$. On reading a key $k_i$ and value $u_i$, the scalar mass
$m_i(q)=[\exp(\operatorname{score}(q,k_i))]$ updates $D_q$ by $D_q\oplus m_i(q)$. The vector
summand for $V_{q,a}$ is the rounded product of $[m_i(q)/a]$ with $u_i$.
After the update, the actual query $q_i$ reads $V_{q_i,D_{q_i}}$.
This is the prescribed operation order, since each coefficient is rounded before value aggregation.
Process the layers of position $i$ in order before advancing the position. The construction is the
query-indexed streaming simulation of Nowak et al.~\cite{NowakEtAl26} with an extra index for
the hypothetical denominator, which is needed when coefficients are normalized before aggregation.

With query and value dimensions at most $w$, the tables use
$B=O(F^w(1+w))$ bits. For $w\le c_0\log_2 N$ this is $O(N^{1-\epsilon}\log N)$
for a constant $\epsilon>0$. The clock can use the known
index to supply its fixed positional embedding. The state $s(X)$ after the first operand (and any fixed
separator at the cut) has at most $2^B$ values and determines the computation on every
formal continuation $(Y,P)$. For division by zero, use the convention that a failed prediction is wrong.
For the upper bound, totalize any failed computation or non-bit output to the bit $0$. One failure flag
suffices in the streaming implementation, and the zero-denominator test uses only actual queries.
\ificml
The simulation stores this same budget throughout its scan, so the bound follows from
\cref{thm:icml-recurrent} with $W=2^B$. The signed prediction of the totalized model
has at least the success probability of the partial model.
\else
The finite-alphabet equality tests and the rounded summations in the circuit simulation of
Li et al.~\cite[Thm.~3.1]{LiEtAl24} also compute this failure flag in constant depth and polynomial size.

\paragraph{Correlation after the cut.}
For each reachable state $s$, fix one representative first operand and hardwire it into the circuit
simulation of the totalized bit output. This gives a sign-valued circuit $C_s(Y,P)$ of constant depth $d$
and polynomial size, and the predictor is $\sum_s1_{\{s(X)=s\}} C_s(Y,P)$. Its correlation with $\hat M_T$
is bounded as follows. On the uniform formal cube of $m=N+T-1$ bits, retain Walsh degrees at most
$r=\lfloor\theta m\rfloor$, choosing fixed $0<\theta<1/2$ with $m h_2(\theta)/2\le DT/4$;
here $h_2$ is binary entropy in bits and $T\ge\alpha N$.
By Parseval and the binomial tail bound the retained Walsh norm is at most $2^{m h_2(\theta)/2}$.
The main theorem bounds the low-degree correlation for each state by $2^{-3DT/4}$.
For the discarded terms, Tal's Fourier tail bound~\cite[Thm.~18]{Tal17} bounds the squared norm on that cube by
$2\cdot2^{-r/\kappa}$, with $\kappa=O_d((\log N)^{d-1})$. On $Y\ge2^{N-j}$ the genuine law of
$(Y,P)$ has density at most $(3/2)2^j$ relative to the uniform cube:
for fixed $Y=y>0$ a given prefix corresponds to an interval of $X$ of length $2^{2N-T+1}/y$,
with at most one extra integer. Choose $j=\lfloor r/(2\kappa)\rfloor$. By Cauchy--Schwarz the tail
correlation for each state on this event is $O(2^{-r/(4\kappa)})$. The removed probability is at most
$2^{-j}$; use the same event in the low-degree terms by multiplying its indicator into the bounded
test of $Y$. Summing over at most $2^B$ states gives the claimed rate since $B=o(r/\kappa)$.
\fi

This statement uses the rounding and iteration order specified above. In particular it requires
the stated upper bound on width. Saturated denominators can support copying all
input bits to a single position when width grows with length, as in Li et al.~\cite{LiEtAl24}.

\fi

\begin{proposition}[A short message can suffice with unrestricted decoding]\label[proposition]{prop:three-bit-decoder}
At $T=N$, a decoder given $Y$, the true prefix, and $X\bmod 8$ can attain
success probability at least $7/8$ for every $N\ge2$.
\end{proposition}
\begin{proof}
Write $p$ for the integer encoded by the $N-1$ prefix bits. For $Y>0$, the input $X$ lies
in the half-open interval
$[p2^{N+1}/Y,(p+1)2^{N+1}/Y)$, intersected with $[0,2^N)$.
If $Y\ge2^{N-2}$ the interval has length at most eight, so it contains at most
one integer in the transmitted residue class. Recover $X$ and compute $M_N$.
This event has probability $3/4$. On its complement a fair independent guess succeeds
with probability $1/2$; one of the two fixed guesses does at least as well there.
\end{proof}
The decoder can read the whole continuation with unrestricted computation.
A scan must process it once within its persistent storage budget.

\paragraph{Previously disclosed numerical bounds.}
This extended source preserves the earlier diffuse-softmax and
ordered-scan arguments. The companion on arithmetic robustness develops
the numerical bounds and derives the midpoint scan from prefix-table rank.
\icmlsoftmaxsection
\section{Sampling stable softmax}\label[appendix]{app:icml-softmax-sampling}

\subsection{Proof of \cref{prop:icml-sharp-diffuse}}
Consider a fixed-depth causal softmax model on the true prefix at
$\alpha N\le T\le N$, for fixed $\alpha>0$. The operands each appear
in MSB-first order and can be interleaved before the prefix. The layout is
fixed and the initial row depends only on its token and position. Scores
use the query and source rows, and values use only the source row.
Each update uses only its incoming row and head outputs. The
readout uses the final unmarked row. At layer
$\ell$, require $\|v_j\|_2\le B_\ell$ and $p_{ij}\le\rho_\ell/i$.
Write $A_\ell$ for the full layer's Lipschitz constant in the maximum
Euclidean row norm, $U_\ell$ for the update's Lipschitz constant in the
maximum Euclidean head-output norm, and $R$ for the per-logit readout
Lipschitz constant. These conditions apply globally, including on the
hybrid states used below. Set
\[
G=R\sum_{\ell=1}^{L}U_\ell B_\ell\rho_\ell\prod_{j>\ell}A_j .
\]

\begin{proof}
We first couple inputs with the same true prefix and opposite targets.
Let $m=2^N$, $d=2^{N-T}$ and $q=\lfloor xy/(md)\rfloor$. Its parity
is $M_T$ and $\lfloor q/2\rfloor$ is the prefix. For any fixed
$\eta>0$, we construct a matching covering $1-\eta$ of the inputs,
where each edge changes at most $2r$ operand bits, all at row indices
at least $T-r$, for a constant $r=r(\eta)$.

For small $\epsilon>0$, the mass of inputs within $2\epsilon$ of an
integer on the $xy/(md)$ scale is $O(\sqrt\epsilon)+o(1)$,
uniformly for $T\ge\alpha N$. For $d\le\epsilon^{-1/2}$, use
uniform residues $xy\bmod m$ over multiples of $\gcd(x,m)$ for
fixed $x$. The mean gcd is $N/2+1$, including $x=0$. The probability
of distance at most $u$ to an integer on the $xy/m$ scale is therefore
at most $2u+(N+2)/m$. Here $u=2\epsilon d$. For $d>\epsilon^{-1/2}$,
discard $y<\epsilon m$. The remaining normalized products at fixed
$y$ lie on a grid of spacing $y/(md)$ in an interval of length
$A=y/d\ge\epsilon 2^T$. At most $A+2$ integer boundary strips
intersect the interval. Counting at most one extra grid point per strip
bounds their mass by $4\epsilon(1+2/A)+(A+2)/m=O(\sqrt\epsilon)+o(1)$.

Partition each operand into dyadic blocks of length $dQ$, $Q=2^r$.
At a block origin $(x_0,y_0)$ put $u_0=x_0/m$, $v_0=y_0/m$.
Almost every $(u_0,v_0)$ in the unit square admits integers $s,t$
with $|sv_0+tu_0-1|<\epsilon$ because the subgroup generated by $u_0,v_0$
is dense when their ratio is irrational. Restrict $|s|,|t|\le K$
for a fixed $K$ large enough to lose at most $\eta/8$ of area.
The finite union of strips has boundary of measure zero, so the
fraction of good block origins converges to this area as $2^T/Q$
grows. Choose one shift $(sd,td)$ per good block. Moving by this shift
changes $xy/(md)$ by $sy/m+tx/m+st d/m$, within
$\epsilon+(2KQ+K^2)/2^T$ of one. The reverse shift has the
corresponding bound about minus one. Match each even $q$ to
its translate if both inputs are in the block and the quotient
increases by one. These edges are disjoint by injectivity and parity.
Outside the boundary strips, and at distance at least $Kd$ from
the block edges, every input in a good block is matched by the
forward or backward shift. The block border mass is at most $4K/Q$.
Choose $\epsilon$ small, then $K$, then $Q$ large, then $N$ large.
An edge preserves $\lfloor q/2\rfloor$, and changes only the $r$ bit
positions starting at offset $\log_2 d$ in each operand. Their row
indices are at least $T-r$ by the layout assumption.

For an edge let $S$ be its changed rows. In layer $\ell$, first
replace the incoming rows in $S$ from one trajectory by those of
the other. At an unmarked query the conditional law over unmarked
keys is unchanged. For masses $p,p'$ on $S$ the head means differ
by at most $2B_\ell\max(p,p')$, using the common background mean.
Only rows $i\ge T-r$ can be affected, so $\max(p,p')\le
\rho_\ell|S|/(T-r)$. The remaining change costs at most
$A_\ell D_{\ell-1}$, where $D_{\ell-1}$ is the maximum incoming row
difference outside $S$. Thus
\[
D_\ell\le A_\ell D_{\ell-1}
   +2U_\ell B_\ell\rho_\ell |S|/(T-r),\qquad D_0=0.
\]
The logit difference on the matched pair is at most $C_{\alpha,\eta}G/N$
in sup norm. Two confident correct labels on an opposite-label
pair need logit difference at least $\gamma$. A sampled opposite-label
pair has average success at most $1/2+\operatorname{TV}/2$, and
softmax at temperature $\tau$ has $\operatorname{TV}\le
\|\Delta\mathrm{logit}\|_\infty/(2\tau)$. Indeed along the logit
segment the probability derivative is $p_j(w_j-\mathbb E_p w)$
with $w=\Delta\mathrm{logit}/\tau$, and its $\ell^1$ norm is at most
$\|w\|_\infty$ by the variance bound. Choose $\eta<a/2$ and
include the exceptional-gap mass for greedy decoding. This proves
both gain lower bounds.

For the error-radius bound, keep one endpoint exact. At the other
endpoint, force all unmarked rows to match it layer by layer by
replacing each unmarked head mean by the reference mean. The same
conditional-law estimate bounds the required error by
$2 B_\ell\rho_\ell|S|/(T-r)$, with the unmarked inputs now identical.
The final predicted distribution is shared, so a matched pair
has robust success at most one half. This argument needs no
Lipschitz or margin assumption.

For sharpness, a uniform head encodes the $d=2N$ operand bits as
$z_j=x_j/n$ in distinct coordinates. Let $y=nz$ and let
$\operatorname{clip}$ truncate to $[0,1]$. Set
$b_j=\operatorname{clip}(4y_j-3/2)$,
$u_j=\operatorname{clip}(8(\min(y_j,1-y_j)-1/4))$,
and $k=\lceil\log_2(2d)\rceil$.
Let $g$ interpolate $(j/(2k),2^j)$ and $h$ interpolate $(2^j,j)$
for integers $0\le j\le2k$. Define
\[
s=d^{-1}\sum_j g(u_j),\qquad q=[1-h(s)/k]_+ .
\]
If any $u_j=1$ then $s\ge2^{2k}/d\ge2^k$ and $q=0$.
Thus on $q>0$ all $b_j$ are locally constant bits. If $y$ is
within $1/4$ in sup norm of a bit vector then $q=1$ and $b$ is
that vector. On linear pieces, $g'\le2k g$ and $h'\le2/s$,
so the gradient of $q$ in the $u$ variables has $\ell^1$ norm at
most $4$. The mask is therefore $32$-Lipschitz in $y$ in sup norm.

Use a constant-depth polynomial-size majority circuit for
multiplication~\citep{HAB2002}, replacing a threshold at integer $r$
by $\operatorname{clip}(\sum_j b_j-r+1)$ on its gate inputs.
The resulting ReLU circuit $C$ returns the desired bit on binary
inputs and takes values in $[0,1]$. The output $[q+C(b)-1]_+$
is zero where $q=0$ and equals either $q$ or zero on each region
where $q>0$. Its continuous piecewise-linear gradient is bounded
by that of $q$ on every piece, giving global Lipschitz constant
$32n$ in $z$ in sup norm, hence also in Euclidean norm.
The interpolants use $O(d\log d)$ additional ReLU units at constant
depth. The threshold circuit and interpolants together have polynomial
parameter count. Center and scale the result to give the desired logit gap.
The averaging layer has bounded remaining parameters, and errors
at most $1/(4n)$ preserve the operand bits. Identity row updates
extend the construction to any fixed attention depth.
\end{proof}

\subsection{Arbitrary attention profiles}
\label{sec:icml-softmax-arbitrary}
We also give bounds under local, reference-state assumptions that
allow arbitrary weight profiles and free-running decoding. The
architecture is the same. Assume uniform Lipschitz bounds on the
layer maps, row updates, and readout, a fixed positive-radius tube
around the reference trajectory, and a bound on the attention-weighted
value variance $\sum_j p_j\|v_j-\sum_b p_bv_b\|_2^2$ in the tube.
The update bound must apply within the chosen tolerance of the exact
head outputs in the tube. Greedy decoding requires a fixed positive
top-two logit gap on the readouts being certified.

\paragraph{A common sampling simulation.}\label{par:icml-softmax-simulation}
For each length, layer, head, row, replicate and eligible key, draw
an independent unit-rate exponential $E_j$. One replicate selects the
maximum of $s_j-\ln E_j$. The clocks $E_j e^{-s_j}$ have rates $e^{s_j}$,
so the first clock is $j$ with probability $p_j$
\citep{maddison2014astar}. Fixing all clocks gives deterministic
pairwise scores and hence an ordinary hard-attention model with $Hk$
top-$1$ heads if there are $k$ replicates. Each replicate returns the
selected value. The row update combines the replicates for each head
before applying the original update.

For a uniform tolerance $\epsilon>0$, partition the draws into groups
of size $m=\max\{1,\lceil144\sigma^2/\epsilon^2\rceil\}$. Write $\mu=\sum_jp_jv_j$
and let $\sigma^2$ bound the weighted variance. The group mean has
expected squared distance to $\mu$ at most $\sigma^2/m$, so for
$m\ge144\sigma^2/\epsilon^2$ it lies within $\epsilon/3$ with
probability at least $15/16$. Take the group mean with smallest radius
containing a strict majority of the group means, using fixed ties.
When a strict majority are good, that radius is at most $2\epsilon/3$,
and the selected neighborhood includes a good mean. The chosen mean is
then within $\epsilon$ of $\mu$. With $k\ge2m$, independent groups give failure
$q(k,\epsilon)\le C\exp(-c_*k\epsilon^2)$ for $0<\epsilon\le1$
by a binomial tail bound. Unused draws can be dropped. This deterministic aggregation is part of the
simulated model's row update.

On a fixed literal tape, condition on the tables through the previous
layer. The head estimates at the current layer are independent.
Let $A_\ell$ be the full layer's Lipschitz bound in the maximum row norm,
and $U_\ell$ the row update's Lipschitz bound in the head outputs.
While all earlier head estimates are good, the maximum row error obeys
$r_0=0$ and $r_\ell\le A_\ell r_{\ell-1}+U_\ell\epsilon$, taking
$U_\ell$ for the maximum norm over heads. Choose $\epsilon$ so these radii stay in the tubes. The good-table
event for a certified literal tape has probability at least
$(1-q(k,\epsilon))^{LHn}$, and its complement has probability
at most $LHnq(k,\epsilon)$. The tape is fixed before the clocks,
and every readout's logit error is $O(\epsilon)$ on this event.

\begin{proposition}[Stable ordinary softmax]\label[proposition]{prop:icml-stable-softmax}
Fix $L,H$ and the uniform variance, tube, Lipschitz and positive greedy
margin bounds above. For arbitrary softmax weight profiles, a correct
teacher-forced bit in the range of \cref{thm:topk-depth} has probability
at most $1/2+\beta+\exp[-\Omega(n^{1/(2^L-1)}/\ln(2n))]$.
There are $c,c'>0$ such that free-running generation within $cN$
deletions on each side has probability at most
$\beta_{\mathrm{edit}}+e^{-c'N}$.
Here $\beta$ is the mass of uncertified true contexts and
$\beta_{\mathrm{edit}}$ is the probability of success without a
certified witnessing tape. A witnessing tape is the literal tape of an
admissible total length whose generated string has a successful alignment.
\end{proposition}

\begin{proof}
Fix $\epsilon>0$ small enough that the logit error of the common
simulation is below half the decoding gap. Each certified greedy
readout then agrees on the good-table event. At this fixed tolerance
write $q(k)=q(k,\epsilon)$.
For teacher forcing apply \cref{rem:topk-explicit} to each
fixed table with $a=1+Hk$ and take $k$ a small constant times
$n^{1/(2^L-1)}/\ln(2n)$, rounded down. Adding the certificate and
sampling failure probabilities gives the single-bit claim.

For free generation we need to compare the replay probability with
the hard-attention success bound as $k$ grows. In the recurrence of
\cref{lem:position-union} put $w_r=\max(\eta,x_r,y_r)$ and use
$\Phi(t)\le C\sqrt t$. While $aw_r\le1$ one has
$z_r\le C'(aw_r)^{1/4}$ and $w_{r+1}\le C''(aw_r)^{1/8}$.
Iterating gives $w_r\le C_L a^{1/7}\eta^{8^{-r}}$.
Thus $\eta=c_{L,\delta}a^{-2\cdot 8^{L-1}}$ with a sufficiently small
$c_{L,\delta}$ keeps $aw_r$ small at all stages and makes the bottom
Walsh norm cost at most $\delta/4$ per position. In the proof of the
edit bound (\cref{app:icml-cot-edit}) take
$\xi=c_L'\eta^2/a^L$. The log list count per position is
at most $K(\xi)=L\Phi(a^L\xi)+a^L\xi\ln2$ by
\eqref{eq:icml-general-alignment}. The same entropy bound gives
$K(\xi)+2\Phi(\xi)\le\eta\ln(2)/32$ and $\xi<\eta/16$.
The fixed-table success probability is at most
$e^{-\eta n\ln(2)/32}$ for all sufficiently large $n$ at each fixed
$k$. This follows by inserting these constants in the mask sum with
$b-\lfloor\xi n\rfloor\ge\eta n/8$.

For each certified successful input choose one witnessing length and
its dense tape before drawing the clocks. Agreement at all readouts on
that tape forces the sampled model to generate it by causal induction.
The hard edit bound holds uniformly for every fixed table, including
the tables for every candidate length. Averaging therefore bounds the
mass of certified success by
\[
 e^{-\eta n\ln(2)/32}(1-q(k))^{-LHn}.
\]
Choose one sufficiently large constant $k$ so that
$LH[-\ln(1-q(k))]\le\eta\ln(2)/64$, which is possible since $q(k)$
decays exponentially and $\eta$ only polynomially in $k$.
This gives the claimed exponential bound at the positive linear
deletion radius.
\end{proof}

\subsection{Sampled categorical readouts}
For every total-length model use the ordinary categorical output
$p(z\mid h)=\exp(\ell_z(h)/\tau)/\sum_w\exp(\ell_w(h)/\tau)$ on its
finite vocabulary, with a position-dependent but state-independent $\tau\ge\tau_0>0$.
Draw independently at each
generation step conditional on the literal prefix. Vocabulary size may
grow. Assume the weighted value-variance, layer and update Lipschitz,
positive tube radius, and uniform per-logit Lipschitz bounds of
\cref{prop:icml-stable-softmax}, on every tape declared certified, uniformly across lengths and input sizes.
No output gap or attention-score gap is required. Let $R$ be the
individual-logit Lipschitz bound. For a single true context, write
$\beta$ for its probability of failing this stability certificate.
For free runs, $\beta_{\rm edit}$ is the joint probability over operands
and dense decoder draws that some allowed length succeeds but no
successful allowed length has a certified generated tape. The union
bound below permits any coupling across total lengths having the
specified within-length autoregressive laws.

\begin{proposition}[Stable softmax with sampled decoding]
\label[proposition]{prop:icml-sampled-softmax}
Fix $L,H$ and the uniform stability bounds above. For a true context
of length $n$ and a target in the range of \cref{thm:topk-depth},
put $d=2^L-1$ and $e_n=n^{-1/(2d)}\ln(2n)$.
There is $C>0$ such that sampled next-token accuracy is at most
$1/2+\beta+Ce_n$. The expected natural-log loss on the true token
is at least $-\ln(1/2+\beta+Ce_n)$. There are $c,c'>0$ such that the
probability of any free-running output within $cN$ deletions on each
side is at most $\beta_{\rm edit}+e^{-c'N}$.
\end{proposition}

\begin{proof}
\emph{Per-tape simulation and the single-bit bounds.}
Use the common simulation with per-head tolerance $\varepsilon$
small enough for the tubes and $k\ge2m$. Its good-table event $G$
on a fixed certified literal tape has failure at most $LHnq$,
where $q\le C\exp(-c k\varepsilon^2)$, and the lower bound
$\Pr_{\mathcal T}(G)\ge(1-q)^{LHn}$. With
$C_0=0$ and $C_j=A_jC_{j-1}+U_j$, the final row error is at
most $C_L\varepsilon$.
Uniformly over readout positions
and output symbols, on $G$ we have
\[
 |\ln p_t(z)-\ln \widetilde p_{\mathcal T,t}(z)|
 \le \gamma,\qquad \gamma=2R C_L\varepsilon/\tau_0.
\]
Indeed, log-sum-exp is $1$-Lipschitz in the sup norm, and the same
bound applies to each scaled numerator logit. In particular the
probability of any one symbol changes by at most $e^\gamma-1$.

At the teacher-forced readout take
$k=\lfloor c_0 n^{1/d}/\ln(2n)\rfloor$ for sufficiently small fixed
$c_0>0$ and $\varepsilon=C_1\sqrt{\ln(2n)/k}$ for sufficiently large
fixed $C_1$. Then $LHnq=O(n^{-2})$, $\varepsilon=O(e_n)$, and the
simulation fits inside the tubes eventually. Freeze the tables and an
independent decoder uniform at this position. The hard single-bit
bound, \cref{rem:topk-explicit}, applies to the resulting deterministic
top-$1$ network with $Hk$ heads. Taking token $1$ as sign $+1$ and all others as $-1$ can only
increase exact-token accuracy. Its excess accuracy is exponentially
small in $n^{1/d}/\ln(2n)$ uniformly in the tables and the uniform.
Averaging, bounding the cost on uncertified inputs by $\beta$, and
using the preceding probability difference proves the accuracy claim.
Finally Jensen gives $\mathbb E[-\ln p_T(M_T)]\ge
-\ln\mathbb E[p_T(M_T)]$. This needs no lower bound on individual
token probabilities.

\smallskip\noindent\emph{Testing only selected transitions.}
Fix an admissible output length $m$ and deletion masks $U\subseteq[m]$
and $V\subseteq[2N]$ with equal retained lengths. Write $n=5N$,
$b=\lfloor\eta n/4\rfloor$ and $B$ for the raw slots paired with the
retained target digits in $\{N-b+1,\ldots,N\}$. The masks are fixed
in this paragraph. For operands $x$ and an arbitrary list $a$ of
symbols at $U$, write $z=z(x,a)$ for the full raw output tape that
places $a$ at $U$ and the corresponding true digit at every retained
slot. (The input operands and separators are always prefixed.) At
fixed $x$, these are distinct tapes. Their dense generation masses are
$\prod_{t=1}^m p_t(z_t\mid x,z_{<t})$. A certified successful tape
contributes only if its indicator $I(x,z)$ of tube certification is one.
Put $S=U\cup B$. For each fixed $x,a$ separately, suppress the prefix
in the notation and use $p_t(z_t)\le1$ to obtain
\[
 I(x,z)\prod_{t=1}^m p_t(z_t)
 \le \frac{e^{\gamma |S|}}{(1-q)^{LHn}}
       \mathbb E_{\mathcal T}\prod_{t\in S}
                        \widetilde p_{\mathcal T,t}(z_t).
\]
For $I=1$ the right-hand product on $G$ is at least
$e^{-\gamma |S|}\prod_{t\in S}p_t(z_t)$, and the good-table lower bound
holds for this literal tape prior to the table draw. For $I=0$ the
inequality is immediate. One may now sum over all $a$ and masks and
average over $x$. This order neither selects a tape after inspecting
the tables nor requests one table that is good on all literal tapes.

For a fixed hard table, the sum over $a$ of the expected hard product
has a direct experiment. Use independent decoder uniforms indexed by
slot. At $U$ sample a token and feed it back. At retained slots always
feed the paired true digit, including at $B$. At $B$ also sample a
separate prediction with that slot's uniform and test whether it is
exactly the required digit. The probability that all these tests pass
is $\sum_a\prod_{t\in S}\widetilde p_t(z_t)$ given $x$ and the table.
This identity follows by fixing the generated list at $U$. Sampling a
prediction for a test does not change any later literal prefix.

Freeze all decoder uniforms as well as the hard tables. The deleted
lists from this experiment have a common universe over all inputs
of size at most $\exp(nK(\xi))$, for $|U|\le\xi n$ and
$K(\xi)=L\Phi((1+Hk)^L\xi)+(1+Hk)^L\xi\ln2$.
Indeed, the proof of \cref{lem:marked-position-code} recovers the
outputs generated at $U$ only from their hard-attention ancestor
marks, the marked binary or fixed tokens at other raw positions, and
the earlier decoded designated tokens. It never requires the tokens
outside $U$ to be produced by the model. Thus true digits forced at
all retained slots, including $B$, satisfy precisely its hypotheses.
Fixed position-indexed decoder uniforms are part of the deterministic
readout maps. For each possible list $a$, drop the requirement that it
actually occurred at $U$. This leaves the midpoint tests on fixed
literal prefixes. As in \cref{app:icml-cot-edit}, accept token $1$ as
sign $+1$ and every other token as sign $-1$; this only enlarges the
exact-match event. The block equality expansion and position-union
lemma give, uniformly in $a$, tables and uniforms, probability over
$x$ at most $2^{-|B|}+2e^{-\eta n}$ for this enlarged event. For fixed
deleted list, each tested prefix uses only operands and earlier true
digits in addition to its fixed tokens, as required for discrepancy.

Consequently, for $r=\lfloor\xi n\rfloor$, summing the hard hybrid
probabilities over lengths and masks and then restoring the uniform
likelihood comparison factor gives the following upper bound for
certified dense success:
\[
 e^{\gamma(b+r)+LHn[-\ln(1-q)]}\,(2r+1)
 e^{n[K(\xi)+2\Phi(\xi)]}
 \bigl(2^{-(b-r)}+2e^{-\eta n}\bigr).
\]
Here the hard marked-list count is paid only after fixing the tables
and decoder uniforms, so no dense-vocabulary cardinality enters.
One may use $\eta=c_{L,\delta}(1+Hk)^{-2\cdot 8^{L-1}}$ as in
\cref{prop:icml-stable-softmax}. First choose a fixed $\varepsilon>0$
within the tube and with $\gamma\le\ln(2)/16$. Next choose $k$ a large
constant with $LH[-\ln(1-q)]\le\eta\ln(2)/128$, possible since $q$
decays exponentially in $k$ at this fixed tolerance. Finally choose
$0<\xi<\eta/16$ small enough that $K(\xi)+2\Phi(\xi)\le\eta\ln(2)/32$.
Then $b-r\ge\eta n/8$ for large $n$, and
$\gamma(b+r)\le5\eta n\ln(2)/256$. The exponents from comparisons,
masks and deleted lists total at most $15\eta n\ln(2)/256$, whereas
the slower term in parentheses consumes at least $32\eta n\ln(2)/256$.
The polynomial front factor is absorbed for all sufficiently large $n$.
The union over allowed lengths and masks includes input- or
run-dependent witnesses. Adding $\beta_{\rm edit}$ proves the claim.
The proof also applies to other output laws with a uniform
$O(\varepsilon)$ log-probability error bound at head tolerance $\varepsilon$.
\end{proof}

\subsection{A gain bound uniform in scratchpad length}

\begin{proposition}[Diffuse sampled generation at any prescribed length]
\label[proposition]{prop:icml-diffuse-transcript}
Use a fixed-depth causal model with a growing context and full causal
attention. At every layer assume the diffuse mass and value bounds
$p_{ij}\le\rho_\ell/i$ and $\|v_j\|_2\le B_\ell$.
Assume the scaled scores are separately Lipschitz in the query and
source row, the value maps are source-Lipschitz, and the row updates
are Lipschitz in the incoming row and the head tuple. These constants
are uniform on convex row domains containing the literal trajectories
and their interpolations, as in the bounded dot-product setting below.
The readout has per-logit Lipschitz constant $R$ and samples at a
state-independent temperature at least $\tau>0$. For a constant $K_L$ depending only
on the layer bounds, the probability of an exact product suffix,
after any prescribed number of auxiliary tokens, is at most
\[
 2^{-N}+\exp\left[-c_L\frac{N}{1+(R K_L/\tau)^2}\right]
\]
when the exponent grows, with $c_L>0$ depending only on depth.
At bounded $R K_L/\tau$ the bound is still exponentially small if
each side may delete $cN$ symbols, for sufficiently small $c>0$.
A union over $F$ specified length models and suffix boundaries
multiplies the second term by $F$.
\end{proposition}

\begin{proof}
Fix $Y\ne0$ and compare inputs differing in one of the last
$\lfloor N/2\rfloor$ bits of $X$, excluding the final prompt row.
The changed raw row $p$ has index at least $\lfloor N/2\rfloor$.
Compare the models at the same literal generated history, so all
other input and generated embeddings agree. Let $D_\ell(i)$ be
the norm of the difference at an unmarked row $i$ after layer $\ell$.
For $i<p$ it is zero by causality. At $i>p$, transplanting only
the marked incoming row costs at most $2U_\ell B_\ell\rho_\ell/i$
in the row update, using the common conditional background mean.
Here $U_\ell$ is the update's Lipschitz constant in its head tuple.

For the rest of the change, keep the marked row common and
interpolate the other rows. For a head mean
$\mu=\sum_j p_{ij}v_j$, its derivative along this path is
$\sum_jp_{ij}v'_j+\sum_jp_{ij}(v_j-\mu)s'_j$, where $s_j$
is the scaled score. The query derivative costs at most
$2B_\ell C_{q,\ell}D_{\ell-1}(i)$, with $C_{q,\ell}$ the query
Lipschitz bound. The source derivatives cost at most
$\rho_\ell(D_{v,\ell}+2B_\ell C_{k,\ell})
 \sum_{p<j\le i}D_{\ell-1}(j)/i$, where $D_{v,\ell}$ and
$C_{k,\ell}$ bound the value and score as functions of a source row.
The derivative bounds hold almost everywhere and integrate along
the segments. Thus for finite layer constants $A,B,C$,
\[
 D_\ell(i)\le A/i+B D_{\ell-1}(i)
                       +(C/i)\sum_{p<j\le i}D_{\ell-1}(j).
\]
Take common maxima of these constants over the layers and
$K_0=0$, $K_\ell=A+(B+C)K_{\ell-1}$. Since
$\sum_{p<j\le i}1/j\le\ln(i/p)$, induction gives
$D_L(i)\le (K_L/i)(1+\ln(i/p))^{L-1}$. This is the
causal, row-dependent version of the single-symbol influence estimate
\citep[Lemma~5]{Hahn20}. The readout indices satisfy $i\ge2N$.
Log-sum-exp is $1$-Lipschitz in the sup norm, so the conditional
log-probability difference for any output token at that step is
bounded uniformly over eligible bits by
$d_i=(2R K_L/\tau)(1+\ln(i/\lfloor N/2\rfloor))^{L-1}/i$.
The squares have sum $S\le C_L(RK_L/\tau)^2/N$ even over
all $i\ge 2N$, by grouping indices into dyadic intervals.

Now leave $b$ of the eligible operand bits free and freeze
the rest. Let $P_x$ be the law of the generated tape at the
prescribed length on the $b$ fair bits $x$ and let $Q=P_0$.
At each step the conditional log likelihood ratio $f_i$
of $P_x$ to $Q$ lies in $[-bd_i,bd_i]$ by flipping bits on the
same literal history. Under the reference step law
$\mathbb E_Q e^{f_i}=1$, hence $\mathbb E_Q f_i\le0$.
Hoeffding's lemma gives $\mathbb E_Qe^{2f_i}\le e^{2b^2d_i^2}$.
Iterating the conditional bounds gives a tape likelihood ratio
second moment at most $e^{2b^2S}$. If each tape is consistent with
at most $V$ choices of $x$ for success, Cauchy--Schwarz under
independent uniform $x$ and reference tape gives probability at
most $\sqrt{V\,2^{-b}e^{2b^2S}}$.

For exact multiplication $V=1$, since $Y\ne0$ makes $XY$
injective in $X$. With up to $r$ deletions on each side, a given
suffix is consistent with at most $2^r S(2N+2r,r)^2$ products,
using the subset-count notation of \eqref{eq:icml-general-alignment}.
Choose the deleted positions and fill the missing binary digits.
Take $b$ to be the integer part of the smaller of
$\lfloor N/2\rfloor-1$ and $\ln(2)/(8S)$, with the first choice when
$S=0$. The exponent before the $\sqrt V$ factor is at most
$-3b\ln(2)/8$. At bounded gain $b=\Theta(N)$ and small linear $r$
preserves this decay. The exceptional event $Y=0$ has mass $2^{-N}$
globally, even after a union over $F$ candidates.
\end{proof}

\paragraph{Uniform guarantees for globally stable families.}
Fix the depth and head count. Suppose the full-layer, row-update,
and per-logit Lipschitz bounds and the attention-weighted value
variance bound hold globally with constants independent of length.
For greedy decoding also assume a uniform positive top-two logit
gap at the relevant readouts. For sampled decoding instead use
a fixed positive temperature floor. Then the exceptional terms
in the next-token and edit bounds vanish if these assumptions hold
on all relevant readouts of all literal tapes. To remove the
next-token exceptional term alone, it suffices to certify the true
contexts at the tested step. The construction of \cref{rem:averaging} uses a gain of $8N$ to recover
the operand bits from normalized averages. This gain keeps its gap
fixed. A shrinking output gap is another way a greedy model can leave
the guarantee.

\paragraph{Checking the stability constants in dot-product blocks.}
Let $Q_0,K_0,V_0$ bound the Euclidean norms of projected
queries, keys and values on a convex row ball. Write
$Q_1,K_1,V_1$ for the operator norms of the corresponding
linear maps. Measure array changes by the maximum Euclidean
row change. For head dimension $d_h$, the eligible score vector
has sup-norm Lipschitz constant at most
\[
 C_s=(Q_1K_0+Q_0K_1)/\sqrt{d_h}.
\]
Indeed, expand the difference of the query--key products
while keeping the same causal mask. Along any score path
with tangent $u$, the softmax derivative has coordinates
$p_j(u_j-\sum_b p_bu_b)$. Its $\ell_1$ norm is at most
$\|u\|_\infty$ by Cauchy--Schwarz and the variance identity.
It follows that a head mean has row-norm Lipschitz constant
at most $V_1+V_0 C_s$. Its weighted value variance is at
most $V_0^2$, since the variance is
$\sum_jp_j\|v_j\|_2^2-\|\sum_j p_jv_j\|_2^2$.

If a row update is $J_\ell$-Lipschitz in its incoming row
and $U_\ell$-Lipschitz in the maximum head-mean norm,
the full layer has
$A_\ell\le J_\ell+U_\ell\max_{\rm heads}(V_1+V_0 C_s)$.
A fixed-depth affine/ReLU block with bounded operator norms
and biases supplies these constants on bounded balls. For
layer normalization with variance offset $\varepsilon_{\rm LN}>0$,
centering is an orthogonal projection and the derivative of
$x\mapsto x/\sqrt{\|x\|_2^2/d+\varepsilon_{\rm LN}}$ has
operator norm at most $1/\sqrt{\varepsilon_{\rm LN}}$.
Coordinate gains therefore multiply this bound by their maximum
magnitude. Bounds on a ReLU update or a normalized update must
still hold uniformly in the sequence length and any growing width.

Take a fixed enlargement of the reference row balls for the
stability tubes. Suppose embeddings and row radii are uniformly
bounded over all valid tokens and lengths, and the displayed
constants and individual logit Lipschitz constants are uniform
on these enlarged balls. The sampled
certificate then holds on every literal tape, with
$\beta=\beta_{\rm edit}=0$ at a fixed temperature floor.
For greedy decoding the separate margin condition is still needed.
No estimate of these quantities on a trained model is used here.

\paragraph{A limitation of the single-bit rate.}
A one-layer uniform-attention example excludes a stretched-exponential
error term for sampled accuracy or log loss, even with a uniform value
bound and no uncertified contexts. At raw position $j$ set the value
to $b_j e_j$, with $b_j$ the token's bit or zero for a fixed separator;
use zero scores and the identity head update. The positional value maps
are $1$-Lipschitz, the full layers are uniformly Lipschitz in the
maximum of the row norms, and values stay uniformly bounded in a
fixed tube. At the final row the code is $(b_j/n)_{j=1}^n$. Distinct
binary contexts have distance at least $1/n$. On true-context codes
assign $g=(2M_T-1)/(2n)$; extend to a $1$-Lipschitz real function on
the full Euclidean space by the McShane extension. With binary unit-
temperature logits $(\ell_0,\ell_1)=(0,g)$ the correct token has
probability $(1+e^{-1/(2n)})^{-1}=1/2+\Theta(1/n)$ and its log loss
is $\ln(1+e^{-1/(2n)})=\ln 2-\Theta(1/n)$. Greedy decoding is exact.

\paragraph{Precision and support.}
Weights on a normalized grid of step $2^{-p}$ have at most $2^p$
positive entries. Small support alone leaves a dependence on the dense normalizer. If rounding
follows normalization, the tail can still change a retained coefficient. For example, keep the first $K$ masses equal to $1$
and $1/(10K)$ for the other $K-1$ entries. Set each of $4K$ tail masses
to $1/(100K)$ or $1/(11K)$. The top $K$ records stay fixed, but the
first normalized mass crosses $3/4$. Nearest rounding to
$\{0,1/2,1\}$ retains just that key with coefficient respectively
$1$ or $1/2$. Set its value to one and all others to zero. A support
count alone therefore cannot establish the proposed precision threshold.
The conclusion depends on the normalization and rounding conventions.

\section{One-pass finite-state models}\label{sec:icml-scan}

A width-$W$ scan has at most $W$ persistent configurations at each token boundary.
Its transitions and readout may depend on position, $N$, and $T$. The schedule reads operand bits
in at most $r$ blocks, each using only one operand, followed by $M_1,\ldots,M_{T-1}$ in order.
The schedule is fixed, with arbitrary bit orders within the blocks and fixed separators. This includes finite-precision recurrent networks and state-space
models when all persistent scalars, buffers, and control flags are counted jointly.

\begin{theorem}[Finite-state prediction]\label{thm:icml-recurrent}
For independent uniform $N$-bit operands, $N\ge8$, and $1\le T\le N$, the bit sign $A$ output by
a width-$W$ scan satisfies, for an absolute constant $C$,
\[
 \big|\E[A(-1)^{M_T}]\big|\le C(N+1)\big(W^{(r+1)/4}2^{-T/8}+2^{-N}\big).
\]
The bound holds for input-independent randomization whose fixed tapes give such scans.
Inverse-polynomial advantage requires $(r+1)\log_2 W\ge T/2-O(\log N)$.
For $X$ then $Y$ then the prefix, at $T=N$ this is $\log_2 W\ge N/6-O(\log N)$.
\end{theorem}

The proof in \cref{app:icml-recurrent} groups the windows of the Mellin integral by the
state at the beginning of the window. The operand blocks partition the input square into
rectangles, so the discrete transfer can charge its smoothing error once.

Storing both operands uses $2N+O(1)$ bits and suffices for exact prediction with arbitrary
readout. Thus the state needed at the midpoint for inverse-polynomial advantage is
$\Theta(N)$ bits. In particular, if all persistent configurations fit in $d$ words of
$p$ bits, then $dp\ge N/6-O(\log N)$. Persistent buffers, flags and counters enter
this budget. Auxiliary output also fits the model if generated from the state without
rereading an external scratchpad, and if all input-dependent timing and stopping information
that can affect later transitions is included in the state. With those assumptions,
compose the between-input transitions to obtain the same scan bound.

The order restriction matters. With unrestricted decoding, a three-bit message from
$X$ suffices for success at least $7/8$ given $Y$ and the true prefix at $T=N$
(\cref{prop:three-bit-decoder}). A finite-state scan must also process that continuation
within its persistent memory budget. The extended supplement treats rounded softmax with logarithmic width
and a fixed scalar format.

For suitably interleaved operands, $O(N/r+\log N)$ bits give error $N^{-\Omega(1)}$
(\cref{prop:icml-scan-order-upper}). The formal memory guarantee thus depends on the
operand order. For $p$ complete passes, the extended supplement gives
an order-$N/p$ midpoint tradeoff through $p\le N/\log_2(N+1)$.

Writing the whole product after the scan requires $2N-o(N)$ bits
for constant exact success, even with an unrestricted decoder
(\cref{prop:icml-output-memory}). Storing both operands attains the
leading constant. The proof counts the possible outputs and uses the
subexponential number of factorizations of a nonzero product.

\paragraph{Read-once branching programs.}
For exact computation from the operands alone, ordered binary decision diagrams require exponential size for the lower middle product bit ($M_{N+1}$) and for the most significant bit ($M_1$), over all variable orders~\cite{Bryant91,Woelfel05,BolligMSB11}.
Ablayev and Karpi\'nski~\cite{AK03} give lower bounds for randomized ordered programs with bounded error on every input, including a hard output bit that can depend on the variable order.
Sauerhoff and Woelfel~\cite{SW03} give branching-program lower bounds for the lower middle bit at small error and a polynomial-size read-once approximation with inverse-polynomial error under uniform inputs, using an order that interleaves the operands.
Our result (\cref{thm:icml-recurrent}) bounds average-case prediction advantage at a specified bit $M_T$ for $T\le N$, even after the scan reads the true product prefix.
It applies to the order that reads all of $X$ before $Y$, with arbitrary fixed bit orders inside each operand.

\section{One-pass recurrent predictors}\label[appendix]{app:icml-recurrent}

\begin{proof}[Proof of \cref{thm:icml-recurrent}]
Fix the randomness. For each state $s$ entering the prefix, let $a_s$ be its sign readout after
processing a formal prefix and put
$w_s(u)=a_s(b_1(u),\ldots,b_{T-1}(u))(-1)^{b_T(u)}$, using terminating binary expansions.
We first bound its Mellin transform by $C\sqrt W\,2^{-T/4}$ uniformly on the critical line.

\emph{Sampling.}
If $|G|\le1$ on $[0,1]$, $0\le a\le1$, and the real frequencies $\gamma_m$ are $1/4$-separated,
then $\sum_m |U_a(\gamma_m)|^2\le C$ for $U_a(\gamma)=\int_0^a G(v)e^{i\gamma v}\,dv$.
Indeed, bound $|U_a(\gamma_m)|^2$ by its average over $[\gamma_m,\gamma_m+1/4]$ plus the
integral of $\big|(|U_a|^2)'\big|$ there. Summing over disjoint interiors costs at most
$4\|U_a\|_2^2+2\|U_a\|_2\|U'_a\|_2$. Both norms are at most $\sqrt{2\pi}$ by Plancherel.

\emph{Shells and windows.}
On substituting $u=2^{-j}x$ for $0\le j<T$ and $1/2\le x<1$, write $D=2^{T-j}$ and
$J=\int_{1/2}^1 w_s(2^{-j}x)x^{-1/2+it}\,dx$. We claim $|J|\le C\sqrt W D^{-1/4}$.
For $|t|\le4$ the weight has bounded derivative. The test has opposite values on the two
children of each level-$(T-j-1)$ cell, so pairing intervals gives $O(D^{-1})$ for $D\ge4$.
Bounded $D$ is covered by the trivial estimate.

For $|t|>4$, take the least power of two $H\ge\sqrt{|t|}$ and write $x=(m+v)/H$,
$H/2\le m<H$, $0\le v<1$. If $H<D$, the $j$ leading zeros and the $\log_2 H$ bits of
$m$ determine a prefix state $q_m$; the remaining digits of $v$ determine an inner test
$F_{q_m}(v)$. There are at most $W$ such tests. For $H\ge D$ every window is constant,
giving at most two tests. Up to a scalar of modulus $H^{-1}$ each window contributes
\[
 \int_0^1 F_{q_m}(v)e^{i\gamma_m v}d_m(v)\,dv,\qquad
 \gamma_m=t/m,\qquad
 d_m(v)=((m+v)/H)^{-1/2}e^{it(\log(1+v/m)-v/m)} .
\]
Here $\|d_m\|_\infty+\|d'_m\|_\infty\le C$ since $|t|/m^2\le4$, and adjacent frequencies
are separated by at least $|t|/H^2>1/4$. Denote the truncated transform of $F_q$ by
$U_{q,a}$. Sampling once per class and applying Cauchy--Schwarz gives
$\sum_m|U_{q_m,a}(\gamma_m)|\le C\sqrt{HW}$ for each $a$. Integration by parts
against $d_m$ uses this estimate at $a=1$ and integrated over $0\le a\le1$.
After division by $H$ it gives $|J|\le C\sqrt{W/H}$, proving the claim when $|t|\ge D$.

For $4<|t|<D$ we have $H\le D/2$ when $D\ge16$. Each $F_q$ has paired children of
width $\delta=H/D$. Retain its values on the first child of each pair and set it to zero
elsewhere, including outside $[0,1)$, obtaining $G_q$. Then
$F_q(v)=G_q(v)-G_q(v-\delta)$. If $V_{q,a}$ is the truncated transform of $G_q$, translation gives
$|U_{q,a}(\gamma)|\le |1-e^{i\gamma\delta}|\,|V_{q,a}(\gamma)|+\delta$.
Since $|\gamma_m|\delta\le2|t|/D$, sampling $V_{q,a}$ in the same integration by parts gives
\[
 |J|\le C\big(\sqrt{W/H}\,|t|/D+H/D\big)\le C\sqrt W\,D^{-1/4}.
\]
The shells carry factors of modulus $2^{-j/2}$. Summing the claim and adding the absolute
integral $2^{1-T/2}$ over $[0,2^{-T})$ gives the asserted Mellin bound.

\emph{Joint integer transfer.}
The transcript of the $r$ operand-block exits partitions the operand square into
$K\le W^r$ rectangles $A_R\times B_R$. Once the entrance and exit states are
fixed, a block constrains only its own operand. Each rectangle fixes a prefix
test $w_R$, whose continuum operator norm is at most
$\mu=C\sqrt W\,2^{-T/4}$ by \cref{lem:hankel}.
Put $m=2^N$, $q=2^T$, and $p_R=|A_R||B_R|/m^2$.
Give each integer $x$ the interval $[x/m,(x+\epsilon)/m)$, $0<\epsilon\le1$,
as in \cref{app:finite-fourier}. Extend each operand indicator by multiplying
its values by $1/\epsilon$ on those intervals and setting it to zero elsewhere.
The product of the two $L^2$ norms on rectangle $R$ is $\sqrt{p_R}/\epsilon$.
The sum of the adjusted correlations is at most
$\mu\sum_R\sqrt{p_R}/\epsilon\le\mu\sqrt K/\epsilon$.
Every $w_R$ changes only on the same $q$-cell grid and every operand pair
belongs to just one rectangle. The common boundary count in \cref{app:finite-fourier}
therefore costs at most $4\epsilon(1+q/m)\le8\epsilon$ in total.
Take $\epsilon=(\mu\sqrt K)^{1/2}$ when it lies in $(0,1]$ and use the trivial
bound if it exceeds one. At zero take the limit. This gives
$O(W^{(r+1)/4}2^{-T/8})$, which implies the stated bound.
Averaging over fixed random tapes preserves it.
\end{proof}

\subsection{Choosing the operand order for a target bit}

\begin{proposition}\label[proposition]{prop:icml-scan-order-upper}
For every $1\le T\le N$, integer $r\ge2$, and fixed $c>0$, there is a fixed order of the
$2N$ operand bits with at most $r$ single-operand runs in which a deterministic one-pass
scan uses $O_c(N/r+\log(N+1))$ persistent bits and predicts $M_T$ with error $O_c(N^{-c})$
under independent uniform operands. The scan ignores the supplied prefix.
\end{proposition}

\begin{proof}
Number operand bits from zero at the least significant end. The target has index $k=2N-T$.
Set $L=\min\{k,\lceil(2c+8)\log_2(N+1)\rceil\}$ and retain the partial sum
$S=\sum_{k-L\le i+j\le k}x_i y_j2^{i+j}$. Higher terms are divisible by $2^{k+1}$.
The sum of lower terms is less than $h=N2^{k-L}$, since each column has at most $N$ terms.
If bit $k$ of $S$ is wrong, adding the lower terms crosses a multiple of $Q=2^k$, so
$XY\bmod Q<h$. If $L=k$, the answer is exact.

Otherwise put $D=2^N$ and $d=k-N$. For $d\le L/2$, \cref{lem:residues} bounds this event's
probability by $(3+1.9N)h/D=O(N^2 2^{-L/2})$. For $d>L/2$, fix $X=x>0$. At most
$Dx/Q+1$ intervals $[aQ,aQ+h)$ can contribute, each containing at most $h/x+1$ values
of $xy$ as $y$ varies. Summing over $x$ and including $x=0$ gives
\[
 \Pr[XY\bmod Q<h]\le h/Q+D/Q+h(1+\log D)/D^2+2/D .
\]
Here $h/Q=N2^{-L}$, $D/Q=2^{-d}<2^{-L/2}$, and $h/D^2=N2^{-L-T}$.
The choice of $L$ gives the claimed error in both ranges.

At times $t=0,\ldots,k$, process $x_t$ and then $y_{k-t}$, taking missing bits as zeros.
Keep the most recent $L+1$ bits of $X$ (initially zeros) and an accumulator modulo $2^{L+1}$.
At time $t$ add $y_{k-t}\sum_{j=0}^L x_{t-j}2^{L-j}$, the contribution with second index
$k-t$ to $S/2^{k-L}$. The accumulator's top bit gives the prediction. To use at most $r$ runs,
split the $k+1\le2N$ times as evenly as possible into $\min\{k+1,\lfloor r/2\rfloor\}$
consecutive batches. In each batch read and buffer its valid $X$ bits first, then perform
the same updates as its valid $Y$ bits arrive. Fixed zero positions need no input and can
be processed within the adjacent transitions. Each input bit is read once. The batch buffer
costs $O(N/r+1)$ bits in addition to the window and accumulator.
\end{proof}

\ificmlextended\subsection{Repeated full passes over operands and prefix}

Write $P=(M_1,\ldots,M_{T-1})$. In each of $p$ passes the scan reads $X$, then $Y$,
then $P$, retaining its state between passes. Each prefix copy is in most-significant-bit
order. Each operand copy may have any fixed internal order, including most-significant-bit order.

\begin{theorem}[Repeated full passes]\label{thm:icml-scan-multipass}
Let $N\ge8$, $1\le T\le N$, and $p\ge1$. For independent uniform $N$-bit operands,
the output sign $A$ of a width-$W$ scan satisfies
\[
 \big|\E[A(-1)^{M_T}]\big|
 \le C(N+1)\big(W^{(5p-2)/4}2^{-T/8}+2^{-N}\big)
\]
with an absolute constant $C$. Input-independent randomization is allowed if each fixed
tape gives such a scan. Inverse-polynomial advantage requires
$(5p-2)\log_2 W\ge T/2-O(\log N)$.
\end{theorem}

\begin{proof}
Fix the random tape. Record the exit state of each of the $3p$ blocks except the last.
A formal transcript $R$ fixes the entrance to every block. Its operand checks define a
rectangle $A_R\times B_R$, since each check uses just the bits of its own operand.
There are at most $K=W^{3p-1}$ transcripts. Each $(x,y)$ belongs to at most
$J=W^{p-1}$ of their rectangles, counted with multiplicity. Indeed, choosing the first
$p-1$ prefix exits determines every operand exit in order once $(x,y)$ is fixed.

For a formal prefix $q$, let $h_R(q)$ vanish if any of its first $p-1$ copies fails
its claimed exit check. Otherwise take the sign readout of its final copy. A width-$W^p$
scan computes $h_R$ in one pass over $q$ by running all $p$ copies in parallel from
their recorded entrances. Thus the test
\[
 w_R(u)=h_R(b_1(u),\ldots,b_{T-1}(u))(-1)^{b_T(u)}
\]
has critical-line Mellin transform bounded by $\mu=CW^{p/2}2^{-T/4}$, by the
prefix estimate in the proof of \cref{thm:icml-recurrent}. That estimate admits
readouts in $\{-1,0,1\}$ since paired children still cancel, and constant windows
use at most three tests.

For each fixed $(x,y)$ and each $u\in[0,1)$, the deterministic scan has a unique
transcript on the formal input $(x,y,b_1(u),\ldots,b_{T-1}(u))$ repeated $p$ times.
Consequently the sum of $w_R(u)$ over rectangles containing $(x,y)$ equals its signed
readout times $(-1)^{b_T(u)}$. At $u=xy/4^N$ this is the desired integrand. Because
the sum has modulus one and is constant on the common level-$T$ cells, linearity of the
common smoothing in \cref{lem:smooth} charges the exceptional residue event just once
per operand pair. Including the axes, the error is $O((N+1)(\rho+2^{-N}))$ at smoothing
width $2^{-N}\rho$, $0<\rho\le1$.

For the smoothed terms, use the measure $\lambda_\Omega$ of the joint integer transfer
in \cref{app:icml-recurrent}. Multiplicity gives
\[
 \sum_R\lambda_\Omega(A_R)\lambda_\Omega(B_R)
 \le J\,\lambda_\Omega(\{0,\ldots,2^N-1\})^2 .
\]
Cauchy--Schwarz over transcripts therefore replaces $\sqrt K$ by $\sqrt{KJ}$ in
that transfer. Its dyadic frequency sum bounds the smoothed correlation by
$C(N+1)\sqrt{KJ}\,\mu(1+\rho^{-1})$.
Take $\rho=(\sqrt{KJ}\,\mu)^{1/2}$ when this is at most one and use the trivial bound
otherwise. Since $\sqrt{KJ}\,W^{p/2}=W^{(5p-2)/2}$, this proves the claim on each
tape and hence after averaging.
\end{proof}

For fixed $c>0$, $p$ full passes suffice for error $O_c(N^{-c})$ with
$O_c(T/p+\log(N+1))$ persistent bits, even for prescribed operand orders.
To see this, take the band $k-L\le i+j\le k$ of \cref{prop:icml-scan-order-upper},
where $k=2N-T$ and $L=O_c(\log(N+1))$. Every contributing operand index is at least
$a=\max(0,k-L-N+1)$, since its partner is at most $N-1$. Thus only virtual times
$a\le t\le k-a$ are needed, with the preceding window initialized to zero.
There are $k-2a+1\le T+2L$ such times. Partition them evenly into at most $p$
consecutive batches, one per pass. Buffer that batch's valid $x_t$ and $y_{k-t}$ bits
in their designated slots as they arrive, then process the stored pairs in virtual-time
order and carry the window and accumulator into the next pass. All other input symbols
can be skipped, including every prefix copy. Each batch holds $O((T+L)/p+1)$ operand
bits. If $T/\log(N+1)\to\infty$ and $1\le p\le T/\log_2(N+1)$, this upper
bound and \cref{thm:icml-scan-multipass} give matching order $\Theta(T/p)$.
\fi

\subsection{Memory for writing the product}\label[appendix]{app:icml-output-memory}
\begin{proposition}[Full-product storage]\label[proposition]{prop:icml-output-memory}
Consider a scan that finishes reading the operands before it emits any
output and retains at most $s$ bits at that cut. Its decoder emits a
$2N$-bit answer using only this state and input-independent randomness.
For independent uniform operands, the probability of exact success is
at most $2^{1-N}+2^{s-2N+o(N)}$. For $0\le\rho\le1/2$, success within
$2N\rho$ Hamming errors has probability at most
$2^{1-N}+2^{s-2N(1-h_2(\rho))+o(N)}$.
\end{proposition}
\begin{proof}
Fix the random tape. Each state at the cut determines at most one output,
so at most $2^s$ output strings occur. The zero product has probability
at most $2^{1-N}$. For each positive $z<2^{2N}$ there are at most
$\tau(z)$ ordered operand pairs with product $z$, where $\tau$ counts
the positive divisors. Uniformly on this range, $\tau(z)=2^{o(N)}$:
for any $\epsilon>0$, each prime $p\ge2^{1/\epsilon}$ raised to exponent
$a$ contributes at most $2^a\le p^{\epsilon a}$ to the divisor count.
The finitely many smaller primes contribute at most a fixed power
of $2N+1$. This gives $\tau(z)\le(2N+1)^{O_\epsilon(1)}2^{2\epsilon N}$.
Counting positive products in the decoder's range proves the first bound.
For Hamming error, each output covers at most $2^{2Nh_2(\rho)}$
products by the binary entropy bound. Average over tapes to finish.
\end{proof}
Thus exact success bounded away from zero requires $2N-o(N)$ bits.
Storing both operands gives the matching leading constant. This cut
requires that persistent state include any input-dependent control or
timing used by the decoder, and gives it no later product prefix.

\section{One operand and the genuine prefix}\label[appendix]{app:icml-one-operand-bayes}

\begin{proposition}[One-operand Bayes limit and two-layer capacity]\label[proposition]{prop:icml-one-operand-bayes}
Let $X,Y$ be independent uniform $N$-bit operands, let $P=M_{<N}$,
and let $Q=2^N$. Write $a_N$ for the optimal advantage over $1/2$
using $(X,P)$. With $h_x=\lfloor Q/x\rfloor$ and $r_x=Q\bmod x$,
\[
 2Q^2a_N=Q+\sum_{x=1}^{Q-1}
 \left[\min(r_x,x-r_x)+\mathbf1_{\{x\ {\rm odd}\}}
                   (h_x-\mathbf1_{\{2r_x>x\}})\right].
\]
Consequently $a_N=3/4-\ln 2+O((N+1)2^{-N})$. For a fixed head bound,
two causal top-$k$ layers receiving just these two blocks, with
one digit per position, have capacity threshold $\Theta(\sqrt n)$
for each fixed advantage in $(0,3/4-\ln 2)$. The same statements hold
with $Y$ in place of $X$.
\end{proposition}

\begin{proof}
Fix $x>0$. The half-cell index $j=\lfloor xy/Q\rfloor$ ranges over
$0\le j<x$. Its occupancy is
$n_j=\lceil(j+1)Q/x\rceil-\lceil jQ/x\rceil=h_x+b_j$,
where the binary residuals $b_j$ sum to $r_x$ and $b_{x-1}=0$.
Majority vote in each $(x,P)$ cell is optimal. The sum of absolute
imbalances in paired half-cells is $r_x$ if $2r_x\le x$, since then
adjacent ones in $(b_j)$ are impossible. If $2r_x>x$, adjacent
zeros are impossible, and this sum is $x-r_x-\mathbf1_{\{x\ {\rm odd}\}}$.
Both claims follow by telescoping two consecutive ceiling increments.
When $x$ is odd, the unpaired last half-cell contributes $h_x$.
The zero row contributes $Q$. Dividing by $2Q^2$ yields the identity.

Write $d(u)=\operatorname{dist}(u,\mathbb Z)$. The odd-row correction
is $O(Q\log Q)$, and $\min(r_x,x-r_x)=x\,d(Q/x)$. Thus the advantage
converges to $\tfrac12\int_0^1 t\,d(1/t)\,dt$ with error $O(\log Q/Q)$.
Indeed on $[1/Q,1]$ the integrand has total variation bounded by the
integral of $1/2+1/t$, and on $[0,1/Q]$ it is at most $t/2$.
Substituting $u=1/t$ and integrating by parts on each unit interval gives
\[
 \frac12\int_1^\infty d(u)u^{-3}\,du
 =\frac14\sum_{j\ge1}
 \left(\frac1j+\frac1{j+1}-\frac2{j+1/2}\right)
 =\frac34-\ln 2,
\]
where grouping the even and odd harmonic sums gives the last equality.

For the upper construction, fixed routing gathers both visible blocks
in two layers at $k=O(\sqrt n)$ by collecting contiguous blocks and
reading their summaries at the final row.
The final update applies the Bayes rule. We prove the corresponding
lower bound for any adaptive scores. For a given $k\le c\sqrt n$, set
$K=\max(k,\lceil n^{1/3}\rceil)$ and $z=n/K^2$. It suffices to bound
the correlation by $C e^{-b z}$ with absolute positive constants
depending on the fixed head bound. We may restrict to sufficiently
small $c$ so that $z\ge1$.

Clip the first-layer records using the following argument, also used
for the comparison theorem. For each query symbol, retain the first
$\ell=\lceil C_1(K+z+\ln(2n))\rceil$ possible position--symbol pairs
in score and priority order. At the query use its hypothesized symbol,
and at a fixed position use its fixed symbol. Use the exact head when
the list fits. Otherwise a miss implies fewer than $K$ present pairs
among the retained pairs. A feasible absence pattern fixes at least
$\ell-K+1$ distinct binary coordinates. There are at most
$(e\ell/K)^K$ such patterns. At least half of the fixed bits are
in one of the two visible words. The operand bits are fair,
and a prescribed pattern on $r$ prefix bits has probability at most
$(2+r\ln 2)2^{-r}$ by the interval count in the proof of
\cref{thm:finite-two}. A union bound over rows, heads and query symbols
therefore gives miss probability at most $e^{-z}$ for sufficiently
large $C_1$. On a miss assign a fixed
fallback state. Every resulting record reads $O_H(K)$ fixed bits since
$z\le K$.

The clipped predictor is a width-$w$ decision list on the formal
visible cube, with $w=O_H(K^2)$. A rule guesses the final winner sets
and the raw bits supporting those records and the query. Each rule
tests its bit assignment and carries the corresponding bounded readout.
Order rules by their concatenated winner score and priority vectors
in descending lexicographic order. All matching rules have the same
query state. The maximal matching rule therefore has the true winner
lists. Each rule uses at most $O_H(K^2)$ bits regardless of list length.

Let $U$ be uniform on the visible cube and $p$ the Walsh projection
of the clipped predictor to degree $r=\lfloor\rho n\rfloor$.
The decision-list switching lemma \citep[Theorem~4]{Rossman2021DL}
gives $\|f-p\|_{L^2(U)}\le C_2e^{-b_2 r/w}$ and $\|p\|_{L^2(U)}\le1$.
For this consequence, leave each bit free with probability
$\lambda=1/(4C_3w)$, where $C_3\ge1$ is the switching constant.
The expected squared restricted Fourier mass at degrees at least $t$
is at most $4^{-t}$ by the decision-tree depth bound. Orthogonality
on the assigned bits expresses that expectation as the sum of
$\widehat f(A)^2\Pr[\mathrm{Bin}(|A|,\lambda)\ge t]$ over supports $A$.
Taking $t=\lfloor r\lambda/2\rfloor$ bounds the tail by
$O(e^{-\Omega(r/w)})$ by the binomial lower-tail bound.
The bounded $r/w$ case follows from $\|f\|_{L^2(U)}\le1$.

On $G=\{X\ge Q2^{-J}\}$ with $J=\lfloor \eta z\rfloor$, the
subprobability law of $(X,P)$ has density at most $3\cdot2^J$ relative
to $U$. For each fixed $x$ and prefix value, the interval of allowed
$Y$ has length $2Q/x$, giving this bound. The discarded probability
is at most $2^{-J}+1/Q$. Thus for small enough $\eta>0$,
the expected error of the projection on $G$ is $O(e^{-b_3 z})$.
For the projected correlation, each Walsh term times $\mathbf1_G$
is a bounded operand test times a prefix parity. With the target
included, its expectation is at most $2^{-DN}$ in modulus by
\cref{thm:main}. The coefficient mass of the projection is at most
$(\sum_{i\le r}\binom ni)^{1/2}$, by Parseval. Choose $\rho$ small
enough that this factor costs at most $2^{DN/2}$.
The clipping, projection and discarded probability bounds combine
to give $C e^{-b z}$. For any fixed target advantage, $c$ can be
chosen small enough to exclude that advantage for all large $N$.
\end{proof}

\subsection{Finite two-layer scratchpad bounds}\label{app:finite-two-scratchpad}

The initial operand gives an exact part of the second-layer search. Every
first-layer record in the initial $X$ segment depends only on $X$, so
rectangle discrepancy lets us verify the best keys in this segment exactly.
For the remaining keys, we filter the ranks of all auxiliary and product
readouts together. A nonnegative filter turns exact generation into a
one-sided event bound.

\begin{proposition}[Two-layer generation at a finite width]\label[proposition]{prop:finite-two-scratchpad}
Use the layout $X,\#,Y,\#$ with independent uniform $N$-bit operands.
At length $t$, take a causal two-layer model of \cref{def:topk} with
$H$ heads in layer one, $J$ in layer two, and top-$k$ at both layers,
$1\le k\le N$. It generates its own $t$ auxiliary symbols before writing
the product. Fix $1\le b\le T\le N$ and $\ell\ge 2k$.
Put $I=\{T-b+1,\ldots,T\}$, $n=2N+2+t+T-1$, and
$\Delta=\max_{\max S\in I}\disc_N(S)$. Define
\[
 B=2\biggl(\sum_{v=0}^{k}\binom{\ell}{v}\biggr)^H,\qquad
 C=(2n)^{kH},\qquad U=(B+1)/2 .
\]
For $k=1$ take $C=(n+1)^H$; when $H=k=1$ one may also take
$U=2\ell/3+5/2$. Let $M=\lceil4U+2\rceil$ and set
\[
 A=C^t\bigl[B(nN B^2)^{kJ}\bigr]^{t+b}.
\]
For each auxiliary readout $a=1,\ldots,t$ put $m=N+a$, and for each
digit readout $i\in I$ put $m=N+t+i$. For every such $m$, append the
$k$ integers $m,m-1,\ldots,m-k+1$ to a list once per second-layer head.
Denote the resulting list by $(m_v)_{v=1}^p$, $p=kJ(t+b)$. For
$0\le d<\min_v m_v$ let
\begin{align*}
 w_{m,j}&=\frac{2j+1}{m+1+j}\prod_{h=0}^{j-1}\frac{m-h}{m+1+h},\\
 S_d&=\sum_{\substack{\alpha\in\mathbb Z_{\ge0}^{p}\\|\alpha|_1\le d}}
                   \prod_v w_{m_v,\alpha_v}.
\end{align*}
With $e=\ell-k+1$ and $u_e$ as in \cref{thm:finite-two}, set
\[
 E=2H\binom{\ell}{k-1}\bigl[(2N+2)2^{-e}+(T-1)u_e\bigr].
\]
The probability of generating the full product is at most
\begin{equation}\label{eq:finite-two-scratchpad}
 E+\frac{2^{-b}+\Delta A M^{2d}}{S_d}.
\end{equation}
For operand-selected $0\le t\le s$, sum this bound over $t$, allowing
different models and witnesses at each length.
\end{proposition}

\begin{proof}
\emph{First-layer records.} For an ordinary position (an operand,
separator or genuine prefix digit), fix its raw query symbol. Each head
orders the pairs (eligible nonauxiliary position, possible raw symbol).
Keep the first $\ell$ entries, or the whole list if shorter. Take the
first $k$ present entries in this list, or all present if fewer, and
merge them with the known auxiliary raw keys to obtain the layer-one
record. Use a fixed default state if there are too few keys for the
specified update. This clipping agrees with the original whenever
the list is uncut or at least $k$ of its first $\ell$ entries are present.
If clipping can change a record, some $e$ of those entries are absent.
In any feasible absence pattern the positions are distinct binary slots.
At an operand or separator its probability is at most $2^{-e}$; at a
digit it is at most $u_e$ by the prefix-pattern estimate in the proof of
\cref{thm:finite-two}. A union bound over the $\binom{\ell}{k-1}$
patterns, ordinary queries, heads and possible raw query symbols gives
$E$. The list here omits all auxiliary keys before truncation, so this bad
event is common to every auxiliary history of the given length.

With auxiliary raw symbols fixed along a replay branch, the ordinary
record code consists of the query symbol and each head's retained ranks.
There are at most $B$ codes. Each code event is a cylinder: it specifies
the retained entries and the necessary prior absences. This includes all
absences after the last retained entry when fewer than $k$ were present.
The auxiliary raw keys are then known and determine the merged lists.
An indicator of any subset of codes has Walsh norm at most $U=(B+1)/2$,
using either the subset or its complement as in \cref{thm:finite-two}.

For the improved $H=k=1$ estimate, condition the ordinary query symbol
algebraically by multiplying by its literal. For each fixed symbol, any
Boolean predicate of the code is a decision list of at most $\ell$ fresh
literals. An already determined test can be skipped or ends the list.
If the first return value is $a\in\{0,1\}$ and the remaining list has
mean $x$ on the independent formal cube, the mean of the list is
$y=(a+x)/2$. Grouping Walsh coefficients by the last tested literal in
their support gives nonconstant mass equal to the sum of $|a-x|/2$
over the list: the current sign coefficient is $\pm(a-x)/2$, gated by
the earlier failure cylinder of norm one, and different last literals
have disjoint Fourier supports. With $V(x)=2(x-1/2)^2$,
\[
 |a-x|\le 2/3+V(x)-V(y).
\]
For $a=0$ the slack is $3(x-2/3)^2/2$; reflection gives $a=1$.
Telescoping, using $0\le V\le1/2$ and adding the constant coefficient,
bounds each list's norm by $\ell/3+5/4$. The two query literals give the
stated $U$.

Immediately after generating an auxiliary symbol, guess its exact
first-layer record for use as a key or a query. At its position the
raw symbol and all eligible auxiliary raw keys are now known. Per head,
guess the ordered top $k$ nonauxiliary (position, raw symbol) pairs.
Their validity is a conjunction of raw literals, and merging with the known
auxiliary keys determines the exact update. This costs at most $C$ choices
per symbol. For $k=1$ a feasible first-present entry is among the first
$n+1$ ranks: all preceding absences must be at distinct bit positions.
Thus an auxiliary key's record, once guessed, is fixed on its branch.
No alphabet-size factor is needed.

\emph{Replaying second-layer ranks.} At a readout, guess its first-layer
record and, for each head, the ordered top $k$ keys from the first $N$
positions, with their records. Given the guessed query record, checking
this entire initial segment is a Boolean test of $X$ alone, by causality.
Independently propose $k$ distinct indices outside the segment, together
with their records. In succession assign each proposed index its rank
among the remaining outside keys. Freeze the earlier indices when
computing each such rank. It counts the outranking keys among $m_v$ others:
each comparison tests one ordinary record (norm at most $U$) or a known
auxiliary record (a constant). Merge these $k$ proposed keys with the
initial segment's top $k$, sort them in model order and return the best
$k$. When all proposed ranks are zero, the true global list is recovered.

At every readout there are at most $B(nN B^2)^{kJ}$ index and record
choices, since an already guessed auxiliary key or query needs no
additional record choice. The cylinder tests for the anchor records
and the $X$-only tests enforce these guesses. For each formal input
and each vector of ranks in $\prod_v\{0,\ldots,m_v\}$ exactly one replay
is consistent: in chronological order take the specified rank among the
remaining keys, emit each proposed auxiliary token, and then recover its
first-layer record. All digit queries use the genuine earlier digits,
irrespective of their proposed readouts. The grid sides therefore depend
only on the query positions.

\emph{A joint endpoint filter.} Use the Gram polynomials $G_{m,j}$
from \cref{thm:finite-two}. Their value at zero is one, their squared
norm under counting measure is $1/w_{m,j}$, and
$A(G_{m,j}(r))\le M^j$ when $r$ counts $m$ predicates of Walsh norm at
most $U$, by the Newton expansion in that proof. Write
$w_\alpha=\prod_v w_{m_v,\alpha_v}$ and
\[
 q(r)=\frac1{S_d}\sum_{|\alpha|_1\le d} w_\alpha
                  \prod_v G_{m_v,\alpha_v}(r_v).
\]
Then $q(0)=1$ and orthogonality on the grid gives $\sum_r q(r)^2=1/S_d$.
On the good clipping event the genuine zero-rank trajectory is among the
replays. Hence the sum, over all replays, of $q(r)^2$ times the indicator
that the $b$ signed answers agree with the true digits majorizes exact
generation there. A token other than $0$ or $1$ may be mapped to either
sign for this upper bound.

Expand the agreement indicator into $2^b$ signed moments.
The empty moment has weight $1/S_d$. For a nonempty set $D\subseteq I$,
keep all auxiliary ranks and only digit ranks belonging to $D$.
Summing over the omitted ranks leaves the signed readouts unchanged:
their predicted digits never enter the genuine context. In the square
of the Gram expansion, summation in an omitted coordinate $v$ replaces
two factors by $\mathbf1_{\alpha_v=\beta_v}/w_{m_v,\alpha_v}$.
Thus the marginal kernel has nonnegative expansion coefficients on
products of Gram factors of total degree at most $2d$. Their sum is
its value at zero on the retained coordinates, at most $1/S_d$ since
this is a partial sum of the nonnegative $q^2$.

Now only retained readouts are replayed. For each fixed branch the
signed readout is known, the anchor cylinders have joint norm at most
one, and all initial-segment checks combine into a single unit-bounded
function of $X$. Apart from that factor the marginal kernel and the
anchor tests have Walsh norm at most $M^{2d}/S_d$. At $u=\max D$,
multiply by the earlier true signs from $D$ and expand in formal
characters. Each term is a product-bit parity ending at $u$ times a
bounded test of each operand, including the initial-segment check on
$X$. Its expectation has absolute value at most $\Delta$ by the
definition of rectangle discrepancy. Counting branches by $A$ bounds
each nonempty moment by $\Delta A M^{2d}/S_d$. The $2^{-b}$
normalization and the clipping error yield \eqref{eq:finite-two-scratchpad}.
\end{proof}

\begin{center}
\begin{tabular}{rcrrr}
\toprule
 & & \multicolumn{3}{c}{Input-selected scratchpad budget $s$}\\
$N$ & Discrepancy & $0$ & $1$ & $2$ \\
\midrule
32,768 & Hand & .37075 & .98526 & \\
32,768 & Certified & .31117 & .82669 & \\
65,536 & Hand & .20178 & .53931 & \\
65,536 & Certified & .15276 & .39476 & .78957 \\
131,072 & Certified & .05417 & .14324 & .28546 \\
\bottomrule
\end{tabular}
\end{center}
These are upper bounds on exact full-product success with one top-$1$
head per layer, rounded upward. At $N=131{,}072$, the certified bounds
for $s=3,4$ are $.51766$ and $.90792$. The hand rows use the analytic
discrepancy in \cref{thm:rates} at $T=\lfloor12N/13\rfloor$. The certified
rows use the three-term estimate in its proof at
$T=\lfloor89N/100\rfloor$. Choosing an interior block gives both a shorter
rank grid and a favorable discrepancy exponent. The integer checker
\texttt{check\_finite\_two\_scratchpad.py} lists all witnesses and rounds the
convolutions defining $S_d$ downward.

\section{Exact accuracy at small operand widths}\label[appendix]{app:exact-head}
We optimize over all one-layer, one-head top-$1$ models in \cref{def:topk} at $T=N$.
For each $N$ the context is the binary string for $X$, a separator, the string for $Y$,
a second separator, and the true product prefix. The last token is the query bit.
The computation uses all $4^N$ ordered operand pairs.
\begin{center}
\begin{tabular}{rrrr}
\toprule
$N$ & Operand pairs & Bayes optimum using prefix only & Exact one-head optimum \\
\midrule
2 & 16 & 13 & 16 \\
3 & 64 & 42 & 58 \\
4 & 256 & 156 & 194 \\
5 & 1024 & 565 & 659 \\
\bottomrule
\end{tabular}
\end{center}

Fix the query bit. Scores then order the possible pairs (key position, token value).
Only one token value is present at each position. The head selects the first present
pair, and its readout is constant on each winner. Constant positions can be
represented by a default. After failure of a literal on a binary position, the other
literal on that position is certain. Therefore every head is represented by an ordered
literal decision list with a default.
Conversely, a head can give the list literals strictly decreasing scores above
one, a separator score one, and other pairs score zero. The general scoring and
update rules allow separate lists and readouts for the two values of the query bit.

For the population subset $D$ remaining after failed tests, let $m(D)$ be its majority count.
The optimum $F(D)$ maximizes $m(D)$ and, over all binary coordinate splits with
both sides nonempty, $m(D_{j=b})+F(D_{j\ne b})$. Induction on $|D|$ proves that this
recurrence is exact. Summing the two query-group optima gives the table.
The supplement reruns the recurrence and replays the optimal heads on every
input. The results bound bilinear UHATs from above as a subclass. They
describe the four specified widths and do not estimate an asymptotic decay rate.

\section{Fixed-width tokenization}\label[appendix]{app:extensions}
\ifdefined\ZENODODOI\fi

\subsection{Fixed-width input and output tokens}
Fix $w,L,H,k$ and $\alpha>0$. Allow a fixed layout of at most $4N$ input positions, each an arbitrary
function of at most $w$ designated operand or earlier product bits. The designated sets may overlap.
The length and the sets depend only on $N$ and the prediction step. The output token encodes a chunk
of $r\le w$ consecutive bits wholly within $[\alpha N,N]$. Its true context ends before the chunk.
The probability of predicting the entire chunk is at most
\[
 2^{-r}+\exp\!\bigl(-\Omega(N/(\ln N)^{2^L-2})\bigr).
\]

To see this, use $m_0=4N$, $a=1+Hk$, $K=4a^L\ln(2^{w+1}m_0)$ and $\omega=4c^2$.
The alphabet count in \eqref{eq:depth-alph} has initial bound $2^w$ in place of $2$.
The proofs of \cref{lem:jointtr} and parts (a) and (b) of \cref{lem:onelayerback} still apply: tests are
conjoined per input word, and the coefficient count is bounded by $e^{Kd}$ using the new $K$.
Use $m_0$ as the upper bound on the number of positions throughout \eqref{eq:depth-rec}.
After all $L$ approximate pullbacks the error is at most $e^{-v}/2$ and the record degree at most
$\omega m_0 K(v/(m_0K^2))^{2^{-L}}$. Each record reads at most $w$ designated bits, so substitution
multiplies ordinary degree by at most $w$, even if supports overlap.

Choose $0<\theta<1/2$ with $h(\theta)\le D\alpha\ln 2/3$, where $h$ is binary entropy in nats, and
$\lambda=\min\{1/2,\theta/(2w\omega),\sqrt{D\alpha\ln2/8}\}$. Take
$v=m_0\lambda^{2^L}/K^{2^L-2}$; this is at least $K+\ln(2L)$ for all sufficiently large $N$.
Then the degree on the formal bit cube is at most $\theta m$ for any formal
dimension $m=2N+u-1$ at an index $u$ in the chunk. By entropy counting, Parseval, and the uniform bound
on the approximant, its Walsh norm is at most $(3/2)\exp(mh(\theta)/2)$.
In the equality indicator of the predicted and true chunks, expand over all subsets of chunk indices.
For a nonempty subset with largest index $u$, approximate the corresponding parity of the predicted
chunk as one bounded function of the top state. Multiply that approximant by the character on the earlier
selected true bits, adding unused formal coordinates if needed. This multiplication preserves Walsh norm.
By \eqref{eq:walsh} the term is at most $e^{-v}/2+(3/2)\exp(-D\alpha N\ln2/2)\le2e^{-v}$
in absolute expectation. The empty term contributes $2^{-r}$. Mapping invalid output tokens to a fixed
word can only increase the success probability. The hidden constant in the exponent can depend on depth.

\section{Effective support of a softmax predictor}\label[appendix]{app:icml-softmax-effective}

Fix $k$ and run the same truncated network on every context. At each head it
selects the $\min(k,i)$ highest-scoring causal keys, with fixed tie priority, and
renormalizes their softmax weights. This is a model of \cref{def:topk}:
the update can recompute every selected score from its query and selected
records. The following certificate uses the discarded weights and values
on this network's own trajectory. It applies at native temperature and
allows ties at the cutoff.

\begin{proposition}[A certificate for greedy softmax predictions]
\label[proposition]{prop:icml-softmax-effective}
On a context $x$, write $\widetilde Y^\ell$ for the array of truncated states at
layer $\ell$. Use the maximum norm over positions and coordinates.
Let $F_\ell$ be the full softmax layer and $A_\ell$ a Lipschitz bound for it.
Initialize $e_0=0$ and compute
\[
 d_\ell(x)=\|F_\ell(\widetilde Y^{\ell-1})-\widetilde Y^\ell\|_\infty,
 \qquad e_\ell=A_\ell e_{\ell-1}+d_\ell(x).
\]
Suppose each last-position output logit is $R$-Lipschitz. Write $m_k(x)$ for the difference
between the two largest truncated output logits. Whenever $m_k(x)>2Re_L$,
the dense and truncated greedy outputs agree. If $b_k$ bounds the prediction
advantage of every top-$k$ model of this architecture under the true-context
law, and $p_k=\Pr_x\{m_k(x)\le 2Re_L(x)\}$, then
\[
 \Pr\{\text{dense output}=M_T\}\le \tfrac12+b_k+p_k .
\]
For fixed $L,H,k$ in the target range of \cref{thm:topk-depth}, one may take
$b_k=e^{-c_0n}$ for all sufficiently large $n$. For growing parameters,
\cref{rem:topk-explicit} gives $b_k=e^{-t}$ under its finite parameter condition.
\end{proposition}

\begin{proof}
At layer $\ell$, insert $F_\ell(\widetilde Y^{\ell-1})$ between the full
and truncated outputs. The triangle inequality and induction bound their
distance by $e_\ell$. Each logit therefore moves by at most $Re_L$,
so the strict margin preserves the winning token. Success can increase by
at most the probability of greedy disagreement.
\end{proof}

\paragraph{Computing the tail contribution.}
At the truncated incoming states of a head, let $q$ be the full softmax law,
$S$ its retained set, $t=q(S^c)$, and $v_j$ its values. The exact error in
this head at these common incoming states is
\[
 t\left\|\mathbb E_q[v\mid S^c]-\mathbb E_q[v\mid S]\right\|_\infty ,
\]
with value zero if $t=0$. Its maximum over positions and heads, multiplied by
the Lipschitz constant $U_\ell$ of the positionwise update in its tuple of
arguments, bounds $d_\ell$. Thus a large tail may contribute zero:
with equal scores, a single retained value zero and discarded values in
balanced pairs $+1,-1$ give identical head outputs.
For an explicit $A_\ell$, suppose the scaled scores and values have Lipschitz
bounds $D_\ell,V_\ell$, respectively, and all values have norm at most $B_\ell$.
Then one may take $A_\ell=U_\ell\max\{1,V_\ell+B_\ell D_\ell\}$.
Indeed, along a logit displacement $w$ the derivative of a softmax law $q$
has entries $q_j(w_j-\mathbb E_q w)$. Its half $\ell_1$ norm is at most
$\|w\|_\infty/2$ by Cauchy--Schwarz. Changing the values costs at most
$V_\ell$ times the incoming distance, and changing the weights costs at most
$B_\ell D_\ell$ times that distance.

\paragraph{Positionwise bounds from the same trace.}
The certificate can also propagate distinct radii for each position.
Start with radius zero at every input. For a head at position $i$, bound
the deviation of each scaled score from its truncated-trace value by
$s_j$ and that of each value vector by $u_j$, throughout the incoming
state boxes. For example, incoming radii $r_i,r_j$ permit
$s_j=D_\ell\max(r_i,r_j)$ and $u_j=V_\ell r_j$ for a key-only value map.
Let $g=\mathbb E_q[v\mid S]$. For each coordinate $c$ and sign
$\sigma\in\{-1,1\}$ put $b_j=\sigma(v_{jc}-g_c)+u_j$. Take the root $z$
of
\[
 \sum_{j\le i} q_j\max\{e^{-s_j}(b_j-z),\ e^{s_j}(b_j-z)\}=0 .
\]
The left side decreases strictly from positive to negative and is
piecewise linear with breakpoints $b_j$. Indeed, the dense probabilities
are proportional to $q_j\lambda_j$ for positive factors
$e^{-s_j}\le\lambda_j\le e^{s_j}$. The root is the largest weighted mean
$\sum_j q_j\lambda_j b_j/\sum_j q_j\lambda_j$ over these intervals:
subtract $z$ in the numerator and maximize each term separately.
It bounds the signed coordinate error of the dense head relative to $g$.
The maximum root over coordinates and signs bounds the head error.
Multiplying $U_\ell$ by the maximum of the previous query radius and all
its head error bounds gives a valid new radius at $i$. Induction now permits
the final position's radius in place of $e_L$ in
\cref{prop:icml-softmax-effective}. All quantities use the same fixed $k$
throughout the computation and across contexts.

\paragraph{Free-running coupling.}
For each fixed total length, let the truncated model generate its own
raw tape. At each of its prefixes, evaluate the certificate above
for the logits of that length's dense and truncated models on the same
literal context. If the certificate holds at every step along that tape,
the dense model generates the identical tape. This follows by induction:
the two prefixes coincide up to the first possible disagreement, and the
margin certificate rules out a disagreement there.

Let $E$ be the event that at least one certificate fails along at least one
truncated run of a candidate output length in the range of
\cref{thm:cot-avg}. On $E^c$, every corresponding dense run has the same
edit distance to the product as its truncated run. Thus the probability
of dense generation within the short edit budget is at most
$e^{-c_1N}+\Pr(E)$. This extension conditions the conclusion on a
certificate along generated trajectories.

\section{Discrepancy down to the least significant bit}\label[appendix]{app:icml-bottom}

\begin{theorem}\label{thm:icml-bottom}
There is an absolute $c>0$ such that for every $N\ge2$, every $N<T\le2N$,
and every mask $S$ with $\max S=T$,
\[
 \disc_N(S)\le 2^{-c\min\{\,2N-T+1,\ \sqrt{N/\log_2(2N)}\,\}} .
\]
\end{theorem}

Together with \cref{app:explicit-d}, this covers every product position. The exponent is linear in distance
from the low end throughout the last $\sqrt{N/\log_2(2N)}$ positions. The stronger pointwise estimate
in \cref{app:midpoint} applies in its overlapping range.

\begin{proof}
Write $H=2^N$, $R=\sqrt{N/\log_2(2N)}$ and $B=\{2N-t:t\in S\}$, so its indices count from the low end.
Put $m=\min B$, $k=|B|$ and $w_B(z)=\prod_{b\in B}(-1)^{\lfloor z/2^b\rfloor}$.
We first work at large $N$. All averages and $L^2$ norms below use normalized counting measure.

\emph{A complete residue table.}
If $q=2^u$ and $A\subseteq\{0,\ldots,u-1\}$ contains $u-1$, then
\[
 |\mathbb E_{x,y\bmod q} F(x)G(y)w_A(xy)|
 \lesssim 2^{-u/4}\|F\|_2\|G\|_2 .
\]
Indeed, on the units modulo $2^s$ a kernel $W(ab)$ has norm at most
$\max_\chi|\mathbb E_z W(z)\chi(z)|$ by multiplicative Fourier expansion and Parseval.
For a Walsh sign containing digit $s-1$, $s\ge2$, characters factoring modulo $2^{s-1}$
cancel by pairing $z$ with $z+2^{s-1}$. For any remaining character take
$h=\lceil s/2\rceil$, $d=s-h$. On $1+2^h v$ the character has the form
$e(cv/2^d)$, $e(t)=\exp(2\pi i t)$, since $2h\ge s$. Here $c$ is odd by nontriviality at
$1+2^{s-1}$. Writing $z=a+2^h v$ splits the Walsh sign into factors in $a,v$.
The $v$-average is a Fourier coefficient of the high factor at $\pm ca^{-1}\bmod 2^d$,
up to a unit scalar. As $a$ ranges over odd residues modulo $2^h$, this frequency is uniform
on odd residues modulo $2^d$. Cauchy--Schwarz and Parseval bound the average modulus by
$\sqrt{2/2^d}\lesssim2^{-s/4}$. Use the trivial bound for $s=1$.
For full residues condition on the operand valuations $i,j$, capped at $u$.
When $s=u-i-j>0$ the odd parts project independently to uniform units modulo $2^s$,
and the reduced mask contains $s-1$. When $s\le0$ the product vanishes modulo $q$.
Both cases cost $O(2^{-u/4}2^{(i+j)/4})$ times the conditional $L^2$ norms.
Summing by Cauchy--Schwarz proves the assertion since $\mathbb E 2^{i/2}=O(1)$.

\emph{Masks avoiding the middle band.}
Fix $\beta=1/128$. If $B\cap[\beta N,N)=\varnothing$, its low part is nonempty.
Let $u-1$ be its highest index, $q=2^u$, and $T_0$ the largest MSB-first index of its
upper part (zero if empty). Thus $m+1\le u\le\beta N+1$ and $T_0\le N$.
For $T_0\le N/2$, tile the operand square by $q$-by-$q$ squares. The upper sign is constant
on each square except those meeting a boundary $xy=j2^{2N-T_0}$, $0<j<2^{T_0}$.
Each descending curve crosses $O(H/q)$ squares by counting grid-line intersections.
The exceptional area fraction is $O(2^{T_0+u-N})$. On every other square the complete-table
bound applies to the low part, giving total discrepancy $O(2^{-u/4})$.
If $T_0>N/2$, instead restrict the tests to each of the $q^2$ residue pairs modulo $q$.
The low sign is constant there, and \cref{thm:main} bounds the total by
$q^2 2^{-T_0/8}\le 2^{2u-N/16}=O(2^{-u/4})$.

\emph{Sparse masks meeting the middle band.}
Suppose $k<c_0R$ for a sufficiently small absolute $c_0$. Starting at a selected index in
$[\beta N,N)$, the gaps between consecutive selected indices, including the final gap to $2N$,
sum to at least $N$. Choose a gap $G\ge N/k$ starting at $b_*\ge\beta N$.
Put $r=\lfloor\min(b_*,G)/16\rfloor\gtrsim N/k$ and $L=2^r$.
For $b<r$, the exact Fourier expansion of the digit sign modulo $2^{b+1}$ uses only odd
harmonics $|h|\le2^b<L$ by half-period antisymmetry. Its coefficients are $O(1/|h|)$
by geometric summation, including the sole harmonic at $b=0$.
For $b\ge r$, take the Fej\'er mean of order $L$ of the periodic square wave evaluated at
$xy/D$, $D=2^{b+1}$. It is bounded by one, uses only odd $0<|h|<L$, and has the same coefficient bound.
Outside distance $\delta=L^{-1/2}$ of the two jumps its error is $O((L\delta)^{-1})$ by the Fej\'er kernel tail.
Since $D\mid H^2$, the exceptional strips contain $O(H^2(\delta+D^{-1}))$ integers in $[1,H^2)$.
Each positive product has at most $d(n)$ representations. Now
$\sum_{n\le H^2}d(n)^2\le\sum_{n\le H^2}d_4(n)\le H^2(1+\log H^2)^3$
by the prime-power inequality and three harmonic sums. Cauchy--Schwarz, the zero-product mass $\le2/H$,
and telescoping give total smoothing error $O(k(1+N)^{3/2}2^{-r/4}+k/H)$.

Each surviving frequency $\theta=\sum_{b\in B}h_b/2^{b+1}$ has partial sum through $b_*$
of exact denominator $2^{b_*+1}$ by oddness. Its remaining tail is at most $2^{r-b_*-G}$.
Hence for $q\le2^r$ it differs from any $p/q$ by at least $2^{-b_*-r-2}\ge2^{-2N+15r-2}$.
Dirichlet approximation at $Q=H^2 2^{-2r}$ therefore gives a reduced $a/q$ with
$2^r<q\le Q$ and $|\theta-a/q|\le1/q^2$.
Cauchy--Schwarz in $x$ and geometric summation bound the squared bilinear form for $e(\theta xy)$ by
$H^{-1}+2H^{-2}\sum_{0<j<H}\min\{H,(2\|j\theta\|)^{-1}\}$.
In blocks of $\lfloor q/2\rfloor$ consecutive $j$ the points are $1/(2q)$-separated modulo one,
so each block costs $O(H+q\log(2q))$. With $O(1+H/q)$ blocks this gives
$O((1+N)2^{-r})$ for the squared form. The total absolute Fourier weight is at most
$[C(1+r)]^k\le2^{r/8}$ by choosing $c_0$ small, since $k^2\log_2(2N)<c_0^2 N$.
Together with the smoothing error this gives $2^{-\Omega(N/k)}$.

\emph{Large masks.}
For $k\ge c_0R$, remove operands below $2^{N-d}$, $d=\lfloor k/16\rfloor$, at probability cost
$\le2^{1-d}$. Partition the rest into dyadic boxes with lower endpoints $2^a,2^b$, swapping
operands so $N-d\le a\le b<N$. Delete indices $\ge a+b+2$, which vanish there.
Replace $y$ by $v=2y$, zero its coefficient on odd $v$, and increment the retained indices
by one to form $A$. This preserves the sum and ensures the second range is strictly longer.
In Bourgain's notation take $\mu=a,\ \nu=b+1,\ \lambda=a+b+3$, so the dyadic scales
are $2^\mu\le2^\nu$, their product is comparable to $X_0=2^\lambda$, and $A\subseteq[0,\lambda)$.
Apply \cref{lem:icml-interior-typeii}, using its uniform cutoff
$M_1=M$ because $\rho$ may be sublinear in $\lambda$.
Choose $\epsilon$ reciprocal-integral as there and $\rho$ the largest
multiple of $1/\epsilon$ below $\kappa k$ for sufficiently small absolute
$\kappa>0$. Here $\mu\ge7N/8$, $\mu/\lambda\ge1/3$ for large $N$,
and $|A|\ge7k/8$. The upper restriction $\rho<\mu/100$ follows for
$\kappa<7/1600$, and $k\gtrsim\sqrt{N/\log_2(2N)}$ ensures the
lower restriction $\epsilon\rho\gg(\log\lambda)^2$. The same
choice of $\kappa$ as in the dense-box argument makes each term
save $2^{-\Omega(k)}$.
Since $X_0$ is eight times the original box area, summing and restoring the strips yields $O(2^{-\Omega(k)})$.

These cases give $C2^{-c'\min(m+1,R)}$. To remove $C$ at bounded exponents, the complete
table has bounded-test discrepancy at most $1-\delta_0$ for some absolute $\delta_0>0$.
Its norm bound suffices for large $u$. For the finitely many smaller tables the rows $0,1$
and columns $0,2^{u-1}$ have negative four-entry product. By compactness, equality at one
would force every signed term to have the same phase, contradicting this minor.
Thus the tiling case inherits a uniform gap, as its exceptional fraction is $o(1)$.
All other bounds tend to zero with $N$. The gap absorbs $C$ after decreasing $c'$.
For the finitely many smaller $N$ the same minor with columns $0,2^m$ gives strict
discrepancy below one for every mask, completing the proof.
\end{proof}

\section{Two layers below the midpoint}\label[appendix]{app:icml-bottom-topk}

\begin{lemma}[a hidden operand block after the midpoint]\label[lemma]{lem:icml-hidden-block}
Fix circuit depth $d$, size exponent $b$ and $D>0$. For sufficiently large $N$ there is a
multiple of eight $\ell\le C_{d,b,D}(\log(2N))^{c_d}$, where $c_d$ is independent of $D$,
with the following property. Suppose $N<T\le2N$, $a\ge T-N$ and $a+\ell\le N$.
Hide bits $a+1,\ldots,a+\ell$ of $X$ in MSB-first indexing. The remaining operand bits
may select any depth-$d$, size-$N^b$ Boolean AND/OR circuit on the whole true prefix,
with negated inputs allowed. Its mean absolute conditional advantage for $M_T$,
averaged over the exposed bits, is at most $N^{-D}$.
\end{lemma}

\begin{proof}
Put $Q=2^N$, $R=2^\ell=\lambda^8$, $t=T-a+1$, $\nu=2^t$, and
$S=2^{2N-a+1}$. Then $R\le Q$ and $R\le\nu\le2Q$. The exposure
$\xi=(x_0,y)$ zeros the hidden bits of $X$ and fixes $Y=y$, so
$X=x_0+2^{N-a-\ell}U$ with $U$ uniform on $\{0,\ldots,R-1\}$.
Writing $Z(u)=x_0y/S+yu/(2QR)$, the first $t$ fractional bits of $Z(U)$
are $M_a,\ldots,M_T$. For $0\le u<R$ the integer $\lfloor2Z(u)\rfloor$ is
$c$ or $c+1$ for $c=\lfloor2Z(0)\rfloor$. Its parity is the first local bit.
Thus every prefix bit before $a$ is a constant or a literal of that bit.
For $V=U+V_0$, with $V_0$ independent uniform on $[0,1)$, the fractional part
of $Z(V)$ is uniform on an arc of length $y/(2Q)$ for $y>0$.
The reference local word $j'=\lfloor\nu Z(V)\rfloor\bmod\nu$ has density at most
$2Q/y$ relative to the uniform $t$-bit word. Its last bit has prediction advantage
at most $2Q/(y\nu)$ in absolute value against any function of the preceding bits:
only the two end pairs of adjacent even/odd cells can contribute to the bias.

\emph{Comparing digit moments.}
Write $\mathrm e(z)=\exp(2\pi i z)$, $K=\nu\lambda^2$, and for integer $h$ set
\[
 G_y(h)=\E_U\mathrm e(hyU/(2QR)),\qquad
 \Phi_y(h)=\E_{V_0}\mathrm e(hyV_0/(2QR)).
\]
For $0<|h|\le K$ one has $\E_y|G_y(h)(1-\Phi_y(h))|\lesssim\lambda^{-2}$.
Indeed, for $|h|\le\lambda^6$ use $|1-\Phi_y(h)|\le\pi |h|/R$.
For $\lambda^6<|h|\le K$ put $\delta=\lambda^{-4}$.
The fraction of $y$ with $\|hy/(2QR)\|\le\delta$ is at most
\[
 |h|/(2QR)+2/Q+2\delta+8\delta R/|h|\lesssim\lambda^{-2}.
\]
There are at most $|h|/(2R)+2$ relevant integers, with at most
$1+4\delta QR/|h|$ points of the equally spaced $y$-grid near each.
Here $K/(2QR)\le\lambda^{-6}$. For all other $y$, geometric summation gives
$|G_y(h)|\le(2R\delta)^{-1}$. Thus $\E_y|G_y(h)|\lesssim\lambda^{-2}$
throughout the high-frequency range, as also used next.

For $W=Z(U)$ or $Z(V)$ one has $\Pr[\|\nu W\|\le1/\lambda]\lesssim1/\lambda$.
The Fej\'er kernel $F_l(z)=l^{-1}|\sum_{j=0}^{l-1}\mathrm e(jz)|^2$,
$l=\lambda/2$, satisfies $F_l(\nu W)\ge4l/\pi^2$ on the indicated event.
Every nonzero frequency $h$ of $F_l(\nu\,\cdot)$ has $\lambda^6<|h|<K$.
The high-frequency estimate and the at most $2l$ coefficients of modulus at most one
give $\E F_l(\nu W)\le1+O(l\lambda^{-2})$.

Let $\omega_I(z)$ be a Walsh character on any $I$ of the $t$ local bits of $z$.
On $\mathbb Z/\nu$ each single digit sign has additive Fourier norm $O(\log(2\nu))$:
over its fundamental period geometric summation bounds the nonzero coefficients by
$O(1/\min(h,\text{period}-h))$, and period repetition preserves the norm.
Convolution of coefficients gives norm at most $[C(N+1)]^{|I|}$ for the product.
Extending the word to a step function on the circle multiplies its normalized discrete
coefficient at $h\bmod\nu$ by $\nu\int_0^{1/\nu}\mathrm e(-hz)\,dz$.
The absolute values of these factors sum to $O(1+\log K)=O(N+1)$
in each residue class over $|h|<K$, using $\min(1,\nu/(\pi|h|))$ for $h\ne0$.
Hence $F_K*\omega_I$ has Fourier norm at most $[C(N+1)]^{|I|+1}$.
Outside $\|\nu W\|\le1/\lambda$, shifts of magnitude below $1/(\nu\lambda)$
preserve the word; the Fej\'er kernel bound $F_K(z)\le1/(4K\|z\|^2)$
then gives error $O(\nu\lambda/K)=O(1/\lambda)$. On the exceptional
set the error is at most two. For each remaining harmonic the difference
of its two conditional averages is
$\mathrm e(hx_0y/S)G_y(h)(1-\Phi_y(h))$. Therefore
\[
 \E_\xi\left|\E[\omega_I(Z(U))-\omega_I(Z(V))\mid\xi]\right|
 \lesssim \lambda^{-1}+\lambda^{-2}[C(N+1)]^{|I|+1}.
\]

\emph{From moments to circuits.}
The success indicator on the local digit cube has depth at most $d+2$ and
size $2N^b+3$, after substituting the exposed constants and literals and testing
agreement with the last bit. For $\eta=N^{-D-3}$, bounded independence
fools all these indicators with error $\eta$ at order
$p\le C_{d,b,D}(\log(2N))^{O(d+1)}$ \cite[Cor.~14]{HarshaSrinivasan19}.
It follows that each success indicator has degree-$p$ upper and lower
sandwich polynomials of one-sided uniform mean gap $\eta$:
the finite linear program maximizing its expectation over laws matching the
uniform moments through degree $p$ has as its dual the minimum uniform mean
of a degree-$p$ majorant; apply the same argument to its complement.
Every coefficient has modulus at most $1+\eta$ by its uniform $L^1$ norm.
The majorant is nonnegative, and the minorant's negative mass is bounded by its gap.
These polynomials may depend on $\xi$. There are at most $(t+1)^p$ coefficients,
so the moment estimate applies term by term after taking their absolute bounds.
The reference density bounds each one-sided reference gap by $2Q\eta/y$.
Since $Q^{-1}\sum_{y=1}^{Q-1}Q/y=O(N+1)$, including the mass $1/Q$ at zero,
the mean conditional advantage in absolute value is at most
\[
 Q^{-1}+C(N+1)(\eta+\nu^{-1})
    +\lambda^{-1}[C'(N+2)]^{2p+2}.
\]
Take $\ell/8$ to be the least integer exceeding
$(2p+2)\log_2(C'(N+2))+(D+4)\log_2(2N)$, with $C'$ enlarged if necessary.
Then $\nu^{-1}\le2^{-\ell}$ and the last display is at most $N^{-D}$.
The exponent in the claimed bound on $\ell$ is independent of $D$.
\end{proof}

\begin{theorem}\label{thm:icml-bottom-topk}
Fix $H,k$. There is $C>0$ such that for every $A>0$, for all sufficiently large $N$,
every two-layer top-$k$ model of \cref{def:topk} with at most $H$ heads per layer has absolute
signed correlation at most $N^{-A}$, and hence prediction advantage at most $N^{-A}/2$,
on the true context whenever
\[
 N<T\le 2N-(\log(2N))^C .
\]
\end{theorem}

\begin{proof}
Write $m=2N-T$, $B=(-1)^{M_T}$ and $F\in\{-1,0,1\}$ for the signed output.
We bound $|\E BF|$. Number operand bits from the most significant end.

For the fixed circuit depth, size exponent and error exponent $D$ used below,
take $\ell$ from \cref{lem:icml-hidden-block}. Starting at $a=T-N$ gives
$J=\lfloor m/\ell\rfloor$ disjoint admissible blocks.

\emph{Retaining verifiable first-layer states.}
A guessed first-layer transcript at one position specifies its query token and the ordered
selected position/token pairs for each head. By \cref{lem:jointtr}, its event is empty or
a conjunction of raw literals. Intersect all tests at a coordinate, discarding
tautologies. There are at most $2(2n)^{Hk}$ guesses per position.
Discard every transcript fixing more than $w$ distinct bits of $X$.
Since $X$ is uniform, the probability that any actual transcript is discarded is at most
$2n(2n)^{Hk}2^{-w}$.

For a subset $E$ of the $J$ blocks also discard transcripts containing any $X$ literal
in those blocks. At each position keep its state only if a surviving transcript holds.
Run the final heads on the retained positions with the original scoring and tie order.
Let $F(E)$ be the resulting signed output, set to zero if the query is absent or
a head has too few candidates for its original list length.
Thus $\E|F-F(\varnothing)|\le4n(2n)^{Hk}2^{-w}$.
For fixed $E$, each of the two sign indicators of $F(E)$ has bounded depth and size
$n^{O_{H,k}(1)}$. Indeed, enumerate the retained transcripts for the final query and
each proposed ordered list of winners. Require their literal conjunctions and forbid
every retained nonselected transcript outranking its head's cutoff. The disjunction
of consistent guesses with the given readout is the indicator. Sharing is unnecessary
for the polynomial size bound. For $E\ne\varnothing$ this circuit ignores every bit
in any one selected block. After fixing the exposed operands it is a circuit on the
true prefix, so the hidden-block bound gives $|\E BF(E)|\le 2N^{-D}$.
Here $F(E)=\mathbf1_{F(E)=1}-\mathbf1_{F(E)=-1}$ and the difference of the
corresponding prediction signs is $2F(E)$ up to sign.

\emph{Several erased blocks.}
Fix an input. Every retained position has a fixed state and disappears precisely when a
further erasure hits its transcript. At a set $E$ with a present query and sufficient
candidates, let $W(E)$ consist of the block indices touched by the actual query and
winner transcripts. It has size at most $s=(1+Hk)w$. If $U$ misses $W(E)$, these
records survive and the remaining competitors can only disappear, so $F(E\cup U)=F(E)$.
At an absent query or insufficient list let $W(E)=\varnothing$, since the default persists.

For $S$ disjoint from $E$ put $\mathcal D_S(E)=\sum_{U\subseteq S}(-1)^{|U|}F(E\cup U)$.
In any available pool disjoint from $E$ there are at most $(2s)^r$ sets $S$ of size
$r$ for which $\mathcal D_S(E)\ne0$. This holds for $r=0$. For $r>0$ such an $S$
must meet $W(E)$, since otherwise its alternating sum is constant times zero.
Choose $j$ in the intersection. Since
$\mathcal D_S(E)=\mathcal D_{S\setminus\{j\}}(E)-\mathcal D_{S\setminus\{j\}}(E\cup\{j\})$,
there are at most $s$ choices for $j$, two recursive cases and $(2s)^{r-1}$ smaller
sets in the pool with $j$ removed. Thus, for a uniform $r$-subset $S$ of all $J$ indices,
\[
 \E_S|\mathcal D_S(\varnothing)|\le 2^r(2s)^r/\binom Jr\le (4rs/J)^r.
\]
All nonempty terms have $|\E BF(U)|\le2N^{-D}$ for fixed $U$. Expanding the
difference and taking $S$ independently of the input gives
\[
 |\E BF|\le 4n(2n)^{Hk}2^{-w}+2^{r+1}N^{-D}+(4rs/J)^r .
\]
Take $r=\lceil(A+3)\log_2(2N)\rceil$,
$w=\lceil(A+Hk+6)\log_2(2n)\rceil$, and $D=2A+10$.
The circuit depth and size exponent are fixed independently of these choices.
For $C=c_d+3$, the hypothesis gives $J/(rs)\to\infty$ uniformly in $T$
at each fixed $A,H,k$, so eventually the last term is at most $2^{-r}$.
Since $n=O(N)$, the sum is less than $N^{-A}$ for large enough $N$.
\end{proof}

At the other end of the range, one head per layer suffices for exact prediction
if $2^{2(m+1)}\le T-2$. Assign an earlier prefix position to each possible pair
of low $(m+1)$-bit operand words. Its first-layer head gives score one to an
operand token that disagrees with the assigned pair and zero to every other token.
The update records whether the selected token was a mismatch and the assigned answer.
The final head retrieves the unique matching position. Its answer is exact because
the target is digit $m$ of the product, counted from zero at the low end.

\section{Two-layer sharpness on genuine and modified prefixes}
\label[appendix]{app:icml-two-layer-sharpness}

The two results below distinguish a restriction on the attention scores
from a restriction on the distributional information used in the proof.
Here $n=\Theta(N)$ is the raw context length, with binary operand and
prefix positions and a fixed number of separators. The head bound $H$
is fixed, and ties use the model's positional priority.

\paragraph{The norm of a losing comparison.}
For a real function on the formal cube, let $\pi(F)$ be the infimum
of $\sum_j|a_j|$ over representations
$F=\sum_j a_j f_j(X)g_j(Y)\chi_{C_j}(P)$ with real $|f_j|,|g_j|\le1$.
Products of two such atoms are atoms, so $\pi$ is submultiplicative.
It is bounded by the raw Walsh norm, and \cref{thm:main} gives
$|\mathbb E[BF]|\le2^{-DT}\pi(F)$ for the true target sign $B$.

\begin{proposition}[Losing comparisons of bounded mixed norm]\label[proposition]{prop:icml-factored-losing}
Fix $\alpha>0$ and $C_0\ge0$, and use the genuine context for
$\alpha N\le T\le N$. In a two-layer top-$k$ model, fix any
last-layer head, query state, cutoff position and cutoff state.
Assume every other eligible position's losing predicate has a real
extension on the formal cube with $\pi$ at most
$W=\exp(C_0\sqrt{k+1})$, agreeing with the predicate on all
inputs realizing the specified states. There are $c,b>0$ such that
for all sufficiently large $N$ and $1\le k\le c\sqrt n$ the absolute
signed correlation is at most $\exp[-bn/(k+1)]$.
The matching capacity scale for constant advantage is $\Theta(\sqrt n)$.
\end{proposition}

\paragraph{Examples.}
Every intersection of constraints
$\chi_C(P)=a(X)b(Y)$, where $a,b$ are arbitrary sign functions,
has indicator of norm at most one. Each constraint contributes a
factor $(1+a(X)b(Y)\chi_C(P))/2$. This includes the intersection
of arbitrary operand tests and affine constraints, as well as
nonaffine graphs such as $P_1=X_1X_2$. Conversely, these intersections
are all zero-one indicators of norm at most one. The unit ball
is the convex hull of sign atoms by expressing each bounded operand
test as a mixture of signs. In a representation of an indicator in
that hull, every atom in the support is $+1$ wherever the indicator
is one, and at least one is $-1$ wherever it is zero.

\begin{proof}
Write $F$ for the signed readout.
We first truncate the first-layer records for use as payloads. For every
head and each hypothetical query bit, list the eligible
(position, raw symbol) pairs in score order, with the prescribed
tie priority. At the query include only its hypothesized symbol,
and at a deterministic position list only its fixed symbol.
On lists of at most $\ell$ pairs use the exact model rule, including
its prescribed short-list size. On longer lists keep the first
$\ell$ pairs. A truncated head misses when these pairs contain fewer
present entries than its prescribed selection size. Use a fixed
fallback state if any head misses. The resulting state at each
position depends on at most $1+2H\ell$ fixed raw coordinates.

A miss requires fewer than $k$ of the first $\ell$ pairs to be present.
For a fixed pattern of present pairs, every absent pair fixes a
distinct raw coordinate, since the two symbols at one position
cannot both be absent. A fixed separator and the hypothesized query
pair cannot be absent on this event. With $\ell\ge128k$ there are at
most $(e\ell/k)^k\le 2^{\ell/10}$ patterns. By \cref{lem:conj} each
feasible pattern has probability at most
$(N+2)2^{-(\ell-k+1)/3}$ on the true context. A union bound over
positions, heads, and query symbols gives probability at most
$e^{-v}$ of any miss for $\ell=\lceil C(k+v+\ln(2n))\rceil$, with $C$
large enough and $v\ge1$.

For a head selecting $p$ of $n_h$ keys, define the rank of a proposed
set to be the number of outsiders above its worst member. There are
$\binom{r+p-1}{p-1}$ sets of rank $r$ in any strict order, for
$0\le r\le M=n_h-p$. We construct a polynomial $q$ with $q(0)=1$,
degree $O(\sqrt{n(k+v)})$, and total squared weight at the false
ranks at most $e^{-v}$, including these multiplicities.

Set $m=\lceil p+v+2\rceil$. If $M\le m$, use
$\prod_{i=1}^M(1-r/i)$, including the case $M=0$.
Otherwise start with $\prod_{i=1}^m(1-r/i)$. For every dyadic
$t=2^jm\le M$, multiply by
\[
 \frac{T_{d_t}((M+t-2r)/(M-t))}{T_{d_t}((M+t)/(M-t))},
 \qquad d_t=\lceil5m\sqrt{M/t}\rceil ,
\]
where $T_d$ is the Chebyshev polynomial. At $t=M$ use $1-r/M$.
Each factor equals one at zero, is bounded in magnitude by one on
$[0,M]$, and is bounded by $e^{-9m}$ on $[t,M]$. For $t<M$
the last bound follows from
$\operatorname{arcosh}((M+t)/(M-t))\ge2\sqrt{t/M}$.
In the shell $2^jm\le r<2^{j+1}m$ the first product has magnitude
at most $(e2^{j+1})^m$, and the rank multiplicity is at most
$(e2^{j+2})^m$. There are at most $2^jm$ integers in the shell.
The squared filtered weight of the shell is at most
$e^{-11m(j+1)}$, by the $j+1$ damping factors. Summing proves the
desired weight estimate, and summing the degrees gives
$O(\sqrt{nm})$.

Sum the proposed readouts with weights $\prod_h q_h(r_h)^2$,
one filter per head, using clipped records for the payload and
the true records for the rank tests. On any input the correct
tuple has weight one, and the total weight of all other tuples
is at most $(1+e^{-v})^H-1$. Including clipping, the resulting
function $Q$ differs from $F$ by $O_H(e^{-v})$ in mean absolute
error.

For a fixed guess, specify the worst selected member per head
and the exact first-layer transcripts of these members and the query.
There are $\exp[O_H((k+1)\ln(2n))]$ guesses. Their transcript
indicators are conjunctions of raw literals with norm at most one.
Multiply by the tests that the other selected members precede the
specified cutoffs, at a cost at most $(1+W)^{Hk}$. The bounded
clipped payload uses at most $O_H((k+1)\ell)$ raw coordinates,
with Walsh norm at most $2^{O_H((k+1)\ell)}$ by Parseval.

For each head, expand the squared filter in Walsh characters of
its $n_h-p$ outsider indicators. The rank polynomial on this
Boolean cube is bounded by one because every feasible rank occurs
in a strict order, with total false weight at most $e^{-v}$. Its
degree is at most $d=O(\sqrt{n(k+v)})$, padding the coefficient
count by $n$ variables. Its coefficient mass is at most
$(\sum_{j\le d}\binom nj)^{1/2}$. Substitution costs at most $1+2W$
per outsider sign. The extensions are used only under the matching
transcript guard, where their values are Boolean. Thus
\[
 \begin{split}
 \ln\max(1,\pi(Q))\le{}&
 O_H((k+1)\ln(2n)+(k+1)\ell+k\ln(1+W))\\
 &+\tfrac H2\ln\sum_{j\le d}\binom nj+Hd\ln(1+2W).
 \end{split}
\]
Take $v=\eta n/(k+1)$ with $\eta$ small and then $c$ small enough
that $k\le v$ for all large $n$. Then $d/n=O(\sqrt{\eta/(k+1)})$,
$d\ln(1+2W)=O_{C_0}(\sqrt\eta\,n)$, and
$(k+1)\ell=O((\eta+c^2)n)+o(n)$. The binomial logarithm divided
by $n$ tends uniformly to zero as $\eta\downarrow0$ by the binary
entropy bound. Hence $\pi(Q)\le2^{DT/2}$, since $T=\Omega(n)$.
Discrepancy and the mean error give the claimed rate.
Fixed routing uses constant losing predicates and gathers the
operands at $k=O(\sqrt n)$ by the two-layer fixed-routing upper bound.
\end{proof}

\begin{proposition}[A limit of fixed-target discrepancy]\label[proposition]{prop:icml-edited-prefix-barrier}
For independent uniform $N$-bit operands and all large $N$, let $B$ be
the sign of the actual midpoint product bit. At
$k=\lceil C_0N^{1/3}\rceil$ one can replace a terminal block of the
prefix by a deterministic function of the operands so that a
two-layer model with one top-$k$ head per layer predicts $B$ with
probability greater than $3/4$. Yet for the resulting prefix $P'$,
\[
 \sup_{|f|,|g|\le 1,\ \mathcal C\subseteq[N-1]}
 |\mathbb E[B f(X)g(Y)\chi_{\mathcal C}(P')]| \le 2^{-c_0N}.
\]
Here $C_0,c_0>0$ are constants. For $k=\lceil N^{1/3+\varepsilon}\rceil$,
$0<\varepsilon<1/6$, the terminal block can have length
$O(N^{1-3\varepsilon})$. All earlier digits are genuine.
\end{proposition}

\begin{proof}
Use $0<D\le1$ from \cref{thm:main}, and write $l=\lfloor DN/8\rfloor$ and use the lowest $l$ bits $x,y$ of the
operands as data. Reserve a disjoint block of $m$ bits $x^*,y^*$ in each
operand, where $m=\lfloor\epsilon_0l\rfloor$ for a small constant
$\epsilon_0>0$. On $m$ bits let $G(u)=2\mathbf1[|u|\ge k]-1$.
The approximate-degree bound of \citet{Paturi1992} and duality of
the sup norm give a signed mass $\psi$ with $\sum|\psi|=1$,
$\sum\psi G>2/3$, and $\sum_u\psi(u)\chi_S(u)=0$ for $|S|<d$,
where $d=\lfloor c\sqrt{mk}\rfloor$ for a small absolute $c>0$.
Here $k\le m/2$. Affine rescaling turns error $2/3$ for the signed
threshold into the error $1/3$ in Paturi's unsigned bound. Set $\nu=|\psi|$.

Choose an $m$ by $l$ binary matrix $A$ such that each row has at most
$(k-2)/2$ ones and each sum of at least $d$ distinct rows has Hamming
weight at least $l/4$. To see existence, take independent entries
with mean $\theta=k/(8l)$. The row bound fails with probability at
most $m e^{-\Omega(k)}$. For any set of at least $d$ rows the sum
has independent coordinates of mean at least
$[1-(1-2\theta)^d]/2>0.49$ once $\theta d\ge2$.
Hoeffding's inequality and a union over the $2^m$ row sets give
the second bound for sufficiently small $\epsilon_0$. We can make
$\theta d\ge2$ by increasing $C_0$. For the stated larger $k$, take
$m=\lceil K N^2/k^3\rceil$ with a large constant $K$, which also
ensures $k\le m/2$ and $m\le\epsilon_0l$ for all large $N$.

Conditional on the operands, sample $u$ with mass $\nu(u)+B\psi(u)$.
This is a probability measure since $\sum\psi=0$ and $B\in\{-1,1\}$.
At terminal position $i$ of the editable block write
$u_i\oplus x_i^*\oplus y_i^*\oplus\langle A_i,x\odot y\rangle$,
with all operations over $\mathbb F_2$ and coordinatewise product
$x\odot y$. Other prefix symbols stay fixed at their genuine values.
At every editable position the first layer retrieves the two pad bits
and the operand bits in the support of its row, using at most $k$
reads. The raw query token then determines $u_i$. At the final
query the second head ranks editable positions with $u_i=1$ first,
followed by the remaining editable positions and then the other
positions. This yields $G(u)$, since the top $k$ positions all belong
to the first group precisely when $|u|\ge k$. The causal mask includes
the final query position. The signed correlation equals
$\sum\psi G+\mathbb E[B]\sum\nu G>2/3-2^{-DN}$.

For a mask split into an edited part $S$ and a retained part $R$,
let $w=A^\top\mathbf1_S$ and
$I_w=(-1)^{\langle w,x\odot y\rangle}$. Absorb the pad characters
into $f,g$. The correlation is the sum of
\[
 \nu_S\mathbb E[Bfg\chi_R I_w]
 \quad\text{and}\quad
 \psi_S\mathbb E[fg\chi_R I_w],
 \qquad
 \nu_S=\sum_u\nu(u)\chi_S(u),\quad
 \psi_S=\sum_u\psi(u)\chi_S(u),
\]
where $\chi_R$ is of the genuine prefix. Both coefficients have
modulus at most one. Expanding $I_w$ by the $x$ bits on
$\operatorname{supp}(w)$ gives at most $2^l$ products of bounded
operand functions. The first expectation is at most $2^{l-DN}$
in modulus. If the last index in $R$ exceeds $N/2$, using it as the
target bounds the second expectation by $2^{l-DN/2}$.

Otherwise set the low $l$ bits of each operand to zero when computing
$\chi_R$. This changes the normalized product by at most
$\Delta=2^{1+l-N}$. A digit changes only within distance $\Delta$
above one of at most $2^{N/2}+1$ boundaries. The event
$X<2^{7N/8}$ has probability at most $2^{-N/8}$, and for other fixed
$X$ the normalized product takes values on a grid of spacing
$X/2^{2N}$. Counting grid points in each boundary interval gives a
total error $2^{-\Omega(N)}$. After truncation the genuine mask is
constant conditional on the higher bits. The second coefficient
vanishes unless $|w|\ge l/4$, by the dual condition and the matrix
property. On the remaining low bits the normalized inner-product
matrix on $\operatorname{supp}(w)$ has operator norm
$2^{-|w|/2}$ by Walsh orthogonality. This bounds the second
expectation after conditioning and proves exponential discrepancy.

Finally sample an editable word independently for every operand pair
and fix the resulting table. For each mask and each pair of sign
tests, Hoeffding's inequality for the $2^{2N}$ independent table
entries bounds a deviation greater than $2^{-N/4}$ by
$2\exp(-2^{3N/2}/2)$. The number of sign-test pairs and masks has
logarithm $O(2^N)$, so a union bound preserves all estimates and
the predictor's correlation. Extremizing over sign tests suffices
for real bounded tests. Complex tests follow by splitting into real
and imaginary parts and decreasing $c_0$. This gives a deterministic
modified prefix. Thus the existence of a positive discrepancy exponent
alone cannot imply vanishing advantage at cube-root capacity. The
unrestricted threshold on the genuine prefix remains open.
\end{proof}

\subsection{Adaptive scores that read the late product prefix}

\begin{proposition}[The near square-root threshold for late-prefix scores]
\label[proposition]{prop:icml-late-scores}
Let $\alpha N\le T\le N$ for fixed $\alpha>0$. In a two-layer
top-$k$ model with fixed head bound, suppose that after fixing any
last-layer query state, each candidate score on the formal context
cube is invariant under changes in the first $J$ product-prefix
digits. The score maps are defined on the product of query and
candidate state alphabets. The query and selected payloads may use
the entire genuine prefix. For any fixed $A>0$, if
$k=o(\sqrt{n/\log n})$ there is $J=o(n)$ for which the signed
correlation has magnitude at most $n^{-A}$. In particular, for
$k=\lceil n^\gamma\rceil$ and fixed $0<\gamma<1/2$,
taking $J=O(k^2\log^2n)$ suffices for correlation
$O(\exp(-c k))$ for some $c>0$.
\end{proposition}

The restriction still allows jointly input-dependent scoring on
both operands and the whole remaining suffix of the genuine prefix.
Fixed routing is in this class and attains capacity $O(\sqrt n)$.
The unrestricted two-layer gap remains open.

\begin{proof}
Write $B=(-1)^{M_T}$ and let $A(f)$ denote the Walsh $\ell_1$ norm.
By \cref{thm:main},
$|\mathbb E[Bf]|\le 2^{-DT} A(f)$ for every formal function $f$.
Also, every nonconstant character on the scoring cube
$(X,Y,M_{J+1},\ldots,M_{T-1})$ has genuine expectation of
magnitude at most $2^{-D(J+1)}$. Operand-only characters have
mean zero. For other characters, use the greatest selected product
index as the target and apply the same discrepancy theorem there.
Write $\mu$ and $U$ for the genuine and uniform laws on this
scoring cube. Thus
$|\mathbb E_\mu p-\mathbb E_U p|\le 2^{-D(J+1)}A(p)$.

Apply the truncation in the proof of
\cref{prop:icml-factored-losing} twice: to the genuine cube and
to the cube with the top $J$ digits forced to zero. In the
second array those positions are deterministic. An absent pair
in a feasible miss pattern still prescribes a distinct unforced
digit, so the same genuine conjunction bound applies. For a target
failure at most $e^{-v}$ across both arrays, take
$\ell=\lceil C(k+v+\log(2n))\rceil$ with a sufficiently large
absolute constant, and let $b_0=1+2H\ell$ bound the number of
fixed raw digits read per truncated record. Use the ordinary
array for the query and for retrieving the selected payloads.
Use the forced array for scoring candidates with that same
query. The score hypothesis shows that the resulting signed
output $\widetilde F$ equals the original except on the clipping
failure. Use an attainable fallback query and extend the final
update boundedly to any unattainable record tuple.

For each head guess its winning set and its last member, then
guess the ordinary query's support assignment and the forced
support assignment of each proposed cutoff. Empty lists require
no cutoff. There are at most
$M=n^{Hk}(k+1)^H2^{(H+1)b_0}$ guesses. After fixing a guess $g$,
each candidate comparison with the fixed cutoff is a function of
at most $b_0$ scoring-cube bits. Prohibit every forbidden assignment
with one clause to encode the winner-membership test as a CNF
$C_g$ with at most $s=\max(2,Hn2^{b_0})$ clauses.
Let $G_g$ be the guess-assignment guards times the readout,
sorting the guessed winners by their forced scores and using
their ordinary payloads. It is bounded by one and depends on
at most $b=(1+2Hk)b_0$ raw bits. Hence $A(G_g)\le 2^{b/2}$.
The positional tie rule makes precisely one guess valid for
each formal input, giving $\widetilde F=\sum_g G_g C_g$.

CNF fooling and linear-programming duality give pointwise
sandwiching polynomials $l_g\le C_g\le u_g$
of degree at most $r=\lceil C_1\log(2s)\log(2s/\epsilon)\rceil$,
with $\mathbb E_U(u_g-l_g)\le 2\epsilon$ and
$\mathbb E_U u_g\le 1+\epsilon$ \cite[\S4.2]{Tal17}.
Use degree at most $n$ by taking the function itself if necessary.
Both $u_g$ and the gap $u_g-l_g$ are nonnegative, so their absolute
Walsh coefficients are bounded by their uniform expectations.
For $L=\sum_{i=0}^{\min(r,n)}\binom ni$ this gives
\[
\begin{split}
|\mathbb E[BF]| \le {}& 2e^{-v}
 + 2M\epsilon(1+2^{-D(J+1)}L)\\
 &+ M 2^{-DT+b/2}(1+\epsilon)L .
\end{split}
\]
Indeed, the middle term uses $|G_g|\le1$ and the small-bias
bound on the nonnegative sandwich gap. The last term uses the
target discrepancy and submultiplicativity of the Walsh norm.
No independence of $G_g$ and the scores is needed.

Take $v=(A+3)\ln(2n)$, $\epsilon=e^{-v}/(8M)$, and
$J=\lceil(\log_2 L)/D\rceil$. In the stated subcritical range,
$r=O((k+\log n)(k+1)\log n)=o(n)$, $b=o(n)$, and
$\log M=o(n)$. The binomial entropy bound gives $\log L=o(n)$,
so $J=o(n)$ and the final term decays exponentially in $T$.
For fixed $\gamma\in(0,1/2)$ instead use $v=k$.
Then $\log L=O(k^2\log^2 n)$ and the other estimates still hold,
giving the asserted rate.
\end{proof}

\section{Operand distributions with min-entropy deficits}\label[appendix]{app:icml-entropy}
Suppose $X,Y$ are independent and have min-entropies at least $N-d_X$, $N-d_Y$ in bits.
Let $U,V$ be independent uniform $N$-bit operands. The functions
$a(x)=2^{N-d_X}\Pr[X=x]$ and $b(y)=2^{N-d_Y}\Pr[Y=y]$ take values in $[0,1]$.
For bounded $f,g$, the correlation under $X,Y$ equals
$2^{d_X+d_Y}\E[f(U)a(U)g(V)b(V)\chi_S(U,V)]$.
Thus the discrepancy in \cref{thm:main} grows by at most the factor $2^{d_X+d_Y}$,
giving the bound $\min(1,2^{d_X+d_Y-DT})$ in each stated linear-exponent range.
In particular the fixed-model advantage is still exponentially small whenever the
combined deficit fits the slack in its exponential bound. More directly, the density
of $(X,Y)$ relative to $(U,V)$ is at most $2^{d_X+d_Y}$, so the exponentially small
full-product and partial-accuracy event probabilities in
\cref{thm:cot-avg,cor:icml-cot-partial} grow by at most this factor.
This probability argument yields an event bound in the nonuniform setting.

Conditioning each operand to have a leading one costs two bits of total
min-entropy. Thus exactly $N$-bit integers retain the exponential
estimates. Independent biased bits also admit a Shannon-entropy
criterion. For fixed biases $p_X,p_Y\in(0,1)$, trim each operand to
self-information at least $N(h_2(p)-u)$ bits, with fixed $u>0$.
The excluded mass is $e^{-\Omega(N)}$ by concentration of the sum
of bounded information increments. The two retained subprobability
marginals have combined deficit
$N(2-h_2(p_X)-h_2(p_Y)+2u)$ in the rectangle bound. The excluded mass
adds only $e^{-\Omega(N)}$ to each bounded parity correlation.
Accordingly the attention and record-transfer estimates remain exponential
if $2-h_2(p_X)-h_2(p_Y)<\liminf DT/N$.
For example, $p_X=p_Y=3/5$ works throughout $N/2\le T\le N$:
$2-2h_2(3/5)<1/12$ and \cref{thm:main} gives $D=1/6$ for large $N$
there. These biased-source estimates use the stated entropy margin;
the event bounds for short generation above use min-entropy.

\fi
\subsection{Prediction resources}\label[appendix]{app:icml-prediction-resources}
At fixed head count and fan-in, inverse-polynomial advantage needs
depth $\Omega(\log\log n)$. For fixed $L$, the necessary selection
budget is $1+Hk\ge n^{1/(2^L-1)-o(1)}$ in the range of
\cref{thm:topk-depth}. Fixed routing suffices for exact prediction
with exponent $1/L$. \Cref{fig:icml-k-gap,rem:topk-explicit}
show the open gap and the explicit resource bounds. A small linear
number of final-layer records also reveals exponentially little
about a binary interior target (\cref{cor:icml-records}).
\begin{figure}[!t]
\centering
\resizebox{.56\columnwidth}{!}{%
\begin{tikzpicture}[font=\small]
\begin{scope}[x=0.84cm,y=3.55cm]
\fill[blue!9] (1,0) -- plot[domain=1:6,samples at={1,2,3,4,5,6}](\x,{1/(pow(2,\x)-1)}) -- (6,0) -- cycle;
\fill[orange!16] (1,1) -- plot[domain=1:6,samples at={1,2,3,4,5,6}](\x,{1/\x}) -- (6,1) -- cycle;
\draw[->] (0.83,0) -- (6.23,0) node[right] {$L$};
\draw[->] (1,0) -- (1,1.09) node[above] {$\beta$};
\foreach \l in {1,...,6}{
  \draw (\l,0) -- ++(0,-.025) node[below] {\l};
}
\foreach \b/\lbl in {0/0,.5/{1/2},1/1}{
  \draw (1,\b) -- ++(-.07,0) node[left] {\lbl};
}
\draw[very thick,blue!65!black] plot[domain=1:6,samples at={1,2,3,4,5,6}](\x,{1/(pow(2,\x)-1)});
\draw[very thick,orange!70!black,dashed] plot[domain=1:6,samples at={1,2,3,4,5,6}](\x,{1/\x});
\node[anchor=west,align=left,text=blue!65!black] at (1.06,.095) {Negligible\\advantage};
\node[anchor=west,text=orange!70!black] at (2.15,.58) {Exact, fixed routing};
\fill[blue!65!black] (3,{1/7}) circle (1.3pt);
\node[font=\scriptsize,text=blue!65!black,anchor=south] at (3,{1/7+.015}) {$1/7$};
\node at (4.2,.151) {Open gap};
\end{scope}
\begin{scope}[shift={(1.55cm,2.41cm)},font=\scriptsize]
\node[align=left,anchor=south west,fill=white,draw=black!30,rounded corners=1pt,inner sep=3pt]
  at (0,0) {Depth with fixed $H,k$\\Adaptive lower: $\Omega(\log\log n)$\\Fixed routing (exact): $\Theta(\log n)$\\Adaptive upper (small error): $O(\log n/\log\log n)$};
\end{scope}
\end{tikzpicture}}
\caption{Fan-in exponents for interior prediction. The blue lower bound
holds in either digit order and every fixed radix.
With a fixed head count and $k=n^\beta$, the blue curve shows the necessary exponent
$1/(2^L-1)$ for inverse-polynomial advantage with adaptive routing, up to $o(1)$
(\cref{rem:topk-explicit}). The dashed curve shows the sufficient exponent $1/L$ for
exact prediction with fixed routing, attained when $k+1\ge n^{1/L}$. The gap leaves
room for both adaptive routing and weaker accuracy requirements. The adaptive upper bound in the depth inset uses
the binary MSB-first construction in the first half in \cref{prop:icml-adaptive-depth-upper}.}
\label{fig:icml-k-gap}
\end{figure}
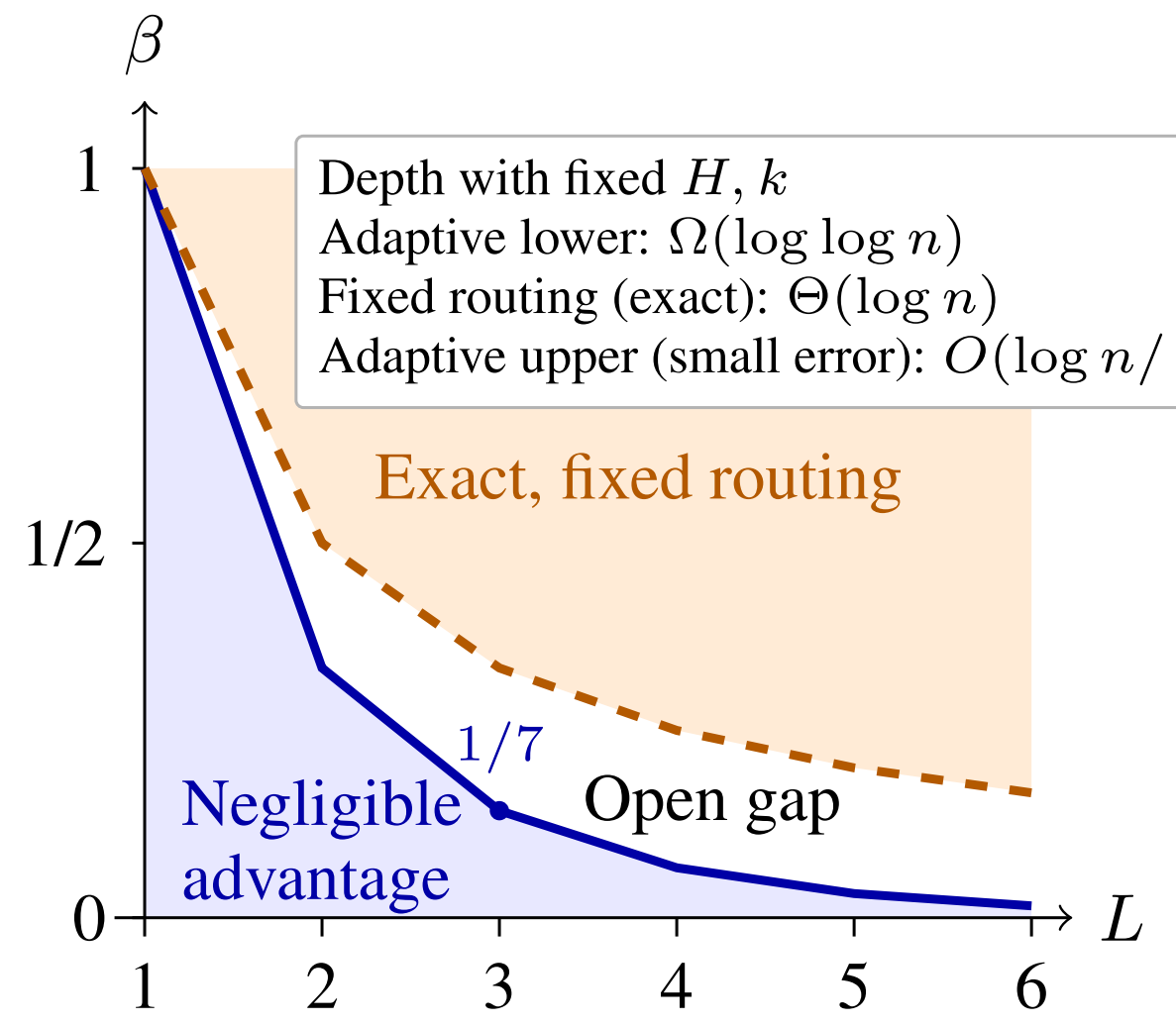

\paragraph{An upper bound on the required $k$.}
For fixed depth $L$, one head per layer suffices for exact prediction when
$(k+1)^L\ge n$ in \cref{def:topk}. Let $b=k+1$. At layer $j$, position $i$
selects the distinct positive positions $i-hb^{j-1}$ for $1\le h\le k$,
using fixed scores and ignoring any fillers. Each state merges these
position-tagged records with its own. After layer $j$, position $i$ holds the
whole interval $[\max(1,i-b^j+1),i]$ by induction. The final update can
therefore compute the next product digit from the full context. This uses unrestricted
states and computation in the update. For $L>1$ it rules out a near-linear
necessary $k$ per head in this model.

\paragraph{A receptive-field bound for fixed routing.}
For binary MSB-first prediction, if fixed routing makes a signed readout a function of at most $m$ formal input bits,
Parseval and Cauchy--Schwarz give Walsh norm at most $2^{m/2}$. Thus \eqref{eq:walsh}
bounds advantage by $\tfrac12\,2^{m/2-DT}$. This also applies to sliding-window,
dilated, and block-sparse softmax with fixed support: the data-dependent weights are
functions of the states inside that support. For $L$ layers with $H$ lists of at most
$k$ fixed positions per query, including residual edges gives $m\le(1+Hk)^L$, so
\begin{equation}\label{eq:fixed-routing-lower}
 \Pr[\text{correct}]-\tfrac12 \le 2^{(1+Hk)^L/2-DT-1}.
\end{equation}
Here $D$ is the applicable discrepancy exponent from \cref{thm:main}. In the range of
\cref{thm:topk-depth}, inverse-polynomial advantage requires $(1+Hk)^L=\Omega(n)$:
for fixed $H,k$ this gives $L\ge\log_{1+Hk}n-O(1)$, and for fixed $L$ it gives
$1+Hk=\Omega(n^{1/L})$. For either output order and every fixed radix,
the sparse-mask estimate gives the same necessary receptive-field size
$m=\Omega(n)$. A bounded character of the readout on $m$ slots has
degree at most $m$ and coefficient mass at most $q^{m/2}$. When $m/n$
is sufficiently small, the uniform sparse correlation bound absorbs this mass.

\icmlrelatedappendix

\ificmlextended\else

\icmlrecordproof

\fi
\end{document}